%% file: main.tex
\documentclass[hdr, twoside, final]{unswthesis}
\usepackage{graphicx}
\usepackage{mystyle} % My packages
\makeatletter
\fancypagestyle{noHeading}{
        \fancyhead{}
        \renewcommand{\headrulewidth}{0pt}
}
\usepackage[final]{pdfpages}

\include{mythesisDetails} % Thesis details e.g. Title
\include{definitions}

\DeclareGraphicsExtensions{.pdf,.jpeg,.png}
\graphicspath{{4-cache/}{6-gpu-abea/}{5-minimap/}{5-minimap/img/}{2-intro/}{3-literature/}{8-io-opti/}{9.5-appendices/}}
\usepackage[normalem]{ulem}
\usepackage{afterpage}

\begin{document}
    
    \renewcommand{\bibname}{References}
    %% pages in the ``frontmatter'' section have roman numeral page number
    \frontmatter  
    \maketitle
    %\includepdf[pages=1]{1-pre/GRIS.pdf}
    %\includepdf[pages=1]{1-pre/GRIS2.pdf}
    % \afterpage{\blankpage}
    %\pagestyle{plain}

    \include{1-pre/abstract}
    \include{1-pre/ack}

    \include{1-pre/publications}
    
    %\cleardoublepage
    %\setcounter{tocdepth}{4}
    \tableofcontents
    %\cleardoublepage
    %\addcontentsline{toc}{chapter}{\listfigurename}
    \listoffigures  % if required
    %\cleardoublepage
    %\addcontentsline{toc}{chapter}{\listtablename}
    \listoftables  % if required
    \include{1-pre/abbreviations}
    \mainmatter
    \pagestyle{fancy}
        \fancyhf{}
        \fancyhead[LE]{\leftmark}
        \fancyhead[RO]{\rightmark}
        \fancyfoot[C]{\thepage}
        \renewcommand{\headrulewidth}{1pt}
        \setcounter{secnumdepth}{3}

    \include{2-intro/main}

    \include{3-literature/main}

    \include{4-cache/cache}

    \include{5-minimap/main}

    \include{6-gpu-abea/main}

    \include{7-integration/main}
    \include{8-io-opti/main}

    \include{9-conclusion/main}

    \clearpage
    \appendix
    \include{9.5-appendices/1-minimap}

    \include{9.5-appendices/2-f5c}
    \include{9.5-appendices/3-f5c-misc}
    \include{9.5-appendices/4-nanocluster-f5p}
    \include{9.5-appendices/5-phone}
    \include{9.5-appendices/5.5-misc-open-source}
    \include{9.5-appendices/6-io}

    \clearpage
    
    \backmatter
    
    \pagestyle{noHeading}
    \bibliographystyle{IEEEtran}
    \bibliography{main}
    %\bibliography{cache/ref,gpu-abea/ref,minimap/ref, minimap/links,intro/ref}
    %cat */*.bib > ref.bib
    %bibtool -r biblatex -s ref.bib -d > refcat.bib  

\end{document}

%% file: mythesisDetails.tex
\thesistitle{Computer Architecture-Aware Optimisation of DNA Analysis Systems}
\thesisschool{School of Computer Science and Engineering}
\thesisauthor{Hasindu Gamaarachchi}
\thesisZid{z5136909}
\thesisdegree{Doctor of Philosophy}
\thesisdate{November 2020}
\thesissupervisor{Sri Parameswaran}% for undergrad theses only

%% file: 1-pre/abstract.tex
\chapter{Abstract}

DNA sequencing---the process that converts the massive amount of chemically encoded data in DNA molecules into a computer-readable form---is revolutionising the field of medicine through a variety of applications such as precision medicine, accurate diagnostics and identifying disease predisposition. DNA sequencing also has many other applications in areas such as epidemiology, forensics and evolutionary biology. DNA sequencers, the machines which perform DNA sequencing, have evolved from the size of a fridge to that of  a mobile phone over the last two decades. The cost of sequencing a complete human genome has remarkably reduced from billions of dollars to hundreds of dollars over this time. The size of a DNA sequencer is expected to become even smaller and the sequencing cost per genome  is expected to be even more affordable in the future. Thus, DNA tests are likely to be performed as routinely and cost-effectively as today's blood tests.  Despite the reduction in size and cost, DNA sequencers output hundreds or thousands of gigabytes of necessary data to account for errors made during the sequencing process. This data must be analysed on computers to discover meaningful information (for instance, mutations and epigenetic modifications) that have biological implications. Unfortunately, the analysis techniques have not kept the pace with rapidly improving sequencing technologies. Consequently, even today, the process of DNA analysis is performed on high-performance computers, just as it was a couple of decades ago. Such high-performance computers are not portable, unlike mobile phone-sized ultra-portable sequencers. Consequently, the full utility of an ultra-portable sequencer for sequencing in-the-field or at the point-of-care is limited by the lack of portable lightweight analytic techniques. 

A primary reason for this lag between the two technologies is because sequence analysis software tools written by computational biologists with the focus on higher accuracy of the results are un-optimised to efficiently utilise computational resources (i.e. software does not map well to the architecture of computers). This thesis proposes computer architecture-aware optimisation of DNA analysis software. DNA analysis software is inevitably convoluted due to the complexity associated with biological data. Modern computer architectures are also complex. Performing architecture-aware optimisations requires the synergistic use of knowledge from both domains, (i.e,  DNA sequence analysis and computer architecture). Computer architecture knowledge helps the efficient mapping and exploitation of existing hardware resources, while the understanding of DNA sequence analysis ensures that the final accuracy of the results is intact. In a nutshell, this thesis aims to draw the two domains together.

In this thesis, gold-standard DNA sequence analysis workflows (a workflow is a few software tools executed sequentially where each software tool is a complex system of dozens of algorithms) are systematically examined for algorithmic components that cause performance bottlenecks. Identified bottlenecks are resolved through architecture-aware optimisations at different levels, i.e., memory level, cache level, register level and processor level. Some example optimisations are: 1, the cache-friendly optimisation of de Bruijn graph construction that is a time-consuming core-component in a branch of software tools called variant callers (2X performance improvement); 2, memory capacity optimisation of reference indexes for the process called read alignment (from 16GB up to 2GB); 3, memory and processor level optimisation (for CPU-GPU heterogeneous systems) of an important time-consuming algorithm called adaptive banded event alignment used for the latest nanopore sequencing technology (3-5X performance improvement). Instead of merely performing algorithmic optimisations, those optimised versions are integrated back to the software and it is demonstrated that there is global efficiency and the accuracy is unaffected.  Finally, the optimised software tools are used in complete end-to-end analysis workflows and their efficacy is demonstrated by running on prototypical embedded systems. The embedded systems are not only fully functional, but the performance is also comparable to an unoptimised workflow on a high-performance computer. The practicality of these embedded systems has been demonstrated by integrating into the sequencing facility at the Garvan Institute of Medical Research in  Sydney.  Such low cost, energy-efficient, sufficiently fast and portable embedded systems enable complete DNA analysis at the point-of-care or in-the-field. Work conducted under this thesis also contributes to the bioinformatics community through contributions to popular bioinformatics tools (i.e. \textit{Platypus}, \textit{Minimap2} and \textit{Nanopolish}) and the design and development of novel open-source bioinformatics software (\textit{f5c}).

%% file: 1-pre/ack.tex
\chapter{Acknowledgement}

I wish to express my deepest gratitude to my supervisors \textbf{Prof Sri Parameswaran}, \textbf{Dr Martin A. Smith} and \textbf{Dr Aleksandar Ignjatovic} for the amazing supervision. Their enthusiasm, encouragement, advice and attitude were too spectacular that I do not have enough words to explain. Due to their great supervision, the time during the PhD was very productive, leading to significant outcomes, at the same time being enjoyable.

I am indebted to \textbf{Hassaan Saadat}, my fellow lab mate at UNSW, for the unwavering support, ingenious suggestions and encouragements. It was thanks to Hassaan that I participated in the ACM SRC that I eventually became a grand finalist.  I am extremely grateful to \textbf{James Ferguson}, my fellow lab mate at Garvan Institute, for countless insights and generously sharing unparalleled knowledge. 

I am also grateful to \textbf{Dr Warren Kaplan} and \textbf{Prof John Mattick} for identifying my talent and providing the opportunity to collaborate with the Garvan Institute of Medical Research, which was a valuable turning point in the PhD. 

I would like to extend my sincere thanks to \textbf{Arash Bayat} and \textbf{Vikkitharan Gnanasambandapillai} who were fellow PhD candidates at UNSW Sydney and also \textbf{Dr Bruno Gaeta} at UNSW for the initial induction to the genomics field. I am also grateful to my progress review panel who provided constructive advice and encouragement. 

Many thanks to all current and former colleagues in the embedded systems research group at UNSW and Genomics Technologies group at the Garvan Institute for the support provided at multiple occasions, especially, \textbf{Dr Darshana Jayasinghe}, \textbf{Dr Jorgen Peddersen}, \textbf{Hsu-Kang Dow}, \textbf{Dr Tuo Li}, \textbf{Shaun Carswell} and \textbf{Dr Ira Deveson}.  Thanks should also go to Data-Intensive Computer Engineering (DICE) group at Garvan Institute for helpful advice and practical suggestions.
I would like to express my deepest appreciation to \textbf{Dr Roshan Ragel}, my undergraduate-project supervisor, who played a decisive role in selecting Prof Sri Parameswaran as my PhD supervisor and also a great amount of assistance during the whole PhD application process. I would like to extend my sincere thanks to all the lecturers at the Department of Computer Engineering of the University of Peradeniya for setting a solid  foundation for my career.

I gratefully acknowledge the assistance from \textbf{Dr Heng Li}, the author of \textit{Minimap2}, and \textbf{Dr Jared Simpson}, the author of \textit{nanopolish}, in  understanding the code, providing with valuable insights and suggestions.

I very much appreciate the invaluable contributions to the software repositories from undergraduate students: \textbf{Chun Wai Lam} (UNSW), \textbf{Gihan Jayatilaka} (University of Peradeniya), \textbf{Hiruna Samarakoon} (University of Peradeniya) and \textbf{Thomas Daniell} (UNSW).

Last but not least, I thank my parents, my brother, relatives, all my former teachers and all my friends.

Funding: I acknowledge the UNSW Tuition Fee Scholarship and UNSW conference funding (Postgraduate Research Student Support and CSE HDR Student Travel)\footnote{I would have been more grateful had the UNSW offered me a prestigious IPRS scholarship as the Australian National University did. Appraisal of Prof Sri Parameswaran by his former students for his astounding supervision was the major factor in selecting UNSW that I witnessed my self with no regret}. I also would like to acknowledge the travel bursaries from Oxford Nanopore Technologies and the ACM SRC Travel Award. I also appreciate the NVIDIA corporation for donating the Jetson TX2 and Tesla K40 GPUs used for experiments in this thesis.

%% file: 1-pre/publications.tex
\chapter{Publications and Presentations}

\section*{List of Publications}

This thesis has led to the following first author journal publications and they are included in lieu of chapters. 
 
\begin{itemize}
    \item \textbf{H. Gamaarachchi}, A. Bayat, B. Gaeta, and S. Parameswaran, “Cache Friendly Optimisation of de Bruijn Graph based Local Re-assembly in Variant Calling,” IEEE/ACM transactions on computational biology and bioinformatics, 2018. DOI: \url{https://doi.org/10.1109/TCBB.2018.2881975}
    \item \textbf{H. Gamaarachchi}, S. Parameswaran, and M. A. Smith, “Featherweight long read alignment using partitioned reference indexes,” Scientific Reports 9, 4318 (2019). DOI: \url{https://doi.org/10.1038/s41598-019-40739-8}   
    \item \textbf{H. Gamaarachchi}, C. W. Lam, G. Jayatilaka, H. Samarakoon, J. T. Simpson, M. A. Smith, and S. Parameswaran, “GPU Accelerated Adaptive Banded Event Alignment for Rapid Comparative Nanopore Signal Analysis,” BMC Bioinformatics 21, 343 (2020). DOI: \url{https://doi.org/10.1186/s12859-020-03697-x}\footnote{Also available as a pre-print in bioRxiv, 2019, DOI: \url{https://doi.org/10.1101/756122}}, 2020.
    \item \textbf{H. Gamaarachchi}, H. Saadat, S. Parameswaran, "Optimisation of Nanopore Sequence Analysis Software for Many-core CPUs", prepared for submission [in progress], 2020.
\end{itemize}

This thesis has also led to the following article in the ACM SRC Grand Finals.

\begin{itemize}
    \item “ESWEEK: G: Real-time, Portable and Lightweight Nanopore DNA Sequence Analysis using System-on-Chip”, ACM SRC Grand Finals, 2020. URL: \url{https://src.acm.org/binaries/content/assets/src/2020/hasindu-gamaarachchi.pdf} --- \emph{\textbf{third place Grand Finalist}}
\end{itemize}

Collaborative research conducted in close relation to the work presented under this thesis has led to the following publications and pre-prints which are not included in the thesis. 

\begin{itemize}
    \item H. Samarakoon, S. Punchihewa, A. Senanayake, J. M.  Hammond, I. Stevanovski, J.M. Ferguson, R. Ragel, \textbf{H. Gamaarachchi} and I. W. Deveson, “Genopo: a nanopore sequencing analysis toolkit for portable Android devices,” Communications biology 3, 538 (2020) DOI: \url{https://doi.org/10.1038/s42003-020-01270-z}
    \item R. P. Mohanty, \textbf{H. Gamaarachchi}, A. Lambert, and S. Parameswaran, “SWARAM: Portable Energy and Cost Efficient Embedded System for Genomic Processing,” ACM Transactions on Embedded Computing Systems (TECS) 18.5s (2019). DOI: \url{https://doi.org/10.1145/3358211}
    \item A. Bayat, \textbf{H. Gamaarachchi}, N. P. Deshpande, M. R. Wilkins, and S. Parameswaran, “Methods for de-novo genome assembly,” Preprints 2020, 2020, DOI: \url{https://doi.org/10.20944/preprints202006.0324.v1}
    \item A.F. Laguna, \textbf{H. Gamaarachchi}, X. Yin, M.Niemier, S. Parameswaran and X. S. Hu, “Seed-and-Vote based In-Memory Accelerator for DNA Read Mapping”, ICCAD 2020 [accepted]
\end{itemize}

\section*{List of Presentations}

\subsection*{Oral presentations:} 

\begin{itemize}
\item "Performance Optimisation of Nanopore DNA Analysis Software: A Computer Architecture Aware Approach," Australasian Leadership Computing Symposium (ALCS), 2019. URL: \url{https://opus.nci.org.au/display/Help/Genomics+Stream?preview=/48497246/50236166/ALCS_Genomics_Gamaarachchi_released.pdf}
\item "Lightweight, Portable and Real-time Embedded Computing Systems for Downstream Nanopore Data Processing", London Calling, 2020. URL: \url{https://www.youtube.com/watch?v=hvXSpqnIB1c}
\item "Real-time, Portable and Lightweight Nanopore DNA Sequence Analysis using System-on-Chip", ACM SRC second-round at ESWEEK, 2019. --- \emph{\textbf{First place and entry into the SRC Grand Finals}}
\end{itemize}

\subsection*{Poster presentations:}

\begin{itemize}
    \item "Portable Real-time Genomic Data Processing: Harmonising Bioinformatics Software to Exploit Hardware", Australasian Genomic Technologies Association Conference (AGTA) 2019. --- \emph{\textbf{Best student poster award}}
    \item "Real-time, Portable and Lightweight Nanopore DNA Sequence Analysis using System-on-Chip", ACM SRC first-round at ESWEEK, 2019. --- \emph{\textbf{Entry into the second-round}}
\end{itemize}

\chapter{Awards}

The work conducted under this thesis has received the following awards.

\begin{itemize}
\item Grand Finalist (third place winner), ACM SRC Grand Finals graduate category, 2020
\item First Place, ACM SIGBED SRC at ESWEEK, 2019
\item Best Student Poster Award, Australasian Genomic Technologies Association Conference (AGTA), 2019
\item Runner-up, UNSW 3 Minute Thesis School level, 2019
\end{itemize}

%% file: 1-pre/abbreviations.tex
%\chapter{Abbreviations}

%%cat */*.tex | grep -wo "[A-Z]\+\{2,10\}" | sort | uniq -c | awk '{print $2}'

\nomenclature{ABEA}{Adaptive Banded Event Alignment}
\nomenclature{ACM}{Association for Computing Machinery}
\nomenclature{ADB}{Android Debug Bridge}
\nomenclature{AIO}{Asynchronous I/O}
\nomenclature{API}{Application Programming Interface}
\nomenclature{ASCII}{American Standard Code for Information Interchange}
\nomenclature{ASIC}{Application-Specific Integrated circuit}
\nomenclature{ASIP}{Application-Specific Instruction-set Processor}
\nomenclature{AVX}{Advanced Vector Extensions}
\nomenclature{BGZF}{Blocked GNU Zip Format}
\nomenclature{BLASR}{Basic Local Alignment with Successive Refinement}
\nomenclature{BLAST}{Basic Local Alignment Search Tool}
\nomenclature{BWA}{Burrows-Wheeler Aligner}
\nomenclature{BWT}{Burrows-Wheeler Transform}
\nomenclature{CCS}{Circular Consensus Sequencing}
\nomenclature{CIGAR}{Concise Idiosyncratic Gapped Alignment Report}
\nomenclature{CLR}{Continuous Long Reads}
\nomenclature{COVID}{COrona VIrus Disease}
\nomenclature{CPU}{Central Processing Unit}
\nomenclature{CRAM}{Compressed Alignment File Format}
\nomenclature{CSE}{Computer Science and Engineering}
\nomenclature{CUDA}{Compute Unified Device Architecture}
\nomenclature{DC}{Direct Current}
\nomenclature{DICE}{Data Intensive Computer Engineering}
\nomenclature{DNA}{Deoxyribonucleic acid}
\nomenclature{DP}{Dynamic Programming}
\nomenclature{DRAM}{Dynamic Random-Access Memory}
\nomenclature{DTW}{Dynamic Time Warping}
\nomenclature{ENA}{European Nucleotide Archive}
\nomenclature{ESWEEK}{Embedded Systems Week}
\nomenclature{FAQ}{Frequently Asked Questions}
\nomenclature{FPGA}{Field-Programmable Gate Array}
\nomenclature{GATK}{GenomeAnalysisToolkit}
\nomenclature{GB}{GigaByte}
\nomenclature{GNU}{GNU's Not Unix}
\nomenclature{GPGPU}{General-Purpose Computing on Graphics Processing Units}
\nomenclature{GPU}{Graphics Processing Unit}
\nomenclature{GZ}{G-Zip}
\nomenclature{HDD}{Hard Disk Drive}
\nomenclature{HDF}{Hierarchical Data Format}
\nomenclature{HDL}{Hardware Description Language}
\nomenclature{HDR}{Higher Degree Research}
\nomenclature{HIV}{Human Immunodeficiency Viruses}
\nomenclature{HLS}{High-Level Synthesis}
\nomenclature{HMM}{Hidden Markov Model }
\nomenclature{HP}{Hewlett-Packard}
\nomenclature{HPC}{High-Performance Computing/Computer}
\nomenclature{HPE}{Hewlett Packard Enterprise}
\nomenclature{HTS}{High-Throughput Sequencing }
\nomenclature{ID}{Identifier}
\nomenclature{IEEE}{Institute of Electrical and Electronics Engineers}
\nomenclature{IGV}{Integrative Genomics Viewer}
\nomenclature{IOPS}{Input/output Operations Per Second}
\nomenclature{IP}{Internet Protocol}
\nomenclature{IPC}{Inter-Process Communication}
\nomenclature{IT}{Information Technology}
\nomenclature{JNI}{Java Native Interface}
\nomenclature{KB}{KiloByte}
\nomenclature{LTR}{Long Terminal Repeat}
\nomenclature{LTS}{Long-Term Support }
\nomenclature{MAPQ}{MAPping Quality}
\nomenclature{MB}{MegaByte}
\nomenclature{MIMD}{Multiple Instruction Multiple Data }
\nomenclature{MPSoC}{Multi-Processor System on a Chip}
\nomenclature{NAS}{Network-Attached Storage }
\nomenclature{NAT}{Network Address Translation}
\nomenclature{NDK}{Native Development Kit}
\nomenclature{NFS}{Network File System}
\nomenclature{NGS}{Next Generation Sequencing}
\nomenclature{NRE}{Non-Recurring Engineering }
\nomenclature{NVCC}{Nvidia CUDA Compiler }
\nomenclature{NW}{Needleman–Wunsch}
\nomenclature{OEM}{Original Equipment Manufacturer}
\nomenclature{ONT}{Oxford Nanopore Technologies}
\nomenclature{OOM}{Out Of Memory}
\nomenclature{OS}{Operating System}
\nomenclature{PAF}{Pairwise mApping Format}
\nomenclature{PC}{Personal Computer}
\nomenclature{PCI}{Peripheral Component Interconnect}
\nomenclature{POSIX}{Portable Operating System Interface}
\nomenclature{RAID}{Redundant Array of Independent  Disks}
\nomenclature{RAM}{Random-Access Memory}
\nomenclature{RNA}{RiboNucleic Acid}
\nomenclature{SAM}{Sequence Alignment Map}
\nomenclature{SARS}{Severe Acute Respiratory Syndrome}
\nomenclature{SATA}{Serial Advanced Technology Attachment}
\nomenclature{SBC}{Single Board Computer}
\nomenclature{SDK}{Software Development Kit}
\nomenclature{SFP}{Small Form-factor Pluggable}
\nomenclature{SGA}{String Graph Assembler}
\nomenclature{SIGBED}{Special Interest Group on Embedded Systems}
\nomenclature{SIMD}{Single Instruction, Multiple Data}
\nomenclature{SINE}{Short Interspersed Nuclear Element}
\nomenclature{SK}{Suzuki-Kasahara}
\nomenclature{SMRT}{Single Molecule, Real-Time}
\nomenclature{SNP}{Single Nucleotide Polymorphisms}
\nomenclature{SNV}{Single Nucleotide Variant}
\nomenclature{SOAP}{Short Oligonucleotide Analysis Package}
\nomenclature{SoC}{System on a Chip}
\nomenclature{SRA}{Sequence Read Archive}
\nomenclature{SRAM}{Static Random-Access Memory}
\nomenclature{SRC}{Student Research Competition}
\nomenclature{SSD}{Solid State Drive}
\nomenclature{SSE}{Streaming SIMD (Single Instruction Multiple Data) Extensions}
\nomenclature{SSH}{Secure Shell}
\nomenclature{SW}{Smith-Waterman}
\nomenclature{TB}{TeraByte}
\nomenclature{TCP}{Transmission Control Protocol}
\nomenclature{TSV}{Tab-Separated Values}
\nomenclature{UCSC}{University of California, Santa Cruz}
\nomenclature{UNSW}{University of New South Wales}
\nomenclature{USA}{United States of America}
\nomenclature{USB}{Universal Serial Bus}
\nomenclature{USD}{United States Dollar}
\nomenclature{VCF}{Variant Call Format}
\nomenclature{WGS}{Whole-Genome Sequencing}

\printnomenclature[5em]

%% file: 2-intro/main.tex
\chapter{Introduction}
Humankind technologically advanced to read or `sequence' their own DNA---nature's blueprint of a living organism---only a few decades ago. This technological advancement was a turning point for healthcare and medicine and led to a new era of medicine that is more precise and tailored to an individual than traditional evidence-based medicine. Sequencing of the first-ever human DNA started as an international effort called the human genome project in 1990 and took 13 years to complete in 2003 \cite{chial2008dna}, at a cost of 2.7 billion USD \cite{initaequence}. Since then, DNA sequencing technologies advanced at a remarkable pace over the last few decades up to a point where today the human genome can be sequenced in just two days at a cost less than 1000 USD \cite{currentcost}. This rapid pace of advancement is continuing, and this sequencing process is becoming possible within a few hours \cite{futuretime} at a cost of less than 100 USD \cite{futurecost}. Thus, DNA sequencing tests are likely to become routine as of today's blood tests in the near future.

Precision medicine considers the variability in genes, environment and lifestyle amongst different individuals to guide the use of effective and safe treatments tailored to a particular individual \cite{precision-medicine}.  This is in contrast to the one-size-fits-all approach in traditional evidence-based medicine that targets the average person \cite{precision-medicine}. Clinical genomics that considers the information encoded in the DNA is being increasingly incorporated into precision medicine protocols \cite{ashley2016towards}. One of the ten highest-grossing drugs (in the USA), \textit{rosuvastin}---a statin used to lower blood cholesterol---is shown to benefit only 1 in 20 \cite{schork2015personalized}. The best benefit ratio for any of the top 10 grossing drugs is 1 in 4, which is still considerably low \cite{schork2015personalized}. In traditional evidence-based medicine, selecting the most effective drug out of available subtypes of drugs for a particular patient is a somewhat trial-and-error process, which is inefficient \cite{ginsburg2001personalized}. In precision medicine, genetic information gathered from the sequencing of individuals' DNA is being increasingly used to determine the most effective drug. For instance, the latest anti-cancer drugs such as \textit{crizotinib} that treat anaplastic lymphoma kinase (ALK) positive lung cancer already require genetic testing of the patient \cite{shaw2011crizotinib}. While genetic testing today is mainly performed for critical cases, it is expected to become more and more frequent in the future with the cost and availability of DNA sequencing improving rapidly.

Another use of DNA sequencing is for implementing a more proactive approach,  "prevention is better than cure". The genetic information of an individual can determine the predisposition to a number of diseases, thus making it possible to implement preventive measures. A very popular example is the actress Angelina Jolie who underwent a double mastectomy in 2013 after a genetic test that revealed a significantly higher chance of developing breast cancer. While the cost of a genetic test in 2013 was probably not affordable for everyone, today they are becoming more and more realistic and common.

DNA sequencing is also beneficial in epidemiological applications. In the recent past, during the Ebola virus outbreak in West Africa (2013–2016) and Zika virus outbreak in Brazil (2015-2016) DNA sequencing has been utilised for viral surveillance.  Today, the utility of DNA sequencing is evident than ever before, due to the ongoing COVID-19 pandemic. DNA sequencing of the viral sequence allows identification of mutations that facilitates the tracking of the disease spread and provides insights into the virus evolution that are useful in vaccine development.

In addition to the above, DNA sequencing is also applicable in several other fields such as forensics, evolutionary biology and agronomy.

DNA sequencing alone is of limited utility if not for the sequence analysis, a very heavy computational analysis that follows the actual sequencing. DNA sequencers---the machines that sequence the DNA---read the DNA sequence in small fragments called `reads'. Computational analyses must be performed to put these reads together to achieve a draft sequence assembly close as possible to the original DNA sequence or to compare against an existing reference of the original DNA sequence. This two-step process, (a) DNA sequencing and (b) sequence analysis are depicted in Fig. \ref{f:seq-vs-analysis}.

\begin{figure}[ht]
\centering
\includegraphics[width=0.5\linewidth]{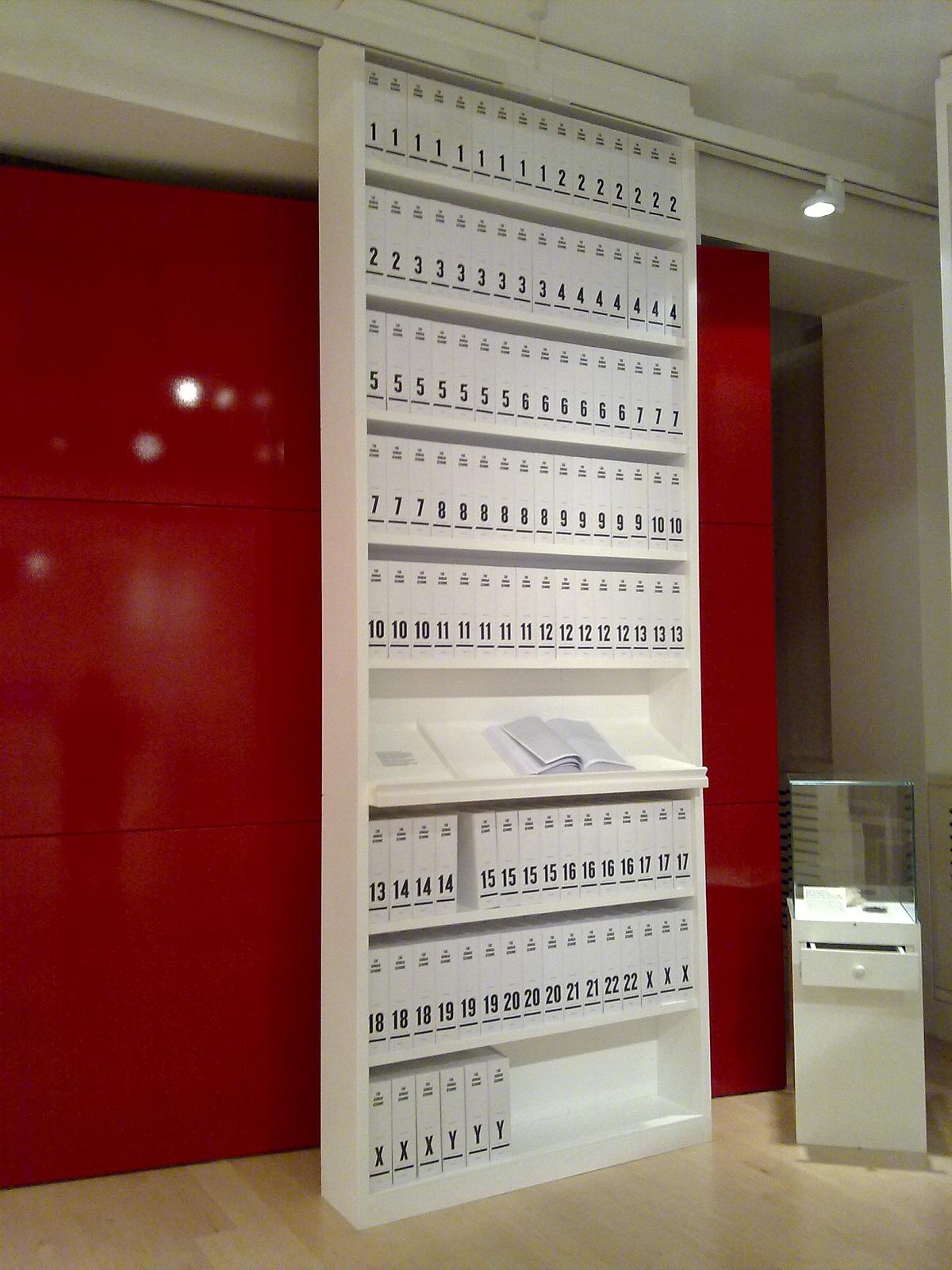}
\caption[Human DNA sequence printed as a series of books]{Human DNA sequence printed as a series of books displayed at Wellcome Collection, Euston Square, London. Photograph from \cite{humangenome-printed} licensed under  Creative Commons Attribution-ShareAlike 2.0 Generic (CC BY-SA 2.0).}
\label{f:bookshelf}
\end{figure}

The input to the DNA sequencing process (Fig \ref{f:seq-vs-analysis}) is a tissue sample of a living organism (e.g., blood). Such a sample contains billions of cells and each and every cell contains a homologous copy of the DNA sequence that is millions of molecular bases long\footnote{Identical copies if cells are normal, i.e., unless cancer cells.}. The full DNA sequence inside a single cell of a human is 3.1 billion bases long\footnote{6.2 billion bases long if copies inherited from both mother and father are considered.} and when printed is a series of books that accommodate a whole bookshelf (Fig. \ref{f:bookshelf}). This long DNA sequence is tightly packed inside the cell and the sample preparation process that unpacks the DNA sequence breaks the fragile DNA strand into small fragments (Fig. \ref{f:seq-vs-analysis})\footnote{Unintended fragmentation in nanopore sequencing or intended fragmentation in Illumina sequencing.}. This prepared sample containing trillions of fragments of DNA from multiple cells are read by an array of sensors in the DNA sequencer and are output as a series of data points that represents the biological sequence (Fig. \ref{f:seq-vs-analysis}).

The reads output by the sequencer---tiny fragments coming from multiple copies of the full DNA sequence---are in random order. The sequence analysis process (Fig. \ref{f:seq-vs-analysis}) that assembles these tiny pieces to obtain the original DNA sequence or compares differences in the reads to a reference of the original DNA sequence is typically challenging and computationally intensive, mainly due to the following reasons:

\begin{figure}[ht]
\centering
\includegraphics[width=\linewidth]{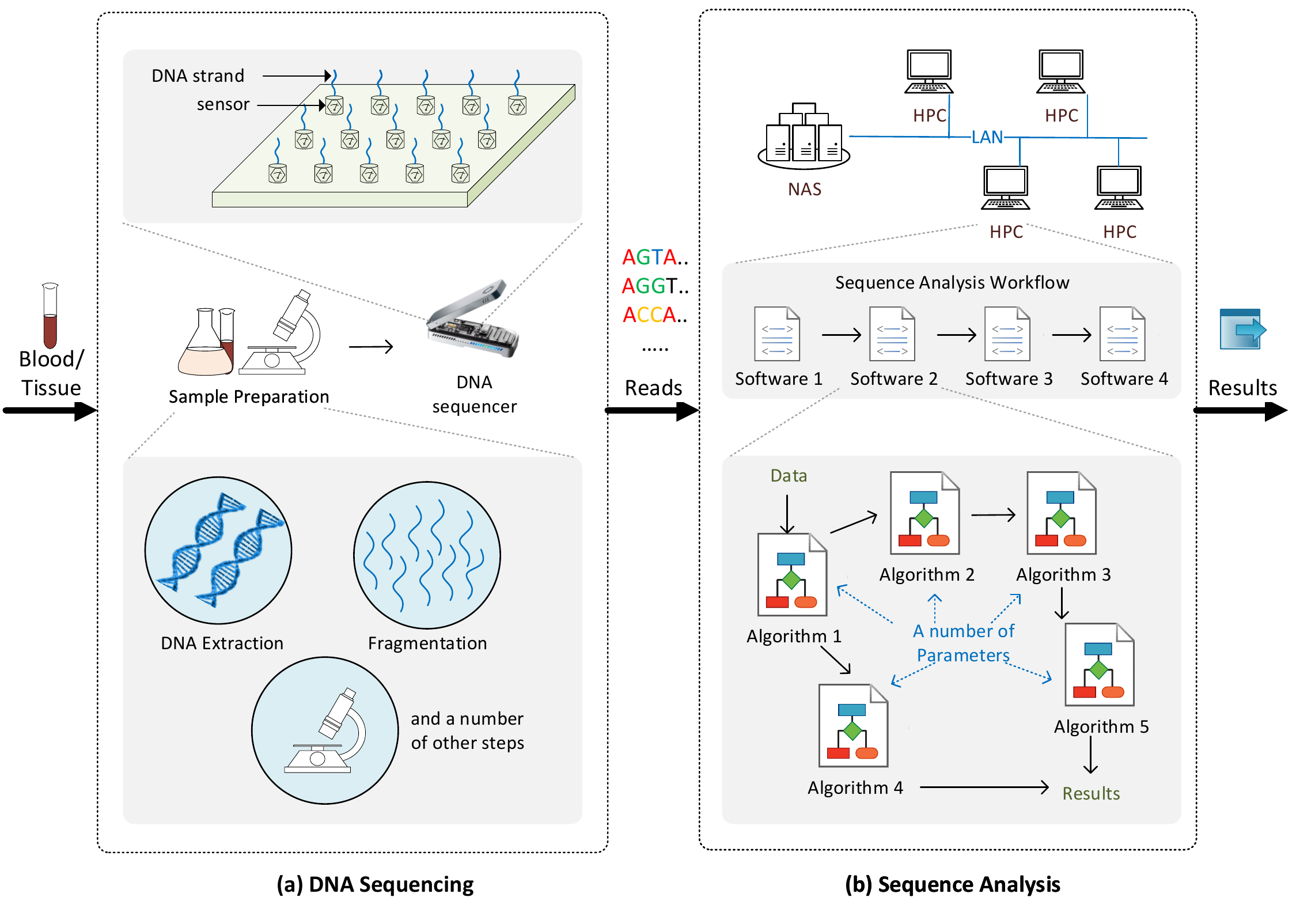}
\caption{DNA sequencing and sequence analysis}
\label{f:seq-vs-analysis}
\end{figure}

\begin{enumerate}
\item Reads are tiny compared to the original full DNA sequence (reads are around 75-500 bases in second-generation sequencers and around 1,000-100,000 bases in third-generation sequencers where the full DNA sequence is typically millions of bases long).
\item The reference sequence used for comparison is somewhat different to the DNA sequence in a sample (around 0.5\% difference in humans due to genetic variation between two humans \cite{levy2007diploid}) and the error caused by sequencers when reading the DNA is comparatively large (around 0.1\%-1\% in second-generation sequencers and around 0.5\%-13\% in third-generation sequencers).
\item The complexity of the human genome (i.e., more than 50\% in the human DNA sequence are repeat regions \cite{de2011repetitive,treangen2012repetitive} and there are numerous types of repeat regions with distinct characteristics).
\item The large data volume output by the sequencers (can be as high as hundreds of gigabytes or even a few terabytes).
\end{enumerate}

Over the last two decades, a plethora of workflows that perform DNA sequence analysis has emerged. A sequence analysis workflow is very sophisticated that a single workflow is a pipeline of different software tools run one after the other (Fig. \ref{f:seq-vs-analysis}) and each single software tool is a collection of numerous algorithms and heuristically determined parameters. Computational biologists or bioinformaticians have attempted to improve the accuracy of the sequence analysis workflows increasingly over the years, and as a result, the workflows have become more sophisticated.

\section{Sequencers vs Computers - Gap Between Technologies} \label{s:thegap}

\begin{figure}[ht]
\centering
\includegraphics[width=\linewidth]{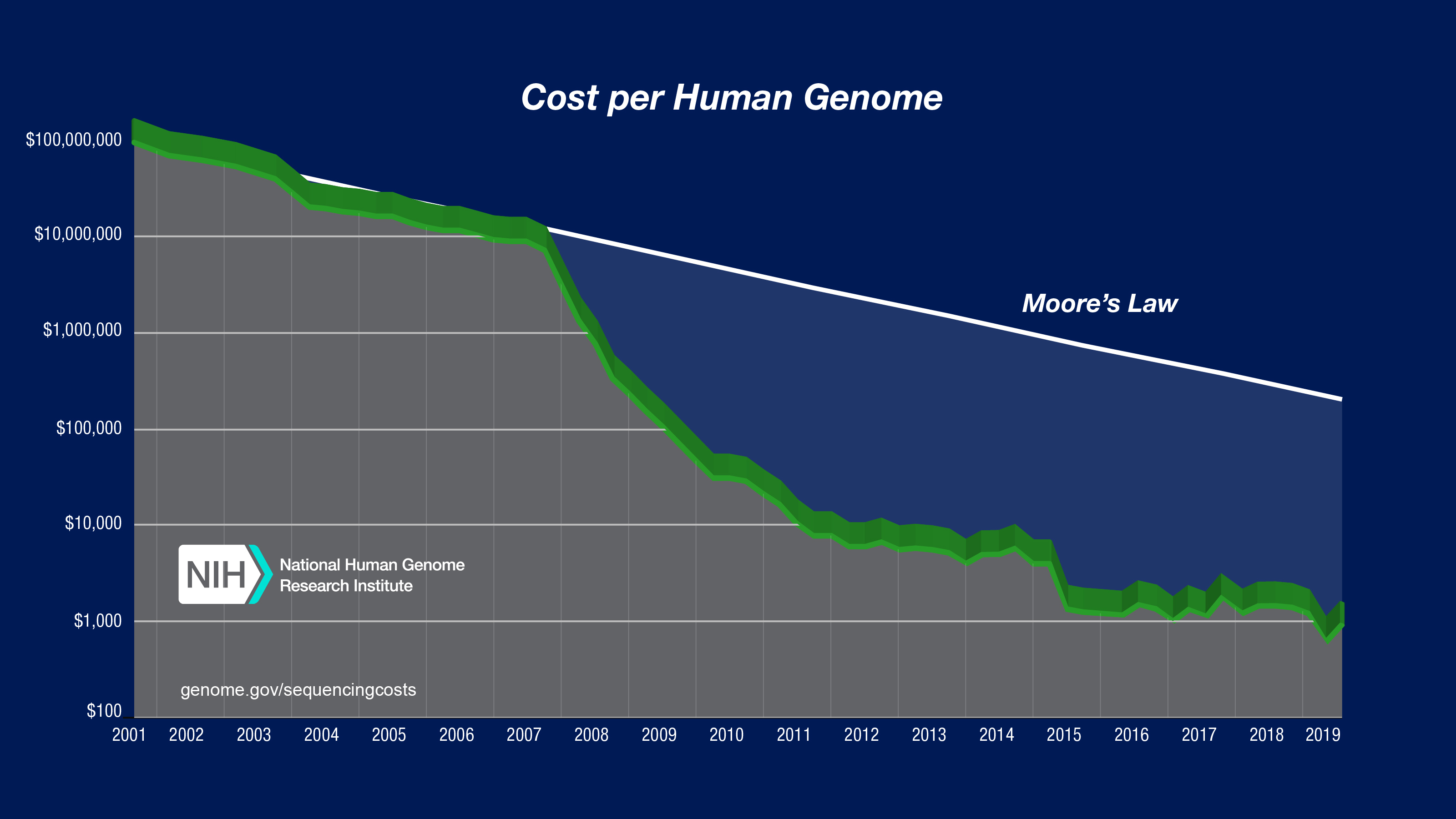}
\caption[Drop of DNA sequencing cost per human genome over the period from 2001 to 2019]{Drop of DNA sequencing cost per human genome over the period from 2001 to 2019. The Figure is from \cite{nhgri}. }
\label{fig:drop-of-sequencing-cost}
\end{figure}

The rapid improvement in DNA sequencing technologies in terms of sequencing cost over the last two decades is depicted in the graph in Fig. \ref{fig:drop-of-sequencing-cost}. The hypothetical line that depicts Moore's law is to compare how fast the sequencing technologies have improved. From 2001 to 2007, the cost of sequencing per human genome has reduced at a rate similar to Moore's law. Then from 2007 to 2019, the drop in sequencing cost has been faster than Moore's law, from 10 million USD to 1000 USD per human genome. Illumina, a leading sequencing company, has announced that their upcoming technology will bring down this cost to 100 USD in the future \cite{futurecost}. This rapid improvement is expected to continue and the sequencing machines which were limited to high-end research facilities are slowly arriving into pathology labs. 

\begin{figure}[ht]
\centering
\includegraphics[width=\linewidth]{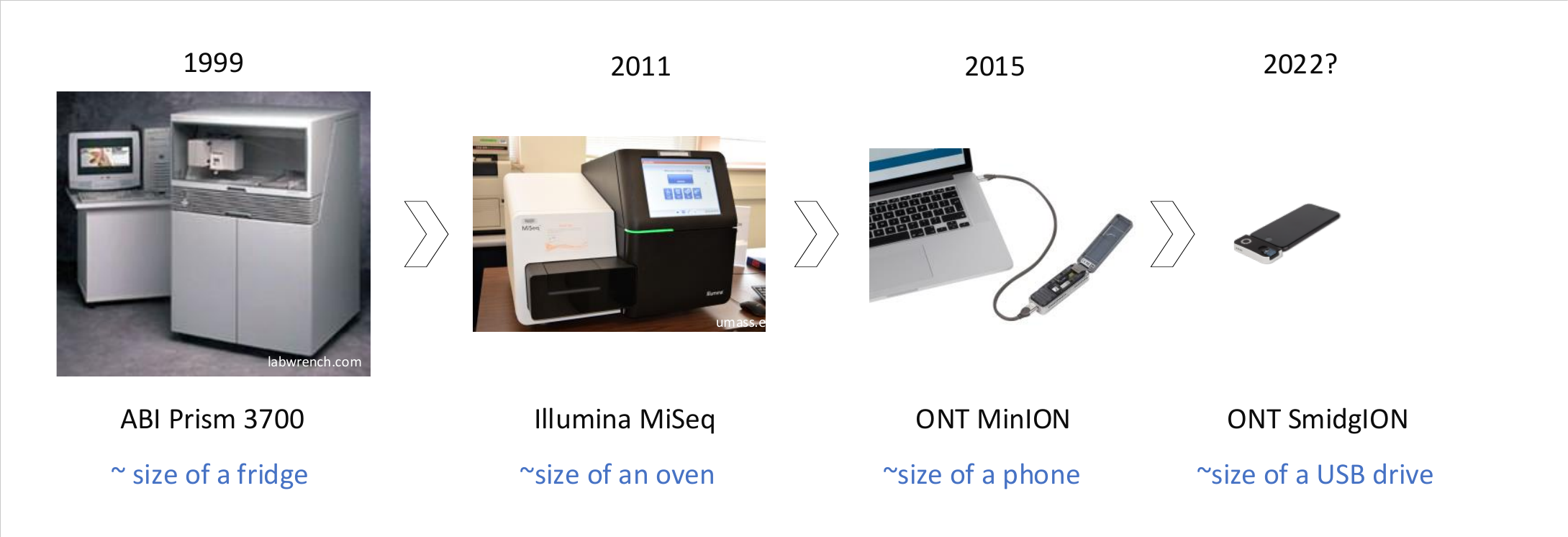}
\caption{Evolution of sequencing machines}
\label{fig:seq-machine-evolve}
\end{figure}

In addition to the improvement of DNA sequencing in terms of cost, the physical size and weight of the DNA sequencers also have improved.  Fig. \ref{fig:seq-machine-evolve} shows the evolution of the sequencing machines over the last two decades. ABI Prison 3700 DNA sequencer, a first-generation sequencer released in 1999, was similar to the size of a fridge (dimensions 134.62 cm x 76.2 cm x 74.93 cm) with a weight over a hundred kilograms. Illumina MiSeq sequencer, a second-generation sequencer released in 2011 was the size of an oven (of the dimensions 68.6 cm × 56.5 cm × 52.3 cm) with a weight of  57.2 kg. The size further shrank with the Oxford Nanopore Technologies (ONT) MinION sequencer, a third-generation sequencer released in 2015 that is the size of a mobile phone (dimensions 10.5 cm x 2.3 cm x 3.3 cm and of weight 87 g). The size is expected to shrink further. Oxford Nanopore Technologies has already announced their upcoming product, ONT SmidgION, a tiny USB thumb drive sized DNA sequencer that could be directly plugged on to a mobile phone.
%ABI Prism 3700 DNA Analyzer  
% 134.62 cm x 76.2 cm x 74.93 cm
% 1999
%[https://americanhistory.si.edu/collections/search/object/nmah_1297334]

%Illumina HiSeq 2000, releaed in 2010 dimesions 118.6 cm × 76.0 cm × 94.0 cm, 312 kg.

Although DNA sequencing has advanced rapidly, sequence analysis technologies are considerably lagging. As a result, even today, sequence analysis is typically performed on clusters of high-performance computers, the same situation as twenty years ago. There are multiple reasons for this lag, which are stated below.

\begin{enumerate}
    \item General-purpose computers have not improved in terms of performance as fast as the DNA sequencers.
    \item Data volume output by DNA sequencers has kept increasing despite the decrease in size.
    \item The sophistication of the sequence analysis software tools also has increased over the years as a result of biologists improving the accuracy of the results.
    \item Sequence analysis software tools written by computational biologists or bioinformaticians with the focus on higher accuracy of the results are un-optimised to efficiently utilise computational resources, i.e. software does not map well to the architecture of computers.
\end{enumerate}

\section{Need for Reducing the Gap}

Sequence analysis steps consume days to weeks if done on a commodity laptop, or not possible at all in certain cases due to limited memory (RAM) in laptop computers. Hence, clusters of high-performance computers are currently being used, yet the process takes hours to complete. Such super-computers are very costly and massive and are typically available in high-end research facilities. As involved data set sizes are massive (can be up to several terabytes), cloud computing that relies on the internet is not ideal. 

The utility of an ultra-portable DNA sequencer such as the MinION is currently limited due to the analysis process being performed on non-portable high-performance computers. There are a number of examples where scientists took these MinION sequencers into the field to perform sequencing. During the 2013-2016 Ebola virus outbreak in West Africa, scientists performed in-the-field sequencing using MinION sequencers \cite{quick2016real}. However, the sequence analysis had to be performed offsite on high-performance computers in Europe. Gigabytes of data were transferred through a mobile internet connection which was expensive and slow. Analysis technologies to perform the analysis in-the-field would have been valuable in such circumstances, not only to reduce the cost but also for a quick turn-around time of results. 

Another similar example is the use of the MinION during the Zika virus outbreak in Brazil \cite{Faria2016}, which the utility was again limited due to limitations in analysis technologies. Going beyond rural areas, scientists have performed sequencing using the MinION in jungles, arctic \cite{goordial2017situ} and even on the international space station \cite{castro2017nanopore}, which all of them would have benefited by efficient sequence analysis technologies. Currently, the ultra-portability of the MinION sequencer is being used to facilitate sequencing of the SARS-CoV-2 in smaller decentralised laboratories around the world \cite{nanopore-covid}. Rapid epidemiological data sharing from places all over the world is a key to a better public health response. Having ultra-portable analysis technologies will further benefit in such circumstances.

Better analysis technologies will not only benefit portable applications such as the above but also in decentralising DNA tests in the future. DNA sequencers that are limited to high-end research facilities today will soon arrive in pathology labs and even doctor's office. Having better sequence analysis technologies will support the data processing in situ without the need to transfer data to centralised high-performance computing facilities. Better sequence analysis technologies can also benefit large scale sequencing studies where the processing is performed in centralised high-performance computing facilities, by reducing the computing cost and the turnaround time of the results.

Considering all of the above factors, improving analysis technologies to match DNA sequencing technologies is a timely need. 

\section{Possible Solutions to Fill the Gap}

Possible solutions to fill the gap between sequencing and analysis technologies are explored below in the context of the four reasons that were stated in the later part of section \ref{s:thegap}.

Performance of sequencers has evolved faster than computers that used to follow Moore's law (Fig. \ref{fig:drop-of-sequencing-cost}). However, general-purpose single-core processor performance only improved by 3\% in the year 2017---much slower than Moore's law and future improvements should focus on application-specific hardware, as pointed by pioneers in computer architecture John Hennessy and David Patterson who received the Turing award in 2018 \cite{hennessy2018new}. Application-Specific Integrated Circuits (ASIC) or custom circuit chips designed specifically for sequence analysis would be a potential solution to reduce the gap, which had already been applied as solutions in other domains such as digital signal processing. However, the field of genomics being still immature and thus the workflows rapidly evolving, designing custom hardware is challenging.  Even little changes in algorithms require redesigning the custom hardware and this design cost is millions of dollars. An option that provides better flexibility than ASIC would be to design Application-Specific Instruction-set Processors (ASIP), which are in-between versions of general-purpose processors and ASICs in terms of flexibility. However, fabricating such ASIPs would still incur billions of dollars. Field Programmable Gate Arrays (FPGA) could be used as `breadboards` to prototype ASICs or ASIPS, however, full sequence analysis workflows are too complicated to be made fully functional on a typical FPGA with limited resources.

The high data volume output by the sequencers is a cause for an increased amount of computations, yet is beneficial to account for errors introduced by the sequencers, i.e,  errors can be normalised when one region of a DNA string is covered by multiple reads. If sequencers become more and more accurate, the amount of data required to assemble a single DNA sequence will reduce. However, production of such accurate sensors that function at nano- and pico-scale measurements is far ahead in the future.

The complexity of the human genome is inevitable, i.e. there are seven categories of repeated sequences, each category has subcategories, each subcategory has a number of different families, each family has subfamilies and these subfamilies have distinct characteristics \cite{jurka2011repetitive}. Also, more than 50\% of the human genome is composed of repeated sequences \cite{de2011repetitive,treangen2012repetitive}. Certain regions (e.g. Telomere, Centromere) in the human genome have been too intractable to the existing technologies and are yet being resolved at the time of writing \cite{miga2019telomere}.  Analysis workflows that work on such complex genomes are thus inevitably sophisticated. What is meant by sophisticated here is not the algorithmic time-complexity, instead, the number of idiosyncratic cases that deviates from the general model. When processing, each of these deviated cases require to be separately handled. For instance, each family of repeated sequences would need to be processed using different algorithms and/or heuristic parameters, leading to a large number of code paths. A sequence analysis workflow as a whole would thus remain sophisticated, however, the time-complexity of each and every algorithm
inside the workflow can be improved. Designing better algorithms with lesser time complexity has been and will be one of the most effective ways to improve performance. Over the past decades, plenty of work has been done in designing better algorithms and this will continue to happen.

%A sequence analysis workflow as a whole would thus remain sophisticated, however, the time-complexity of each and every algorithm inside the workflow can be improved. In fact, plenty of work has been done over the last decades and most algorithms in today's workflows are of linear time-complexity. Thus, constant values typically ignored when determining the time-complexity (coefficients that are ignored in big O notation, i.e if time is $kn$, $k$ is ignored to give $O(n)$) are of importance for further performance improvements and this has been considered to a certain level in today's workflows.  

Sequence analysis software tools are typically designed and developed by computational biologists or bioinformaticians whose major focus is to develop methods that are predicated on answering a research question or producing a specific outcome. Typically those computational biologists or bioinformaticians have access to near unlimited computational resources in their research environment---clusters of high-performance servers with hundreds of gigabytes of RAM. Their focus is not on maximal optimisation of the software, which requires detailed knowledge of computational systems. Consequently, sequence analysis software tools are typically severely un-optimised to efficiently run on computing systems with limited resources such as laptops or desktops. In other words, sequence analysis software tools severely lack computer architecture-aware optimisations that consider the knowledge of underlying hardware architecture. Note that these architecture-aware optimisations are not to be confused with algorithmic time-complexity optimisations which have already been done to a considerably adequate level in current sequence analysis software. Consider a hash table versus a contiguous array in memory. Despite accessing both the hash table and the array having the same time-complexity, contiguous accesses to an array are tens of times faster than random accesses to a hash table in a modern computer due to the presence of memory caches. Such optimisations that map existing sequence analysis software components to efficiently map with complex architectural features in modern computer systems are henceforth referred to as computer architecture-aware optimisations.

Out of the solutions discussed above, the most timely solution is to perform  architecture-aware optimisations on existing sequence analysis software. Such optimisations are cost-effective and practical, yet rarely applied to sequence analysis software. The focus of this thesis is such architecture-aware optimisations on existing DNA sequence analysis software. In addition to the provision of efficient performance on general-purposes computers, such optimisations would complementary benefit any future-focused ASIP design efforts.

\section{This Thesis}

\subsection{Philosophy}

\begin{figure}[ht]
\centering
\includegraphics[width=\linewidth]{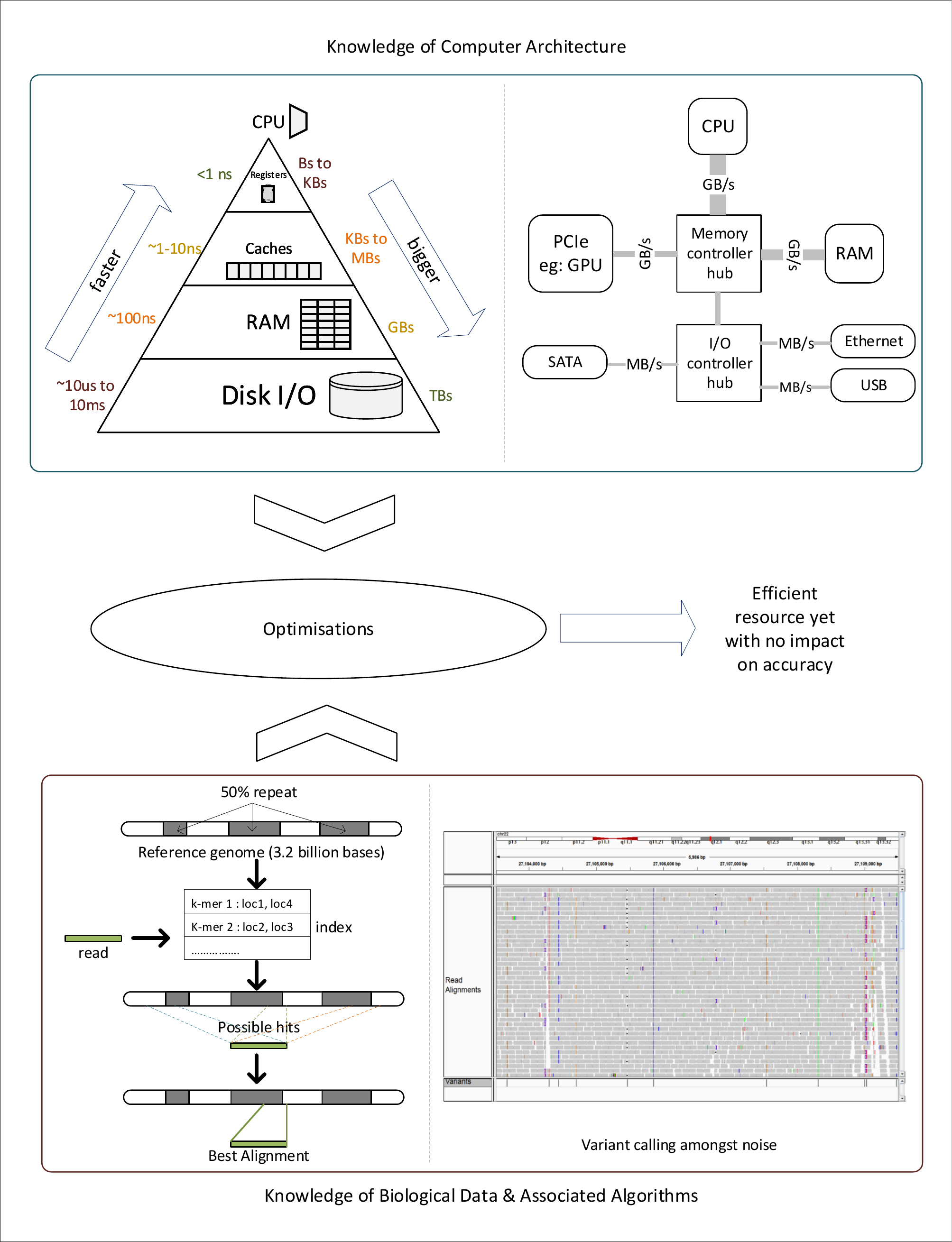}
\caption{Synergistic use of knowledge from both computer architecture and biology}
\label{fig:arch-hierarchy}
\end{figure}

The architecture of modern computer systems is complex. Understanding such complex architectures requires the knowledge of a number of topics such as:

\begin{itemize}
    \item the memory hierarchy (Fig. \ref{fig:arch-hierarchy});
    \item interfacing  between the processor, memory and co-processors (Fig. \ref{fig:arch-hierarchy});
    \item internal details of processors and co-processors such as multiple cores, instruction scheduling and instructions set architecture; and,
    \item low-level software such as operating system (processes, threads, scheduling, disk caches, virtual memory, etc.) and device drivers.
\end{itemize}

Simultaneously, the field of DNA analysis is also utterly complex and understanding such analysis tools require knowledge of a number of topics such as:

\begin{itemize}
    \item basic molecular biology involving the structure and function of DNA, chromosomes, genome, etc.;
    \item features of DNA sequences such as various types of repetitive sequences;
    \item different generations of DNA sequencing technologies;
    \item characteristics of data produced from sequencing technologies such as the read lengths and error rate; and,
    \item sequence analysis algorithms such as sequence alignment, variant calling and methylation calling.
\end{itemize}

Developing efficient software that conforms to hardware architectures requires the knowledge of all these computer architecture related topics.
Developing sequence analysis software that produces accurate results require
the knowledge of all above DNA analysis related topics. Thus, architecture-aware optimisation of sequence analysis software requires the simultaneous use of knowledge from both computer architecture and DNA analysis (Fig. \ref{fig:arch-hierarchy}).

DNA sequence analysis software tools are sophisticated and modern computer architectures are also sophisticated. Computer architecture knowledge helps to efficiently utilise resources. At the same time, knowledge of characteristics of biological data and associated algorithms ensures that the accuracy of the final results is unaffected. The domain knowledge from both the fields is utilised for the optimisations (Fig. \ref{fig:arch-hierarchy}). This thesis attempts to bridge the two interdisciplinary fields--computer architecture and DNA analysis---to produce sequence analysis software that can efficiently utilise existing resources in a modern computer system.

\subsection{Chapter Outline}

This thesis is about architecture-aware optimisations to existing DNA sequence analysis software. We present architecture-aware optimisations at different levels: Processor level, register level, cache level, RAM level and disk I/O level. The outline of the rest of the thesis is as follows.  

In chapter \ref{c:literature}, the background required to understand the technical chapters and a detailed literature review of the state-of-the-art are given. First, the background of DNA and DNA sequencing is presented in a simplified fashion for a reader from a non-biological background. Then, the background and the state-of-the-art of sequencing analysis workflows are presented. Finally, previous efforts of architecture-aware optimisation of DNA analysis workflows are presented.

In chapter \ref{c:cacheopti}, cache and register level optimisations to a popular variant calling software called \textit{Platypus} are presented. A major time-consuming component of this software---de Bruijn Graph construction---was improved by a  using cache and register level optimisations without any impact on the accuracy.

In chapter \ref{c:minimap}, memory (RAM) size optimisation of a popular sequence alignment software called \textit{Minimap2} is presented.  The peak memory usage in \textit{Minimap2} was reduced  through a divide and conquer strategy, most importantly, without compromising the accuracy. This work enabled performing sequence analysis in low memory systems such as mobile phones, which was otherwise not possible. 

Chapter \ref{c:gpuabea} discusses RAM level, cache level and processor level optimisations to a core algorithm component in analysing data produced from Oxford Nanopore sequencers called the Adaptive Banded Event Alignment (used in the popular Nanopore analysis toolkit Nanopolish). This includes how the algorithm was parallelised for CPU-GPU heterogeneous architectures. Importantly, the impact on the accuracy of the final results is negligible.

Chapter \ref{c:integration} discusses how the optimisations proposed in chapters \ref{c:cacheopti},\ref{c:minimap},\ref{c:gpuabea} were integrated to develop fully functional embedded system prototypes for a popular nanopore sequence analysis workflow. It is shown the performance of the prototypical embedded systems employed with proposed optimisations is surprisingly comparable to the performance on the same workflow (unoptimised version) run on a high-performance server.

Then, going beyond embedded systems, chapter \ref{c:ioopti} presents the identification of the primary bottleneck in nanopore sequence analysis workflows that seriously affect high-performance servers. Solutions for alleviating this bottleneck are also presented.

Finally, the thesis is concluded in chapter \ref{c:conclusion} with a discussion of  future directions.

\subsection{Publications}

Chapter \ref{c:cacheopti} is published in IEEE/ACM transactions on computational biology and bioinformatics \cite{gamaarachchi2018cache}. Chapter \ref{c:minimap} is published in Nature Scientific Reports \cite{minimap2arm} and has received global attention amongst the community (altimeter score 89, picked up by 8 news outlets). Chapter \ref{c:gpuabea} is available as a pre-print in bioRxiv \cite{gamaarachchi2019gpu} and a modified version is accepted for publication BMC Bioinformatics. Chapter \ref{c:integration} may be adapted for publication in the future. Chapter \ref{c:ioopti} is prepared to be submitted to an IEEE/ACM journal or conference proceedings. 

In addition, collaborative research conducted in close relation to the work presented under this thesis has led to second author publications and pre-prints in \cite{mohanty2019swaram,samarakoon2020f5n,bayat2020methods}. However, none of the content from these second author publications is claimed as a part of this thesis.

\subsection{Open-source Contributions}

This thesis benefits the community through a number of contributions to existing open-source bioinformatics software and the development of new open-source bioinformatics software. 
Those existing open-source software tools that were contributed are \textit{Platypus} variant caller (see chapter \ref{c:cacheopti}), popular sequence aligner \textit{Minimap2} (see chapter \ref{c:minimap}, appendix \ref{a:minimap-supps} and appendix \ref{a:opensource}) and popular nanopore signal analysis toolkit \textit{Nanopolish} (see appendix \ref{a:opensource}). The new bioinformatics software developed under this thesis are \textit{f5c} (see chapter \ref{c:gpuabea}, appendix \ref{a:f5c-documentation} and appendix \ref{a:f5c-supps}) and \textit{f5p} (see chapter \ref{c:integration} and appendix \ref{a:nanocluster}). Also, the design and the associated software for the prototype embedded systems are released as open-source (see chapter \ref{c:integration} and appendix \ref{a:nanocluster}).

\subsection{Miscellaneous}

The research conducted in support of this thesis has won third place in the grand final of ACM Student Research Competition 2020, amongst competitors from 22 major ACM conferences. The entry to the ACM SRC Grand finals was through winning the first place in ACM SRC at ESWEEK 2019 conference. Research conducted under this thesis has received the best poster award in Australasian Genomic Technologies Association Conference 2019.

%% file: 3-literature/main.tex
\chapter{Literature Review}\label{c:literature}

In this chapter, the background of DNA sequencing is discussed in section \ref{s:sequencing-back}. Then, the background of sequence analysis and associated data structures and algorithms are discussed in section \ref{s:sequence-analysis}. In section \ref{s:related-work}, related work that has focused on computational optimisation of sequence analysis software is discussed.

\section{DNA Sequencing} \label{s:sequencing-back}

\subsection{Terminology and Basics of DNA}

In this subsection, the terminology in DNA sequencing and basic concepts of DNA sequencing are introduced.

\subsubsection{DNA}

Deoxyribonucleic Acid (DNA) is the blueprint of life. DNA is a molecule that encodes the structure and the function of a living organism \cite{felsenfeld1985dna}. A closer analogy from computer science is a computer program. A computer program is composed of instructions and data to achieve a particular outcome, whereas DNA is composed of instructions and data to make a living organism from scratch and to maintain its function. Instructions and data in a computer program are encoded in binary (base-2), whereas instructions and data in DNA are encoded in quaternary (base-4). The four bases in the DNA alphabet are Adenine (A), Cytosine (C), Guanine (G) and Thymine (T), which are molecules called nucleotides. 

A long chain of nucleotide bases connected through chemical bonds forms a \textit{DNA strand}.  Two such strands that are coiled around each other, forming the double helix-shaped DNA molecule (Fig. \ref{f:basics-of-dna}). Both strands contain the same information and having two such strands facilitates DNA replication, the process by which DNA copies itself during cell division. The two strands are complementary to each other and are held together by hydrogen bonds between G-C and A-T base pairs, i.e., a base `G' is complementary to base `C' (and vice versa) and a base `A' is complementary to base `T' and vice versa).

\subsubsection{Chromosome}

\begin{figure}
    \centering
    \includegraphics[width=\linewidth]{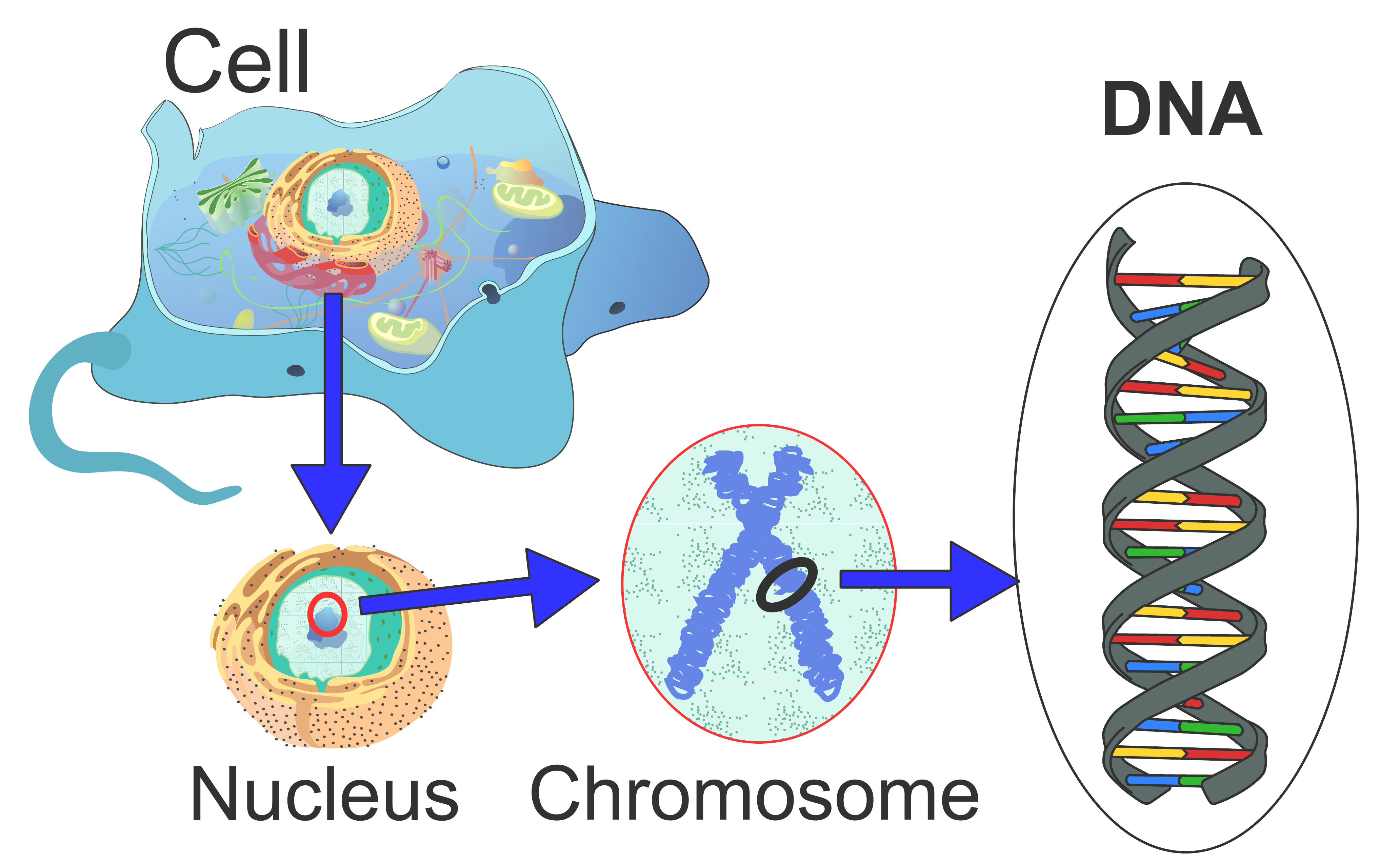}
    \caption[The DNA molecule and a chromosome]{The DNA molecule and a chromosome. Chromosomes are present inside the nucleus of a cell of an organism. The DNA molecule that is the major component of a chromosome is double-stranded. A DNA strand is composed of four nucleotide bases A, C, G and T depicted in blue, yellow, red and green, respectively.
    The figure is from \url{https://commons.wikimedia.org/wiki/File:Eukaryote_DNA.svg} licensed under Creative Commons Attribution-Share Alike 3.0 Unported license.}
    \label{f:basics-of-dna}
\end{figure}

A DNA molecule is tightly coiled many times and packaged with proteins to form a structure called a \textit{chromosome}. Inside the nucleus of every cell of a human being, there are 23 pairs of such chromosomes (Fig. \ref{f:basics-of-dna}). Those chromosome pairs are named chromosome 1 to chromosome 22 and the 23\textsuperscript{rd} chromosome pair determines the sex. This 23\textsuperscript{rd} chromosome pair in females contains two X chromosomes and in males contains
an X chromosome and a Y chromosome. In humans, the largest chromosome is chromosome 1 ($\sim$247 million bases) and the smallest is chromosome 21 ($\sim$47 million bases).
In each chromosome pair, one chromosome is inherited from the mother and the other from the father.

The number of chromosomes varies amongst organisms. It can be just one chromosome or even thousands of chromosomes. The \emph{ploidy}---whether the chromosomes exist in pairs, single or a higher number of sets---also varies amongst different organisms.

\subsubsection{Genome}

The complete nucleotide sequence of all the chromosomes within a cell is called the \textit{genome}. The size of a genome is measured using the number of nucleotide bases. Following the metric prefixes, thousands of bases, millions of bases and billions of bases can be called kilobases, megabases and gigabases, respectively. Biologists typically use kb, Mb and Gb as symbols for those units, but this thesis uses the symbols kbases, Mbases and Gbases to prevent any confusion with kilobytes, megabytes and gigabytes.

The sizes of the genome of various organisms are listed in Table \ref{t:genome-sizes}. Viral genomes are typically the smallest with a size of several thousand bases. 
Bacterial genomes can vary from several hundred thousand bases to a few million bases ($\sim$100 kbases to $\sim$15 Mbases). Insect genomes are in the order of hundreds of millions of bases ($\sim$100 Mbases to $\sim$900 Mbases). The genome size of complex organisms is billions of bases long. For instance, the human genome is 3.1 Gbases (6.2 Gbases if both chromosomes in a pair are considered). The largest genome found so far is of a rare Japanese flower plant called \textit{Paris japonica} and is 149 Gbases long.

\begin{table}[!ht]
    \centering
    \caption{Genome size of various organisms}\label{t:genome-sizes}
    \begin{tabular}{|l|r|}
        \hline
        \textbf{Organism} & \textbf{Genome size} \\  \hline
        HIV (virus) & 10 kbases \\  \hline
        H1N1 (virus) & 14 kbases \\  \hline 
        SARS-CoV-2   (COVID-19 virus) & 30 kbases \\  \hline
        \textit{Helicobacter pylori} (bacteria) & 1.7 Mbases\\ \hline 
        \textit{Escherichia Coli} (bacteria) & 4.6 Mbases \\ \hline 
        Yeast & 12.1 Mbases \\ \hline
        Fruit fly (insect) & 140 Mbases \\ \hline
        Mouse & 2.5 Gbases \\ \hline
        Cow & 3 Gbases \\ \hline
        Human & 3.1 Gbases \\\hline
        Wheat & 17 Gbases \\ \hline
        Marbled lungfish & 130 Gbases \\ \hline
        \textit{Paris japonica} (plant) & 149 Gbases \\ \hline
    \end{tabular}
\end{table}

\begin{figure}
    \centering
    \includegraphics[width=\linewidth]{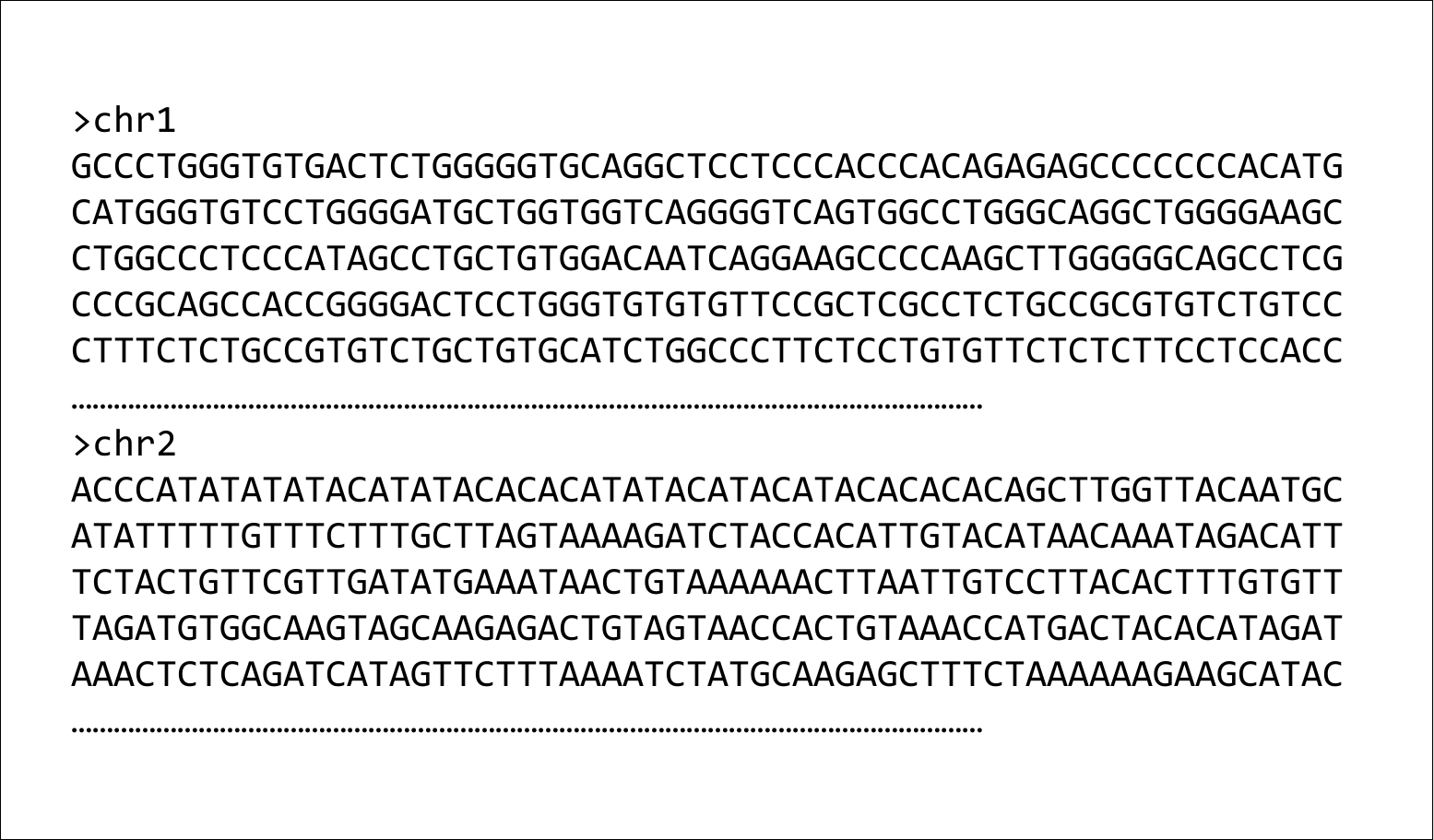}
    \caption[An example of \textit{FASTA} file format]{An example of \textit{FASTA} file format. The dotted lines are to indicate that the long chromosome sequences  continue, i.e., dotted lines are not actually present in a \textit{FASTA} file. Note that the sequences here are hypothetical and are not representative of a particular species.}
    \label{f:fasta-format}
\end{figure}

The widely used file format for storing a genome on a computer is the \textit{FASTA} (\textit{.fa}) format. \textit{FASTA} is a simple text-based format where the characters `A', `C', `G' and `T' denote the nucleotide bases. An example genome stored in the \textit{FASTA} format is given in Fig. \ref{f:fasta-format}. The line that starts with a `>' character contains the name of the chromosome (may contain other metadata) and the subsequent lines contain the actual DNA sequence (Fig. \ref{f:fasta-format}).

To save space, \textit{FASTA} can be compressed using the extended \textit{gzip} format called \textit{bgzip} that allows random access to genomic locations in the compressed file at the expense of a slightly lesser compression  ratio than \textit{gzip}.

A representative example of the genome of a particular species if known as the \textit{reference genome}. Species such as humans have a high-quality reference as a result of the human genome project that produced a draft assembly, which was subsequently improved by scientists over the years. The latest version of the human genome is named as GRCh38 (Genome Reference Consortium Human Build 38).

%\todo{can Why text instead of bits, if required.}

%The human genome consists of 3.1 billion nucleotide bases \cite{lander2001initial}.
\subsubsection{Repeats}
%https://asia.ensembl.org/info/genome/genebuild/assembly_repeats.html

%DNA sequences that occur repeatedly in multiple locations in a genome are referred to as \emph{repeats}, \emph{repeated sequences} or \emph{repetitive elements}. 

Repeats are quite common in genomes, for instance, more than 50\% of the human genome is composed of repeats \cite{de2011repetitive,treangen2012repetitive}. \emph{Repeats} are also known by terms such as \emph{repeated sequences}, \emph{repetitive elements} or \emph{repeat regions}. 
Repeats have always introduced complications to the sequence analysis process, due to reads coming from such regions that are non-unique being extremely challenging to be accurately aligned \cite{treangen2012repetitive}.

Repeats have been classified into several types based on the characteristics of the sequence, for instance, satellite repeats,  simple repeats, tandem repeats, transposons, etc. \cite{Ensembl-repeats,jurka2011repetitive}.

\subsubsection{Genes}

The exact interpretation of the genome is not fully understood yet. However, scientists have interpreted the genome to a considerable extent. Millions of regions in the DNA called genes are individually or collectively responsible for features and functions of a living organism.

\subsubsection{Variants}\label{s:vcf}

About 99.5\% of the genome of all humans is the same \cite{levy2007diploid}. The 0.5\% difference encompasses human genetic variation. The differences in the genome of a particular individual to the reference genome of the particular species are called variants. 

\begin{figure}
    \centering
    \includegraphics[width=\linewidth]{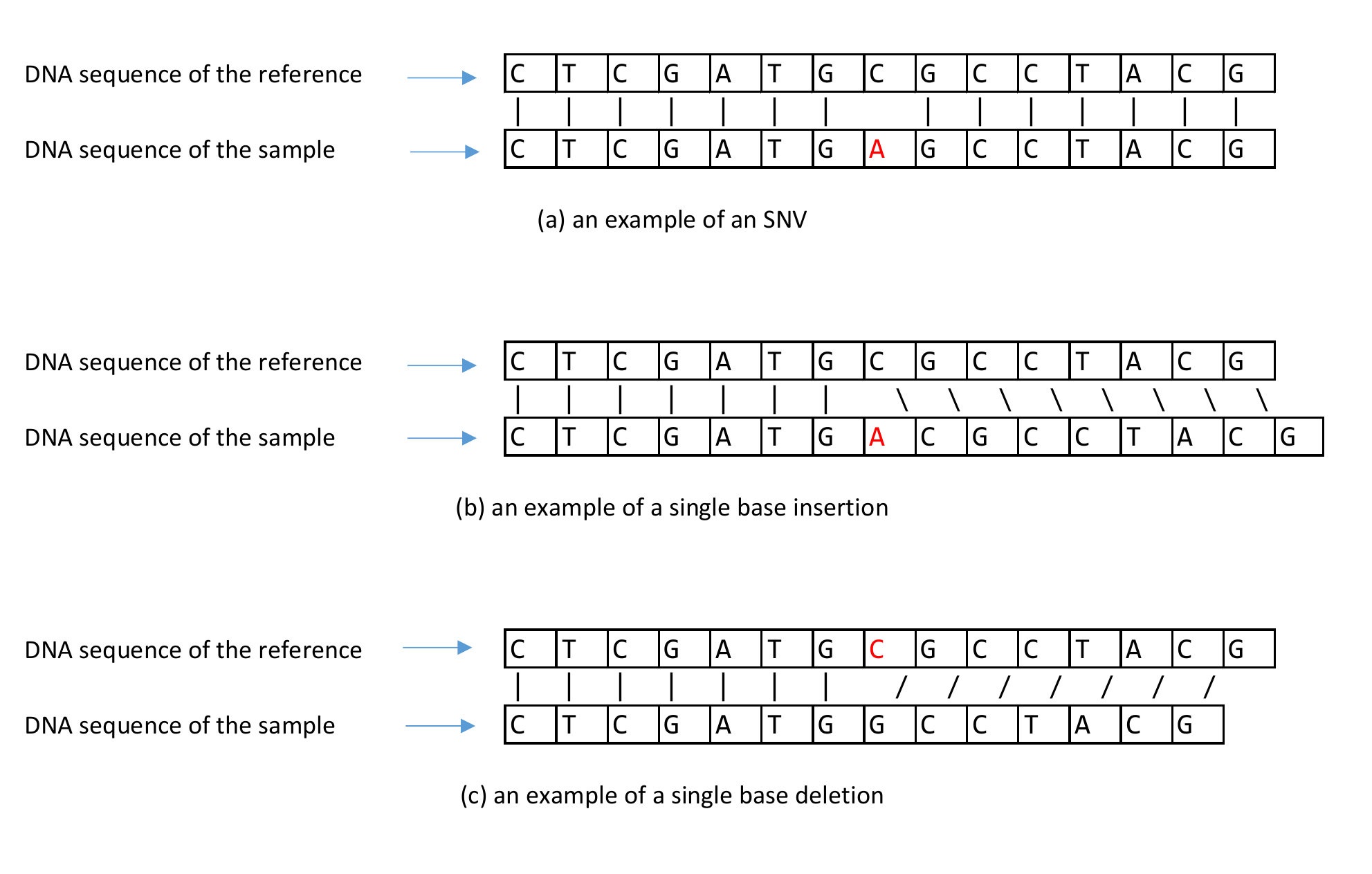}
    \caption{Elaboration of SNV and Indels}
    \label{f:variants}
\end{figure}

Different types of variants exist. A variation of a single nucleotide base is called a Single Nucleotide Variation (SNV). An SNV that is prevalent amongst a sufficiently large fraction of the population is referred to as Single Nucleotide Polymorphism (SNP).
Insertion or deletion of one or more contiguous bases is called an Indel \cite{kondrashov2004context}. Examples of these three types of variants SNV, insertions and deletions are shown in Fig. \ref{f:variants}. Indels can be small as one or two bases or large as ten thousand bases. Variants that are 50 or more bases (50 is the typical value and this number can be arbitrary) are known as structural variants \cite{mahmoud2019structural}. Structural variants include many different sub-types such as long insertions, long deletions,  copy number variants and inversions.

Most of these variants cause natural differences among individuals. However, some of the variants are responsible for various diseases. For instance,  diseases such as sickle cell anaemia \cite{ingram1956specific} and beta-thalassemia\cite{chang1979beta} are directly associated with SNVs. Diseases such as Asthma and Allergic Rhinitis are caused by a  complex contribution of both genetic and environmental factors \cite{ober2011genetics}. A large number of genetic variants that contribute to various diseases have been discovered.  \textit{ClinVar} is a public database containing such medically significant genetic variants  \cite{landrum2013clinvar}. More and more novel variants and their correspondence to various conditions are being readily discovered. 

\begin{figure}
    \centering
    \includegraphics[width=\linewidth]{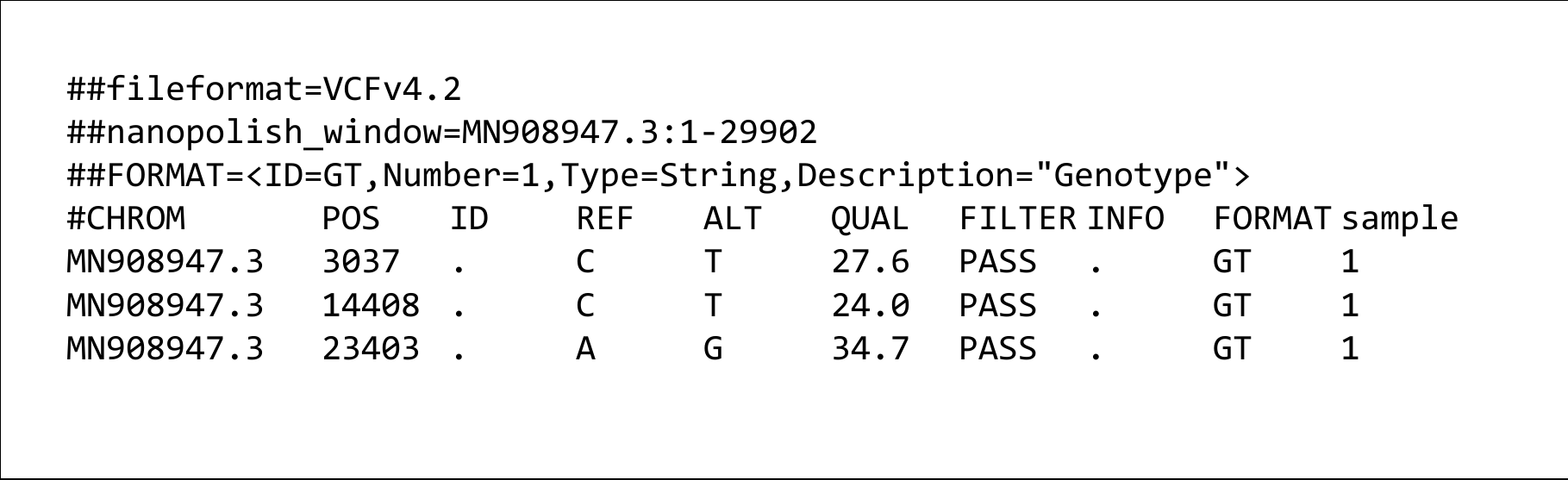}
    \caption{An example of VCF file format}
    \label{f:vcf-format}
\end{figure}

Detected variants are typically stored in the file format called Variant Call Format (VCF) \cite{team2015variant}, which is text-based format exemplified in Fig. \ref{f:vcf-format}. The header contains lines starting with `\#' character and describes metadata. Then, the details of variants are listed as tab-separated values with one variant per line.

\subsubsection{Epigenome}

Nucleotide bases in the DNA can naturally undergo biochemical modifications when  chemical compounds are attached to nucleotide bases. Such nucleotide bases with additional chemical compounds attached are known as modified bases. To date, more than 17 different types of base modifications have been identified in DNA \cite{xu2019recent}. The set of base modifications undergone by every base in the genome when taken together is called the \textit{epigenome}. A common type of base modification in humans is the addition of methyl groups to nucleotide bases, which is known as DNA methylation. 

DNA methylation is known to be a regulator of the genome, i.e., the expression of genes can be regulated by base modifications. DNA methylation is also known to affect development and tissue differentiation. DNA methylation is altered by environmental factors, but those alterations can be passed onto the next generations.

% \subsubsection{Application of Genomic and epi-genomic Research}

% \todo{this section is probably unnecessary}

% The genome can reveal the physical traits, disease predisposition and the drug response of a particular individual. The knowledge of DNA has given rise to new medical treatment methods. Latest drugs such as crizotinib that treat anaplastic lymphoma kinase (ALK) positive lung cancer requires genetic testing of the patient \cite{shaw2011crizotinib}. Currently, the most effective antidepressant to treat depression of a particular person is found out by trial and error basis. Research is being conducted to find the best drug by genetic testing \cite{winner2013psychiatric,depression}. 

% \todo{Explain about virus outbreaks and phylogeny}

% Understanding of the genome has not only benefited healthcare, but also to a number of fields ranging from forensics \cite{homer2008resolving}, evolutionary biology, and anthropology \cite{kaestle2002ancient}.

\subsubsection{DNA Sequencing}

The DNA molecule exists inside a human cell in a very compact form (scale of nanometres) with the DNA strand coiled many times, which if uncoiled would be a few metres long. To read the DNA strand, it has to be extracted from the cell and uncoiled. The DNA strand being very fragile when uncoiled, reading the full DNA strand from one end to another accurately is still a technological challenge. The best available technology today can only read this DNA strand in fragments of contiguous bases. This is due to the fragile DNA strand breaking into fragments at random locations during the DNA extraction process from the cell, uncoiling and even during the reading process. The process of reading the DNA sequence is termed \emph{DNA sequencing} \cite{heather2016sequence} and the machines that perform this sequencing are called \emph{DNA sequencers}.

\subsubsection{Reads}

The sequencing machine takes a tissue sample of an organism, for instance, blood (more accurately a prepared sample out of tissue where the DNA strands have been extracted), and outputs the order of the bases in a digital form. The DNA strands break into fragments and the sequencing machine reads these fragments of DNA strands\footnote{Fragmentation can be intentional in certain technologies such as Illumina.}. The resultant series of data points denoting bases of a DNA fragment is called a \emph{read}. The reads are output in random order by the sequencer. This is mainly due to the DNA fragments floating randomly in the liquid solution of the sample.

\subsubsection{Coverage}

\begin{figure}
    \centering
    \includegraphics[width=\linewidth]{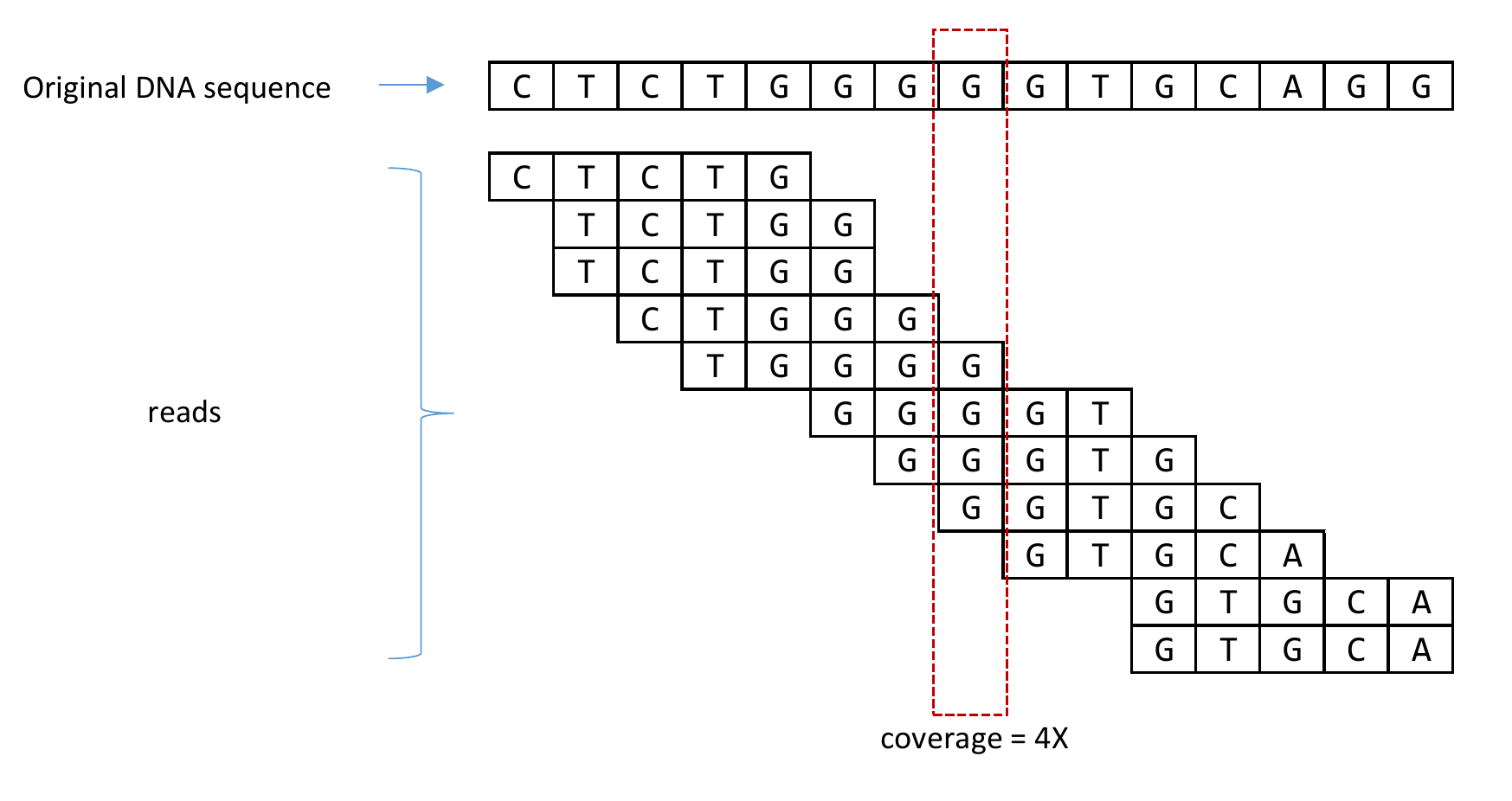}
    \caption{Elaboration of the concept of coverage in sequencing. }
    \label{f:coverage}
\end{figure}

A sample prepared for sequencing contains fragments of DNA that originated from nearly identical DNA molecules (each cell has a copy of the DNA and there are millions of cells in a sample).  Fragmentation of DNA occurs at random locations on the DNA strand. The Sequencer  randomly sequences a subset of these DNA fragments floating in the solution and outputs them as reads. Consequently, a single position in the genome is covered by multiple reads. The number of reads that overlap a particular position on the genome is known as the \textit{depth} or \textit{coverage}. Fig. \ref{f:coverage} elaborates the coverage using an example. In the example, the coverage of the marked base position is 4$\times$ because the particular position on the original sequence is covered by four reads.

\subsubsection{Base-calling}\label{s:fastq}

The process of converting direct or indirect measurements of the nucleotide bases (in the DNA strand) captured by sensors in the sequencer into ASCII reads is called \textit{base-calling}. The base-calling process is not 100\% accurate due to the presence of noise in measurements, sensor limitations and restrictions of the software involved in  base-calling. One or more bases in a read can be incorrectly base-called and such errors are known as base-calling errors or sequencing errors. 

The volume of data output by a sequencer or the sequencing yield is typically measured using the total number of bases in all the reads generated during a sequencing run (the duration in which the sequencing machine is operated). Modern sequencers can generate reads that sum to billions of bases and thus the common unit used for yield is Gbases. 

\begin{figure}
    \centering
    \includegraphics[width=\linewidth]{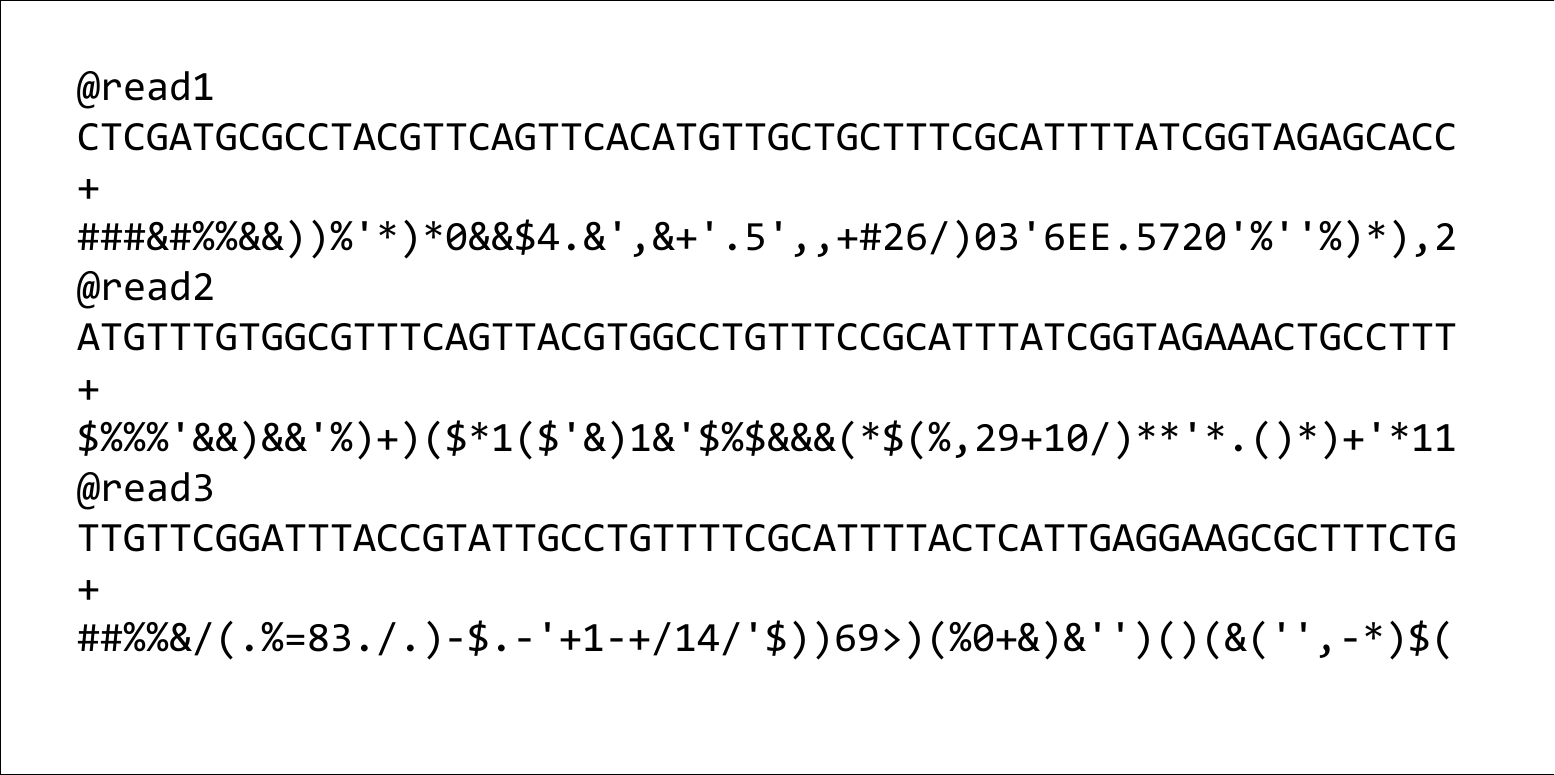}
    \caption{An example of \textit{FASTQ} file format}
    \label{f:fastq-format}
\end{figure}

Base-called reads are typically stored in the file format called \textit{FASTQ} \cite{cock2009sanger}, a text-based file format extended from the previously discussed \textit{FASTA} format. In \textit{FASTQ} format, a single read takes four lines  (Fig. \ref{f:fastq-format}): the first line is the read name (read identifier and optional metadata) that starts with an `@' character; the second line is the actual read sequence in ACGT characters; the third line is always a `+' character; and, the fourth line is the per-base phred quality score encoded in ASCII (phred quality score $Q$ is given by $Q = -10 log_{10}(P)$ where $P$ is the  base-calling error probabilities  \cite{ewing1998base}).

\subsection{DNA Sequencing Technologies}

As of today, there have been three generations of DNA sequencing technologies. They are detailed below.

\subsubsection{First-Generation Sequencing}

Sanger et al. used a method called the plus-minus system to sequence the first complete DNA which was of a virus in 1977 \cite{sanger1977nucleotide}. The introduction of the chain termination method (also known as Sanger Sequencing) \cite{sanger1977dna} was a breakthrough in sequencing technologies due to its accuracy and convenience. With various improvements to this method, automated DNA sequencers were produced that were capable of sequencing complex genomes. These automated Sanger sequencers are known as first-generation sequencing machines. 

\begin{figure}
    \centering
    \includegraphics[width=\linewidth]{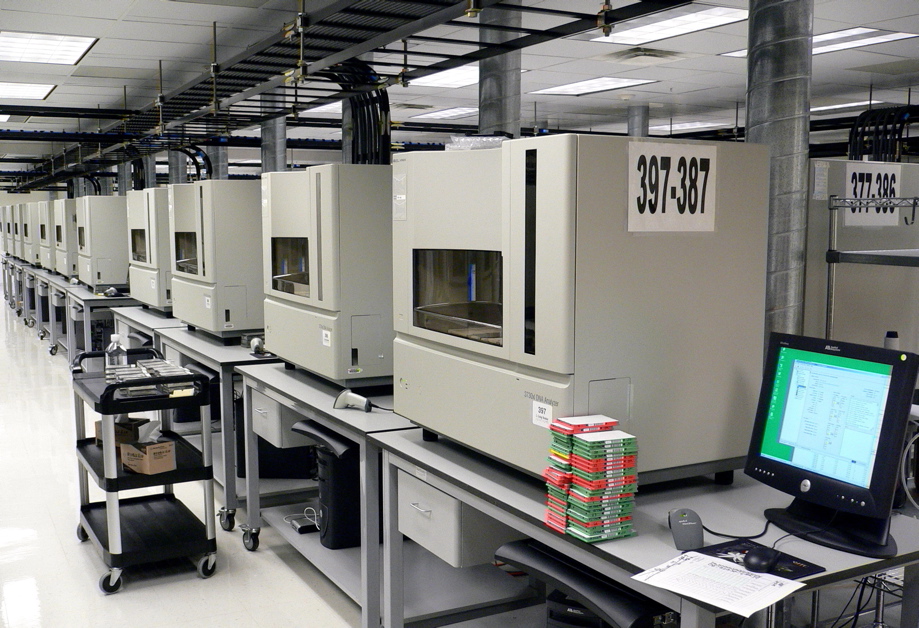}
    \caption[First-generation sequencers]{First-generation sequencers. A row of Applied Biosystems 3730xl DNA Analyzer machines. Photo is from \cite{photofirstgen} licensed under CC BY 2.0.
    The weight of a machine is 180 kg and the dimensions are 100 cm (W) x 73 cm (D) x 89 cm (H)  \cite{thermofisher}.}
    \label{f:first-gen-seq}
\end{figure}

First-generation sequencers can produce high-quality (accurate) long-reads at the expense of high cost and low throughput. For instance, Applied Biosystems 3730xl first-generation sequencer in Fig. \ref{f:first-gen-seq} could output reads at around 99\% accuracy and 400 to 900 bases length. However, a single sequencing run that spans over a duration of 20 minutes to 3 hours generates only 1.9-84 kbases \cite{liu2012comparison}. In fact, the human genome project that started in 1990 mainly used first-generation sequencers \cite{venter2001sequence}. The human genome project took 13 years to complete at an expense of billions of dollars. Today, first-generation sequencers are infrequently used.

\subsubsection{Second-Generation Sequencing}

In 1985, a different technique to that used in first-generation sequencers was introduced \cite{nyren1985enzymatic} and the eventual improvements in the 1990s led to the \textit{second-generation sequencing} technology. In literature, the term \textit{next-generation sequencing} has been used instead of second-generation sequencing, which is no longer appropriate due to the emergence of the third-generation. Therefore, this thesis will continuously use the term second-generation sequencing.

Second-generation sequencers are capable of sequencing multiple DNA fragments (up to billions of fragments) in parallel and thus are also referred to using terms such as  \textit{high-throughput sequencing} or \textit{massively parallel sequencing}. The sample preparation step for second-generation sequencing involves the intentional fragmentation of DNA strands into short pieces. The read lengths produced by second-generation sequencers are around 75-500 bases and these reads are referred to as \textit{short-reads}. Second-generation sequencing has enabled sequencing complete genomes at an extremely low cost at a much faster rate when compared to first-generation sequencers. For instance,  the Illumina X Ten sequencer was the first to achieve whole-genome sequencing (WGS) for 1000 USD in less than 3 days \cite {illumina}.

\begin{figure}
    \centering
    \includegraphics[width=\textwidth]{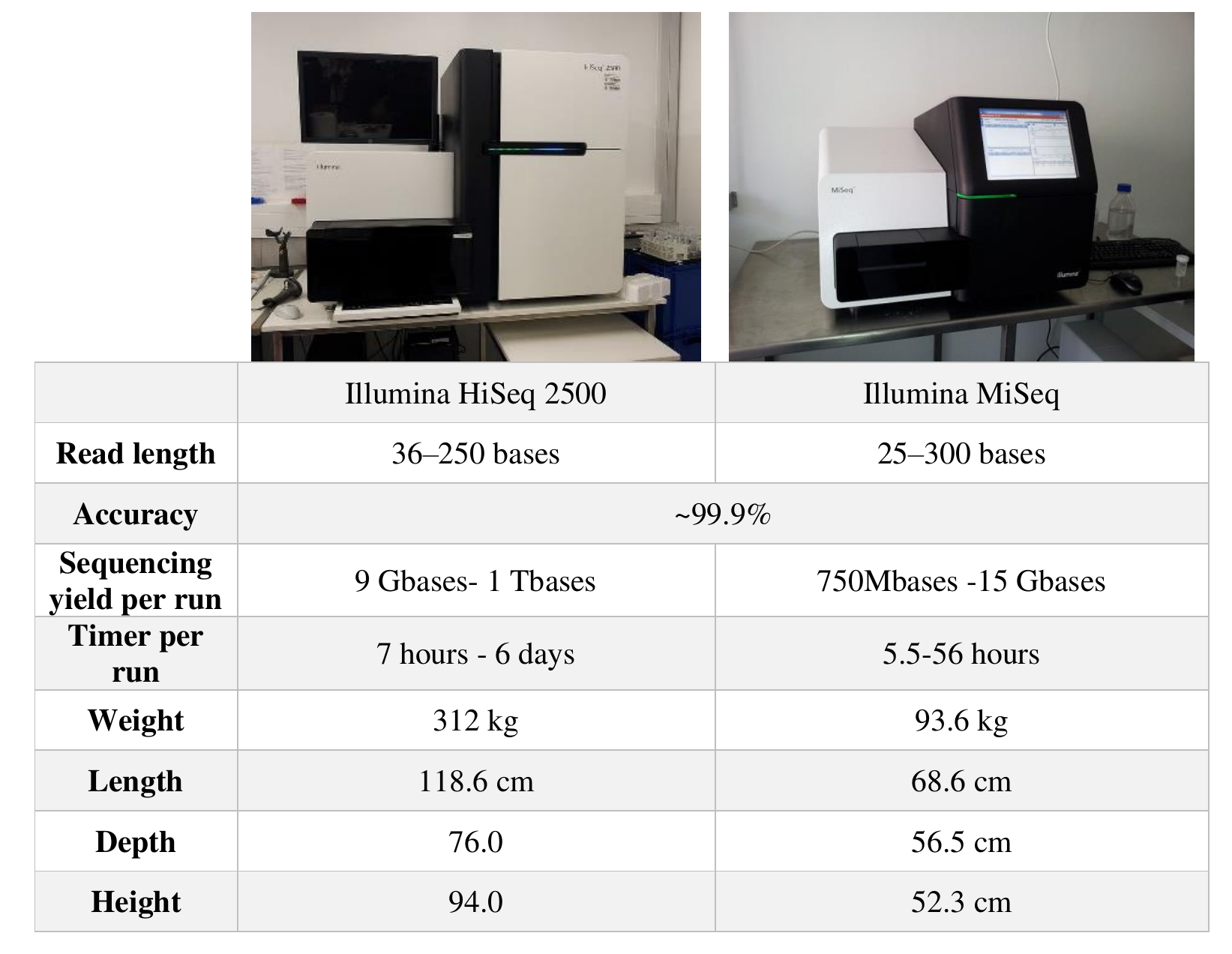}
    \caption[Illumina second-generation sequencers]{Illumina second-generation sequencers. Photograph of Hiseq sequencer is from \url{https://commons.wikimedia.org/wiki/File:Illumina_HiSeq_2500.jpg} and Miseq is from \url{https://en.wikipedia.org/wiki/File:Illumina_MiSeq_sequencer.jpg}, both licensed under CC0 1.0. Note that sequencing yield. Note that values for sequencing yield are to give a rough idea and may change based on a number of factors.  Time of a sequencing run is given as a range since the exact value differs based on the configured value for the read length during sequencing.}
    \label{f:iilumina-sequencers}
\end{figure}

Illumina has become the dominant company in the production of second-generation sequencers. Illumina machines have an error rate of around 0.1\%-1\% per each base sequenced \cite{lou2013high}. Fig. \ref{f:iilumina-sequencers} depicts two different Illumina sequencing machines, HiSeq 2500 used in large-scale sequencing centres and Miseq that is a relatively smaller benchtop device.

Second-generation sequencers are widely used at present. Due to the low cost of sequencing with good accuracy, second-generation sequencers are suitable for SNV and short indel detection. However, the primary limitation of second-generation sequencing is that variants occurring in repeat regions of the genome cannot be easily resolved. This is because reads coming from such repeat regions usually align to multiple locations of the reference genome. Also, structural variants that are longer than the short-read lengths cannot be easily identified using  second-generation sequencing.

Chapter \ref{c:cacheopti} in this thesis is about software used to analyse second-generation sequencing data.

\subsubsection{Third-Generation Sequencing}\label{s:3rdgenseq}

Sequencing approaches that are different from the second-generation sequencing appeared in the late 2000s and eventually led to the third-generation of sequencing technology \cite{hayden2009genome}. Third-generation sequencers produce much longer reads with lower accuracy when compared to second-generation sequencers  \cite{schadt2010window}. Reads produced by third-generation sequencers are known as long-reads. Similar to second-generation sequencers, third-generation sequencers are also capable of sequencing thousands of reads in parallel and thus fall under the category of high-throughput sequencers. Currently, two major companies produce third-generation sequencers. These are: Pacific Biosciences (PacBio); and, Oxford Nanopore Technologies (ONT). Third-generation sequencing technologies are under active development and are not as matured as second-generation sequencers. The read lengths and the accuracy are continually improving with time, and the values given here are to give a rough idea.

\begin{figure}
    \centering
    \includegraphics[width=0.5\textwidth]{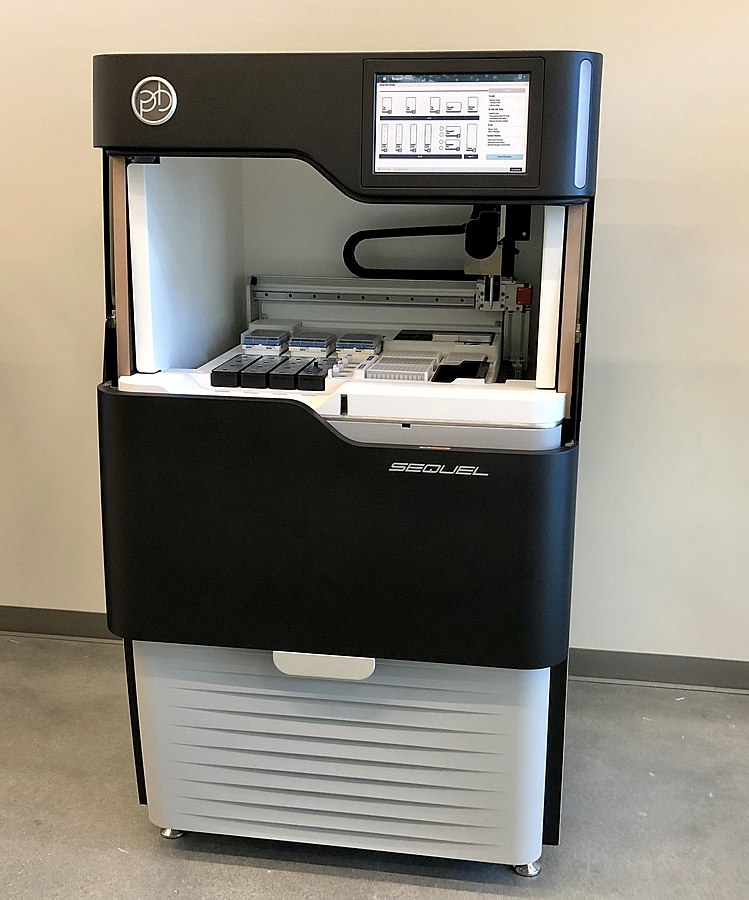}
    \caption[Pacific Biosciences Sequel Sequencer.]{Pacific Biosciences Sequel Sequencer. Photograph from \url{https://en.wikipedia.org/wiki/File:SequelSequencer.jpg} licensed under CC BY-SA 4.0. Dimensions are 92.7 x 86.4 x 167.6 cm}
    \label{f:pacbio}
\end{figure}

PacBio uses a technology known as Single-Molecule Realtime Sequencing (SMRT). Fig. \ref{f:pacbio} depicts one of their sequencers called Sequel. PacBio sequencers can produce reads under two distinct modes. These are Continuous Long-Reads (CLR); and, Circular Consensus Sequencing (CCS) reads. CLR are much longer (up to around 40 kbases) at the expense of lower accuracy ($\sim$87\%), while CCS reads are more accurate ($\sim$99\%), at comparatively shorter lengths (<2.5 kbases) at the time of writing\cite{wei2018npbss}.

Oxford Nanopore Technologies (ONT) uses a technology known as nanopores. Nanopores are nanometre scale biological (protein) pores. Nanopore sequencers measure the ionic current when a DNA strand passes through a nanopore. The produced ionic current is in the range of pico-amperes and this instantaneous current varies based on the nucleotide bases inside the nanopore. Nanopore sequencers have hundreds of such nanopores and thus DNA strands are sequenced in parallel. The measured ionic currents are referred to as \textit{raw signals} and are used during the base-calling process to deduce nucleotide sequences. Thus, nanopore sequencers are capable of directly measuring the actual DNA strand, unlike other sequencing technologies (second-generation Illumina or third-generation PacBio) that perform \textit{sequencing by synthesis}.

The average length of reads produced by nanopore sequencers is typically 10-20 kbases, and the exact value of the length depends on fragmentation during sample preparation and the library preparation protocol. Ultra-long-reads that are longer than 1 Mbases have been recorded.  The accuracy of raw base-called reads of Nanopore sequencers is $\sim$90-95\%  \cite{wick2019performance} and is constantly improving. Nanopore sequencers also output the raw signal data in addition to the base-called reads. This signal data can be used later during the sequence analysis process to reach  a final consensus accuracy of $\sim$99.8\% \cite{jain2018nanopore}. 

\begin{figure}
    \centering
    \includegraphics[width=\textwidth]{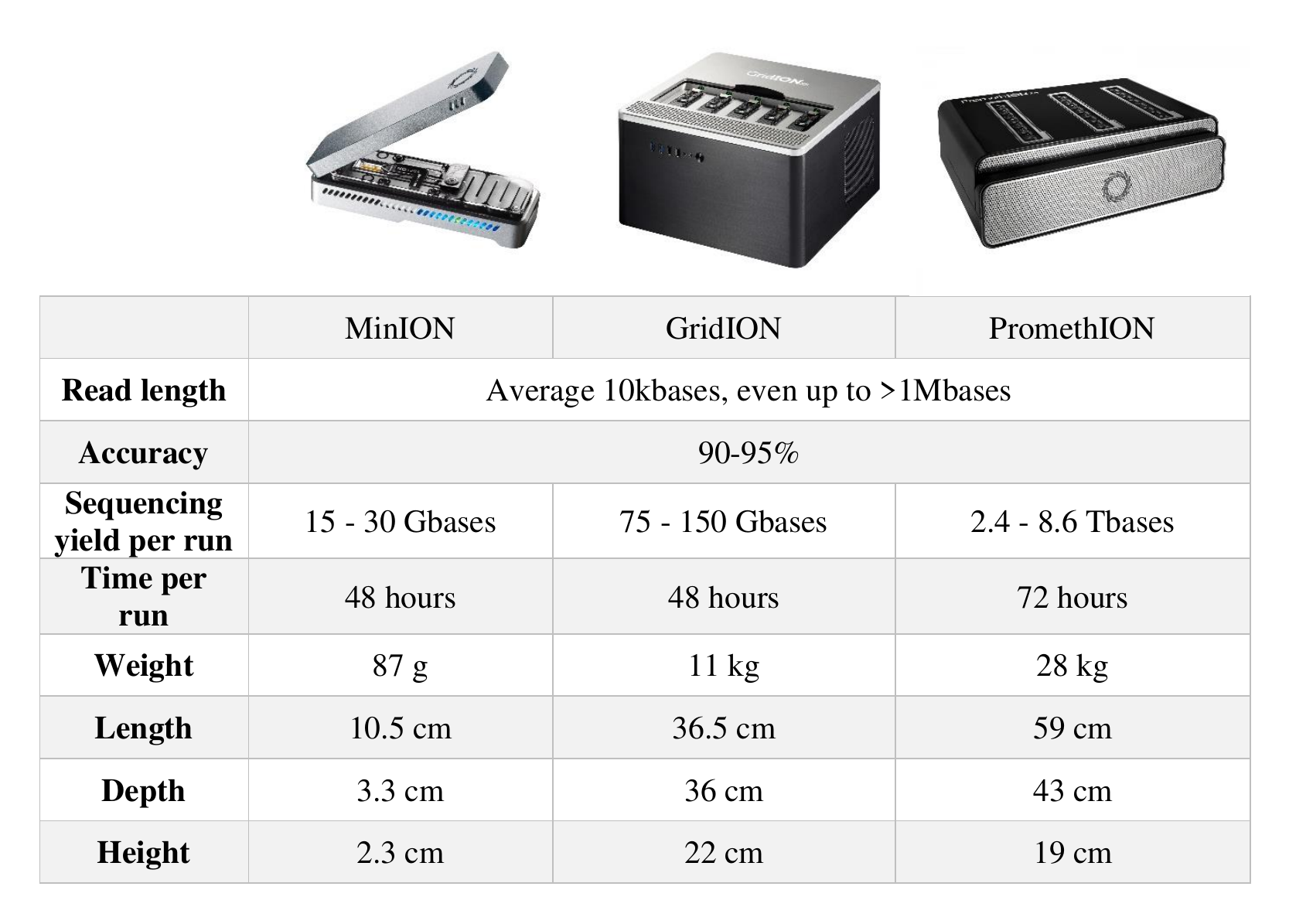}
    \caption[Nanopore third-generation sequencers]{Nanopore third-generation sequencers. The photographs are from Nanopore \url{https://nanoporetech.com/about-us/for-the-media}. The read lengths and the sequencing yields values are for the purpose of giving a rough idea and may change on a variety of different factors.}
    \label{f:nanopore-sequencers}
\end{figure}

Currently, ONT produces three different sequencers as depicted in Fig. \ref{f:nanopore-sequencers}. MinION is the mobile phone-sized ultra-portable sequencer for in-the-field sequencing. GridION is a bench-top sequencer and is equivalent to five MinIONs in sequencing capacity. PromethION is for massive scale sequencing facilities and is capable of sequencing up to about 48 human genomes in parallel. 

\begin{figure}[!ht]
  \centering
\begin{subfigure}[!ht]{0.33\linewidth}
  \centering
    \includegraphics[width=\textwidth]{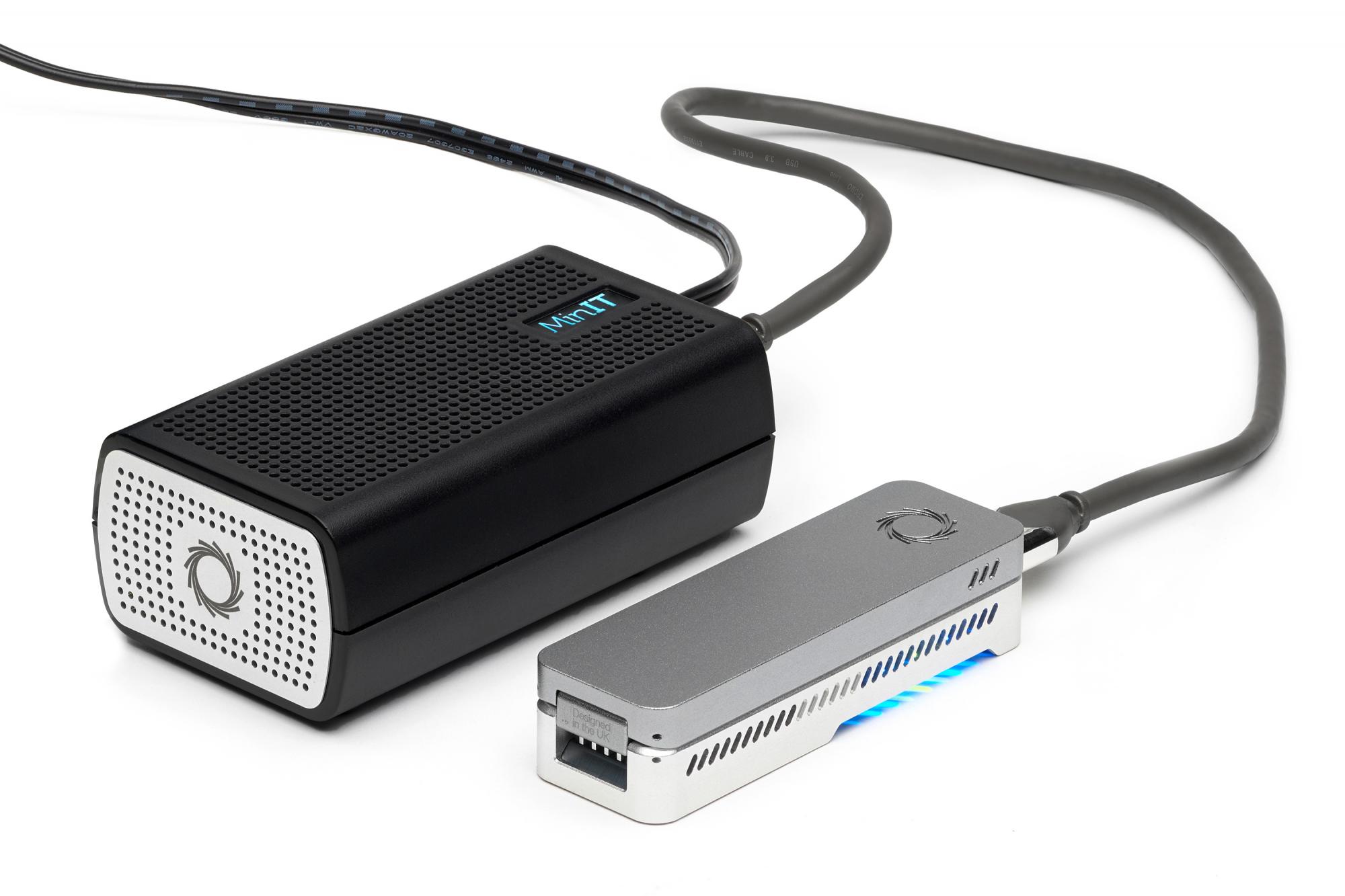}
    \caption{MinION sequencer (right) connected to the MinIT base-calling unit (left)} 
    \label{f:minit}
\end{subfigure}
\begin{subfigure}[!ht]{0.3\linewidth}
  \centering
    \includegraphics[width=\textwidth]{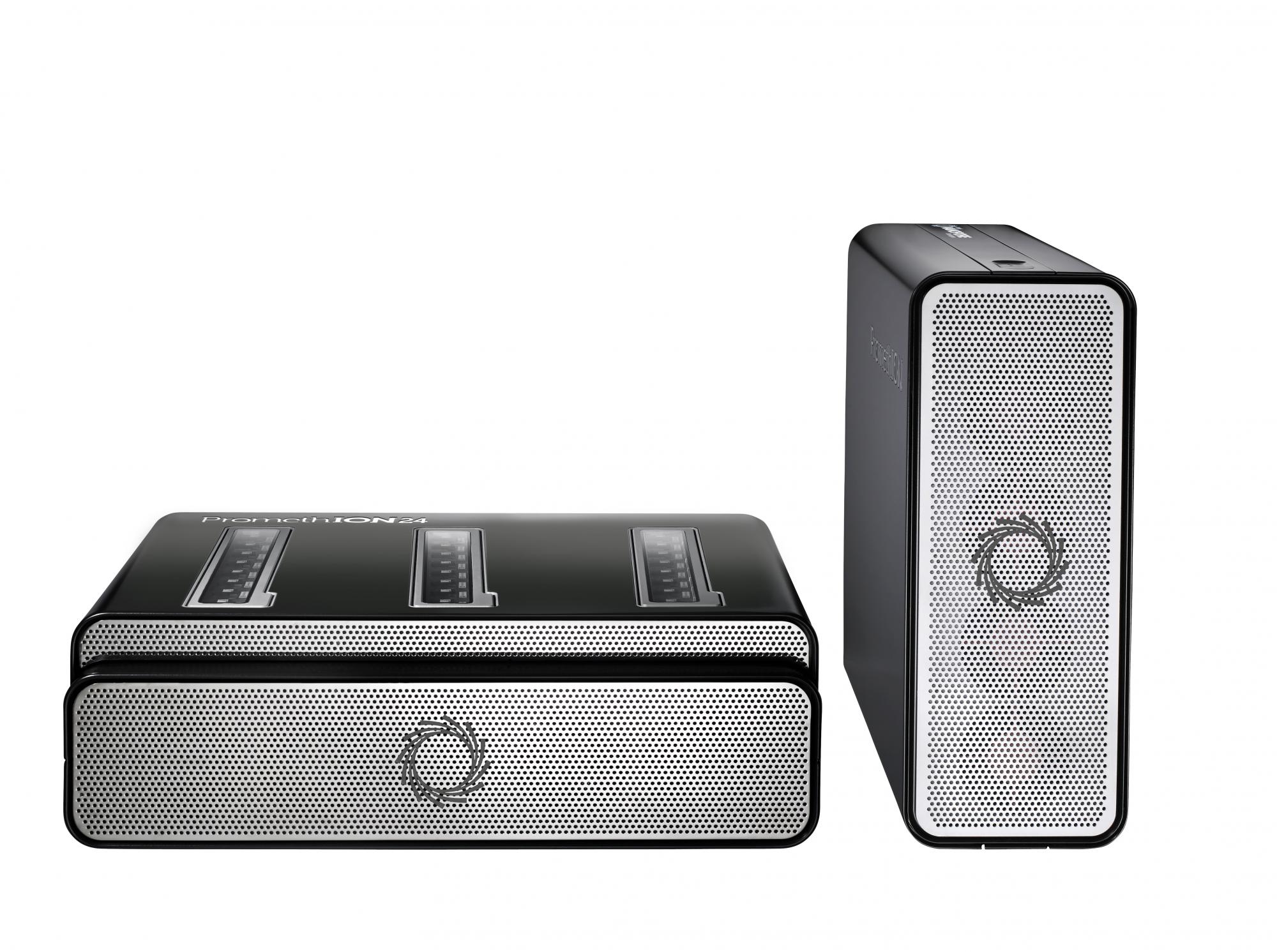}
    \caption{PromethION sequencer (left) and its compute tower (right)} 
    \label{f:prom-compute}
\end{subfigure}
\begin{subfigure}[!ht]{0.3\linewidth}
  \centering
    \includegraphics[width=\textwidth]{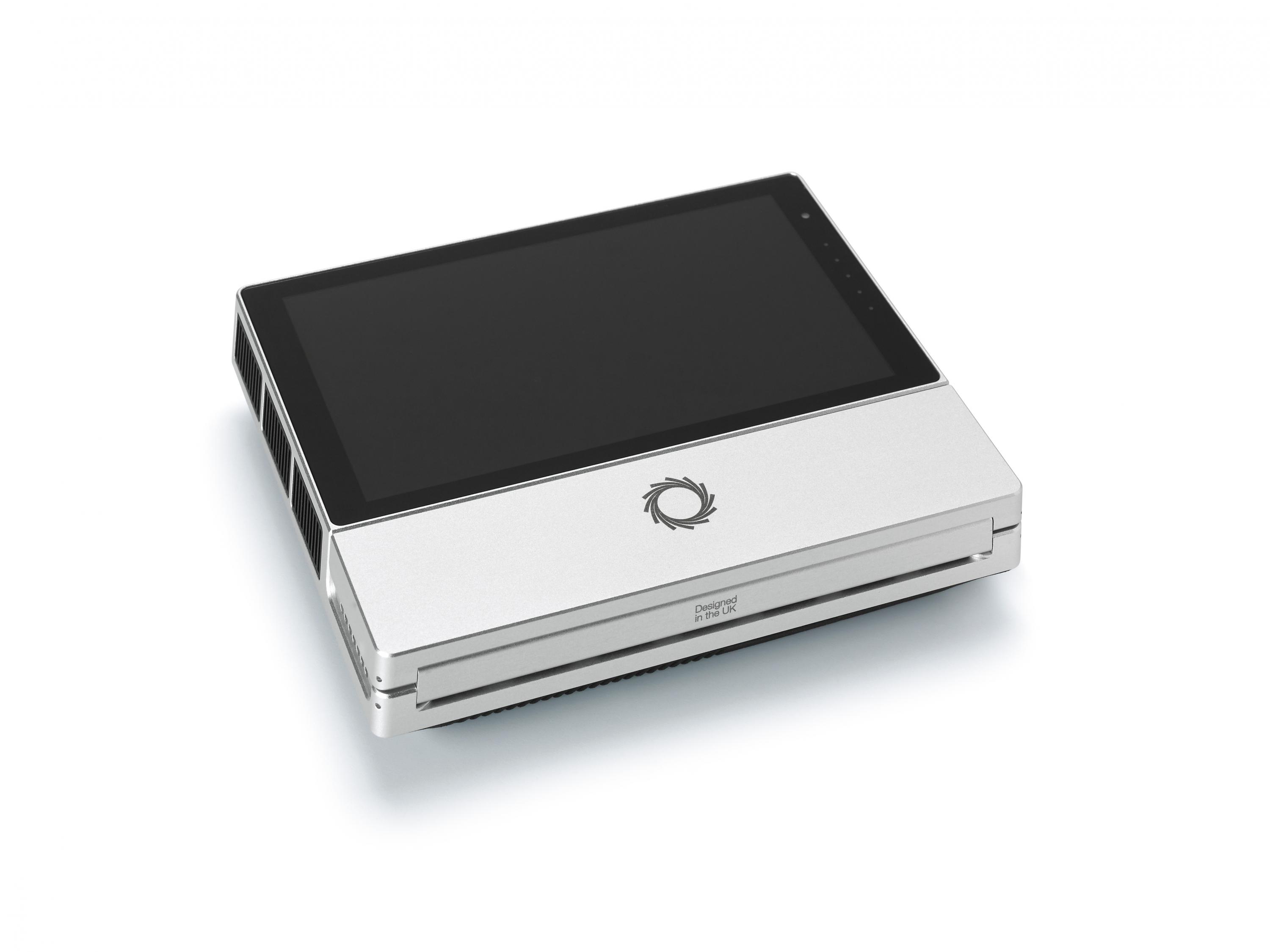}
    \caption{MinION Mk1C sequencer with integrated base-calling unit} 
    \label{f:mk1c}
\end{subfigure}
    \caption[ONT MinIT, PromethION compute tower and MinION Mk1C]{ONT MinIT, PromethION compute tower and MinION Mk1C. The photographs are from Nanopore \url{https://nanoporetech.com/about-us/for-the-media}.} 
    \label{f:ont-misc}
\end{figure}

The MinIONs sequencer in Fig. \ref{f:nanopore-sequencers} does not have a built-in computing unit for base-calling. Thus, the base-calling had to be performed on a workstation or a laptop. ONT recently released (in 2018) an ultra-portable compute module called MinIT that is directly pluggable to the ultra-portable MinION (Fig. \ref{f:minit}) to make the base-calling process ultra-portable. In addition, ONT very recently (in 2019) released the next version of the MinION sequencer called MinION Mk1C that has an integrated base-calling compute module (Fig. \ref{f:mk1c}). GridION (Fig. \ref{f:nanopore-sequencers}) has a built-in computing unit composed of GPUs for base-calling.  The PromethION sequencer comes with a computer tower (high-end workstation) to be used for base-calling (Fig. \ref{f:prom-compute}).

Third-generation sequencers are mainly used by researchers for detecting structural variants and resolving complex and highly repetitive regions that were not possible with the read lengths of previous generation sequencers. For instance, 29 unresolved regions of the X chromosome of the human genome reference were only resolved very recently using third-generation sequencing technology \cite{miga2019telomere}.

Unlike other technologies, Nanopore sequencers can stream data in real-time which facilitates  data analysis on-the-fly (while the sequencer is operating). Also, Nanopore sequencers such as the MinION are ultra-portable, and they are in harmony with the intention of this thesis to construct embedded systems for sequence analysis.

%\todo{Need to add that streaming is not possible in other gen. As per Martin it may be not theoretically impossible. But companies have not exposed such feature.}

\section{Sequence Analysis} \label{s:sequence-analysis}

The goal of sequence analysis is to: assemble the reads into the actual DNA sequence in the sample (or the genome); or, to compare differences in the reads to a reference genome (e.g., to detect variants or epigenetic modifications). The former is performed when a high-quality reference genome is not available and thus the assembly has to be performed from the scratch (known as \textit{de novo assembly}). The latter performed When a high-quality reference genome is available (referred to as \textit{reference-guided sequence analysis}). This thesis focuses on reference-guided sequence analysis. For well-known species like humans, scientists have spent years compiling a high-quality reference sequence. Therefore, for most practical purposes involving humans, reference-guided sequence analysis is adequate.

%This can be achieved through  or \textit{denovo assembly}. , this reference genome can be used as a guide during the assembly process and is known as \textit{reference guided assembly}. Conversely, when no reference genome is available, the assembly is performed from the scratch and is known as \textit{denovo assembly}.

While the reference-guided sequence analysis has some similarities between second-generation and third-generation sequencing, there are important differences. Section \ref{s:workflow2ndgen}, describes the typical reference-guided sequence analysis workflow for second-generation sequencing and section \ref{s:workflow3rdgen} for third-generation sequencing. Despite being not required for the thesis, a brief account of de novo assembly is given in section \ref{s:denovo} for the sake of completeness.

\subsection{Reference-guided second-generation workflow} \label{s:workflow2ndgen}

\begin{figure}[!ht]
\centering
\includegraphics[width=\linewidth]{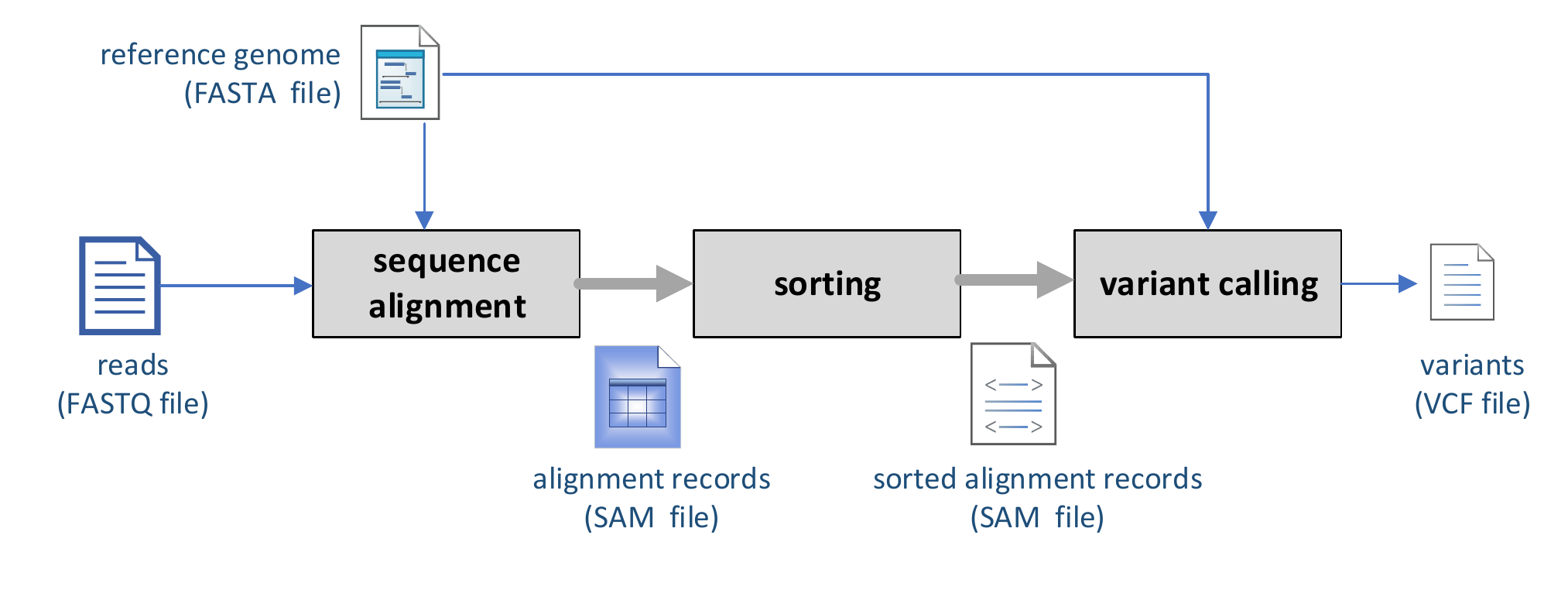}
\caption{Simplified second-generation workflow }
\label{f:2ndgenworkflow}
\end{figure}

\begin{figure}[!ht]
\centering
\includegraphics[width=\linewidth]{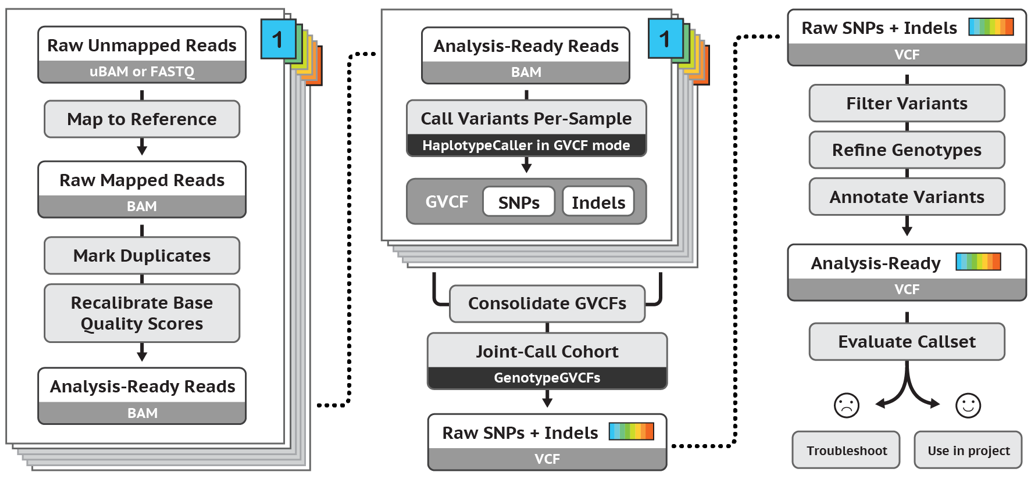}
\caption[GATK Best Practices pipeline.]{GATK Best Practices pipeline. Image from \url{https://gatk.broadinstitute.org}}
\label{f:gatkpipeline}
\end{figure}

%For well known species like humans, scientists have spent years compiling a high quality reference sequence. This reference can be used as a guide when performing computational analysis to put the reads in order. However, the DNA sequences of two individuals slightly differ (around 1\%) and this computational analysis involves approximately aligning reads to the reference. This process of aligning a read to a reference is known as \emph{read alignment} or \emph{read mapping}. The differences in the DNA sequence (assembled using the reads) to the reference are known as \emph{variants}. The computational process of finding these variants is known as \emph{variant calling}. These variants can reveal important information about the individual such as disease predisposition and drug response. However, variant calling is quite challenging as variants should be identified amongst differences in the reads to the reference due to  sequencing errors and alignment artefacts.
A simplified second-generation bioinformatics workflow is given in Fig. \ref{f:2ndgenworkflow}. Certain workflows may contain additional steps such as filtering and calibration (i.e. GATK Best Practices pipeline from Broad institute in Fig. \ref{f:gatkpipeline}). However, the most important and computationally challenging steps are the ones shown in Fig. \ref{f:2ndgenworkflow}.

The reads, typically in \textit{FASTQ} format (discussed previously in section \ref{s:fastq}), are first aligned to the reference genome (step one in Fig. \ref{f:2ndgenworkflow}). This process is known by various terms such as \textit{sequence alignment}, \textit{read alignment} or \textit{read mapping}. Sequence alignment process produces the alignment records for every read (whether the read was successfully mapped, mapping coordinates, mapping quality, etc.), in a file format called sequence alignment/map format (SAM) \cite{li2009sequence}. Tools and associated algorithms for sequence alignment are detailed in section \ref{s:seqaln}.  

The alignment records in the SAM file are then sorted (step two in Fig. \ref{f:2ndgenworkflow}) based on genomic coordinates. That is, sorting first by chromosome order and then by base position in each chromosome. The sorted alignment records are typically stored in a file format called BAM, which is a binary version of SAM format with BGZF compression support \cite{li2009sequence}. BAM allows random accesses to alignment records for a given genomics region through an index called the BAM index. The most popular tool for sorting is \textit{samtools} \cite{li2009sequence} written in C programming language, which is reasonably optimised for performance. Other tools such as \textit{Picard} \cite{picard} written in Java programming language and \textit{Sambamba} \cite{tarasov2015sambamba} written in D programming language  can also be used for sorting.

The next step is the identification of variants amongst sequencing errors and alignment artefacts, and this process is known as \emph{variant calling} (step three in Fig. \ref{f:2ndgenworkflow}). The variant calling step takes the sorted alignment records (BAM file) and the reference genome (\textit{FASTA} file) and outputs the identified variants in VCF file format (discussed previously in section \ref{s:vcf}). These variants can reveal important information about the individual, such as disease predisposition and drug response. However, variant calling is quite challenging as variants should be differentiated correctly from sequencing errors and alignment artefacts. Tools and associated algorithms for variant calling are detailed in section \ref{s:varcall}.

\subsubsection{Sequence Alignment} \label{s:seqaln}

\begin{figure}[ht]
\centering
\includegraphics[width=\linewidth]{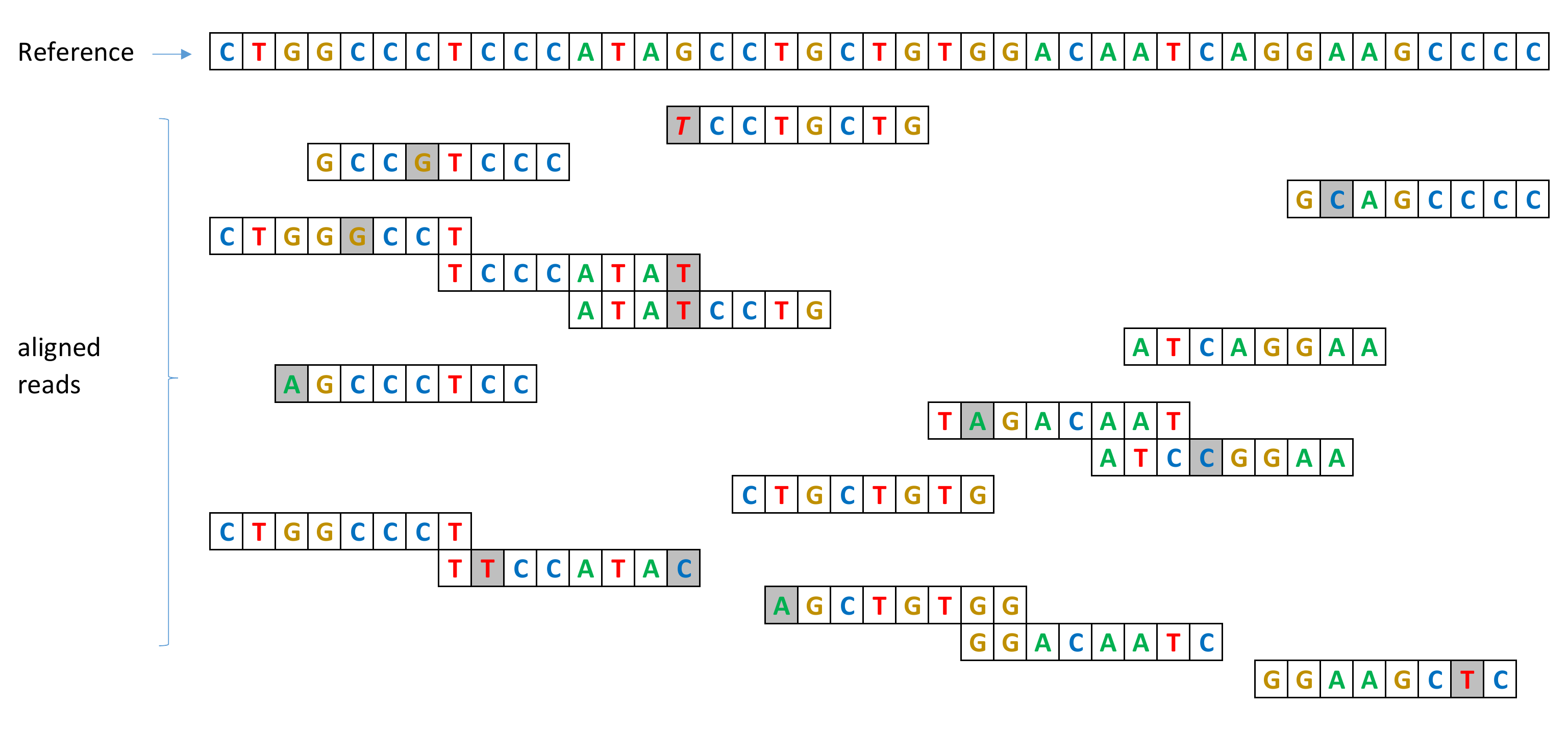}
\caption{Simplified illustration of aligned sequence reads to a reference}
\label{fig:mapping}
\end{figure}

Fig. \ref{fig:mapping} is a simplified elaboration of sequence alignment. Sixteen reads with read lengths of 8 bases have been aligned to the reference. The differences in the reads to the reference (due to sequencing errors or actual variants) have been shaded in grey. Note that, only single base mismatches are in this demonstration, where in reality there can be insertions and deletions. The average number of reads that overlaps a particular nucleotide position is called coverage. Terms such as depth and depth of coverage are also interchangeably used \cite{sims2014sequencing}. The required coverage depends on the application \cite{sims2014sequencing}, for instance, a coverage of 30X or more is recommended \cite{ahn2009first} for detection of SNV and indels.

To date, a large number of sequence alignment tools have been published \cite{flicek2009sense}. Modern sequence alignment tools typically perform the alignment in two steps: first, potential mapping locations of a given read on the reference genome are searched using an index (e.g. hash table); and second, the read is aligned at base-level to those potential locations in the reference using dynamic programming-based alignment algorithms to identify the optimal alignment.

Use of an index is required to reduce the search space in a large genome. Performing base-level alignment of a read on to the whole reference genome is impractical due to computational and memory complexity when the reference genome is large. Locating a few locations on the reference genome (for instance 5-10) using an index is thus vital. The two common indexing approaches use hash tables and the Burrows-Wheeler transform (BWT) \cite{burrows1994block}. 

Earlier short-read alignment tools used the hash table-based approach. The alignment tool \textit{MAQ} \cite{li2008mapping} builds a hash table out of the reads and iterate through the reference sequence to find potential mappings. In contrast, alignment tools such as \textit{SOAP} \cite{li2008soap} and \textit{BFAST} \cite{homer2009bfast} build the hash table using the reference genome and iterate through the reads to find potential mappings. 

Modern short-read alignments tools typically rely on a BWT-based index called an FM-index \cite{ferragina2000opportunistic}.  An FM-index is constructed by taking the BWT of the reference genome, which effectively compresses the data while allowing sub-string indexing at the same time. The FM-index-based approach has gained popularity due to its superiority to hash tables in terms of both performance and memory footprint. Alignment tools such as BOWTIE \cite{langmead2009ultrafast,langmead2012fast}, BWA \cite{li2009fast,li2010fast,li2013aligning} and SOAP2 \cite{li2009soap2} use this approach.

After potential mapping locations are identified quickly using an index, more accurate base-level alignment algorithms are dispatched to find the optimal alignment.
%The dynamic programming algorithm called Needleman-Wunch \cite{needleman1970general} and its successor called Smith-Waterman \cite{smith1981identification} introduced in 1981 are undeniably the heart of all modern base-level alignment algorithms. 
These algorithms to determine the optimal alignment between two biological sequences typically utilise dynamic programming (DP). Very first of such algorithms, the Needleman-Wunsch (NW)  algorithm dates back to the 1970s \cite{needleman1970general}. NW and its variant, the Smith-Waterman (SW) algorithm \cite{smith1981identification}  are of quadratic time and space complexity. Both NW and SW were used extensively to perform fine alignment of DNA sequences with high quality. However, due to its extended time consumption, several heuristic improvements have been proposed by researchers to improve the speed of alignment without losing quality. 

\begin{figure*}[!ht]
  \centering
\begin{subfigure}[t]{0.65\textwidth}
    \centering
    \includegraphics[trim=40 580 320 40,clip,width=\textwidth]{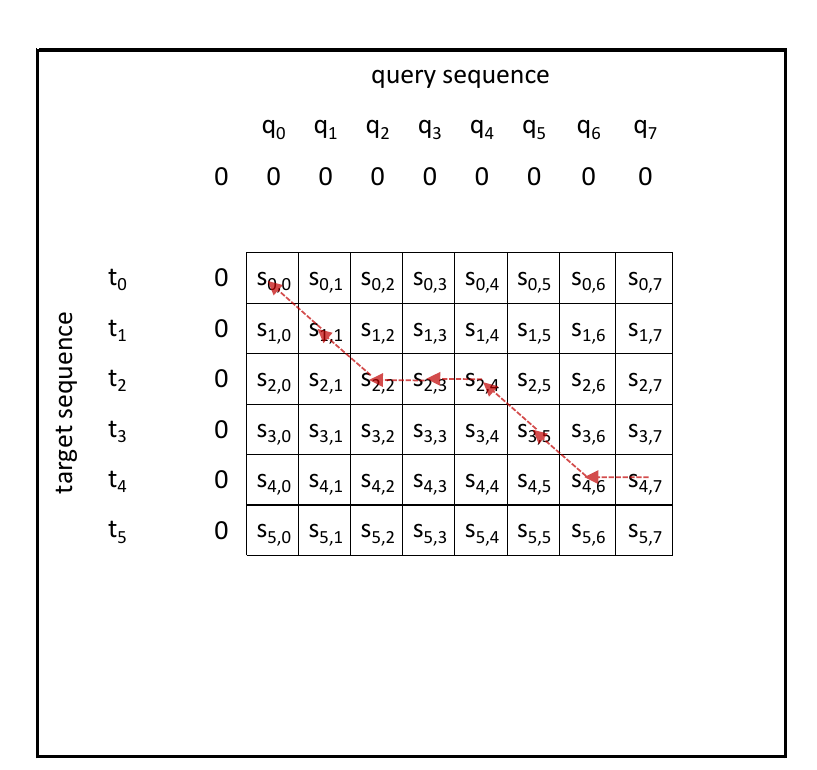}
    \caption{optimal sequence alignment} 
    \label{f:sw}
\end{subfigure}

\begin{subfigure}[t]{0.65\textwidth}
  \centering
    \includegraphics[trim=40 580 320 40,clip,width=\textwidth]{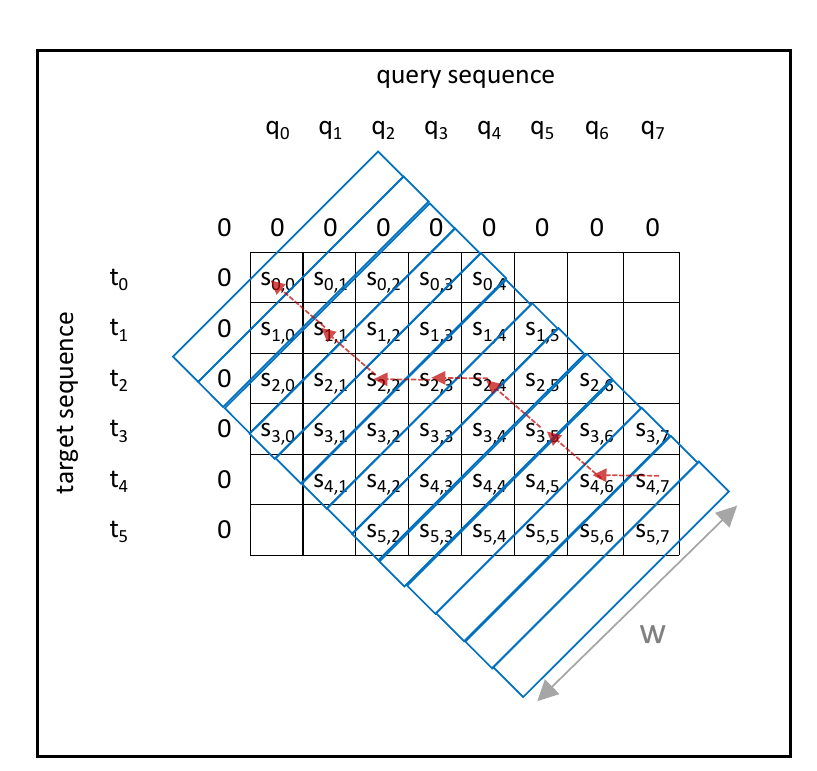}
    \caption{Banded sequence alignment (band-width=4)} 
    \label{f:bandedswbw4}
\end{subfigure}
\caption{Dynamic programming based sequence alignment}
\end{figure*}

Fig. \ref{f:sw} exemplifies an original SW based alignment (no heuristic) between two sequences, \textit{target sequence t\textsubscript{0}t\textsubscript{1}t\textsubscript{2}t\textsubscript{3}t\textsubscript{4}t\textsubscript{5} (6 bases long), and \textit{query sequence q\textsubscript{0}q\textsubscript{1}q\textsubscript{2}q\textsubscript{3}q\textsubscript{4}q\textsubscript{5}q\textsubscript{6}q\textsubscript{7}}} (8 bases long). The DP table (scoring matrix) contains  6x8 cells as shown. First, the initial values are set (shown as 0 in the figure); second, the score for each cell (s\textsubscript{x,y}) is computed based on a scoring scheme; and third, the trace-back (backtracking denoted by red arrows on the figure) starting from the highest-scoring cell and ending at a cell with 0 score, outputs the optimal alignment that yields the highest score (please refer \cite{durbin1998biological} for a detailed explanation of SW).

In the case of short-read alignment, the sequences to be aligned are small (typically 75-500 bases). Two sequences (each sequence \textasciitilde100 bases long) can be aligned by filling \textasciitilde10\textsuperscript{4} cells. While a single such alignment can be quickly handled by a modern computer, it is computationally demanding when the number of alignments to be performed scales up to hundreds of millions and billions, which is the case for short-reads. To reduce the number of computations, banded alignment approaches were introduced \cite{chao1992aligning}, where only the cells in the DP table along the left diagonal band are computed as shown in Fig. \ref{f:bandedswbw4}. The underlying assumption is that the sequences that are aligned to each other are essentially similar, thus the alignment (the trace-back arrows) should lie close to the left diagonal. Note that in the figure, only the cells in a band of width (W) four have been computed. This computation is sufficient since the band contains the alignment. 

X-drop in BLAST (Basic Local Alignment Search Tool) \cite{altschul1997gapped} is another notable heuristic to SW that terminates the computation when the drop in the alignment score reaches a threshold. An extended version of X-drop called Z-drop is used in the modern alignment tool BWA MEM \cite{li2013aligning}.

In addition to computing the alignment and the alignment score for each read, modern alignment tools also compute an important quantity called the Mapping Quality (MAPQ). The concept of mapping quality was introduced in MAQ aligner \cite{li2008mapping}. MAPQ is computed per read as: $−10 log_{10}(P)$ rounded off to the nearest integer, where $P$ is the probability of the mapping position being incorrect. This probability value is heuristically determined through different formulas in different software but essentially considers both the alignment score and the number of sub-optimal mappings of the read. A higher number of sub-optimal mappings means that the read is likely to be from a repeat sequence and thus the chance of being incorrect is high. MAPQ is an important score for the variant calling step, i.e., to avoid false-positive variants.

\subsubsection{Variant Calling}\label{s:varcall}

\begin{figure}[ht]
\centering
\includegraphics[width=\linewidth]{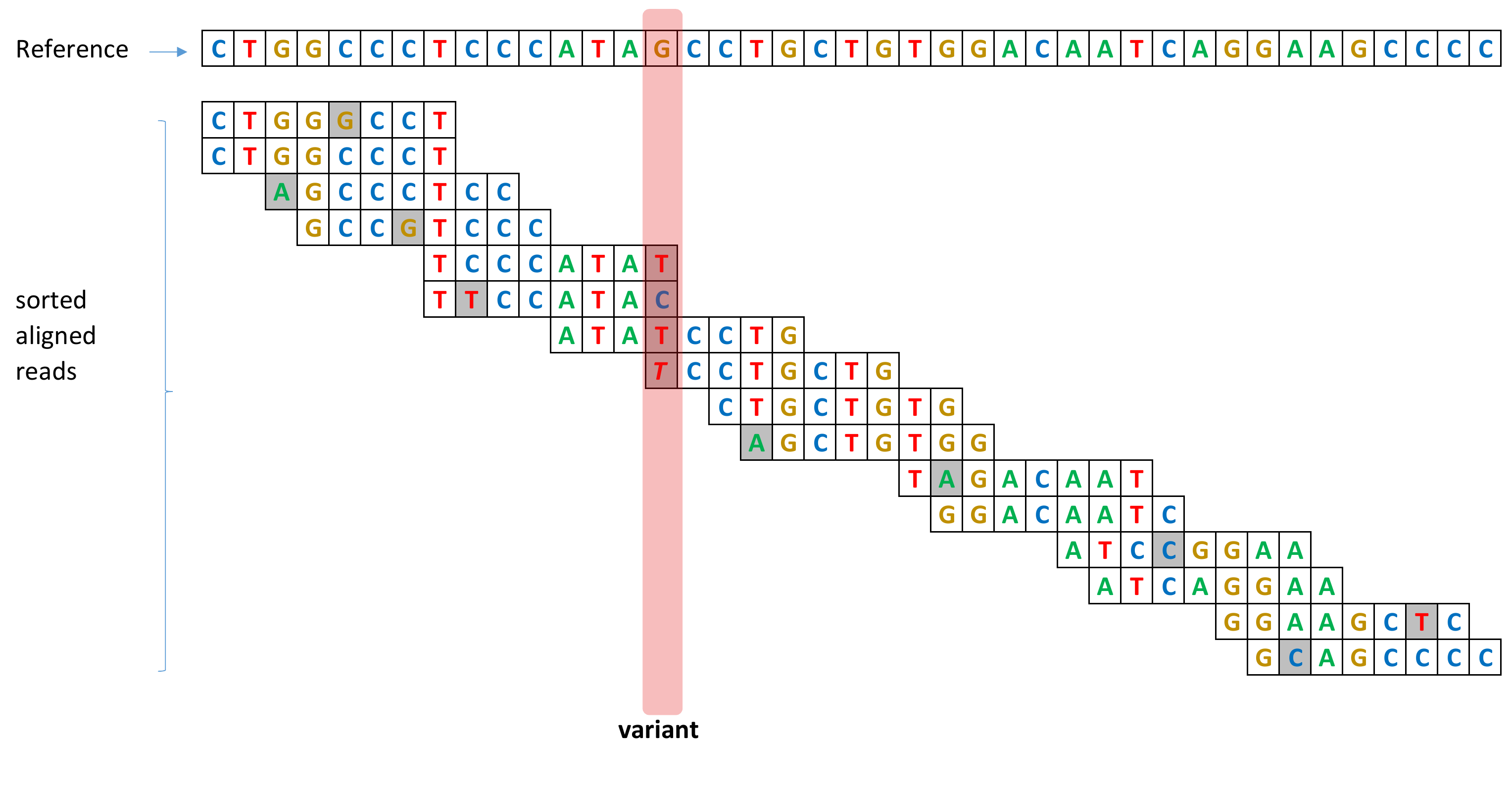}
\caption{Simplified elaboration of variant calling}
\label{f:varcall}
\end{figure}

Variant calling is the process of identifying the variants amongst sequencing errors and alignment artefacts. One of the simplest possible examples illustrating the variant calling process in Fig. \ref{f:varcall}, which is based on the same reads and the reference used in the previous example (Fig. \ref{fig:mapping}). Note that in Fig. \ref{f:varcall}, the reads have been sorted based on genomic coordinates and the marked variant is simply based on the majority vote. However, such a simple strategy will not be adequate for accurately identifying variants in real genomic data (to minimise both false positives and false negatives) and numerous sophisticated variant calling software tools have been introduced.

More than 40 open source tools have been released in the last decade \cite{sandmann2017evaluating}. Most tools utilise a probabilistic framework (Bayesian approach is the most common) and popular variant calling tools such as GATK UnifiedGenotyper \cite{depristo2011framework}, GATK HaplotypeCaller \cite{poplin2018scaling} (the  of UnifiedGenotyper), FreeBayes \cite{garrison2012haplotype}, SAMtools package (\textit{samtools} and \textit{bcftools}) \cite{li2011statistical}  and Platypus \cite{Rimmer2014} are some examples.  
In contrast to the probabilistic methods, certain tools such as VarScan rely on heuristic approaches \cite{wei2011snver,koboldt2012varscan}. 

Past variant callers (e.g, GATK UnifiedGenotyper) solely relied on the read alignment performed by the aligning tool. However, alignment artefacts due to indels were found to affect the accuracy of the variants calling results \cite{alkan2011genome}. Thus, separate pre-processing tools such as GATK IndelRealigner were introduced to perform local re-alignment in the affected regions \cite{homer2010improved} before executing the variant caller. Modern variant calling tools such as GATK HaplotypeCaller and Platypus have a built-in local de novo assembly step to address the aforementioned issue, making GATK IndelRealigner redundant. In local de novo assembly, the genome is broken into small regions and de novo  assembly is performed separately in these regions. For local de novo assembly, Platypus uses a variant of Bruijn graphs called coloured de Bruijn graphs \cite{Iqbal2012}, while GATK HaplotypeCaller also uses a de Bruijn like graph \cite{haplotypeCallerAlgo}. 

In the past variant callers (e.g, GATK UnifiedGenotyper), each base position on the genome was considered independently when calculating probabilities. However, recent variant callers such as GATK HaplotypeCaller and Platypus breaks the genome into overlapping haplotypes\footnote{A haplotypes is a group of variants that tend to occur together} based on initially identified variations. They perform probability calculation on these haplotypes by mapping reads to each haplotype. GATK HaplotypeCaller uses pair Hidden Markov Model (pairHMM) \cite{pachter2002applications} and Platypus uses Needleman-Wunch for mapping reads to haplotypes. Haplotype-based approaches have increased the accuracy of variant calls \cite{10002010map}. 

Sandmann et al. \cite{sandmann2017evaluating} evaluated the accuracy of eight variant calling tools including GATK, Platypus and SAMtools. None of the variant callers could detect all the variants in their data sets. They also observed that increased sensitivity decreases precision. Further, the accuracy of different tools varied with different data sets. Hence, modern variant calling tools are being frequently updated to gradually improve accuracy.

Variant calling is a time-consuming step that takes hours on a high-performance computer. Despite this, many variant callers  such as VarScan, FreeBayes, SNVer \cite{wei2011snver} and VarDict \cite{lai2016vardict} do not support multi-threading. GATK HaplotypeCaller does support multi-threading. However, multi-threaded executions of GATK HapplotypeCaller frequently crash and thus are not recommended to be used as stated in the manual \cite{HaplotypeCaller}. Even during instances that do not crash, the multi-threaded execution of GATK HaplotypeCaller marginally improves the run-time due to inefficient multi-core utilisation. Further, multi-threaded execution could not reproduce the same result as single-threaded execution as observed by Sandmann et al. \cite{sandmann2017evaluating}.  Platypus variant caller is capable of efficiently utilising multi-CPU cores through its in-built multi-processing.

% are amongst the few variant callers that support multi threading, however, 
% Meanwhile, there .
% Hence, multi threaded implementations require improvements for increasing the reproducibility. Tools such as GATK are written using Java. Though Java is application independent, bioinformatics applications written in Java are slower and less memory efficient when compared to those written in C/C++ \cite{fourment2008comparison}. 

%Further, preliminary tests done by us revealed that the implementations done in tools such as Platypus are very naive and a considerable number of optimisations for performance and memory efficiency are possible.

Chapter \ref{c:cacheopti} describes  memory optimisation algorithms associated with variant calling. Specific details of the underlying algorithms are discussed in the background of that chapter.

\subsubsection{Characteristics of data}

\textbf{Read length}: In a second-generation sequencing dataset, the lengths of all the reads in the dataset are typically the same (at least for Illumina Sequencing that dominates the second-generation sequencing market). The read length can be initially configured to a particular value between around 50 and 500 bases at the start of a sequencing run (depending on the sequencing machine) and all the reads generated from that sequencing run would of that configured length.

\begin{figure}[ht]
\centering
\includegraphics[width=\linewidth]{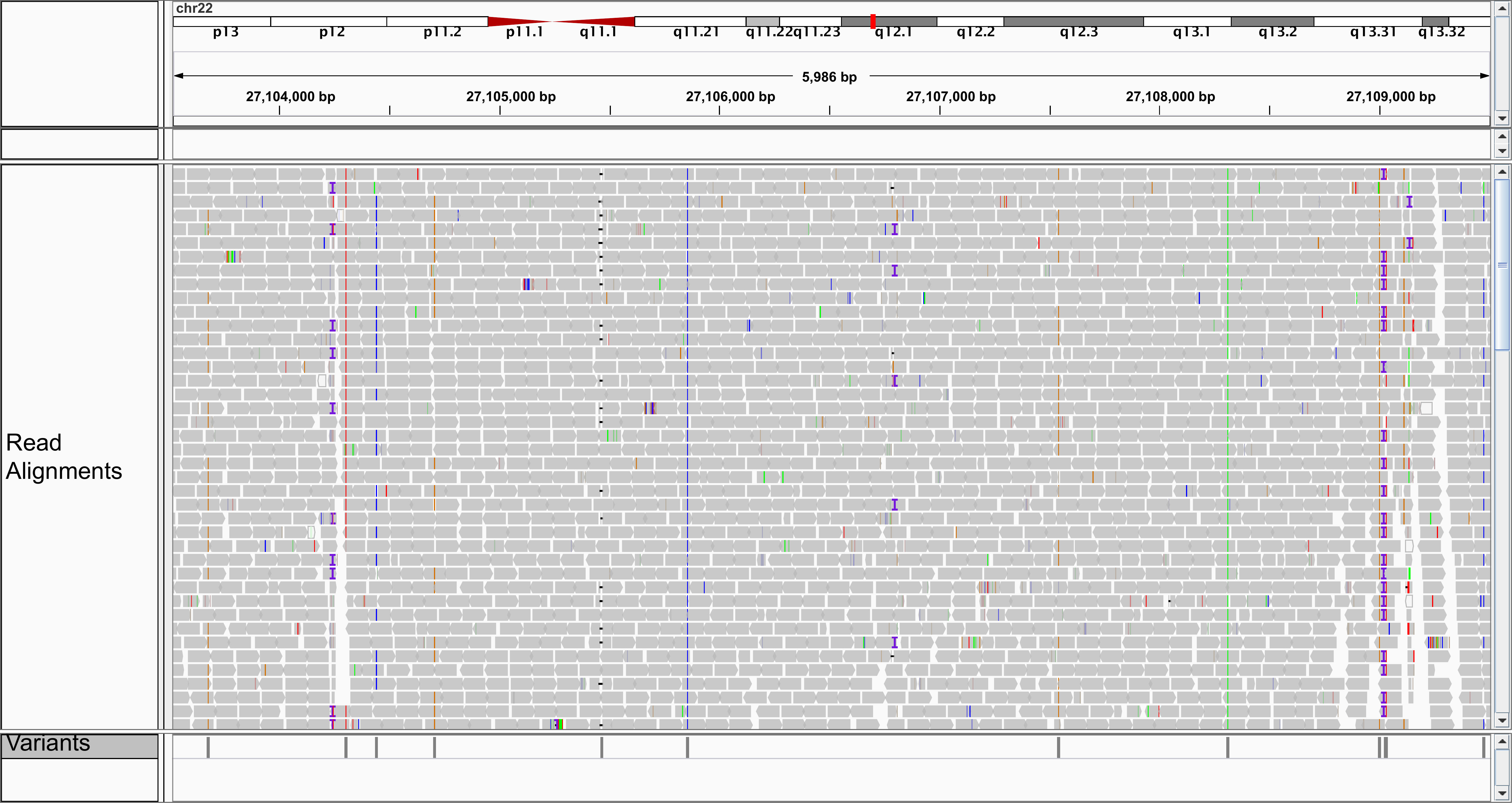}
\caption[A screenshot from IGV from an NA12878 dataset]{A screenshot from IGV for the region chr22:27,103,514-27,109,534 from an NA12878 dataset aligned to the human genome}
\label{f:illumina-errors}
\end{figure}

\textbf{Error rate:} An example demonstrating the error rate of second-generation sequencing is in Fig. \ref{f:illumina-errors}. This example uses Illumina short-reads from a real dataset (NA12878 dataset from 1000 genomes project)\footnote{NA12878 is a well-studied human genome sample from a particular Utah woman} aligned to a reference (human genome). Fig. \ref{f:illumina-errors} is a screenshot of a $\sim$6 kbase region in chromosome 22 taken through the Interactive Genome Viewer (IGV). The grey colour horizontal blocks on Fig. \ref{f:illumina-errors} represent the reads and other colours represent differences in those reads to the reference. The bottom panel shows the variants that are present in this region.

\textbf{Data size:} The human genome is 3.1 Gbases and the \textit{FASTA} file (uncompressed) is around 3.1 GB. If the human genome is sequenced at an average coverage of 30X, the yield is around 96 Gbases. If the read length is assumed to be 100, the dataset would contain around 960 million reads. A \textit{FASTQ} file (uncompressed) storing such a dataset is around 200-250 GB. The generated result from the alignment step stored in a SAM file (uncompressed)  is around 250-300 GB. The sorted alignments stored as a BAM file (BGZF compressed) is around 30-40 GB. The VCF file generated from the variant calling step is around 1 GB.

%-rw-rw-r--  1 hasindu hasindu 872M Apr 24 20:05 output_platy.vcf
% -rw-rw-r--  1 hasindu hasindu  36G Apr 24 19:06 tmp.bam
% -rw-rw-r--  1 hasindu hasindu 7.2M Apr 24 19:26 tmp.bam.bai
% -rw-rw-r--  1 hasindu hasindu 214G Apr 24 12:44 tmp.fastq
% -rw-rw-r--  1 hasindu hasindu 258G Apr 24 16:40 tmp.sam

\subsection{Reference-guided third-generation workflow for nanopore data} \label{s:workflow3rdgen}
 %The downside is the massive size of data. At current sampling rate of 4096 and the DNA movement of 450 bases per second, one base is XX samples. Each sample consumes around 2bytes and the output of XX Gbases of a Minion leads to YY and Promethion ZZ.

\begin{figure}[!ht]
\centering
\includegraphics[width=\linewidth]{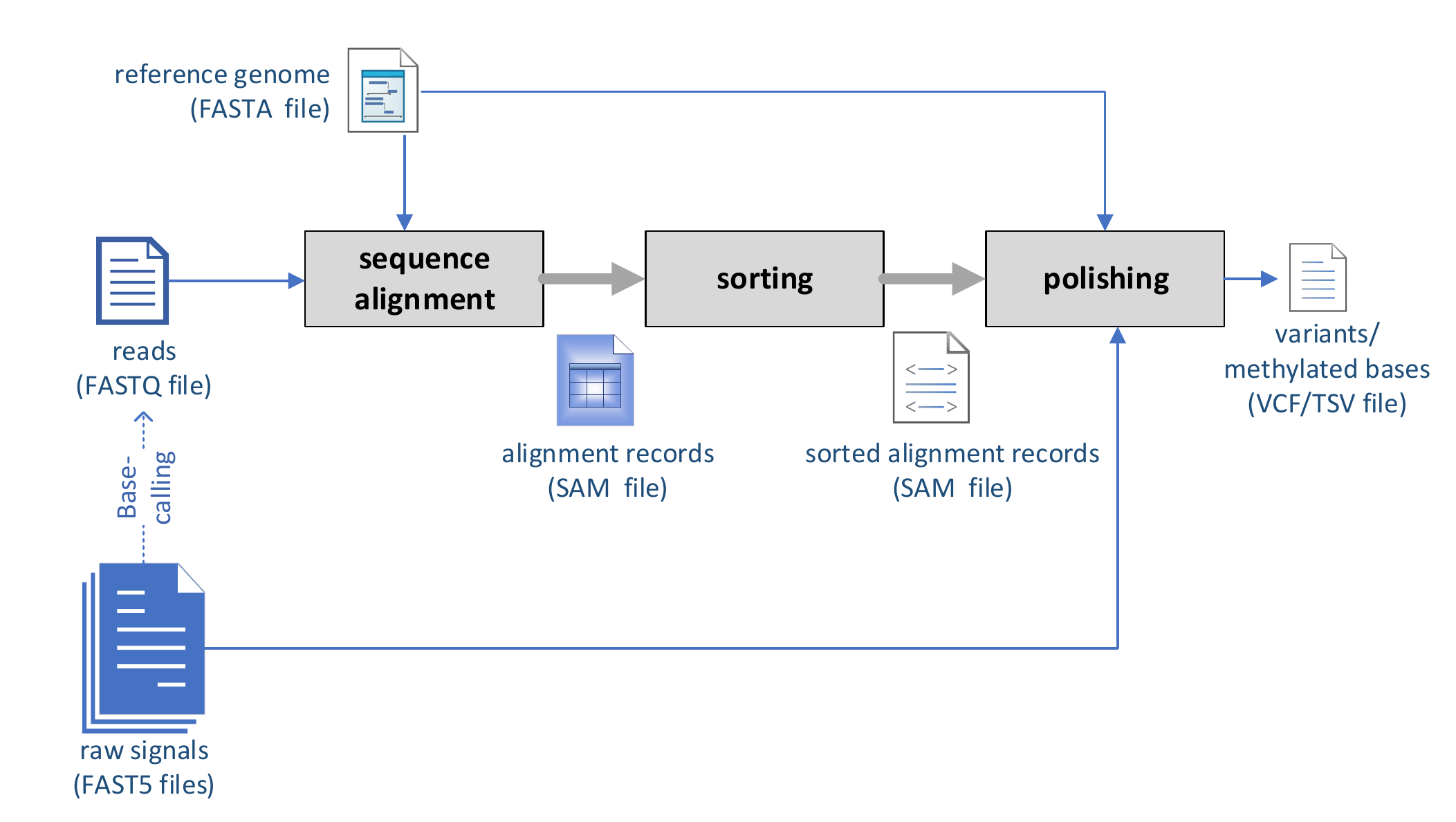}
\caption{Simplified third-generation nanopore workflow}
\label{f:3rdgenworkflow}
\end{figure}

Third-generation sequencing technology is currently under active development and no standard or best practises workflow exists at present (as opposed to the second-generation). Third-generation sequencing workflows are not stable and are constantly evolving. Fig. \ref{f:3rdgenworkflow} shows the typical workflow for nanopore data processing at the time of writing.

The reads (in \textit{FASTQ} file format) are first aligned to the reference genome (step one in Fig. \ref{f:3rdgenworkflow}). The alignment is conceptually similar to that of the second-generation workflow. However, software tools used for aligning third-generation sequencing have distinct characteristics which are different from the previous aligners and are detailed in section \ref{s:3rdgen-aln}. After the alignment step, the aligned reads are sorted (step two in Fig. \ref{f:3rdgenworkflow}). The sorting step is identical to that of the second-generation workflow and the most popular sorting tool remains \textit{Samtools}. The next step (step three labelled as polishing in Fig. \ref{f:3rdgenworkflow}) now can be either variant calling or detection of epigenetic base modifications (e.g., methylation calling). Variant calling or detection of epigenetic base modifications is a challenging process where true variants and/or base modifications must be identified amongst highly erroneous reads (currently 5\%-10\%). Thus, this step typically uses the raw signals (raw sensor output from the sequencer) in addition to the base-called reads. Associated software tools for  variant calling and detection of epigenetic base modifications are detailed in section \ref{s:3rdgen-polish}. 

As stated in section \ref{s:3rdgenseq}, a raw signal is the ionic current measurement when a DNA strand passes through a protein nanopore. Nanopore sequencers output these raw signals in a file format called \textit{fast5}. \textit{Fast5} format is essentially the Hierarchical Data Format 5 (HDF5) \cite{hdf5}, with a specific scheme determined by ONT to store raw signal data and metadata. Before 2018, a single raw signal (corresponds to a single read) was stored as a single fast5 file, which is currently referred to as a \textit{single-fast5 file}. However,  millions of files generated from a sequencing run were difficult to manage and now a fast5 file contains a batch of raw signals (by default 4000 reads). Such fast5 files containing multiple reads are called \textit{multi-fast5 files}. HDF5 is a versatile file format with numerous features (including compression). However, HDF5 is a very complicated file format of a monolithic design and a lengthy specification. Consequently, HDF5 files must be accessed through the only existing library provided by the HDF5 group, which has limitations such as lack of efficient multi-threading access. Chapter \ref{c:ioopti} of this thesis explores the impact of this limitation on efficient raw signal access and presents alternate solutions to circumvent the limitation.

%The base-calling step (conversion of raw sensor data to bases) in second generation sequencing was a step exclusively performed inside the sequencer and does not belong to sequence analysis. 

As nanopore sequencers output raw signals, base-calling can be optionally performed externally on a general-purpose computer. However, the latest nanopore sequencers either come with an internal compute-module or support an externally attachable dedicated base-calling module running ONT's proprietary base-callers. Thus, base-calling will not be considered under sequence analysis in this thesis.

\subsubsection{Sequence Alignment}\label{s:3rdgen-aln}
 
When examined from a higher level, long-read aligners also use a two-step approach similar to previous aligners: finding potential mapping locations using an index; and, applying accurate dynamic programming algorithms to obtain the optimal alignment. However, when looked microscopically, long-reads aligners have major differences in underlying algorithms and parameters to handle distinct characteristics of long-reads. Numerous long-read aligners have been published over the last decade, for instance,  BWA MEM \cite{li2013aligning}, BLASR \cite{chaisson2012mapping}, GraphMap \cite{sovic2016fast}, Kart \cite{lin2017kart}, NGMLR \cite{Sedlazeck2018}, LAMSA \cite{liu2017lamsa} and Minimap2\cite{minimap2}. Note that BWA MEM is an extended version of previously discussed BWA (BWA was initially designed for short-reads), that supports long-reads up to a certain degree. Minimap2\cite{minimap2} is the most popular long-read aligner amongst all the other long-read aligners, due to its superior performance, accuracy and robustness. Minimap2 \cite{minimap2} employs a hash table-based genome index to quickly locate potential mappings and is both fast and accurate compared to the FM-index-based approach in BWA MEM \cite{li2013aligning}. However, the RAM requirement is higher for a hash table-based index compared to FM-index. For instance, The hash table data structure itself consumes about 8 GB of memory (RAM) in \textit{Minimap2}. The typically RAM consumption of \textit{Minimap2} is around 12 GB on average when memory is allocated for internal data structures (i.e. dynamic programming tables). However, the peak RAM for the human genome can occasionally exceed 16 GB depending on the characteristics of data, such as the length of the reads. Chapter \ref{c:minimap} of this thesis focuses on memory optimisations to \textit{Minimap2} to reduce peak RAM.

Banded versions of dynamic programming algorithms such as SW and NW (fig) used for short-reads are not directly suitable for long-reads. In contrast to short-reads, the long-reads which emanate from Nanopore, PacBio etc, have lengths which are 10 to 10000 orders of magnitude bigger than short-reads, are noisier (with a greater number of errors) and are typically not suitable for such small static bands. The 10\% base-calling error rate would result in the alignment significantly deviating from the diagonal (diagonal mentioned in section \ref{s:seqaln}). A major advantage of long-reads is the detection of long indels (insertions and deletions occasionally spanning lengths longer than short-reads themselves). When aligning such reads, the alignment path deviates significantly from the diagonal. The high errors and the large indels require the bands to be of large width if they are to be static.  High bandwidth requirement causes processing times to be extremely high when aligning millions of reads. To improve the speed of this processing,  Suzuki-Kasahara (SK) heuristic algorithm \cite{suzuki2018introducing} was introduced in 2017. SK utilises an adaptive band scheme, letting a smaller band to contain such an alignment within the band, which is exemplified as below.

\begin{figure*}[!ht]
  \centering
% \begin{subfigure}[t]{.475\textwidth}
%     \centering
%     \includegraphics[trim=40 580 320 40,clip,width=\textwidth]{img/sw.pdf}
%     \caption{optimal sequence alignment} 
%     \label{f:sw}
% \end{subfigure}\hfill
% \begin{subfigure}[t]{.475\textwidth}
%   \centering
%     \includegraphics[trim=40 580 320 40,clip,width=\textwidth]{img/bandedsw.pdf}
%     \caption{Banded sequence alignment (band-width=4)} 
%     \label{f:bandedswbw4}
% \end{subfigure}\vspace{0.7cm}

\begin{subfigure}[t]{.65\textwidth}
  \centering
    \includegraphics[trim=40 580 320 40,clip,width=\textwidth]{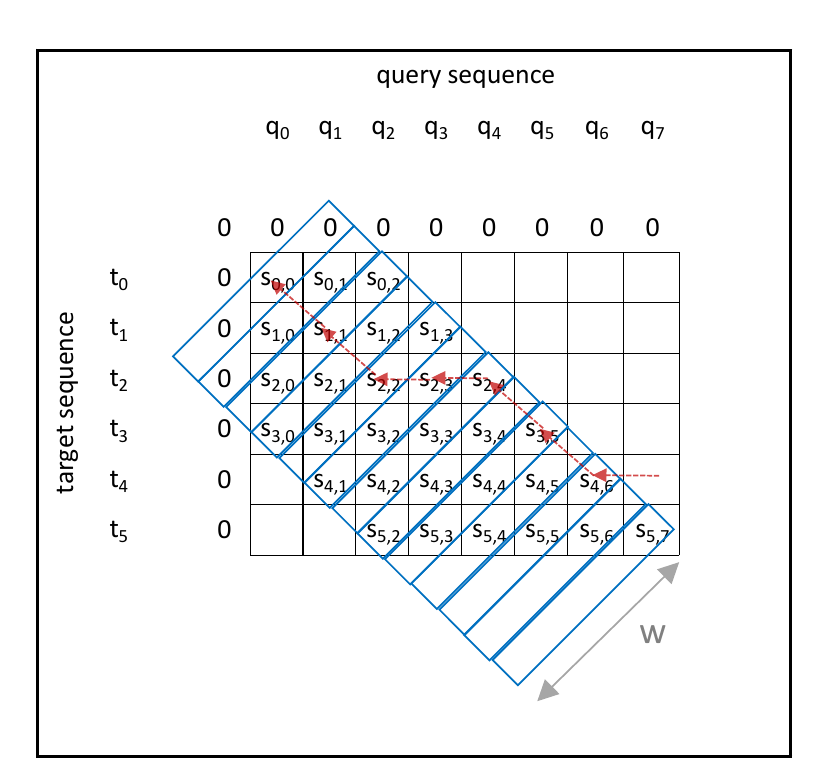}
    \caption{Banded sequence alignment (band-width=3)} 
    \label{f:bandedsw}
\end{subfigure}

\begin{subfigure}[t]{.65\textwidth}
  \centering
    \includegraphics[trim=40 580 320 40,clip,width=\textwidth]{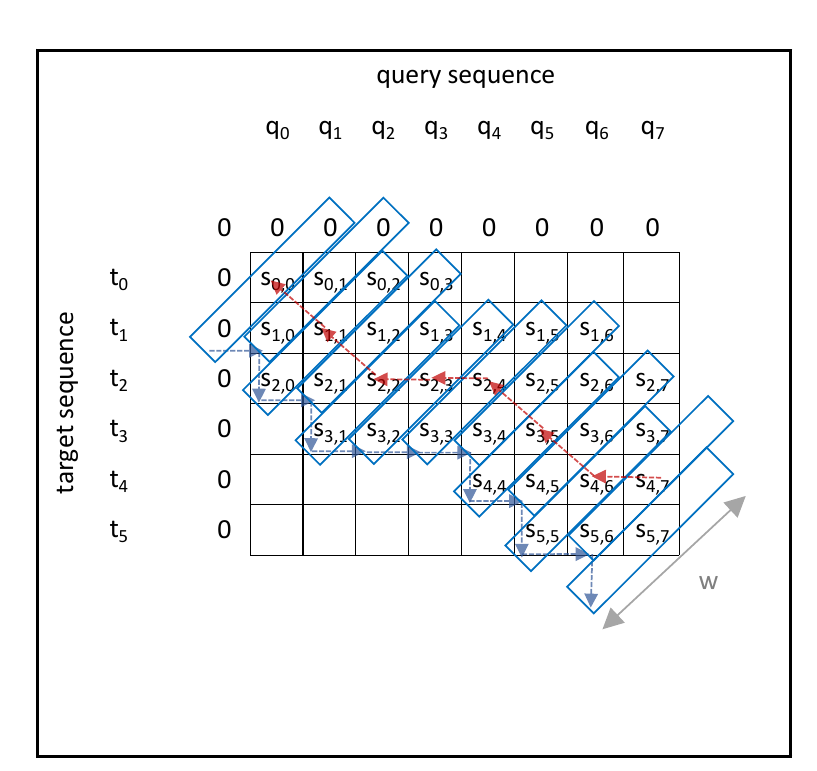}
    \caption{Adaptive banded sequence alignment} 
    \label{f:bandedadaptive}
\end{subfigure}
\caption{Evolution of dynamic programming-based sequence alignment}
\end{figure*}

Consider the same example in Fig. \ref{f:bandedswbw4} (performed previously with a static band of size 4) is now performed only with a band-width of size 3, as shown the Fig. \ref{f:bandedsw}. Observe that the band is no longer sufficient to contain the whole alignment, i.e. the cell s\textsubscript{4,7} which previously contained the maximum score is no longer computed, thus the trace-back would begin from the maximum value within the band, which leads to a non-optimal alignment. This is remedied using an \textit{adaptive band} in Fig. \ref{f:bandedadaptive}. The band moves either down or to the right (the band dynamically adapts) as determined by the Suzuki-Kasahara heuristic, which is illustrated by blue arrows. Observe how the alignment is possible to be contained inside a band of width 3 which was previously infeasible using a static band.

\subsubsection{Variant calling / detection of epigenetic base modifications}\label{s:3rdgen-polish}

As stated before, variant calling or detection of epigenetic base modifications is a downstream processing step which utilises both base-space alignments and raw signals. This step reuses the raw signals to recover the lost biological information during base-calling. Previous research \cite{jain2018nanopore,loman2015complete} has shown that the identification of genetic variants can be improved up to an accuracy of more than 99\% by using raw signal data from multiple overlapping nanopore reads.  It has also been shown that methylated C bases can be differentiated from non-methylated C bases by the use of signal data, using algorithms such as the one implemented in the software package \emph{Nanopolish} \cite{simpson2017detecting}. Thus, the downstream analysis that reuses raw signal data could also detect modified nucleotide bases.

At the time of writing \textit{Nanopolish} \cite{simpson2017detecting} is the most popular software package amongst the nanopore community for variant calling and detection of epigenetic base modifications.
\textit{Nanopolish} takes the reads, their alignments to the reference genome and the raw signal of each read as the input. Initially, the raw signal is segmented in the time domain based on sudden jumps in the signal and these segments are known as \emph{events}. The events are then aligned to a hypothetical signal model using an algorithm called \emph{Adaptive Banded Event Alignment (\textit{ABEA})}. The output of ABEA and alignment details of reads to the reference genome are sent through a Hidden Markov Model (HMM) to detect variants or base modifications. Nanopolish is written in C/C++ and supports multi-threading through openMP. Chapter \ref{c:gpuabea} of this thesis is about the optimisation of \textit{Nanopolish} (ABEA algorithm in particular) and details of the algorithm are discussed in the background of that chapter. 

Tombo is another software for detection of modified bases which also uses raw signals for the process. Tombo has been developed in Python programming language. Recently, a few neural network-based variant callers also have been released, for instance, Medaka\cite{medaka}, Clairvoyante \cite{luo2019multi} and Clair \cite{luo2020exploring} (successor of Medaka Clairvoyante). These neural network-based variant callers have been developed in Python programming language and use Tensorflow in the backend. Unlike \textit{Nanopolish}, these neural network-based variant callers only rely on base-called reads and are incapable of using raw signal data. 
%https://github.com/nanoporetech/medaka
%medaka includes a basic workflow for aggregating Guppy basecalling results for Dcm, Dam, and CpG methylation. The workflow is currently very preliminary and subject to change and impr
%medaka is a tool to create a consensus sequence from nanopore sequencing data. This task is performed using neural networks applied from a pileup of individual sequencing reads against a draft assembly. It outperforms graph-based methods operating on basecalled data, and can be competitive with state-of-the-art signal-based methods, whilst being much faster.

% Tombo
% modified base detection
%https://github.com/nanoporetech/tombo
%Stoiber, M.H. et al. De novo Identification of DNA Modifications Enabled by Genome-Guided Nanopore Signal Processing. bioRxiv (2016).
%Tombo is a suite of tools primarily for the identification of modified nucleotides from nanopore sequencing data. Tombo also provides tools for the analysis and visualization of raw nanopore signal.

\subsubsection{Characteristics of data}

\begin{figure}[ht]
\centering
\includegraphics[width=\linewidth]{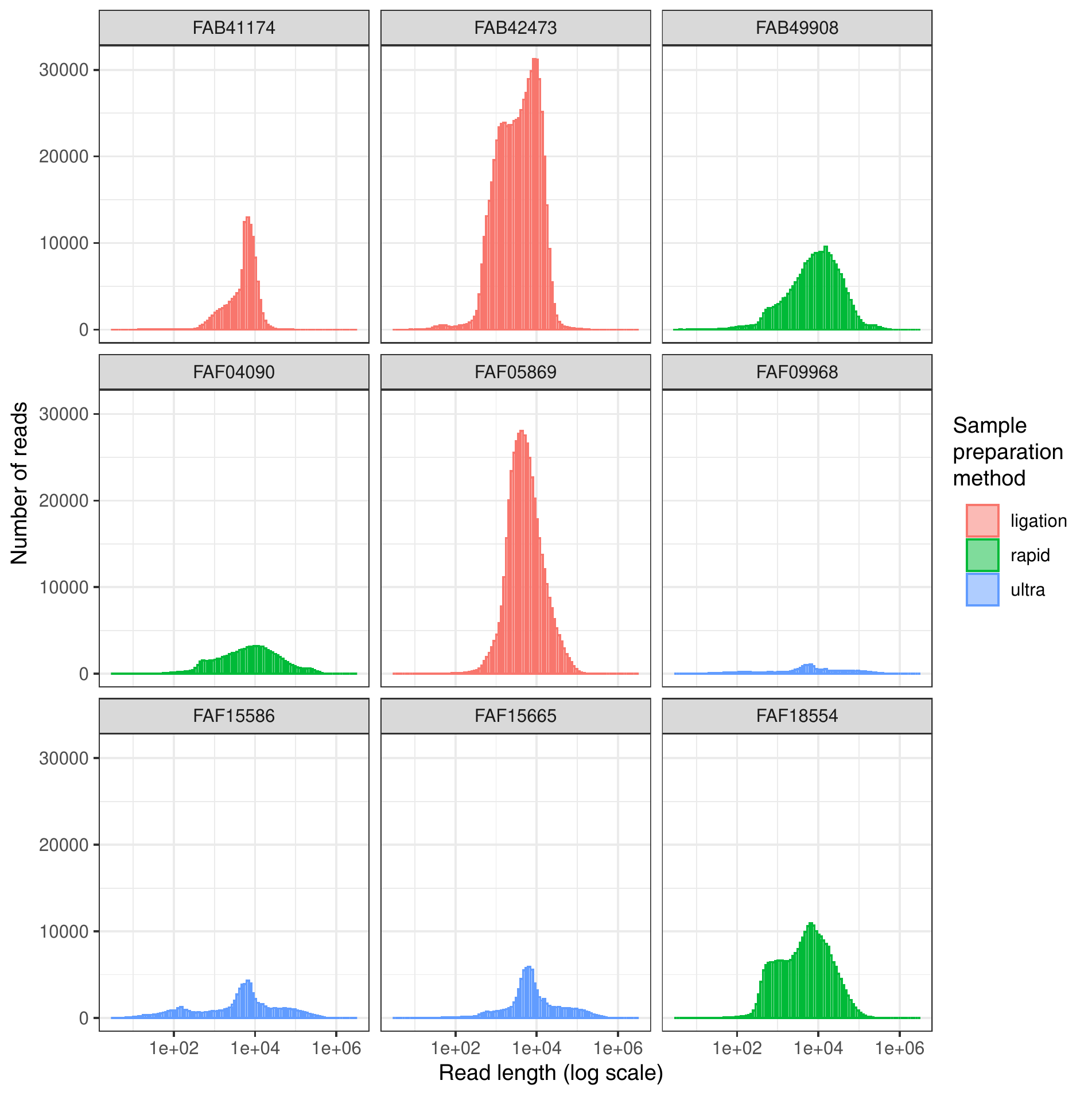}
\caption{Read length distribution nanopore consortium}
\label{f:3rdgenrlen}
\end{figure}

\textbf{Read length}: In third-generation sequencing, read lengths can significantly vary within a dataset. For instance, read lengths can be from a few hundred bases to >1 Mbases in nanopore sequencing. Fig. \ref{f:3rdgenrlen} shows read length distributions of nine datasets, out of the 53 publicly available NA12878 datasets (different datasets produced at different sequencing run of the NA12878 sample) from the nanopore consortium \cite{jain2018nanopore}. The library preparation method is a major factor that affects the read lengths. Currently, there are three library preparation method for nanopore, \textit{ligation} and \textit{rapid}, which are officially from Oxford Nanopore, and \textit{Ultra} which is community-developed \cite{jain2018nanopore}. Fig. \ref{f:3rdgenrlen} shows three datasets from each of those library preparation methods and demonstrates that the read length distributions vary not only among different library preparation methods but also among different datasets from the same library preparation method.

\begin{figure}[ht]
\centering
\includegraphics[width=\linewidth]{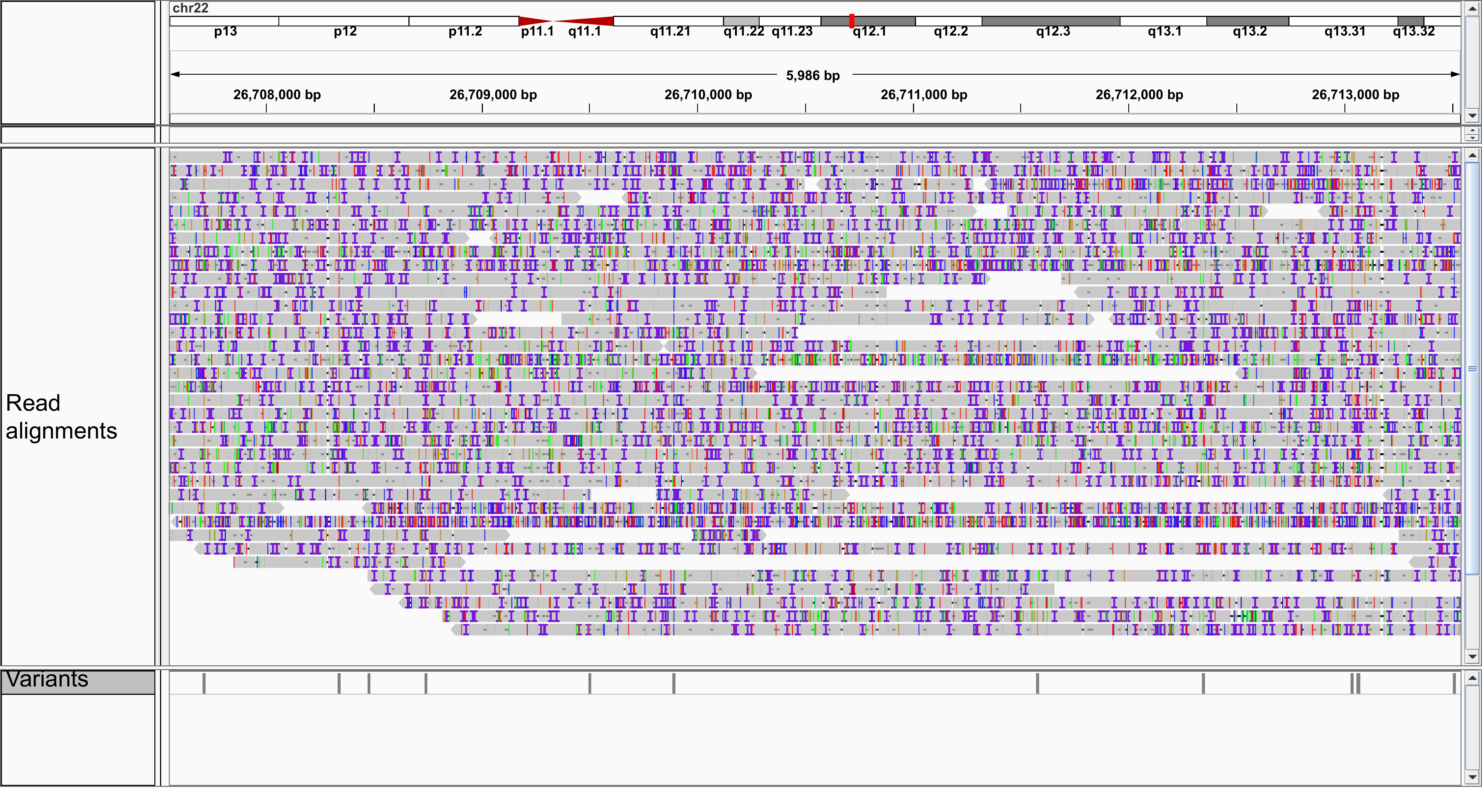}
\caption[A screenshot from IGV from an NA12878 nanopore dataset]{A screenshot from IGV for a region from NA12878 sample nanopore consortium genome project}
\label{f:nanoporeerrors}
\end{figure}

\textbf{Error rate:} The error rate of nanopore third-generation sequencing is demonstrated in Fig. \ref{f:nanoporeerrors} using reads from a real dataset (all NA12878 data from nanopore consortium \cite{jain2018nanopore}) aligned to a reference (human genome). Fig. \ref{f:nanoporeerrors} is a screenshot of a $\sim$6 kbase region in chromosome 22 taken through the Interactive Genome Viewer (IGV). Similar to Fig. \ref{f:illumina-errors} for second-generation sequencing, the grey colour horizontal blocks on Fig. \ref{f:nanoporeerrors} represent the reads and other colours represent differences in those reads to the reference. The bottom panel shows the variants that are present in this region. Observe how high the error rate in third-generation sequencing (Fig. \ref{f:nanoporeerrors}) is, compared to second-generation sequencing (Fig. \ref{f:illumina-errors}).

\textbf{Data size:} When all NA12878 datasets from the nanopore consortium are aggregated, the total is around 132.931 Gbases. This is equivalent to about 40X coverage of the human genome. The number of reads is 15.667 million. The \textit{FASTQ} (uncompressed) file containing these read is 250GB in size. The SAM file (uncompressed) generated by aligning these reads using \textit{Minimap2} is 280 GB. The sorted BAM file (compressed) is around 150 GB. Per-read methylation calls generated from \textit{nanopolish} which are in TSV format (uncompressed) consume around 70 GB. The VCF file generated from the variant calling step is around 1 GB. Raw signals corresponding to the aforementioned reads stored in the latest fast5 format (multi-fast5 with compression) consume around 2.2 TB. Note that this used to be 46 TB a few years ago when stored as single-fast files (mostly due to redundant data such as the event table).

%provides around 40X coverage

%We give figures base don publicly available NA12878 consortium whole dataset. 

% The human genome sequenced at around 40X coverage is about . FASTQ contains $\sim$132.931 gbases 132.931reads and is 250GB. SAM file 280 GB. BAM file is around 150 GB. Raw methylation calls are around 70 GB. Frequencies counted are around 1 GB. The VCF file generated from the variant calling step is around 1 GB.

% Fast5 are around 2.2 TB
% Used to be around 46 TB

% fast5 2371.419976 GB
% used to be 46163 GB 

\subsection{De novo assembly} \label{s:denovo}

 If it is the first time that the DNA of the particular species is sequenced, then the reads must be assembled without any reference, only using the information in the reads. This process is known as de novo assembly. 
 
 Early de novo assemblers such as SEQAID \cite{peltola1984seqaid} were based on greedy algorithms. Modern assemblers rely on graph-based techniques. Short-read assemblers mostly use de Bruijn graphs \cite{zerbino2008velvet} and examples are ABySS \cite{simpson2009abyss}, Velvet \cite{zerbino2008velvet}, Spades \cite{bankevich2012spades} and Cortex \cite{iqbal2012novo} use de Bruijn graphs. However, SGA \cite{simpson2012efficient}  which is also a short-read assembler uses overlap graphs (a type of overlap called string graphs \cite{myers2005fragment}). Almost all long-read assemblers use overlap graph-based methods. Examples of long-read aligners are miniasm \cite{li2016minimap}, flye \cite{kolmogorov2019assembly}, canu \cite{koren2017canu} and  wtdbg2 \cite{ruan2020fast}. 
  
 Currently, there are three de novo assembly workflows: 1, using only short-reads; 2, using only long-reads; and, 3, using both called hybrid assembly. For more information of de novo assembly readers may refer to \cite{bayat2020methods}.

\section[Related Work]{Architecture-aware optimisation of sequence analysis algorithms}\label{s:related-work}

Studies in the field of DNA analysis have predominantly focused on improving the accuracy of algorithms. Such improvements have further increased the computing power required for the analysis. Compared to the plethora of studies focusing on accuracy,  studies attempting to reduce the gap between DNA sequencing and analysis technologies (architecture-aware optimisation of sequence analysis algorithms) are minimal. This section presents those architecture-aware optimisation  studies under four categories: work that optimises sequence analysis algorithms for general-purpose CPU, HPC, cloud computing and distributed computing in section \ref{s:archaware-opti-cpu}; work on GPU-based optimisations in section \ref{s:archaware-opti-gpu}; FPGA-based optimisations in section \ref{s:archaware-opti-fpga}; and, specialised hardware design for sequence analysis in section \ref{s:archaware-opti-asic}.

\subsection{CPU/HPC/Cloud}\label{s:archaware-opti-cpu}

Core sequence alignment algorithms for second-generation sequencing such as SW and NW (discussed in section \ref{s:seqaln}) have been optimised to efficiently utilise Single-Instruction Multiple-Data (SIMD) instructions in modern Intel CPUs (SSE and AVX). Libraries such as libssa \cite{rognes2011faster},  Parasail \cite{daily2016parasail}, SeqAn \cite{reinert2017seqan}, SSW \cite{zhao2013ssw} and SWPS3 \cite{szalkowski2008swps3} are some examples of such SIMD-based optimisations. SK, the core alignment algorithm for third-generation sequencing (discussed in section \ref{s:3rdgen-aln}) has also been accelerated using SIMD instructions in a library called libgaba  \cite{suzuki2018introducing}. The most popular second-generation read alignment tool, BWA MEM (discussed in section \ref{s:seqaln}), has been very recently optimised for better cache, memory and SIMD instruction utilisation by researchers from the parallel computing lab of Intel, yielding 2.4X improvement in performance. This work has been released as open-source software named BWA MEM 2 \cite{vasimuddin2019efficient}. 

The GATK best practices pipeline for second-generation sequencing (discussed in section \ref{s:workflow2ndgen} has been optimised for HPC environments in \cite{banerjee2016efficient} to efficiently utilise available multiple cores and bandwidth. Another work called ADAM which is a library and a command-line tool enables the use of Apache Spark to efficiently parallelise genomic data analysis across cluster/cloud computing environments \cite{massie13,nothaft15}.  

Attempts to utilise cloud computing for DNA analysis have been made \cite{langmead2009searching, schatz2009cloudburst, o2013big}. However, transferring DNA data which are hundreds of gigabytes in size over the Internet is not efficient as the data transfer itself may consume more time than the analysis. Additionally,  uploading sensitive DNA information has privacy concerns \cite{schatz2010cloud}.

\subsection{Graphics Processing Units (GPU)}\label{s:archaware-opti-gpu}

Suitability of massively parallel Graphics Processing Units (GPU) for DNA sequence analysis has been investigated. The core alignment algorithm SW, has been accelerated using GPU in examples such as \cite{liu2006gpu,manavski2008cuda,Liu2009}. GPU-accelerated aligners such as SOAP3 \cite{liu2012soap3}, BarraCUDA \cite{klus2012barracuda} and MUMmerGPU \cite{schatz2007high} have been released for second-generation sequencing. GPU-accelerated variant calling tools for second-generation sequencing such as BALSA \cite{luo2014balsa} are also available for use. 

GPU-acceleration efforts have been made for third-generation sequencing as well. Nanopore base-calling software known as \textit{Guppy} exploits NVIDIA GPUs for fast processing \cite{wick2019performance}.  \textit{Guppy} is a proprietary software provided by ONT that uses deep neural networks. Design details of \textit{Guppy} are not known due to the program being closed source. \textit{Guppy} likely benefited by the plethora of work focusing on GPU optimisations in the neural network domain. Minimap2, the popular open-source base-space aligner for long-reads (discussed in section \ref{s:3rdgen-aln}) has been recently accelerated with the simultaneous use of a GPU and an Intel Xeon Phi co-processor \cite{feng2019accelerating}.  However, the source code for this accelerated Minimap2 is not openly available. Recently, NVIDIA corporation has shown an interest in developing open-source libraries such as \textit{Clara Genomics}\cite{clara} for accelerating long-read data analysis on their GPUs. \textit{Clara Genomics} library contributes to the nanopore data analysis domain through the acceleration of core algorithmic components such as all-vs-all read mapping and partial order alignment for performing de novo assembly. 

%Third-generation nanopore DNA sequence analysis is a relatively new field (first nanopore MinION emerged in 2015). Despite the short time, many software tools have been rapidly developed by biologists for nanopore DNA sequence analysis. However, works on accelerating those software or even those exploring computational bottlenecks are very limited. 

%However, GPU programming is challenging than CPU programming due to the requirement of knowledge on  complex GPU architectures. Further,  GPU implementations should be followed by massive optimisation effort, in order to gain speed up. Meanwhile, analysis of human genome would require high-performance GPUs such as Tesla K40, which are costly and must be fixed on servers through PCI express. Mobile graphic chips on laptops are not powerful enough and do not have sufficient memory for human genome analysis. Hence, a portable DNA analysis cannot be achieved through GPU based acceleration as well. 

\subsection{Field Programmable Gate Arrays (FPGA)}\label{s:archaware-opti-fpga}

The utility of Field Programmable Gate Arrays (FPGA) for accelerating key computational kernels in second-generation sequence analysis has been explored by researchers. The SW alignment algorithm has been accelerated in work such as \cite{benkrid2009highly,houtgast2017high}. Edit distance-based alignment has been accelerated by 40-60 times in \cite{banerjee2018asap}. Pair-HMM alignment, a major bottleneck in the GATK HaplotypeCaller, has been accelerated in studies such as  \cite{huang2017hardware} (487 times performance improvement) and \cite{banerjee2017accelerating} (14.85 times throughput improvement). The reported speedups for FPGA-accelerated key algorithms are impressive. However, the overall speedup when such components are integrated into an end-to-end analysis workflow is yet to be explored.

FPGA-based commercial accelerators also exist for second-generation sequence analysis. Examples are DeCypher \cite{decipher} and Dragen \cite{dragon}. DeCypher \cite{decipher} is from a company called Timelogic. Their proprietary FPGA cards are fixed to servers using Peripheral Component Interconnect (PCI) Express interface. Alignment algorithms such as BLAST, SW and HMM are supported on these cards. Timelogic claims that their {\small FPGA} card is equivalent to 860 generic CPU cores \cite{decipher}. Dragen is from a company called Edico genome (recently acquired by the sequencing giant Illumina). Edico genome claim that the whole genome analysis pipeline including sequence mapping and variant calling can be completed within 22 minutes \cite{dragon2}. A major drawback of these commercial systems is that they are proprietary, and the users are restricted to the few algorithms provided by the company. These commercial systems are also prohibitively expensive when compared to purchasing general-purpose servers.

To date, the use of FPGA for rapidly evolving third-generation sequence analysis is not explored.  Traditional implementations for FPGA that are done using Hardware Descriptive languages (HDL) are not very flexible and a slight algorithmic change requires considerable modification in the implementation. In the future, when third-generation sequence analysis algorithms are relatively stable, FPGA-based acceleration of such algorithms are anticipated.

We believe that ongoing advancements in high-level synthesis (HLS) would further increase the FPGA-based acceleration efforts in the future. HLS attempts to improve flexibility while achieving performance similar to hand-optimised HDL. The OpenCL framework is increasingly becoming popular for FPGA acceleration. Preliminary attempts that use open-CL framework for accelerating genomic kernels have been made in work such as \cite{rucci2018swifold,rucci2017accelerating}.

\subsection{Application-specific Hardware} \label{s:archaware-opti-asic}

There have been rare attempts to design custom hardware for sequence analysis.  MESGA \cite{gnanasambandapillai2018mesga} is a Multiprocessor system on a chip (MPSoC) architecture based on embedded processors for short-read alignment.  DARWIN \cite{turakhia2018darwin} is a co-processor for long-read alignment. Large speedups have been reported for these custom hardware, as anticipated.  However, these systems have been evaluated only using simulations, potentially due to the impractical fabrication cost unless mass produced. 

Though custom hardware can provide extremely fast performance with a smaller footprint, the design flow is complex and the non-recurring engineering (NRE) cost is very high. DNA analysis algorithms improve rapidly and new algorithms are frequently introduced, especially for new sequencing technologies that are ever-improving. Thus, custom hardware for genomics processing is unlikely to become mainstream in the near future. In 1997, a company named Paracel introduced GeneMatcher, a specialised genome analyser based on  Application-Specific Integrated Circuits (ASIC) \cite{genematcher}. The second version, GeneMatcher 2 was equipped with more than 27000 processors.  Unfortunately, ASIC based Genematcher systems did not thrive.

%\section{Summary}

%Together, these studies indicate that there is a considerable gap between DNA sequencing and DNA analysis technologies. Even though several studies have been successful in accelerating the analysis process, yet the process is performed centrally in server computers. The recent introduction of portable DNA sequencers indicates that methods to decentralise the analysis process should be investigated. To date, according to best of our knowledge, there is no portable computer system that is capable of performing DNA analysis. We will be building a low cost, efficient and portable DNA analyser that fully analyses the human DNA in a reasonable time. Our system will be using easily programmable general purpose processors, making the system flexible for future algorithmic changes. 

%% file: 4-cache/cache.tex
\chapter[Cache Friendly Variant Calling]{Cache Friendly Optimisation of de Bruijn Graph based Local Re-assembly in Variant Calling} \label{c:cacheopti} 

\rule{\textwidth}{0.4pt} 
This chapter has been published in IEEE/ACM Transactions on Computational Biology and Bioinformatics. \textcopyright 2018 IEEE. Reprinted, with permission, from \textbf{H. Gamaarachchi}, A. Bayat, B. Gaeta, and S. Parameswaran, “Cache Friendly Optimisation of de Bruijn Graph based Local Re-assembly in Variant Calling,” IEEE/ACM transactions on computational biology and bioinformatics, 2018. DOI: \url{https://doi.org/10.1109/TCBB.2018.2881975} \cite{gamaarachchi2018cache}.\\ 
\rule{\textwidth}{0.4pt}

A variant caller is used to identify variations in an individual genome (compared to the reference genome) in a genome processing pipeline.  
%Variant calling is an important but time consuming process in genome analysis. These applications usually have a number of steps, such as: 
For the sake of accuracy, modern variant callers perform many local re-assemblies on small regions of the genome using a graph-based algorithm. However, such graph-based data structures are inefficiently stored in the linear memory of modern computers, which in turn reduces computing efficiency. Therefore, variant calling can take several CPU hours for a typical human genome. 
We have sped up the local re-assembly algorithm with no impact on its accuracy, by the effective use of the memory hierarchy. The proposed algorithm maximises data locality so that the fast internal processor memory (cache) is efficiently used. By the increased use of caches, accesses to main memory are minimised. 
The resulting algorithm is up to twice as fast as the original one when executed on a commodity computer and could gain even more speed up on computers with less complex memory subsystems.

\section{Introduction}

The capability of Next Generation Sequencing technology (NGS) has grown faster than Moore's law in the past few years \cite{heather2016sequence}. Both the time and the cost of sequencing a genome have dropped to an affordable level and is expected to drop further \cite{alkan2011genome}. However, the capability of the computing technology has not kept up with the pace of improvement in sequencing technology \cite{Muir2016}. Hence, it is an increasing challenge for computers to process such massive amount of data. 
%in the  speed of analysing genomes 

%Personalised medicine in an emerging trend where patients are genetically tested to improve diagnosis and treatment. Furthermore, genomic research helps the discovery of mutations that causes certain diseases. 

The most commonly used NGS technologies produce a large number of short DNA fragments known as reads. The reads are aligned to a reference genome using sequence aligners such as BWA \cite{li2009fast} and Bowtie \cite{langmead2009ultrafast}. % This process of alignment is referred to as sequence alignment or sequence mapping \cite{alkan2011genome}. 
%Sequence alignment is challenging due to the presence of actual variants and sequencing errors, where certain reads will not exactly match the reference genome. 
After sequence alignment, a process called variant calling \cite{nielsen2011genotype} identifies the actual genomic variations, amongst the artefacts generated by sequencing machines (sequencing errors).
%Variant calling is a process to identify genomic variations amongst artefact generated by sequencing machines (sequencing errors). 
Traditional variant callers such as GATK UnifiedGenotyper \cite{depristo2011framework} fully rely on the alignment produced by sequence aligners. Sequence aligners only perform pairwise alignment between an individual read and the reference genome. However, a more accurate alignment can be obtained by considering all the reads that are aligned to a particular region of the genome. Tools such as GATK IndelRealigner \cite{homer2010improved} were designed to improve the accuracy of alignments using information from all the reads mapped to a region.  Such tools are run prior to the use of a traditional variant calling algorithm. %As a result all the reads should be aligned to the genome first so that one could extract those that are mapped to a particular region. 

Modern variant callers, such as GATK HaplotypeCaller \cite{depristo2011framework}, Platypus \cite{Rimmer2014}, SOAPindel \cite{li2013soapindel} and Scalpel \cite{narzisi2014accurate} take a different approach compared to traditional variant callers. These modern variant callers utilise \textit{de-Bruijn} graph-based \textit{de-novo} local re-assembly, which assemble the genome locally (only a small region) using all the reads mapped to that region. The local re-assembly results in greater accuracy in variant identification \cite{Rimmer2014}. The typical workflow of a modern variant caller is given below (though there are slight differences in each variant calling tool, the basic workflow remains the same).
%
%Early variant callers such as GATK UnifiedGenotyper \cite{depristo2011framework} directly relied on the output from the aligner. Detection of structural variants such as indels had limited success with these early tools \cite{alkan2011genome} .
%Modern variant callers such as GATK HaplotypeCaller \cite{depristo2011framework}, Platypus \cite{Rimmer2014}, SOAPindel \cite{li2013soapindel} and Scalpel \cite{narzisi2014accurate} do not fully rely on the output from aligner. Instead, they perform local re-assembly as the first step. 

\begin{itemize}
\item{\textbf{Local re-assembly}: A \textit{de-Bruijn} graph \cite{pevzner2001eulerian} is formed using the reads aligned to a particular region of the genome. The graph is then traversed to detect candidate variant sites. This graph construction and graph traversal are performed for a region at a time \cite{Rimmer2014}. 
}
\item{\textbf{Aligning reads to haplotypes}:
Haplotypes are formed using candidate variant sites identified through local re-assembly. Then, the reads in the corresponding region are aligned to each haplotype using a pairwise alignment algorithm such as Needlemann-Wunch \cite{needleman1970general} or pair-HMM \cite{pachter2002applications}.
}
\item{\textbf{Finding statistically significant variants} :
Statistical approaches are applied to find the most probable variants using reads aligned to the haplotypes.}% Finally, additional filtering steps might be applied to identify the best variant \cite{Rimmer2014}}
\end{itemize}

%Though sequence mapping and variant calling are the two major steps that, additional steps (for instance GATK best practises) are recommended for better results. However sequence mapping and variant calling are the most compute intensive steps in the pipeline. 

Thus far, a number of studies have been performed to improve the variant calling process. % and  sequence alignment.  
%Though numerous improvements to sequence alignment have targeted both accuracy and performance, 
 However, most research on variant calling has been restricted to improving accuracy. A very little attention has been paid to optimising core components of modern variant calling algorithms, such as \textit{de-Bruijn} based local re-assembly.
%
%In this paper, we present algorithmic improvements to local re-assembly process that can accelerat using computer architecture knowledge on the memory hierarchy. The proposed method improves performance without affecting the accuracy of the variant calling results.

\begin{figure}[!t]
  \centering
    \includegraphics[width=\columnwidth]{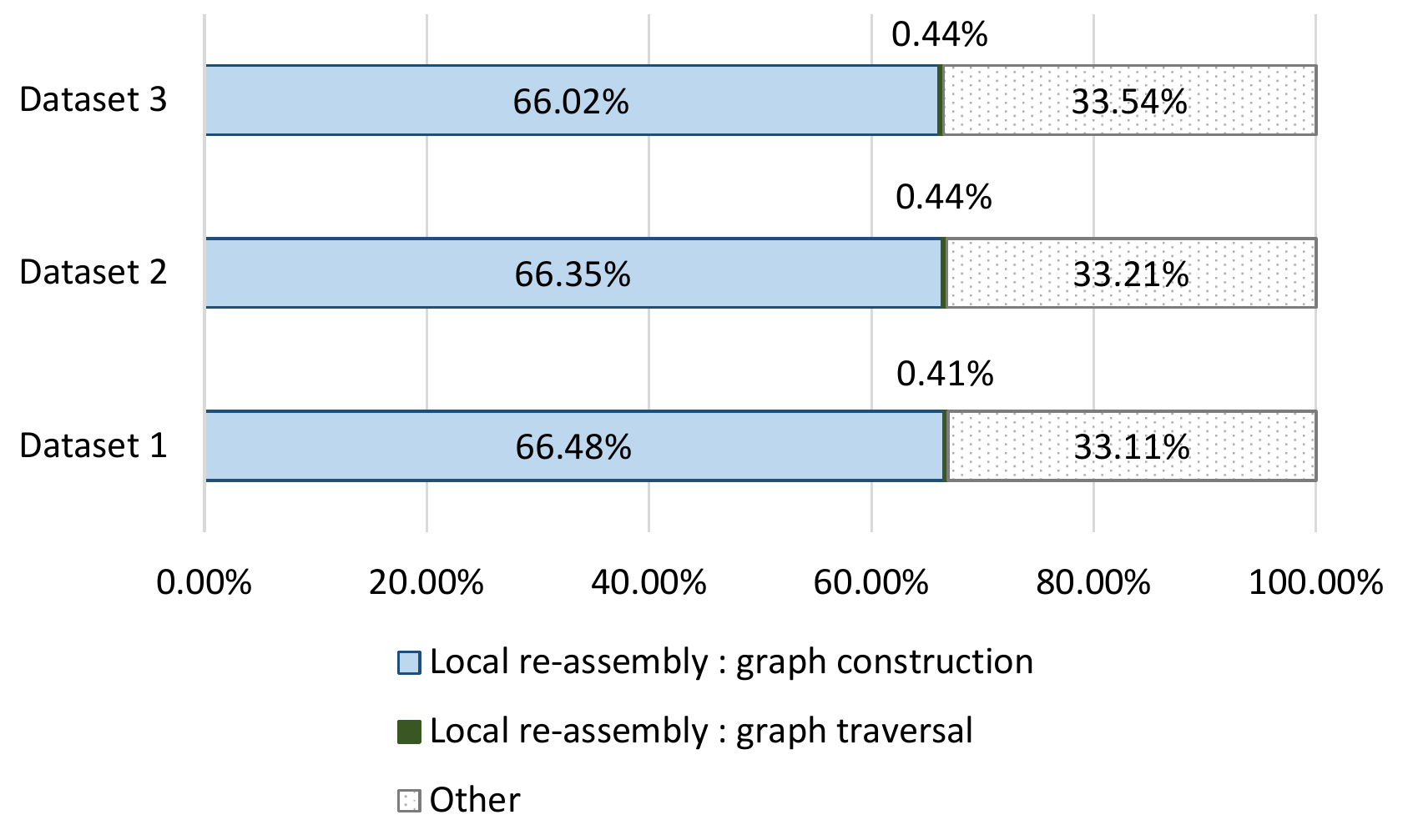}
    \caption{Distribution of execution time for Platypus variant caller}  
    \label{f:platytime}
\end{figure}

Modern variant callers are compute and memory intensive compared to traditional variant callers. The main reason for this additional computation is the local re-assembly step.
%For instance GATK HaplotypeCaller is several times slower than the GATK UnifiedGenotyper \cite{hasan2015performance}.
%Initially, the performance bottleneck was aligning reads to the haplotypes. However, latest variant callers such as Platypus \cite{Rimmer2014} and newer versions of GATK HaplotypeCaller utilise SIMD implementations of alignment algorithms.Consequently, the alignment step has become faster and now the memory intensive local re-assembly step has become the bottleneck.
%In order to support our claim, an experiment has been performed where three datasets were processed using the modern variant caller Platypus   
This is illustrated in Fig. \ref{f:platytime} that shows the time requirement for different steps of variant calling using Platypus variant caller (refer to Section \ref{s:results} for information on the datasets and the machine configuration where the experiment was performed). De Bruijn graph-based assembly involves two major steps; graph construction and graph traversal. For the three datasets in Fig. \ref{f:platytime}, graph construction took around  66\% of the total time.  Graph traversal took less than 0.5\% of the total time. All other tasks including reads to haplotype alignment, probability computation and disk accesses added up to around 33\% of the total time. 
This experiment suggests that improving the performance of graph construction will result in a significant increase in overall performance. The  graph construction is time-consuming since it  requires random accesses to memory which reduce locality of memory accesses. When there is locality in accesses to memory, most accesses can be handled by the fast internal processor memory (cache). If data exists in the cache (cache hit) it can be accessed quickly. If data does not exist in the cache (cache miss) it has to be loaded from main memory (RAM) which usually takes a much longer time.

%The execution time for each processing step is reported in Fig. \ref{f:platytime}. 
%As shown in Fig. \ref{f:platytime}, 
%\textit{de-Bruijn} graph construction takes 60\% of the total execution time.
%Note that \textit{other} in Fig. \ref{f:platytime} includes all other steps including the disk access. Profiling was done using Platypus for three real data sets (refer \hyperref[s:results]{``Results''} section for information on data sets and machine configuration). 
%Fig. \ref{f:platytime} also reveals that graph traversal takes negligible amount of time when compared to graph construction.

%{\textbf{Random access to memory is costly} - Computer systems have a memory hierarchy to tradeoff cost and performance. RAM is high in capacity but slow. An access to the RAM takes more than 100 CPU clock cycles in commodity computer.  Cache memories are smaller in capacity but faster. An L1 cache access takes only about 10 CPU clock cycles. Registers which are very low in capacity are much faster (generally only 1 CPU clock cycle). Temporal and spatial locality in memory accesses are required for effective utilisation of this memory hierarchy. Random memory accesses will always end up in the RAM making the performance lower. Hence, the algorithms must be tuned to maximise efficient usage of the memory hierarchy. 

%However, the use of optimised version for alignment algorithms (such as SIMD versions) has improved the performance. 

%However, de Bruijn based assembly is a memory intensive step which has now become the bottleneck. 

In this paper, we introduce several improvements to the original \textit{de-Bruijn} graph-based local re-assembly algorithm such that locality in memory accesses is increased and the processor cache is utilised efficiently. One of the most efficient improvements we apply is to exploit existing alignment information in the graph construction process. Our proposed improvements result in the algorithm being about twice as fast as the original algorithm. In order to test the proposed algorithm, we ported it into the Platypus variant caller. The modified Platypus implementation is available at \cite{ourcode}.

The rest of the paper is organized as follows.  Section \ref{s:related} discusses related work. Then in Section \ref{s:existing} we explain the de-bruijn based local assembly algorithm. Then in Section \ref{s:proposed} we present our optimization techniques. Next, in  Section \ref{s:results} we present the experimental setup and results. After that, Section \ref{s:discussion} elaborates the future directions. Finally, we conclude in  Section \ref{s:conclusion}.

\section{Background of deBruijn Graph based Local Re-assembly} \label{s:related} 

Modern variant callers such as GATK HaplotypeCaller \cite{depristo2011framework} and Platypus\cite{Rimmer2014}, consist of  local re-assembly, reads to haplotypes alignment and variant identification.  In the GATK Haplotypecaller, aligning reads to the haplotypes (using the Pair-HMM algorithm)  takes much of the processing time  \cite{huang2017hardware}. However, by the use of Intel's Advanced Vector Extension (AVX) instructions and FPGAs, this processing time can be considerably reduced (by 720X using Intel AVX and 3,857X  using FPGA \cite{altera}). The latest versions of GATK HaplotypeCaller support Intel AVX and FPGA \cite{GATKIntel} (the latter as an experimental feature)\footnote{Additionally, Sentieon Inc. company has optimised GATK without specialised hardware, however, it is commercial.}.
Furthermore, tools such as Avacado \cite{nothaft2015scalable} consist of algorithms with lower time complexity for aligning reads to haplotypes. 
%GATK first identifies regions with variation (called active regions). De Bruijn assembly is performed for only those regions. Next pair-HMM is used for aligning reads to haplotypes. 
% Pair-HMM algorithm used to align reads to haplotypes was the bottleneck in HaplotypeCaller \cite{huang2017hardware}.
% However, it is no longer the case with the introduction of accelerated pair-HMM using Intel Advanced Vector Extension (AVX) instructions and FPGA. 
% For instance  the Intel AVX based pair-HMM is 720X times faster \cite{altera} and is already integrated to GATK HaplotypeCaller \cite{GATKIntel}.
%OpenCL based FPGA implementation of pair-HMM is claimed to be 3,857 times faster \cite{altera} and GATK HaplotypeCaller now supports FPGA (as an experimental feature \cite{GATKIntel}). %Thus, after local re-assembly is accelerated, de Bruijn based local reassembly would be the present bottleneck in GATK haplotypeCaller which warrants optimisation.
%  
%FPGA implementation in \cite{huang2017hardware} is 487X time faster.
Platypus is a newer variant caller which is faster, with better indel accuracy than the widely used GATK HaplotypeCaller
\cite{hasan2015performance,Rimmer2014}\footnote{Note that, the indel accuracy of GATK HaplotypeCaller is higher than Platypus today. However, at the time of publishing the manuscript it was not so.}. Platypus aligns reads to haplotypes using a Single Instruction Multiple Data (SIMD) accelerated Needleman-Wunch algorithm. Hence the alignment phase is already fast. However, as discussed above more than 60\% of the time of Platypus is spent on \textit{de Bruijn} graph-based local re-assembly.

\textit{De Bruijn} graphs were first used for \textit{de novo} assembly, where assembling is done solely using the reads when a reference genome is unavailable.  Several researchers have proposed techniques for \textit{de Bruijn} graph optimisation, for \textit{de novo} assembly, where the inputs to the assembler are unaligned reads.  However, in the case of aligned reads for local re-assembly, the utility of the optimisations used for \textit{de novo} assembly are limited. The graph for local re-assembly is much smaller: only a few megabytes for local re-assembly compared to several gigabytes for \textit{de novo} assembly. This is because the whole genome is considered at once for \textit{de novo} assembly, whereas the local assembly is performed in small regions  (For instance a region is 1500 bases in Platypus). Consequently, optimisation techniques for \textit{de novo} assembly have focused on processing large graphs through memory size optimisation or parallelising. Work such as Cortex \cite{iqbal2012novo}, SOAPdenovo2 \cite{luo2012soapdenovo2} and \cite{chikhi2013space} have compressed the graph in size, while ABySS \cite{simpson2009abyss} and \cite{georganas2014parallel} have parallelised the graph across clusters. However, as the graph is small for local re-assembly, techniques used for large graphs are superfluous.

%However, de novo assembly for a human consumes giga bytes of memory, to construct the graph for the whole genome. Therefore, optimisation of de Bruijn graphs in the context of de novo assembly has been with respect to large data structures. Conversely, local re-assembly is performed on a small region of the genome at a time. For instance a region is 1500 bases in Platypus. Hence, the number of nodes in a graph are in the order of several thousands and the data structures are few megabytes. 

%EULER \cite{pevzner2001eulerian}, Velvet \cite{zerbino2008velvet}, ABySS \cite{simpson2009abyss} and cortex \cite{iqbal2012novo} are examples for de Bruijn graph based de novo assemblers.
 %ABySS is optimized to work across a cluster. Cortex which uses coloured de Bruijn graphs consists of an efficient hash table data structure that encodes the graph. SOAPdenovo2 \cite{luo2012soapdenovo2}  includes methods to reduce memory consumption. \cite{chikhi2013space} uses bloom filters for better memory usage. \cite{georganas2014parallel} introduced an algorithm that parallelises de Bruijn graph creation and traversal over large clusters. Correspondingly, there is a number of studies that propose optimisation techniques for de Bruijn based de novo assembly. Therefore, optimisation techniques in de novo assembly are not very suitable local re-assembly. 

In summary, the present bottleneck in modern variant callers is the local re-assembly process. 
To the best of our knowledge no one  has focused upon optimizing \textit{de Bruijn} graph-based local re-assembly by utilising the information from aligned reads for superior cache usage.  

%Edico genome has claimed that they are providing FPGA based acceleration for variant calling pipeline. However, methods are not  published. 

%\subsection{Optimisation of de Bruijn based Graphs for local re-assembly}

\section{Baseline Algorithm} \label{s:existing}

\begin{figure}[!t]
  \centering
    \includegraphics[width=\columnwidth]{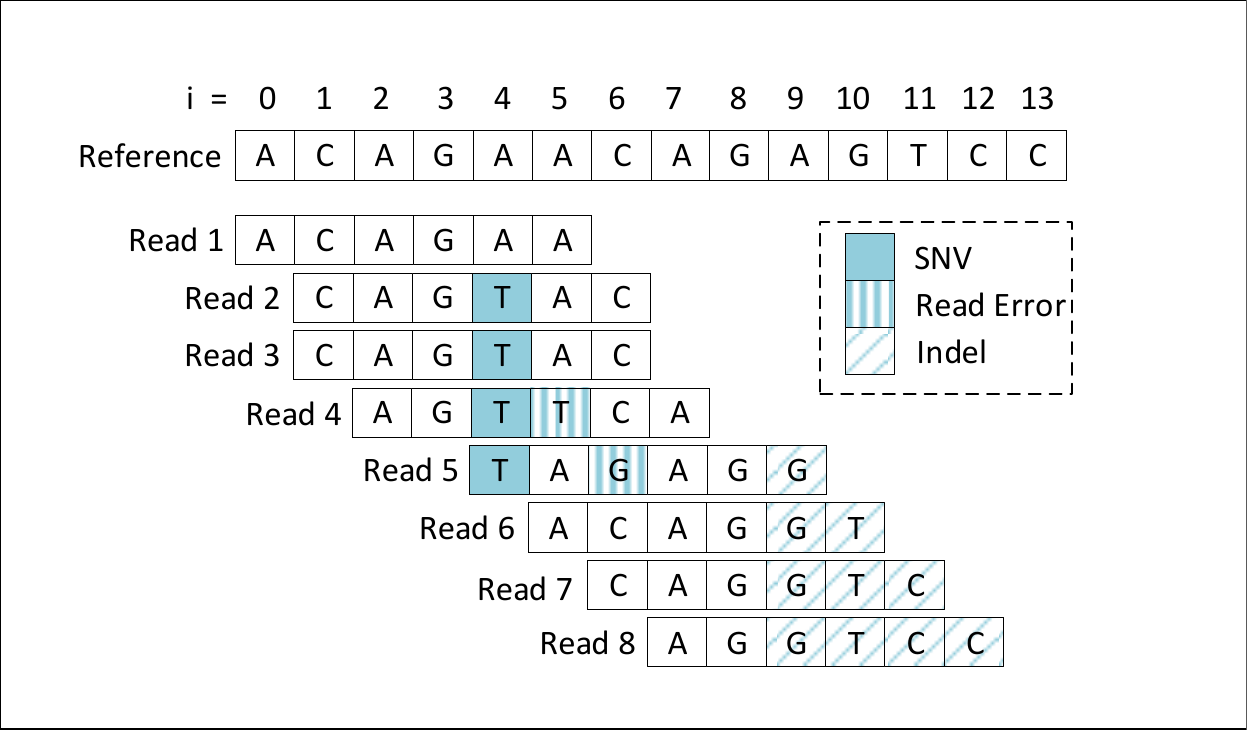}
    \caption{A region of the reference and few mapped reads} 
    \label{refdem}
\end{figure}

\begin{figure}[!t]
  \centering
    \includegraphics[width=\columnwidth]{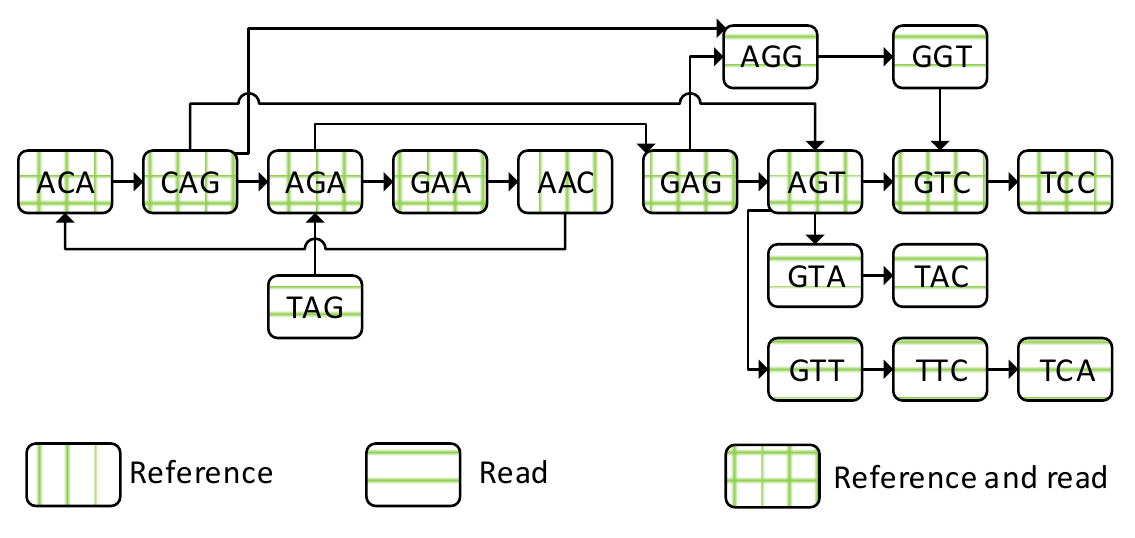}
    \caption{De Bruijn graph for the region} 
    \label{debruj}
\end{figure}

In local re-assembly, the reference and the reads are used to construct a \textit{de Bruijn} graph. Typically, a region  of several thousands of bases is considered at a time.
The reference and the reads are broken into k-mers, such that adjacent k-mers overlap by k-1 bases (see example in Fig. \ref{refdem} and Fig. \ref{debruj}). Each unique k-mer forms a node in the graph, and the edges of the graph link adjacent k-mers.

This section gives a brief account of a typical \textit{de Bruijn} graph construction algorithm for local re-assembly. The example region of the reference, and the mapped reads to the region in Fig. \ref{refdem} and its \textit{de Bruijn} graph in Fig. \ref{debruj} will be used throughout the algorithm explanation. In this simplified example, the read size and the k-mer size are  6 and 3 respectively (this is for demonstration purpose only and in reality, they are around 100 and 15 respectively).  Bases corresponding to a probable single nucleotide variant (SNV), read errors and an indel are shaded as per the legend in Fig. \ref{refdem}. The \textit{de Bruijn} graph in Fig. \ref{debruj} is constructed out of all the k-mers in the region in Fig. \ref{refdem}.
Nodes can be either shared by both reference and reads or belong to only one, as shown in Fig. \ref{debruj}. In addition note that a node for a certain k-mer is unique, despite the number of occurrences of the k-mer in the reference or reads. The graph is typically stored in computer memory using dynamically allocated nodes where memory pointers represent the edges.

%By traversing the graph based on edges with high weights, assembly is performed. De novo assembly process is very memory intensive and time consuming.

%Coloured de Bruijn graph method as in Platypus is discussed. The algorithm is generally similar in all software, except minor differences. For instance, in GATK there is no colours for nodes. Instead all nodes from reads and reference are considered equal. 
%

% The typical approach for denoting the de Bruijn graph in computer memory is explained using Fig. \ref{edge}. This figure represents two nodes (labelled as start node and end node) that are connected by an edge. The node data structure stores required data about the node (labelled as node data) such as the k-mer and position depending on the implementation.  The slots labelled as *edge are memory pointers for edge data structures. In this figure 4 such pointers are included, which is usually adequate for local re-assembly. The edge data structure stores data about the edge data (labelled as edge data) such as the weight. The slot labelled as *node is a memory pointer to a node structure. Observe how the start node stores the memory pointer to the edge and the edge stores the memory pointer to the end node.

% \begin{figure}[!t]
%   \centering
%     \includegraphics[width=\columnwidth]{edge.pdf}
%     \caption{Representation of a graph edge in memory}
%     \label{edge}
% \end{figure}

\begin{figure}[!t]
  \centering
    \includegraphics[width=0.7\columnwidth]{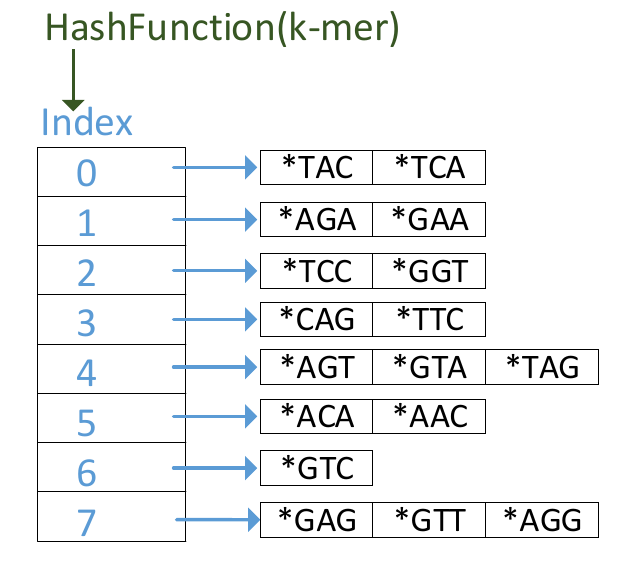}
    \caption{The hash table} 
    \label{hash}
\end{figure}

A hash table data structure is required so that repeated accesses to the same node can locate this node quickly for performance reasons during construction of the graph. As an example, consider the `CAG' k-mer in Fig. \ref{debruj} which is shared by both reference and reads.  Fig. \ref{refdem} shows that `CAG' is repeated twice in the reference and five times in the reads. During graph construction, a node will be created at the first occurrence,  but it must be found within the created graph for each subsequent occurrence. Instead of exhaustively searching  all nodes, the hash table is used for fast search. A hash table for the example in Fig. \ref{debruj} is shown in Fig. \ref{hash}. The index of the hash table for a particular k-mer is found by applying a hash function to the k-mer. The simple hash function used in this example is the addition of the ASCII values of the characters in the k-mer, subjected to the modulo operator of the number of hash table entries. For instance, there are 8 entries in the hash table and the index for the k-mer `ACA' can be found by (`A'+`C'+`A')\%8 which is 5. Hence The address to the node containing `ACA' is in index 5 in Fig. \ref{hash} as denoted by `*ACA'. %Note that each index links to a bucket (implemented as a linked list or array list) comprising of space for pointers to handle collisions. 

Algorithm \ref{ref1} and Algorithm \ref{reads1} show how the graph is constructed. Note that these algorithms are not full implementations, but are the necessary components to explain our methodology.  Algorithm \ref{ref1} shows how the reference is loaded to the \textit{de Bruijn} graph. Algorithm \ref{reads1} illustrates how reads are then loaded  to the same graph. 

In Algorithm \ref{ref1}, \textit{reference} is an array that contains the reference genome.
The algorithm iterates through the reference while adding edges between adjacent k-mers. 
For the example in  Fig. \ref{refdem} and Fig. \ref{debruj}, the algorithm will add edges in the order, ACA-CAG, CAG-AGA, AGA-GAA etc.

%In Platypus $REGION\_SIZE$ and $KMER_SIZE$ are 1500 and 15 by default respectively. 

\begin{algorithm}[ht]
\caption{Load the reference to de Bruijn Graph}\label{ref1}
\begin{algorithmic}[1]
\Function{\textit{loadReference}}{\textit{reference}}
\For{\textit{i = 0 : (region\_size - kmer\_size)}}
    \State \textit{kmer1}$\gets$ \textit{reference[i : i+kmer\_size}]
    \State \textit{kmer2}$\gets$ \textit{reference[i+1 : i+1+kmer\_size]}
    \State \textit{addEdge(kmer1,kmer2)}
\EndFor
\EndFunction
\end{algorithmic}
\end{algorithm}

In Algorithm \ref{reads1}, \textit{readsInWindow} is a buffer consisting of all the mapped reads in the region sorted using the mapped position. For each read, edges are added between adjacent k-mers. In our example, ACAGAA is the first read. Edges are added in the order ACA-CAG, CAG-AGA, AGA-GAA for this read. Similarly, all other reads will be processed. 

\begin{algorithm}[ht]
\caption{Load reads to de Bruijn Graph}\label{reads1}
\begin{algorithmic}[1]
\Function{\textit{loadReads}}{\textit{readsInWindow}}
\For{\textit{read} \textbf{in} \textit{readsInWindow}}
	\For{\textit{i = 0 : (read\_length-kmer\_size)}}
    	\State \textit{kmer1} $\gets$ \textit{read[i : i+kmer\_size}]
    	\State \textit{kmer2} $\gets$ \textit{read[i+1 : i+1+kmer\_size]}
    	\State \textit{addEdge(kmer1,kmer2)}
    \EndFor
\EndFor
\EndFunction
\end{algorithmic}
\end{algorithm}

Algorithm \ref{edge1} shows the implementation of \textit{addEdge} function used in Algorithms \ref{ref1} and \ref{reads1}. The \textit{hashTableLookUpOrInsert} function in Algorithm \ref{edge1} locates the corresponding node for a given k-mer, using the hash table. If no memory pointer exists in the hash table for a node containing the k-mer, then memory is allocated for a new node and the hash table is updated. Finally, \textit{hashTableLookUpOrInsert} returns the memory pointer to the node. %In case of null pointer, a new node is allocated in memory and added to the hash table as shown in Algorithm \ref{edge1}. 
The \textit{addEdge} function calls \textit{hashTableLookUpOrInsert}  on both \textit{kmer1} and \textit{kmer2}. Finally, \textit{createLink} actually adds the connection between the nodes by storing the pointer to the second node \textit{ptr2} in the first node\footnote{if the edge already exists, the weight parameter of the edge (details not discussed as not required to understand the cache behaviour) is updated}.  
% Parameters such as node data and edge data can be inserted or modified depending on the application. 
% \textbf{This part is confusing as the explanation for node was removed}

\begin{algorithm}[ht]
\caption{Add an edge connecting kmer1 and kmer2}\label{edge1}
\begin{algorithmic}[1]
\Function{\textit{addEdge}}{\textit{kmer1,kmer2}}
	\State \textit{ptr1} $\gets$ \textit{hashTableLookUpOrInsert(kmer1)}
    %\If{$ptr1==NULL$}
    %	\State $ptr1 = createNode(kmer1)$
    %    \State $hashTableAdd(ptr1,kmer1)$
    %\EndIf
    \State \textit{ptr2} $\gets$ \textit{hashTableLookUpOrInsert(kmer2)}
    %\If{$ptr2==NULL$}
    %	\State $ptr2 = createNode(kmer2)$
    %    \State $hashTableAdd(ptr2,kmer2)$
    %\EndIf   
	\State \textit{createLink(ptr1,ptr2)}
\EndFunction
\end{algorithmic}
\end{algorithm}

 %Hence, features such as the colour of the node are not shown in the Algorithms. 

\section{Methodology}\label{s:proposed}

% \subsection{Hash table optimisation}\label{hashopti}

%As mentioned earlier, hash tables are not cache friendly due to random accesses to the RAM.  
In this section, We show how the algorithms in Section \ref{s:existing} are modified to minimise accesses to the RAM. First, we give a simplified overview of our methodology. Then, we present additional technical information so that our method can be replicated in any \textit{de Bruijn} graph-based variant caller.

\subsection{Simplified Overview}

The hash table produced during graph construction (explained in Section \ref{s:existing}) is too large to totally reside in cache. Hence it must reside in RAM. Accesses to a hash table are random accesses and therefore cache misses will occur very frequently. 
We propose techniques to minimize these random memory accesses, by exploiting the following two factors:
\begin{itemize}
\item{\textbf{The input reads to the variant caller are already aligned to the reference} - 
The input to a variant caller is a set of reads which are already aligned to the reference. Information from the already aligned reads can be exploited for efficient utilization of the memory hierarchy, minimizing random accesses to the RAM. For instance, alignment information can be used to predict the majority of memory locations. Such predictions can minimise random accesses to the RAM.}
\item{\textbf{Variants and sequencing artefacts are rare} - In cases of variants or sequencing artefacts, memory accesses will deviate from the expected pattern. For instance, predictions   that bypass the hash table can be incorrect, requiring random accesses to the RAM. However, these events are very low as we show in Section \ref{s:results}. }
\end{itemize}

Algorithm \ref{ref1} is modified such that a cache friendly array is filled during the construction of k-mers for the reference genome. Then, Algorithm \ref{reads1} is modified to utilise this filled cache friendly array, to reduce the memory accesses to the RAM. Furthermore, Algorithm \ref{edge1} is introduced with an additional variable which is register-friendly. This register-friendly variable  accelerates the construction of graph edges by eliminating redundant accesses to the cache and the RAM.

\subsection{Algorithm in Depth}

\begin{figure}[!t]
  \centering
    \includegraphics[width=\columnwidth]{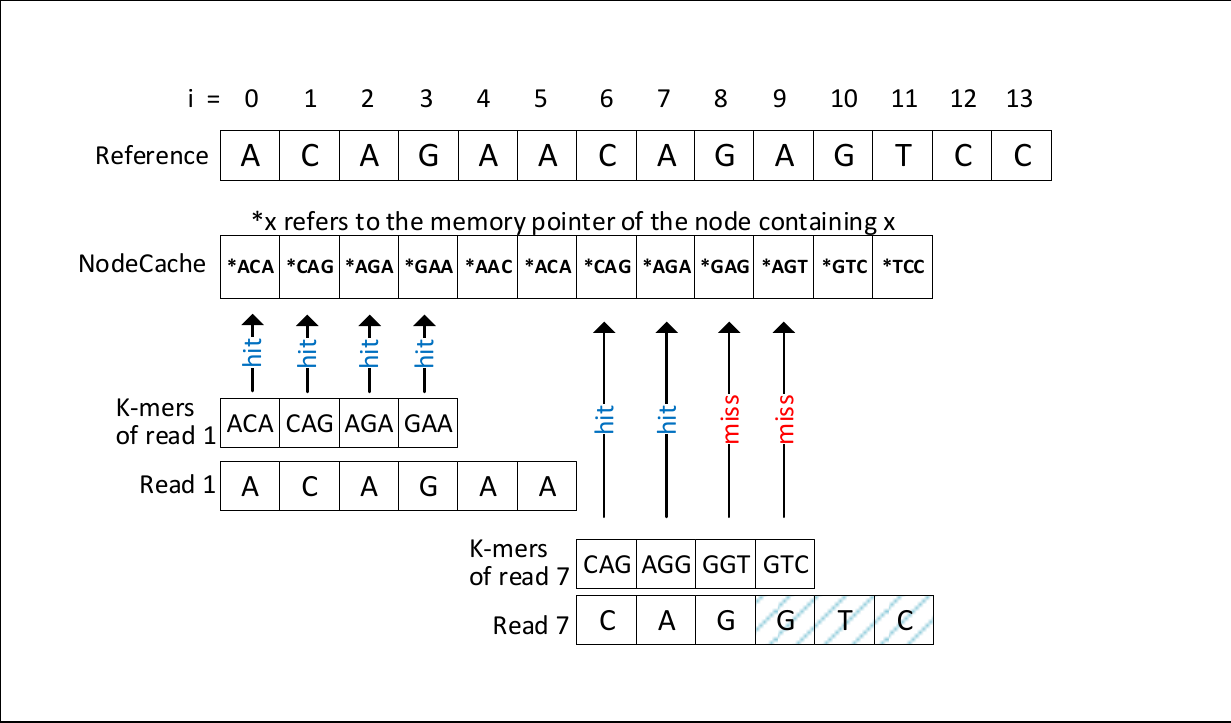}
    \caption{Elaboration of how mapping information is used for improving cache performance}
    
    \label{f:ref2}
\end{figure}

In Fig. \ref{refdem}, for position 1 (i=1), CAG is the k-mer in the reference as well as the first three reads. Similarly, if there were no variants (such as SNV, Indel) or read errors,  the k-mer in the reference and the reads should be the same for each position. However, variants and read errors cause k-mers in the reads to differ from that of the reference.
For instance, the k-mers (at position 4) for the second to fifth reads are TAC, TAC, TTC and TAG while the k-mer on the reference for that particular position is AAC. This difference 
is due to the SNV at position 4 and the read errors at positions 5 and 6 in the fourth and fifth reads.  However, variants and read errors are less frequent. Variation between a particular human genome and the reference human genome is about 0.5\% \cite{levy2007diploid}, and the read errors in modern sequencing machines are about  0.1\% - 1\% \cite{lou2013high}. Thus, in about 99\% of the time, the k-mers in the reads would be identical to the k-mers in the reference at a particular position.  
This high probability can be used to predict the memory addresses of nodes, minimizing accesses to the hash table. This is accomplished by a pointer array referred to as the \textit{nodeCache} (see Fig. \ref{f:ref2}), which is populated when loading the reference to the graph. 

Fig. \ref{f:ref2} elaborates on how the \textit{nodeCache} is used. The reference and  the  two reads in Fig. \ref{f:ref2} used for illustration  are taken from Fig. \ref{refdem}. The \textit{nodeCache} in Fig. \ref{f:ref2} has been populated 
when loading the reference to the graph, by storing the node address that corresponds to position  \textit{i}. For instance, position 0 on the reference forms the k-mer ACA and the memory address for the node containing this k-mer is stored at index 0 of the \textit{nodeCache}.
This \textit{nodeCache} is utilised when loading the reads to the graph. 
Consider the read ACAGAA and its k-mers in Fig. \ref{f:ref2}. This read is mapped to position 0 on the reference and hence the four k-mers ACA, CAG, AGA and GAA map to positions 0, 1, 2 and 3 respectively. When loading these k-mers to the graph, the corresponding location in the \textit{nodeCache} is looked up as shown using arrows in Fig. \ref{f:ref2}. As this read is exactly the same as the reference, all the accesses are hits to the \textit{nodeCache}. However, in the other read on Fig. \ref{f:ref2}, two misses have occurred due to the marked indel on the read. As explained earlier, variants and read errors are less frequent and therefore, the misses are fewer. In the case of a hit to the \textit{nodeCache}, the node can be directly accessed.  The hash table has to be accessed only in case of a miss to the \textit{nodeCache}. If the \textit{nodeCache} was not used then  all these accesses would have to go through the hash table, and thus access memory randomly.  

Although this \textit{nodeCache} is an array originally residing in the memory, it is cache friendly due to the spatial and temporal locality of accesses to the array. For instance, 
in  Fig. \ref{f:ref2}, observe how the accesses to the \textit{nodeCache} (marked using arrows) for consecutive k-mers in a read exhibit spatial locality.  Temporal locality is observed for accesses to the \textit{NodeCache} due to consecutive reads. In Fig. \ref{refdem}  note how consecutive reads overlap to each other. 
Hence, locations in the \textit{nodeCache} corresponding to a read are re-accessed for the next read, exhibiting temporal locality. For instance, the k-mer CAG  which is accessed for the first read ACAGAA has to be accessed again when processing the second read CAGTAC.  Due to both this spatial and temporal locality among accesses, the \textit{nodeCache} array is  cache friendly. %\textcolor{blue}{ The lack of programmer's control over caches, in general purpose processor, limits the the \textit{nodeCache} to be an array that originally resides in the RAM. However in an ASIP based system,  this \textit{nodeCache} can be made an exclusive cache during custom processor design such as in ASIP.}

In addition to the above, the end node of an edge will always be the start node of the next edge, throughout the read. For instance, the edges will be added in the following order: 1. ACA-CAG; 2. CAG-AGA; and 3. AGA-GAA for the read ACAGAA. Note how CAG and AGA which are the end nodes  for edges 1 and 2, are the starting nodes for edges 2 and 3, which causes repeated accesses to memory locations. %As the reads are generally about 100-150 bases and one read will have a number of edges that in turn will increase the number of double accesses.
This pattern is observed for the reference as well. Though these accesses are already cache friendly due to the temporal locality, they can be made even faster by using a register (which we refer to as \textit{lastAccess}). This \textit{lastAccess} register is used to store the memory pointer to the end node of the current edge,  to be used when loading the start node of the next edge. In an implementation for a general purpose processor, \textit{lastAccess} can be a globally declared variable
%It is possible to exclusively allocate a register for \textit{lastAccess} when using a programming language such as C or assembly. 

Implementation for the method above is given in Algorithm \ref{ref2} and Algorithm \ref{reads2}. Algorithm \ref{ref2} shows how the reference is loaded to the \textit{de Bruijn} graph. A globally accessible variable \textit{lastAccess}  is used for storing the end node of each edge (line 1 of Algorithm \ref{ref2}). A globally accessible array \textit{nodeCache} is initialized with NULL pointers (line 2 of Algorithm \ref{ref2}). The k-mers are extracted from the reference as previously. Function \textit{addEdge} called at line 7 of Algorithm \ref{ref2} now returns the pointer (\textit{ptr1}) to \textit{kmer1} which  is then stored in the \textit{nodeCache} (at line 8 of Algorithm \ref{ref2}).

\begin{algorithm}[ht]
\caption{Load the reference to de Bruijn Graph}\label{ref2}
\begin{algorithmic}[1]
\State \textit{global} \textit{lastAccess = NULL} 
\State \textit{global} \textit{nodeCache[region\_size] = \{NULL\}} 
\Function{\textit{loadReference}}{\textit{reference}}
\For{\textit{i = 0 : (region\_size-kmer\_size)}}
    \State \textit{kmer1} $\gets$ \textit{reference[i : i+kmer\_size]}
    \State \textit{kmer2} $\gets$ \textit{reference[i+1 : i+1+kmer\_size]}
    \State \textit{ptr1} $\gets$ \textit{addEdge(kmer1,kmer2,i,i+1)}
    \State \textit{nodeCache[i]} $\gets$ \textit{ptr1}
    
\EndFor
\EndFunction
\end{algorithmic}
\end{algorithm}

Algorithm \ref{reads2} shows how reads are loaded.  Function \textit{getMappedPosition} at line 3 retrieves the position on the reference to which the read is mapped. The positions of the k-mers are then computed and provided as arguments to \textit{addEdge} (line 7-9 of Algorithm \ref{reads2}).

\begin{algorithm}[ht]
\caption{Load reads to de Bruijn Graph}\label{reads2}
\begin{algorithmic}[1]
\Function{\textit{loadReads}}{\textit{readsInWindow}}
\For{\textit{read} \textbf{in} \textit{readsInWindow}}
	\State \textit{readMapping = getMappedPosition(read)}
	\For{\textit{i = 0 : (read\_length-kmer\_size)}}
    	\State \textit{kmer1} $\gets$ \textit{read[i : i+kmer\_size]}
    	\State kmer2 $\gets$ \textit{read[i+1 : i+1+kmer\_size]}
        \State \textit{pos1} $\gets$ \textit{readMapping+i}
        \State \textit{pos2} $\gets$  \textit{readMapping+i+1}
    	\State \textit{addEdge(kmer1,kmer2,pos1,pos2)}
    \EndFor
\EndFor
\EndFunction
\end{algorithmic}
\end{algorithm}

Algorithm \ref{edge2} elaborates the \textit{addEdge} function. First, \textit{kmer1} (which is the start node of the current edge) is compared with the k-mer in \textit{lastAccess} at line 2 of Algorithm \ref{edge2}. If identical, there is no requirement to inspect the \textit{NodeCache} or hash table.
Otherwise, \textit{LookUp} will be called which will inspect the \textit{NodeCache} at line 5 of Algorithm \ref{edge2}. Unlike \textit{kmer1}, the end node of the edge \textit{kmer2} is directly checked in the \textit{NodeCache} by calling \textit{LookUp} at line 7 of Algorithm \ref{edge2}. Finally, the end node of the current edge is backed up on \textit{lastAccess} for future use and \textit{createLink} is called (line 8-9 of Algorithm \ref{edge2}). \textit{createLink} is same as described previously in Section \ref{s:existing}.

\begin{algorithm}[ht]
\caption{Add an edge connecting kmer1 and kmer2}\label{edge2}
\begin{algorithmic}[1]
\Function{\textit{addEdge}}{\textit{kmer1,kmer2,pos1,pos2}}
	\If{\textit{lastAccess!=NULL} \textbf{and} \textit{lastAccess.kmer==kmer1}}
    \State \textit{ptr1} $\gets$ \textit{lastAccess}
    \Else	
    \State \textit{ptr1} $\gets$ \textit{LookUp(kmer1,pos1)}
     % \If{$ptr1==NULL$}
     %     \State $ptr1 = createNode(kmer1)$
     %     \State $hashTableAdd(ptr1,kmer1)$
     % \EndIf    
    \EndIf
	
    \State \textit{ptr2} $\gets$ \textit{LookUp(kmer2,pos2)}
    %\If{$ptr2==NULL$}
    %	\State $ptr2 = createNode(kmer2)$
    %    \State $hashTableAdd(ptr2,kmer2)$
    %\EndIf   
    \State \textit{lastAccess} $\gets$ \textit{ptr2}
 
	\State \textit{createLink(ptr1,ptr2)}
    \State \Return \textit{ptr1}
\EndFunction
\end{algorithmic}
\end{algorithm}

The function \textit{LookUp} in Algorithm \ref{cache} which is called at lines 5 and 7 of Algorithm \ref{edge2}, attempts to locate the node for the k-mer in the \textit{nodeCache}. If it is a hit to the \textit{nodeCache}, the pointer to the node can be immediately returned (line 3-4 of Algorithm \ref{cache}). The hash table is accessed only in case of a miss (line 6 of Algorithm \ref{cache}).

\begin{algorithm}[ht]
\caption{First lookup in the nodeCache and then in hash table}\label{cache}
\begin{algorithmic}[1]
\Function{\textit{LookUp}}{\textit{kmer,pos}}
\State \textit{cacheItem = nodeCache[pos]}
\If {\textit{cacheItem!=NULL} \textbf{and} \textit{cacheItem.kmer==kmer}}
	\State \textit{ptr} $\gets$ \textit{cacheItem}
\Else
	\State \textit{ptr} $\gets$ \textit{hashTableLookUpOrInsert(kmer)}
\EndIf
\State \Return \textit{ptr}
\EndFunction
\end{algorithmic}
\end{algorithm}

\begin{figure}[!t]
  \centering
    \includegraphics[width=0.7\columnwidth]{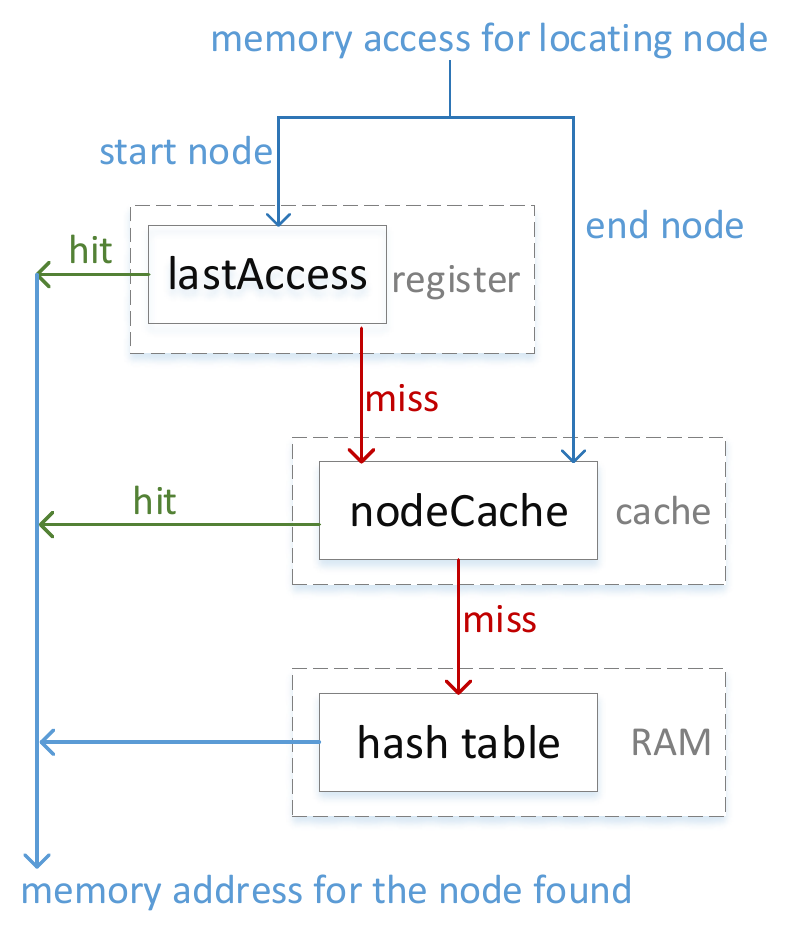}
    \caption{Summary of the outcome of the proposed method} 
    \label{f:summary}
\end{figure}

Fig. \ref{f:summary} summarises how memory accesses are allocated. A memory access to lookup the memory address of a node in the graph corresponds either to a start node or an end node of an edge in the graph (as explained previously). If the access is for a start node, the \textit{lastAccess} register will be inspected as shown in Fig. \ref{f:summary}. If a hit occurs when looking up  the \textit{lastAccess} register, the lookup process completes at the cost of only reading that register. In case of a miss to the \textit{lastAccess} register, the node will be looked up in the \textit{nodeCache} as shown in the figure.  In case of a hit to the \textit{nodeCache}, the process ends there, at the cost of accessing cache memory.  In case of a miss to the \textit{nodeCache}, the hash table has to be accessed as shown. The cost of accessing the hash table residing in the RAM is high, but misses that end up in the hash table are rare (as mentioned previously). For end nodes, the inspection starts directly from the \textit{nodeCache} as shown in  Fig. \ref{f:summary}. In summary, if not for the \textit{lastAccess} register and the \textit{nodeCache} all the accesses would directly go to the hash table causing frequent accesses to RAM.

\section{Results}\label{s:results}

All experiments were performed on a server with four Intel Xeon X7560 processors (total of 32 cores / 64 CPU threads and 256 GB of RAM). The Platypus variant caller (downloaded from \cite{platyoriginal}) that implements the algorithm in Section \ref{s:existing} is referred to as the \textit{baseline implementation}. The modified version of Platypus  based on the method in Section \ref{s:proposed} is referred to as the \textit{optimised implementation}. Three real datasets from the 1000 genomes project (same whole genome sequencing data used in \cite{Rimmer2014} for performance assessment of Platypus, downloaded from \cite{data}) were used for experiments. The three datasets are aligned 75-86 X 100-bp paired-end Illumina HiSeq 2000 reads (BAM files) for the parent-offspring trio NA12878, NA12891 and NA12892. Platypus was run using the default parameters with \textit{de Bruijn} based assembly turned on. Variants were called on all chromosomes including X and Y.
%\textbf{We changed the hash function to 101}

%As explained earlier, Fig. \ref{f:platytime} shows how time consumption for different parts of the variant caller are distributed. De Bruijn graph based assembly involves two major steps; constructing the graph and traversing the graph. For the three datasets,  graph construction takes around  66\% of the total time.  Graph traversal takes about 0.5\% of the total time. All other tasks including aligning reads to haplotypes, computing probabilities and disk accesses  takes around 33\%. Therefore, constructing the graph is the bottle neck which is addressed using the method in \hyperref[s:proposed]{``Methodology''} section. To get the time measurements in Fig. \ref{f:platytime},
%the baseline implementation was run with 64 threads. Times reported by each thread were summed together. 

% \begin{table}[ht]
% \centering
% \caption{Percentage of time spent for de Bruijin based assembly}
% \label{t:percentage}
% \renewcommand{\arraystretch}{1.3}
% \begin{tabular}{|c|c|c|c|}\hline
% {} & dataset 1 & dataset 2 & data set 3 \\\hline
% De Bruijn & 66.89\% & & \\\hline
% Other & 33.11 & & \\\hline

% \end{tabular}
% \end{table}

% \begin{table}[ht]
% \centering
% \caption{Percentage of time spent for each step}
% \label{t:percentage2}
% \renewcommand{\arraystretch}{1.3}
% \begin{tabular}{|c|c|c|c|}\hline
% {} & dataset 1 & dataset 2 & data set 3 \\\hline
% Creating the Graph & 99.38\% & & \\\hline
% Traversing the Graph & 0.62\% & & \\\hline
% \end{tabular}
% \end{table}

\begin{figure}[!t]
  \centering
    \includegraphics[width=\columnwidth]{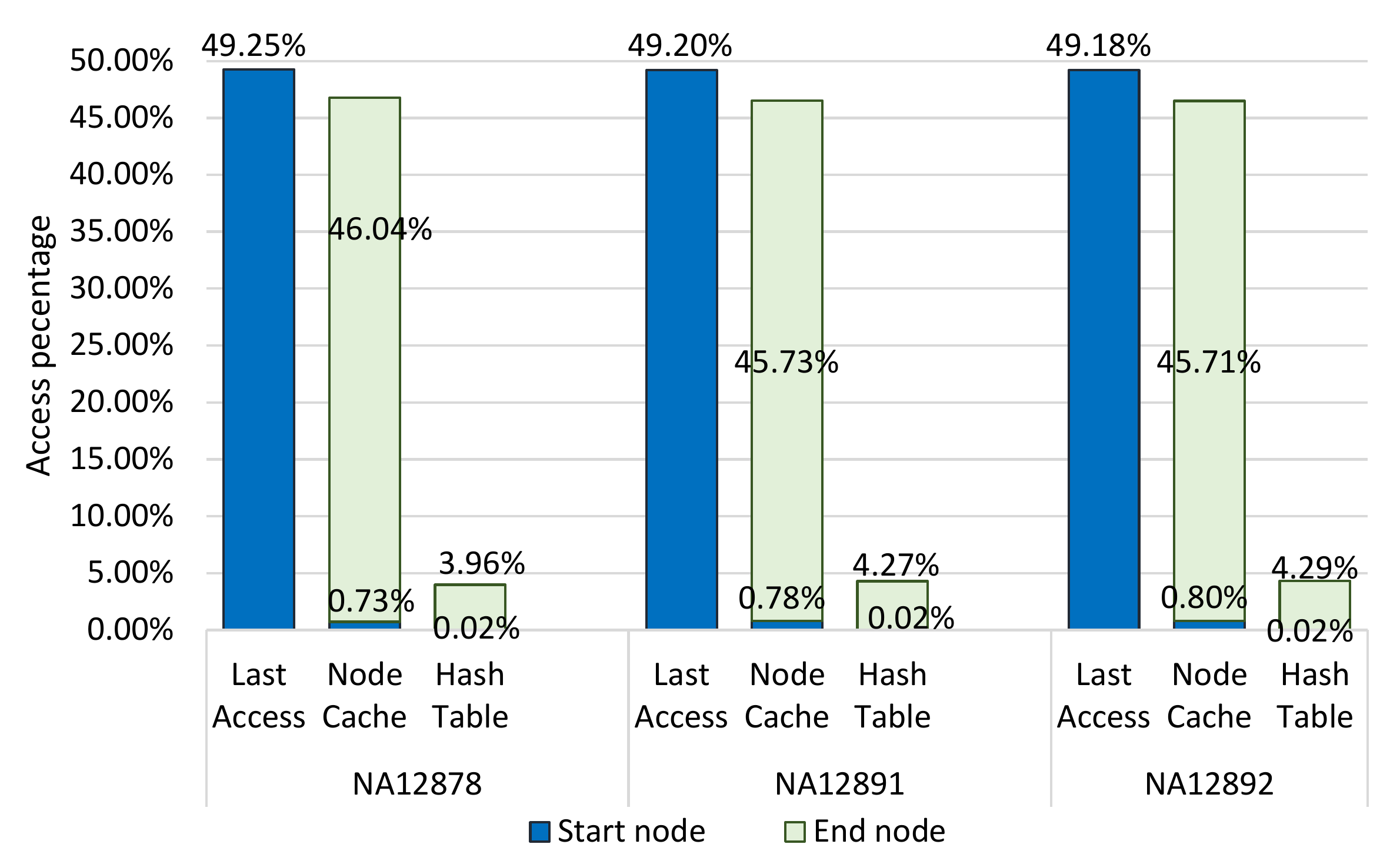}
    \caption{Distribution of memory accesses in the optimised implementation} 
    \label{f:hits}
\end{figure}

Section \ref{s:proposed} described how accesses to the hash table are minimised by using a register that stores the previous node (referred as \textit{lastAccess}) and a cache friendly array (referred as \textit{NodeCache}).
Fig. \ref{f:hits} shows how memory accesses to \textit{lastAccess}, \textit{nodeCache} and the hash table are distributed. The X-axis in Fig. \ref{f:hits} denotes the dataset and the type of memory access. Y-axis shows the access percentage for each item on the X-axis.  The data used to compute the percentages in  Fig. \ref{f:hits} are given in Table \ref{t:memaccess}. These data in the table were obtained by running the \textit{optimised implementation} with software counters introduced to count different memory accesses. The first column of Table \ref{t:memaccess} is the dataset and the second column is the memory access type. The third column contains the number of memory accesses occurred when locating start nodes of the edges in the \textit{de Bruij}n graph. Similarly, the fourth column is for end nodes. The last column is the total number of accesses which is the sum of columns three and four. Note that the numbers are given in x10\textsuperscript{9}. The number of hits to the \textit{lastAccess} register is equal to the number of times the program reaches line 3 in Algorithm \ref{edge2}. Only the accesses to start nodes are responsible for \textit{lastAccess} hits. Therefore, \textit{lastAccess} hits due to end nodes are 0 as shown in the Table. Similarly, \textit{Nodecache} accesses and hash table accesses map to lines 4 and 6 respectively in Algorithm \ref{cache}. The function \textit{LookUp} in Algorithm \ref{cache} called at line 5 of Algorithm \ref{edge2} corresponds to start nodes.  Similarly, \textit{LookUp} called at line 7 of Algorithm \ref{edge2} corresponds to end nodes. The data in Table \ref{t:memaccess} includes accesses occurred when loading both the reference and the reads to the \textit{de Bruijn} graph. The percentage value for each item in Fig. \ref{f:hits} is calculated out of the total memory accesses for that data set. For instance, the total memory accesses for dataset NA12878 is the sum of the three values 317.36,301.35 and 25.66 in the last column of Table \ref{t:memaccess}. The values 317.36, 4.71, 296.64, 0.11 and 25.55 for dataset NA12878 when expressed as a percentage of the above sum equates the percentages 49.25\%,0.73\%, 46.04\%, 0.02\% and 3.96\% respectively in Fig. \ref{f:hits}.

\begin{table}[!t]
%\centering
\caption{Memory access distribution in the optimised implementation}
\label{t:memaccess}
\renewcommand{\arraystretch}{1.3}
\begin{tabular}{|c|p{3.5cm}|p{2cm}|p{2cm}|p{2cm}|}\hline
Dataset & Memory access type & Start nodes ($\times$10\textsuperscript{9}) & End nodes ($\times$10\textsuperscript{9}) & Total ($\times$10\textsuperscript{9}) \\\hline\hline
\multirow{4}{*}{NA12878} & Last Access &  317.36 & 000.00 & 317.36 \\\hhline{~----}
{} & Node Cache & 004.71  & 296.64 & 301.35\\\hhline{~----}
{} & Hash Table & 000.11  & 025.55 & 025.66\\\hline\hline
\multirow{4}{*}{NA12891} & Last Access & 288.57 & 000.00 & 288.57\\\hhline{~----}
{} & Node Cache & 004.59 & 268.26 & 272.85\\\hhline{~----}
{} & Hash Table &  000.11 & 025.02 & 025.13\\\hline\hline
\multirow{4}{*}{NA12892} & Last Access & 284.67  & 000.00 & 284.67\\\hhline{~----}
{} & Node Cache & 004.61 & 264.54 & 269.15\\\hhline{~----}
{} & Hash Table & 000.11 & 024.86 & 024.97\\\hline
\end{tabular}
\end{table}

In Fig. \ref{f:hits}, observe that for all datasets about 49\% of total accesses are hits to the \textit{lastAccess} register. Note that all hits to \textit{lastAccess} are for start nodes. Then about 46.5\% of accesses are hits to the \textit{nodeCache}. The majority are from end nodes, as most of the start nodes have already been resolved through the \textit{lastAccess} register. Observe that only about 4.5\% of the accesses must go to the hash table. The implications of this figure are that only 4.5\% of the accesses are misses to the RAM and therefore the techniques presented in Section \ref{s:proposed} have enabled efficient usage of the memory hierarchy.
However, 4.5\% is higher than the percentage anticipated in Section \ref{s:proposed} mainly because the values in Fig. \ref{f:hits} also include the memory accesses when loading the reference to the graph. Additionally, alignment artefacts would cause mismatches between k-mers in reference and reads, which in turn would also have increased the percentage.

% \begin{table}[ht]
% \centering
% \caption{Memory access distribution for start nodes}
% \label{t:memaccessstartt}
% \renewcommand{\arraystretch}{1.3}
% \begin{tabular}{|c|c|c|c|}\hline
% {} & dataset 1 & dataset 2 & data set 3 \\\hline
% LastAccess hits & 98.50\% & &\\\hline
% NodeCache hits & 1.46\%  & & \\\hline
% Access to hash table &  0.03\% & & \\\hline 
% \end{tabular}
% \end{table}

% \begin{table}[ht]
% \centering
% \caption{Memory access distribution for end nodes}
% \label{t:memaccessend}
% \renewcommand{\arraystretch}{1.3}
% \begin{tabular}{|c|c|c|c|}\hline
% {} & dataset 1 & dataset 2 & data set 3 \\\hline
% NodeCache hits & 92.07\%& & \\\hline
% Access to hash table & 7.93\%& & \\\hline 
% \end{tabular}
% \end{table}

% \begin{table}[ht]
% \centering
% \caption{Memory access distribution for end nodes}
% \label{t:memaccessend}
% \renewcommand{\arraystretch}{1.3}
% \begin{tabular}{|c|c|c|c|}\hline
% {} & dataset 1 & dataset 2 & data set 3 \\\hline
% NodeCache hits & 92.07\%& & \\\hline
% Access to hash table & 7.93\%& & \\\hline 
% \end{tabular}
% \end{table}

\begin{figure}[!t]
  \centering
    \includegraphics[width=\columnwidth]{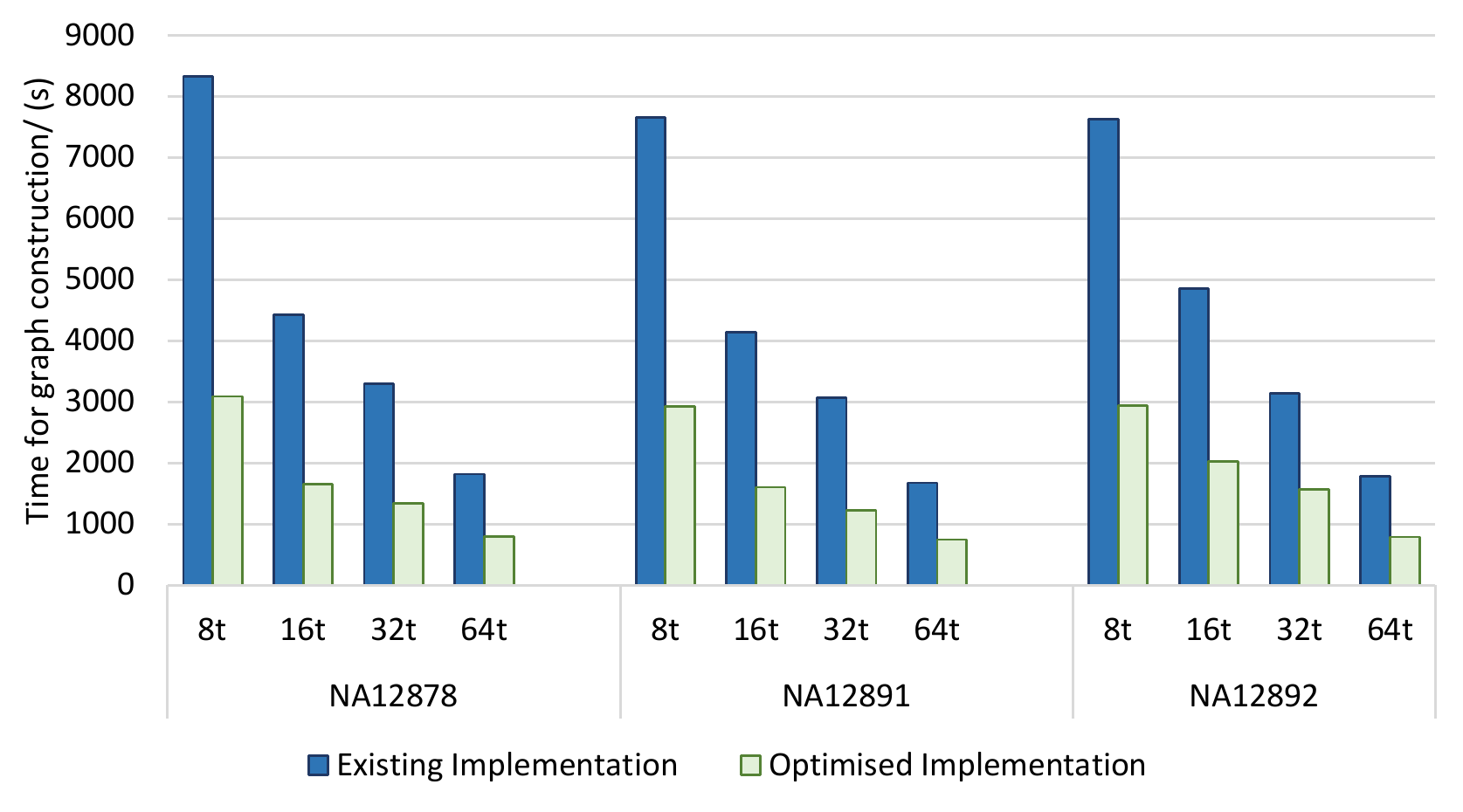}
    \caption{Execution time for the baseline implementation and the modified implementation} 
    \label{f:speedup}
\end{figure}

Fig. \ref {f:speedup} compares the time taken for graph construction by the baseline implementation and the optimised implementation. For each dataset, each implementation was run using 8, 16, 32 and 64 threads, which are shown as 8t, 16t, 32t and 64t along the X-axis. The Y-axis shows the runtime for graph construction for each case, in seconds. In all cases, the optimised implementation was at least twice as fast as the baseline implementation\footnote{The overall speedup of Platypus with our optimised implementation integrated was around 1.4-1.6 times}. %The times used for Fig. \ref {f:speedup} are tabulated in Table \ref{t:timeconsumption}. 
%It is observed that the speed up has slightly decreased with the increase in the number of threads. The competition for the shared L3 cache and the RAM by the threads would have been the reason for this slight decrease.
Since the execution times for Platypus is considerable for the three large datasets used, the execution times given in Fig. \ref {f:speedup} are an average of three repetitions. All three repetitions consistently produced values which were significantly similar on a general-purpose server (Supplementary Table S1). To further validate the claims on the speed-up, we performed two experiments with small datasets so that each test can be repeated a large number of times, with the intention to test the following: 1. the variability of the execution time for the same dataset (randomness due to the operating system scheduling); and 2. the variability in the speedup for different datasets.

In the first experiment we repeatedly ran the baseline implementation and the optimised implementation for a single data set 100 times for each implementation (Supplementary Table S2). Only the chromosome 1 of the NA12878 dataset is considered and executed with 64 threads for 100 repetitions. The two distributions (baseline implementation and optimised implementation) are near normal (Supplementary Figure S1 and S2 - the two outliers are explained in the figures). We performed an independent random sample t-test on log-transformed data to test the null hypothesis that the data in the two distributions come from populations with equal means. The log-transformed values were preferred, as the exponent of the difference between two means provides the speed-up. 
The mean speed-up was 2.047, with a 95\% confidence interval of 2.042-2.053. The null hypothesis could be rejected at the 5\% significance level with a p-value \textless 0.0001. Hence, we may conclude that the observed speed-up is not due to random variations.
%The null hypothesis could be rejected at the 5\% significance level with a p-value of 9.9519x10\textsuperscript{-309}. The confidence interval of the speed-up was 2.0416-2.0532 at a mean speed-up of 2.0474. Hence, it can be concluded that the observed speed-up is not due to random variations.

In the second experiment we executed the baseline implementation and the optimised implementation on 69 different datasets (Supplementary Table S3). Each chromosome (chr 1-22 and chr X) of the three datasets NA12878, NA12891 and NA12892 were considered as a separate dataset (thus the 69 different datasets). A paired sample t-test was performed on the log-transformed data (X - log transformed times for baseline implementation and Y - log transformed times for optimised implementation) to test the null hypothesis that the mean of X-Y is equal to 0 (speed-up is 1). The distribution of X-Y was near normal (Supplementary Figure S3). 
The mean speed-up was 2.028, with a 95\% confidence interval of 2.017-2.039. The null hypothesis could be rejected at the 5\% significance level with a p-value \textless 0.0001. Hence, the speed-up is evident across different datasets.
%The null hypothesis could be rejected at the 5\% significance level with a p-value of 4.7981x10\textsuperscript{-104}. The mean speed-up was 2.0279. The confidence interval was 2.0171-2.0388. Hence, the speed-up is evident across different datasets.

% \begin{table}[!t]
% %\centering
% \caption{Time consumption for the baseline implementation and the optimised implementation}
% \label{t:timeconsumption}
% \renewcommand{\arraystretch}{1.3}
% \begin{tabular}{|c|p{1.5cm}|p{2cm}|p{2cm}|}\hline
% Dataset & No. of CPU threads & Time for the baseline implementation(s) & Time for the optimised implementation(s) \\\hline
% \multirow{4}{*}{NA12878} & 8 &	8328.87 &	3093.34 \\\hhline{~---}
% {} &	16	& 4427.14 &	1657.93 \\\hhline{~---}
% {} &	32	& 3306.32 &	1337.35	\\\hhline{~---}
% {} &	64 &	1817.18 &	0795.15	\\\hline
				
% \multirow{4}{*}{NA12891} &	8 &	7664.28	& 2929.74 \\\hhline{~---}
% {} &	16	& 4143.99 &	1605.01 \\\hhline{~---}
% {} &	32	& 3070.96 &	1232.29	\\\hhline{~---}
% {} &	64	& 1678.63 &	0744.21	\\\hline
				
% \multirow{4}{*}{NA12892} &	8	& 7627.56	& 2945.08 \\\hhline{~---}
% {}&	16 &	4862.93	& 2025.90 \\\hhline{~---}
% {}&	32 &	3137.50	& 1570.20 \\\hhline{~---}
% {}&	64 &	1781.77	& 0790.46 \\\hline
% \end{tabular}
% \end{table}

\section{Discussion}\label{s:discussion}

According to profiling done on the baseline implementation, approximately about:  
\begin{enumerate}[label=(\alph*)]
\item{70\% of the memory accesses  to the RAM (during graph construction) are due to the hash table}, and
\item{30\% are other memory accesses not due to the hash table}. 
\end{enumerate}

Our optimisation technique reduced only (a) memory accesses. The speed up for the whole graph construction process was about 2X. 
However, note that this  speed up was obtained purely by modifying the algorithm that runs on a general purpose processor. On general purpose processors, the programmer's control of the caches and the registers is limited. Hence, \textit{nodeCache} is implemented as an array that originally resides in the RAM, and \textit{lastAccess} is implemented as a global variable. 
In contrast, it is possible to have an exclusive cache for the \textit{nodeCache} and an exclusive register for \textit{lastAccess}  when building a custom processor.
Therefore, the proposed algorithm opens the door to building custom processors such as Application Specific Instruction Set Processors (ASIP), where the baseline algorithm is not suitable. In such a case, the observed speed up would be higher.  
Furthermore, the proposed algorithm can lead to efficient local re-assembly implementations for any other system having a memory hierarchy such as Graphics Processing Units (GPU) and Field Programmable Gate Arrays (FPGA),

RAM accesses in (b) are due to other memory accesses that occur during graph construction, such as,
\begin{enumerate}
\item{reading the reference genome in Algorithm \ref{ref1} (line 3 and 4) },
\item{reading the reads in Algorithm \ref{reads1} (line 4 and 5) }, and
\item{writing to the located nodes to add connection between k-mers inside \textit{createLink} function (called at line 4 of in Algorithm \ref{edge1})}
\end{enumerate}

Out of these, 1 and 2 are already cache friendly due to the spatial locality of access. Accesses to the RAM are still caused when refilling cache lines. In contrast, 3 is not cache friendly due to the large size of the data structure that stores a node. A node contains space for fields such as memory pointers to adjacent nodes and the total node size is even larger than the size of a cache line of a general purpose processor. For instance, the typical cache line size of a CPU is 64 bytes and the size of a node in Platypus is 65 bytes. Hence at least one access to the RAM is required for each node access to fill a cache line. A large cache line size that fits several nodes to one cache line can be implemented during custom processor construction. In addition, strategies such as cache pre-fetching (fetching the next adjacent cache line from the RAM before it is actually required) would be helpful to boost the performance.

%Though the observed speed up is around 2 times, the anticipated speed up according to the memory access distribution in Fig. \ref{f:hits} would have been higher. For instance, if the number of clock cycles for access to the RAM, cache and resisters are estimated to be 100, 10 and 1 respectively, the access percentages in  Fig. \ref{f:hits} gives an average access time of 9.64 clock cycles (100*4.5\%+10*46.5\%+1*49\%). Without the \textit{nodeCache}  and \textit{lastAccess}, the access time will be always close to 100 as all the accesses are directly to the hash table causing random accesses. Hence, a speed-up around 10 times would have been expected. 

%the observed speed up for the whole graph construction process is 2 times. A main reason for not getting 10 times speed up would be the 
%However other cache unfriendly memory accesses such as when accessing a located node (at line 3 of Algorithm \ref{edge1}). This is  cache unfriendly because the structure that holds a node is large in capacity due to numerous fields such as pointers to other nodes. When a structure is close to the size of a cache line (typical size of a cache line is 64 bytes), spatial locality features of accesses cannot be utilised, and hence are not cache friendly. 

\section{Summary}\label{s:conclusion}

The \textit{de Bruijn} graph construction during the local re-assembly step of modern variant callers consumes more than 60\% of the total variant calling time. We have shown how the existing algorithm can be modified such that the locality of memory accesses are improved, which in turn improves the efficient usage of faster cache memories. The results show that these changes improve the performance of \textit{de Bruijn} graph construction by a factor of around two when implemented on a general purpose processor. The modified algorithm opens the door to much higher acceleration of local re-assembly on GPU, FPGA and ASIP. The implementation of the algorithm which is integrated into the Platypus Variant Caller is publicly available at \cite{ourcode}.

%% file: 5-minimap/main.tex
% \newif\ifsupp
% %\supptrue 
% \suppfalse
% \newif\ifmain
% \maintrue
	
% \ifsupp
\newcommand{\suppref}[2]{\ref{#1}} 
% \else
%\newcommand{\suppref}[2]{{#2}} 
% \fi
\newcommand{\revise}[1]{\textcolor{black}{#1}}
\newcommand{\revisestart}{ \color{black} }
\newcommand{\reviseend}{ \color{black} }

\chapter[Featherweight Long Read Alignment]{Featherweight Long Read Alignment using Partitioned Reference Indexes} \label{c:minimap} 

\rule{\textwidth}{0.4pt} 
This chapter is published in Nature Scientific Reports under Creative Commons CC BY license at \textbf{H. Gamaarachchi}, S. Parameswaran, and M. A. Smith, “Featherweight long read alignment using partitioned reference indexes,” Scientific Reports 9, 4318 (2019). DOI: \url{https://doi.org/10.1038/s41598-019-40739-8} \cite{minimap2arm}\\
\rule{\textwidth}{0.4pt}

The advent of Nanopore sequencing has realised portable genomic research and applications. However, state of the art long read aligners and large reference genomes are not compatible with most mobile computing devices due to their high memory requirements. We show how memory requirements can be reduced through parameter optimisation and reference genome partitioning, but highlight the associated limitations and caveats of these approaches. We then demonstrate how these issues can be overcome through an appropriate merging technique. We \revise{incorporated multi-index merging into} the Minimap2 aligner and demonstrate that long read alignment to the human genome can be performed on a system with 2GB RAM with negligible impact on accuracy.

%%%%%%%%%%%%%%%%%%%%%%%%%%%%%
%		INTRODUCTION		%
%%%%%%%%%%%%%%%%%%%%%%%%%%%%%

\section{Introduction}
Long read sequencing has revolutionised genome research by facilitating the characterisation of large structural variations, repetitive regions, and de-novo assembly of whole genomes. Pacific Biosciences (PacBio) and Oxford Nanopore Technologies (ONT) are leading manufacturers that produce long read sequencers. In particular, ONT manufacture sequencers smaller than the size of a mobile phone that can nevertheless output more than 1TB of data in 48 hours. Such highly portable sequencers have realised the possibility of performing genome sequencing in the field. For instance, ONT's MinION sequencer has been used for Ebola virus surveillance in Guinea \cite{quick2016real}, mobile Zika virus surveillance in Brazil \cite{Faria2016}, and for experiments on the International space station \cite{castro2017nanopore}.\\

The advent of highly portable DNA sequencers raise the need for local data processing on devices such as mobile phones, tablets and laptops. Facilitating genomic data analysis on mobile devices avoids the need for high speed internet connections and enables real-time genomic tests and experiments. For Nanopore sequencers, a pico-ampere ionic current signal is produced for each DNA read, which is subsequently converted to nucleotide bases via applied machine learning models. Until recently, a high performance workstation (Quad-core i7 or Xeon processor, 16GB RAM, 1TB SSD) was required for live base calling, the process of converting ionic signal to nucleotide sequences. \\

Most genomic analyses depend on base calling as an initial step, which can be efficiently performed through GPGPU software implementations on graphics cards or, quite conveniently, on dedicated portable hardware (ONT manufacture one such device, termed `MinIT'). Next, base called reads are typically aligned/mapped to a reference, in case of reference guided assembly, or aligned to themselves in case of de-novo assembly. Subsequent analyses (i.e. consensus sequence generation, variant calling, methylation detection, etc) should follow this alignment step. Therefore, an alignment tool that can run on portable devices such as mobile phones, tablets and laptops is the next step in realising the full portability of the whole Nanopore processing pipeline. \\

Minimap2 \cite{minimap2} is a general purpose mapper/aligner that is compatible with both DNA and RNA sequences. Minimap2 can align both long reads and short reads, either to a reference or an assembly contig. Minimap2 first employs hashing followed by chaining for coarse grain alignment. Then it performs an optional base level alignment using an optimised implementation of the Suzuki-Kasahara DP formulation \cite{suzuki2018introducing}. Minimap2 stands out as the current aligner of choice for long reads, among other long read aligners such as BLASR \cite{chaisson2012mapping}, GraphMap \cite{sovic2016fast}, Kart \cite{lin2017kart}, NGMLR \cite{Sedlazeck2018} and LAMSA \cite{liu2017lamsa}; not only is it ~30 times faster than existing long read aligners, but its accuracy is on par or superior to other algorithms \cite{minimap2}. Hash table based approach in Minimap2 has shown to be effective against long reads. In contrast, FM-index \cite{ferragina2000opportunistic} based  short read aligners such as BWA \cite{li2009fast} and Bowtie \cite{langmead2009ultrafast} have shown to fail with ultra long reads (i.e. several hundred kilobases or more) \cite{jain2018nanopore}.\\

Most alignment tools build an index of reference sequences that is stored in volatile memory. Whilst this is manageable for small genomes such as individual bacteria ($\mathtt{\sim}$5Mb), fungi ($\mathtt{\sim}$50Mb) or insects ($\mathtt{\sim}$400Mb), most vertebrates and some plant species require large amounts of memory, as their genomes are in the 1-100 Gb range. \revise{For example, building an index for the GRCh38 human genome reference requires over 11.2GB of volatile memory, and at the very least 8.8GB to map nanopore reads to a pre-computed index with Minimap2.} \\

\revise{To accelerate the development and uptake of real-time genomic applications in field research and point of care medical testing, long read sequence alignment should be performed on ultra-portable computing devices. These can include mobile phones, microcomputer boards, Field-Programmable Gate Arrays (FPGAs), and other embedded systems, such as ONT’s “MinIT” and ``Mk1c MinION'' devices. Such hardware rarely have more than 4GB of RAM, therefore more expensive and less portable equipment is typically required for sequence alignment (high-end laptops, internet connectivity, power generation, etc).} \\

\revise{ Here, we describe strategies for long read alignment to large reference genomes (or collections of genomes) using low amounts of memory. We present an efficient approach to achieve this by splitting a genome index into smaller partitions. Although partitioned indexes are not a novel concept \cite{mohamadi2015dida, doi:10.1093/bioinformatics/bty567, gnanasambandapillai2018mesga}, we expose the caveats of their use on the accuracy of entailing alignments. We present a solution to these issues by merging multi-part alignments via serialisation of internal data structures of the Minimap2 aligner, and demonstrate how this strategy produces alignments that are almost indiscernible from a classical single index using simulated long reads, Nanopore NA12878 reference human genome sequencing data, and a 470kb long chromothriptic read from a human cancer cell line. }

%%%%%%%%%%%%%%%%%%%%%%%%%%%%%
%		RESULTS				%
%%%%%%%%%%%%%%%%%%%%%%%%%%%%%
\section{Results}

\subsection{Effect of parameters on memory usage}

With default options, Minimap2 requires more than 11GB of memory to \revise{create an index from the human reference genome sequence and align Nanopore reads against it (Table \suppref{t:memusage}{S1}). Although the pre-calculated index can be saved to disk, 7.7GB are nonetheless required to subsequently load the index into memory, and between 8.8 and 11.3GB are required when intermediate data structures during alignment are included. This exceeds the average RAM capacities of high-end mobile phones and mid-range laptops.} Hence, running Minimap2 on human data with default options on a typical laptop with 8GB memory or a typical mobile phone with 2GB of memory is not \revise{feasible}. \\

\begin{table}[!ht]
\centering
\caption[Memory usage of Minimap2 for default parameters]{
{\bf Memory usage of Minimap2 for default parameters}}
\begin{tabular}{|p{8cm}|r|r|}\hline
{} & \textbf{PacBio} & \textbf{Oxford nanopore}\\ \hline
\textbf{Index construction} & 8.67 GB & 11.32 GB\\ \hline
\textbf{Index residence} & 6.33 GB & 7.71 GB\\ \hline
\textbf{Mapping with base-level alignment (SAM output)} & 8.56 GB & 11.30 GB \\ \hline
\textbf{Mapping without base-level alignment (PAF output)} & 7.14 GB &8.80 GB \\ \hline
\end{tabular}
\begin{flushleft} \footnotesize Minimap2 was run with default parameters. Pre-set profiles \textit{map-pb} and \textit{map-ont} were used for PacBio and Oxford Nanopore, respectively. The peak memory usage for each event is presented. Index construction refers to the building of the index and then serialising the index to a file. Index residence is the memory required only for the index to reside in memory, such as when loading a pre-built index. 
\end{flushleft}
\label{t:memusage}
\end{table}

We therefore tested the relative effect of alignment parameters on peak memory usage in Minimap2 (see Materials and methods) \revise{ to investigate if parameter optimisation alone can significantly reduce the memory requirements without compromising alignment quality. For this purpose, we used Sequins---synthetic DNA spike-in controls that are designed from the reverse or `mirrored' human genome sequence \cite{deveson2016representing}. This chirality reproduces diverse properties of the human genome, such as nucleotide frequencies, complexity, repetitiveness, somatic variation, etc. As detailed in Materials and methods, we aligned Nanopore sequencing data from Sequins to both native and reversed (not complemented) human reference genomes to compare the relative impact of Minimap2 parameters on alignment accuracy. Specifically:}
\begin{itemize}
\item \textit{k} the minimiser k-mer length (default = 15 for ONT data);
\item \textit{w} the minimiser window size (default = 10);
\item \textit{t} the number of threads (default = 4);
\item \textit{K} the number of query bases loaded into memory at a time (default = 500M). 
\end{itemize} 

\begin{figure}[!htp]
\centering
\includegraphics[width=\linewidth]{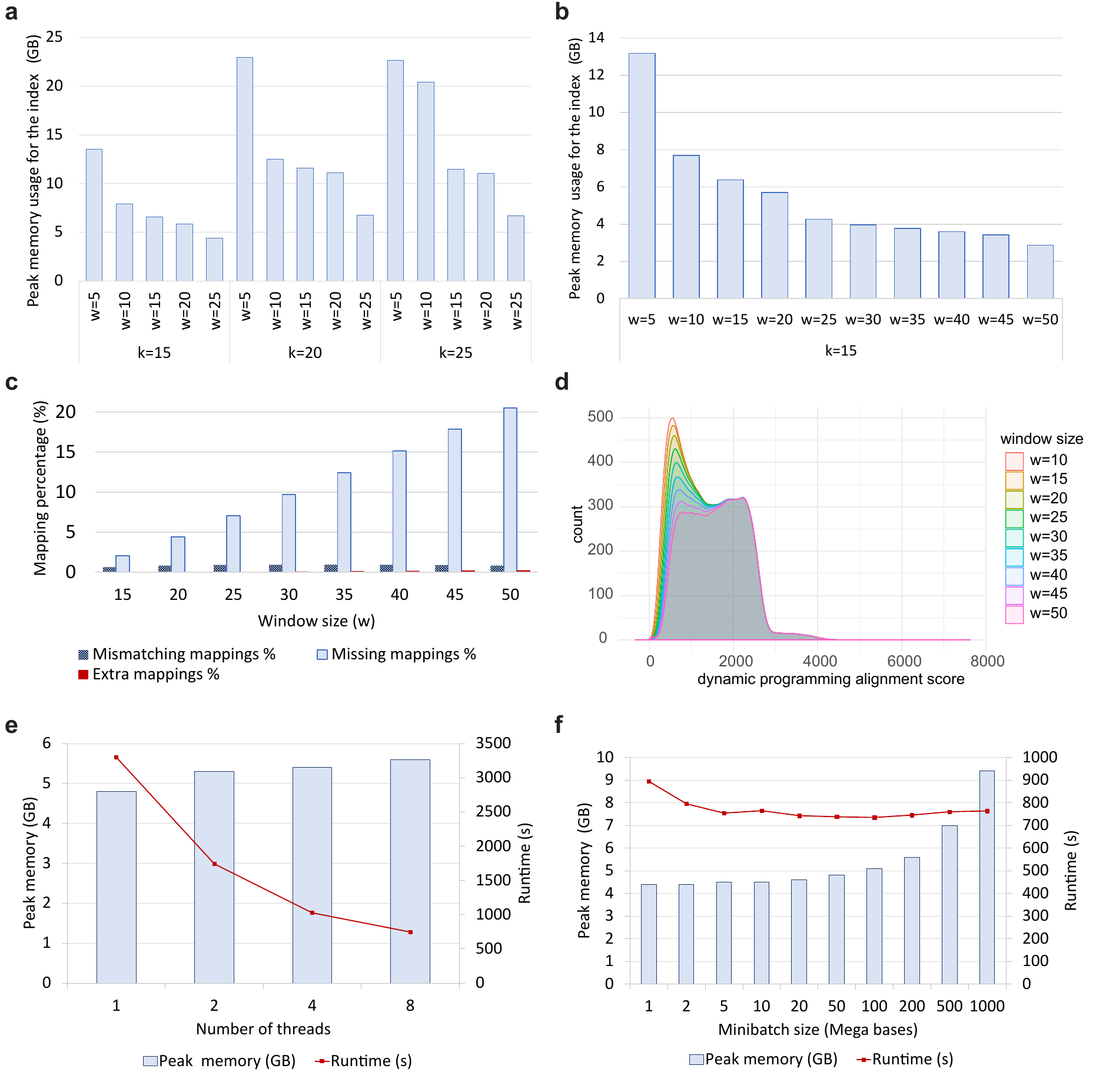}
\caption[Effect of parameters on  memory usage, performance and accuracy]{{\bf Effect of parameters on  memory usage, performance and accuracy.}\\ \footnotesize (\textbf{a}) Peak memory usage of the index for different combinations of \textit{k} and \textit{w}. (\textbf{b}) Peak memory usage of the index for a large range of \textit{w} with \textit{k}=15. (\textbf{c}) \revise{The effect of \textit{w} on sensitivity and error relative to the default window size.} The x-axis is the minimiser window size (\textit{w}). The k-mer size is held constant at 15 for all values of \textit{w}. The y-axis shows the number of missing mappings / mismatches or extra mapping (compared to the mappings from default \textit{w}=10) as a percentage of the number of reads. (\textbf{d}) \revise{Distribution of the dynamic programming alignment score for different minimiser window sizes (\textit{w}). The x-axis is the score and the y-axis is the smoothed number of mappings for a particular score. Note that the distribution is smoothed to show the trend.}
%The mismatches in C are further classified. About 80\% of the mismatches are chromosome mismatches and the remaining 20\% are different loci in the same chromosome.
(\textbf{e}) Effect of the number of threads on memory and performance. The parameters \textit{k}, \textit{w} and \textit{K} were held constant at 15, 25 and 200M respectively while changing the number of threads. Both the peak memory usage and the runtime were measured on a PC with an Intel i7-6700 CPU and 16GB of RAM. (\textbf{f}) Effect of the number of query bases loaded at a time. The parameters \textit{k}, \textit{w} and \textit{t} were held constant at 15, 25 and 8 respectively. }
\label{f:fig1}
\end{figure}

Parameters \textit{k} and \textit{w} considerably affect the peak memory usage for holding the index in memory (Figure \ref{f:fig1}a). For an index without homo-polymer compression, \textit{k} =  15 consumed the least amount of memory out of the values tested, as expected. In fact, the default k-mer size for \revise{ONT data} in Minimap2 (pre-set command line parameter \textit{map-ont}) is 15.

Unsurprisingly, parameter \textit{w} has the most prominent impact on memory usage, which decreases considerably when increasing \textit{w} (Figure \ref{f:fig1}b). At \textit{w}=50, memory usage is capped at 3GB, but the sensitivity \revise{(see Materials and methods) is substantially reduced compared to the default value of parameter \textit{w} (missing mappings } in Figure \ref{f:fig1}c). A larger \textit{w} of 50 reduces sensitivity compared to the default \revise{value of} \textit{w} by 20\%, \revise{whereas} a \textit{w} of 25 entails an \revise{apparent} reduction in sensitivity of about 7\% \revise{while nonetheless requiring} more than 4GB of memory. \revise{Although} sufficient for a computer with 8GB of RAM, this is still too high for smaller devices.\\

\begin{figure}[!ht]
\begin{center}
\includegraphics[width=\textwidth]{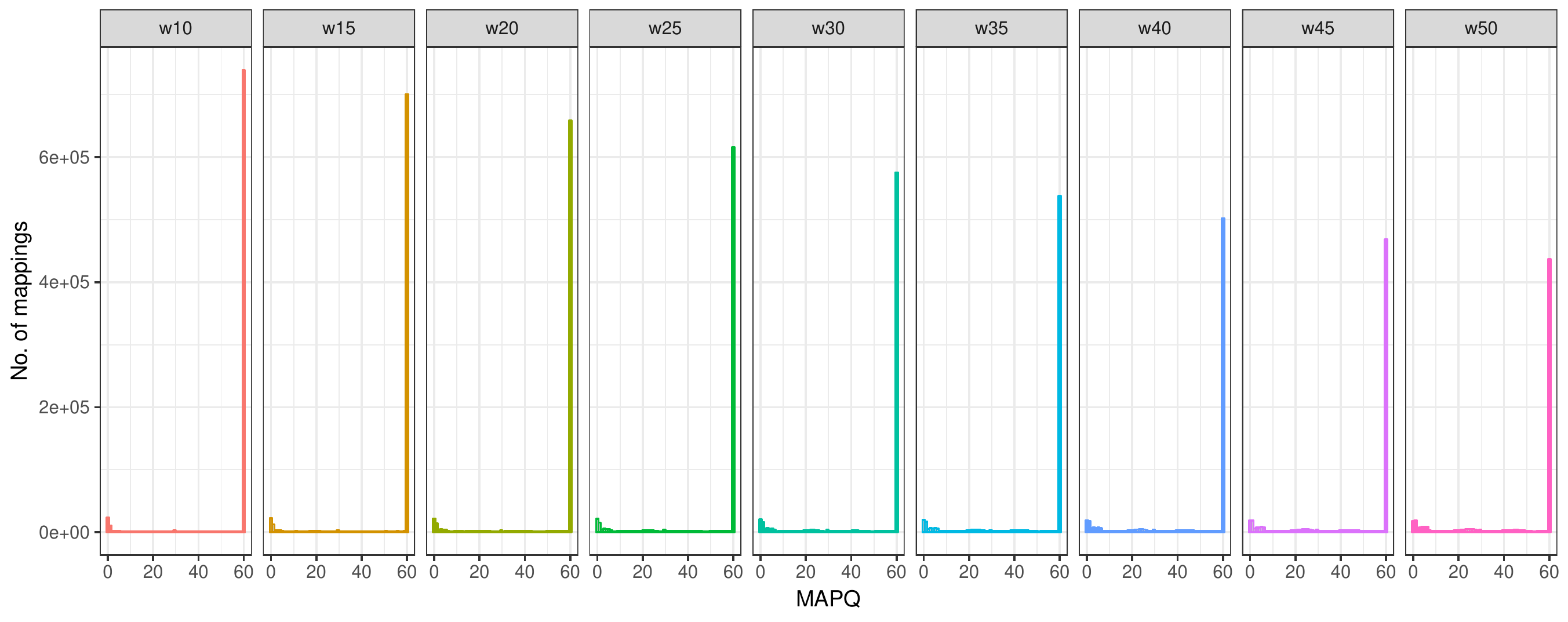}
\end{center}
\caption[Effect of the window size parameter on the MAPQ distribution for synthetic spike-in controls]{{\bf Effect of the window size parameter \textit{w} on the distribution of mapping qualities (MAPQ) for synthetic spike-in controls.}  }
\label{f:mapq_distrib}
\end{figure}

Importantly, the \revise{amount of mismatched mappings in} mapped reads are not significantly affected by the \textit{w} parameter (mismatches in Figure \ref{f:fig1}c), \revise{nor are the high-quality alignments as demonstrated by their dynamic programming (DP) alignment score distribution (cf. DP score > 2000 in Figure \ref{f:fig1}d). However, lower ($\approx$1000) DP scores are less frequent with increasing window size, as are alignments with high MAPQ scores, while those with low MAPQ scores are more prevalent (Figure \suppref{f:mapq_distrib}{S1}). The effect of the window size is also apparent with simulated PacBio reads, where both the sensitivity and error-rate of alignments are negatively affected (Figure \suppref{f:synth_window}{S2})}.\\

The number of threads marginally increases peak memory usage   (Figure \ref{f:fig1}e). Only about 0.8GB of additional memory was consumed when moving from 1 to 8 threads, while producing a 6-fold gain in speed. Hence, reducing the number of threads for the sake of reduced memory usage is not an efficient solution. \\

Intuitively, the number of query bases loaded to the memory at once (also known as the mini-batch size) heavily impacts peak memory usage. This affects the size of the internal data structures used for mapping, but does not affect the index size. Hence, this parameter does not affect alignment accuracy or the sensitivity. A lower mini-batch size reduces the peak memory usage, at the cost of reduced multi-threading efficiency (Figure \ref{f:fig1}f). The runtime drops when changing the mini-batch size from 1M to 5M. However, the runtime is relatively stable from a mini-batch size of 5M onwards. It is important to note that the values in Figure \ref{f:fig1}f are only valid for 8 threads. A large number of threads would require a large mini-batch size for optimal performance. \revise{} \\

\revise{Although parameter adjustments (small minimiser window size value and mini-batch sizes from 5M-20M in particular) can be suitable for systems with limited RAM (for 8 CPU threads or less), tuning parameters alone cannot bring down the memory usage to a value lesser than 4GB due to a substantial loss of sensitivity. As a consequence, this inspired us to investigate the use and suitability of partitioning (or splitting) the reference sequences into distinct indexes.}

\subsection{Caveats of naive partitioned indexes}

\begin{figure}[!htp]
	\centering
		\centering
		\includegraphics[width=0.8\linewidth]{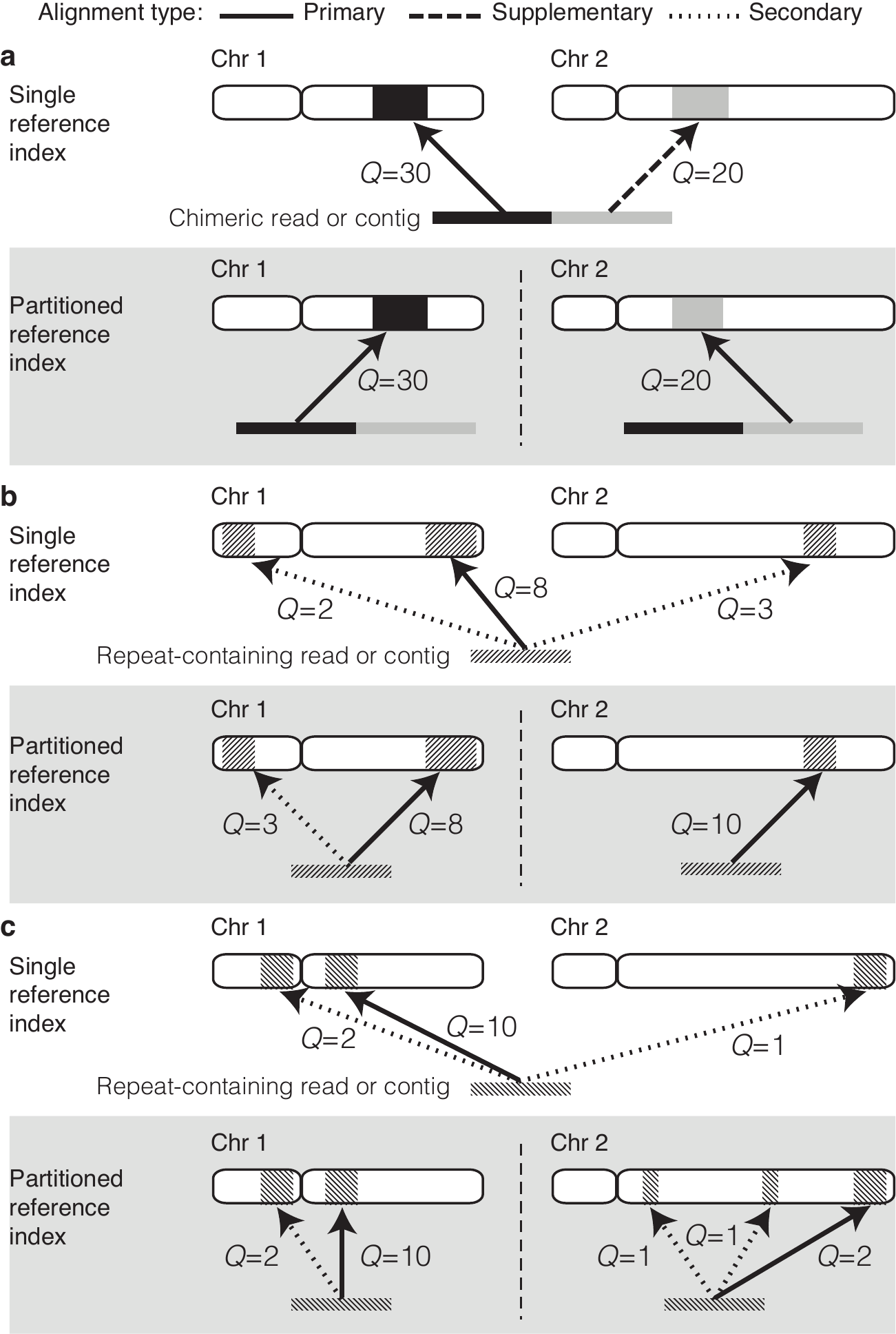}
        \caption[Effect of aligning sequences to single vs partitioned indexes]{\textbf{Effect of aligning sequences to single vs partitioned indexes.}
        
\footnotesize Uniquely mapping chimeric reads (\textbf{a}) can be reconstructed from a partitioned reference index with relative ease. However, sequences (or sub-sequences) that are difficult to map (i.e. low complexity regions, repetitive elements, etc) can cause artefacts when aligning to a partitioned reference index. (\textbf{b}) An example where one partition (chr2) contains less homologous sequences to the query sequence, producing the situation where the best alignment when using a single reference is not achieved. (\textbf{c}) An example where a partitioned reference introduces several additional low quality mappings that would be dismissed with a single reference index. \textit{Q}: mapping quality score.}
        \label{f:multissue}
\end{figure}

Minimap2 allows the reference index to be split by a user specified number of bases through the option \textit{I}, effectively dividing a reference into smaller indexes of comparable size. This facilitates parallel computation and, in theory, enables lower peak memory requirements. However, this feature is not ideal for mapping single reads to large references, mainly because global contiguous information about the reference is unavailable. As a result, several mapping artefacts can occur, as listed below and in Figure \ref{f:multissue} (N.B. these may not be as prominent when overlapping reads--the application for which index partitioning in Minimap2 was originally developed).

\begin{enumerate}
\item The mapping quality is incorrect.\\
The mapping quality estimated in Minimap2 is accurate as it deliberately lowers the mapping quality for repetitive hits. However, this is not possible when only a fraction of a whole genome is present in the index (see supplementary materials of Li, H. \cite{minimap2}). In a partitioned index, if the same repeat lies across different partitions, the mapping quality will be overestimated (Figure \ref{f:multissue}b.)

\item Incorrect alignment flags.\\
For a chimeric read where different sub-sequences map to different chromosomes, the supplementary mappings would be marked as primary mappings (Figure \ref{f:multissue}a). A repeat containing read that maps to multiple locations across different partitions will have multiple primary alignments instead of a single primary alignment (Figure \ref{f:multissue}b and Figure \ref{f:multissue}c).

\item Large output files.\\
A spurious unmapped record will be printed for each partition of the index where a particular read does not map to.  Furthermore, if a maximal amount of secondary alignments are specified, that number of secondary alignments would be output for each partition (Figure \ref{f:multissue}c). Hence, the more partitions used, the larger the output files will be. Such large outputs not only waste disk space, but they are also time consuming to parse or sort.

\item Multiple hits of the same query may not be adjacent in the output \cite{minimap1git}\\
This causes difficulties to analyse or evaluate mapping results. For instance, the \textit{Mapeval} utility in \textit{Paftools} (a tool bundled with Minimap2 for evaluating alignment accuracy) is not compatible with such outputs. Sorting by the read identifier would fix the issue, but requires significant computations for large files.

\item{Incomplete headers in the sequence alignment/map (SAM) output}\\
For a partitioned index, Minimap2 suppresses the reference sequence dictionary (SQ lines) in the SAM header. Users must manually add SQ lines to the header for compatibility with downstream analysis tools.
\end{enumerate}
 
\noindent We resolved these issues by \revise{serialising and storing} the internal state of Minimap2 while mapping reads, then merging the output and processing the result \textit{a posteriori} (see Materials and methods). The accuracy of this technique is discussed below. \\

\subsection{Effect of using a partitioned index on alignment accuracy}

We compared the alignment accuracy between a single reference index and a partitioned index, with and without merging the output. The following acronyms will be used in the subsequent text (see Materials and methods for more details):
\begin{itemize}
\item \textit{single-idx}: Aligning reads to a single reference index;
\item \textit{part-idx-no-merge}: Aligning reads to a partitioned index without merging the output;
\item \textit{part-idx-merged}: Aligning reads to a partitioned index while applying our merging technique.
\end{itemize}

\subsubsection{Synthetic long reads}

\begin{figure}[!htp]
	\centering
    \includegraphics[width=\linewidth]{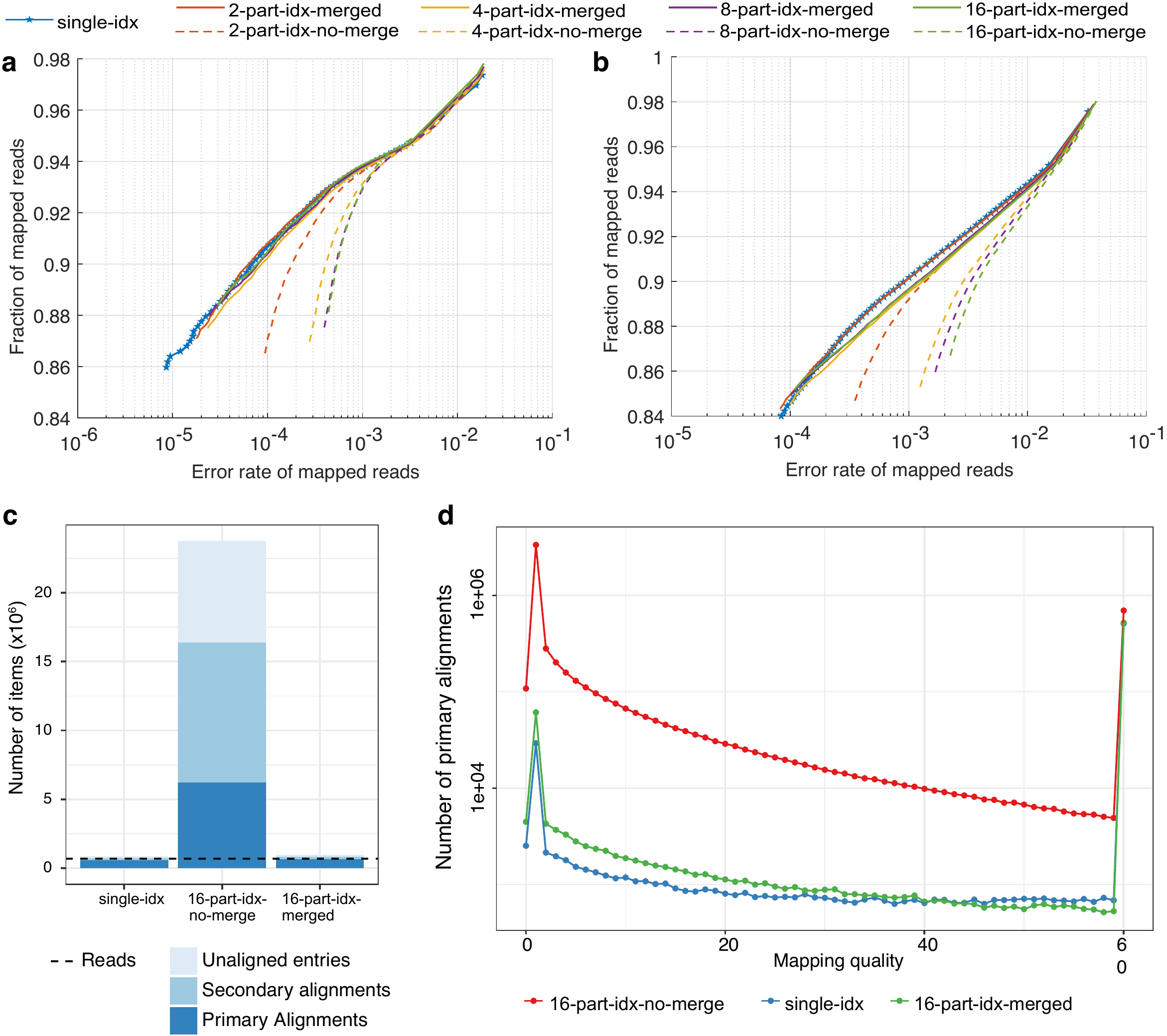}
\caption[Effect of using partitioned indexes versus a single reference index on alignment quality]{\textbf{Effect of using partitioned indexes versus a single reference index on alignment quality.}\\ \footnotesize
\revise{(\textbf{a})} Base-level  and \revise{(\textbf{b})} locus- or block-level  alignment accuracy from synthetic long reads. The x-axis shows the error rate \revise{of alignments in log scale} (see Materials and methods). The y-axis shows the fraction of aligned reads out of all input reads. Each point in the plot corresponds to a mapping quality threshold that varies from 0 (top right) to 60 (bottom left). \revise{These plots are akin to precision-recall plots with the x-axis inverted}. (\textbf{c}) and (\textbf{d}): Alignment statistics for Nanopore whole genome sequencing data from NA12878 \cite{jain2018nanopore} using a 16-part index. (\textbf{c}) The number of total entries (primary+secondary+unaligned) for \textit{single-idx}, \textit{16-part-idx-merged}, and \textit{16-part-idx-no-merge}, in log scale. The dotted horizontal line represents the number of reads. (\textbf{d}) Number of primary mappings in function of Minimap2 mapping quality (log scale).}
\label{f:fig3}  
\end{figure}

Synthetic long reads were used as a ground truth for the evaluation of alignment accuracy (see Materials and methods). The accuracy of \textit{part-idx-merged} is similar to \textit{single-idx}, despite employing significantly more partitions (Figure \ref{f:fig3}a and \ref{f:fig3}b) as exemplified by the overlap of their respective curves.

In contrast, the results of \textit{part-idx-no-merge} are considerably less accurate, in particular for larger quantities of index partitions. A lower error rate is observed for \textit{part-idx-merged} when compared to \textit{single-idx} for the lowest mapping quality values, but this effect is marginal and is associated with low sensitivity. \\

\subsubsection{For real Nanopore NA12878 reads}

As no ground truth is available for biological data, we evaluated alignment accuracy by comparing the number of primary/secondary alignments and unmapped reads across single and multi-partition indexes. When using \textit{single-idx}, Minimap2 outputs 12.1GB of base-level alignment data in SAM format, whereas \textit{part-idx-no-merge} generates much larger output (180GB). However, \textit{part-idx-merged} generates 12.4GB of data--proportional to the output produced with \textit{single-idx}. Hence, \textit{part-idx-merged} reduces disk usage by about 14-fold compared to \textit{part-idx-no-merge}. Peak disk usage is also minimised in \textit{part-idx-merged} as only intermediate alignments are \revise{serialised} as temporary binary files. The resulting size of temporary files generated with \textit{part-idx-merged} is 29.2GB, thus achieving maximal disk usage of 41.6GB, 4 times less than \textit{part-idx-no-merge}. The increased output produced by \textit{part-idx-no-merge} is due to redundant unmapped entries and spurious mappings (Figure \ref{f:fig3}c and Table \suppref{t:stats}{S2}). \\

\begin{table}[!ht]
\centering
\caption[Statistics for alignment outputs for 689,781 reads from NA12878]{
{\bf Statistics for alignment outputs for 689,781 reads from NA12878}}
\begin{tabular}{|p{4cm}|r|r|r|}\hline

{} & \textbf{single-idx} & \textbf{16-part-idx-no-merge} & \textbf{16-part-idx-merged}\\\hline
\textbf{File size (SAM file)} &12.1 GB &180 GB &12.4 GB\\\hline
\textbf{No of SAM entries}&862,427&23,749,310&969,223\\\hline
\textbf{No of unaligned entries}&127,177&7,365,979&120,775\\\hline
\textbf{No of  aligned entries} & 735,250&16,383,331& 848,448 \\\hline
\textbf{No of primary alignments}&592,748&6,228,567&654,302\\\hline
\textbf{No of secondary alignments}&142,502&10,154,764&194,146\\\hline
\end{tabular}
\label{t:stats}
\end{table}

The number of total entries (primary + secondary + unaligned) for \textit{single-idx} and \textit{part-idx-merged} are comparable to the number of input reads (689,781), while \textit{part-idx-no-merge} generates abundant---presumably spurious---hits. Furthermore, the distribution of mapping qualities for primary alignments between \textit{part-idx-merged} and \textit{single-idx} are quite similar (Figure \ref{f:fig3}d). Interestingly, \textit{part-idx-merged} produces slightly more primary alignments with lower mapping quality scores than \textit{single-idx}, a likely consequence of sampling less repetitive regions in partitioned indexes. All the strategies produce almost the same amount of mappings with quality = 60. In contrast, \textit{part-idx-no-merge} has a very high number of spurious mappings for mapping qualities between 0 to 59. \\

\revisestart

\begin{figure}[!ht]
\begin{center}
\includegraphics[width=\textwidth]{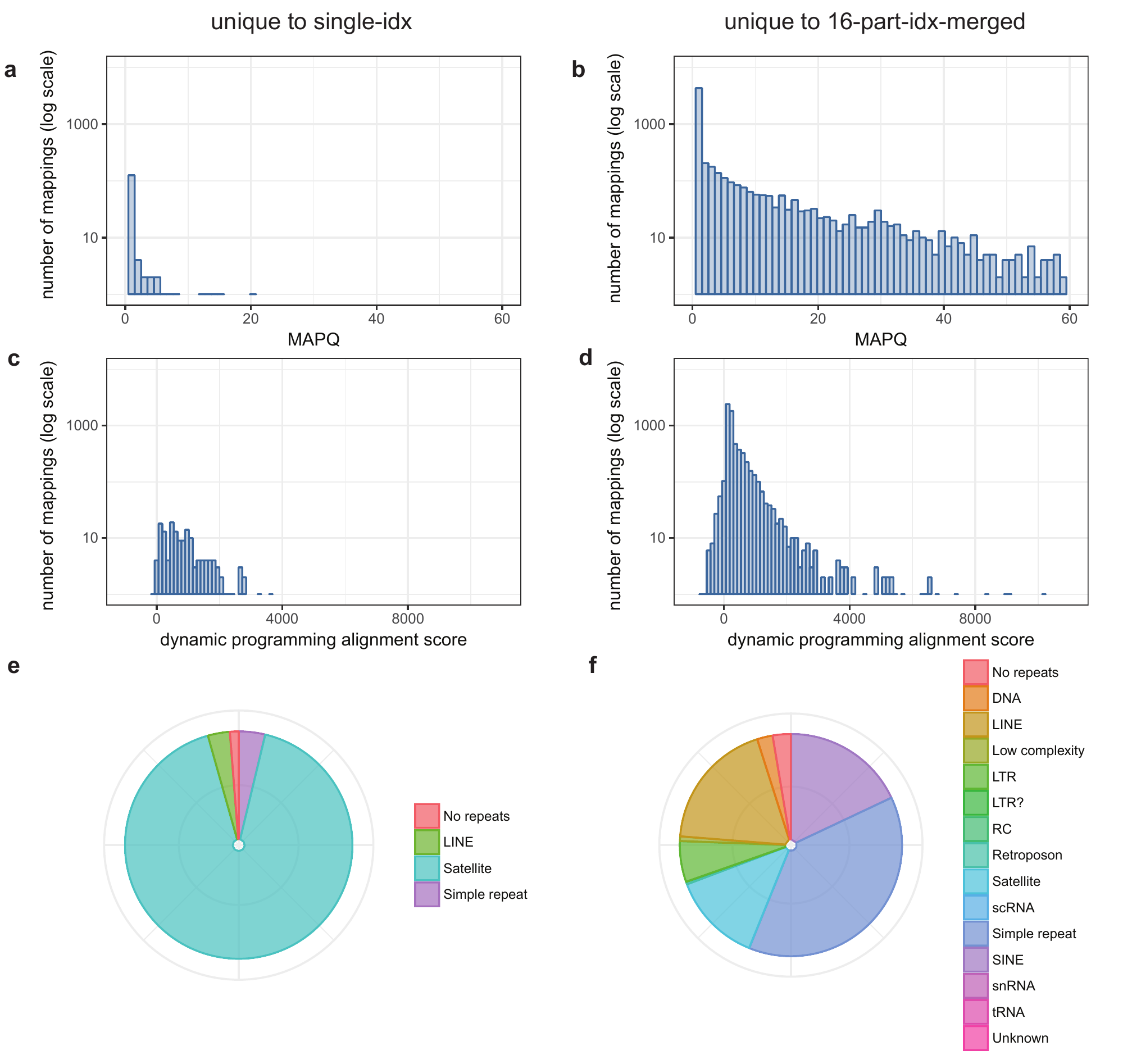}
\end{center}
\caption[Features of alignments that were discordantly mapped in \textit{single-idx} and \textit{16-part-idx-merged} indexing strategies]{{\bf Features of alignments that were uniquely mapped in single (left column) and multiple (right column) partition indexing strategies.\\}
\footnotesize \textbf{(a)} and \textbf{(b)} are the distribution of mapping quality scores (MAPQ). \textbf{(c)} and  \textbf{(d)} are  the distribution of dynamic programming alignment scores. \textbf{(e)} and \textbf{(f)} are the proportion of reads that map to regions of the human genome annotated as repeat elements (Repeat masker track of GRCh38 UCSC genome browser). }
\label{f:unique_sm}
\end{figure}

\begin{figure}[!ht]
\begin{center}
\includegraphics[width=\textwidth]{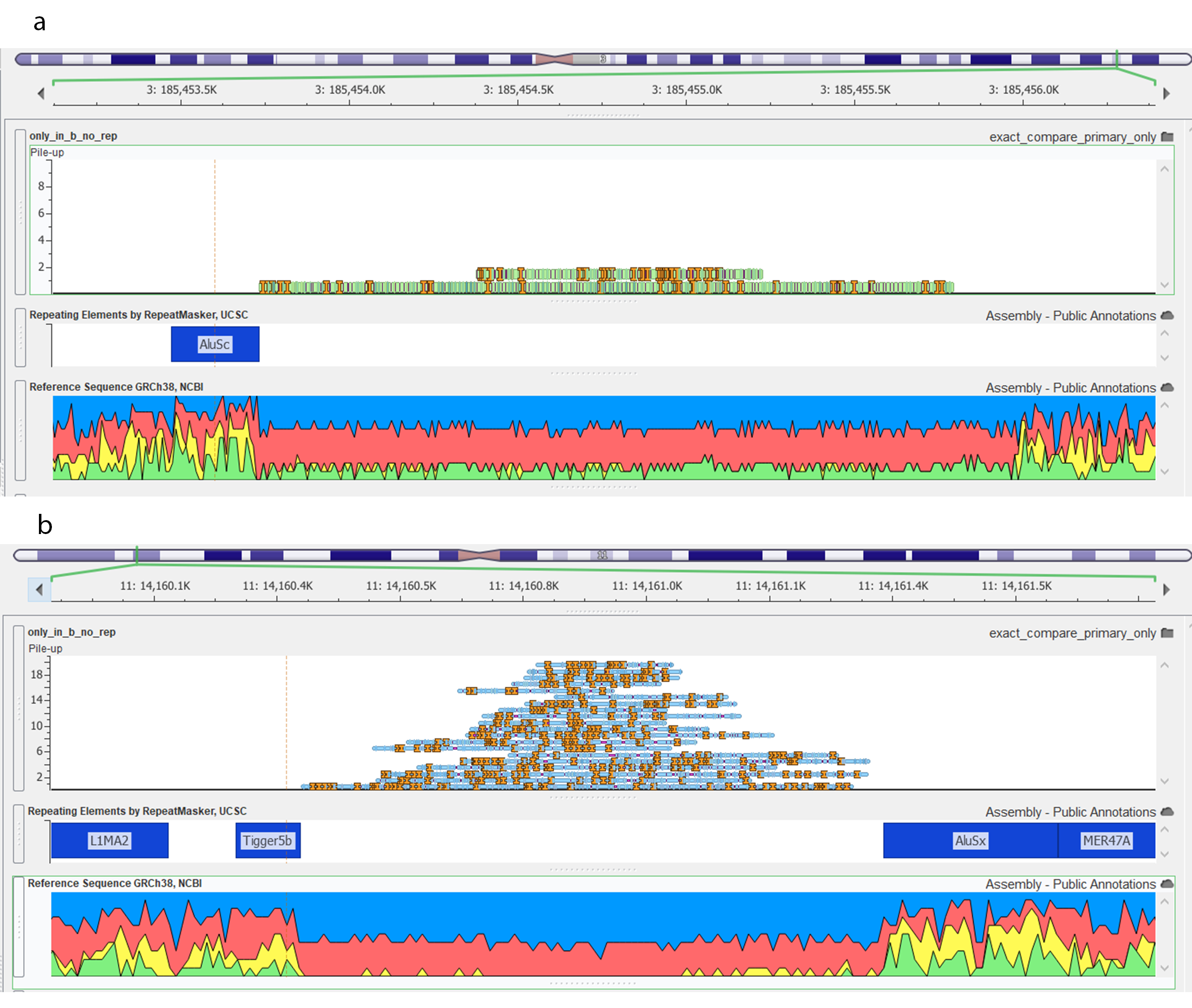}
\end{center}
\caption[Genome browser screenshots of alignments unique to the \textit{16-part-idx-merged} that do not overlap annotated repeats]{{\bf Genome browser screenshots of alignments unique to the 16-part index that do not overlap annotated repeats.} 
\footnotesize
Out of the 281 alignments only found in \textit{16-part-idx-merged} with no overlap with repeat regions, the two highest scoring alignments were found in the regions
(a) chr3:185,453,170-185,456,442 and (b) chr11:14,160,389-14,161,287 respectively. The screen shots are from the Golden Helix GenomeBrowse [http://goldenhelix.com/products/GenomeBrowse]. The first track of each screen shot shows the pile-up of the alignments for the genomic region. The second track shows the repeat elements from the UCSC Repeatmasker track for GRCh38. The third track visualises the relative sequence composition of the GRCh38 reference for the particular region (A - red, C - yellow,  G - green and T - blue).}
\label{f:misseq}
\end{figure}

A more detailed comparison of 689,781 ONT reads aligned using \textit{single-idx} and \textit{16-part-idx-merged} revealed the following:
120,623 (17.49\%) reads were unmapped in both;
 152 (0.02\%) reads mapped only in \textit{single-idx};
 6,554 (0.95\%) reads mapped only in \textit{16-part-idx-merged};
 562,452 (81.54\%) reads mapped in both. 
Less than 1\% of all reads presented discordant mappings when using a single or a multi-part index. Of those discordant mappings, 6,423 (95.8\%) overlapped regions in the human genome annotative as repeat elements or low-complexity sequences, the majority of which were satellites and ALR/Aplha repeats from centromeres (Figure \suppref{f:unique_sm}{S3} and \suppref{f:misseq}{S4}). 
Furthermore, most (97.7\%) of these index-specific unique alignments stem from the multi-part index, which suggests that a reduced search space can help Minimap2 map less complex sequences, presumably through more frequent recourse of the dynamic programming step in Minimap2.\\

% Among the 152 reads that mapped only in \textit{single-idx}
% 150 contains known repeat regions in repeat masker. Out of 150, 141 had ALR/Alpha repeats. The 2 reads that did not involve repeats had very low dynamic programming alignments scores (106 and -26) and MAPQs (2 and 1). 
% 6,273 of 6,554 reads mapped only in \textit{part-idx-merged} involved repeat regions. 1712 of 6,273 contained (AGAAT)n repeats and 1353 contained  ALR/Alpha repeats. Out of the ones that did not involve repeat regions one with maximum DPA score 1567 had a MAPQ of 1. The one with maximum MAPQ 32 had a DP score of only 92.

Among the 562,452 reads that mapped in both \textit{single-idx} and \textit{16-part-idx-merged}, 545,306 (96.95\%) had the exact same primary mappings (same chromosome, strand and position).
Out of the remaining 17,146 aligned reads (3.05\%): 
2,748 (16.02\%) of mapping coordinates overlapped by at least 10\% in both sets;
 952 (5.55\%) were classified as supplementary mappings in \textit{16-part-idx-merged} and the primary mapping in \textit{single-idx}; 
 3,891 (22.69\%) were classified as secondary mappings in \textit{16-part-idx-merged} and primary mapping in \textit{single-idx}.   \\

\begin{figure}[!ht]
\begin{center}
\includegraphics[width=\textwidth]{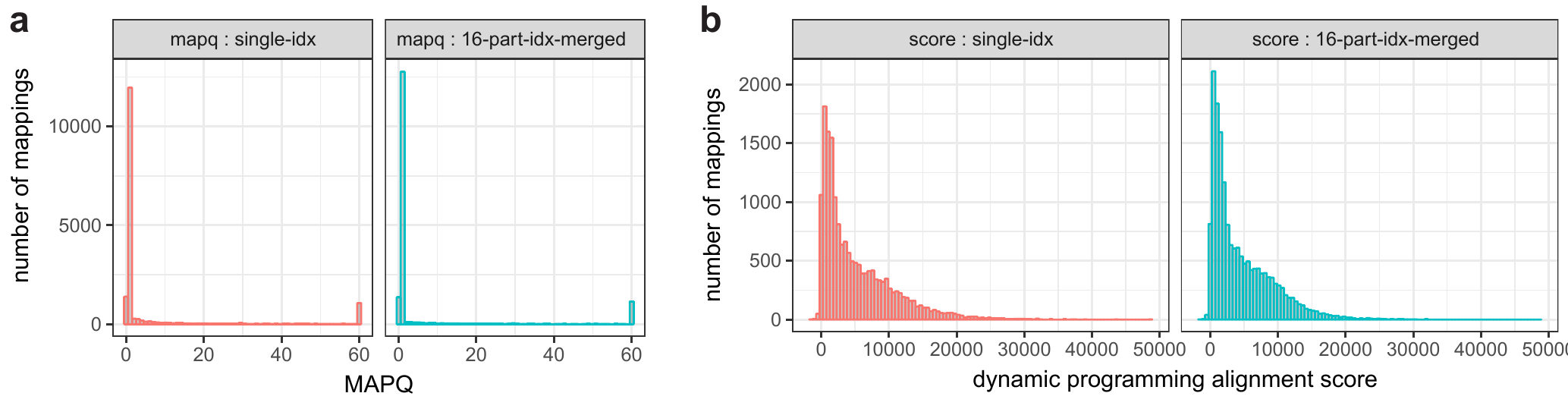}
\end{center}
\caption[Distribution of MAPQ and DP alignment score for the disparate mappings (different by at least one base position) between \textit{single-idx} and \textit{16-part-idx-merged}]{\textbf{ Distribution of (a) mapping quality and (b) dynamic programming alignment score for the disparate mappings (different by at least one base position) between \textit{single-idx} (left) and \textit{16-part-idx-merged (right)}.}

\footnotesize Disparate mappings here refer to the 17,146 aligned NA12878 reads with different primary mappings (different by at least one base position) between the two partition strategies. }
\label{f:mishis}
\end{figure}

\begin{figure}[!ht]
\begin{center}
\includegraphics[width=0.8\textwidth]{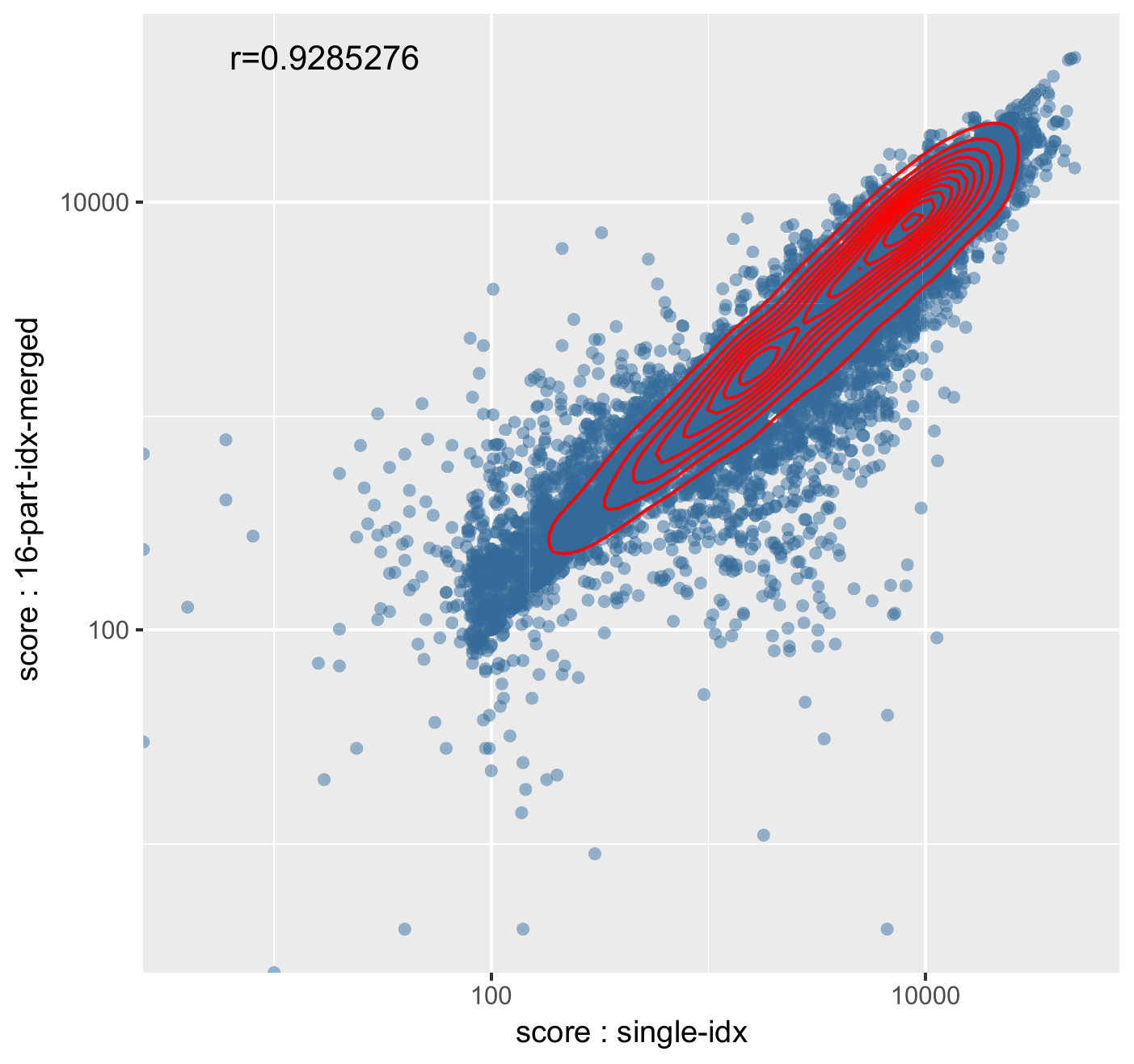}
\end{center}
\caption[Scatter plot of alignment scores of \textit{single-idx} vs \textit{16-part-idx-merged} for disparate alignments (different by at least one base position)]{\textbf{Scatter plot of alignment scores of \textit{single-idx} vs \textit{16-part-idx-merged} for disparate alignments (different by at least one base position).}

\footnotesize The scatter plot contains 17,146 points representing each disparate mapping - different by at least one base position (Pearson's correlation (r) of 0.9285). The x and y axes are in log scale. 
50.5\% had higher dynamic programming alignment scores for \textit{single-idx}, while 42.9\% had higher scores in the \textit{16-part-idx-merged}. }
\label{f:miscorr}
\end{figure}

\begin{figure}[!ht]
\begin{center}
\includegraphics[width=\textwidth]{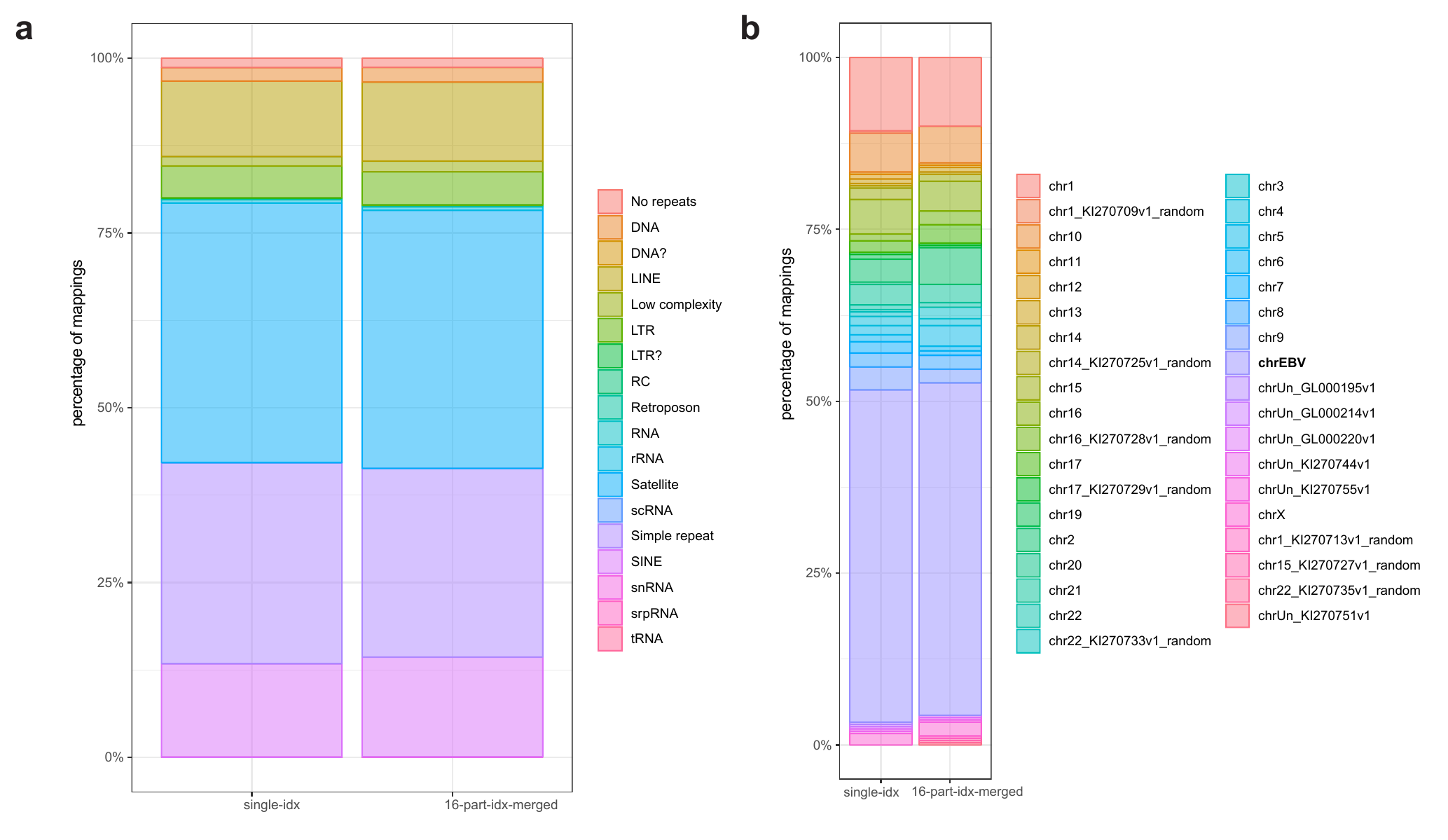}
\end{center}
\caption[Distribution of genomic features of disparate alignments (different by at least one base position) between \textit{single-idx} and \textit{16-part-idx-merged}]{{\textbf{Distribution of genomic features of disparate alignments (different by at least one base position).}}

\footnotesize \textbf{(a)} The distribution of repeat diversity in the disparate alignments (different by at least one base position) between \textit{single-idx} and \textit{16-part-idx-merged}. (\textbf{b}) Distribution of genomic targets for the 456 (left) and 472 (right) disparate alignments (that did not overlap with annotated repeat regions) between \textit{single-idx} and \textit{16-part-idx-merged}, respectively.}
\label{f:misrep}
\end{figure}

\begin{figure}[!htp]
\begin{center}
\includegraphics[width=0.9\textwidth]{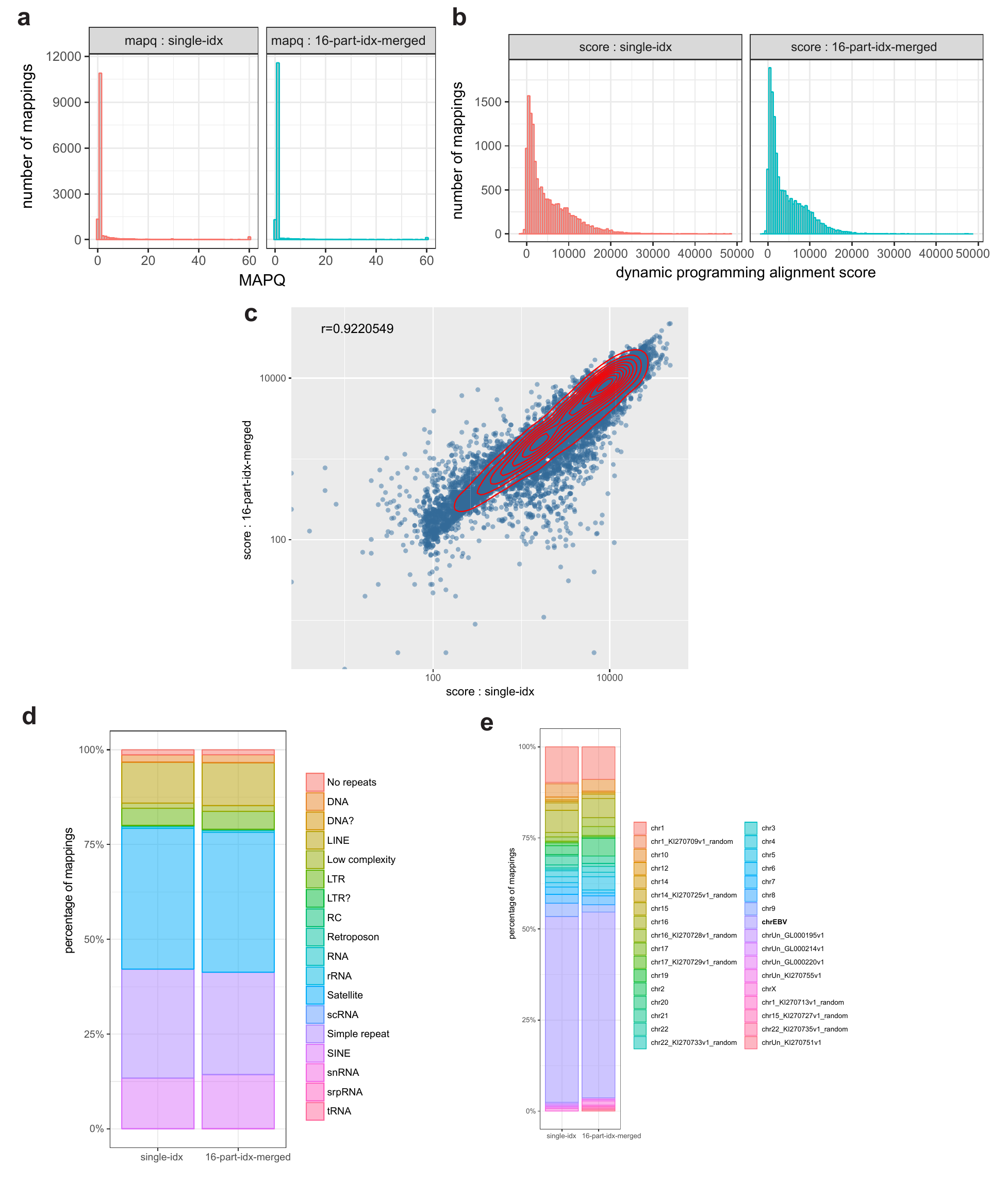}
\end{center}
\caption[Statistics and genomic features of disparate mappings (mapping positions not overlapping at least by 10\%) between \textit{single-idx} and \textit{16-part-idx-merged}]{{\bf Statistics and genomic features of disparate mappings (mapping positions not overlapping at least by 10\%) between \textit{single-idx} and \textit{16-part-idx-merged}}\\
\footnotesize Out of the 17,146 mappings that were different by at least one base position, 14,398 did not even overlap by 10\% or more with their mapped locations on the reference. For those 14,398 mappings
the distribution of the \textit{(a)} mapping qualities and \textit{(b)} alignment scores, \textit{(c)} the scatter plot between the alignment scores, \textit{(d)} the distribution of repeat diversity and \textit{(e)} the mapping location of the mappings that did not contain repeats were almost identical to respective plots for disparate alignments --different by at least one base position.
}
\label{f:olpmismatch}
\end{figure}

Of the 17,146 reads with disparate mappings, 50.5\% had higher DP alignment scores for the single index, while 42.9\% had higher scores in the 16-part index (Pearson’s correlation = 0.93, Figures \suppref{f:mishis}{S5} and \suppref{f:miscorr}{S6}). This effect was similarly observed for the MAPQ scores, with 15.2\% and 12.8\%, respectively, suggesting that alignments are of marginally better quality when generated with a single index. Again, these disparate mappings are largely composed of repetitive and viral sequences (Figure \suppref{f:misrep}{S7}). when the 2,748 overlapped reads were removed from the 17,146 disparate mappings, the trend of the DP scores, MAPQs and the repeat distributions are very similar to when those are not removed (Figure \suppref{f:olpmismatch}{S8}). \\

%if we just say overlapping by 10% it is vague isn't it? Shouldn't we say: the "mapping coordinates overlapped" rather than just "overlapped"?

\reviseend

\subsubsection{For an ultra-long chromothriptic read}

\begin{figure}[!htp]
	\centering
    \includegraphics[width=\linewidth]{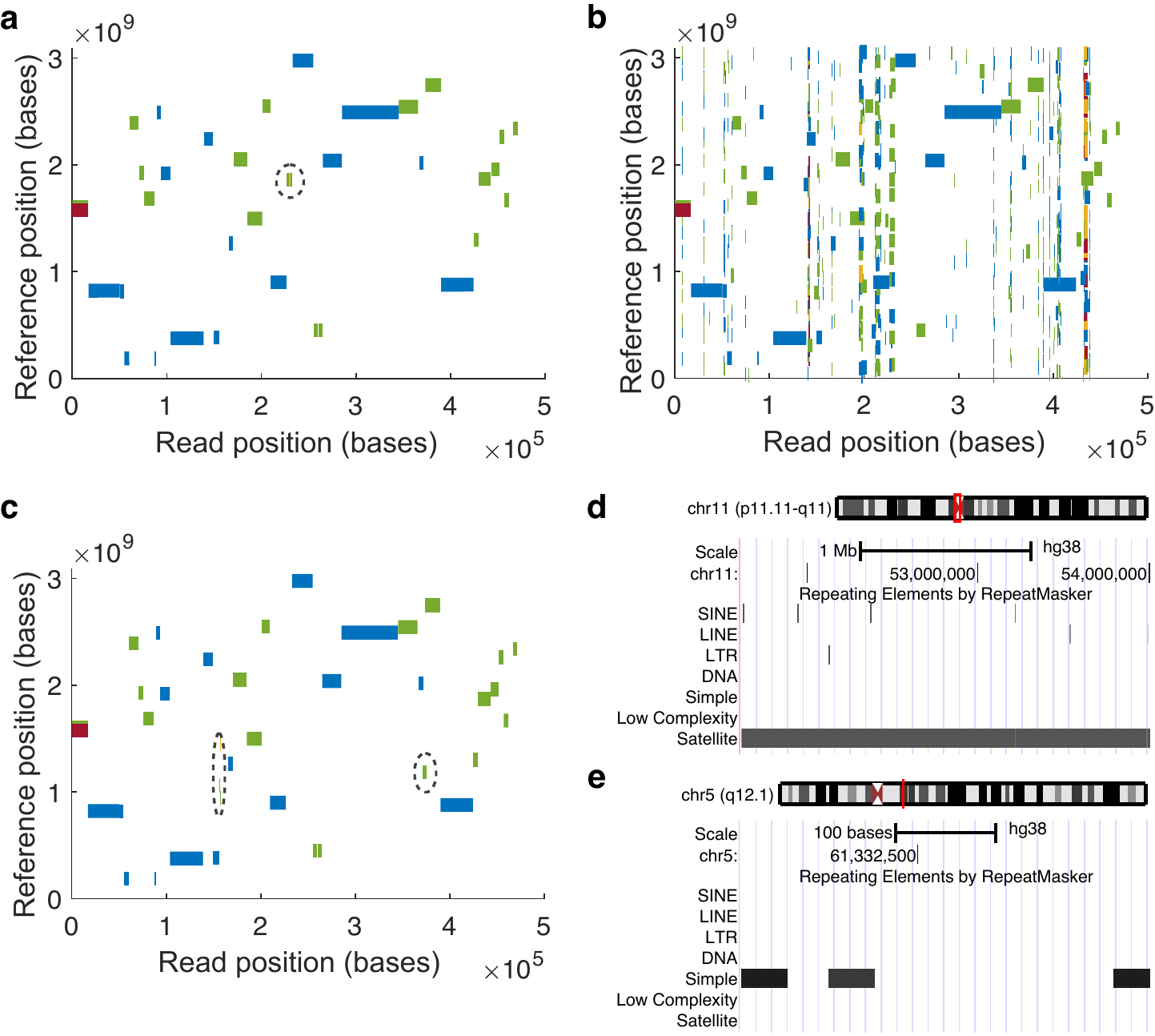}

\caption[Alignment of an ultra-long Nanopore read from a chromothriptic region]{{\bf Alignment of an ultra-long Nanopore read from a chromothriptic region.}\\ 
\footnotesize Mapping coordinates in the entire human reference genome (y-axis) in function of the position in the read, showing where sub-sequences of the chimeric read map to in the genome for \revise{(\textbf{a})} \textit{single-idx} , \revise{(\textbf{b})} \textit{part-idx-no-merge} , and \revise{(\textbf{c})} \textit{part-idx-merged}. The y-axis begins with chromosome 1 at 0 and ends with chromosome X, Y, and the mitochondria at the top. The length of rectangles along the x-axis are in the correct scale to the length of the read. However, the length along the y-axis are exaggerated to a fixed value so that it is clearly visible. In (\textbf{a}) and (\textbf{c}), the areas with dotted circles contain the differences between unique mappings for each alignment strategy. Circled regions in (\textbf{a}) map to a genomic locus harbouring the satellite repeat displayed in (\textbf{d}). Out of the 6 unique mappings in (\textbf{c}), the segment with the highest mapping quality (6) maps to the simple repeat containing region displayed in (\textbf{e}).}
\label{f:chimeric}  
\end{figure}

To evaluate how chimeric reads will be affected by aligning them to partitioned indexes, we tested this case on an ultra-long (473kb) chromothriptic Nanopore read from a patient-derived liposarcoma cell line \cite{garsed2014architecture}. Chromothripsis is a genetic phenomenon often associated with cancer and congenital diseases. It is caused by several rounds of breakage-fusion-bridge, which produce complex and localised genomic rearrangements in a relatively short segment of DNA. The \textit{single-idx} produced 41 (36 primary + 5 secondary mappings) mappings (Figure \ref{f:chimeric}a). However, \textit{part-idx-no-merge} (16 partitions) produced 688 (608 primary + 80 secondary) mappings (Figure \ref{f:chimeric}b), while mapping with \textit{part-idx-merged} resulted in 47 (42 primary + 5 secondary) mappings (Figure \ref{f:chimeric}c).\\

In \textit{single-idx} and \textit{part-idx-merged}, 34 mappings were the same. Interestingly, there were 7 mappings unique to \textit{single-idx} and 6 unique to \textit{part-idx-merged} (Table \suppref{t:chemericdiff}{S3}). All 7 alignments unique to \textit{single-idx} map to the centromeric region of the chromosome 11 (Figure \ref{f:chimeric}d), which is composed of large arrays of repetitive DNA (also known as satellite DNA). The alignments that are unique to \textit{part-idx-merged} map to simple repeats (e.g. GAGAGAGA).

\begin{table}[!ht]
\centering
\caption[Mappings of the chromothriptic read which are different in \textit{single-idx} and \textit{16-part-idx-merged}]{\bf Mappings of the chromothriptic read which are different in single-idx and 16-part-idx-merged}
\label{t:chemericdiff}
\includegraphics[width=0.8\textwidth]{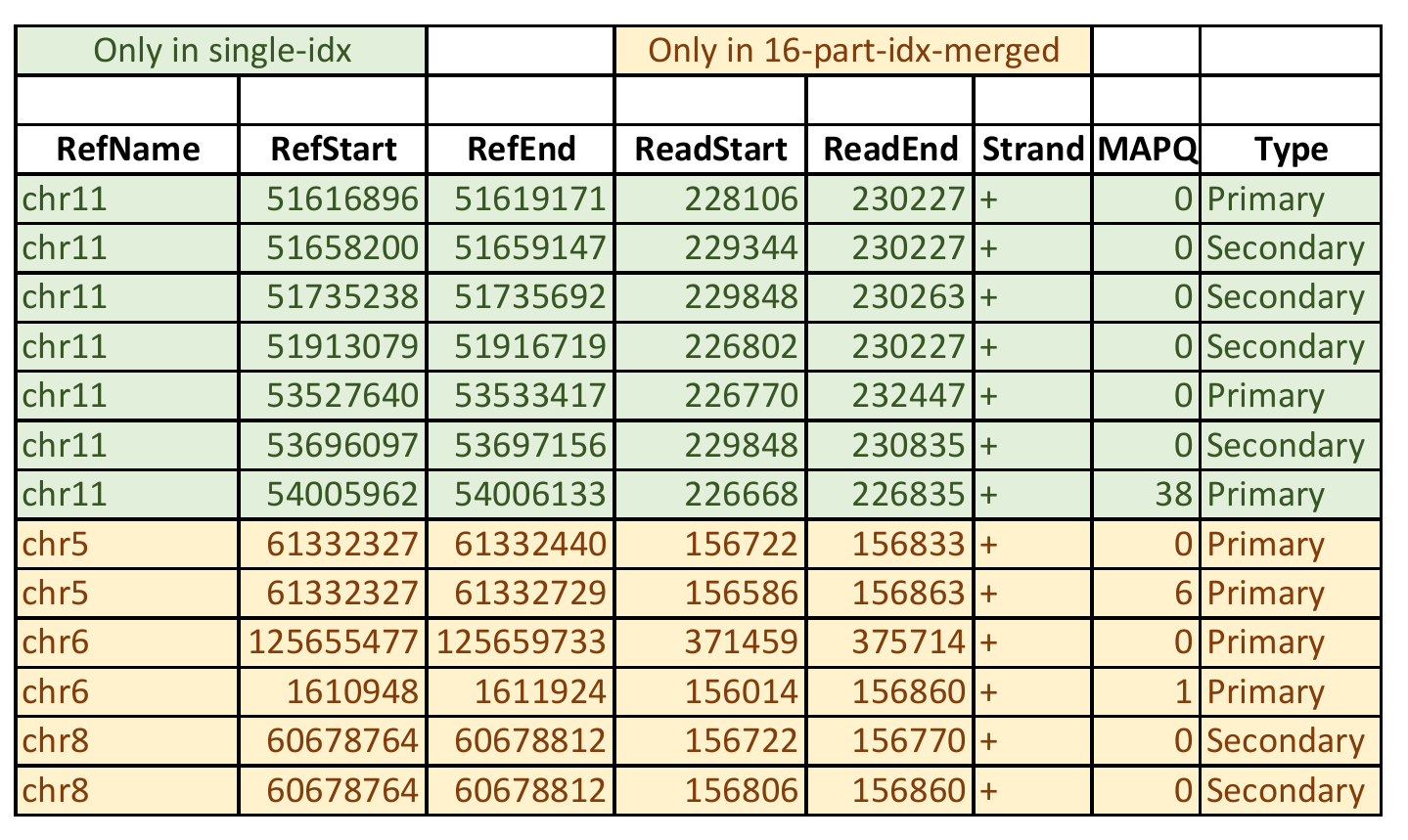}
\begin{flushleft} \footnotesize The alignments which were only found in the \textit{single-idx} mapped to locations in the range chr11:51616896-54006133. The ones unique to \textit{16-part-idx-merged} mapped to repetitive regions in chr5, chr6 and chr8.
chr5:61332327-61332729, chr6:1610948-1611924 and chr8:60678764-60678812 contained simple repeats.
chr6:125,655,477-125,659,733 had simple repeats,SINE repeats and LTR repeats
\end{flushleft}
\end{table}

\subsection{Memory usage and runtime of partitioned indexes}
 
\begin{figure}[!htp]
\centering
\includegraphics[width=0.9\textwidth]{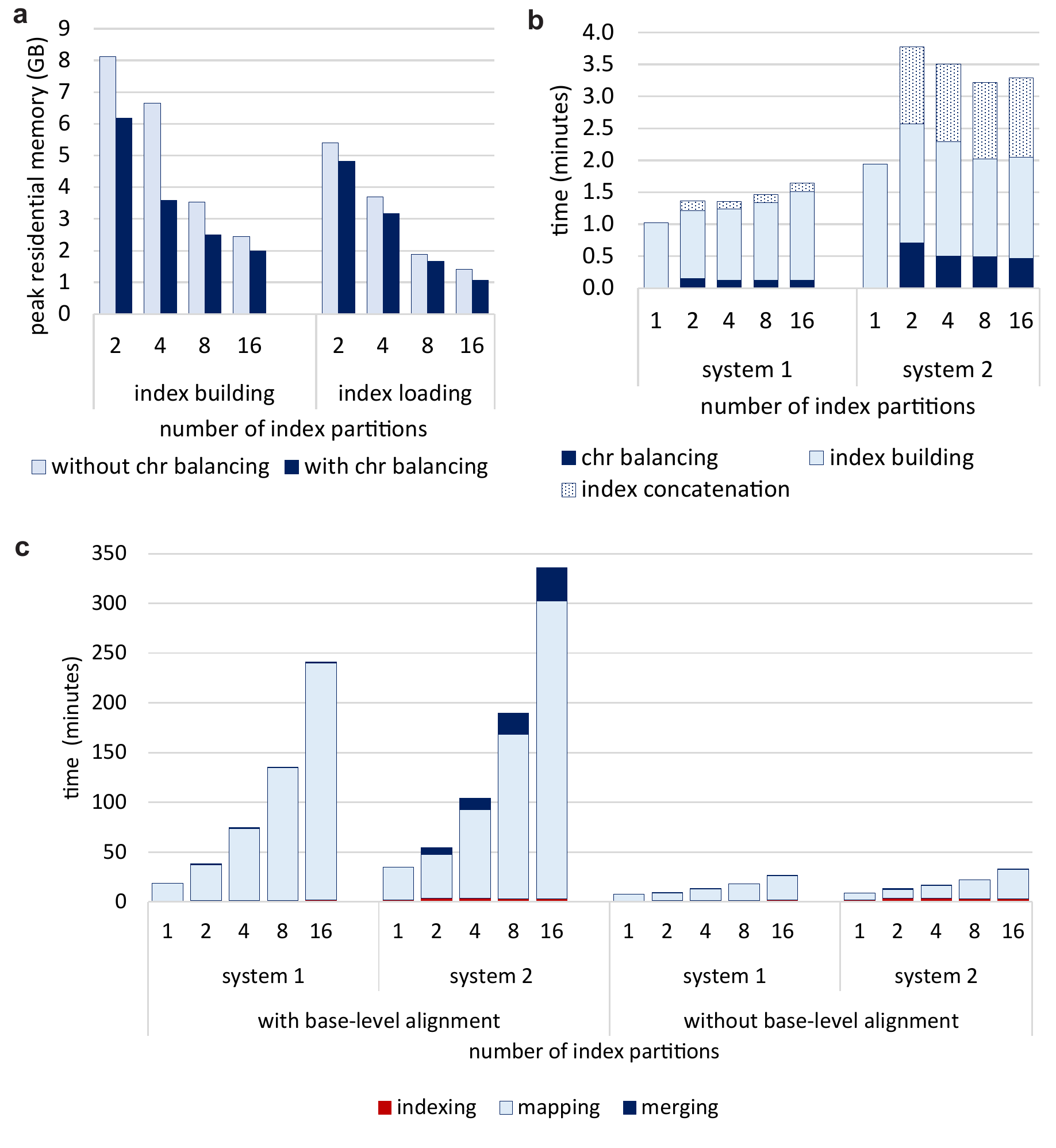}
        %\label{f:indexloadmem}
%\label{f:indexbuildmem}
\caption[Peak memory usage and runtime for a partitioned index of the GRCh38 human reference genome]{\revise{ \textbf{Peak memory usage and runtime for a partitioned index of the human genome.}}

\footnotesize
\revise{(\textbf{a}) Peak memory usage in function of the number of partitions for \textit{Ab initio} index generation (left) and loading a pre-built index (right). Dark bars represent memory usage when performing chromosome balancing as described herein, whilst light bars represent default iterative partition distribution as implemented in Minimap2.
(\textbf{b}) Detailed runtime metrics for index building across two computational systems. \textit{System 1} is a laptop with flash memory (Intel i7-8750H processor, 16GB of RAM and Toshiba XG5 series NVMe SSD) while \textit{system 2} is a workstation  with a mechanical hard disk (Intel i7-6700 processor, 16GB of RAM and Toshiba DT01ACA series HDD). The total indexing time has been broken down into three steps; 
\textit{chr balancing}, \textit{index building} and \textit{index concatenations}. \textit{Chr balancing} includes the overhead for chromosome sorting, partitioning and writing each partition to a separate file. 
(\textbf{c}) Runtime for base-level alignment (left) and block/locus level mapping (right). System 1 and 2 are as described in (\textbf{b}). Alignment was performed on the NA12878 data (see Materials and methods) with the \textit{map-ont} pre-set in Minimap2 using 8 threads. Runtime statistics are composed of \textit{indexing} (index generation including the overhead), \textit{mapping} (total time for aligning reads to each partition), and \textit{merging} using the method described herein. Runtimes include file reading and writing.}}
\label{f:indeximprove}
\end{figure}

\revise{In addition to comparable quality of alignments, using a partitioned index yields impressive reductions on peak memory usage during indexing. About 7.7GB of memory is required to hold a single reference index, whereas only 1.5GB is needed for a partitioned index with 16 parts (Figure \ref{f:indeximprove}a). Peak memory usage can be further reduced by distributing or 'balancing' chromosomes across partitions based on their size (see Materials and methods). These indexing approaches combined with a mini-batch size between 5-20 Mbases (Minimap2 parameter \textit{K}) enables alignment of long reads to the human genome with less than 2GB of RAM. Although generating an index \textit{ab initio} requires more memory than loading a pre-built one, this only needs to be done once for a given reference and can be performed \textit{a priori}, if required.}\\

The reduced peak memory usage of a partitioned index comes with an inherent sacrifice in computing speed. \revise{Alignment requires significantly more time than the balancing, indexing and merging steps when generating an index \textit{ab initio}, which we observed to be relatively constant across different partition ranges (Figure \ref{f:indeximprove}b). Less than 10\% of the total compute time (5.7h) for base-level alignment is dedicated to balancing, indexing, and merging when using mapping to 16 partitions with eight CPU threads and a mechanical hard drive. When using flash memory, the overheads have very minor impact (3\% of the total compute time). Chromosome balancing, for instance, required less than 1 minute and merging required less than 2 minutes.} \\

  \revise{We observed that the alignment time increased less than linearly with the number of partitions in the index (Figure \ref{f:indeximprove}c). Since our motivation is to reduce the memory requirements of mapping to large references, this will inevitably impact speed. However, this also facilitates parallelisation of alignments, where several small index partitions can be distributed across an array of low-memory devices (e.g. microcomputing boards such as Raspberry Pi). It also enables the use of mobile computing devices, such  mobile phone or inexpensive laptops, which would otherwise be impossible. Considering that ONT’s MinION sequencer generates about 80\% of all data in the first 24h, using 16 partitions would enable real-time mapping on a system with 2 GB of RAM in parallel to data acquisition, whereas a system with 4GB RAM would only require 4 partitions and less than 1h of compute to align a typical ONT MinION dataset.}\\

%%%%%%%%%%%%%%%%%%%%%%%%%%%%%
%		DISCUSSION			%
%%%%%%%%%%%%%%%%%%%%%%%%%%%%%
\section{Discussion}
\revise{ This work details two ways to reduce memory requirements for performing alignments on large genomes, or collections of genomes, using Minimap2. By tuning alignment parameters, peak memory usage can be lowered marginally, although with non-negligible impact to the accuracy of alignments. We demonstrated this effect by sequencing and mapping a diverse and representative set of synthetic spike-in controls, which can be used as a ground truth to assess sequencing and alignment accuracy. However, these data were not used to benchmark the alignment accuracy of Minimap2, but to demonstrate the comparative and relative impact of alignment parameters on memory usage. In this regard, we show that partitioning a large reference into smaller indexes upstream of an appropriate merging process drastically reduces the peak memory usage without compromising alignment accuracy.\\ }

\revise{ Previous studies have described indexing strategies to improve computational efficiency. DIDA \cite{mohamadi2015dida} and DREAM-Yara \cite{doi:10.1093/bioinformatics/bty567} use bloom filters to distribute sequencing reads to the most appropriate index partitions. However, these works employ methods dependent on indexes generated with the Burrows-Wheeler transform algorithm---which are ideal for short, less noisy reads generated by second generation sequencing platforms. 
%For instance, the accuracy of mapping reads with DIDA was evaluated by simply considering the total number of alignments, with no analysis of mapping qualities or the use of independent controls. 
These strategies focus on indexing enhancements centred on reducing the alignment search space by delivering reads to the most suitable index partition. Our work focuses on multi-index alignment merging, which is independent and complementary to these strategies. We reveal the problems derived from mappings to multiple partitions such as the accuracy of mapping reads and the overestimation of the  mapping quality. We show that our merging technique circumvents those issues by the analysis of mapping qualities and the use of independent controls.}
%Similarly, DREAM-Yara is a distributed framework based on Yara (which is only compatible with short reads) and focuses on maximising sensitivity without considerations for specificity. 
The partitioned reference approach has also been previously used for reducing the memory usage of BWA, a popular short read alignment program \cite{gnanasambandapillai2018mesga}. However, the final output consists of the indiscriminate concatenation of the alignments from all the partitions, raising several of the caveats exposed in Figure 2. We have demonstrated that performing appropriate merging of alignment output is required to eliminate many mapping artefacts, thus improving overall accuracy. \\

We also showed that \textit{part-idx-merged} can provide a better result than a simple strategy of filtering out results with low mapping quality in \textit{part-idx-no-merge}. This is supported by the results from synthetic reads, where the accuracy of alignments with mapping quality 60 in \textit{part-idx-no-merge} is lower than those from \textit{part-idx-merged}. Furthermore, a simple strategy to remove all short mappings from \textit{part-idx-no-merge} is also less than ideal. In fact, \textit{Paftools} (which was used for evaluating the synthetic read alignments) considers the longest primary mapping when multiple primary mappings exist to assess alignment accuracy.\\
 
However, \textit{part-idx-merged} can sometimes generate non-identical alignments to that of \textit{single-idx}. This is like a consequence of slight variations in highly abundant k-mers observed when the index is build. Overall, this affects only a few reads which would nonetheless have low mapping qualities--an issue that has previously been reported by the author and users of Minimap2 (see the public code repository associated with Li, H. \cite{minimap2}. Further, the reported alignments might differ in long low-complexity regions, as Minimap2 may generate suboptimal alignments in long low-complexity regions (see supplementary materials of Li, H. \cite{minimap2}).\\
 
Although a partitioned index reduces peak memory usage, the runtime is proportionately higher. This is because all the reads should be repeatedly mapped to each partition of the reference. However, this strategy lends itself well to distributed computing, in particular when many smaller, less expensive computing devices are available. \\

A limitation of this method also lies in the maximal number of  partitions an index can be split into, which currently depends on the longest chromosome or contig. We have not yet investigated the impact of splitting chromosomes into fragments, although we anticipate this would not drastically affect results (as exemplified from the chromothriptic read example above). Furthermore, we have not tested the impact of this strategy for RNA sequencing read alignment, which implements different alignment scoring metrics. \\ 

In addition to capability of mapping long reads to large genomes on devices with a small memory footprint, our extension to Minimap2 could potentially be useful for the following applications:
\begin{itemize}
\item \textit{Mapping to huge reference genome databases}. Meta-genomic databases can be hundreds of gigabytes in size. Hence, holding the index for the whole database would be challenging even for high-specification servers. Especially when multiple species with similar genomes are present, an accurate mapping quality with correct flags, headers, and reduced output file sizes is always appreciated.
Alternatively, mapping genome assembly contigs, or a select amount of long reads, to a large public sequence repository (akin to a BLASTN nucleotide databse query) could benefit from our approach. However, the effect of merging output from such large queries has yet to be investigated.
\item \textit{Mapping with a lower window size for increased sensitivity}. Minimap2 runs on a default minimiser window size of 10. However, reducing this value improves the mapping sensitivity, but increases the memory consumption. For application where high sensitivity may be preferred, for instance when confronted low coverage sequencing data, our method can be beneficial.
\end{itemize}

\noindent While preparing this manuscript, our method was integrated into the source code of the original Minimap2 software repository. In Minimap2 version 2.12, the option \textit{-{}-split-prefix} can be used to align to a partitioned index. The developer of Minimap2 has expanded our implementation to support paired end short reads and multi-threading for the merging process. The original version we implemented for conducting the above experiments is available in the associated \emph{github} repository \cite{gitrepo} and can be useful for understanding the underlying algorithm. The partitioned index functionality can be invoked with the option \textit{-{}-multi-prefix}. \revise{Instructions to run the tools are detailed in the Supplementary Note \ref{s:minimap2-supp3}}\\

%%%%%%%%%%%%%%%%%%%%%%%%%%%%%
%			METHODS			%
%%%%%%%%%%%%%%%%%%%%%%%%%%%%%
\section{Materials and methods}

\subsection{Exploration of parameters that affect memory usage in Minimap2}

For measuring peak memory usage and runtime, publicly available NA12878 Nanopore reads \cite{jain2018nanopore} were aligned to the human genome reference (GRCh38) with Minimap2 \cite{minimap2}. Peak memory usage and  runtime were measured by using the GNU command line \textit{time} utility with the \textit{-v} option. \\

Sensitivity and error rate calculations for different window sizes (Minimap2 parameter \textit{w}) were performed using \revise{Sequins, synthetic human genome spike-in controls and synthetic PacBio reads (see below). By reversing (not complementing) the sequences from regions of interest, these spike-in controls reproduce most features of the human genome, including nucleotide frequencies, somatic variation, low-complexity regions, and repeats. Given their chiral or `mirror' design, Sequins do not align to the native reference sequence but will align to a mirror copy of the human reference genome. They can thus be used to benchmark alignment accuracy when spiked-in to a normal sample, although they were sequenced in isolation for this study. The particular Sequins design we employed was unpublished at the time this manuscript was prepared (Deveson et al., under review), but it is conceptually similar to what is reported by Deveson et al. \cite{deveson2016representing}. 1 $\mu$g of Sequins DNA was sequenced on a ONT R9.4.1 flow cell, using the LSK108 sample preparation kit and the results were base called with ONT's proprietary Albacore software (version 1.2.6).} Reads were mapped to the reverse human genome, using Minimap2 under the pre-set \textit{map-ont} for different window sizes. \\

\revisestart
 We leveraged the chiral design of Sequins to qualify any mapping to the normal reference genome as a false positive. True positive Sequin alignments should display the exact mapping positions on the mirrored human genome, as intended by their design. However, given stochastic variations in sequencing (base calling idiosyncrasies, involuntary library fragmentation, sample degradation, etc) the primary mappings derived from the default window size parameter (\textit{w} = 10) in Minimap2 were used as a reference to assess the relative effect and impact of parameter tuning. Then, for a \revise{given} window size:
\begin{itemize}
\item \textit{Mismatching mappings} refer to primary mappings that had different positions to the mappings with reference parameters;
\item \textit{Missing mappings} refer to primary mappings that were not observed in empirical alignments, but were observed in alignments with reference parameters;
\item \textit{Extra mappings} refer to primary mappings that were observed in empirical alignments, but were not observed in alignments with reference parameters.
\end{itemize}
\noindent The above counts were expressed as a percentage of the total number of reads. The sum of mismatch and extra mapping percentages were taken as an approximation of the \revise{relative} error rate. The \revise{relative} sensitivity was approximated by subtracting the percentage of missing mappings from 100.\\
\reviseend

\subsection{Merging of mappings from a partitioned Index}

We extended the partitioned index approach of Minimap2 to eliminate alignment artefacts as described below. The index partitioning in Minimap2 is inherited from the first version of Minimap \cite{li2016minimap}. This feature is for finding long read overlaps to be used with assembly tools such as Miniasm \cite{li2016minimap}. As overlap computing requires all-vs-all mapping of reads, the index is built for chunks of 4 Gbases (can be overridden with the \textit{-I} argument) at a time, effectively partitioning the alignment index to keep the maximum memory capped at around 27GB. For each part of the index, Minimap attempts to map all the reads. The concatenated alignments from all the parts is the final output.\\

We modified Minimap2 to \revise{serialise and store} the software's internal state during the alignment process. The internal state is \revise{serialised} in binary format to reduce disk usage. The internal state includes: (i) mapped positions, chaining scores and other mapping statistics for each alignment record;  (ii) DP score, CIGAR string, and other base-level alignment statistics for each alignment record (when base-level alignment is specified); and (iii) sum of region length of read covered by highly repetitive k-mers for each read (referred to as repeat length). These data form the \revise{serialised} binary files, one for each partition of the index.\\ 

When an alignment process has completed, we simultaneously open all the \revise{serialised binary} files together with the queried sequence file. For each queried read (or contig), the previously \revise{serialised} internal states of all  alignments for the given read (resulting from all the index partitions) are loaded into memory. If no base-level alignment has been requested, the alignments are sorted based on the chaining score in descending order. Otherwise, the sorting is based on the DP alignment score in descending order. The classification of primary and secondary chains is re-iterated as implemented in Minimap2. This corrects the primary and secondary flags in the output. Then, the secondary alignment entries are filtered based on a user requested number of secondary alignments, and the requested minimum primary to secondary score ratio, effectively removing spurious secondary alignments. If a SAM output has been requested, the best primary alignment is retained as the primary alignment and all other primary alignments are classified as supplementary alignments. An unaligned record is printed only if the read is not mapped to any part of the index.\\

\revise{ The length of the read covered by repeat regions in the whole genome (repeat length) is one of the parameters required to estimate an ideal mapping quality (MAPQ). The MAPQ produced by Minimap2 is a globally computed heuristic that depends on a large number of parameters, including this repeat length}.
We estimate this global repeat length by taking the maximum of the previously \revise{serialised} repeat lengths (for each partition of the index) for that particular read. The Spearman correlation between this estimated repeat length and the global repeat length is 0.9961. In theory, it would possible to exactly calculate this value by \revise{serialising and storing} the positions of repeats within the read. However, as the MAPQ is itself an estimation and the accuracy of mappings was adequate in our initial tests, we simply took the maximum. Hence, the computed MAPQ during merging of a partitioned index is not exactly the same as for a single reference index, but very similar overall. This computed MAPQ is more accurate than a MAPQ computed only from the repeat length for a single part of the index. \\

% For space efficiency, dumped data contains numeric identifiers for references (the same used by internal Minimap2 data structures) rather than the reference name. This numeric identifier is determined per index partition and is fixed during the merging using an emulated single reference index. When each index partition is loaded/constructed during mapping, the meta data of reference sequences (sequence name, length, etc) in the index are dumped along with their respective numeric identifiers. At the beginning of merging, these meta-data are reloaded and the numeric identifier is rectified by adding a necessary offset to form an emulated single reference index. This allows generation of a correct SAM header. The numeric identifiers for dumped alignments are also fixed in the same way at the time of loading. \\

Merging is performed in the order of input read sequences, and mappings for a particular read ID will be adjacent in the output. \revise{As the  serialised data are loaded into memory for each read (or a batch of few reads) at a time, the memory usage of merging is only a few megabytes.} %N.B. reads could be batched to improve performance. \revise{That is, rather than performing merging one read by one read, a batch of reads (and the internal states for mapping for that read) can be loaded at once, and merged using multiple threads.}
For a detailed explanation of the merging algorithm refer the supplementary Note \ref{s:minimap2-supp1}. 
%and the source code at [\url{https://github.com/hasindu2008/minimap2-arm/blob/v0.1-alpha/merge.c}].
\\

\subsection{ Chromosome balancing }

\revise{The construction of partitioned indexes in Minimap2 (specified by the \textit{-I} option) processes reference sequences iteratively, which does not distribute reference sequences (i.e. chromosomes) evenly when using multiple partitions. We implemented a simple sorting and binning algorithm to mitigate this effect. First, a command line parameter describing the number of desired partitions is considered. Then, the reference sequences (or chromosomes) are sorted in descending order of size. Next, a chromosome is assigned to the bin (or partition) with the lowest sum of bases, and the sum of that bin is then incremented by the chromosome size. This effectively distributes the chromosomes to roughly balanced partitions in $\mathcal{O}(n\log{}n)$ time complexity--adding negligible overhead to the overall indexing process (Table \suppref{t:runtimedetails}{S4}). We output the reference sequences belonging to the each bucket in a separate file. Finally we launch the Minimap2 indexer on each file and concatenate the indexes. Refer to Supplementary Note \ref{s:minimap2-supp2} for a detailed explanation. This approach is available under \textit{misc/idxtools} in the \emph{github} repository \cite{gitrepo} and the instructions to run the tool are detailed in the Supplementary Note \ref{s:minimap2-supp3}.} %[\url{https://github.com/hasindu2008/minimap2-arm/tree/master/misc/idxtools}]. 

\begin{table}[!ht]
\centering
\caption[Detailed runtime for partitioned indexes]{
{\bf Detailed runtime for partitioned indexes}} 
\label{t:runtimedetails}
\includegraphics[width=\textwidth]{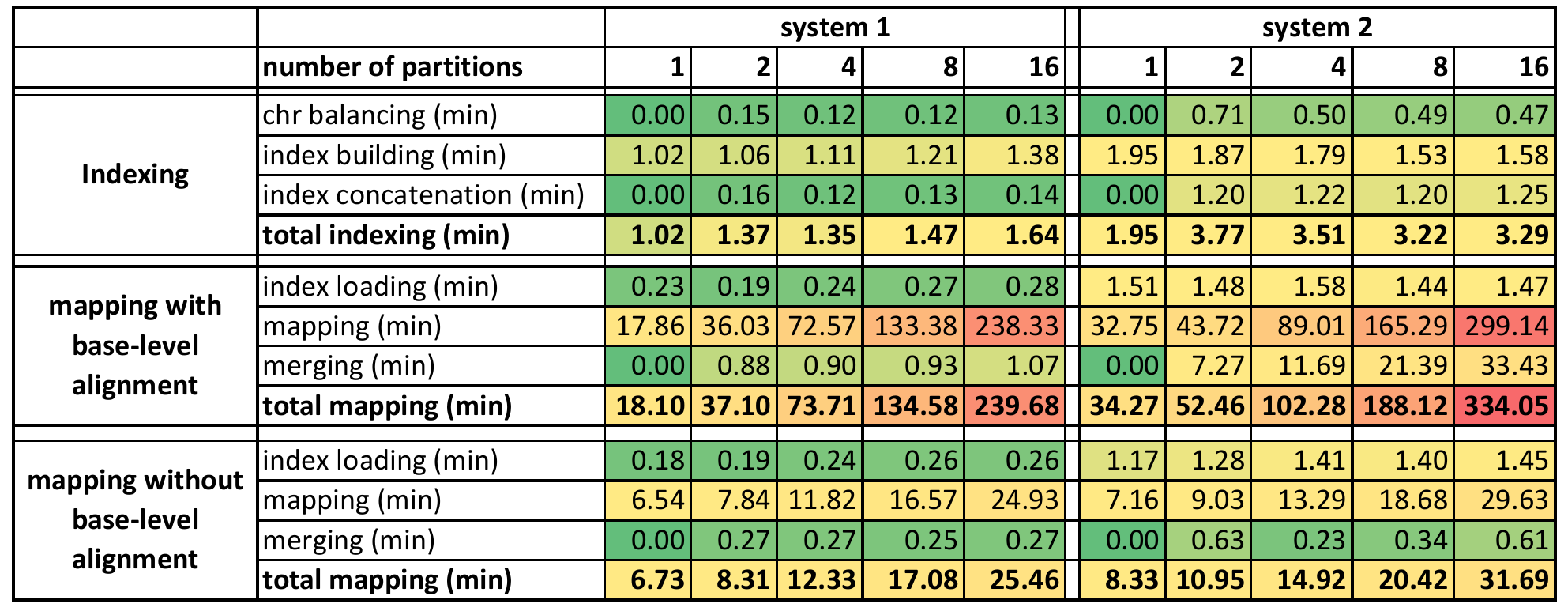}
\begin{flushleft} \footnotesize \textit{System 1} is a laptop with flash memory (Intel i7-8750H processor, 16GB of RAM and Toshiba XG5 series NVMe SSD) while \textit{system 2} is a workstation  with a mechanical hard disk (Intel i7-6700 processor, 16GB of RAM and Toshiba DT01ACA series HDD). Alignment was performed on the NA12878 Nanopore data with the \textit{map-ont} pre-set in Minimap2 using 8 threads. 
\end{flushleft}
\end{table}

\subsection{Datasets and evaluation methodology}

All experiments were performed using the human genome as a reference (GRCh38 with no ALT contigs). The scripts and tools written for performing the experiments are available under \textit{misc/idxtools/eval} in the \emph{github} repository \cite{gitrepo}.
%[\url{https://github.com/hasindu2008/minimap2-arm/tree/master/misc/idxtools/eval}].

\subsubsection{Synthetic reads}

Mapping accuracy was evaluated using synthetic long reads. We generated about 4 million PacBio reads using PbSim \cite{ono2012pbsim} \revise{ under "Continuous Long Read" mode (long reads with a high error rate)}. The minimum, maximum and the mean read length were set to be 100 bases, 25 kbases and 3 kbases respectively with a standard deviation of 2300. The minimum, maximum and the mean accuracy of bases were set to 0.75, 1.00 and 0.78 respectively with a standard deviation of 0.02. The ratio between substitution:insertion:deletion was 10:60:30. \\

\begin{figure}[!ht]
\begin{center}
\includegraphics[width=\textwidth]{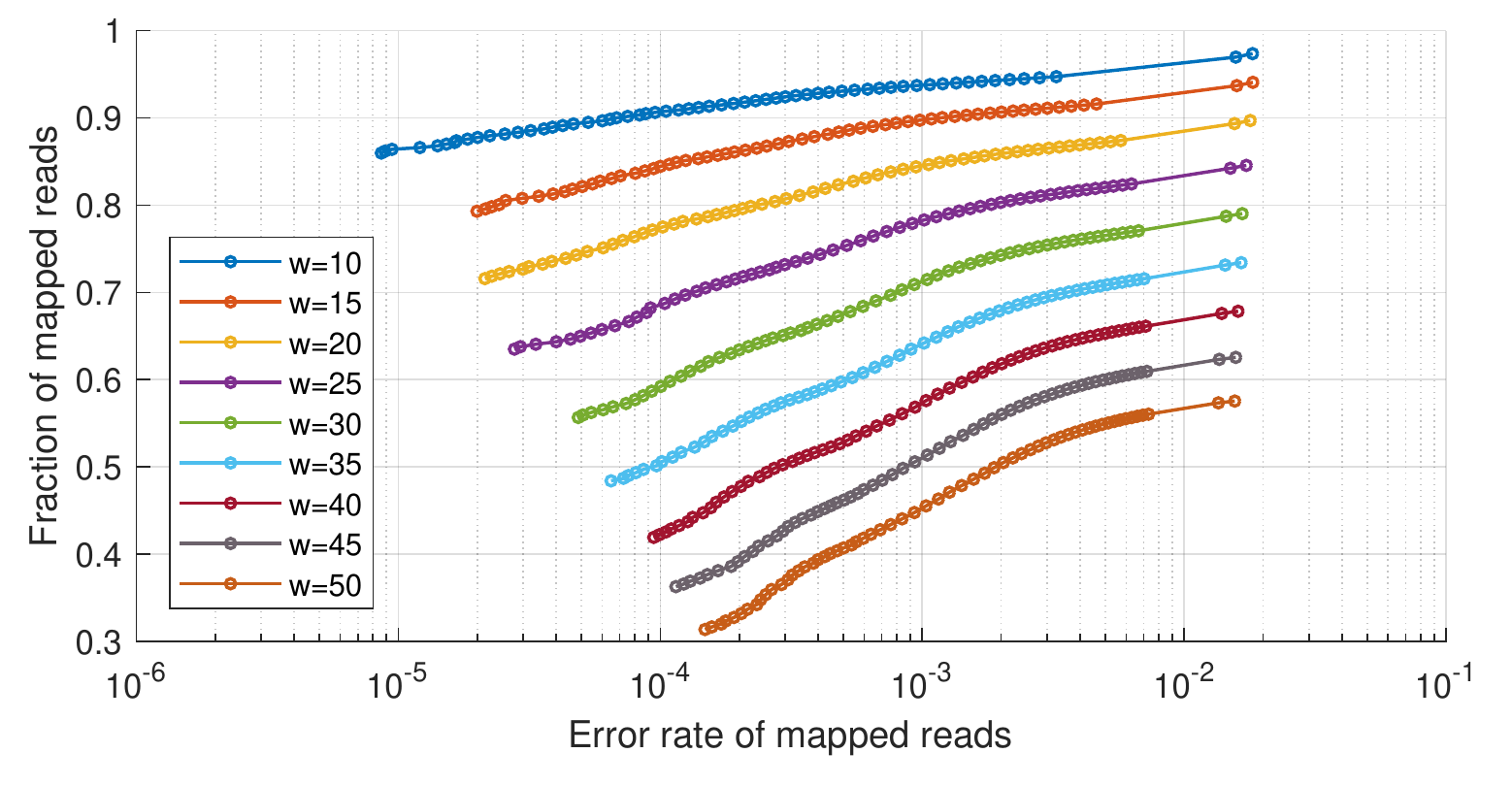}
\end{center}
\caption[Effect of the window size on the error rate and  sensitivity for simulated reads]{{\bf Effect of the window size parameter \textit{w} on the error rate and  sensitivity for simulated reads.}
\footnotesize Effect of window size on the proportion of mapped reads and the associated error rate (log scale) for 4 million simulated long reads (see Materials and Methods). A single curve contains points for each mapping quality threshold (MAPQ score), one point for each mapping quality threshold from 60 (leftmost) to 0 (rightmost).}
\label{f:synth_window}
\end{figure}

\revise{In the context of parameter tuning (Figure \suppref{f:synth_window}{S2}), the reads were mapped using Minimap2 with different window sizes while keeping other parameters constant. Then the accuracy evaluation was performed using the \textit{Mapeval} utility in \textit{Paftools}---part of the Minimap2 software package---where a read is considered correctly mapped if the mapping coordinates of its longest alignment overlaps with the true reference coordinates with an overlap length of 10\% or higher.}\\

In the context of multi-part index accuracy, simulated long reads were aligned using Minimap2 with single reference index (\textit{single-idx}), partitioned index without merging (\textit{part-idx-no-merge}) and partitioned index with merging (\textit{part-idx-merged}). Partitioned indexes with 2, 4, 8 and 16 parts were tested. For each instance, we evaluated base-level alignments (default SAM output) as well as locus- or block-based alignment (default PAF output without CIGAR information). To evaluate alignment accuracy, \textit{Mapeval} utility in \textit{Paftools} was used with default options, which consider only the longest primary alignment for a read. However, Paftools assumes that all alignments for a particular read reside contiguously. Hence, for \textit{part-idx-no-merge}, we first sorted the alignments based on the read ID. The output from Paftools contains the accumulative mapping error rate and the accumulative number of mapped reads for different mapping quality thresholds \cite{paftoolslink}. The fraction of mapped reads is taken as a measure of sensitivity.\\

\subsubsection{Nanopore sequencing data}

We could not find a suitable Nanopore simulator. Published Nanopore simulators explored at the time of writing had issues such as dependence on Minimap2 (would cause a bias), unavailability of trained models for human genome, being unstable or unavailability of source code. (For instance DeepSimulator \cite{deepsim} and NanoSim \cite{yang2017nanosim} are dependent on Minimap2, SNaReSim \cite{faucon2017snaresim} code was not available.) Hence we used a a dataset from the publicly available NA12878 sample (rel3-nanopore-wgs-84868110-FAF01132 \cite{jain2018nanopore}). The dataset had 689,781 reads with about 5.5 Gbases. We aligned this dataset to the human genome using a 16-part index with merging (\textit{part-idx-merged}) and without merging (\textit{part-idx-no-merge}) with base-level alignment. Then we compared those outputs by generating some alignment summary metrics with the result from a single reference index (\textit{single-idx}). We initially attempted to perform an extensive comparison using tools such as \textit{CompareSAMs} utility in \textit{Picard} \cite{picard} and \textit{qProfiler} utility in \textit{AdamaJava} \cite{adamajava}.  They crashed probably because they are designed to be worked with short reads.  \revise{Hence, we first obtained simple summary metrics using \textit{samtools}\cite{li2009sequence} together with custom Linux shell scripts. Then we performed an extensive read by read comparison between the SAM outputs from \textit{single-idx} and \textit{part-idx-merged} using a custom tool written in C. The tool sequentially reads two SAM files while loading all the alignments for a particular read to the memory at a time. For a particular read, it compares and then outputs the alignment entries when the mappings positions between the two sets are disparate or if mapped only in one set (discordant). On these disparate and discordant mappings we used \textit{bedtools}\cite{quinlan2010bedtools} to find overlaps with the UCSC repeatMasker track.}\\ 

\revise{The same above NA12878 dataset was used to measure the runtime of partitioned indexes. The runtime and the peak memory usage were measured using the GNU \textit{time} command line utility. }\\

The ultra-long chromothriptic read was sourced from an unpublished patient-derived dataset generated in house (see Garsed et al\cite{garsed2014architecture} for more details on the cancer cell line). The data was generated on a MinION MkI sequencer (MN16218) with MinKNOW version 1.1.17 on a first generation R9 flowcell (MIN105, no spot-on loading, flow cell ID FAD24075) using the SQK-RAD001 library preparation kit from ONT. The raw data for the read was live base called with MinKNOW 1.1.17 and produced an average fastq score of 7.8.

%%%%%%%%%%%%%%%%%%%%%%%%%%%%%
%		CONCLUSION			%
%%%%%%%%%%%%%%%%%%%%%%%%%%%%%
\section{Summary}

Aligning long reads generated from third generation high-throughput sequencers to large reference genomes is possible on computers with limited volatile memory. Parameter optimisation alone cannot substantially reduce memory usage without considerably sacrificing alignment quality. Partitioning an alignment index, saving the internal state, and merging the output \textit{a posteriori} substantially reduces memory usage. This strategy reduces the memory requirements for aligning Nanopore reads to the human reference genome from 11GB to less than 2GB, with minimal impact on accuracy. 

\section{Data Availability}

% The synthetic read set and the Nanopore chromothriptic read is available in \cite{synth} and \cite{chromo}. The data generated from the experiments (along with the scripts used for the experiments) are available at \cite{experimentdata}. The NA12878 Nanopore reads were downloaded from the publicly available repository at \url{https://github.com/nanopore-wgs-consortium/NA12878}. 

The datasets generated and analysed during the evaluation are available in the \emph{figshare} repository \cite{alldatainone} [\url{https://doi.org/10.6084/m9.figshare.6964805.v1}].

%\clearpage

%\clearpage

%\setcounter{secnumdepth}{0}

%% file: 6-gpu-abea/main.tex
\chapter[GPU Accelerated Adaptive Banded Event Alignment]{GPU Accelerated Adaptive Banded Event Alignment for Rapid Comparative Nanopore Signal Analysis} \label{c:gpuabea}

\rule{\textwidth}{0.4pt} 
This chapter is available as a pre-print in bioRxiv under the CC-BY-NC 4.0 International license at \cite{gamaarachchi2019gpu}. An adapted version of this chapter is published in BMC Bioinformatics: \textbf{H. Gamaarachchi}, C. W. Lam, G. Jayatilaka, H. Samarakoon, J. T. Simpson, M. A. Smith, and S. Parameswaran, “GPU Accelerated Adaptive Banded Event Alignment for Rapid Comparative Nanopore Signal Analysis,” BMC Bioinformatics 21, 343 (2020). DOI: \url{https://doi.org/10.1186/s12859-020-03697-x}. \\
\rule{\textwidth}{0.4pt} 

%Advances in DNA sequencing and analysis are revolutionising precision medicine.  A human genome---the sum total of all DNA in a cell---is typically characterised via high-throughput sequencing machines and subsequently analysed using high-performance computing. H

Nanopore sequencing has the potential to revolutionise genomics by realising portable, real-time sequencing applications, including point-of-care diagnostics and in-the-field genotyping. Achieving these applications requires efficient bioinformatic algorithms for the analysis of raw nanopore signal data. For instance, comparing raw nanopore signals to a biological reference sequence is a computationally complex task despite leveraging a dynamic programming algorithm for Adaptive Banded Event Alignment (ABEA)—a commonly used approach to polish sequencing data and identify non-standard nucleotides, such as measuring DNA methylation. Here, we parallelise and optimise an implementation of the ABEA algorithm (termed \textit{f5c}) to efficiently run on heterogeneous CPU-GPU architectures. By optimising memory, compute and load balancing between CPU and GPU, we demonstrate how \textit{f5c} can perform \textasciitilde3-5$\times$ faster than the original implementation of ABEA in the \textit{Nanopolish} software package. We also show that \textit{f5c} enables DNA methylation detection on-the-fly using an embedded System on Chip (SoC) equipped with GPUs. Our work not only demonstrates that complex genomics analyses can be performed on lightweight computing systems, but also benefits High-Performance Computing (HPC). The associated source code for \textit{f5c} along with GPU optimised ABEA is available at \url{https://github.com/hasindu2008/f5c}.

\section{Introduction}

Advances in genomic technologies have given rise to a new era in biomedical sciences, improving the feasibility and accessibility of rapid species identification, accurate clinical diagnostics, and specialised therapeutics, amongst other applications. Whole genome sequencing involves ‘reading’ the entire DNA sequence of a cell, revealing the genetic variation that underlies biological diversity and the onset of disease. A human genome encompasses two copies of ∼3.2 billion DNA nucleotides, or ‘letters’. Therefore, analysing the data generated by contemporary high-throughput sequencing technologies typically requires  high-performance computing support. 

\begin{figure}[!ht]
  \includegraphics[width=\textwidth]{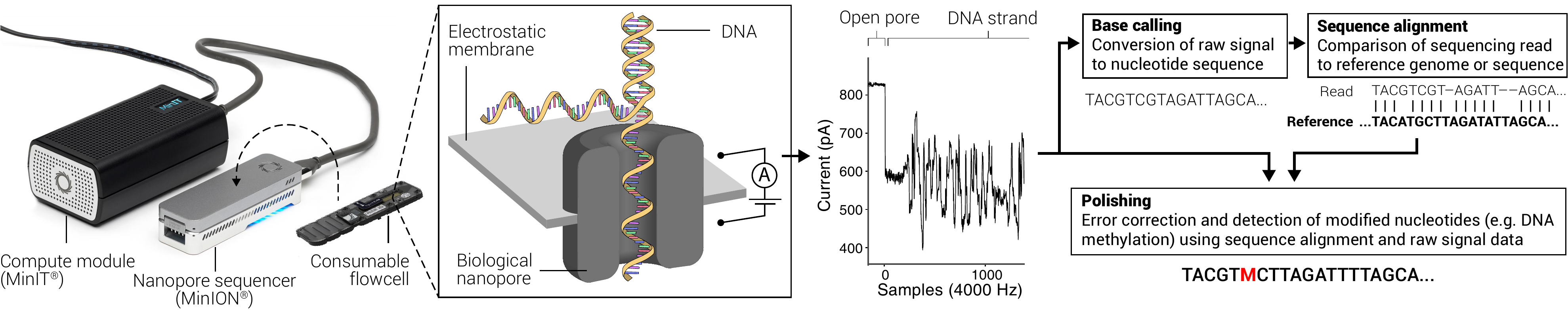}
  \caption{Nanopore portable sequencer and associated data analysis}
  %\Description{description}
  \label{fig:teaser}
\end{figure}

The latest generation (third generation) of sequencing technologies can generate ultra-long DNA `reads' from single molecules in real-time. In particular, Oxford Nanopore Technologies (ONT) manufacture a pocket-sized sequencer called MinION (Fig. \ref{fig:teaser}), a relatively inexpensive and portable sequencing device capable of sequencing in-the-field (e.g. remote area with no network connectivity) or at the point-of-care (e.g. hospital, clinic, pharmacy). 

In contrast to `second' generation sequencers, which produce highly accurate short reads through enzymatic synthesis of the complementary strand of DNA, nanopore sequencing measures characteristic disruptions in the electric current (referred to hereafter as \textit{raw signal}) when DNA passes through a biological nanopore (Fig. \ref{fig:teaser}).  A consumable flowcell containing an array of hundreds or thousands of such nanopores is loaded into the sequencing device (e.g. MinION), which is coupled to a generic (e.g. laptop) or dedicated (e.g. MinIT) compute module to acquire sequence data and base-call (the process of converting the raw signal to nucleotide characters) in parallel.

Nanopore sequencing offers several benefits over other technologies, including ultra-long reads (>1 Mbases), detection of non-standard DNA bases or biochemically modified DNA bases, and real-time analysis, at the expense of a higher error-rate, which is predominantly caused by the conversion of the raw signal into DNA bases via probabilistic models (referred to as `base-calling'). To overcome base-calling errors, the raw signal can be revisited \textit{a posteriori} (see the polishing step in Fig. \ref{fig:teaser}). Such \textit{a posteriori} ‘polishing’ can correct for base-calling errors by aligning raw signal to a biological reference sequence, thus identifying idiosyncrasies in the raw signal by comparing observed signal levels to expected levels at all aligned positions. This process can also reveal base substitutions (i.e. mutations) or base modifications such as 5-methylcytosine (5mC), a dynamic biochemical modification of DNA that is associated with genetic activity and regulation. Detecting 5mC bases is important for the study of DNA methylation in the field of epigenetics.

A crucial algorithmic component of polishing is the alignment of raw signal--a time series of electric current to a biological reference sequence.  One of the first raw nanopore signal alignment is implemented in the popular tool \textit{Nanopolish} \cite{simpson2017detecting}, which employs a dynamic programming strategy referred to as  Adaptive Banded Event Alignment (ABEA)\footnote{Recently, deep learning based tools such as \textit{Deepsignal} and \textit{Deepmod} have been released for methylation calling \cite{xu2019recent}. However, those tools have been developed using Python and are dependent on a large number of bulky libraries including Tensorflow. Thus, those tools are not very practical to be optimised for embedded systems with limited resources. Conversely, Nanopolish that utilises traditional alignment and Hidden Markov Model based methods, developed using C/C++, was a good candidate to be optimised for embedded systems.}. ABEA is one of the most time consuming steps when analysing raw nanopore data. For instance, when performing methylation detection with \textit{Nanopolish}, the ABEA step consumes \textasciitilde70\% of the total CPU time. Consequently, it is important to investigate strategies to reduce the runtime of ABEA to improve the turnaround time of certain nanopore sequencing applications, such as real-time polishing or methylation detection. 

In this study, we dissect the ABEA algorithm to optimise and parallelise its use on diverse hardware platforms, including Graphics Processing Units (GPUs). Adapting this ABEA algorithm for the GPU is not a straight forward task due to three main factors: (i) Read lengths vary significantly (from \textasciitilde100 to >1M bases), thus requiring millions to billions of dynamic memory allocations---an expensive operation in GPUs. (ii) Inefficient memory access patterns which are not ideal for the GPUs having relatively less powerful and smaller caches (compared to CPUs) result in frequent instruction stalls. (iii) Varying read lengths cause irregular utilisation of the GPU cores. %(\textasciitilde60\% of clock cycles for GPU based ABEA are wasted on memory dependency stalls - see results). 

 %By exploiting knowledge of computer architecture, bioinformatics algorithms, and biological data idiosyncrasies, optimise it.
 
We overcome the above mentioned challenges by: (i) employing a custom heuristic-based memory allocation scheme; (ii) tailoring the algorithm and the GPU user-managed cache to exploit cache-friendly memory access patterns; and, (iii) using a heuristic based work-partitioning and load-balancing scheme between the CPU and GPU.

We demonstrate the utility of our GPU optimised ABEA by incorporating to a completely re-engineered version of the popular methylation detection tool \textit{Nanopolish}. First, we re-engineered the original \textit{Nanopolish} methylation detection tool to efficiently utilise existing CPU resources, which we refer to as \textit{f5c}. Then, we incorporated our GPU optimised ABEA algorithm into the re-engineered \textit{f5c}. We demonstrate how \textit{f5c} enables DNA methylation detection using nanopore sequencers in real-time (i.e. on-the-fly processing of the output) by using an embedded System on Chip (SoC) equipped with a GPU.  We also  demonstrate how \textit{f5c} benefits a wide range of computing systems from embedded systems and laptops to workstations and high performance servers.

%Improves the turn-around time of results for sequencing facilities.

%We demonstrate how an embedded system equipped with a NVIDIA tegra SoC is capable of processing raw signal and  detect DNA methylation on the fly. Such an embedded system will speedup biological data analysis, reduce data storage and data transfer requirements, as well as accelerating the development of portable genomic applications predicated on nanopore signal analysis.

%\todo{[TODO : In addition to SoC, can be deployed on NVIDIA GPU equipped laptops, desktops and servers.The optimisations not only benefit embedded systems, but also HPC.  Possibility of use in other workflows such as  re-squiggle, variant calling applications etc.]}

The key contributions of this paper are: (i) the first example of GPU acceleration and optimisation of raw signal alignment algorithm; (ii) \textit{f5c}, a re-engineered and optimised version of the popular DNA methylation detection tool \textit{Nanopolish}; and, (iii) real-time detection of DNA methylation using a lightweight and portable embedded system (previously only possible on high-performance servers).

In the rest of the paper, we discuss  the background of nanopore sequencing and ABEA algorithm in Section \ref{background}, related work in Section \ref{relwork}, methodology in Section \ref{method}, results in Section \ref{res}, followed by the discussion and future work in Section \ref{discussion}. The associated tool that includes the GPU-based acceleration is available at \url{https://github.com/hasindu2008/f5c}.

\section{Background} \label{background}

Basic terms and concepts of DNA sequencing and data analysis are given in Section \ref{s:basicsnanopore}.  Section \ref{methcall}  briefly explains methylation calling,  an example nanopore data analysis workflow. Section \ref{adaptivebanded}  explains the Adaptive Banded Event Alignment (ABEA) algorithm, the algorithm which is optimised in this paper for execution on a CPU-GPU heterogeneous architecture. In Section \ref{gpu},  a brief account of GPU architectures and the programming methods for GPUs.

\subsection{Nanopore sequencing and analysis} \label{s:basicsnanopore}
\subsubsection{Whole genome sequencing}
The \emph{genome} is a long sequence composed of four types of nucleotide bases: adenine (A), cytosine (C), guanine (G) and thymine (T). Nucleotide bases will be  simply referred to as \textit{bases} hereafter. The human genome is around 3.2 gigabases (Gbases) long and is composed of 23 pairs of chromosomes (46 chromosomes in total), where each chromosome is a single molecule of continuous deoxyribonucleic acid (DNA) polymer. The process of reading strings of contiguous bases is called \emph{sequencing}, and the resulting strings of bases are called \emph{reads}. In order to be sequenced, DNA molecules must be extracted and purified from cells before being biochemically prepared for sequencing. This \emph{library preparation} process can fragment chromosomes (especially large ones) into smaller segments--either intentionally or incidentally--which are `read' by the sequencer. Given that samples contain multiple cells, and thus several distinct DNA molecules, and that sequencing may introduce errors, it is desirable to generate enough reads to cover a particular position several times. The average number of reads at a given position is termed sequencing \emph{coverage}. High coverage facilitates the characterisation of genetic variation and correct for errors. A human genome sequenced at around 20$\times$ average coverage corresponds to around 64 Gbases of sequencing reads. 
%\todo{[refer XYZ for more info.]}

\subsubsection{DNA methylation}
DNA undergoes naturally regulated biochemical modification through the addition of a methyl group to certain bases. Methylation is reversible and can control the activity of a DNA segment, such as turning the expression of genes on or off, without modifying the genetic code itself---a process called \emph{epigenetic} regulation. DNA methylation is dynamically regulated during normal biological development and in function of environmental factors; it plays an important role in disease aetiology and clinical diagnostics \cite{liyanage2014dna,lewandowska2011dna,fraser2017genomic}. Methylation of cytosine (`C') bases is of particular interest in human biology, where CpG dinucleotides ( `C' base followed by a `G' base) are dynamically methylated in normal development and disease \cite{saxonov2006genome,bird2002dna,gonzalo2010epigenetic}.

\subsubsection{Nanopore sequencing and the raw signal}
Nanopore sequencing is a third generation sequencing technology that involves physical observation of atomic properties of DNA fragments using a nanometer scale biological pore coupled to an ammeter (see Fig. \ref{fig:teaser}). The pore acts as a bottleneck to generate characteristic disruptions in ionic current (in the range of pico-amperes) that are indicative of the molecules passing through the pore. The size and nature of the pore influence the measured instantaneous current and how it is subsequently analysed. Oxford Nanopore Technologies (ONT) sequencing devices measure DNA strand passing through biological nanopores composed of recombinant (or `designer') proteins at an average speed of \textasciitilde450 bases/s while the current is sampled and digitised at \textasciitilde4000 Hz\footnote{these are typical values at present which may vary in the future}. The instantaneous current measured in ONT nanopore depends on 5-6 contiguous bases \cite{lu2016oxford}.  The measured signal also presents stochastic noise due to several factors, such as homopolymers (same base repeating multiple times) which produce constant current levels, contaminants in the sample, entanglement of long DNA strands, depletion of ions, etc \cite{rang2018squiggle}. Additionally, the movement
speed of the DNA strand through the pore can vary, causing the signal to warp in the time domain \cite{rang2018squiggle}. The raw signal is converted into character representations of DNA bases (e.g. A,C,G,T)  using artificial neural networks, generating a typical accuracy >90\% for single reads \cite{wick2019performance}. This conversion process is referred to as \textit{base-calling} and the software tools that perform this conversion are referred to as  \textit{base-callers}.
Please refer to \cite{lu2016oxford} for a detailed discussion of ONT sequencing.

\begin{figure}[!ht]
\begin{subfigure}[!ht]{.72\textwidth}
  \centering
    \includegraphics[width=\columnwidth]{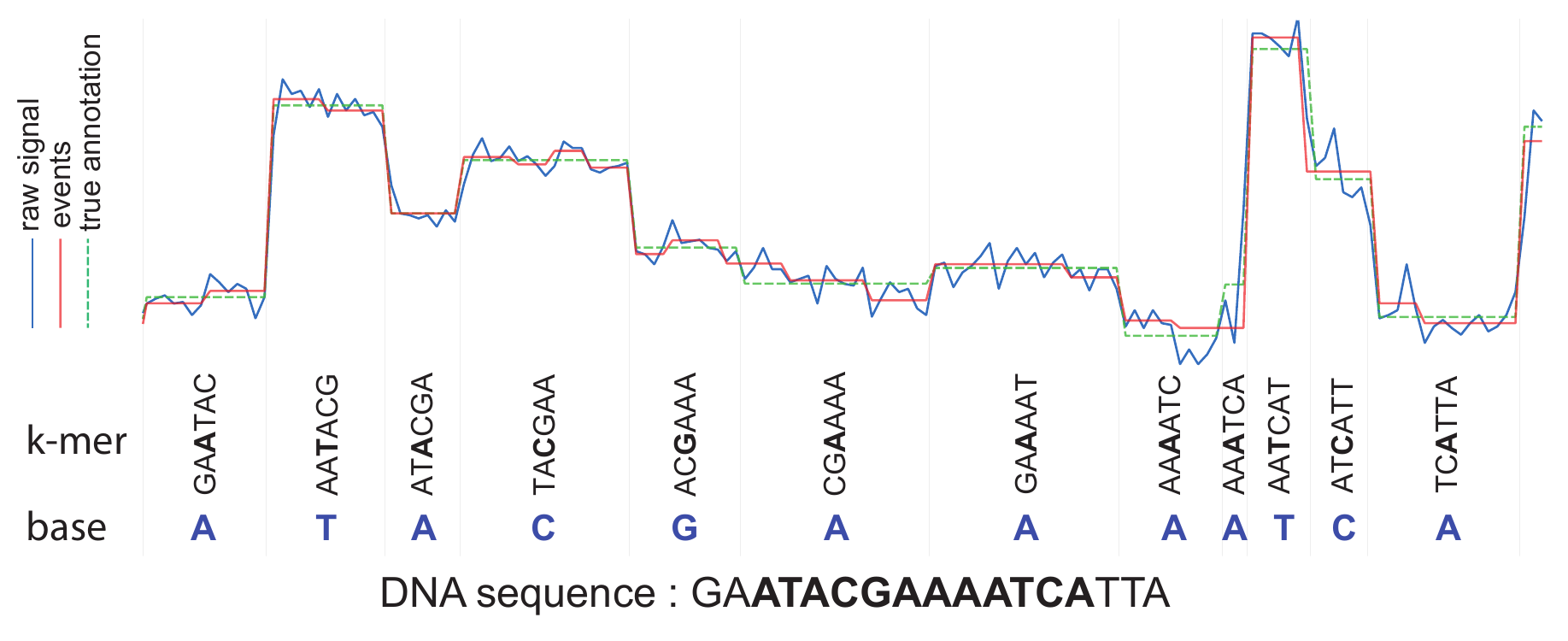}
    \caption{An example nanopore raw signal and events} 
    \label{f:kmers}
\end{subfigure}
    \begin{subfigure}[!ht]{.27\textwidth}
  \centering
    \includegraphics[width=\linewidth]{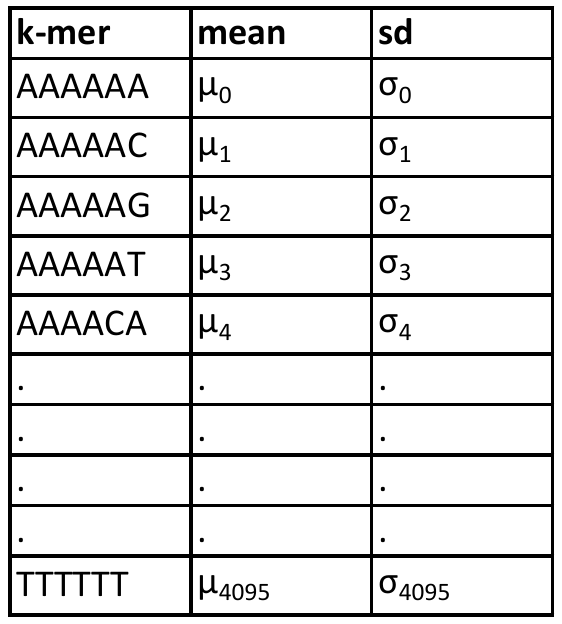}
    \caption{an example \textit{pore-model}} 
    \label{f:kmermodel}
    \end{subfigure}
    \caption{Illustration of a nanopore raw signal, events and \textit{pore-model}} 
\end{figure}

An example of a raw nanopore signal is shown in Fig.  \ref{f:kmers} using blue coloured line. Assume that the signal is generated from the DNA sequence \textit{GAATACGAAAATCATTA} which passed through the nanopore. In this example, the instantaneous current of the signal is affected by a string of 6 contiguous bases, known as a \textit{6-mer} (or a \textit{k-mer} in general).
%)This string of contiguous bases is known as a k-mer, where k is 6 in this case.
%(known as a \textit{6-mer}--a \emph{k-mer} is a string of \textit{k} contiguous bases).  
Let us assume that the annotation of the signal to the corresponding \textit{k-mers} is known (the process of getting this annotation is detailed in Section \ref{methcall}). The \textit{6-mers} in the sequence and the corresponding segments in the raw signal are marked using vertical grey lines in Fig. \ref{f:kmers}. When the DNA sequence \textit{GAATACGAAAATCATTA} moves through the pore, the first \textit{6-mer} is \textit{GAATAC}. Similarly, the subsequent \textit{6-mers} are \textit{AATACG}, \textit{ATACGA}, \textit{TACGAA}, ..., \textit{TCATTA}.
%The first \textit{6-mer} in the sequence is \textit{GAATAC} and the corresponding segment in the raw signal is marked in Fig. \ref{f:kmers} (\textit{GAATAC} in the first column).  The corresponding segments in the raw signals for these k-mers are marked in Fig.  \ref{f:kmers} under the corresponding signal segment. 
\textit{True annotation} (depicted by dotted green coloured step function in Fig.  \ref{f:kmers}) corresponds to the \textit{ideal} average level of current for each \textit{k-mer}. These ideal average values are obtained using the \textit{pore-model} provided by ONT, which is elaborated in Section \ref{methcall}. The red coloured step function corresponds to an \emph{event}---detailed in Section \ref{methcall}. 

To deduce the sequence from the \textit{k-mers}, the base at the centre (3rd base) of each k-mer is taken, as shown on the bottom of Fig. \ref{f:kmers}. For instance, we take \textit{A} from \textit{GAATAC}, \textit{T} from \textit{AATACG}, \textit{C} from  \textit{TACGAA} and etc. Hence, we obtain a sequence \textit{ATACGAAAATCA} which is a part of the original sequence \textit{GAATACGAAAATCATTA}. Note that the beginning and the end of the sequence (GA at the beginning and TTA at the end) are clipped. 

\subsubsection{Nanopore read length distribution}
\begin{figure}[!ht]
  \centering
    \includegraphics[width=\columnwidth]{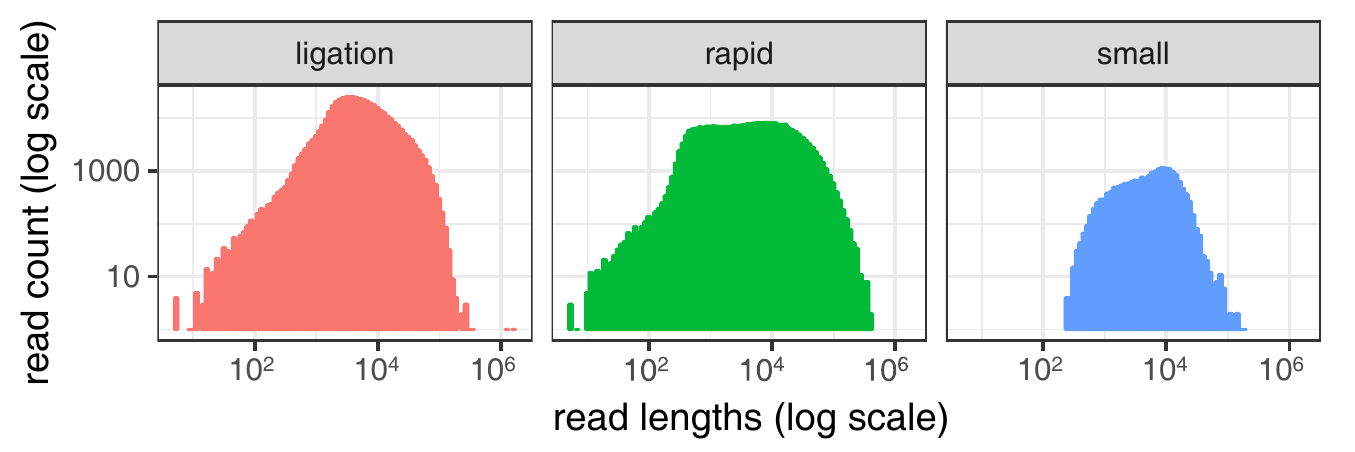}
    \caption{example nanopore read length distributions} 
    \label{f:rlen}
\end{figure}

 The length of the reads generated from nanopore sequencers can vary from several hundred bases to even more than 2 million bases.  A typical sequencing run of a particular sample (which completes after 48-64 hours) generates millions of such reads. The distribution of the read lengths varies in function of DNA integrity, extraction protocols, and sample preparation methods. Example distributions for three different samples are shown in Fig. \ref{f:rlen}, where both x and y axes are in logarithmic scale.  The average read length of a sample typically falls between 8-20 Kilobases. 
 
\subsubsection{Sequence alignment/mapping in the base-space}\label{s:alignment}
Once a nanopore read is base-called, the sequence is aligned to a reference sequence (see Fig. \ref{fig:teaser}). A reference sequence consists of a previously generated consensus sequence (such as the human genome reference). Sequence alignment involves global optimisation algorithms to identify the most similar target and to compare any differences between sequences. 
Compared to biologically occurring variation in individual genomes (<1\% difference to the reference), the error-rate of nanopore sequencing is  relatively high  (5-10\%). Thus, sequence alignments derived from nanopore reads are distinct in nature from previous sequencing technologies (such as highly accurate short reads). Consequently, unique analytic tools must be considered when aligning such reads. Alignment tools such as Minimap2 \cite{minimap2} that employ a hash table based genome index followed by a base-level dynamic programming alignment step can successfully align long and noisy reads.

\subsubsection{Polishing/Downstream processing using raw signal}

The base-space alignment discussed previously in Section \ref{s:alignment} is followed by `polishing', a downstream processing step which utilises both the base-space alignment results and the raw signal (see Fig. \ref{fig:teaser}). The polishing step reuses the raw signal to recover the lost biological information during base-calling. This polishing step can be to correct errors during base-calling or to detect modified nucleotide bases (eg: DNA methylation).

Previous research has shown that identification of genetic variants can be improved up to an accuracy of more than 99\% by using raw signal data from multiple overlapping reads \cite{jain2018nanopore,loman2015complete}. Thus, the downstream analysis that reuses raw signal data could correct for base-calling errors. It has also been shown that methylated C bases can be differentiated from non-methylated C bases by the use of signal data, using algorithms such as the one implemented in the software package \emph{Nanopolish} \cite{simpson2017detecting}. Thus, the downstream analysis that reuses raw signal data could detect modified nucleotide bases.

Signal-space alignment is one of the crucial steps performed in these downstream analyses such as error correction and modified base detection. This signal alignment step is described in the context of modified base detection in the following sections. 

%However, the base-space alignment is still crucial as it reveals the coordinates on the genome to which a given read maps. Thus, the search space is reduced where a signal-space alignment can be efficiently performed - where the raw signal of the read is aligned to a hypothetical model signal.

\subsection{Methylation calling} \label{methcall}

As discussed above, important biological information is lost during base-calling.
Some base-calling models may not accommodate methylated data, either because they are trained on unmethylated sequences, or because they abstract away non-canonical bases. Therefore, these molecules may be erroneously classified as unmethylated bases. The process of identifying methylation  is known as \textit{methylation calling}.

As implemented in \emph{Nanopolish}, methylation calling requires: 1, raw signals; 2, base-called reads; and 3, base-space alignment to a reference genome (output of the sequence alignment step described above). For a given read, the main steps for methylation calling are: 1, event detection; 2, signal-space alignment; and 3, Hidden Markov Model (HMM) profiling. These steps are performed for each individual read in the data set.

\emph{Event detection} is the time series segmentation of the raw signal based on sudden signal level changes. Each segment is called an \emph{event} and is typically denoted using the mean ($\mu_{\bar{x}}$), standard deviation ($\sigma_{\bar{x}}$) and the duration of the raw signal samples ($n_{\bar{x}}$) pertaining to the particular segment. The red step function in Fig. \ref{f:kmers} denotes such detected events by plotting the mean value of the samples ($\mu_{\bar{x}}$) corresponding to the segment. Note that in Fig. \ref{f:kmers}, events (red line) roughly match to the true annotation (dotted green line), nevertheless,  are not exactly the same. Mostly, the signal has been over-segmented (eg: portion corresponding to k-mer \textit{CGAAAA} has been segmented into 3 events) and seldom under-segmented (eg: k-mer \textit{AAATCA}).

%Refer to the associated code in \cite{simpson2017detecting} for more details on event detection.

%Though ideally there should be one event per base, in a real-world noisy signal there can be under segmentation and over-segmentation. As a lower threshold is typically used during event detection, we mostly observe over-segmentation in Fig. \ref{f:kmers}. The event detector in Nanopolish (which is based on the implementation in Scrappie \cite{}) generates an average of around 2.5 events per base.

To obtain the true annotation in Fig. \ref{f:kmers}, the events detected in the event detection step are aligned to a generic k-mer model signal. This generic k-mer model signal is derived from the base-called sequence and a \textit{pore-model} provided by ONT. The \textit{pore-model} corresponds to a table of all possible k-mers matched to their mean signal value and standard deviation (4\textsuperscript{6} k-mers if k is 6, as shown in Fig. \ref{f:kmermodel})\footnote{there can be other values in addition to mean and standard deviation, which are not required for our methylation calling}. For each 6-mer in the base-called read, the corresponding entry in the \textit{pore model} (\textit{mean,sd}) is obtained and these \textit{mean,sd} pairs form the generic k-mer model signal.
\textit{Nanopolish} aligns the events from the event detection step to this generic k-mer model signal by using the algorithm named \emph{Adaptive Banded Event Alignment (ABEA)} explained in Section \ref{adaptivebanded}. 

ABEA above produces the alignment between the events and the k-mers in the base-called read. The base-space sequence alignment then is used to deduce which event corresponds to a given k-mer in the reference genome. Finally, this alignment between the events and the k-mers in the reference genome are subjected to Hidden Markov Model (HMM) profiling to identify if a given base is methylated or not.

% \clearpage

% \par\noindent\textcolor{red}{\rule{\textwidth}{3pt}}

\subsection{Adaptive Banded Event Alignment (ABEA)} \label{adaptivebanded}

Modified versions of the SK (explained in section \ref{s:workflow3rdgen}) algorithm are used for event-space alignment as exemplified in \textit{Nanopolish} and is referred to as \textit{Adaptive Banded Event Alignment (ABEA)}. In ABEA, the events are aligned to the k-mers of the base-called read (as stated in Section \ref{methcall}). As typically there are many more events than k-mers (usually by a factor 1.5-2) due to the frequent over-segmentation of events (discussed in Section \ref{methcall}), event alignment is even more difficult than base-space long read alignment if performed with static banding around the diagonal. Thus, an adaptive band is essential for event alignment.

%The band-width used for signal alignment is typically around (\textasciitilde100).
The scoring function for signal alignment uses a 32 bit floating point data type, as opposed to 8-bit integer data type in sequence alignment. Furthermore, the signal alignment scoring function that computes the log-likelihood (which we elaborate shortly) is computationally expensive.

\begin{figure*}[!ht]
  \centering
\begin{subfigure}[t]{.4\textwidth}
  \centering
    \includegraphics[trim=250 300 130 200,clip,width=\textwidth]{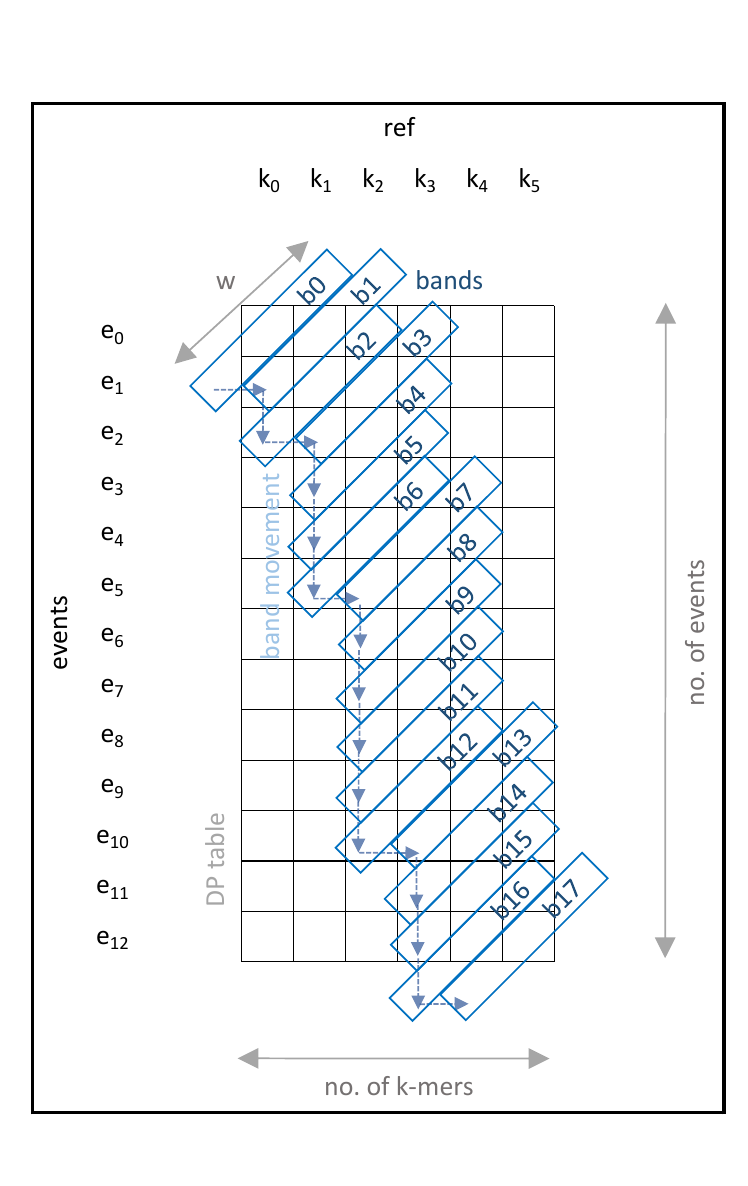}
    \caption{band movement} 
    \label{f:adaptive}
\end{subfigure}
\begin{subfigure}[t]{.4\textwidth}
  \centering
    \includegraphics[trim=250 300 130 200,clip,width=\textwidth]{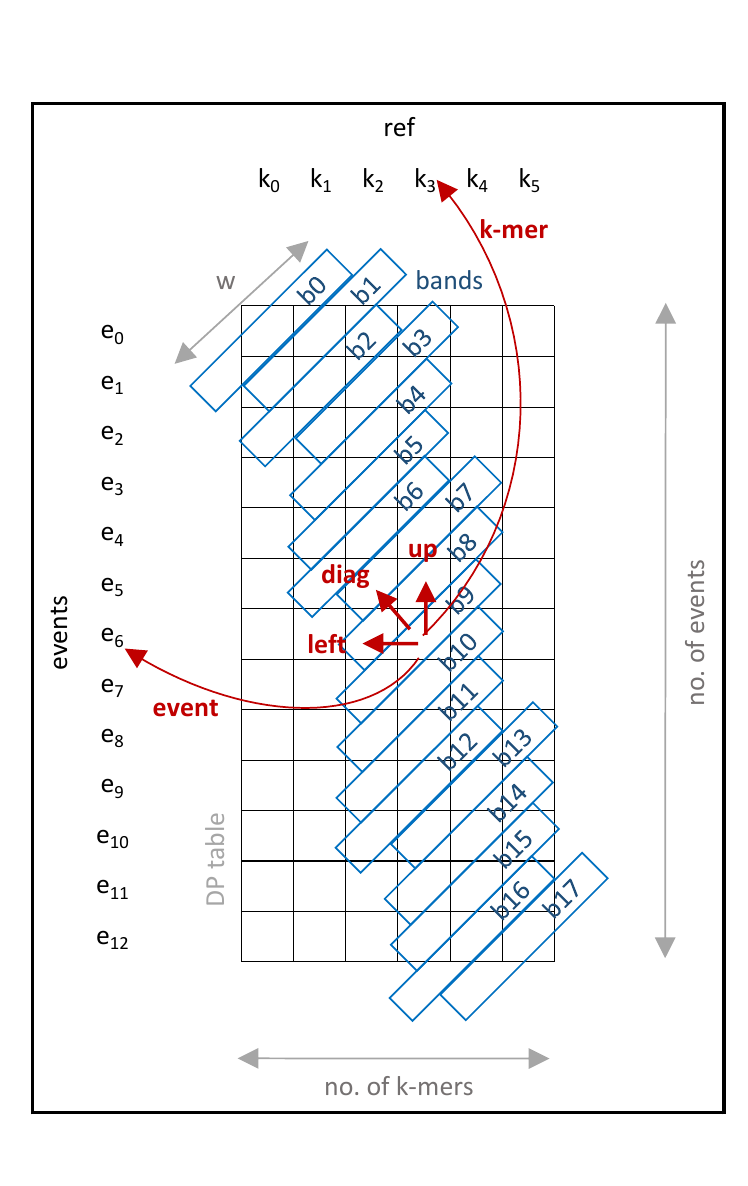}
    \caption{computing a single cell score} 
    \label{f:adaptive2}
\end{subfigure}

\begin{subfigure}[t]{.4\textwidth}
  \centering
    \includegraphics[trim=250 300 130 200,clip,width=\textwidth]{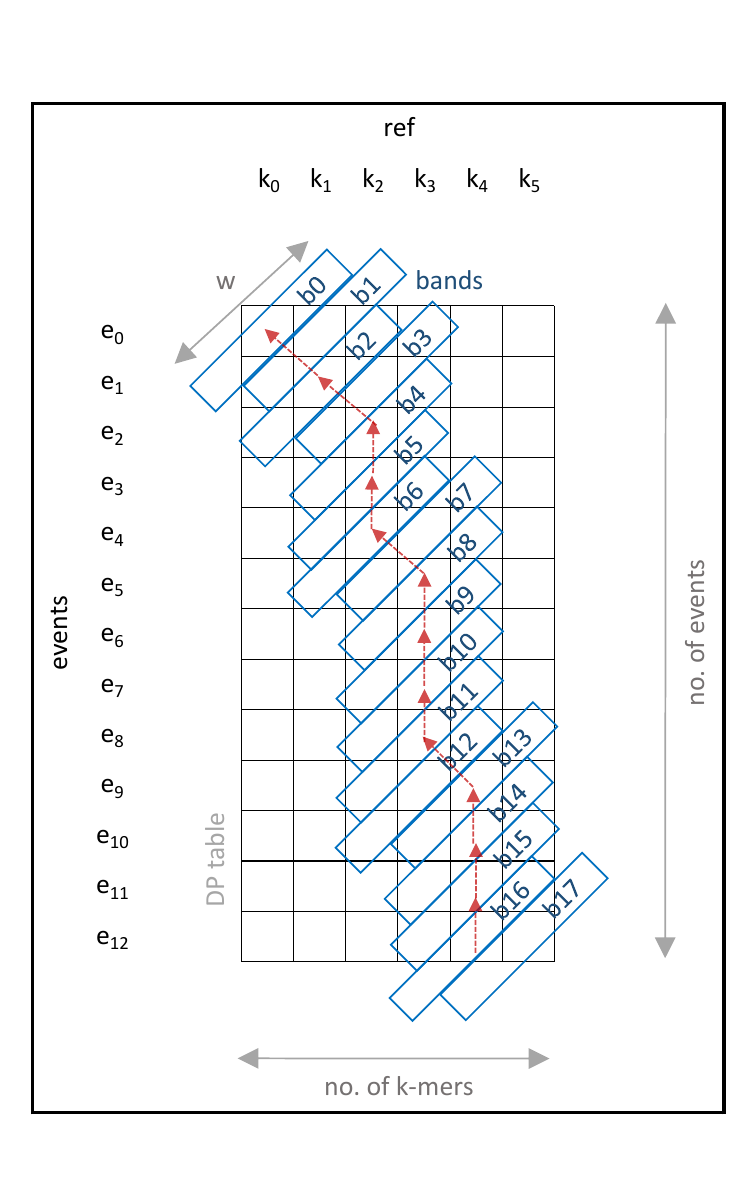}
    \caption{trace-back} 
    \label{f:adaptive3}
\end{subfigure}\hfill
\caption{Adaptive Banded Event Alignment}
\label{f:abea}
\end{figure*}

A simplified example of ABEA is shown in Fig. \ref{f:adaptive}. In Fig. \ref{f:adaptive} the horizontal axis represents the events (results of the event detection step) and the vertical axis represents the \textit{ref} k-mers (k-mers of the base-called read). The dynamic programming table (DP table) in  Fig. \ref{f:adaptive} is for 13 events, indexed from e\textsubscript{0}-e\textsubscript{12} vertically, and the \textit{ref} k-mers, indexed from k\textsubscript{0}-k\textsubscript{5} horizontally.  
 As mentioned previously for computational and memory efficiency, only the diagonal bands (marked using blue rectangles) with a band-width of $W$ (typically $W$=100 for nanopore signals) are computed. The bands are computed along the diagonal from top-left (\textit{b0}) to bottom-right (\textit{b17}). Each cell score is computed in function of five factors: scores from the three neighbouring cells (up, left and diagonal); the corresponding \textit{ref} k-mer; and, the event (shown for the cell e\textsubscript{6}, k\textsubscript{3} via red arrows in Fig. \ref{f:adaptive2}, details of the computation is explained later). Observe that all the cells in the $n$\textsuperscript{th} band can be computed in parallel as long as the $n-1$\textsuperscript{th} and $n-2$\textsuperscript{th} bands are computed beforehand. To contain the optimal alignment, the band adapts by moving down or to the right as shown using blue arrows in  Fig. \ref{f:adaptive}. The adaptive band movement is determined by the Suzuki-Kasahara heuristic rule \cite{suzuki2018introducing}.

Algorithm \ref{adaptivebandedcpu} summarises the ABEA algorithm used in \textit{Nanopolish} \cite{simpson2017detecting} and is explained with the aid of the example in Fig. \ref{f:adaptive}. 

%Gihan%

\begin{algorithm}[!ht]
\caption{Adaptive Banded Event Alignment}\label{adaptivebandedcpu}
{\fontsize{10}{7.5}\selectfont
\begin{flushleft}
\textbf{Input:} \\
\hspace*{\algorithmicindent}  \textit{ref[]} : the base-called read (1D char array) \\
\hspace*{\algorithmicindent} \textit{model} : \textit{pore-model} (Fig. \ref{f:kmermodel}) \\ %(k-mers and their $\mu$,$\sigma$) in Fig.   \\
\hspace*{\algorithmicindent} \textit{events[]} : event table containing \{$\mu_{\bar{x}}$,$\sigma_{\bar{x}}$,$n_{\bar{x}}$\} of each event---1D \{\textit{float},\textit{float},\textit{float}\} array\\
\textbf{Output:} \\
\hspace*{\algorithmicindent} \textit{alignment[]} : alignment denoted by a list of \{\textit{event index},\textit{k-mer index}\}---1D \{\textit{int},\textit{int}\} array \\ 
\textbf{Intermediate:}\\
\hspace*{\algorithmicindent} \textit{score[][]} : scores of the cells in banded area---2D float array \\ 
\hspace*{\algorithmicindent} \textit{trace[][]} : back-track flags of the cells in banded area---2D char array \\ 
\hspace*{\algorithmicindent} \textit{ll\_idx[]} : \{event index,k-mer index\} for each band's lower left cell---1D \{\textit{int},\textit{int}\} array\\ 
%\green{TODO: algorithm 1 take the band-width as a parameter}\\
\end{flushleft}
\begin{algorithmic}[1]

\Function{\textit{align}}{\textit{ref,model,events}}

\State \textit{initialise\_first\_two\_bands(score,trace,ll\_idx)} \Comment \textcolor{gray}{band b0 and b1 in Fig. \ref{f:adaptive}, see line \ref{a:abeacpu:init}} \label{a:abeacpu:init1}

%\State ll[0].event\_idx $\gets$ $\frac{band-width}{2} - 1$
%\State ll[0].kmer\_idx $\gets$ $-1 - \frac{band-width}{2}$
    
% \State \textit{ll[0]} $\gets$ $[\frac{w}{2} - 1,-1 - \frac{w}{2}]$
% \State \textit{ll[1]} $\gets$ move\_down(\textit{ll[0]})

\For{\textit{i $\gets$ 2 to n\_{bands}}} \Comment \textcolor{gray}{Iterate from b2 to b17 in Fig. \ref{f:adaptive}} \label{a:abeacpu:outerloop}
    % \State $ll$ $\gets$ lower left from band array
    % \State $ur$ $\gets$ upper right from band array
    % \If{out\_of\_bounds(\textit{ll,ur}) }
    %     \State \textit{right} $\gets$ is\_odd(\textit{i})
    % \Else
    %     \State \textit{right} $\gets $ $ll < ur$ ?
    % \EndIf
    
    \State \textit{dir} $\gets$ \textit{suzuki\_kasahara\_rule(score[i-1])} \Comment \textcolor{gray}{\textit{score[i-1]} is of the previous band} \label{a:abeacpu:bandmoves}
    \If{$dir == right$}
        \State \textit{ll\_idx[i]} $\gets$ \textit{move\_band\_to\_right(ll\_idx[i - 1])} \Comment  \textcolor{gray}{see line \ref{a:abeacpu:moveright}}
    \Else
        \State\textit{ ll\_idx[i]} $\gets$ \textit{move\_band\_down(ll\_idx[i - 1])} \Comment \textcolor{gray}{see line \ref{a:abeacpu:movedown}}
    \EndIf \label{a:abeacpu:bandmovee}
    
    \State \textit{min\_j,max\_j} $\gets$ \textit{get\_limits\_in\_band(ll\_idx[i])} \Comment \textcolor{gray}{get index bounds in current band\textsuperscript{*}}
    
    \For{\textit{j} $\gets$ \textit{min\_j} to \textit{max\_j} } \Comment \textcolor{gray}{Iterates through each cell in band i} \label{a:abeacpu:innerls}

         \State \textit{s,d} $\gets$ \textit{compute(score[i-1],score[i-2],ref,events,model)} \Comment \textcolor{gray}{see Algorithm \label{a:abeacpu:cellcompute} \ref{adaptivebandedcpucompute}}%\Comment \textit{score[i-1]} and \textit{score[i-2]} are previous and second previous bands
         \State \textit{score[i,j]} $\gets$ \textit{s}
         \State \textit{trace[i,j]} $\gets$ \textit{d}
         
        %\State \textit{trace[i,j]}$\gets$ direction\_to\_backtrack(\textit{band[i-1]})
    \EndFor \label{a:abeacpu:innerle}
\EndFor \label{a:abeacpu:outerloope}
%\State \textit{trace, band} $\gets$ trim()
\State \textit{alignment} $\gets$ backtrack(\textit{score}, \textit{trace}. \textit{ll}) \Comment \textcolor{gray}{the trace-back red arrows in Fig. \ref{f:adaptive3}.} \label{a:abeacpu:backtrack}

\EndFunction
\\
\Function{\textit{initialise\_first\_two\_bands}}{\textit{score,trace,ll\_idx}} \label{a:abeacpu:init}

    \State \textit{score[0,*], trace[0,*]}  $\gets - \infty, 0$ \Comment \textcolor{gray}{Initialise first band b0} \label{a:abeacpu:21}
%     \State \textit{ll\_idx[0].event\_idx} $\gets$ $\frac{W}{2} - 1$ \Comment $W$ is the width of the band
% \State \textit{ll\_idx[0].kmer\_idx} $\gets$ $-1 - \frac{W}{2}$

    \State \textit{score[1,*], trace[1,*]}  $\gets - \infty, 0$ \Comment \textcolor{gray}{Initialise second band b1}\label{a:abeacpu:22}
    
    \State \textit{ll\_idx[0]} $\gets$ \{$ei_0$,$ki_0$\} \Comment \textcolor{gray}{$ei_0=1$ and $ki_0=-1$ in Fig. \ref{f:adaptive}\textsuperscript{**}} \label{a:abeacpu:23}
    \State \textit{ll\_idx[1]} $\gets$ \{$ei_1$,$ki_1$\} \Comment \textcolor{gray}{$ei_1=1$ and $ki_1=0$ in Fig. \ref{f:adaptive}\textsuperscript{**}}\label{a:abeacpu:24}

    \State \textit{score[0,$si_0]$} $\gets$ 0 \Comment  \textcolor{gray}{$si_0$ is 0 is Fig. \ref{f:adaptive}\textsuperscript{***}}\label{a:abeacpu:25}

\EndFunction
\\
\Function{\textit{move\_band\_to\_right}}{\textit{ll\_previous}} \label{a:abeacpu:moveright}
    \State \textit{ll\_current.event\_idx} $\gets$ \textit{ll\_previous.event\_idx + 1}
    \State \textit{ll\_current.kmer\_idx} $\gets$ \textit{ll\_previous.kmer\_idx}
\EndFunction
\\
\Function{\textit{move\_band\_down}}{\textit{ll\_previous}}\label{a:abeacpu:movedown}
    \State \textit{ll\_current.event\_idx} $\gets$ \textit{ll\_previous.event\_idx}
    \State \textit{ll\_current.kmer\_idx} $\gets$ \textit{ll\_previous.kmer\_idx+1}  
\EndFunction

\end{algorithmic}

\begin{flushleft}
{\textcolor{gray}{\textsuperscript{*}For instance, in Fig. \ref{f:adaptive} \textit{min\_j=1,max\_j=1} for b0 and b17; \textit{min\_j=0,max\_j=1} for b1; \textit{min\_j=1,max\_j=2} for b16;  and, \textit{min\_j=0,max\_j=2} for the rest\\
\textsuperscript{**}these initial event and k-mer indices corresponding to the lower left of the band are computed with respect to band-width $W$\\
\textsuperscript{***}the score of cell that corresponds to k-mer index -1 in band b0 is initiliased to 0}}
\end{flushleft}
}
\end{algorithm}

%---eg:[$\mu_{\bar{x1}}$,$\sigma_{\bar{x1}}$],[$\mu_{\bar{x2}}$,$\sigma_{\bar{x2}}$],[$\mu_{\bar{x3}}$,$\sigma_{\bar{x3}}$],....

 The input to the Algorithm \ref{adaptivebandedcpu} are: 1, \textit{ref} (the sequenced read in base-space---eg: \textit{GAATACG...}); 2, \textit{events} (the output of the event detection step mentioned in Section \ref{methcall}); and 3, \textit{model} (\textit{pore-model}---Fig. \ref{f:kmermodel}). As mentioned in Section \ref{methcall}, the ABEA algorithm (Algorithm \ref{adaptivebandedcpu}) attempts to align the events to the generic signal model (produced with the use of \textit{ref} and the \textit{model}) and outputs the alignment as \textit{event-ref} pairs. The algorithm requires three intermediate arrays, namely \textit{score} (2D floating point array), \textit{trace} (2D byte array) and \textit{ll} (1D pointer array) to formulate the intermediate state during alignment computation, which is the DP table shown in Fig. \ref{f:adaptive}).
Note that, \textit{ll} stands for lower-left, which holds the coordinate of the start point of the band.
 
 The initialisation of the first two bands (\textit{b0} and \textit{b1}) in Fig. \ref{f:adaptive} is performed by line \ref{a:abeacpu:init} of Algorithm \ref{adaptivebandedcpu}. Then, the outer loop (starting from line \ref{a:abeacpu:outerloop}) iterates through rest of the bands from top-left to bottom-right of the DP table. The inner loop (lines \ref{a:abeacpu:innerls}-\ref{a:abeacpu:innerle}) iterates through each cell in the current band \textit{bi}. To ensure that only cells within the DP table are computed, the loop counter $j$ iterates from \textit{min\_j} to \textit{max\_j}, instead of 0 to $W-1$. Lines \ref{a:abeacpu:bandmoves}-\ref{a:abeacpu:bandmovee} of Algorithm \ref{adaptivebandedcpu} correspond to the movement of the band (corresponds to the blue arrows in Fig. \ref{f:adaptive}). Band movement is actuated by proper placement of the band in the static 2D arrays, \textit{score} and \textit{trace} via the array \textit{ll} using the functions \textit{move\_band\_right} and \textit{move\_band\_down}.  
 
 Line \ref{a:abeacpu:cellcompute} of the algorithm performs the cell score computation (explained in detail later) and generates a score and a direction flag for subsequent backtracking, which are henceforth stored in the arrays \textit{score} and  \textit{trace}.  When all the cells in the DP table are computed, the final operation is to find the actual alignment (\textit{event-ref} pairs) through the backtracking operation (line \ref{a:abeacpu:backtrack} of Algorithm \ref{adaptivebandedcpu} and red trace-back arrows in Fig. \ref{f:adaptive3}), which uses both the cell scores and the direction flags stored in \textit{trace}.

The \textit{compute} function (called at line \ref{a:abeacpu:cellcompute} of Algorithm \ref{f:adaptive}) is elaborated in Algorithm \ref{adaptivebandedcpucompute}. A number of heuristically determined constants suitable for Nanopore data, which are used during subsequent calculations are listed at the beginning of this algorithm. The first step of this algorithm is the computation of \textit{lp\_emission}, a log probability value (likelihood of the particular signal event being the particular \textit{ref} k-mer), performed using the function elaborated in Algorithm \ref{adaptivebandedcpucomputelog}. This computed \textit{lp\_emission} is used in lines \ref{a:compute:lps}-\ref{a:compute:lpe} of Algorithm \ref{adaptivebandedcpucompute} along with the heuristically determined constants (\textit{lp\_skip,lp\_stay,lp\_step}) to compute three scores from the diagonal, left and up (\textit{score\_d, score\_u, score\_l}). The maximum of the three scores and direction from which the max score came (flags pertaining to diagonal, up or left) are returned as outputs from this function. The line \ref{a:compute:getscore} of Algorithm \ref{adaptivebandedcpucompute}  refers to accessing the scores of the upward, left and diagonal cells which was previously mentioned with respect to cell e\textsubscript{6},k\textsubscript{3} and the red arrows in Fig. \ref{f:adaptive2}. 
 
 The log probability computation in Algorithm \ref{adaptivebandedcpucomputelog} involves  floating point log probability computations. For the k-mer at the specific \textit{ref} position, the \textit{pore-model} table (Fig. \ref{f:kmermodel}) is accessed to obtain the corresponding model values. This \textit{model\_kmer} (mean and the standard deviation of the particular model k-mer) and the mean value of the event is used for the log probability computation as shown in the Algorithm \ref{adaptivebandedcpucomputelog}. Note that for event alignment neither the standard deviation or the duration of the event are used.

 %, thus involving random memory accesses. However, this \textit{model} array is small (typically\textasciitilde32KB) which easily fits in CPU cache.

 %used for intenal book keeping of the band movement - the movement of the band inside the 2D arrays are done by placing the band accordingly using this)
 
 %The Algorithm \ref{adaptivebandedcpu} requires two 2-D arrays \textit{bands} (for storing the computed scores) and \textit{trace} (stores direction for later backtracking) and a 1-D array \textit{ll} (and henceforth not throughly discussed). The 2D arrays are for the cells marked in figure for the band and each array is (row+cols)x$w$.

%The algorithm generates a set of pairs (mappings) from events to references.\\
%The algorithm takes the sequence computed by the previous signal processing. \textit{kmer\_rank} is computed for every character in the sequence. They corresponds to the events axis in figure \ref{adaptivebanded}. They are matched against the models for reference.\\
%The output of the algorithm is stored as a collection of mappings between events and references. The computation is done using the array \textit{ll} which is then backtracked to obtain the final answer.\\

\begin{algorithm}[!ht]
\caption{Adaptive Banded Event Alignment - cell score computation}\label{adaptivebandedcpucompute}
\begin{flushleft}
\textbf{Constants:} \\
\hspace*{\algorithmicindent} $events\_per\_kmer = \frac{n\_events}{n\_kmers}$\\
\hspace*{\algorithmicindent} $\epsilon = 1^{-10}$ \\
\hspace*{\algorithmicindent} \textit{lp\_skip} $= \ln(\epsilon)$  \\
\hspace*{\algorithmicindent} \textit{lp\_stay} $=\ln(1-\frac{1}{events\_per\_kmer+1})$ \\ 
\hspace*{\algorithmicindent} \textit{lp\_step} $= \ln(1.0 - e^{lp\_skip} - e^{lp\_stay})$ \\
\end{flushleft}
\begin{algorithmic}[1]

% \hspace*{\algorithmicindent} \textit{model} : pore-model (k-mers and their $\mu$,$\sigma$)  \\

\Function{computation}{\textit{score\_prev},\textit{score\_2ndprev},\textit{ref},\textit{events},\textit{model}}

\State \textit{lp\_emission} $\gets$ \textit{log\_probability\_match(ref,events,model)} \Comment \textcolor{gray}{see Algorithm \ref{adaptivebandedcpucomputelog}}
\State \textit{up,diag,left} $\gets$ \textit{get\_scores(score\_prev,score\_2ndprev)} \label{a:compute:getscore} \Comment \textcolor{gray}{see red arrows in Fig. \ref{f:adaptive2}}
\State \textit{score\_d} $\gets$ \textit{diag} $+$ \textit{lp\_step} $+$ \textit{lp\_emission}  \label{a:compute:lps}
\State \textit{score\_u} $\gets$ \textit{up} $+$ \textit{lp\_stay} $+$ \textit{lp\_emission}  \label{a:compute:lpe}
\State \textit{score\_l} $\gets$ \textit{left} $+$ \textit{lp\_skip} 
\State \textit{s} $\gets$ \textit{max(score\_d,score\_u,score\_l)}
\State \textit{d} $\gets$ \textit{direction from which the max score came}
\EndFunction

\end{algorithmic}

\end{algorithm}

\begin{algorithm}[!ht]
\caption{Adaptive Banded Event Alignment - log probability computation}\label{adaptivebandedcpucomputelog}
\begin{algorithmic}[1]

\Function{log\_probability\_match}{\textit{ref,events,model}}
\State \textit{event,kmer} $\gets$ \textit{get\_event\_and\_kmer(ref,events)} \Comment \textcolor{gray}{see red arrows in Fig. \ref{f:adaptive2}} \label{a:logp:2}
\State $ x \gets$ \textit{event.mean}
\State \textit{model\_kmer} $\gets$ \textit{get\_entry\_from\_poremodel(kmer,model)} \label{a:logp:4}
\State $\mu \gets model\_kmer.mean$ \label{a:logp:scale}
\State $\sigma \gets model\_kmer.stdv$
\State $z \gets \frac{x-\mu}{\sigma}$
\State \textit{lp\_emission} $\gets \ln(\frac{1}{\sqrt{2\pi}})-\ln(\sigma)-0.5z^2$

% \hspace*{\algorithmicindent} \textit{model} : pore-model (k-mers and their $\mu$,$\sigma$)  \\

\EndFunction

\end{algorithmic}

\end{algorithm}

The above elaboration covers the ABEA algorithm  to a sufficient enough level to explain our GPU implementation and optimisations. Therefore, implementation details of checking out of bound array accesses and the backtracking process were not discussed. Furthermore, the concept of the `trim state' and `event scaling' were not discussed as the control flow of the algorithm are not affected by them. Thus, those details not vital for the elaboration GPU implementation. However, for the sake of completeness, a brief account of this `trim state' and `read-model scaling' are  given below.

The raw signal may contain samples at the beginning/end that may be ignored by the base-caller and hence does not contribute to the base-called sequence. These samples may be open pore signal immediately before or after the DNA molecule is detected (i.e. the electric current when nothing is in the nanopore), or perhaps part of the adaptor (molecules bounds to the ends of the DNA molecules to enable sequencing). The `trim states' allow the alignment to ignore these samples, since such samples should not be considered to be part of the base-called read. 

Due to reasons such as slight variations between different nanopores and characteristic changes of the same nanopore with time, an event will not directly match the \textit{pore-model} in Fig. \ref{f:kmermodel} \cite{david2016nanocall}. Therefore, to account for these variations either the events or the \textit{pore-model} should be scaled on a per-read basis. In \textit{Nanopolish}, two scaling parameters namely \textit{shift} and \textit{scale} are estimated on a per-read basis, prior to ABEA algorithm, using a `Method of Moments' approach \cite{david2016nanocall}. Then, during ABEA, the \textit{pore-model} mean values are scaled using these two parameters.  The scaling should be performed at line \ref{a:logp:scale} of Algorithm \ref{adaptivebandedcpucomputelog} as $\mu \gets model\_kmer.mean \times scale + shift$ instead of directly assigning $model\_kmer.mean$ to  $\mu$.

% The heuristics were mostly empirically determined to work well - in particular applying a large p\_skip penalty was observed to help keep the alignment within the dynamic band. If there is a particular parameter you are interested in I can try to write up a justification for it (if one exists). epsilon and lp\_skip :  A small value that still allows skips to be used in some (rare) cases. I didn't spend much time determining the optimal parameter, this just worked well for the purpose; double lp\_stay ) : empirical determination ; double lp\_step : this is set so the transition probabilities sum to 1; double lp\_trim : 100 events to be trimmed, so this probability is 1/100.

% \clearpage
% \par\noindent\textcolor{red}{\rule{\textwidth}{3pt}}
% \clearpage

\subsection{GPU architecture and programming} \label{gpu} 

Graphics Processing Units (GPUs) were originally designed as co-processors for graphics processing and rendering. Graphics processing and rendering algorithms involve pixel-wise operations which expose fine-grained parallelism, thus GPUs consists of hundreds of compute cores to perform parallel processing. Eventually, the concept of general purpose graphics processing units (GPGPU) emerged where the GPUs were exploited to accelerate compute intensive, yet highly parallelism portions of general purpose algorithms. GPUs are quite popular in scientific computations due to the significant speedup when used for common matrix manipulation which contains fine-grained parallelism. From around a decade ago, GPUs which are explicitly designed for high performance computers are available (e.g., Tesla GPUs from NVIDIA).

GPUs are of Single Instruction Multiple Data (SIMD) architecture (or more accurately Single Instruction Multiple thread, as stated by NVIDIA), where multiple threads run the same stream of instructions in parallel yet on different data. Conversely, CPUs are of Multiple Instruction Multiple Data (MIMD) architecture, where each thread runs its own instruction sequence and own data stream, independent of the others. GPUs have hundreds or even thousands of processing cores while a CPU would maximally have a few dozen  cores. However, the GPU cores are relatively less complex (fewer instructions, smaller caches, no sophisticated branch prediction units etc.) and run at a lower clock speed when compared to a CPU. Due to these significant differences between CPU and GPU architectures, serial algorithms designed and developed for the CPUs are not suitable for execution on GPUs. Such algorithms have to be adapted and parallelised in a way that the GPU architectural features are efficiently used.

NVIDIA provides a programming model/framework for programming their GPUs for general purpose computations, called Compute Unified Device Architecture (CUDA). CUDA includes CUDA C/C++ (extended C/C++ syntax) and an Application Programming Interface (API) to provide a platform to write programs for the NVIDIA GPU. We used this CUDA C/C++ for our GPU implementation of the Adaptive Banded Event Alignment algorithm.

We will now briefly give GPU/CUDA related terms. Readers  are advised to refer to \cite{cudaprog} and \cite{cudabest} for further information.

A GPU \emph{kernel} is a function that is executed on a GPU. A GPU kernel is written from the execution perspective of a single GPU thread. These GPU kernels will run in parallel, based on the parameters specified with the function call, known as the  \emph{thread configuration}.  This thread configuration in CUDA is an abstraction which employs a hierarchy of threads.
In the thread hierarchy, a group of threads are known as a \emph{block}. A group of blocks form a \emph{grid}. Instances of a single kernel are executed in a single grid.   Blocks and grids can be 1 dimensional, 2 dimensional or 3 dimensional. The presence of this thread hierarchy lets the programmer  organise and map the threads conveniently to a grid. These logical threads would be  mapped to the hardware cores automatically by the underlying driver software and hardware.

A thread block consists of one or more \emph{thread warps}. A warp is a group of threads sharing the same program counter. A data dependent conditional branch inside a warp causes the threads to execute each code path while disabling threads that are not in the path, known as \emph{warp divergence}. The warp divergence affects the performance and should be minimised.

The \emph{occupancy} is the percentage of the number of active warps to the maximally supported warps on the GPU. A lesser occupancy leads to under utilisation of GPU resources. Thus, a higher occupancy is preferable for better utilisation of GPU resources.

GPUs also employ a memory hierarchy. Relatively larger but slow Dynamic Random Access Memory (DRAM) that forms the lowest level in the memory hierarchy is known as \emph{global memory}. Global memory is typically allocated using \emph{cudaMalloc()} API function. Memory allocated in this global memory can be exclusively accessed by all the threads in the grid. The next level in the memory hierarchy which is made of relatively fast, yet smaller SRAM is called \emph{shared memory}. Shared memory is allocated on a per-thread-block basis and is shared by all the threads in the block. Shared memory can be called user managed cache (more accurately a programmer managed cache) as the programmer is expected to identify and load frequently accessed data to the shared memory. In addition, there are one or more levels of  SRAM caches managed by the hardware. The registers are the fastest and highest in the hierarchy and are allocated by the compiler on a  per-thread basis.

The global memory can be easily saturated when hundreds of threads compete to access the memory at the same time. Thus, memory accesses should be batched such that contiguous threads access contiguous memory locations. This process is referred to as \emph{memory coalescing} and reduces global memory requests thus reducing the impact on performance compared to scattered memory accesses. Additionally, the programmer could utilise the shared memory to load and store frequently accessed data, which also reduces global memory traffic.

\section{Related work} \label{relwork}

An algorithm to call methylation using the raw signal from ONT sequencers was introduced by Simpson et al. \cite{simpson2017detecting}. The associated C++ based implementation of this algorithm is a sub-module under the open source tool \textit{Nanopolish}. \textit{Nanopolish} was designed to run on high-performance computers and is not lightweight or suitable for  deployment on embedded systems.

The signal-space alignment algorithm, termed \textit{Adaptive Banded Event Alignment (ABEA)}, used in \textit{Nanopolish} is a customised version of the Suzuki-Kasahara alignment algorithm \cite{suzuki2018introducing} for base-level sequence alignment. According to the best of our knowledge, neither of these algorithms (ABEA or Suzuki-Kasahara) have GPU accelerated versions. The root origins of these algorithms are dynamic programming sequence alignment algorithms, such as Smith-Waterman and Needleman-Wunsch. A number of GPU accelerated versions for Smith-Waterman exist in previous research \cite{Liu2009,Liu2010,Liu2013a} \cite{Liu2010} \cite{Liu2013a}. However, the Smith-Waterman algorithm has a compute complexity of $O(n^2)$ and is most practical when the sequences are short, especially when millions of sequences need to be aligned. As nanopore sequencers can produce reads >1 million bases long,  computing the full DP table for such reads using SW would require >10\textsuperscript{12} computations and hundreds of gigabytes of RAM---and even more if aligning raw nanopore signals. 

Heuristic approaches such as banded Smith-Waterman attempt to reduce the search space by limiting computation along the diagonal of the DP table. While the approach is suitable for Illumina short reads, it is less so for noisy long nanopore reads as substantial band-width is required to contain the alignment within the band. The Suzuki-Kasahara algorithm uses a heuristic that allows the band to adapt and move during the alignment, thus containing the optimal alignment within the band but allowing large gaps in the alignment. Modified versions of the adaptive banded alignment algorithm are used for signal-space alignment, as exemplified in \textit{Nanopolish}. The band-width (width of the band) used for signal-space alignment is typically higher (\textasciitilde100) compared to other banded algorithms used for sequence alignment. In addition, the scoring function for signal alignment uses a 32 bit floating point data type, as opposed to 8-bit integers in sequence alignment. Furthermore, the signal alignment scoring function that computes the log-likelihood is computationally expensive. Taken together, these  reasons motivated us to consider using GPUs to speedup the computation of signal-space alignment.

The portable compute module, MinIT, manufactured by ONT is composed of a NVIDIA SoC \cite{minitout} that exploits GPUs for performing live base-calling, which can perform  base-calling at a speed of \textasciitilde150 Kbases per second, thus keeping up with the MinION sequencer's output. In addition, our previous work has optimised  the popular \textit{Minimap2} \cite{minimap2} sequence alignment tool (which typically requires \textasciitilde16GB memory) for reduced peak memory usage, enabling the software to be executed on embedded processors \cite{minimap2arm}. The data processing steps required for methylation calling are thus possible to run on embedded processors, therefore supporting the implementation of a portable, offline DNA methylation detection application that would facilitate such analyses in the field. 

Load balancing between the CPU and GPU for heterogeneous processing has been explored for areas such as fluid dynamics \cite{huismann2017load} and conjugate gradient method \cite{lang2013dynamic}. However, nanopore data have different characteristics compared to aforementioned applications which are predominately based on matrices. Furthermore, the signal-space alignment algorithm is different from linear algebra algorithms used in these fields. We exploit characteristics of Nanopore data and algorithms to perform memory, compute and load balancing optimisations.

\section{Methodology} \label{method}

To optimise the performance on GPUs, we process a batch of reads (original source code processes a read at a time) at a time. Such batch processing minimises data transfer initialisation overhead (between RAM and GPU memory); reduces the GPU kernel invocation overhead; and,  allows  parallelism which sufficiently occupies all available GPU cores. The execution flow follows the typical GPU programming paradigm, which is elaborated in Algorithm \ref{a:exec}. In Algorithm \ref{a:exec}, \textit{gpu\_alignment(...)} refers to the GPU implementation of the Adaptive Banded Event Alignment (CPU algorithm is elaborated in Algorithm \ref{adaptivebandedcpu}). We present our methodology in three steps: parallelisation and compute optimisations in Section \ref{computeopti}; memory optimisation in Section \ref{memopti};  and, the resource optimisation through heterogeneous processing in Section \ref{loadbalmet}.

\begin{algorithm}[bt]
\caption{Outline of execution flow}\label{a:exec}
\begin{algorithmic}[1]
\For{batch of $n$ reads}
\State ... \Comment \textcolor{gray}{CPU processing steps before the Adaptive Banded Event Alignment eg: event detection}
\State $memcpy\_ram\_to\_gpu(...)$  \Comment \textcolor{gray}{copy inputs of the Adaptive Banded Event Alignment to the GPU memory}
\State $gpu\_alignment(...)$ \Comment \textcolor{gray}{Perform the event alignment on the GPU}
\State $memcpy\_gpu\_to\_ram(...)$ \Comment \textcolor{gray}{copy results back to the RAM}
\State ... \Comment \textcolor{gray}{CPU processing steps after the alignment eg: HMM}
\EndFor
\end{algorithmic}
\end{algorithm}

\subsection{Parallelisation and compute optimisations} \label{computeopti}

The GPU implementation of the Adaptive Banded Event Alignment (ABEA) algorithm is broken into three GPU kernels. Breaking down into the three GPU kernels allows for efficient thread assignment based on the workload type, synchronisation of all GPU threads (a GPU kernel execution is inherently a synchronisation barrier \cite{cudaprog}) and minimising warp divergence compared to a big all-in-one GPU kernel.

The three GPU kernels are:

\begin{itemize}
\item \textit{pre-kernel} - Initialising the first two bands of the dynamic programming table (corresponds to line \ref{a:abeacpu:init1} of algorithm \ref{adaptivebandedcpu}) and pre-computing frequently accessed values by the next GPU kernel;
\item \textit{core-kernel} - The filling of dynamic programming table which is the compute intensive portion of the ABEA algorithm (corresponds to line \ref{a:abeacpu:outerloop}-\ref{a:abeacpu:outerloope} of Algorithm \ref{adaptivebandedcpu} composed of nested loop); and,
\item \textit{post-kernel} - Performs backtracking (corresponds to line \ref{a:abeacpu:backtrack} of algorithm \ref{adaptivebandedcpu})
\end{itemize}

 %The three GPU kernels are elaborated in detail in the following sections.

\subsubsection{pre-kernel}\label{s:prekernel}

The \textit{pre-kernel} initialises the first two bands of the dynamic programming table (initialisation performed at line \ref{a:abeacpu:init1} of Algorithm \ref{adaptivebandedcpu} on CPU). The \textit{pre-kernel} also pre-computes the values in a data structure called \textit{kcache}, a newly introduced data structure in the GPU implementation that improves cache hits during the subsequent execution of the  \textit{core-kernel}. 

\begin{figure}[!ht]
  \centering
    \includegraphics[width=.7\linewidth]{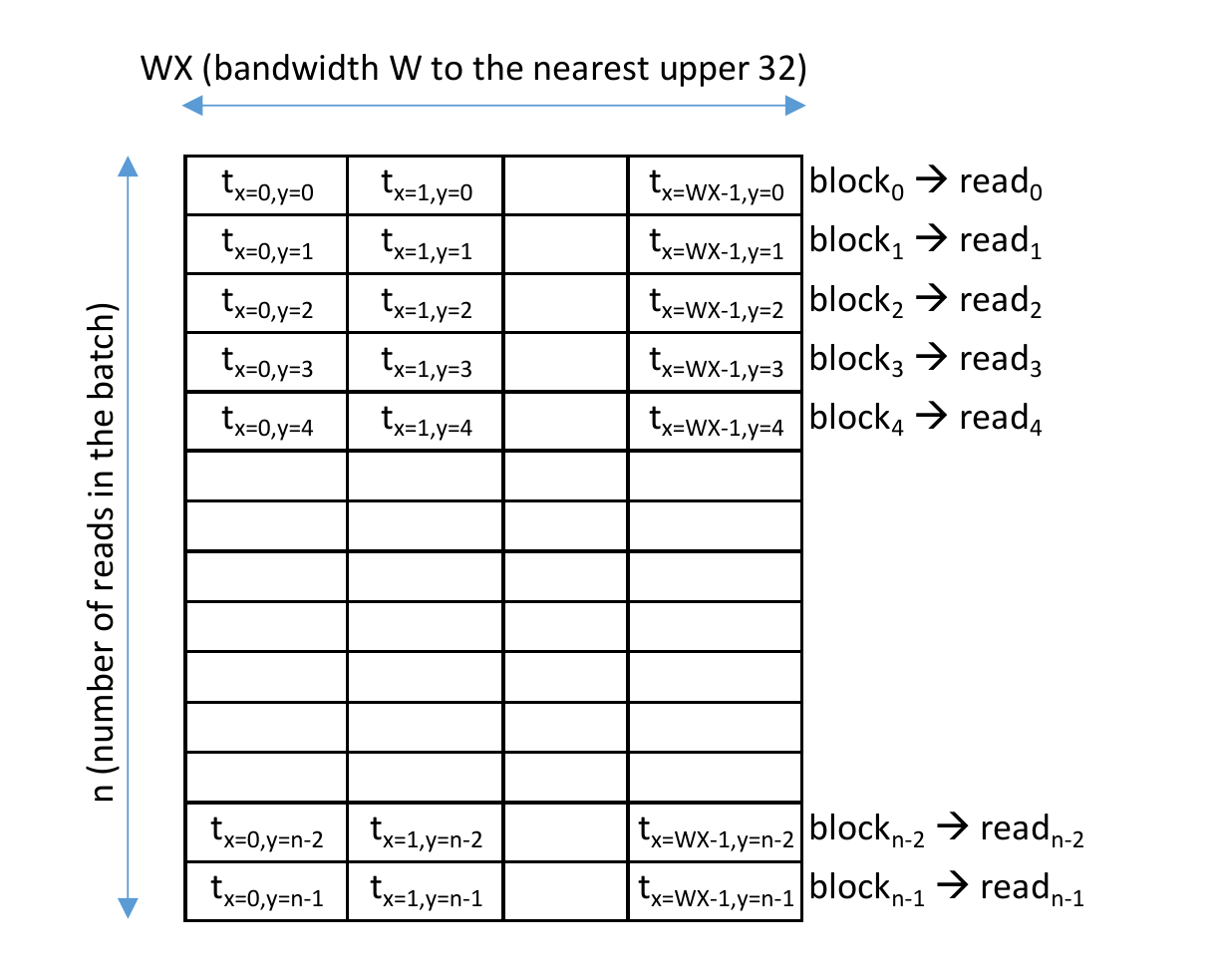}
    \caption{Thread configuration of \textit{pre-kernel}} 
    \label{f:theradconfig}
\end{figure}

A simplified version of the \textit{pre-kernel} is in Algorithm \ref{adaptivebandedcudapre} and  thread configuration for the invocation of the \textit{pre-kernel} is in Fig. \ref{f:theradconfig}. Note that the GPU kernel is presented (as is always the case) from the perspective of a single GPU thread in Fig. \ref{f:theradconfig}.

Each cell in Fig. \ref{f:theradconfig} represents a GPU thread denoted as \textit{t}, where the subscripts \textit{x} and \textit{y} denotes the thread index along the x-axis and the y-axis respectively. The thread grid in Fig. \ref{f:theradconfig} is composed of \textit{n} thread blocks, where \textit{n} is the number of reads in the batch. Each thread block contains \textit{WX} threads where \textit{WX} is the nearest upper ceiling multiple of 32 to the band-width \textit{W} (band-width of the ABEA algorithm); i.e. $WX = (int)\frac{W+31}{32} \times 32$ For instance, if \textit{W=100}, \textit{WX} is 128. The reason for taking  a multiple of 32 is due to performance attributed by a thread block size that a multiple of the warp size (warp size is 32 currently) \cite{cudabest} . As shown in Fig. \ref{f:theradconfig}, a single thread block composed of \textit{WX} threads is assigned to a single read.

\begin{figure}[!ht]
  \centering
    \includegraphics[width=\linewidth]{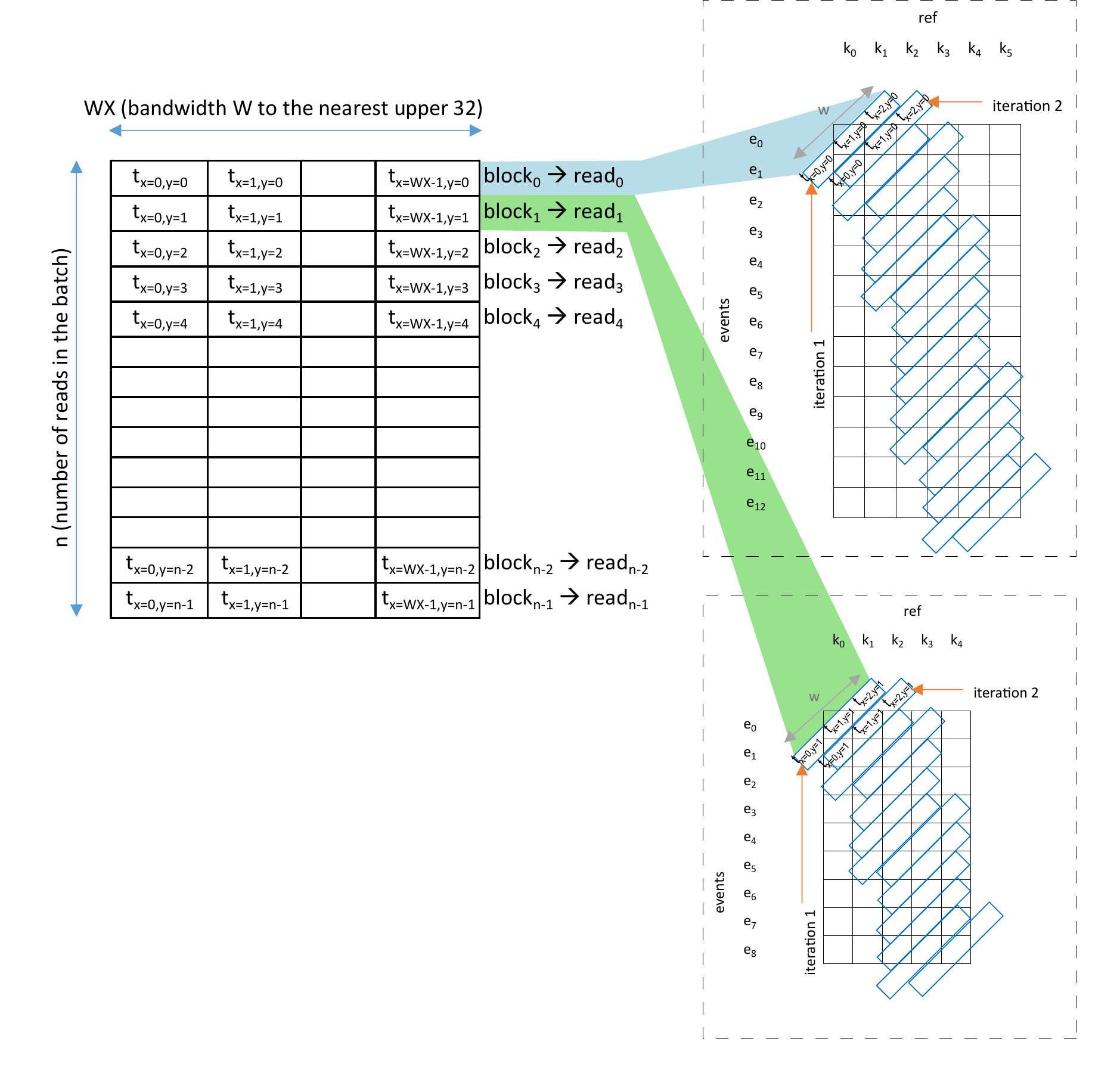}
    \caption[Thread assignment of \textit{pre-kernel}]{Thread assignment of \textit{pre-kernel}. The assignment for  the first two reads are shown.  Each thread block has a read assigned to it (block\textsubscript{0} refers to threads $t_{X=0,y=0}$ to $t_{x=WX-1,y=0}$, and read\textsubscript{0} is processed by all threads in block\textsubscript{0}; similarly, block\textsubscript{1} refers to $t_{X=0,y=1}$ to $t_{x=WX-1,y=1}$ and  read\textsubscript{1} is processed by threads in block\textsubscript{1}).} 
    \label{f:threadassign1}
\end{figure}

  In the Algorithm \ref{adaptivebandedcudapre}, lines \ref{a:pre2}-\ref{a:pre3} get the thread index of the thread being executed, i.e. the thread indices denoted as \textit{x} and \textit{y} in Fig. \ref{f:theradconfig}. Line \ref{a:pre4} obtains the memory pointers of the input array \textit{ref}; intermediate arrays \textit{score} and \textit{trace}; and the \textit{kcache}, the use in which is explained in the memory optimisation Section (Section \ref{memopti}).
%The lines 1-2 of  Algorithm \ref{adaptivebandedcudapre} shows the thread configuration In which the \textit{pre-kernel} is launched, thread block sized [x=$WX$,y=1] and the grid sized [x=1,y=$n$]. 

Lines \ref{a:pre5}-\ref{a:pre8} of Algorithm \ref{adaptivebandedcudapre} initialises the first two bands of the dynamic programming table (which was performed at line \ref{a:abeacpu:init1} of original CPU Algorithm \ref{adaptivebandedcpu}). The kernel is in from the perspective of a single thread and thus  a single cell is initialised by a single thread. The collective execution of all the threads in Fig. \ref{f:theradconfig}, effectively sets a band for all the reads in the batch in parallel, which is illustrated in Fig. \ref{f:threadassign1}. Note that, only the first two reads are elaborated in Fig.  \ref{f:threadassign1}, and in reality each thread block has a read assigned to it. In Fig. \ref{f:threadassign1}, each cell in band\textsubscript{0} (marked as iteration 1) contains the index of the thread which performs the initialisation at line \ref{a:pre6} of Algorithm \ref{adaptivebandedcudapre}.
Similarly, iteration 2 corresponds to line \ref{a:pre7} of Algorithm \ref{adaptivebandedcudapre}.

The \textit{if} condition on line \ref{a:pre5} of Algorithm \ref{adaptivebandedcudapre} is to limit the threads to the width of the band $W$, a consequence of selecting $WX$ which is a multiple of 32 (as stated previously). Note that there is a  1024 thread limit for a block \cite{cudaprog} in current NVIDIA CUDA/GPU architecture, thus our implementation will only work for a maximum band-width of 1024. This limit is more than sufficient for a typical $W$ of 100 in ABEA.

Line \ref{a:pre10}-\ref{a:pre11} of  Algorithm \ref{adaptivebandedcudapre} initialises the index of the lower left band which corresponds to line \ref{a:abeacpu:23}-\ref{a:abeacpu:24} of Algorithm \ref{adaptivebandedcpu}. Note that this initialisation is executed by one thread per read (thread id 0 along y-axis).
Lines \ref{a:pre:13}-\ref{a:pre:16} in Algorithm \ref{adaptivebandedcudapre} initialises \textit{kcache}. As stated previously \textit{kcache} is a  newly introduced array for the GPU implementation to minimise random accesses to the GPU memory during the \textit{core-kernel} and will be explained in Section \ref{s:corekernel}. Note that, this \textit{kcache} initialisation in line \ref{a:pre:13}-\ref{a:pre:16} is also executed by one thread per read (thread id 0 along y-axis). The loop in \ref{a:pre:13}-\ref{a:pre:16} can be further parallelised; however, as the time spent on \textit{pre-kernel} is comparatively negligible (see results), further parallelising this loop  is superfluous.

\begin{algorithm}[!ht]
\caption{Adaptive Banded Event Alignment - \textit{pre-kernel}}
\label{adaptivebandedcudapre}
% \begin{algorithmic}[1]
% \State \textit{block} $ \gets (x=W,y=1) \quad;\quad  \textit{grid } \gets (x=1,y=nreads)$
% \State \textit{align\_kernel\_pre\_kernel} $<<<grid,block>>>$ \textit{(...)} \Comment kernel configuration
% \end{algorithmic}
\begin{algorithmic}[1]
\Function{\textit{align\_pre}}{\textit{...,model}} \Comment \textcolor{gray}{... refers to other arguments which are later explained Section \ref{memopti}}

    \State \textit{j} $\gets$ \textit{thread index along x} \Comment \textcolor{gray}{the x subscript of a thread Fig. \ref{f:theradconfig}}\label{a:pre2}
    %\Comment offset in band and read id 
    \State \textit{i} $\gets$ \textit{thread index along y} \Comment \textcolor{gray}{the y subscript of a thread Fig. \ref{f:theradconfig}} \label{a:pre3} %\Comment offset in band and read id

	\State \textit{(ref,score,trace,ll\_idx,kcache)} $\gets$ \textit{get\_cuda\_pointers(i,...)}  \label{a:pre4}
	\Comment \textcolor{gray}{get memory pointers of the arrays corresponding to read i (explained in Section \ref{memopti})}

 \If {\textit{j} $ < W$} \Comment \textcolor{gray}{Though a block is $WX$ wide (Fig. \ref{f:theradconfig}) only $W$ threads should execute } \label{a:pre5}
    %\For{\textit{k=0 to 1}}
    \State \textit{score[0,j], trace[0,j]}  $\gets - \infty, 0$ \Comment \textcolor{gray}{corresponds to line \ref{a:abeacpu:21} of Algorithm \ref{adaptivebandedcpu}} \label{a:pre6}
    \State \textit{score[1,j], trace[1,j]}  $\gets - \infty, 0$ \Comment \textcolor{gray}{corresponds to line \ref{a:abeacpu:22} of Algorithm \ref{adaptivebandedcpu}} \label{a:pre7}
    %\EndFor
 \EndIf \label{a:pre8}
 
 \If {\textit{j==0}} \Comment \textcolor{gray}{only thread 0 process this Section} \label{a:pre9}
  \State \textit{ll\_idx[0]} $\gets$ \{$ei_0$,$ki_0$\} \Comment \textcolor{gray}{corresponds to line \ref{a:abeacpu:23} of Algorithm \ref{adaptivebandedcpu}} \label{a:pre10}
    \State \textit{ll\_idx[1]} $\gets$ \{$ei_1$,$ki_1$\} \Comment \textcolor{gray}{corresponds to line \ref{a:abeacpu:24} of Algorithm \ref{adaptivebandedcpu}} \label{a:pre11}
    
    \State \textit{score[0,$si_0]$} $\gets$ 0 \Comment  \textcolor{gray}{corresponds to line \ref{a:abeacpu:25} of Algorithm \ref{adaptivebandedcpu}}
    
    \For{\textit{k=0 to numkmers}} \Comment \textcolor{gray}{ Iterate through each kmer in ref from left to right} \label{a:pre:13}
        \State \textit{kmer} $\gets$ \textit{get\_kmer\_at(ref,k)} \Comment \textcolor{gray}{k-mer at position k in \textit{ref}}
        \State \textit{kcache[k] = get\_entry\_from\_poremodel(kmer,model)}
    \EndFor \label{a:pre:16}
    %\State \textit{initialise\_rest(score,trace,ll)} \Comment a few individual slots in score,trace and ll arrays
 \EndIf

\EndFunction
\end{algorithmic}
\end{algorithm}

\subsubsection{core-kernel}\label{s:corekernel}

A simplified version of the \textit{core-kernel} which fills the dynamic programming table in Fig. \ref{f:adaptive} (corresponds to line \ref{a:abeacpu:outerloop}-\ref{a:abeacpu:outerloope} of the original Algorithm \ref{adaptivebandedcpu}) is in Algorithm \ref{adaptivebandedcudacore}. This kernel is executed with the same kernel thread configuration as \textit{pre-kernel} in Fig. \ref{f:theradconfig}. Thus, a batch of reads are processed in parallel with a block of threads assigned to a single read in a similar way to that in \textit{pre-kernel} (Fig. \ref{f:threadassign1}). The only difference in Fig. \ref{f:threadassign1} for the \textit{core-kernel} is that the third band to the last band are processed instead of the first two bands.

\begin{algorithm}[!hht]

\caption{Adaptive Banded Event Alignment - \textit{core-kernel}}
\label{adaptivebandedcudacore}
{\fontsize{10}{9}\selectfont
\begin{algorithmic}[1]
%\State \textit{block} $ \gets (x=W,y=1) \quad;\quad \textit{grid} \gets (x=1,y=nreads)$
%\State \textit{align\_core} $<<<grid,block>>>$ \textit{(...)}

\Function{align\_kernel\_core(...)}{} \Comment \textcolor{gray}{... refers to the arguments which are later explained in Section \ref{memopti}}

    \State \textit{j} $\gets$ \textit{thread index along x} \Comment \textcolor{gray}{the x subscript of a thread Fig. \ref{f:theradconfig}}
    %\Comment offset in band and read id 
    \State \textit{i} $\gets$ \textit{thread index along y} \Comment \textcolor{gray}{the y subscript of a thread Fig. \ref{f:theradconfig}} %\Comment offset in band and read id

	\State \textit{(events,score,trace,ll\_idx,kcache)} $\gets$ \textit{get\_cuda\_pointers(i,...)} 
	\Comment \textcolor{gray}{get memory pointers of the arrays corresponding to read i (explained in Section \ref{memopti}} \label{a:core4}

	\State \textit{n\_bands} $\gets$ \textit{n\_events + read\_len}
		
    \State \textit{\_\_shared\_\_ c\_score[W], p\_score[W], pp\_score[W]} \Comment \textcolor{gray}{allocate space in fast shared memory for scores of  current, previous and 2nd previous bands} \label{a:abeagpu-sm1}
    
    \State \textit{\_\_shared\_\_ c\_ll\_idx, p\_ll\_idx, pp\_ll\_idx} \Comment \textcolor{gray}{allocate space in fast shared memory for indexes of lower left cells of current, previous and  2nd previous bands} \label{a:abeagpu-sm2}

    \If{ \textit{(j<W)} } \Comment \textcolor{gray}{similar behaviour as in \textit{pre-kernel}}

        \State \textit{p\_score[j],pp\_score[j]} $\gets$ \textit{score[1,j],score[0,j]} \Comment \textcolor{gray}{copy initialised b0 and b1 scores}    \label{a:abeagpu-sm3}                      
        %\State \textit{c\_score[j]} $\gets$ \textit{score[2,j]}
        %\State \textit{p\_score[j]} $\gets$ \textit{score[1,j]}
        %\State \textit{pp\_score[j]} $\gets$ \textit{score[0,j]}
         
        \State \textit{p\_ll\_idx,pp\_ll\_idx} $\gets$ \textit{ll[1],ll[0]} \Comment \textcolor{gray}{copy initialised b0 and b1 indexes}  \label{a:abeagpu-sm4}
        
        %\State \textit{c\_ll\_idx} $\gets$ \textit{ll[2]}
        %\State \textit{p\_ll\_idx} $\gets$ \textit{ll[1]}
        %\State \textit{pp\_ll\_idx} $\gets$ \textit{ll[0]}
         
        \State \textit{\_\_syncthreads()} \Comment \textcolor{gray}{synchronise threads in the block} 
   
    \For{\textit{i $\gets$ 2 to n\_{bands}}}   \Comment \textcolor{gray}{similar to Algorithm \ref{adaptivebandedcpu}}
            
            \If{\textit{(j==0)}} \Comment \textcolor{gray}{only thread 0 process this} \label{a:abeagpu-t0}

                \State \textit{dir} $\gets$ \textit{suzuki\_kasahara\_rule(p\_score)} \Comment \textcolor{gray}{similar to Algorithm \ref{adaptivebandedcpu}} \label{a:abeagpu-sm5}
                \If{ \textit{$dir == right$}}
                    \State  \textit{c\_ll\_idx} $\gets$ \textit{move\_band\_to\_right(p\_ll\_idx)}   \Comment \textcolor{gray}{similar to Algorithm \ref{adaptivebandedcpu}}
                    \State \textit{ll[i]} $\gets$ \textit{c\_ll\_idx} \Comment \textcolor{gray}{store to global memory}
                \Else
                    \State \textit{c\_ll\_idx} $\gets$ \textit{move\_band\_down(p\_ll\_idx)}  \Comment\textcolor{gray}{ similar to Algorithm \ref{adaptivebandedcpu}   }
                    \State \textit{ll[i]} $\gets$ \textit{c\_ll\_idx} \Comment \textcolor{gray}{store to global memory}
                \EndIf \label{a:abeagpu-sm6}

                %\State \textit{right} $\gets$ \textit{p\_score[0] < p\_score[W - 1]} \
                %\\ \Comment Suzuki's rule

                %\State \textit{trim\_j} $\gets$ \textit{get\_trim\_j(c\_ll\_idx)}
                %\Comment trim state
                %\State \textit{score,from} $\gets$\textit{ get\_trim\_state(trim\_j,c\_ll\_idx)}
				%\State \textit{c\_score[trim\_j]} $\gets$\textit{ score}
				%\State \textit{trace[i,trim\_j]} $\gets$\textit{ from}
				
            \EndIf
			
            \State \textit{\_\_syncthreads()} \Comment \textcolor{gray}{synchronise threads in the block}

			\State \textit{min\_j,max\_j} $\gets$ \textit{get\_limits\_in\_band(c\_ll\_idx)} \Comment \textcolor{gray}{similar to Algorithm \ref{adaptivebandedcpu}}

            \State \textit{\_\_syncthreads()} \Comment \textcolor{gray}{synchronise threads in the block}
   
            \If{(\textit{j} $\geq$ \textit{min\_j} \textbf{AND} \textit{j} $<$\textit{ max\_j}) } \Comment \textcolor{gray}{fill the cells in band i in parallel} \label{a:abeagpufills}

				%\State \textit{ju,jl,jd} $\gets$ \textit{get\_j(p\_ll\_idx,pp\_ll\_idx)}
				
				%\State \textit{up,left,diag} $\gets$ \textit{p\_score[ju],p\_score[jl],pp\_score[jd]}
                % \State \textit{up} $\gets$ \textit{p\_score[j\_up]}
                % \State \textit{left} $\gets$ \textit{p\_score[j\_left]}
                % \State \textit{diag} $\gets$ \textit{pp\_score[j\_diag]}
                %\State \textit{s,d} $\gets$ \textit{compute (kcache, events, c\_ll\_idx,j,up,left,diag)}

             \State \textit{s,d} $\gets$ \textit{compute(p\_score,pp\_score,kcache,events,model)} \Comment \textcolor{gray}{see Algorithm \ref{adaptivebandedcpucomputegpu}} \label{a:abeagpu-sm7}%\Comment \textit{score[i-1]} and \textit{score[i-2]} are previous and second previous bands
				
                \State \textit{c\_score[j]} $\gets$ \textit{s} \Comment \textcolor{gray}{store score to shared memory} \label{a:abeagpu-sm8}
                \State \textit{trace[i,j]}  $\gets$ \textit{d} \Comment \textcolor{gray}{store backtrack flag directly to global memory}
            \EndIf \label{a:abeagpufille}

            \State \textit{ \_\_syncthreads()} \Comment \textcolor{gray}{synchronise threads in the block}

            \State\textit{ score[i,j]} $\gets$ \textit{c\_score[j]} \Comment \textcolor{gray}{store the scores in global memory} \label{a:abeagpu-sm9}
            
            \State \textit{pp\_score[j], p\_score[j], c\_score[j]} $\gets$ \textit{p\_score[j], c\_score[j],} $-\infty$ \Comment \textcolor{gray}{update band scores for the next iteration} \label{a:abeagpu-sm10}
            % \State \textit{pp\_score[j]} $\gets$ \textit{p\_score[j]}
            % \State \textit{p\_score[j]} $\gets$ \textit{c\_score[j]}
            % \State \textit{c\_score[j]} $\gets$ $\infty$

            \If{\textit{j==0}}
                %\textit{pp\_ll\_idx}$\gets$\textit{p\_ll\_idx}
                \State \textit{pp\_ll\_idx, p\_ll\_idx} $\gets$ \textit{p\_ll\_idx, c\_ll\_idx} \Comment \textcolor{gray}{update band indexes for the next iteration}
            \EndIf \label{a:abeagpu-sm11}

            \State\textit{\_\_syncthreads()} \Comment \textcolor{gray}{synchronise threads in the block }

        \EndFor
    \EndIf
\EndFunction
\end{algorithmic}
}
\end{algorithm}

\begin{algorithm}[!ht]
\caption{Adaptive Banded Event Alignment - \textit{core-kernel} - cell score computation}\label{adaptivebandedcpucomputegpu}
\begin{flushleft}
\textbf{Constants:} \\
\hspace*{\algorithmicindent} $events\_per\_kmer = \frac{n\_events}{n\_kmers}$\\
\hspace*{\algorithmicindent} $\epsilon = 1^{-10}$ \\
\hspace*{\algorithmicindent} \textit{lp\_skip} $= \ln(\epsilon)$  \\
\hspace*{\algorithmicindent} \textit{lp\_stay} $=\ln(1-\frac{1}{events\_per\_kmer+1})$ \\ 
\hspace*{\algorithmicindent} \textit{lp\_step} $= \ln(1.0 - e^{lp\_skip} - e^{lp\_stay})$ \\
\end{flushleft}
\begin{algorithmic}[1]

% \hspace*{\algorithmicindent} \textit{model} : pore-model (k-mers and their $\mu$,$\sigma$)  \\

\Function{computation}{\textit{score\_prev,score\_2ndprev,\textcolor{blue}{kcache},events}}

\State \textit{lp\_emission} $\gets$ \textit{log\_probability\_match(\textcolor{blue}{kcache},events)} \Comment \textcolor{gray}{see Algorithm \ref{adaptivebandedcpucomputeloggpu}} \label{a:computegpu:2}
\State \textit{up,diag,left} $\gets$ \textit{get\_scores(score\_prev,score\_2ndprev)} \Comment \textcolor{gray}{see red arrows in Fig. \ref{f:adaptive2}}
\State \textit{score\_d} $\gets$ \textit{diag} $+$ \textit{lp\_step} $+$ \textit{lp\_emission} 
\State \textit{score\_u} $\gets$ \textit{up} $+$ \textit{lp\_stay} $+$ \textit{lp\_emission} 
\State \textit{score\_l} $\gets$ \textit{left} $+$ \textit{lp\_skip} 
\State \textit{s} $\gets$ \textit{max(score\_d,score\_u,score\_l)}
\State \textit{d} $\gets$ \textit{direction from which the max score came}
\EndFunction
\end{algorithmic}
\begin{flushleft}
\textcolor{blue}{Note: Changes to Algorithm \ref{adaptivebandedcpucompute} are highlighted in blue}
\end{flushleft}
\end{algorithm}

\begin{algorithm}[!ht]
\caption{Adaptive Banded Event Alignment - \textit{core-kernel} - log probability computation.}\label{adaptivebandedcpucomputeloggpu}
\begin{algorithmic}[1]

\Function{log\_probability\_match}{\textcolor{blue}{\textit{kcache}},\textit{events}}
\State \textcolor{blue}{\textit{event} $\gets$ \textit{get\_event(events)}} \Comment \textcolor{gray}{see red arrow in Fig. \ref{f:adaptive2}}
\State $ x \gets$ \textit{event.mean}
\State \textcolor{blue}{\textit{model\_kmer} $\gets$ \textit{get\_entry\_from\_kcache(kcache)}} \label{a:loggpu:4}
\State $\mu \gets model\_kmer.mean$ 
\State $\sigma \gets model\_kmer.stdv$
\State $z \gets \frac{x-\mu}{\sigma}$
\State \textit{lp\_emission} $\gets \ln(\frac{1}{\sqrt{2\pi}})-\ln(\sigma)-0.5z^2$

% \hspace*{\algorithmicindent} \textit{model} : pore-model (k-mers and their $\mu$,$\sigma$)  \\

\EndFunction

\end{algorithmic}
\begin{flushleft}
\textcolor{blue}{Note: Changes to Algorithm \ref{adaptivebandedcpucomputelog} are highlighted in blue}
\end{flushleft}
\end{algorithm}

All the $W$ cells in a given band (Fig. \ref{f:adaptive}) are computed by $W$ number of GPU threads in parallel (lines \ref{a:abeagpufills}-\ref{a:abeagpufille} of Algorithm \ref{adaptivebandedcudacore}), thus the inner loop of Algorithm \ref{adaptivebandedcpu} (lines \ref{a:abeacpu:innerls} and \ref{a:abeacpu:innerle}) is now no longer present. However, the outer loop of Algorithm  \ref{adaptivebandedcpu} cannot be parallelised due to band $n$ depending on $n-1$ and $n-2$ bands as explained the background.
The movement/placement of the band (described in background) is performed by a single thread using the condition given on line \ref{a:abeagpu-t0} Algorithm \ref{adaptivebandedcudacore} that limits the code segment to thread 0. In addition, synchronisation barriers per-thread-block basis (\textit{\_\_syncthreads}) in Algorithm \ref{adaptivebandedcudacore} prevent any data hazards due to multiple threads assigned to a single read.

Another notable difference in the GPU implementation is the use of GPU shared memory \cite{cudaprog} (user-managed cache or more accurately programmer-managed cache) for exploiting the temporal locality in the memory accesses to the dynamic programming table (\textit{n\textsuperscript{th}} band in Fig. \ref{f:adaptive} is computed using bands \textit{n-1} and \textit{n-2}). Shared memory is allocated for three bands (current, previous band and second previous) by line \ref{a:abeagpu-sm1}-\ref{a:abeagpu-sm2} of Algorithm \ref{adaptivebandedcudacore} which are then initialised at lines \ref{a:abeagpu-sm3}-\ref{a:abeagpu-sm4} of Algorithm \ref{adaptivebandedcudacore}. These initialised memory locations are used during band direction computation (lines \ref{a:abeagpu-sm5}-\ref{a:abeagpu-sm6} of Algorithm \ref{adaptivebandedcudacore}) and the cell score computation (lines \ref{a:abeagpu-sm7}-\ref{a:abeagpu-sm8} of Algorithm \ref{adaptivebandedcudacore}), eliminating any accesses to the slow GPU global memory (shared memory-SRAM vs global memory-DRAM). The cell score is written to the global memory at the end of the iteration (line \ref{a:abeagpu-sm9} of of Algorithm \ref{adaptivebandedcudacore}) as scores are later required for backtracking. Finally, current, previous and second previous bands are set for the next iteration (lines \ref{a:abeagpu-sm10}-\ref{a:abeagpu-sm11} of Algorithm \ref{adaptivebandedcudacore}).

\begin{figure*}[!ht]
\begin{subfigure}[t]{\textwidth}
    \centering
    \includegraphics[trim=0 92 0 0,clip,width=\textwidth]{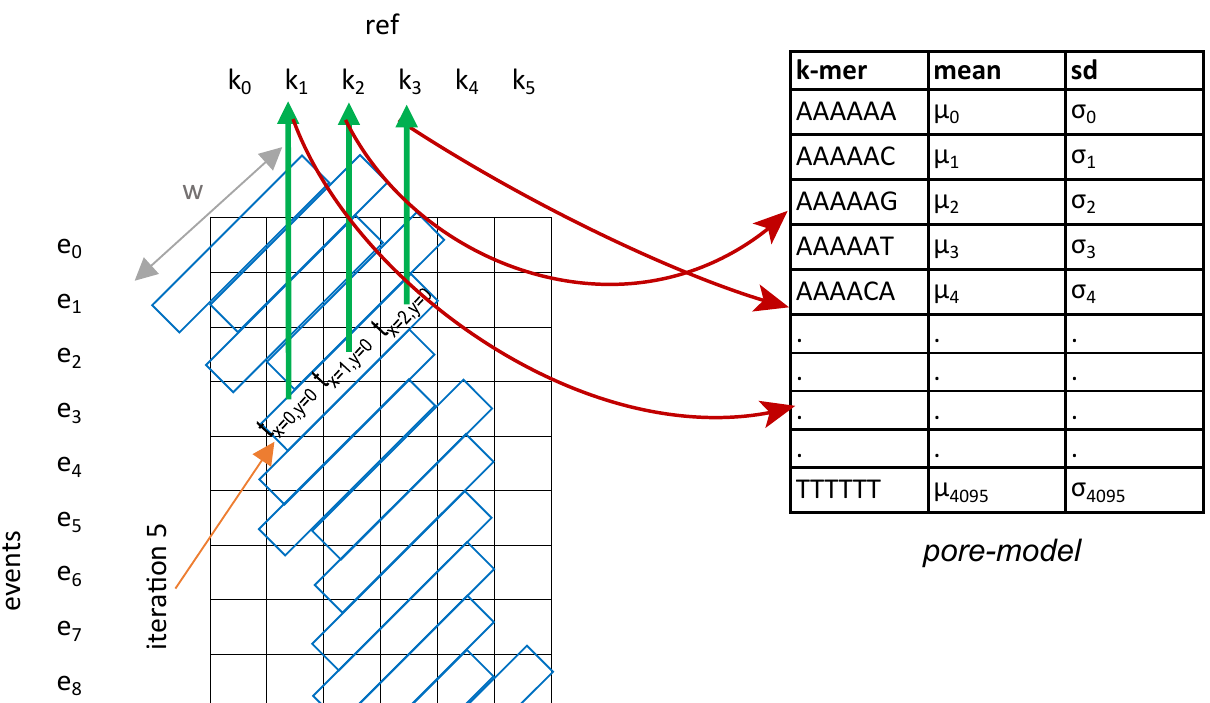}
    \caption{Random accesses to the \textit{model} array (red arrows) when \textit{kcache} is not employed}
    \label{f:coalease1}
\end{subfigure}\vspace{0.7cm}

\begin{subfigure}[t]{.6\textwidth}
    \centering
    \includegraphics[width=\textwidth]{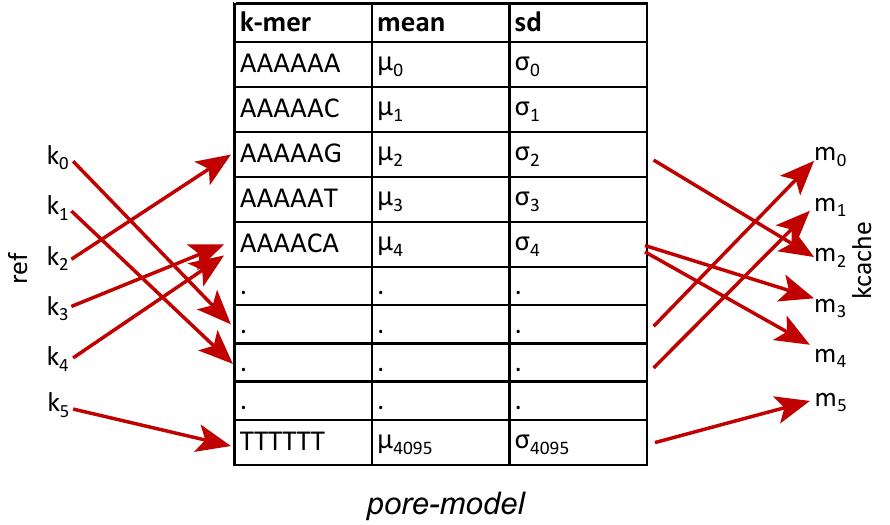}
    \caption{Construction of the \textit{kcache} in \textit{pre-kernel}}
    \label{f:coalease2}
\end{subfigure}\hfill\hfill
\begin{subfigure}[t]{.4\textwidth}
    \centering
    \includegraphics[trim=0 90 20 0,clip,width=\textwidth]{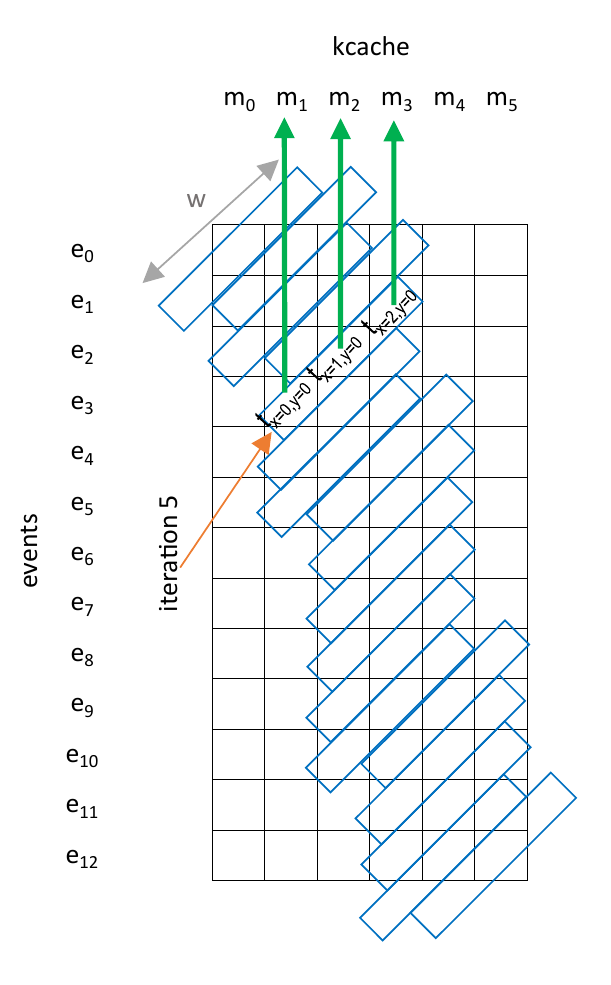}
    \caption{Spatially local memory accesses (green arrows) when \textit{kcache} is employed}
    \label{f:coalease3}
\end{subfigure}
\caption{Utility of \textit{kcache} in the \textit{core-kernel} to improve memory coalescing}
\end{figure*}

As stated under Section \ref{s:prekernel}, the data structure \textit{kcache} introduced to the GPU implementation facilitates memory coalescing by minimising random memory accesses to the \textit{model} array (\textit{pore-model} array in Fig. \ref{f:kmermodel}). If \textit{kcache} did not exist, access pattern by contiguous threads in the \textit{core-kernel} (shown for the iteration 5 of read 0) would look like in Fig. \ref{f:coalease1} where accesses to the \textit{ref} are shown in green colour arrows and the subsequent accesses to the \textit{pore-model} are in red colour arrows. The green arrows (relates to getting the k-mer at line \ref{a:logp:2} of Algorithm \ref{adaptivebandedcpucomputelog} in the CPU version) are spatially local and would facilitate memory coalescing in the GPU. However, red arrows (relates to line \ref{a:logp:4} of Algorithm \ref{adaptivebandedcpucomputelog} in the CPU version) to the \textit{model} array are random accesses. Note that such random accesses would occur during each iteration (iteration 3 to the last band iteration). Such multiple threads accessing random GPU memory locations degrade the performance due to smaller and less powerful GPU caches (compared to CPU), for instance, 32KB \textit{pore model} array is larger than 8KB GPU constant cache \cite{cudaprog}. 

These random accesses are eliminated by the \textit{kcache} constructed in \textit{pre-kernel} (stated under Section \ref{s:prekernel}) which is then passed as an argument to the \textit{compute} function at line \ref{a:abeagpu-sm7} in Algorithm \ref{adaptivebandedcudacore}). This \textit{kcache} is then passed on to the \textit{log\_probability\_match} function (at line \ref{a:computegpu:2} of Algorithm \ref{adaptivebandedcpucomputegpu}) which is then used at line \ref{a:loggpu:4} of Algorithm \ref{adaptivebandedcpucomputeloggpu}. The construction of the caches in the \textit{pre-kernel} requires random accesses to the model as shown in Fig. \ref{f:coalease2}, which happens only once. However, this \textit{kcache} is utilised by the \textit{core-kernel} in every iteration and facilitates memory coalescing (see green arrows in Fig. \ref{f:coalease3}  which are spatially local accesses to the \textit{kcache} by contiguous threads in iteration 5).

%to the \textit{ref} in Fig. \ref{f:adaptive}, where adjacent locations in \textit{ref} are accessed by contiguous cells (contiguous cells are processed by consecutive threads) in a band. This \textit{ref} signal was constructed on-the-fly in original algorithm (see background) through random memory accesses (to the \textit{model} array). 

It is noteworthy to mention that allocating one thread block per read is critical (in the kernel configuration) to: use lightweight  block synchronisation primitives \textit{\_\_syncthreads} (instead of expensive kernel invocations as synchronisation barriers \cite{cudaprog}); minimise warp divergence (otherwise the longest read in the thread block would consume the longest time which corresponds to the band filling loop); and, use shared memory per read (shared memory is allocated per block).

% For the core-kernel
% \begin{itemize}
%     \item flattened row major for mem coaleasing
%     \item other challenge is the number of registers
% \end{itemize}

\subsubsection{post-kernel}
 
The backtracking operation performed by this \textit{post-kernel} (one thread assigned to one read) does not expose fine grained parallelism as in previous kernels and thus not ideal for the GPU. However, performing this on GPU is still advantageous when compared to transferring huge intermediate arrays (\textit{scores} and \textit{trace}---size in order of GB) from GPU to the RAM. In addition, no additional memory in the RAM is required, thus reducing peak RAM usage. 

Allocating one thread block per read (as in \textit{core-kernel} to reduce warp divergence) is not ideal for this \textit{post-kernel} due to the lack of fine grained parallelism (i.e. 1 block having 1 thread), which results in reduced GPU occupancy (occupancy will be limited by the maximum thread blocks that can simultaneously reside in a GPU multi-processor). This is remedied without affecting the warp divergence by allocating a large number of threads per block (eg: 1024) and then limiting only the first thread in the warp (a warp is composed of 32 contiguous threads \cite{cudaprog} and thus thread with indices 0, 32, 64, 96 ... etc) to perform the actual computation (backtracking for a read).

\clearpage

\subsection{Memory optimisation} \label{memopti}

CPU version of the Adaptive Banded Event Alignment (ABEA) algorithm performs dynamic memory allocations (\textit{malloc}) on a per read basis. The number of reads in a dataset is in the order of millions and thus incur millions of \textit{malloc} calls. However, dynamic memory allocations (\textit{malloc} performed inside GPU kernels) are extraordinarily expensive in terms of execution time \cite{cudaprog}. In-fact, our initial GPU kernel implementation which performed such memory allocations was more than 100$\times$ slower than the CPU implementation. An intuitive approach of statically allocating memory at the compile time is not practical as nanopore read lengths vary significantly (\textasciitilde100 bases to >1 Mbases as explained previously) and thus the associated data structures vary from \textasciitilde200 KB to >1.5 GB.  We present a methodology that significantly reduces the number of memory allocations by pre-allocating large chunks of contiguous memory at the beginning of the program to accommodate a batch of reads, which are then reused throughout the life-time of the program. The sizes of these large chunks are determined by the available GPU memory and the average number of events per base (i.e. average value of the number of events divided the by the read length). For a given batch of reads, we assign reads to the GPU until the allocated GPU memory chunks saturate, and the rest of the reads are assigned to the CPU.
 
We describe the memory allocation technique in two steps: in Section \ref{memserialise} how the memory allocation for a batch of reads at a time is performed; and, in Section \ref{heuristicmem}, how the method in Section \ref{memserialise} can be expanded to reuse large chunks of memory, allocated at the beginning of the program.

\subsubsection{Data array serialisation} \label{memserialise}

In the three GPU kernels elaborated in Section \ref{computeopti}, the associated data arrays per each read are \textit{ref}, \textit{kcache}, \textit{events}, \textit{score}, \textit{trace}, \textit{ll\_idx} and \textit{alignment} (final output from the \textit{post-kernel}). If any of these arrays are allocated inside the GPU kernels on a per-read basis, for instance if \textit{score} and \textit{trace} arrays are allocated at line \ref{a:pre4} of Algorithm \ref{adaptivebandedcudapre} using \textit{malloc}), the performance will be degraded. 

We identified that the sizes of all the aforementioned data arrays are dependent only on the read length (known at run-time during file reading) and the number of events for the read (known after event detection described in Section \ref{background}). Thus, the sum of read lengths and the number of events for a batch of $n$ reads (GPU processes a batch of $n$ reads at a time) is used to calculate the sizes of memory allocations required for the particular batch according to the formulation below. 

Let $n$ be the number of reads loaded to the RAM (from the disk) at a time. Let $r[]$ be the read length and  $e[]$ be the number of events for all the reads in batch of $n$ reads. Column 1 of Table \ref{t:datastructperread}, lists the data arrays. The size of arrays \textit{ref} and \textit{kcache} depends only on read lengths $r$;  \textit{events} and \textit{alignment} depend on number of events $e$; and, \textit{score}, \textit{trace} and \textit{ll\_idx} depend on both read length  $r$ and number of events $e$. Based on these dependencies, the arrays are categorised in Table \ref{t:datastructperread} by horizontal separators. The second column of Table \ref{t:datastructperread}  states the data-type size of each array, denoted by constants of the form $c_x$. Typical values of these constants (in our implementation) are given inside the brackets. For instance, the data type for \textit{ref} is \textit{char} and thus $C_{r}$ is 1 byte. The data type for events is a \textit{struct} of size $C_{e}$ that is 20 bytes. Note that, the exact values may depend on the implementation and the underlying processor architecture, nevertheless are constants known at compile time. The third column  of Table \ref{t:datastructperread} shows the size required for the particular array for a single read, i.e. the size for the i\textsuperscript{th} read (assume 0 based index origin) in the batch of $n$ reads.  For instance, \textit{ref} depends on the read length of the particular read and the datatype, thus the size is $C_{r}r[i]$. \textit{Score} depends on read length, number of events, data type  size and band-width ($W$), thus $WC_s(r[i]+e[i])$. The last column of Table \ref{t:datastructperread} is the total size required for a batch of reads (based on sum of $r$ and $e$). For instance, the sum of all the \textit{ref} arrays for the batch is the product of data type size $C_r$ and sum of all read lengths in the batch $\sum_{i=0}^{n-1}{r[i]}$.

\begin{table}
\caption{Data arrays associated with ABEA and their sizes}

\begin{tabular}{|l|l|l|l|}
\hline
\textbf{Array}          & \textbf{Data type size}     &\textbf{Size for read $i$ in batch} &  \textbf{Size per batch}   \\
& \textbf{(bytes) } & & \\ \hline
\textit{ref[]}           & $ C_r$ (1)             & $ C_r r[i] $              &  $ C_r \sum_{i=0}^{n-1}{r[i]} $             \\
\textit{kcache[]}       & $ C_k$ (12)            & $ C_k r[i] $              &  $ C_k \sum_{i=0}^{n-1}{r[i]} $             \\ \hline
\textit{events[]}             & $ C_e$ (20)             & $ C_e e[i] $              &  $ C_e \sum_{i=0}^{n-1}{e[i]} $            \\ 
\textit{alignment[]}        & $ C_a $(8)           & $ 2 C_a e[i] $            &  $ 2 C_a \sum_{i=0}^{n-1}{e[i]} $            \\ \hline
\textit{score[][]}                   & $C_s $ (4)      & $ W C_s (r[i] + e[i]) $   &  $ W C_s \sum_{i=0}^{n-1}{(r[i] + e[i])} $     \\
\textit{trace[][]}                   & $C_t$ (1)     & $ W C_t (r[i] + e[i]) $     &  $ W C_t \sum_{i=0}^{n-1}{(r[i] + e[i])} $    \\
\textit{ll\_idx[]}         & $C_l$(8)    & $ C_l (r[i] + e[i]) $     &  $ C_l \sum_{i=0}^{n-1}{(r[i] + e[i])} $     \\
 \hline
\end{tabular}

\label {t:datastructperread}
\end{table}

Based on the total array sizes in the last column Table \ref{t:datastructperread}, we can allocate seven big chunks of linear contiguous memory in the GPU. Let the base address of those chunks be represented by uppercase letters: \textit{REF}; \textit{KCACHE}; \textit{EVENTS} etc. These memory allocations are performed using \textit{cudaMalloc()} API calls, just before the kernel invocations and are deallocated after the kernels. Note that for now,  we do these allocations and deallocations for each batch of reads.
 
The GPU arrays \textit{REF}, \textit{KCACHE}, \textit{EVENTS} etc, allocated using \textit{cudaMalloc} above are 1D arrays, thus multi-dimensional arrays in the RAM (eg: an array of pointers---each pointer pointing to a string/char array) must be serialised/flattened. One option is to save a series of pointers  associated to each above array during the serialisation and then utilising those pointers for addressing a particular element later. However, this can be performed better by storing only two offset arrays of length $n$ each: \textit{read offset} array $p[]$, which is the cumulative sum of read lengths in the batch ($p[i] =\sum_{j=0}^{i-1}{r[j]}$); and, \textit{event offset} array $q$, which is the the cumulative sum of events in the batch ($q[i] = \sum_{j=0}^{i-1}{e[j]}$). Note that, $r$ and $e$ have the same definitions as before.  These two offset arrays $p$ and $q$ can be used to deduce the associated pointer to a given element when required, by computing the array offset as shown in Table \ref{t:pointercompute}. The first column of Table \ref{t:pointercompute} is the base address of the large GPU arrays we allocated above. The offset of the element pertaining to the i\textsuperscript{th} read (assume 0-indexing) in the particular array is given in the second column of Table \ref{t:pointercompute}.  The definition of constants $C_x$ and $W$ are the same as for the previous Table \ref{t:datastructperread}. These 1D array base addresses in the first column of Table \ref{t:pointercompute} and the two associated offset arrays $p[]$ and $q[]$, are passed as arguments to the GPU kernels (Algorithm \ref{adaptivebandedcudapre} and Algorithm \ref{adaptivebandedcudacore}). These arguments are used for the the memory pointer computation inside the GPU kernels (line \ref{a:pre4} of Algorithm \ref{adaptivebandedcudapre} and line \ref{a:core4} of Algorithm \ref{adaptivebandedcudacore}) based on the second column of Table \ref{t:pointercompute}.

\begin{table}
\caption{GPU data arrays, pointer computation and heuristically determined sizes}
\begin{subtable}{0.499\textwidth}
\caption{Computation of pointer for the read i}
\begin{tabular}{|l|l|}
\hline
\textbf{1D GPU array}          		& \textbf{Offset to element} \\
\textbf{(base address)}          		& \textbf{i in the batch}
\\ \hline
\textit{REF}         & $ C_r p[i] $ 									\\
\textit{KCACHE}        	& $ C_k p[i] $ 									\\ \hline
\textit{EVENTS}        & $ C_e q[i] $ 									\\ 
\textit{ALIGNMENT}       & $ 2 C_a q[i] $ 								\\ \hline
\textit{SCORE}     & $ W C_s (p[i]+q[i]) $ 						\\
\textit{TRACE}     & $ W C_t (p[i]+q[i]) $ 						\\
\textit{LL\_IDX}  & $ C_l (p[i]+q[i]) $ 					 \\ \hline
\end{tabular}
\label {t:pointercompute}
\end{subtable}
\begin{subtable}{0.499\textwidth}
\caption{Heuristic allocation}
\begin{tabular}{|l|l|}
\hline
\textbf{1D GPU array}   & \textbf{Allocated size} \\
\textbf{(base address)}   & \textbf{per batch} 
\\ \hline
\textit{REF}         			& $C_r X$			 \\
\textit{KCACHE}        			& $C_k X$ 			 \\ \hline
\textit{EVENTS}        			& $C_e Y$ 		 \\ 
\textit{ALIGNMENT}       		& $2 C_a Y$ 				 \\ \hline
\textit{SCORE}     				& $ W C_s (X+Y)$ 				 \\
\textit{TRACE}     				& $W C_t (X+Y)$				 \\
\textit{LL\_IDX}  				& $ C_l (X+Y)$ 				 \\ \hline
\end{tabular}
\label {t:actuallyacllocated}
\end{subtable}
\end{table}

% Instead of storing an associated pointer array for each of those arrays, we compute 
%  the memory location of a given element \textit{i} is computed based on the 

% and the actually allocated size of the array (section \ref{heuristicmem}) are in columns 2-6 respectively. All sizes are in bytes, value in brackets in column 2 are the default data-type values and $W$ is the band-width.

Algorithm \ref{a:memalloc1} elaborates how the above mentioned strategy is integrated into the previous execution flow depicted in Algorithm \ref{a:exec}. Lines \ref{a:mem:3}-\ref{a:mem:7} of Algorithm \ref{a:memalloc1}  show how the offset arrays $p$ and $q$ are computed for each batch of reads.
Line \ref{a:mem:8} of Algorithm \ref{a:memalloc1}  performs the serialisation of the multi-dimensional arrays with the use of offset arrays $p$ and $q$.
Line \ref{a:mem:9} of Algorithm \ref{a:memalloc1} allocates GPU arrays based on sizes in last column of Table \ref{t:datastructperread}. Then, the serialised arrays are copied to allocated GPU memory (line \ref{a:mem:10} of Algorithm \ref{a:memalloc1}), GPU kernels (the three kernels discussed in Section \ref{computeopti}) are executed (line \ref{a:mem:11}) and the alignment result is copied back from the GPU (line \ref{a:mem:12}). At the end, the alignment result is converted back to multi-dimesional arrays (line \ref{a:mem:13}) and then the GPU memory (allocated at line \ref{a:mem:9}) is deallocated (line \ref{a:mem:14}).

The offset arrays $p$ and $q$ (and also REF, KCACHE, EVENTS, etc.) are passed onto the GPU kernels and are utilised inside the GPU kernels to compute the memory pointers (line \ref{a:pre4} of Algorithms \ref{adaptivebandedcudapre} and \ref{adaptivebandedcudacore}) through the equations listed on the second column of Table \ref{t:pointercompute}. 

\begin{algorithm}[!ht]
\caption{Memory allocation---data structure serialisation}\label{a:memalloc1}
\begin{algorithmic}[1]
\For{batch of $n$ reads}
\State ... \Comment \textcolor{gray}{CPU processing steps before the ABEA eg: event detection}
\State $rs,es \gets 0,0$ \Comment \textcolor{gray}{cumulative sum of read lengths and no of events} \label{a:mem:3}
\For{each read $i$}
\State $p[i],q[i] \gets rs,es$ \Comment \textcolor{gray}{save current read and event offsets}
\State $rs \gets rs + r[i]$; $es \gets es + e[i]$
\EndFor \label{a:mem:7}
\State $serialise\_ram\_arays(p,q,...)$ \Comment \textcolor{gray}{flatten multi dimensional arrays in RAM to 1D arrays}\label{a:mem:8}

\State  \textit{allocate\_gpu\_arrays(rs,es,...)} \Comment \textcolor{gray}{GPU arrays REF, KCACHE, EVENTS, etc.}\label{a:mem:9}
\State $memcpy\_ram\_to\_gpu(...)$  \Comment \textcolor{gray}{copy inputs of the ABEA to the GPU memory}\label{a:mem:10}
\State $gpu\_alignment(p,q...)$ \Comment \textcolor{gray}{Perform ABEA on the GPU} \label{a:mem:11}
\State $memcpy\_gpu\_to\_ram(...)$ \Comment \textcolor{gray}{copy alignment result back to the RAM}\label{a:mem:12}
\State $deserialise(p,q,....)$ \Comment \textcolor{gray}{convert 1D result array to multi dimensional array} \label{a:mem:13}
\State $free\_gpu\_arrays()$ \Comment \textcolor{gray}{free GPU arrays REF, KCACHE, EVENTS, etc.}\label{a:mem:14}
\State ... \Comment \textcolor{gray}{CPU processing steps after ABEA eg: HMM}
\EndFor
\end{algorithmic}
\end{algorithm}

The limitation of this strategy is the GPU memory allocation and de-allocation (line \ref{a:mem:9} and \ref{a:mem:14} of Algorithm \ref{a:memalloc1}) performed for each batch of reads (which is expensive on certain GPUs, see Section \ref{s:memoptireasons}). This limitation is remedied by the heuristic based pre-allocation strategy explained in the next subsection.

% arrays in Table \ref{t:propk} are proportional to the  $n$. The third column mentions whether each vector is an input or an output to the GPU. 
% The length of each read (in a batch of $n$ reads) is in $r$ which is populated based on input reads. The number of events $e$ are based on the previous computation performed on CPU.  The read offsets $p$ and event offsets $q$ are computed based on the cumulative sums and are required to serialise the reads later into linear arrays. As $n$ is typically around 1000 these arrays are in order of kilo-bytes.

% \begin{table}[!ht]
% \begin{tabular}{|l|l|l|}
% \hline
%  \textbf{vector} & \textbf{description} & \textbf{type} \\ \hline
%  $r$ & read lengths & input \\ \hline
%  $p$ & read offsets : $ p[i] = \sum_{k=0}^{i-1}{r[k]} $ & input \\ \hline
%  $e$ & number of events & input \\ \hline
%  $q$ & event offsets : $ q[i] = \sum_{k=0}^{i-1}{e[k]} $ & input \\ \hline
%  $a$ & number of alignment pairs & output  \\ \hline
% \end{tabular}
% \caption{arrays proportional to $n$}
% \label {t:propk}
% \end{table}

\subsubsection{Heuristic based memory pre-allocation} \label{heuristicmem}

The GPU memory allocations in the previous section which were performed for each batch could be eliminated by pre-allocating all the available GPU memory at the startup of the program and then re-using for subsequent batches of reads). If the sizes of the arrays depended only on the read length, the total read length accommodable into the available GPU memory can be derived.  Then, the available memory can be allocated among the seven large arrays ($REF$, $KCACHE$, $EVENTS$ etc) in correct proportion. However, these array sizes depend both on the read length and the number of events which are unknown at the beginning of the program; thus, memory cannot be partitioned among the data arrays. Therefore, We present a heuristic approach which exploits characteristic of nanopore data to estimate the proportion to maximally utilise the available GPU memory. In summary, we obtain the average number of events per base (average of the number of events divided by read length), use this average to determine the maximum read length that can be accommodated to the GPU, and proportionally allocate the GPU arrays. This approach is formulated as follows.

Sum of all the cells in column 4 of Table \ref{t:datastructperread} is total memory required for a batch of $n$ reads.  This sum simplifies to equation \ref{e:totalsize} (due to the properties of constants) where $C_R = C_r+C_k+WC_s+WC_t+C_l$ and  $C_E = C_e+2C_a+WC_s+WC_t+C_l$.
This sum represents the total size of all array (for adapted banded event alignment algorithm) for a batch of $n$ reads.

\begin{equation} 
\label{e:totalsize}
S = C_R\sum_{i=0}^{n-1}{r[i]} + C_E\sum_{i=0}^{n-1}{e[i]} 
\end{equation}

If $\bar{\mu}$ is the average number of events per base (total number of events divided by the total read length for all reads in the batch), we can write as $\sum_{i=0}^{n-1}e[i] = \bar{\mu} \sum_{i=0}^{n-1}r[i]$.  Now substituting this in equation \ref{e:totalsize} gives  $S = (C_R+ \bar{\mu} C_E) \sum_{i=0}^{n-1}{r[i]} $. We observed  that for a sufficient batch size (>64), $\bar{\mu}$ is stable \textasciitilde 2.5 (on more than 10 datasets we tested). Let this estimated value for $\bar{\mu}$ be represented by the constant $\mu$. Thus, the total memory required for a batch of reads can be estimated using equation \ref{e:mem}.
 
\begin{equation} 
\label{e:mem}
M = (C_R+ \mu C_E) \sum_{i=0}^{n-1}{r[i]}
\end{equation}

Equation \ref{e:mem} can be used to estimate the maximum number of bases (sum of read lengths) that a given amount of GPU memory can accommodate. Let $M$ in equation \ref{e:mem} be the available GPU memory. Then, the approximate maximum number of bases $X$ that fits available GPU memory $M$ can be computed via equation \ref{e:X}. Then, the associated total number of total events $Y$ which the GPU memory can accommodate, is found by equation \ref{e:Y}. 

\begin{equation}
\label{e:X}
 X = floor\left(\frac{M}{C_R+ \mu C_E}\right)
\end{equation}
\begin{equation}
\label{e:Y}
 Y = floor(\mu X) 
\end{equation}

These $X$ and $Y$ allow the available GPU memory to be allocated among the seven large arrays ($REF$, $KCACHE$, $EVENTS$ etc) with approximately correct proportions, as shown in the second column of Table \ref{t:actuallyacllocated}. The values in the second column of Table \ref{t:actuallyacllocated} are obtained by substituting $\sum_{i=0}^{n-1}{r[i]}$ with $X$ and  $\sum_{i=0}^{n-1}{e[i]}$ with $Y$ in the last column of Table \ref{t:datastructperread}.

By incorporating the above heuristic based memory allocation  strategy to Algorithm \ref{a:memalloc1}, we get the execution flow in Algorithm \ref{a:memalloc}. The major changes to the previous Algorithm \ref{a:memalloc1} are highlighted in blue text. Now the GPU memory is allocated at the beginning of the program based on the estimated $X$ and $Y$ on line \ref{a:heuris:1} of Algorithm \ref{a:memalloc}. As $X$ and $Y$ are approximations, the GPU arrays may saturate for certain batches of reads.
Line \ref{a:heuris:6}  of Algorithm \ref{a:memalloc} checks if GPU arrays are saturated and assigns the read to either GPU (line \ref{a:heuris:9}) or CPU (line \ref{a:heuris:11}), accordingly. Only a few reads are assigned to the CPU and these  few reads are processed on the CPU in parallel to the GPU kernel execution, and thus no additional execution time is incurred.

\begin{algorithm}[!ht]
\caption{heuristic memory allocation scheme}\label{a:memalloc}
\begin{algorithmic}[1]
\State  \textcolor{blue}{\textit{allocate\_gpu\_arrays(X,Y)}} \Comment \textcolor{gray}{pre-allocate GPU arrays REF, KCACHE, EVENTS, etc.} \label{a:heuris:1}

\For{batch of $n$ reads}
\State ... \Comment \textcolor{gray}{CPU processing steps before the ABEA eg: event detection}
\State $rs,es \gets 0,0$ \Comment\textcolor{gray}{cumulative sum of read lengths and no of events}

\For{each read $i$}
\If{ \textcolor{blue}{($rs + r[i] \leq X$ \textbf{and} $es + e[i] \leq Y$)}} \Comment \textcolor{gray}{check if GPU arrays have adequate free space}  \label{a:heuris:6}
\State $p[i],q[i] \gets rs,es $ \Comment \textcolor{gray}{save current read and event offsets}
\State $rs \gets rs + r[i] ; es \gets es + e[i]$
\State  \textcolor{blue}{\textit{assign\_to\_gpu(i)}} \Comment \textcolor{gray}{GPU arrays have space, thus assign read to the GPU} \label{a:heuris:9}
\Else 
\State  \textcolor{blue}{\textit{assign\_to\_cpu(i)}} \Comment \textcolor{gray}{a GPU arrays is already full, thus assign the read to the CPU} \label{a:heuris:11}
\EndIf
\EndFor
\State $serialise\_ram\_arays(p,q,...)$ \Comment \textcolor{gray}{flatten multi dimensional arrays in RAM to 1D arrays}
\State $memcpy\_ram\_to\_gpu(...)$  \Comment \textcolor{gray}{copy inputs of the ABEA to the GPU memory}
\State $gpu\_alignment(p,q...)$ \Comment \textcolor{gray}{Perform ABEA on the GPU}
\State  \textcolor{blue}{\textit{process\_rest\_on\_cpu()}} \Comment \textcolor{gray}{execute on the CPU in parallel to the GPU kernels}
\State $memcpy\_gpu\_to\_ram(...)$ \Comment  \textcolor{gray}{copy alignment result back to the RAM}
\State $deserialise(p,q,....)$ \Comment \textcolor{gray}{convert 1D result array to multi dimensional array} 

\State ... \Comment \textcolor{gray}{CPU processing steps after ABEA eg: HMM}
\EndFor
\State  \textcolor{blue}{\textit{free\_gpu\_arrays()}} \Comment \textcolor{gray}{free GPU arrays REF, KCACHE, EVENTS, etc.}
\end{algorithmic}
\begin{flushleft}
\textcolor{blue}{Note: Changes to Algorithm \ref{a:memalloc1} are highlighted in blue}
\end{flushleft}
\end{algorithm}

With the heuristic based memory pre-allocation strategy described in this section, \textit{cudaMalloc} operations are invoked only at the beginning of the program and thus no additional memory allocation overhead during the processing. Note that, our implementation is future proof; i.e. $\mu$ is a user specified parameter (that is initialised to 2.5 by default) in case nanopore data characteristics change in future.

% Each array for the whole data batch is flattened. We have two offsets to  compute the memory pointers in the flattened array. 
% Read\_ptr\_array : cumulative sum of read lengths, event\_ptr\_array : cumulative sum of events. 
% Intermediate DP array 

% No mallocs inside the GPU kernels as intermediate arrays follows the same approach. 

% Heuristic for the allocation of each big array. Eliminates CUDA mallocs for each batch. However, the average number of events per base and a maximum threshold is needed to balance.

%\todo{importance of user parameter B with K}

\subsection{Heterogeneous processing} \label{loadbalmet}

If all the reads were of similar length, GPU threads that process the reads would complete approximately at the same time, and thus GPU cores will be busy throughout the execution. However, as stated in Section \ref{background}, there can be a few reads which are significantly longer than the other reads (we will refer to them as \emph{very long reads}). When the GPU threads process reads in parallel, presence of such \textit{very long reads} will cause all other GPU threads to wait until the GPU threads processing the longest read complete. This thread waiting leads to under utilisation of GPU cores. Thus, we process these \textit{very long reads} on the CPU while the GPU is processing the rest in parallel. However, there can be exceptionally long reads (we will refer to them as \emph{ultra long reads}) which the CPU would take longer time than what the GPU took to process the whole batch. Such reads would lead the GPU to idle until the CPU completes. Thus, \textit{ultra long reads} will be skipped and will be processed separately at the end by the CPU. Similarly, there can be a few \emph{over segmented reads} (i.e. reads with a significantly higher events per base ratio than the others) which cause GPU under utilisation. These over-segmented reads will also be processed on the CPU.

We discuss these problems of \textit{very long reads} and \textit{ultra long reads} in detail with examples in Section \ref{s:vlrulr}, along with the solutions.  Then, in this Section \ref{s:oversegreads}, we discuss the problem of over segmented reads and the respective solution. Then, in Section \ref{s:batchsize}, we discuss another  factor that affects performance, the batch size (number of reads loaded to the RAM at a time). Finally, in Section \ref{s:decision}, we describe a method to detect and prompt the user of any drastic impacts on performance along with suggestions to tune parameters to minimise the impact.

\subsubsection{Very long reads and ultra long reads}\label{s:vlrulr}

Consider a batch of reads where \textasciitilde90\% of the reads are less than 30 Kbases in length. Assume the longest read in the batch is 90 Kbases. Assume that the GPU is processing all the reads (in the batch) in parallel. Suppose that GPU threads processing reads of length <30 Kbases (90\% of the threads) would complete in <300ms while GPU threads processing the longest 90 Kbases read would take 900ms. As a result, the completed GPU threads will have to wait for additional 600ms. Similarly, the few \textit{very long reads} consume a significant time to process on the GPU in comparison to other reads in the batch. Majority of the GPU threads will have to wait and this causes under-utilisation of GPU compute-cores. Furthermore, \textit{very long reads} negatively affects the GPU occupancy by occupying a significant portion of GPU memory. For instance, a read of size \textasciitilde10 Kbases requires only \textasciitilde18 MB of GPU memory while a read with 90 Kbases requires \textasciitilde160MB memory. Hence, \textit{very long reads} occupy a significant portion of GPU memory, limits the number of reads that could be processed in parallel. This reduces the amount of parallelism and the occupancy of the GPU is reduced.

Fortunately,  \textit{very long reads} being few (see the typical read length distribution under results), the CPU (core frequency faster than on GPU) could process those reads while GPU is processing the rest of the reads. In the above example, selecting a static threshold (eg: processing reads of length <30Kbases on GPU and rest on CPU) would give reasonable performance. However, selecting such a static threshold is not ideal due to variations in the read length distributions based on the dataset (see background).  Thus, we use the product of \textit{max-lf} and the average read length in the batch to determine the threshold dynamically, where \textit{max-lf} is a user-parameter that defaults to 5.0. This threshold was empirically determined.

%However, selecting such a static universal threshold is not ideal due to variations in: read length distributions based on the data-set (see background); and, specification of systems (CPU, RAM, GPU memory and GPU compute cores etc.).  Thus, the strategy we propose in section \ref{s:decision} selects the threshold based on the characteristics of data.
Now assume amongst the \textit{very long reads} processed on the CPU, a few \emph{ultra long reads} (eg: read >100 Kbases in a dataset where >99\% of the reads are <100 Kbases). Such \emph{ultra long reads} could cause a severe load imbalance between the CPU and the GPU. For instance, assume that there exists a read which is 1 Mbases in a given read batch. Despite the high core frequency, the CPU will take a few seconds to process such an \emph{ultra long read}. The GPU meanwhile would process the whole batch in less than 1s (see results for empirical evidence). Such \emph{ultra long reads} being <1\%, are skipped during the processing (while being written to a separate file) and are separately processed by the CPU at the end. In our implementation, the threshold for \emph{ultra long reads} is a user defined parameter which defaults to 100 Kbases. There is an additional advantage of processing \emph{ultra long reads} later. \emph{Ultra long reads} usually require a significant amount of RAM (a few gigabytes) and may crash on limited memory systems. In the end, it is possible to process these reads with a limited amount of threads to reduce the peak memory consumption, particularly if the size of the RAM is limited.

\subsubsection{Over segmented reads}\label{s:oversegreads}
Once the \textit{very long reads} and \textit{ultra long reads} are processed as in Section \ref{s:vlrulr}, the performance impact due to the over-segmented events become prominent. While majority of the reads have a number of events per base that is close to the average $\mu(=2.5)$, a few reads can have a very large value. For instance, a few reads with a number of events per base being more than eight times the average $\mu(=2.5)$ can violate the suitability of our partitioning of GPU memory as $X$ and $Y$ ($X$ and $Y$ are derived in equations \ref{e:X} and \ref{e:Y}). These over-segmented reads lead to the GPU arrays that are proportional to $Y$ be full,  while the arrays proportional to $X$ are left under-utilised. For instance, arrays proportional to $Y$ can become 100\% while arrays proportional to $X$ are only filled to <70\%. Hence, over segmented reads lead to under-utilisation of GPU memory and results in limiting the number of reads which are processed in parallel. We process the over-segmented reads on the CPU based on a user specifiable threshold \textit{max-epk} which defaults to 5.0.

On rare occasions, reads with >100 events per base were observed. Such severely over-segmented reads can be processed separately at the end or ignored totally as such rare reads amongst millions of other reads are unlikely to affect the final polishing result.

\subsubsection{Batch size}\label{s:batchsize}

Selection of proper batch size (reads loaded to RAM from the disk at a time) is another important parameter that affects performance. If the batch size is too small compared to what the GPU memory can accommodate, the number of reads to be processed in parallel is limited, thus leads to in-adequate occupancy. Conversely, if the batch is too large to fit the GPU, CPU will have to process many surplus reads that could not be accommodated into the GPU. The batch size in our implementation is determined by two user specified parameters: $K$ which is the maximum number of reads; and, $B$ which is the maximum number of total bases. When  reading from the disk to RAM, the true batch size ($n$-number of reads and $b$-number of total bases are capped by $K$ and $B$) is determined by the first value ($n$ or $b$) reaching the cap ($K$ or $B$) first. Having such a limit $B$ allows to cap peak RAM due to adjacent \textit{very long reads}. The suitable value for $B$ is dependent on the available GPU memory, which can be estimated via the equation \ref{e:X} discussed in Section \ref{memopti}.

%Further, the average events per base should  be modified in case it differs in data.

\subsubsection{Detection of performance anomalies}\label{s:decision}

While we have empirically determined  typical parameters/thresholds (associated with above strategies), an unusual situation (for instance, a big gap between the CPU and GPU specifications or a data set that severely deviates from the heuristics we use) may cause performance anomalies. We employ the following method to detect a severe performance anomaly caused by such an unusual scenario.

\begin{figure}[!ht]
  \centering
\begin{subfigure}[!ht]{\textwidth}
  \centering
    \includegraphics[width=\columnwidth]{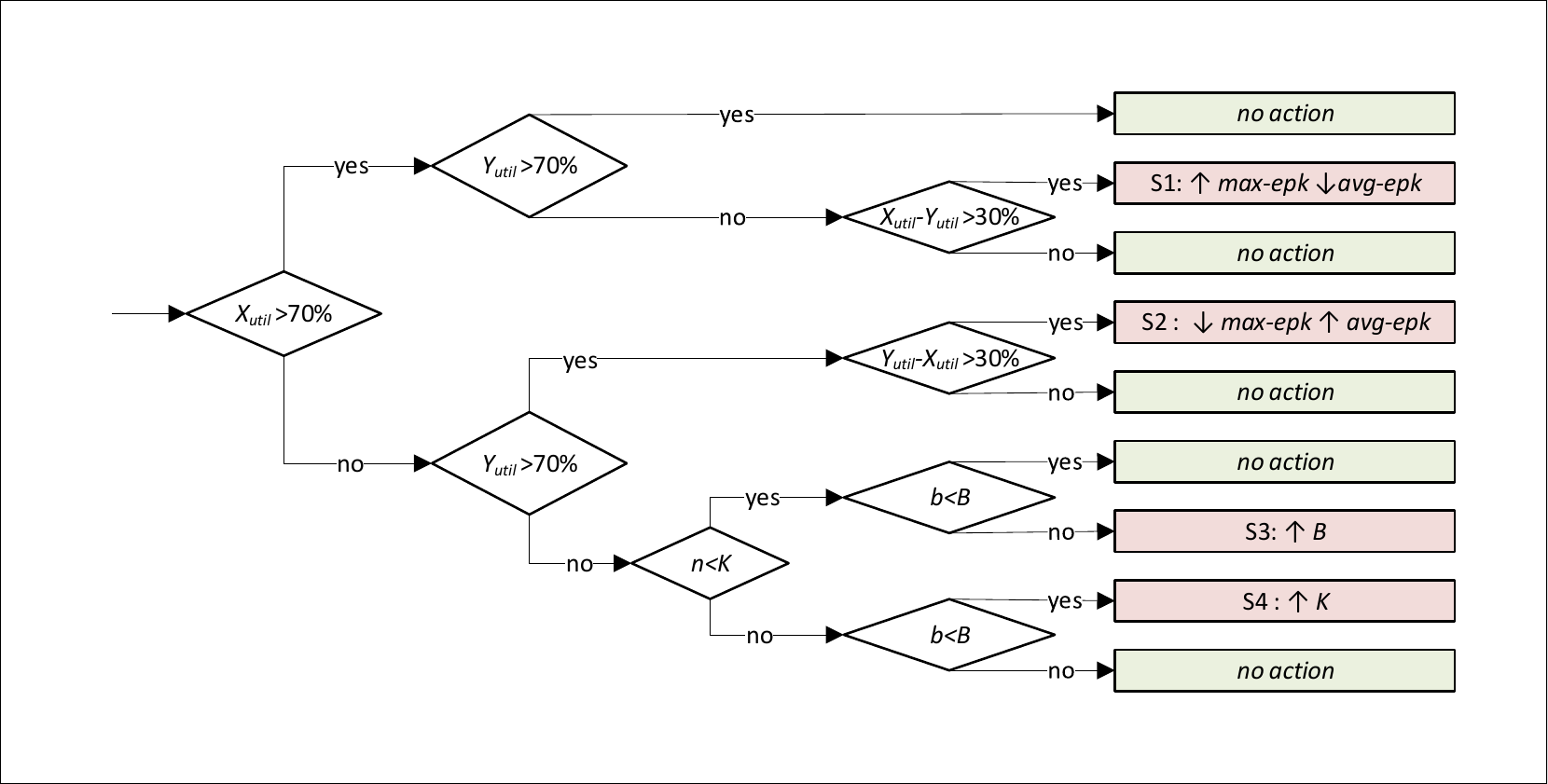}
    \caption{memory balancing} 
    \label{f:mem_balance}
\end{subfigure}

\vspace{0.5cm}
\begin{subfigure}[!ht]{\textwidth}
  \centering
    \includegraphics[width=\columnwidth]{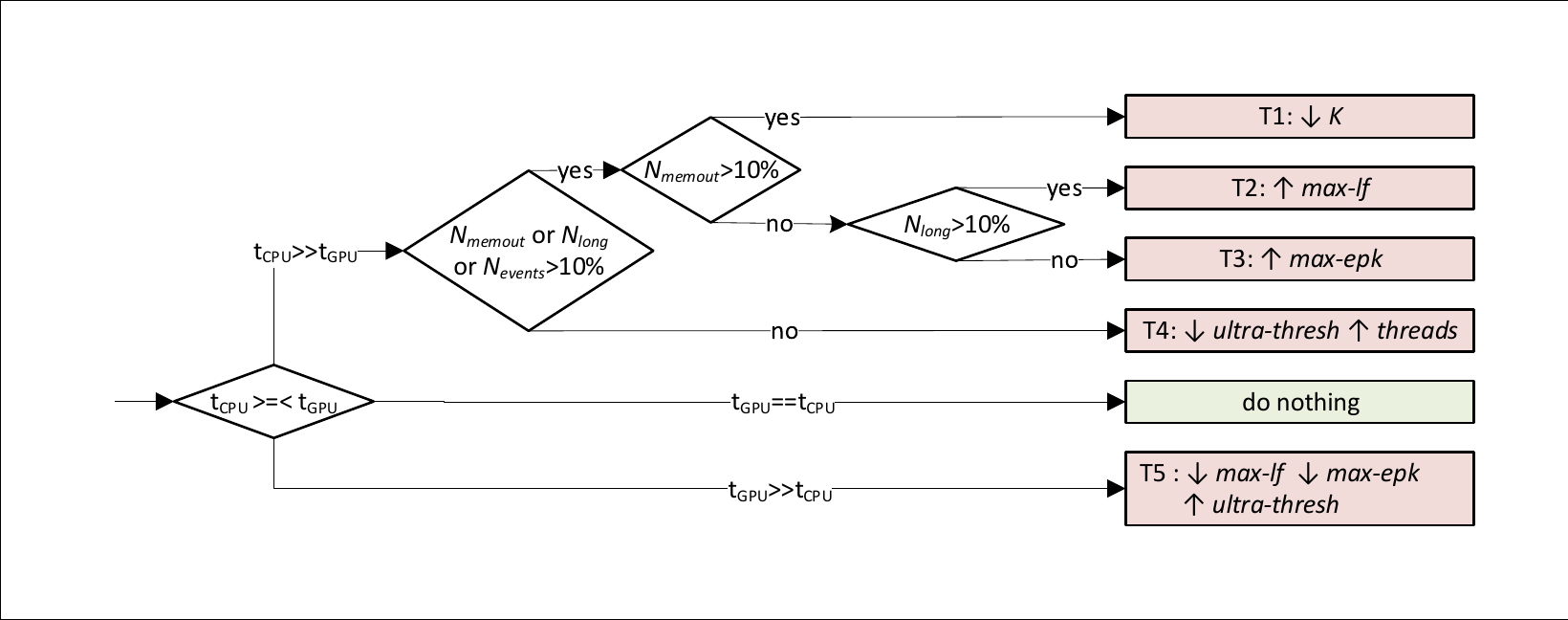}
    \caption{load balancing} 
    \label{f:load_balance}
\end{subfigure}
    \caption{Decision trees for resource optimisation} 
    \label{f:decision_trees}
\end{figure}

We measure the quantities representing resource utilisation during  run time, which are  listed in Table \ref{t:measuredquantities}.  These quantities are measured per  batch of reads loaded to the RAM at a time. We use those measured quantities to determine any severe performance issues and  suggest  suitable parameter adjustments to the user.  The adjustable parameters (or thresholds) that can be tweaked to improve the resource utilisation are defined in Table \ref{t:adjustableparameters}. Determination of performance issues and suggestions are done via two decision trees, one that corresponds to GPU memory usage (Fig. \ref{f:mem_balance}) and another which corresponds to balancing the load between CPU and GPU (Fig. \ref{f:load_balance}).

%These quantities in Table \ref{t:measuredquantities} are used to detect any drastic performance issues and to The detection of performance issues and the suggestion of parameter adjustments are done using the decision trees in Fig. \ref{f:decision_trees}.  

%However, , long reads () are very few and they can be performed on the CPU---. 

%For instance, (based on formulation in section \ref{memopti}) a single base requires x bytes of memory, thus for example an individual 10k, 100k and  1M base reads would requires xx MB, yy MB and zz GB respectively. Further, if a read is very long compared to the other reads in the batch, the longer read would be the critical path.

% \begin{itemize}

%     \item an ideal length threshold is to be selected that keeps the load between the CPU and GPU similar. However Selecting the threshold for the ultra-long reads is tricky, due to the  as well as the presence of multiple CPU GPU combinations.
%     \item 
%     \item This can be made automatic in theory, the method is not difficult, but the implementation is hard due to the number of cases requires to handle. SO future work.
%     \item Load balancing between the CPU and GPU is based on heuristic. At the moment automatically profiles during the run time and gives suggestions on performance improvement through the changing of parameters.
%     \item The approach is explained below.
%     \item 
    
% \end{itemize}

 \begin{table}[!ht]
     \centering
     \begin{tabular}{|l|p{12cm}|}
     \hline
     \textbf{quantity} & \textbf{description} \\ \hline

     t\textsubscript{CPU} & processing time on CPU \\ \hline  
     t\textsubscript{GPU} & processing time on GPU \\ \hline  
     X\textsubscript{util} & utilisation percentage of the arrays proportional to $X$  (\textit{rs} as a percentage of $X$ in Algorithm \ref{a:memalloc}) \\ \hline  
     Y\textsubscript{util} & utilisation percentage of the arrays proportional to $Y$ (\textit{es} as a percentage of $Y$ in Algorithm \ref{a:memalloc}) \\ \hline   
     N\textsubscript{memout} & number of reads assigned to CPU due to GPU memory getting prematurely full (corresponds to line \ref{a:heuris:11} of Algorithm \ref{a:memalloc}) \\ \hline  
     N\textsubscript{long} & number of \textit{very long reads} assigned on to the CPU (corresponds to user parameter \textit{max-lf}) \\ \hline
     N\textsubscript{events} & number of reads with too many events per read assigned onto the the CPU (corresponds to user parameter \textit{max-epk}) \\ \hline  
     $n$ & the number of reads actually loaded to the RAM \\ \hline   
     $b$ & the number of bases actually loaded to the RAM \\ \hline

     \end{tabular}
     \caption{measured quantities}
     \label{t:measuredquantities}
 \end{table}

 \begin{table}[!ht]
     \centering
     \begin{tabular}{|l|p{12cm}|}
     \hline
     \textbf{parameter} & \textbf{description} \\ \hline
     \textit{max-lf} & reads with length $\le$ \textit{max-lf} $\times$ \textit{average\_read\_length} are assigned to GPU and rest to CPU \\ \hline
     \textit{avg-epk} & average number of events per base used for allocating GPU arrays as discussed previously ($\mu$) \\ \hline
     \textit{max-epk} & reads with events per base $\le$ \textit{max-epk} are assigned to GPU, rest to CPU \\ \hline
    $K$ & upper limit of the batch size with respect to the number of reads \\ \hline 
    $B$ & upper limit of the batch size with respect to the number of bases \\ \hline 
    $t$ & number of CPU threads\\ \hline 
    \textit{ultra-thresh}  & threshold to skip \textit{ultra long reads}\\ \hline
     \end{tabular}
     \caption{adjustable user parameters}
     \label{t:adjustableparameters}
 \end{table}

Fig. \ref{f:mem_balance} shows the decision tree that detects any imbalance in the proportions $X$ and $Y$ associated with GPU arrays allocation ($X$ and $Y$ derived in equations \ref{e:X} and \ref{e:Y}). The objective of this decision tree is to detect any GPU memory wastage and to increase the number of reads which the GPU gets to process in parallel. 

As shown in Fig. \ref{f:mem_balance}, if both $X_{util}$  and $Y_{util}$ (\textit{rs} as a percentage of $X$ and \textit{es} as a percentage of $Y$ in Algorithm \ref{a:memalloc}) are more than 70\%, the utilisation of GPU arrays is considered reasonable. Note that 70\% is an empirically determined value that provides adequate performance. If $X_{util}$ is reasonable (>70\%) and $Y_{util}$ is unreasonable (<70\%), we inspect for any significant imbalance between $X_{util}$ and $Y_{util}$ ($X_{util}$-$Y_{util}$>30\%). Such a significant gap suggests an under-utilisation, which should be remedied through the increase of \textit{max-epk} (the threshold at which over-segmented reads are offloaded to the CPU) or reducing \textit{Y} by decreasing \textit{average-epk} (node S1 in Fig. \ref{f:mem_balance}).  In contrast, if $Y_{util}$ is reasonable and the $X_{util}$ is unreasonable, the strategy is the opposite, i.e, either decrease \textit{max-epk} or increase \textit{average-epk} (follow up to the node S2 in Fig. \ref{f:mem_balance}).

 If both $X_{util}$ and $Y_{util}$ are less than 70\%, a likely cause is an inadequate batch  size to fill the GPU memory. The actual batch size ($n$,$b$) is determined by both $K$ and $B$ as stated previously. As shown in Fig. \ref{f:mem_balance}, we check which limit out of $K$ and $B$ was reached first. If both $n<K$ and $b<K$, the currently processed batch being the last batch in the dataset (end of input data reached) is the likely cause. Thus, no parameter tuning action is necessary. If $B$ was reached first ($n<K$ and not $b<B$), $B$ is the limitation and should be increased (S3 in Fig. \ref{f:mem_balance}). If $K$ was reached first (not $n<K$ and $b<B$), $K$ should be increased (S4 in Fig. \ref{f:mem_balance}). 
 %Both $n<K$ and $b<B$ is a very rare case out and is ignored, to be eventually detected by the subsequent batches.

Fig. \ref{f:load_balance} shows the decision tree for CPU-GPU workload balancing. For a particular batch, if the CPU takes significantly more time than the GPU, the decision tree first inspects whether the CPU is assigned with an excessive workload. An excessive workload on the CPU can be attributed by: an extensively over-sized batch size (in comparison to the available GPU memory), which results in a majority of the reads being assigned to the CPU (N\textsubscript{memout}>10\%); excessive number of \textit{very long reads} assigned to the CPU (N\textsubscript{long}>10\%); and, excessive number of over-segmented reads events assigned to the CPU (N\textsubscript{events}>10\%). If N\textsubscript{memout}>10\%, $K$ is reduced (node T1 in Fig. \ref{f:load_balance}); if N\textsubscript{long}>10\%, \textit{max-lf} is increased (T2 in Fig. \ref{f:load_balance}); and, if N\textsubscript{events}>10\%,  \textit{max-epk} is increased (T3 in Fig. \ref{f:load_balance}).

If the cause for higher CPU time is not the aforementioned excessive workload, a likely cause is \textit{ultra long reads}, where a single \textit{ultra long reads} processed on the CPU taking more time than the time taken by GPU for the whole batch. In such an event, \textit{ultra-thresh} threshold is reduced so that more \textit{ultra long reads} are skipped. Another likely cause is that the program was executed with inadequate threads (if the CPU had more hardware threads than the program was launched), which is to be remedied by increasing the number of CPU threads. Another cause might be that the CPU is not sufficiently powerful to match with the GPU and thus no action can be taken (except upgrading the CPU). These actions are denoted by T4 in Fig. \ref{f:load_balance}.

The ideal case is when the CPU and GPU take similar times which requires no intervention. 

Conversely, if the GPU takes significant time than the CPU, the likely causes are  \textit{very long reads} or over-segmented reads. In such event, the thresholds \textit{max-lf} and \textit{max-epk} are decreased so that more \textit{very long reads} and over-segmented reads
are assigned to the CPU. Another likely cause is the \textit{ultra long read} which can be remedied by increasing \textit{ultra-thresh} threshold. Another cause might be an insufficiently powerful GPU (less compute cores or less memory) compared to the CPU and no action is taken (except to upgrade the GPU).

%one reason might be that many reads are processed in the CPU, as a result of GPU memory (X or Y arrays) being full. If it is the case (number of reads processed on the CPU due to GPU memory being full is grater than 5\% of the batch size), we check the utilisation of X and Y arrays. If X arrays are full, while Y arrays are comparatively under-utilised (lesser than a threshold 80\% in the figure), it might be either due to too many long reads occupying the space or the actual average events per base value is lower than the value we used $\mu=2.5$. So if the number of long reads that were assigned to the CPU is higher than a threshold (5\% of the batch size in figure), it can be assumed that the data-set contains longer reads than usual and thus 

To reduce false positives due to incidental under utilisation, a suggestion is provided to the user, only if the same condition (condition that led to the decision in the decision tree, S1 to S4 T1 in Fig. \ref{f:mem_balance} and T1 to T4 in Fig. \ref{f:load_balance}) consecutively repeats more than a few times (eg: >3 times).

Note that the above mentioned strategy is to warn and suggest of potential parameter adjustments in the event of drastic performance degradation, rather than to obtain optimal performance or to determine the exact parameter values.

\section{Results} \label{res}

 Experimental setup is given  in Section \ref{s:exsetup}. In Section \ref{s:reasons}, we present experimental evidence that justify the selection of steps presented in Section \ref{method}. Next in Section \ref{s:adaptiveres}, we compare the GPU implementation of the Adaptive Banded Event Alignment (ABEA) algorithm to its CPU implementation. Finally, we show the overall speedup of the GPU implementation when incorporated into an actual work-flow (i.e. detection of methylated bases).

\subsection{Experimental setup} \label{s:exsetup}

We re-engineered the \textit{Nanopolish} methylation calling tool (existing methylation detection tool discussed in Section \ref{background}) to: one, load a batch of \textit{n} reads from disk to RAM at a time, instead of on-demand loading; two, synchronise CPU threads prior to GPU kernel invocation (\textit{Nanopolish} assigns a thread dynamically to a particular thread, thus each read follows its own code path); and three, optimise the CPU implementation which otherwise would result in an apparent un-fair speedup (when the optimised GPU version is compared to an un-optimised CPU version). Re-engineered \textit{Nanopolish} employs a fork-join  multi-threading model (with work stealing) implemented using C POSIX threads. ABEA algorithm for the GPU was implemented using CUDA C. This re-engineered \textit{Nanopolish} will be hitherto referred to as \emph{f5c}.

% \begin{itemize}
% \item on-demand loading of signal data from files---a CPU thread assigned to the particular read invokes a file access just prior to signal alignment. Conversely we read to RAM and then bulk transfer to GPU memory, a batch of \textit{n} reads at a time. 
% %However, one by one PCI express transfer will bring no advantage. Instead have to transfer in a batch. Hence we had to re-write the Nanopolish processing framework in a such a way that loading and processing of a batch is performed batch wise. 
% \item thread model un-suitable for GPU acceleration---a thread is dynamically assigned to a read using openMP, thus each read has its own code path. However, offloading a batch of reads to the GPU for signal alignment requires code paths of all the reads in the batch to have converged before the GPU kernel is invoked. In addition, accurately measuring time, benchmarking and profiling of individual algorithmic components is hindered by such divergent code paths.
% %\textit{pthread} based approach that interleaves input reading, processing and output.
% \item not optimised for efficientresource  utilisation (eg: marginal performance improvement beyond 16 threads on servers and heavy-weight for embedded systems due to spurious \textit{malloc} calls)---a comparison of such a version with the GPU would result in an apparent high speedup,which is unfair.
% % C/C++
% % OpenMP
% % Lazy loading
% % Malloc
% % MIMD
% \end{itemize}

We used publicly available NA12878 (human genome) Nanopore WGS Consortium sequencing data \cite{jain2018nanopore} for the experiments. The datasets used for the experiments, their statistics (number of reads, total bases, mean read length and maximum read length) and their source are listed in Table \ref{t:datasets}.  D\textsubscript{small} which is a small subset, is used for running on a wide range of systems (all systems in Table \ref{t:systems}: embedded system, low-end and high-end laptops, workstation and high-performance server).  Two complete MinION data sets (D\textsubscript{ligation} and D\textsubscript{rapid}) are only tested on three systems due to large run-time and incidental access to the other two systems. D\textsubscript{ligation} and D\textsubscript{rapid} represent the two existing nanopore sample preparation methods (ligation and rapid \cite{libraryprep}) that affects the read length distribution.

\begin{table}[!t]
\footnotesize
\caption{Information of the datasets}
\begin{tabular}{|l|p{2cm}|p{2cm}|p{2cm}|p{2cm}|p{2.5cm}|}
\hline
\TstrutS
\textbf{Dataset} &   \textbf{Number of reads} & \textbf{Number of bases (Gbases)} & \textbf{Mean read length (Kbases)} & \textbf{Max read length \textbf{(Kbases)}} & \textbf{Source / SRA accession}\\ \hline
\TstrutS
D\textsubscript{small}        &  19275                                 &  0.15 & 7.7 &  196   & \cite{nanopolishmethlink}    \\ \hline
\TstrutS
D\textsubscript{ligation}     & 451020                                  & 3.62  & 8.0 &   1500   &  ERR2184733 \\ \hline
\TstrutS
D\textsubscript{rapid}        & 270189                                  & 2.73 & 10.0 & 386     & ERR2184734    \\ \hline
\end{tabular}
\label{t:datasets}
\end{table}

\begin{table}[!t]
\caption{Different systems used for experiments}
\footnotesize
%\begin{tabular}{|p{0.6cm}|p{1.5cm}|p{0.9cm}|p{0.6cm}|p{0.4cm}|p{0.7cm}|p{0.5cm}|p{0.7cm}|}
\begin{tabular}{|p{1cm}|p{2.5cm}|p{2.5cm}|p{1cm}|p{1cm}|p{1cm}|p{1cm}|p{1cm}|}
\hline
\Tstrut
\textbf{System Name} & \textbf{Info} & \textbf{CPU} & \textbf{CPU cores / threads} & \textbf{RAM (GB)} & \textbf{GPU}& \textbf{GPU mem (GB)} & \textbf{GPU arch}\\ \hline
\TstrutS
SoC	&	NVIDIA Jetson TX2 embedded module	&	ARMv8 Cortex-A57 + NVIDIA Denver2	&	6 / 6	&	8 	&	Tegra	&	shared with RAM	& Pascal / 6.2\\ \hline
\TstrutS
lapL	&	Acer F5-573G laptop	&	i7-7500U	&	2/4	&	8 	&	Gefore 940M	&	4  & Maxwell / 5.0	\\ \hline
\TstrutS
lapH	&	Dell XPS 15 laptop 	&	i7-8750H	&	6/12	&	16 	&	Gefore 1050 Ti	&	4  & Pascal / 6.1	\\ \hline
\TstrutS
%wsL	&	Custom Workstation	&	i7-6700K	&	4/8	&	32 GB	&	Tesla C2075	&	6 GB	\\ \hline
ws	&	HP Z640 workstation	&	Xeon E5-1630	&	4/8	&	32 	&	Tesla K40	&	12 & Kepler / 3.5	\\ \hline
\TstrutS
HPC	&	Dell PowerEdge C4140	&	Xeon Silver 4114	&	20/40	&	376 	&	Tesla V100	&	16  & Volta / 7.0	\\ \hline

\end{tabular}
\label{t:systems}
\end{table}

D\textsubscript{small} dataset was used for experiments under Sections \ref{s:memoptireasons}, \ref{s:computeoptires} and \ref{comparisonsaccorssdevices}. For experiments under Sections \ref{s:bigdatasets} and \ref{s:compnano}, the datasets D\textsubscript{rapid} and D\textsubscript{ligation} were used.

To obtain the results for Section \ref{s:heterorres}, first we grouped the reads in dataset D\textsubscript{rapid} based on their read lengths. We grouped the read into 10 Kbases bins (i.e., 0K-10K,10K-20K...90K-100K). Reads with >100 Kbases were grouped into larger bins (100K bin sizes; 100K-200K, 200K-300K and 200K-300K) as the read count is very little in the range that certain 10K bins would contain no reads at all. Then, we ran \textit{f5c} with only CPU and \textit{f5c} with GPU acceleration on each group of the reads separately. Then, we computed the speedup of ABEA for each group of reads: the kernel only speedup (\textit{GPU kernel time / time on CPU}); and, the speedup with overheads (overheads such as memory copy, data structure serialisation). This experiment was performed on the system \textit{lapH}.

For Sections \ref{s:reasons} and \ref{s:adaptiveres}, time measurements were obtained by inserting \textit{gettimeofday} function invocations directly into the C source code. Total execution time and the peak RAM usage in Section \ref{s:compnano} were measured by running the \textit{GNU time} utility with the \textit{verbose} option.

\subsection{Effect of individual optimisations}\label{s:reasons}

\subsubsection{Compute optimisations}\label{s:computeoptires}

Fig. \ref{f:kerneldistrib} shows the time consumed by the three GPU kernels after applying the compute optimisation techniques discussed in Section \ref{computeopti}. Time taken by each of the three GPU kernels (\textit{pre-kernel}, \textit{core-kernel} and \textit{post-kernel}) is plotted for each different GPU. It is observed that the \textit{core-kernel}, which computes the dynamic programming table (compute-intensive portion), still consumes the majority of the GPU compute time. The \textit{pre-kernel} which performs data structure initialisation consumes much lesser time and shows that there is no  need to further parallelise the loop in Algorithm \ref{adaptivebandedcudapre}  (explained in Section \ref{computeopti}). Despite the lack of fine-grained parallelism in \textit{post-kernel} (which performs backtracking), the elapsed time is still considerably lesser than the \textit{core-kernel}. Thus, any future optimisations should still mainly focus on the \textit{core-kernel}, followed by the  \textit{post-kernel}.

The efficacy of our compute optimisations on the compute intensive \textit{core-kernel} can be elaborated using the reported statistics from the NVIDIA profiler (instruction level profiling---PC sampling in NVIDIA visual profiler \cite{cudaprof}). The profiler reports the percentage distribution of reasons that caused the thread warps to stall, based on the number of clock cycles. The percentage of the number of clock cycles that a warp was stalled due to a memory dependency (waiting for a previous memory accesses to complete), improved from 59.10\% to 44.81\% after the use of GPU shared memory. After exploiting the \textit{kcache} for improving memory coalescing, this percentage further improved to 28.62\%.

%\todo{TODO : with constant cache}
%\todo{how latency samples are improved}

%\todo{Note about the inefficiency  of using the dynamic parallelism features - kernel overhead is not worthwhile compared to the amount of processing. We tested this approach initially and the performance gain was marginal. Hence a 2D thread model with suitable thread limiters is more efficient.}
%\todo{how occupancy is improved} 
%\todo{Final kernel time distributions explain}

%\todo{This table is generated via \laterurl{https://unsw-my.sharepoint.com/:x:/g/personal/z5136909_ad_unsw_edu_au/EUjJ_WMtfv5JqQcje4rh2dABAOM2ccPhc9P2QCvEV7GnaQ?e=LPCPjm}}

\subsubsection{Memory optimisations} \label{s:memoptireasons}

%Fig. \ref{perf1}a shows the effect of data serialisation (section \ref{memserialise}). A speedup of 2-3X is observed in GPU signal alignment (left bars of the graph) compared to the CPU (right bars). Note that this CPU version is optimised, which otherwise would have been much slower. 
As stated in Section \ref{memserialise}, the data array serialisation technique eliminated all memory allocations inside GPU kernels (\textit{malloc}); still, required memory allocations per each batch of reads (\textit{cudaMalloc}). The overhead due to these \textit{cudaMalloc} calls are plotted in Fig. \ref{f:malloceffect} along with the time for kernel execution and data transfer to/from the GPU (using \textit{cudaMemcpy}). Observe that on certain GPUs (Jetson TX2, GeForce 940M and Tesla K40), the overheads due to \textit{cudaMalloc} operations are significant in comparison to the compute kernels (even higher  than the compute kernels in Jetson TX2). Such significant overheads justify why we proposed a heuristic based memory pre-allocation technique (Section \ref{heuristicmem}) which completely eliminates this overhead.

Interestingly, Tesla K40 and Gefore 940M which incurred high \textit{cudaMalloc} overheads are of relatively older GPU architectures in comparison to GeForce 1050 and Tesla V100, where the overheads were minimal. This is probably due to hardware supported memory allocation in latest GPU architectures. 
However, the aforementioned observation seems to be valid only for GeForce GPUs (targeted for gaming on PC/laptops) and Tesla GPUs (targeted for high performance computing). On Tegra GPUs (SoC targeted for embedded devices) the overhead seems to be significant in spite of the latest architectures (Jetson TX2 is the same Pascal architecture as GeForce 1050).
We additionally tested on a Jetson AGX Xavier (the most recent Tegra GPU based SoC --- Volta architecture) and \textit{cudaMalloc} was yet expensive (40s on GPU kernels and 44s on \textit{cudaMalloc}, not shown in figure). Thus, our memory pre-allocation strategy (in Section \ref{heuristicmem}) which totally eliminates this \textit{cudaMalloc} overhead is specifically beneficial for GPU on SoCs.

%Pascal and Volta respectively, still the cudaMalloc takes  Older devices Kepler K40 (CUDA 3.5 - Kepler Architecture) and Gefore 940M (cuda 5.0 - Maxwell architecture) spent considerable time for memory allocation. 

%\todo{Graphs should be generated via \laterurl{https://unsw-my.sharepoint.com/:x:/g/personal/z5136909_ad_unsw_edu_au/EcTvu_PBxh5IgNBEZjVpdT8BN-g5q2s0Q0YJrqvXZITFcA?e=8M1zTX}}

\subsubsection{Heterogeneous processing}\label{s:heterorres}

\begin{figure}[!ht]

\begin{subfigure}[!t]{0.495\textwidth}
  \centering
    \includegraphics[width=\textwidth]{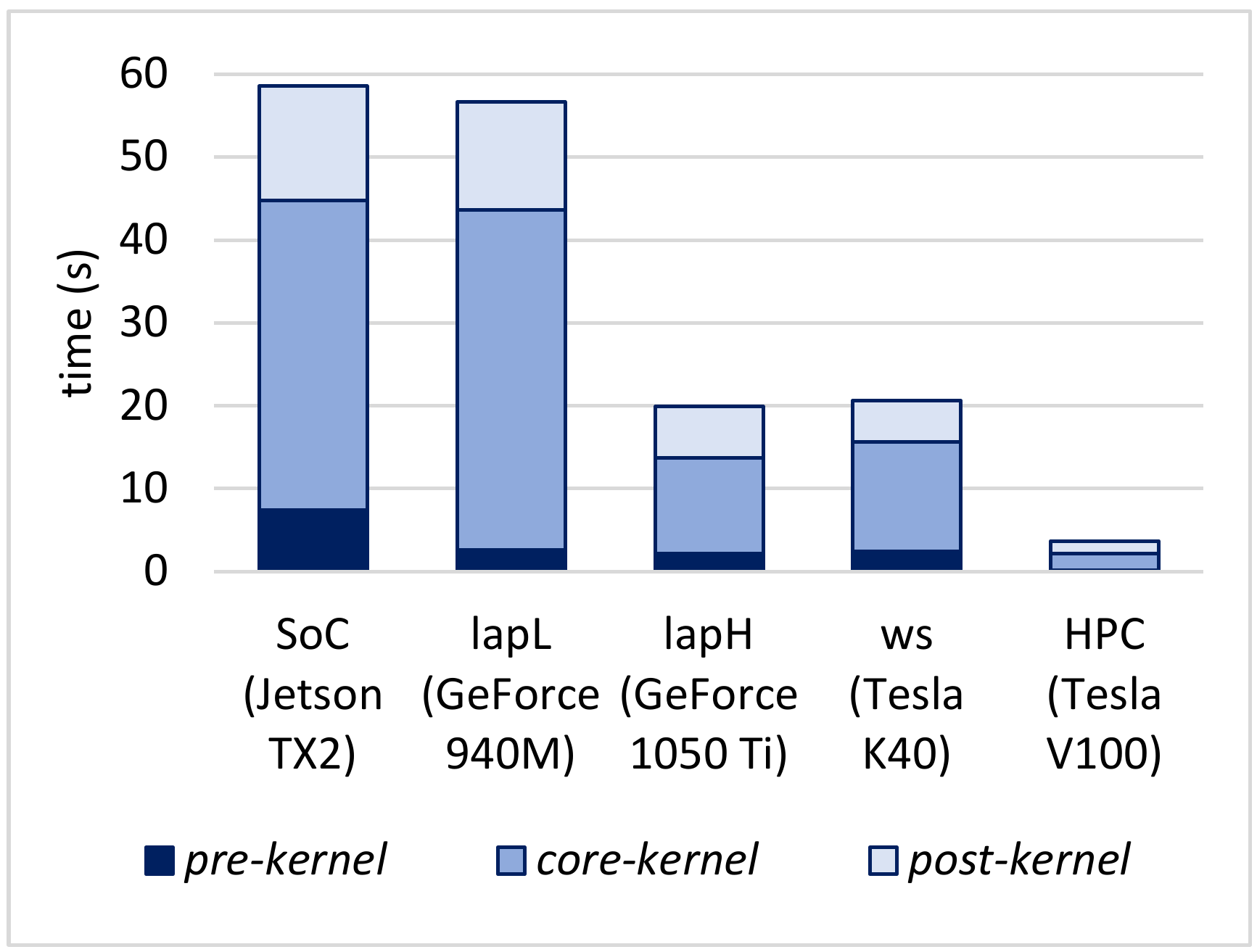}
    \caption{Distribution of GPU kernel execution time}
    \label{f:kerneldistrib}
\end{subfigure}
  \centering
\begin{subfigure}[!b]{0.495\textwidth}
  \centering
    \includegraphics[width=1.02\textwidth]{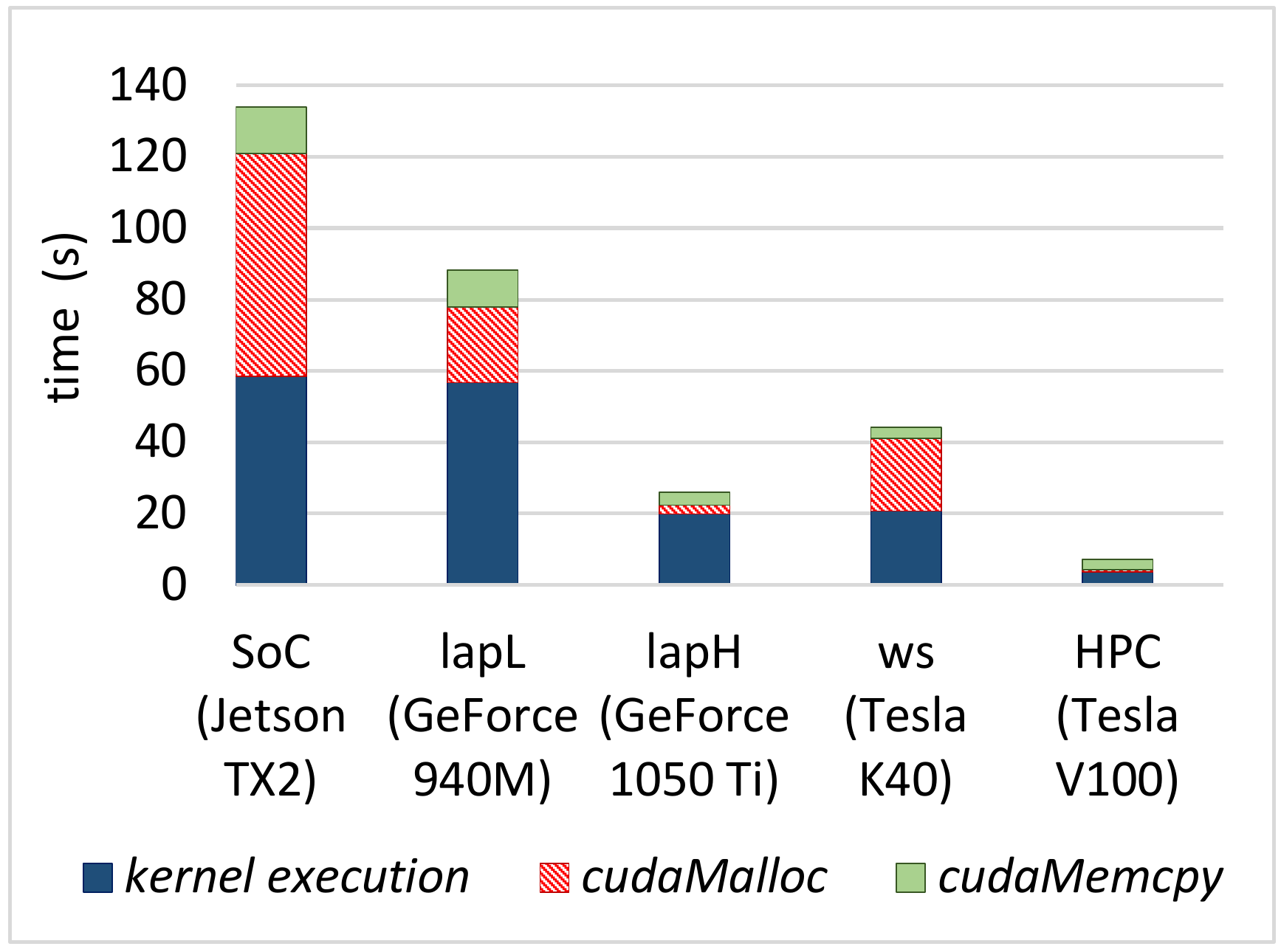}
    \caption{Time for GPU kernels compared \textit{cudaMalloc}} 
    \label{f:malloceffect}
\end{subfigure}

\vspace{0.5cm}

\begin{subfigure}[!ht]{0.495\textwidth}
  \centering
    \includegraphics[width=\textwidth]{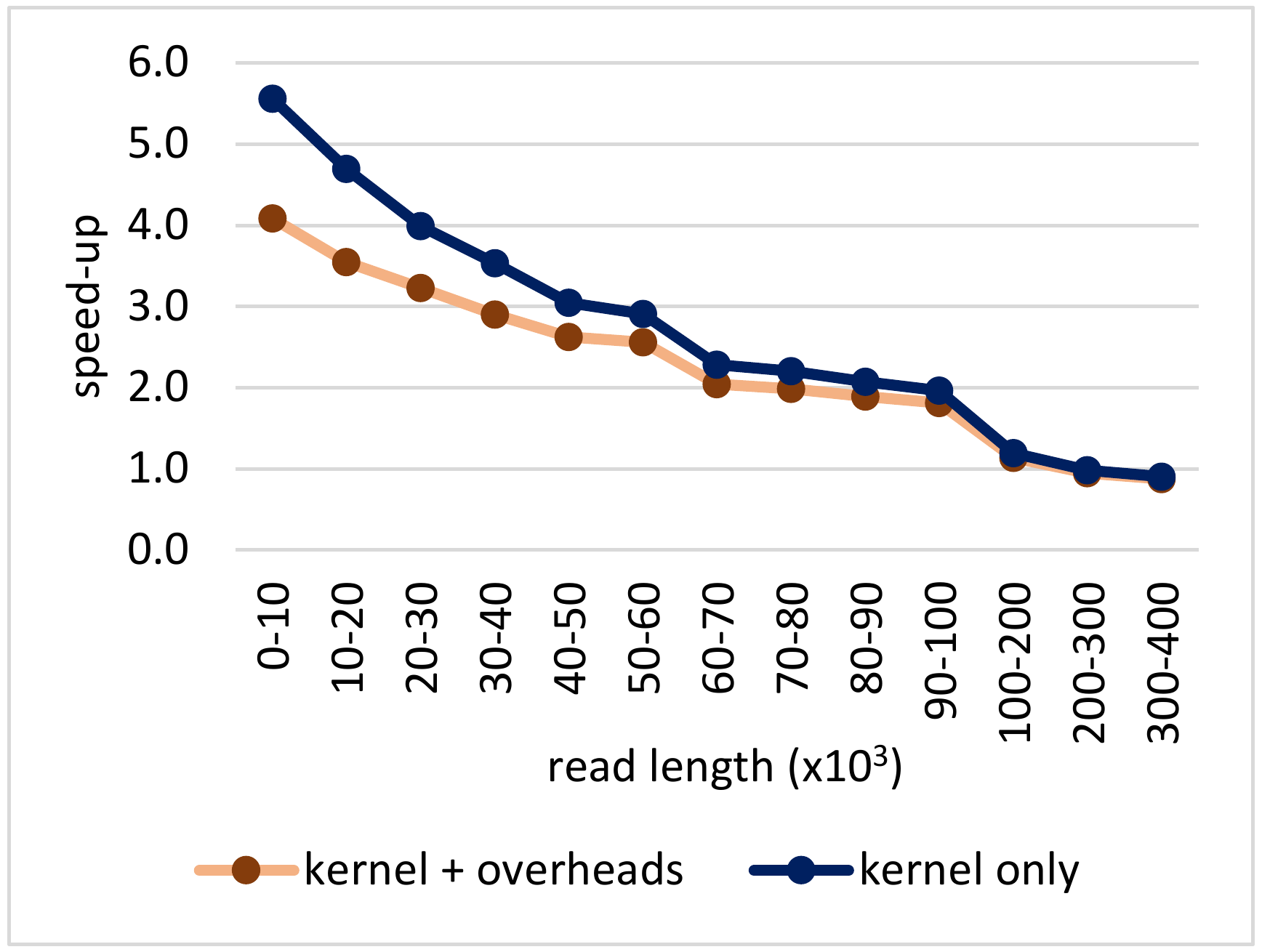}
    \caption{Effect of the read length on the speedup} 
    \label{f:readlen_speedup}
\end{subfigure}
\begin{subfigure}[!ht]{0.495\textwidth}
  \centering
    \includegraphics[width=1.02\textwidth]{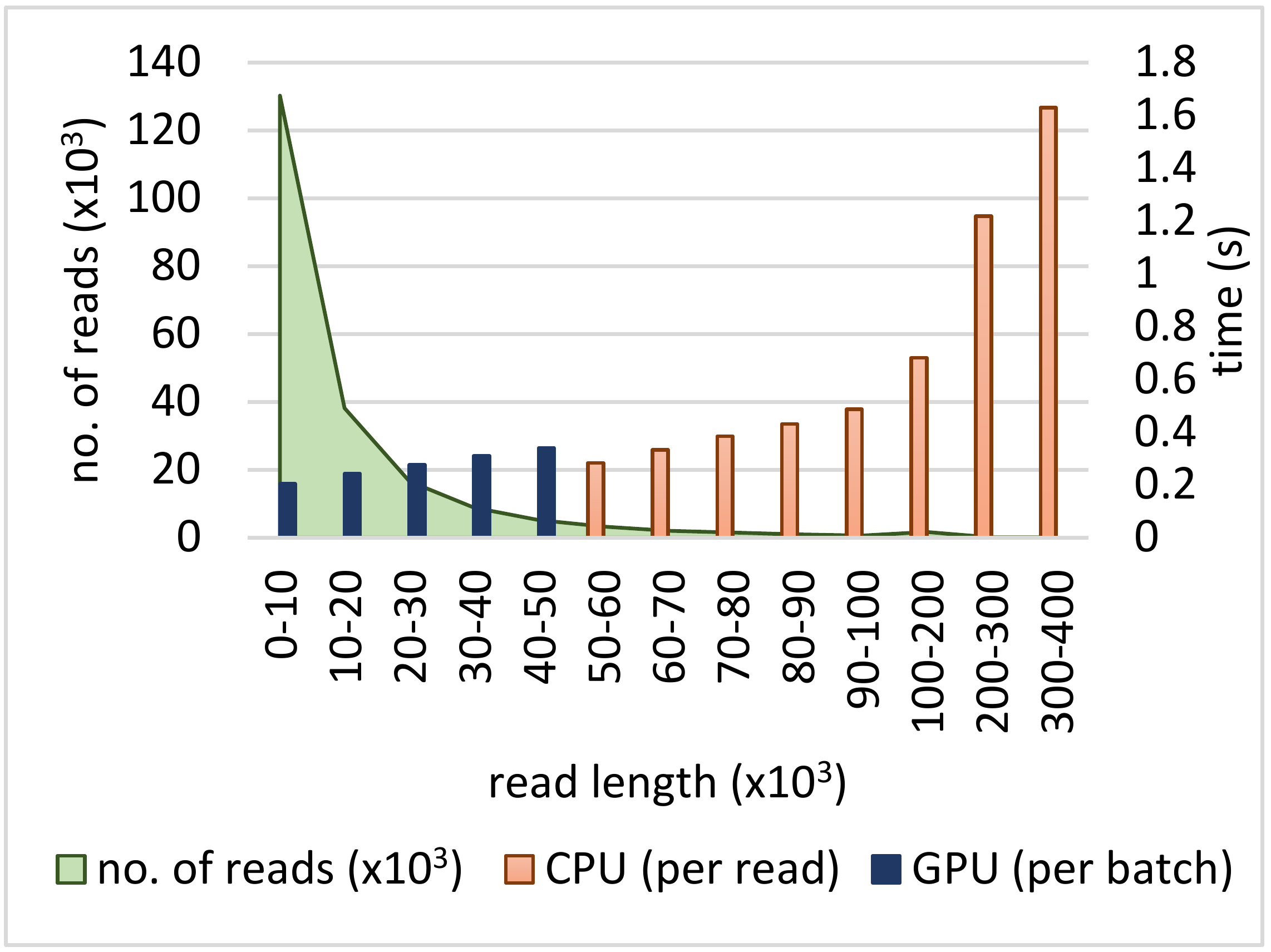}
    \caption{need for load-balancing based on read lengths} 
    \label{f:readlen_kernels}
\end{subfigure}

\caption{Effect of individual optimisations}
\end{figure}

We stated in Section \ref{loadbalmet} that \textit{very long reads} if processed on the GPU, limits the GPU occupancy. 
Fig. \ref{f:readlen_speedup} provides experimental evidence and shows the need to process \textit{very long reads} on CPU (explained in Section \ref{loadbalmet}).
Fig. \ref{f:readlen_speedup} plots the variation of the speedup (GPU compared to CPU for ABEA) as the read length varies. The x-axis labels the range of the read length for which the speedup was computed (explained in the experimental setup). For instance, 0-10 on the x-axis refers to the group of reads with read length 0-10Kbases. Note that in Fig. \ref{f:readlen_speedup} the bins are 100K wide from 100K-200K on-wards, due to less number of reads of those lengths (explained in the experimental setup). The speedup of \textit{computations} (GPU kernel time / CPU time) and the speedup including  \textit{overheads} (GPU kernel time + overheads such as memory copy, data structure serialisation) are plotted in Fig. \ref{f:readlen_speedup}. Speedup of more than 4X was observed for smaller read lengths (0-10K). speedup drops with increasing read-length and is less than 3X from 50K-60K.  The longer the reads are, the lesser number of reads can be processed in the GPU in parallel (reduced occupancy), thus the reduced speedup. Hence, \textit{very long reads}  that significantly affects the performance should be performed on the CPU while the GPU is processing the rest.
%To compute the speedups (y axis of the figure) the reads (in data-set D\textsubscript{rapid}) were partitioned based on the read length to the groups labelled on the x axis of the figure.

%We stated in section \ref{loadbalmet} that \textit{very long reads} if processed on the GPU, would cause GPU cores to idle due to waiting. Fig. \ref{f:readlen_kernels}  provides the evidence. 
 %\todo{0-10k has xxx reads and GPU takes xxs to process a batch of reads. However, with read length increasing the read count decreases, but the GPU time for a batch increases. For instance at 40-50k the read count is xxx, but to process a batch GPU takes yy time. The GPU times for the rest further increases with the rlen (not shown on graph).  In the graph were run on groups of similar lengths. but in a real sample mixed. So the longest read in the batch will determine the GPU time. Hve to select a balance of read counts and performance. let us say 50k < on GPU and > CPU. Now CPU is doing 0k-When compared to the 40k-50k performedon GPU, a single 50k-60k takes less time than that on the GPU. However, for 90k-100k range the CPU takes more time for a single read that a whole GPU batch and become the limiting factor. This

Fig. \ref{f:readlen_kernels}  shows the need for processing \textit{ultra long reads} separately (explained in Section \ref{loadbalmet}). The x-axis in the figure is the read-length (similar to Fig. \ref{f:readlen_speedup}). The blue bars (with reference to the right y-axis) denote the average time consumed by the GPU to process a batch of reads (1.5 Mbases), for each group of read lengths from 0 bases to 50Kbases. The orange bars (with reference to the right y-axis) denote the average time consumed by the CPU (1 thread) to process a single read in the particular group of reads. The read length distribution (left y-axis) is shown  shaded in green colour to depict the abundance of reads in each read length. Observe that CPU takes >1.6s for a single read of 300K-400K length while the GPU completes a whole 40K-50K batch in <0.4s. Thus, the GPU would idle for >1.2s until the CPU completes processing. Hence, such \textit{ultra long reads} (eg : >100 Kbases) must be skipped and processed separately at the end. Note that such \textit{ultra long reads} are very few (green coloured read length distribution in Fig. \ref{f:readlen_kernels}).

% gets worse for ultra-long reads (read >100k). For instance there were
% very few reads in the range 300k-400k (only 20 reads compared to
% 130k for 0k-1k) , where the average time per read on CPU was 1.6s
% (compared to 0.2s for a whole batch on the GPU for 0k-10k range).
% If one single such read was present the whole gain on processing a
% whole batch in the GPU is gone—hence we have to process them
% separately later.

%

% Time for a single read on CPU is less than for a whole batch on GPU before 70k.

% However, a single read on CPU takes more time than a batch on GPU after 70k. This idle time gets worse for ultra-long reads (read >100k), for instance in  thus with no performance gain. 

% 

%\todo{These figures are outdated, update from \laterurl{https://unsw-my.sharepoint.com/:x:/g/personal/z5136909_ad_unsw_edu_au/EZ28TP53PaRLtaTpmwt7vxcBBx9wyWvZnZtrR3FgdymDFg?e=E5qC6o}}

\subsection{Speedup of Adaptive Banded Event Alignment} \label{s:adaptiveres}

In this subsection, we present the performance of the GPU ABEA implementation when all the optimisations  in Section \ref{method} are applied together. Note that we compare this optimised GPU version with optimised CPU version in \textit{f5c} (not the CPU version in original \textit{Nanopolish}). The CPU version was run with maximum supported threads on the system. The optimised CPU version will be hitherto referred to as \textit{CPU-opti} and the optimised GPU version will be referred to as \textit{GPU-opti}. First, we compare the run-time of \textit{CPU-opti} and \textit{GPU-opti} on a wide range of different computer systems in Section \ref{comparisonsaccorssdevices}, and then on the two big datasets in Section \ref{s:bigdatasets}.

\subsubsection{Across different devices}\label{comparisonsaccorssdevices}

\begin{figure}[!ht]
  \centering
\begin{subfigure}[!ht]{.495\textwidth}
  \centering
    \includegraphics[width=\columnwidth]{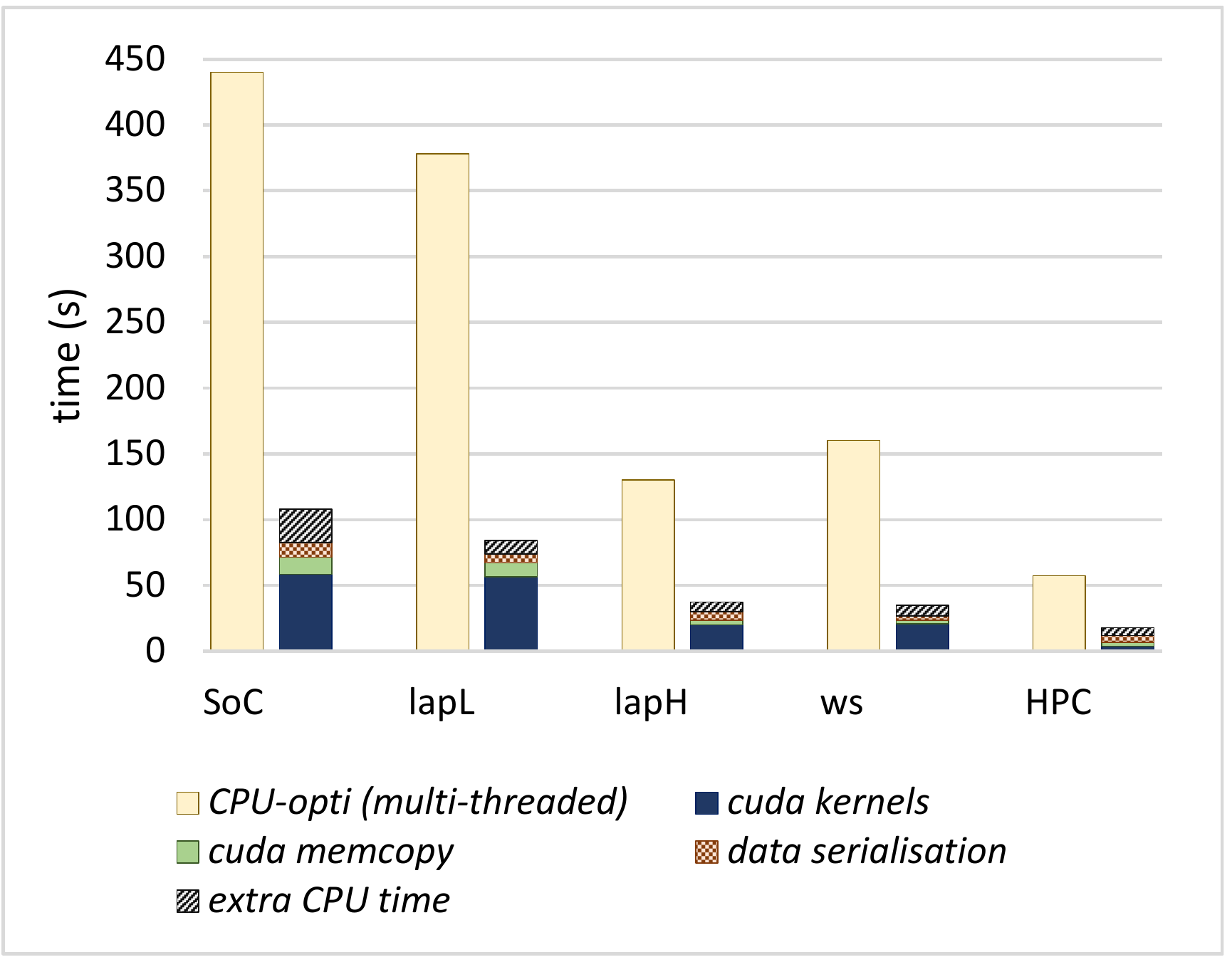}
    \caption{Performance comparison of ABEA on CPU vs GPU for D\textsubscript{small} over a wide range of systems} 
    \label{f:alignment_speedup}
\end{subfigure}
\begin{subfigure}[!ht]{.495\textwidth}
  \centering
    \includegraphics[width=\columnwidth]{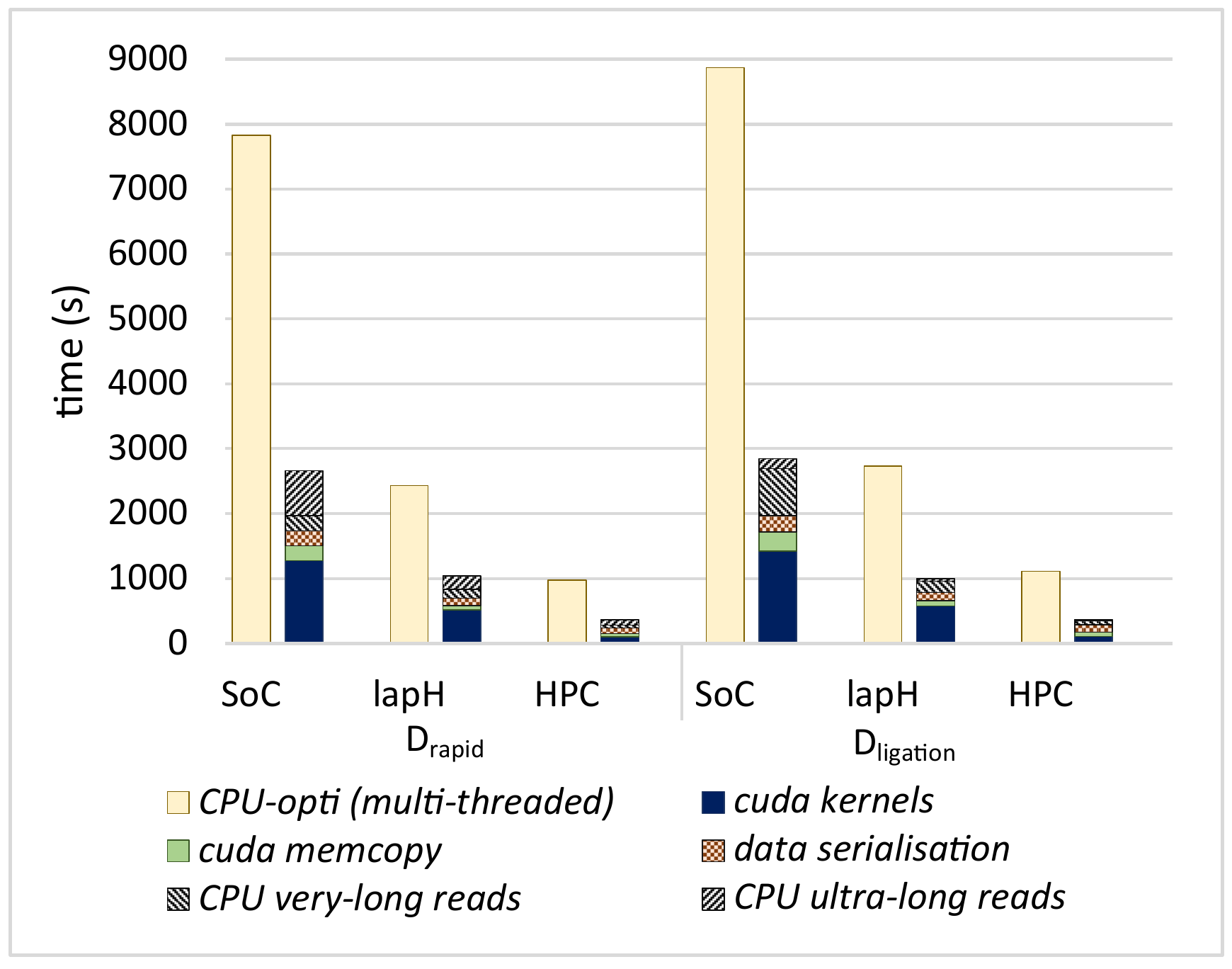}
    \caption{Performance comparison of ABEA on CPU vs GPU across full datasets} 
    \label{f:alignment_speedup_datasets}
\end{subfigure}

\caption{Speedup of ABEA on GPU compared to CPU}

\end{figure}

Fig. \ref{f:alignment_speedup} shows the time for \textit{CPU-opti} (left bars) and the \textit{GPU-opti} (right bars) for the Dataset D\textsubscript{small}, for each system listed in Table \ref{t:systems}. The run-time for the GPU has been broken down in to:  compute kernel time; different overheads (memory copying to/from the GPU, data serialisation time); and, the extra CPU time due to reads processed in the CPU. The compute kernel time includes the sum of time for all the three kernels (\textit{pre-kernel}, \textit{core-kernel} and \textit{post-kernel}). The extra CPU time is the additional time spent by the CPU to process the reads assigned to the CPU (excluding the processing time that overlaps with the GPU execution, i.e. only the extra time which the GPU has to wait after the execution is included). Note that the \textit{ultra long reads} were not separately processed on the CPU as the D\textsubscript{small} contains a minuscule number of \textit{ultra long reads}.

%processing time for\textit{exceptionally long reads} which are performed separately on the CPU  (stated in section \ref{loadbalmet}). 
 Speedups (including all the overheads) observed for \textit{CPU-opti} compared to \textit{GPU-opti} are: \textasciitilde4.5$\times$ on the low-end-laptop and the workstation; \textasciitilde4$\times$ on Jetson TX2 SoC; and \textasciitilde3$\times$ on high-end-laptop and HPC. Note that only \textasciitilde3$\times$ speedup on high-end-laptop and HPC (in comparison to >=4$\times$ on other systems) is due to the CPU on those particular systems having comparatively a higher number of CPU threads (12 and 40 respectively).

%\todo{Graphs to be generated via \laterurl{https://unsw-my.sharepoint.com/:x:/g/personal/z5136909_ad_unsw_edu_au/EUcYy--YKDFBg_Un7BwlBakBHQQNfOH802-Epq4APWhKUg?e=SmRZCM}}

\subsubsection{Benchmark on big datasets}\label{s:bigdatasets}

Time taken for \textit{CPU-opti}  compared to \textit{GPU-opti} for the two big datasets (D\textsubscript{ligation} and D\textsubscript{rapid}) is shown in Fig. \ref{f:alignment_speedup_datasets}.  Experiments were performed only on three systems due to the limited availability of other devices (mentioned previously). The graph is similar to the previous Fig. \ref{f:alignment_speedup}, except the \textit{extra CPU time} has been further broken down to: \textit{CPU} \textit{very-long reads}; and, \textit{CPU} \textit{ultra long reads}. \textit{CPU} \textit{very long reads} refers to the additional time spent by the CPU to process \textit{very long reads} and, \textit{CPU} \textit{ultra long reads} refer to the \textit{ultra long reads} (reads >100 Kbases) processing time performed separately on the CPU. A speedup up of \textasciitilde3$\times$ was observed for all three systems. Due to more \textit{ultra long reads} in D\textsubscript{ligation} and D\textsubscript{rapid} than in D\textsubscript{small}, the overall speedup for \textit{SoC} is limited to around \textasciitilde3$\times$ compared to \textasciitilde4$\times$ for D\textsubscript{small}.

\subsection{Total run-time of \textit{f5c} compared with original \textit{Nanopolish}}\label{s:compnano}

In this section, we demonstrate the overall performance when the GPU accelerated ABEA is incorporated into an actual methylation detection work-flow. As stated in the experimental setup, we re-engineered \textit{Nanopolish} to overcome the limitations of original \textit{Nanopolish}. We compare the total run-time for methylation calling using original \textit{Nanopolish} against \textit{f5c} (both CPU only and GPU accelerated versions).

We refer to original \textit{Nanopolish} (version 0.9) as \textit{nanopolish-unopti}, \textit{f5c} run only on the CPU as \textit{f5c-cpu-opti} and  GPU accelerated \textit{f5c} as \textit{f5c-gpu-opti}.
We executed \textit{nanopolish-unopti}, \textit{f5c-cpu-opti} and \textit{f5c-gpu-opti} for the full datasets D\textsubscript{rapid} and D\textsubscript{ligation}. Note that all  the executions were performed with the maximum number of CPU threads supported on each system.

\begin{figure}[!ht]
  \centering
    \includegraphics[width=\columnwidth]{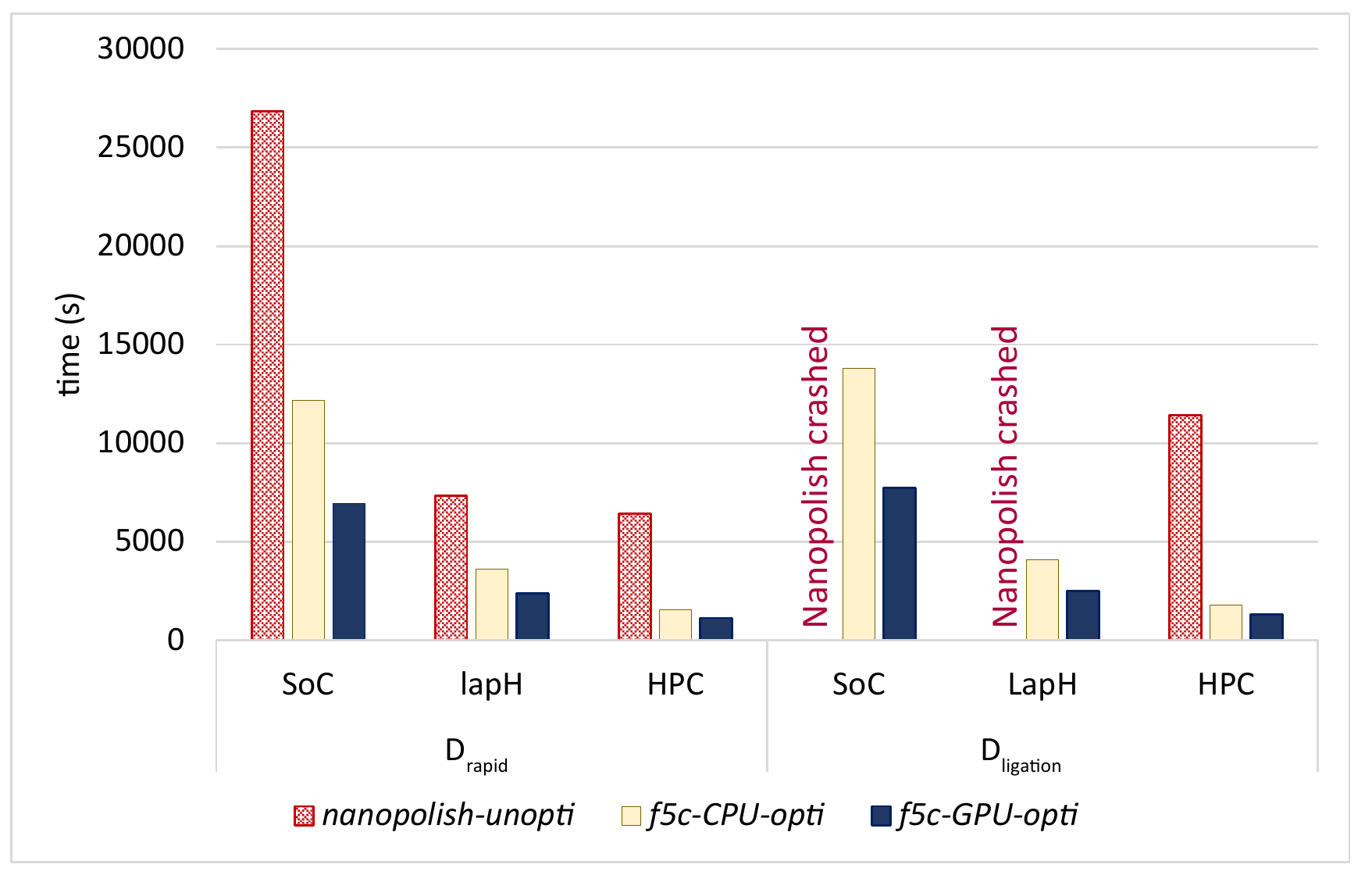}
    \caption{Comparison of \textit{f5c} to \textit{Nanopolish}} 
    \label{f:compare_nano}
\end{figure}

The run-time results are shown in Fig. \ref{f:compare_nano}.
The reported run-times are for the whole methylation calling (all steps mentioned in Section \ref{methcall}) and also includes disk I/O time. As each read executes on its own code path in original \textit{Nanopolish} (as mentioned in the experimental setup) the time for individual components (eg: ABEA) cannot be accurately measured, thus we only compare the total run-times.

\textit{f5c-cpu-opti} for D\textsubscript{rapid} dataset was: \textasciitilde2$\times$ faster on SoC and lapH; and, \textasciitilde4$\times$ faster on HPC. \textit{nanopolish-unopti} crashed on SoC (8GB RAM) and lapH (16GB RAM) when run for D\textsubscript{ligation} dataset due to Linux Out Of Memory (OOM) killer \cite{oomkiller}. When run for D\textsubscript{ligation} on HPC, \textit{f5c-cpu-opti} was not only 6$\times$ faster than original \textit{Nanopolish}, but also consumed only \textasciitilde15 GB RAM opposed to >100 GB by original \textit{Nanopolish} (both run with 40 threads). Hence, it is evident that CPU optimisations alone can do significant improvements. 

As per Fig. \ref{f:compare_nano} for the whole methylation-calling process (including disk I/O),  \textit{f5c-gpu-opti} (only ABEA is performed on GPU) compared to \textit{f5c-cpu-opti} was 1.7$\times$ faster on SoC, 1.5-1.6$\times$ on the lapH and <1.4$\times$ on HPC. On HPC the speedup was limited to <1.4$\times$ due to file I/O being the bottleneck.

When the execution time of \textit{f5c-gpu-opti} for D\textsubscript{rapid} is compared with original \textit{Nanopolish}, it is \textasciitilde4$\times$, \textasciitilde3$\times$ and \textasciitilde6$\times$ faster on SoC, laptop and HPC, respectively. On HPC for D\textsubscript{ligation}, \textit{f5c-gpu-opti} was \textasciitilde9$\times$ faster.

Note that we used \textit{Nanopolish} v0.9 for comparison as the re-engineering was done on this particular version. As we incorporated a number of CPU optimisations identified during the re-engineering into the subsequent version of \textit{Nanopolish} (only those that did not require major code refactoring), latest \textit{Nanopolish} v0.11 should be faster than v0.9 used in this paper.

\section{Discussion} \label{discussion}

\begin{figure}[!ht]
  \centering
    \includegraphics[width=\columnwidth]{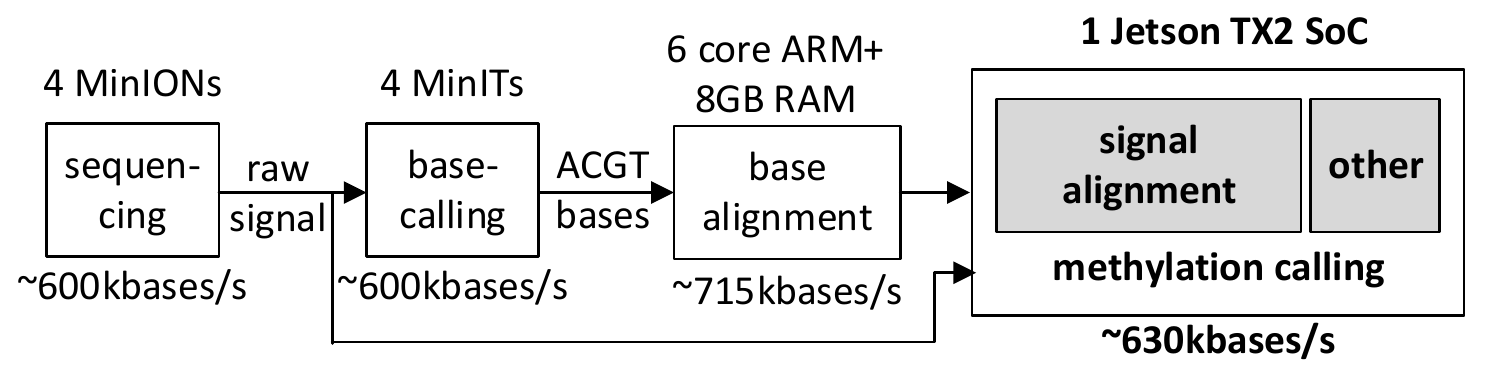}
    \caption{Human genome processing on-the-fly} 
    \label{f:onfly}
\end{figure}

%(a beta stage PromethION flow cells is equivalent to the capacity of around four minIONs and thus a single minION with such a capacity is foreseeable in the near future)
With the method discussed in this paper,  the complete methylation calling of a human genome can now be performed on-the-fly (process in real-time while the nanopore sequencer is operating) on an embedded system (e.g., an SoC equipped with ARM processor and an NVIDIA GPU) as shown in  Fig. \ref{f:onfly} (four Oxford Nanopore MinION devices sequencing in parallel or a single Oxford Nanopore GridION, is capable of sequencing a human genome at an adequate coverage). \textit{f5c} powered by GPU accelerated ABEA can process the output from the rest of the pipeline on a single NVIDIA TX2 SoC, at a speed of (>600 Kbases per second) to keep up with the sequencing output (\textasciitilde600 Kbases per second \cite{minitout}) as shown in Fig. \ref{f:onfly}. Conversely, if the original \textit{Nanopolish} was executed on the NVIDIA TX2 SoC, the processing speed is limited to \textasciitilde256 Kbases per second. Our work will not only reduce the associated costs of Nanopore data processing and data transfer, but will also improve turnaround time of the final test outcome. 

%Note that in  Fig. \ref{f:onfly} four Oxford Nanopore minION devices sequencing in parallel is capable sequencing a human genome at an adequate coverage yielding an output ~600 Kbases per second \cite{}. Base-calling can be performed on the GPU powered portable device Oxford Nanopore MinIT. A minIT connected to each minION can match the sequencing output as shown in Fig.~\ref{f:onfly}. 

In addition to embedded systems, our work benefits  systems with or without GPU. Due to reduced peak memory usage, methylation calling can be performed on laptops with <16GB of RAM. Furthermore, post sequencing methylation calling execution on high performance computers also benefit from a significant speedup in processing.

A limitation of our implementation is that the parameter tuning cannot be performed automatically, which instead prompts the user when an un-optimal parameter is detected. This limitation is expected to be addressed in a future version by automatically tuning the parameters at run-time; or, by the use of pre-set parameter profiles for different types of datasets and/or computer systems. 

%Furthermore, future work includes incorporating the GPU accelerated signal alignment into other Nanopore processing algorithms such as variant calling.  However, as we have shown there likely to be enough room for CPU optimisations and restructuring/re-engineering of software first. GPU optimisations unless followed by such CPU related optimisations would be superfluous. Though these changes and optimisations may not necessarily be sophisticated and cutting edge research, doing such changes to the software requires understanding the whole computer hardware and software stack in conjunction to the algorithms and characteristics of DNA data. 

% server with a large number of threads and performance gain, reduced peak memory usage. Future work : GPU accelerated signal alignment for variant calling, automate parameter tuning during the run time

% 

% The base-space alignment typically performed using the aligner Minimap2 \cite{minimap2} has shown to be capable of running on embedded devices with limited memory \cite{minimap2arm}, thus an ARM hexa-core processor with 8GB RAM can keep up with the sequencing output as shown on Figure. 

%Our version that supports both the CPU and the GPU is available at [\laterurl{https://github.com/hasindu2008/f5c/}]. 
The documentation of \textit{f5c} is in appendix \ref{a:f5c-documentation}. Supplementary material on the design, development and deployment of \textit{f5c} is available in appendix \ref{a:f5c-supps} and appendix \ref{a:potablebin}.

% \todo{Apart from these our re-implementation has many additional advantages}
% \begin{itemize}
% \item I/O and processing are interleaved: the I/O latency is considerably minimised.
% \item Our CPU version alone is around 1.5X-2X faster than the nanopolish call methylation implementation and is very lightweight - suitable for embedded systems : Due to the careful use of  data structures and algorithms
% \item Detect load balance problems between CPU and GPU, and report user with suggestion for appropriate parameters
% \item  Works with package manager's system wide installations of HDF5, hence no need to wait ages for HDF5 to be locally compiled
% Dependency hell has been minimised for both CPU and GPU version
% G++ 4.8 or higher, works even with CUDA toolkit 6.5. We have
% Suggestive error message for troubleshooting issues with GPU. 
% \item Pthread based thread framework  written in C that interleaves I/O with processing can be also used as a starting point for future Nanopore tools
% \item Allows the possibility of benchmarking section by section to identify the bottlenecks in the current state of the art Nanopore processing .
% \item The framework should be suitable for acceleration of core kernels through other methods such as FPGA.
% \item Again the ones that I forgotten by now.
% \item Other tools like tombo
% \item event alignment for var calling
% \item GPU cpy can be performed asynchronously
% \item automate performance
% \end{itemize}

\section{Summary}

Adaptive Banded Event Alignment algorithm is one of the key  components in nanopore data analysis. Despite this algorithm being not  embarrassingly parallel, we presented an approach that makes  this algorithm efficiently execute on GPUs. The high variability of the read lengths was one of the main challenges, which was remedied through a number of memory optimisations and a heterogeneous processing strategy that uses both CPU and GPU. Our optimisations yield around 3-5$\times$ performance improvement on a CPU-GPU system when compared to a CPU. We incorporated the optimised Adaptive Banded Event Alignment algorithm into a methylation detection workflow and demonstrated that an embedded SoC equipped with an ARM processor (with six cores) and NVIDIA GPU (256 cores) is adequate to process data from a portable nanopore sequencer in real-time.

This work not only benefits embedded SoC, but also a wide range of systems equipped with GPUs from laptops to servers. The re-engineered version of the \textit{Nanopolish} methylation detection module, \textit{f5c} that employs the GPU accelerated Adaptive Banded Event Alignment was not only around 9$\times$ faster on an HPC, but also reduced the peak RAM by around 6$\times$ times. The source code of \textit{f5c} is made available at \url{https://github.com/hasindu2008/f5c}.

%\section*{Acknowledgements}

%We thank NVIDIA for providing the Jetson TX2 board (to University of New South Wales) and Tesla K40 GPU (to University of Peradeniya) through the GPU donation programme. We thank Dr. Roshan Ragel and Dr. Swarnalatha Radhakrishnan at University of Peradeniya who assisted the research. We thank our colleagues who provided support, especially, James Ferguson and Shaun Carswell at Garvan Institute of Medical Research; Thomas Daniell, Hassaan Saadat and Darshana Jayasinghe at University of New South Wales; Geesara Pratap at Innopolis University; and, Pim Schravendijk. We also thank the Data Intensive Computer Engineering (DICE) team at Data Sciences Platform, Garvan Institute of Medical Research.
% %We thank Manuel Ballesteros, Derrick Lin and Warren Kaplan at Garvan Institute of Medical Research for providing access to the Tesla V100 equipped server blade.
%\todo{Whoever I forgot.}

%% file: 7-integration/main.tex
\chapter{System Integration} \label{c:integration}

\section{Introduction}\label{s:integration-intro}

This chapter presents how the different optimisations proposed in previous chapters are integrated to construct different prototype embedded systems that perform end-to-end DNA analysis workflows. 

In collaboration with two other PhD candidates in the research group, an embedded system called SWARAM was constructed for performing a variant calling pipeline for second-generation sequencing.  SWARAM consisted of 16 Odroid XU4 single board computers (SBC) interconnected through Ethernet. The optimisations to the \textit{Platypus} variant caller presented in chapter \ref{c:cacheopti} are applied in SWARAM to facilitate fast variant calling. However, the integration details and the architecture of SWARAM have been generously shared to be used in another PhD candidate's thesis and thus not discussed or claimed under this thesis. Refer to the published article at \cite{mohanty2019swaram} for those details. 

Inspired by SWARAM, another embedded system called the nanopore-cluster was constructed, now in a different architecture to SWARAM, to process third-generation nanopore sequencing data.  The optimisations proposed in chapter \ref{c:minimap} and \ref{c:gpuabea} were used in this nanopore-cluster to enable a real-time workflow for nanopore sequencing data. As mentioned in section \ref{s:3rdgenseq}, nanopore is a highly portable technology. Thus, an embedded like the nanopore-cluster system is harmonious with the ultimate goal of such ultra-portable sequencers to enable complete DNA sequencing in-the-field. Further, unlike second generation Illumina sequencers, third-generation nanopore sequencers allows streaming and thus the processing can be performed while sequencing. This streaming capability can be exploited in an embedded system like the nanopore-cluster, to  perform data analysis on-the-fly while the sequencer is operating, intending to produce the result soon after the sequencing run is completed.

\section{System Architecture of Nanopore-cluster} \label{s:nanopore-cluster-arch}

\subsection{Hardware Architecture} \label{s:nanopore-cluster-hw-arch}

The hardware architecture of the proposed system is in Fig. \ref{f:arch-nanopore-cluster}. The system comprised of the DNA Sequencer and the base-caller, Network Attached Storage (NAS) and the computational nodes (head node and the worker nodes) are interconnected using Ethernet via a layer 2 Switch. The system is interfaced with the Internet or the Intranet through a router (layer 3 switch) supporting Network Address Translation (NAT). The function and details of individual components are elaborated below:

\begin{figure}[!ht]
    \centering
    \includegraphics[width=\textwidth]{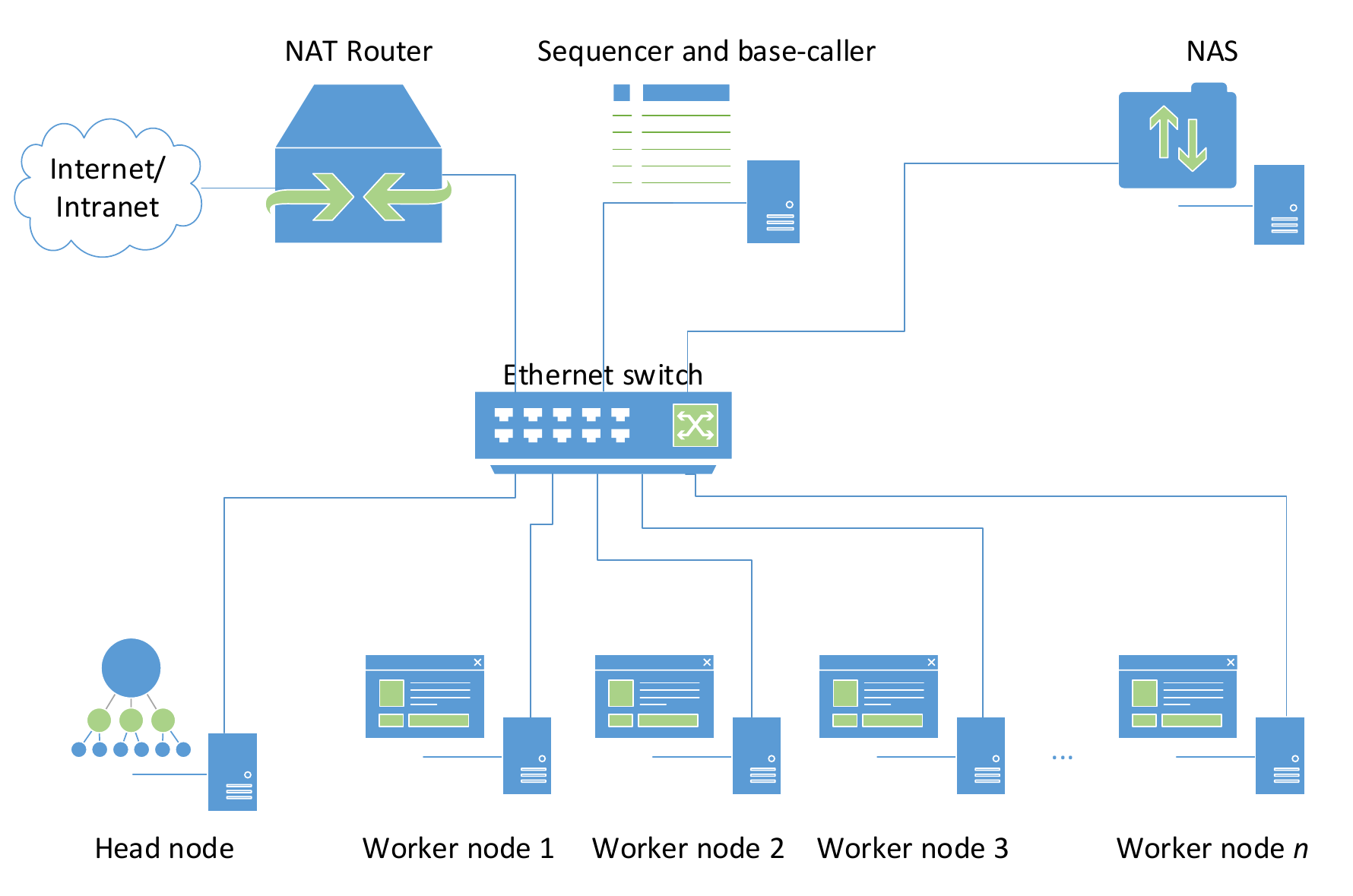}
    \caption{Hardware architecture of the proposed system}
    \label{f:arch-nanopore-cluster}
\end{figure}

\textbf{Sequencer and base-caller:} The sequencer can be one of the available nanopore sequencers - MinION, GridION or PromethION (Fig. \ref{f:nanopore-sequencers}). If the sequencer is a MinION or a PromethION, it must be connected to the corresponding base-calling unit (to the MinIT in case of the MinION or the compute tower in case of the PromethION; see Fig. \ref{f:ont-misc}). It is this base-calling unit  that is connected to the Ethernet switch through the available Ethernet interface on the case-calling unit. If the sequencer is a GridION or a MinION Mk1C, then the base-calling unit is integrated with the sequencer and has a direct Ethernet interface.

\textbf{NAS:} The NAS acts as the storage buffer between the base-caller and the computational nodes (head node and the worker nodes). The sequencer and base-caller produces a batch of data every few minutes or so, which is copied to the NAS. The computational nodes fetch these batches from the NAS into their local storage and process the data. The NAS can also be used as an archive for the sequencing data in case the raw data is later required. The NAS is not necessarily a dedicated hardware NAS, alternately it can be virtual, i.e., the secondary storage of the base-caller or the head node exposed as a network drive.

\textbf{Head node:}
The head node can be an SBC, a laptop  or even a desktop. The head node monitors the NAS and assigns processing jobs to the worker nodes. The head node is also responsible for the administration of worker nodes (controlling, updating software, deploying software). Note that the head node is not expected to be on high CPU load and thus the head node is extremely unlikely to freeze.

\textbf{Worker nodes:}
Worker nodes are SBCs. They are for processing the data and are controlled by the head node.

\textbf{NAT router:}
A router that supports NAT is not mandatory but is recommended. The Ethernet switch can be indeed connected to the local intranet or the Internet, however,
 administrators of centrally managed IT infrastructure may be reluctant due to potential risk of switching loops.  The NAT router streamlines the integration by hiding the Ethernet switch behind NAT.  The NAT router (usually comes with a built-in firewall) has additional benefits in terms of security. It can be configured to only allow limited inbound traffic (for instance, only particular ports on the head node only) while allowing outbound traffic for Internet access.

\subsection{Software Architecture}

\subsubsection{Overview}\label{s:arch-overview}

The NAS is mounted on the base-calling unit, head node and all worker nodes. The sequencer outputs the reads as raw signal data that are acquired  by the base-calling unit. When a batch of reads are accumulated, the base-caller performs base-calling and produces two files -- a multi-FAST5 file containing the raw signals (used to be a directory containing one FAST5 file per each reads last year) and a single-FASTQ containing the base-called reads. The batch size is by default 4000 as set by ONT. When the base-calling of the batch is completed, the multi-FAST5 file (if single-FAST5, a tarball of the directory containing single-FAST5 files) and the FASTQ file is copied to the NAS. 

The head node monitors the NAS for the recently copied data batches. Once such a fresh data batch is found, the worker node assigns the batch to a free worker node to be processed. If multiple worker nodes are free, the assignment is done randomly. If all worker nodes are busy, it will be assigned as soon as a worker node becomes free. 

A sequencing run lasts for 48 hours (MinION or GridION) or 64 hours (PromethION). The base-caller will continuously produce batches from the read data produced by the sequencer. The head node will continuously monitor and assign the work to worker nodes.

At the beginning of the sequencing run, all the pores in the flow-cell of the sequencer are functional and data batches are produced at a faster rate. With time, the pores slowly die and
 the rate of data batches produced will decrease.
 The objective of the proposed architecture is to finish processing soon after the sequencing run completes. At the beginning of the sequencing run when the sequencer outputs faster, there is no strict need for the worker nodes to keep-up processing at the same rate as the data is produced. Later when the sequencing rate decreases, the worker nodes can catch up.

\subsubsection{Challenges}

\textbf{Devices tend to occasionally freeze under high load.} Using an embedded system for nanopore data processing facilitates portability and is potentially cheaper due to the low cost of SBC compared to an expensive server. However, processing on such an embedded system is at the same time challenging due to the low reliability of those SBCs - i.e., they occasionally freeze when under high computational load potentially due to a bug in the operating system\footnote{This was observed for Rock64 devices we used in our experimental setup.}. Most of the time, the freeze is detected by the watchdog timer and the device reboots automatically. In rare cases, the device completely freezes until manually power reset.

\textbf{Dynamic workload and scalability.}
Sequencing yield differs between MinION, GridION and PromethION. Library preparation techniques and the quality of the flow-cell also affects the sequencing yield. The rate of the sequencing output also varies. The embedded system for processing should support such variations. Thus, the optimal number of SBC required to process the data on-the-fly varies and the system should be scalable to add or remove SBCs based on the requirement and the budget. Thus, the scheduling of the processing jobs should dynamically scale with available resources.

\textbf{Flexibility to support evolving workflows.}
Nanopore bioinformatics workflows evolve rapidly. The changes can be small as changing user-specified parameters to the program, moderate replacing of a particular tool with another, considerable as adding additional steps to the workflow or significant as using another workflow. The embedded system for processing should be flexible to support these imminent changes in the future.

The above challenges were overcome by the proposed strategy called \textit{f5p} which is detailed below.

\subsubsection{\textit{f5p} - Lightweight Scheduler and Failure Handler} \label{s:f5p}

\textit{f5p} is a lightweight job scheduler with integrated failure handling capability designed to overcome the above-mentioned challenges in a nanopore data processing embedded system.

\textit{f5p} is composed of two components, namely, \textit{f5pd} and \textit{f5pl} which are explained below.

\textbf{\textit{f5pd}:} \textit{f5pd} is the daemon program that runs on worker nodes. \textit{f5pd} is launched at the startup of the worker node and runs indefinitely while listening on a port. \textit{f5pd} accepts connections from the head node (\textit{f5pl} below) and receives job scheduling commands from the head node. Job scheduling command is the location of a shell script and the location of a data batch as the arguments. The shell script contains the commands for copying the data from the NAS to the local storage, executing the steps in the data processing workflow and copying the results back to the NAS. Once, a job scheduling command is received, \textit{f5pd} executes the job on the node and once the job is completed \textit{f5pd} sends the status (success or failure with an error code) back to the head node. Once the job is completed. \textit{f5pd} will continue to accept another job scheduling command.

\textbf{\textit{f5pl}:} \textit{f5pl} is the launcher that is run by the user on the head node. \textit{f5pl} accepts the pipeline shell script and configuration settings such as the directory path to be monitored and the IP addresses of the worker nodes to be used. First, \textit{f5pl} establishes connections to the worker nodes and copy the pipeline shell script to all the nodes. Then, \textit{f5pl} keeps monitoring the specified directory and when a batch of reads is available, \textit{f5pl} assigns the job to a free worker node.
If the rate of data batches produced by the sequencer is high, \textit{f5pl} will assign until all worker nodes are occupied. Then, \textit{f5pl} waits until a worker node becomes free and assigns the next waiting data batch accordingly. The process repeats until the end of the sequencing run completes. If a worker node is hung  (restarted by the watchdog timer), \textit{f5pl} waits until the worker node is alive and assigns the same data batch again. If N consecutive freezes occur, the worker node will be retired and will not be used for the rest of the sequencing run. In the rare case where a device is totally hung (not restarted by the watchdog), \textit{f5pl} will retire the dead node and will continue with the rest of the worker nodes.

\textit{f5p} is thus capable of handling failures due to unreliable SBC. Also, \textit{f5p} is capable of dynamically assigning the jobs based on the available worker nodes. As \textit{f5p}  accepts a shell script that can be easily customised by the user, it ensures flexibility. 

Administration tasks such as updating software, deployment of new software and configuration management of the worker nodes are done by an existing IT Automation software (e.g. \textit{Ansible})

\section{Experimental Setup}\label{s:integration-exp}

\subsection{Rock64-cluster Hardware Setup}

\begin{figure}[!ht]
    \centering
    \includegraphics[width=\textwidth]{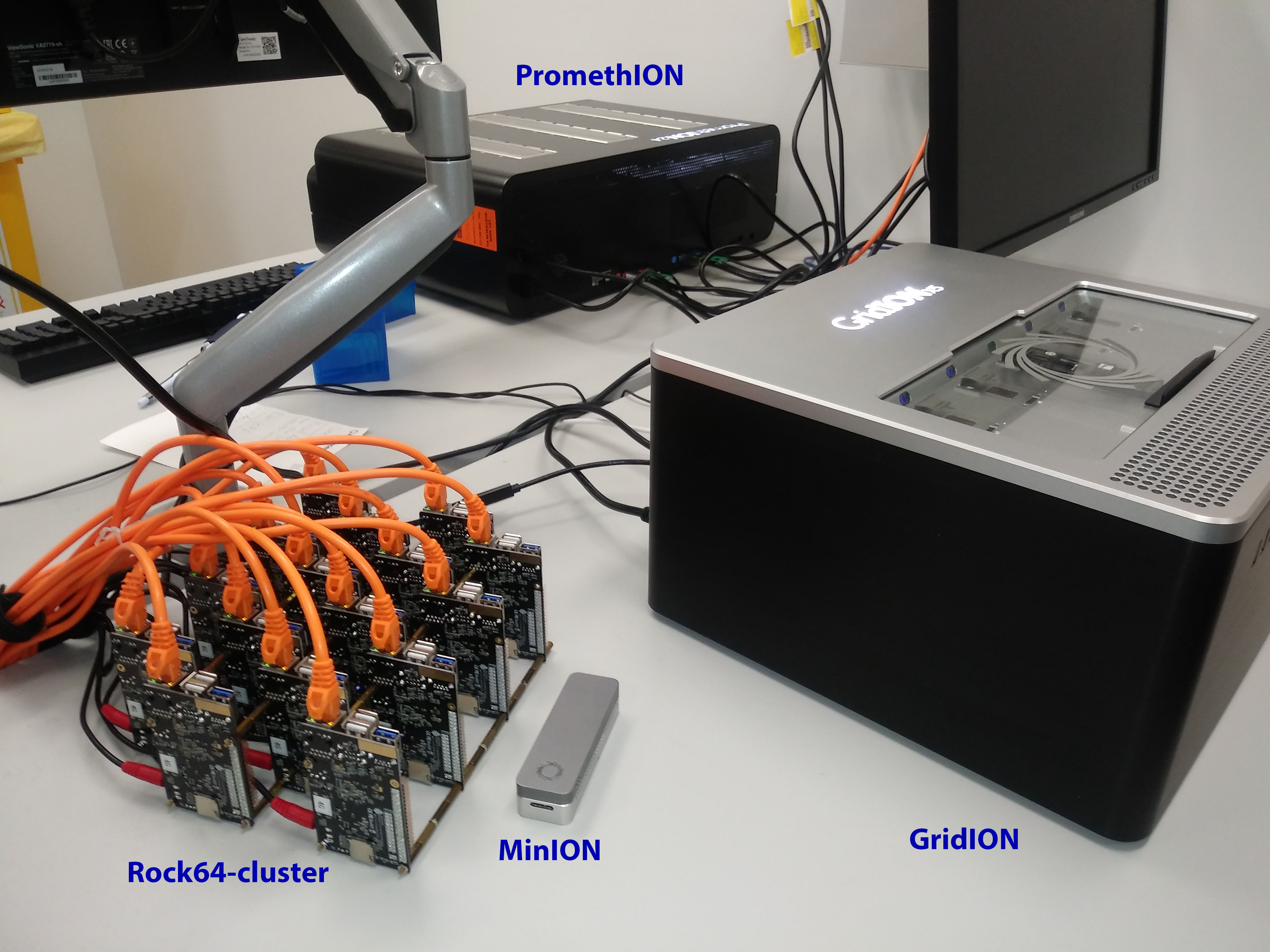}
    \caption{Rock64-cluster placed alongside the nanopore sequencers at Garvan Institute of Medical Research}
    \label{f:rock64-cluster-garvan}
\end{figure}

The architecture proposed in section \ref{s:nanopore-cluster-hw-arch} was realised in the sequencing facility at Garvan Institute of Medical Research using 16 Rock64 SBCs as worker nodes. The cluster of these 16 SBCs is referred to as the Rock64-cluster. Rock64-cluster placed alongside the nanopore sequencers is shown in Fig. \ref{f:rock64-cluster-garvan}.

\begin{figure}
\centering
    \begin{subfigure}[!ht]{0.465\linewidth}
    \centering
    \includegraphics[width=\textwidth]{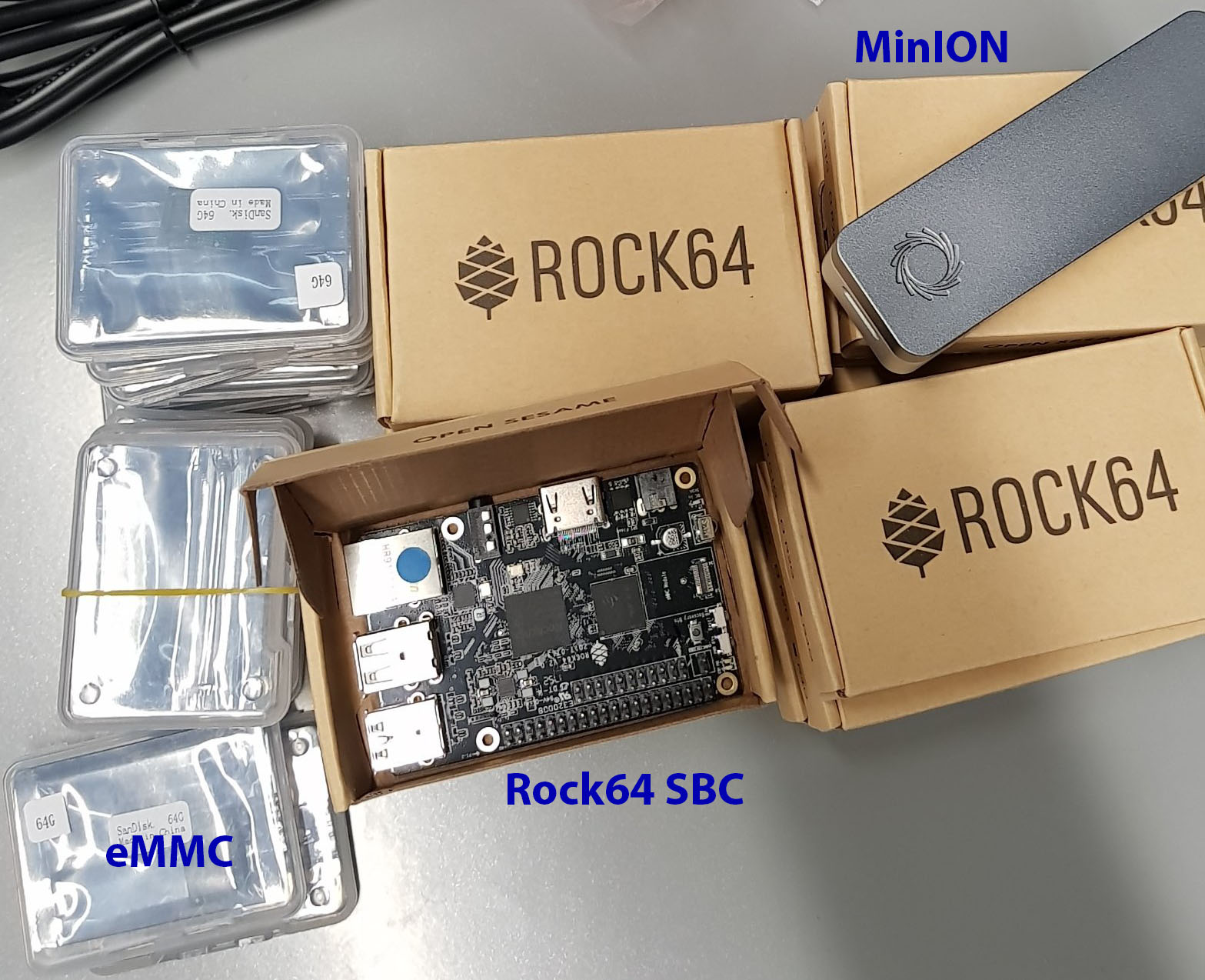}
    \caption{A newly opened Rock64 SBC. Photograph credit: Martin A. Smith.}
    \label{f:newrock64}
    \end{subfigure}
    \begin{subfigure}[!ht]{0.5\linewidth}
    \centering
    \includegraphics[width=\textwidth]{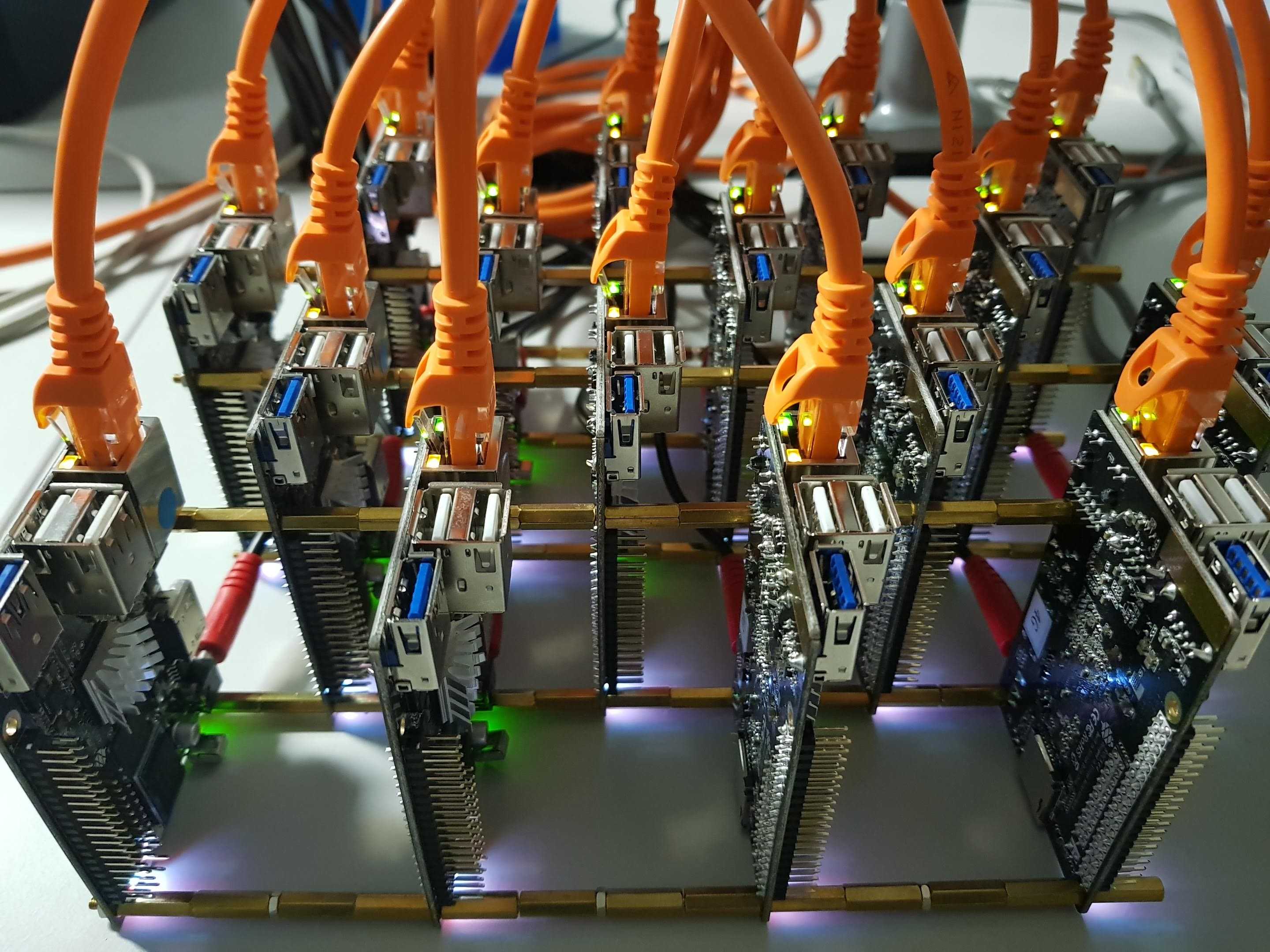}
    \caption{Rock64 SBCs stacked using M2.5 cylinders. Photograph credits: Martin A. Smith.} \label{f:stacked-rock64}
    \end{subfigure}    
    \begin{subfigure}[!ht]{\linewidth}
    \centering
    \includegraphics[width=\textwidth]{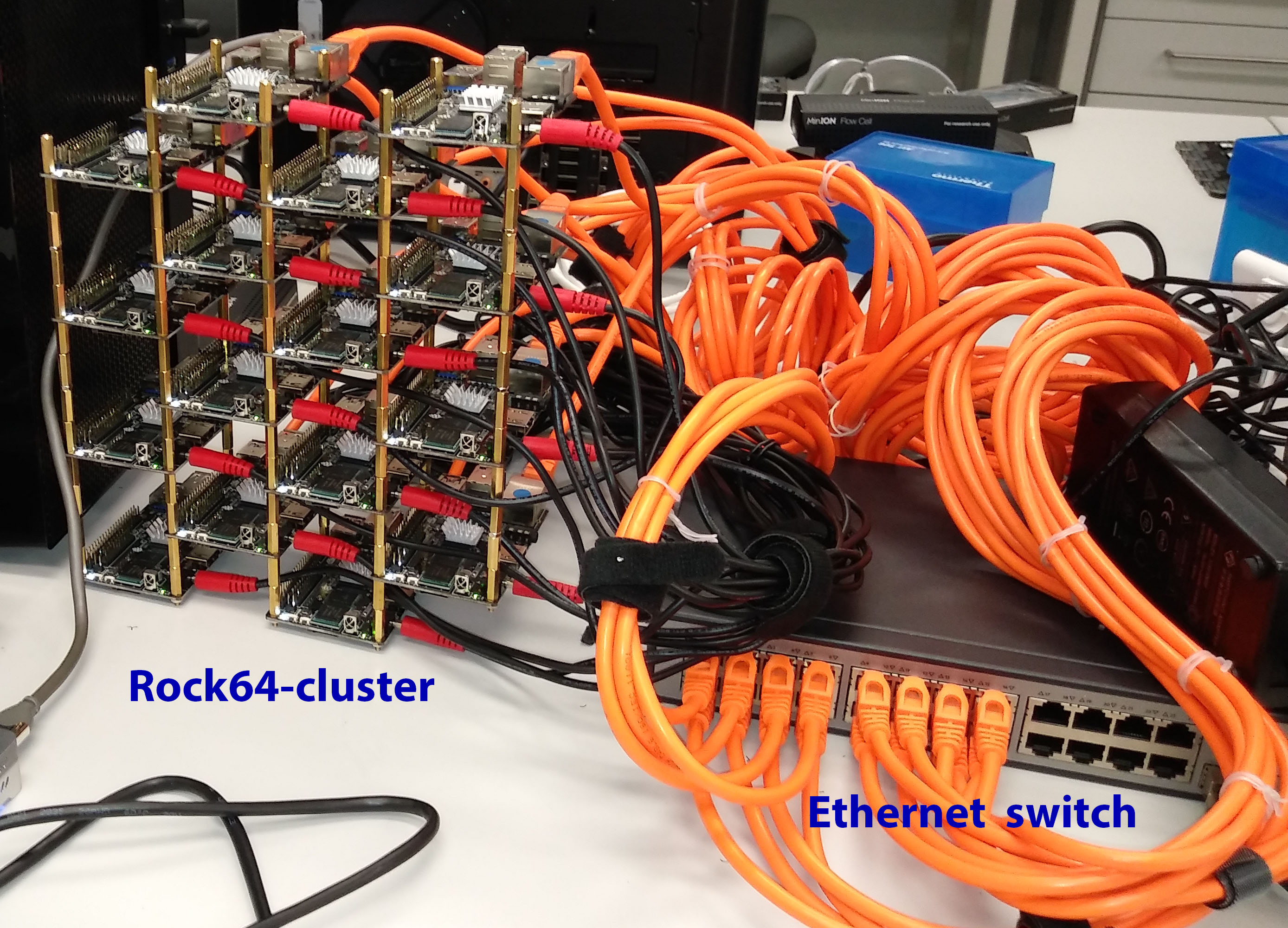}
    \caption{Rock64 SBCs connected using Ethernet} \label{f:ethernet-connected-rock64}
    \end{subfigure}        
    \caption{Construction of the Rock64-cluster.}
    \label{f:rock64-construction}
\end{figure}

Each Rock64 SBC (Fig. \ref{f:newrock64}) composed of a quad-core ARM processor, 4GB of RAM and 64 GB eMMC storage \cite{rock64} was running Ubuntu 16.04 LTS as the operating system. The 16 SBCs were stacked using M2.5 Copper Cylinders (Fig. \ref{f:stacked-rock64}) and were connected using Ethernet on to an HPE OfficeConnect 1950 24G switch (Fig. \ref{f:ethernet-connected-rock64}). A Synology DS3617xs system with 5 TB storage was used as the NAS and a Ubiquiti 10G SFP+ EdgeRouter Infinity was used as the NAT router. A desktop computer with an Intel i7-4790 processor and 16 GB of RAM running Ubuntu 16.04 was used as the head node. Refer to appendix \ref{a:nanocluster} for a step by step guide on building the Rock64-cluster.

\subsection{Rock64-cluster Software Setup} \label{s:rock64-software-implementation}

\textit{f5pd} and \textit{f5pl} proposed in section \ref{s:f5p} were implemented in C programming language. TCP sockets were used for communication between \textit{f5pd} and \textit{f5pl}. Multiple worker nodes were handled in \textit{f5pl} using multiple threads implemented with \textit{pthreads}. \textit{f5pd} was launched at the startup of each worker node and was ensured for continuous running using \textit{systemd}. TCP keepalive feature in the Linux kernel \cite{busatto2007tcp} was used to detect hung worker nodes (by determining if the connection is still up and running or if it has broken).

\begin{figure}[!ht]
    \centering
    \includegraphics[width=\textwidth]{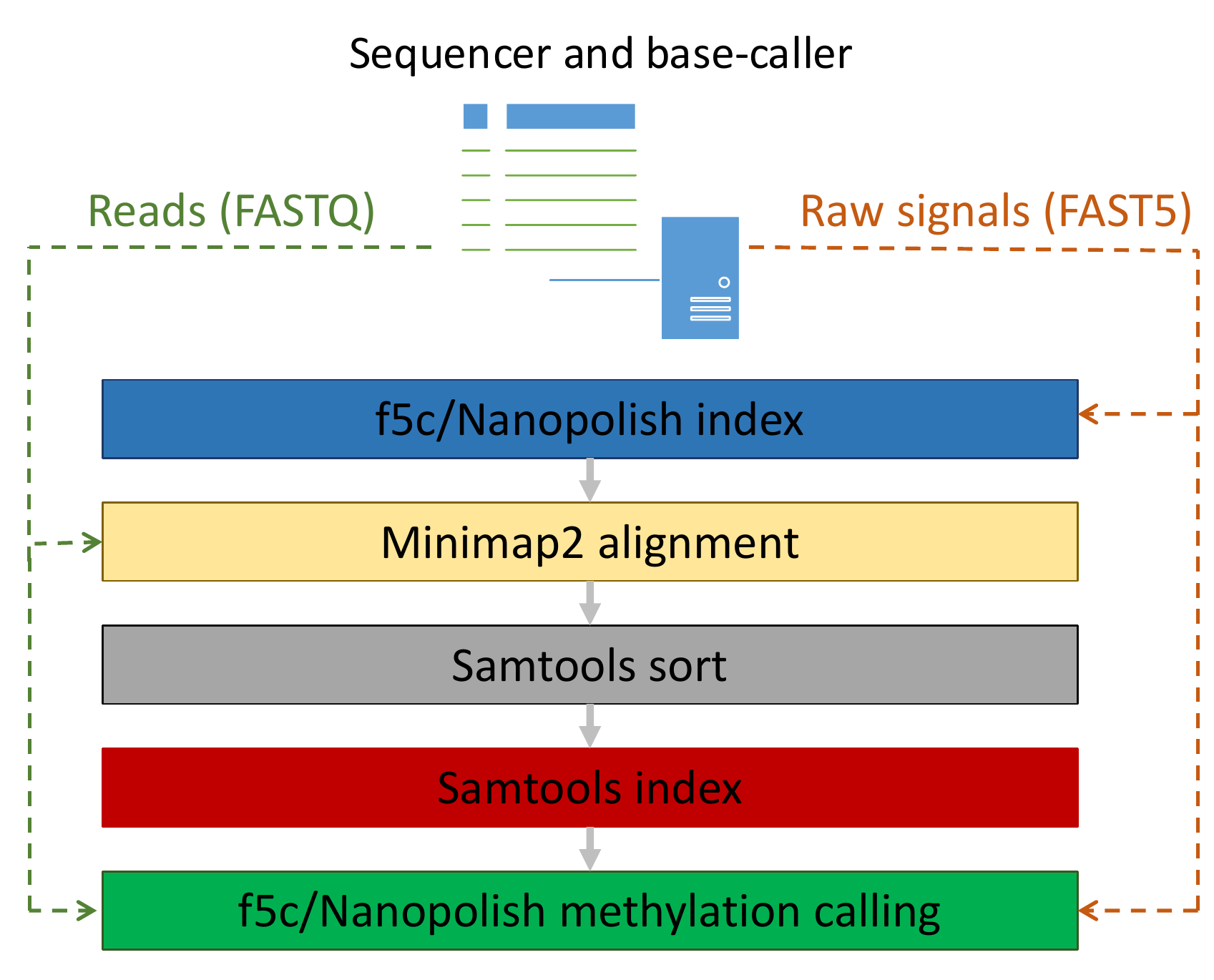}
    \caption{Methylation calling workflow and its software tools}
    \label{f:pipeline-rock64}
\end{figure}

The Rock-64 cluster was evaluated using a state-of-the-art nanopore methylation calling workflow (presented previously in Fig. \ref{f:pipeline-rock64}) consisting of software tools \textit{Minimap2}, \textit{Nanopolish} and \textit{Samtools}. The optimised version of \textit{Minimap2} for efficient memory capacity under chapter \ref{c:minimap} is used on the Rock64-cluster where the original \textit{Minimap2} cannot run due to limited RAM on each node. \textit{Samtools} was compiled for ARM architecture. A modified version of \textit{Nanopolish} was initially used on the Rock-64 cluster, which was eventually replaced with \textit{f5c} developed in chapter \ref{c:gpuabea}. Original \textit{Nanopolish} which did not compile on ARM due to Intel specific SSE instructions had to be modified and a bug that affected ARM architecture had to be fixed to successfully run on ARM architecture. These fixes are now on the original Nanopolish \textit{repository}, see appendix \ref{a:opensource}.  Later \textit{Nanopolish} was replaced with \textit{f5c} as \textit{f5c} for faster performance and memory efficiency. Despite, Rock64 devices not having a GPU, \textit{f5c} was around twice faster compared to \textit{Nanopolish}.

The aforementioned workflow was implemented as a shell script. This pipeline shell script runs on each worker node for each data as mentioned in section \ref{s:f5p}. The shell script first copies data from NAS to local eMMC storage and (extract if a single-fast5 tarball), performs the commands of the aforementioned pipeline in the order presented in Fig. \ref{f:pipeline-rock64} and finally copies the result back to the NAS. 

\begin{figure}
\centering
    \begin{subfigure}[!ht]{\linewidth}
    \centering
    \includegraphics[width=\textwidth]{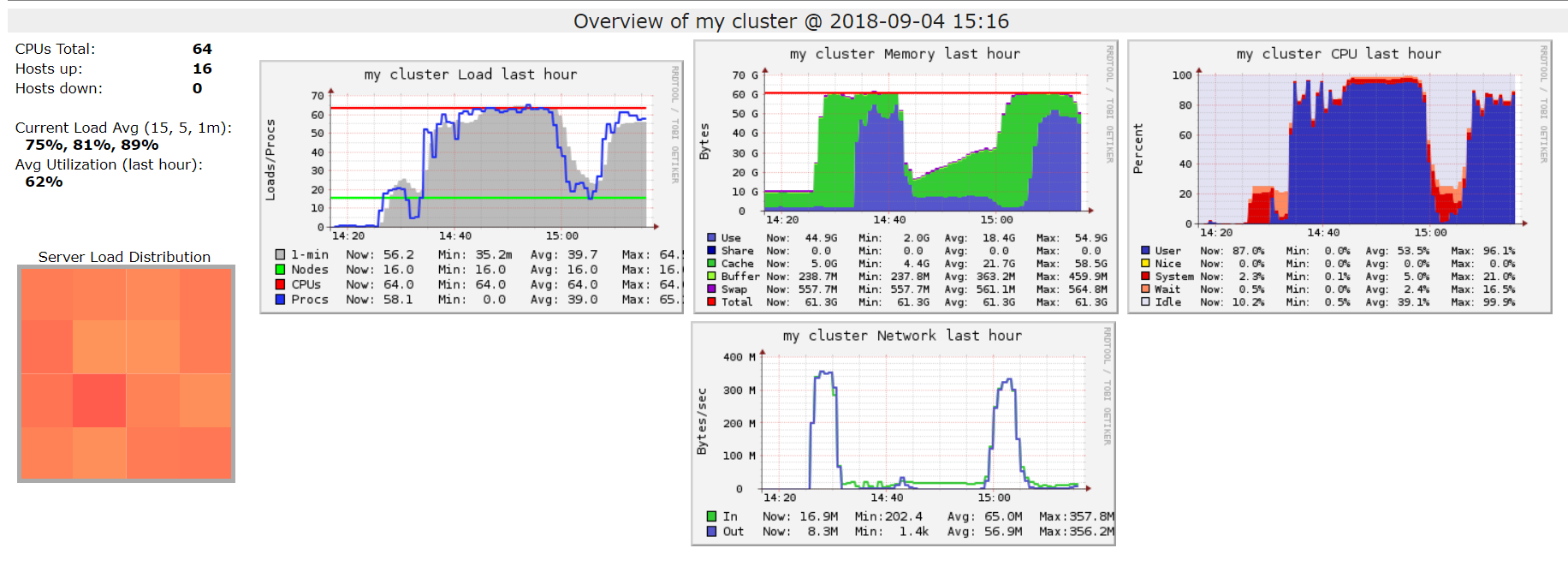}
    \end{subfigure}
    \begin{subfigure}[!ht]{\linewidth}
    \centering
    \includegraphics[width=\textwidth]{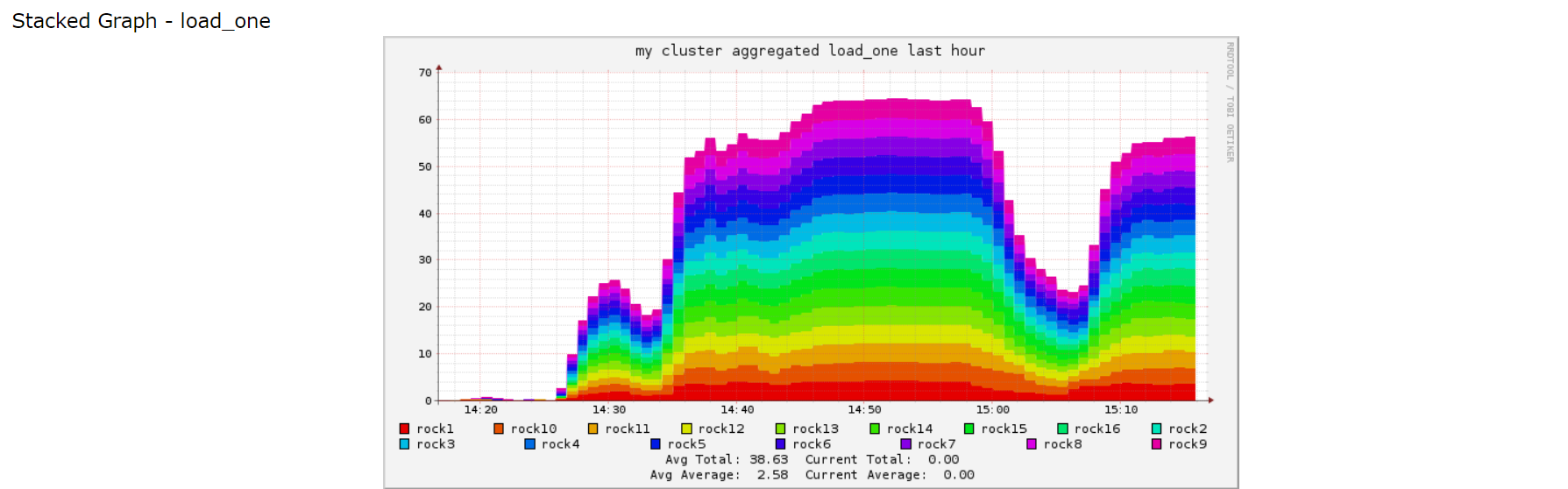}
    \end{subfigure}    
    \begin{subfigure}[!ht]{\linewidth}
    \centering
    \includegraphics[width=\textwidth]{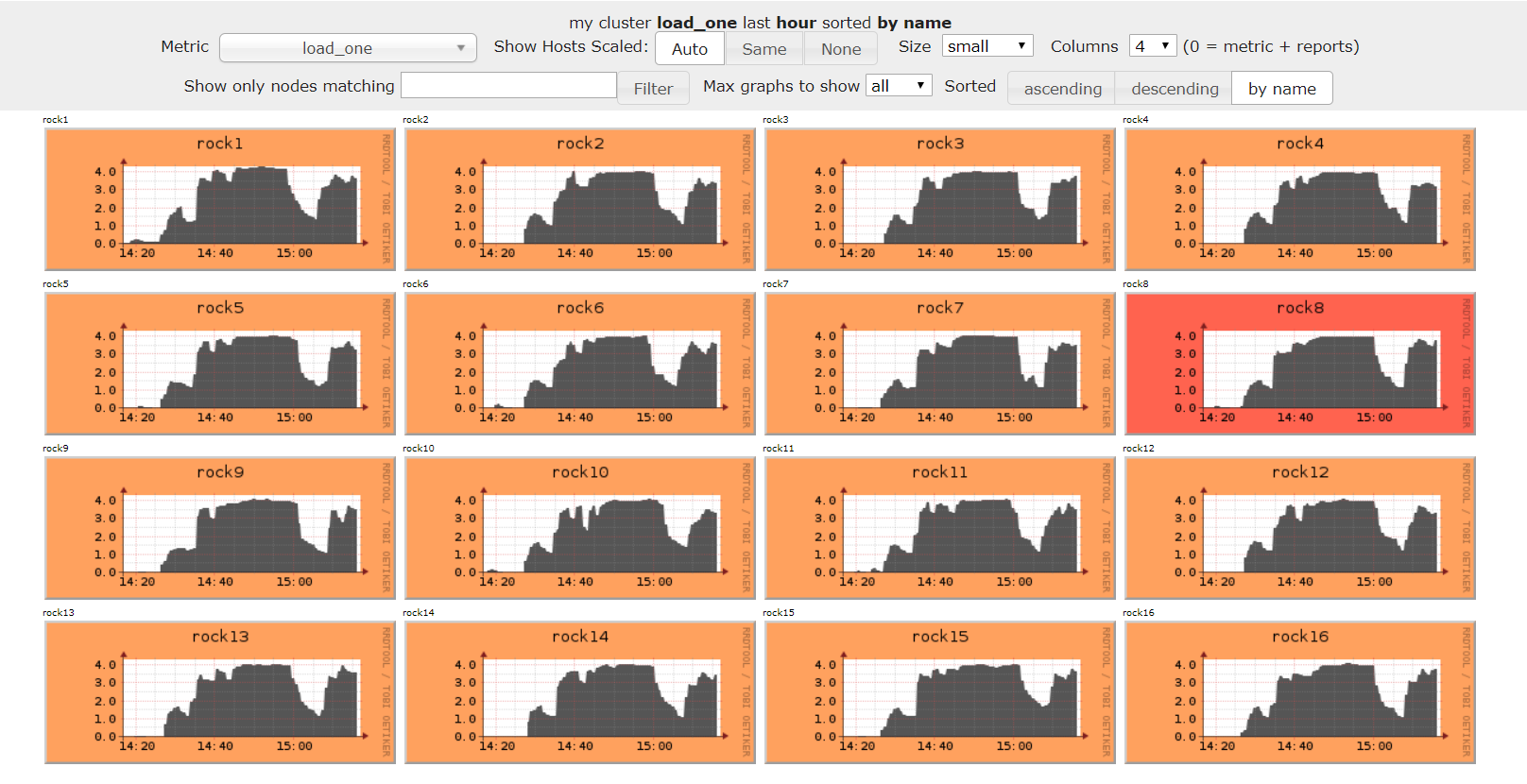}
    \end{subfigure}      
    \caption{Screenshot from Ganglia monitoring system}
    \label{f:ganglia}
\end{figure}

\begin{figure}[!ht]
    \centering
    \includegraphics[width=\textwidth]{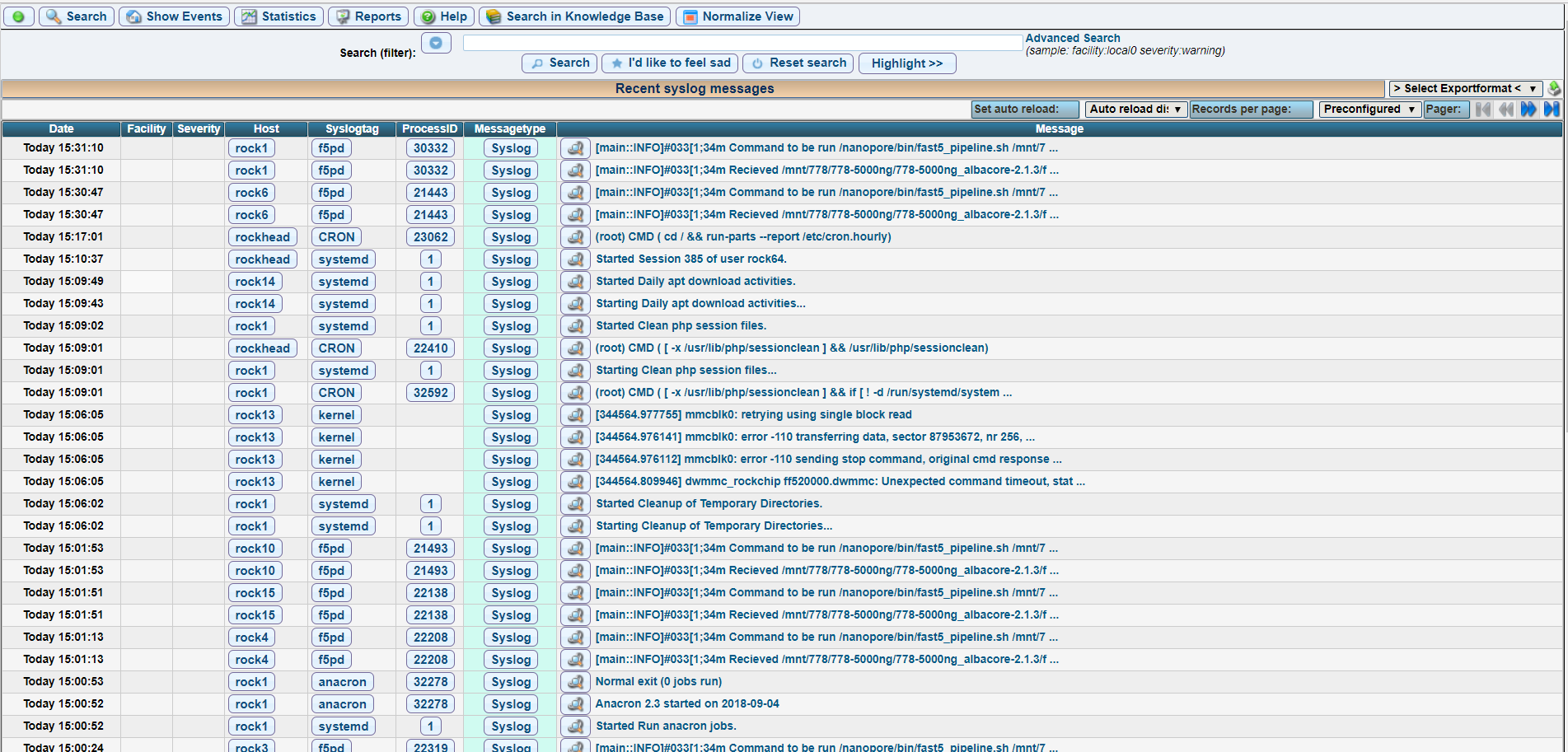}
    \caption{Screenshot from LogAnalyzer}
    \label{f:loganalyser}
\end{figure}

\textit{Ansible} was used to automate administration task such as deploying software updates across worker node, configuring settings across all worker nodes, performing maintenance operations etc. \textit{Ansible} installed on the head node accesses worker nodes through password-less key-based SSH. the ganglia monitoring system \cite{massie2004ganglia} was set up on the nodes to centrally observe the state and the utilisation of worker nodes from the head node (screenshot in Fig. \ref{f:ganglia}). Also, \textit{rsyslog} coupled with loganalyzer \cite{loganalyzer} was configured to centrally view the worker node logs from the head node (screenshot in Fig. \ref{f:loganalyser}).

The detailed steps to install and manage the Rock64-cluster are in \ref{a:nanocluster} and the associated scripts are in the GitHub repository at \url{https://github.com/hasindu2008/nanopore-cluster}.
Source code of \textit{f5p} is available at the GitHub repository \url{https://github.com/hasindu2008/f5p} and more details are in appendix \ref{a:f5p}.

\subsection{NVIDIA Jetson SBCs}

The architecture presented in section \ref{s:nanopore-cluster-hw-arch} is not limited to Rock64 devices, instead can be any other SBC. Two other SBCs, namely Jetson TX and Jetson Nano from NVIDIA, were evaluated as an alternative to Rock64. However, due to prohibitive  cost of building a complete cluster of the Jetson SBCs, this evaluation was performed only using a single worker node, which is adequate as the multi-node architecture was already verified using Rock64 SBCs.

\begin{figure}
\centering
    \begin{subfigure}[!ht]{\linewidth}
    \centering
    \includegraphics[width=\textwidth]{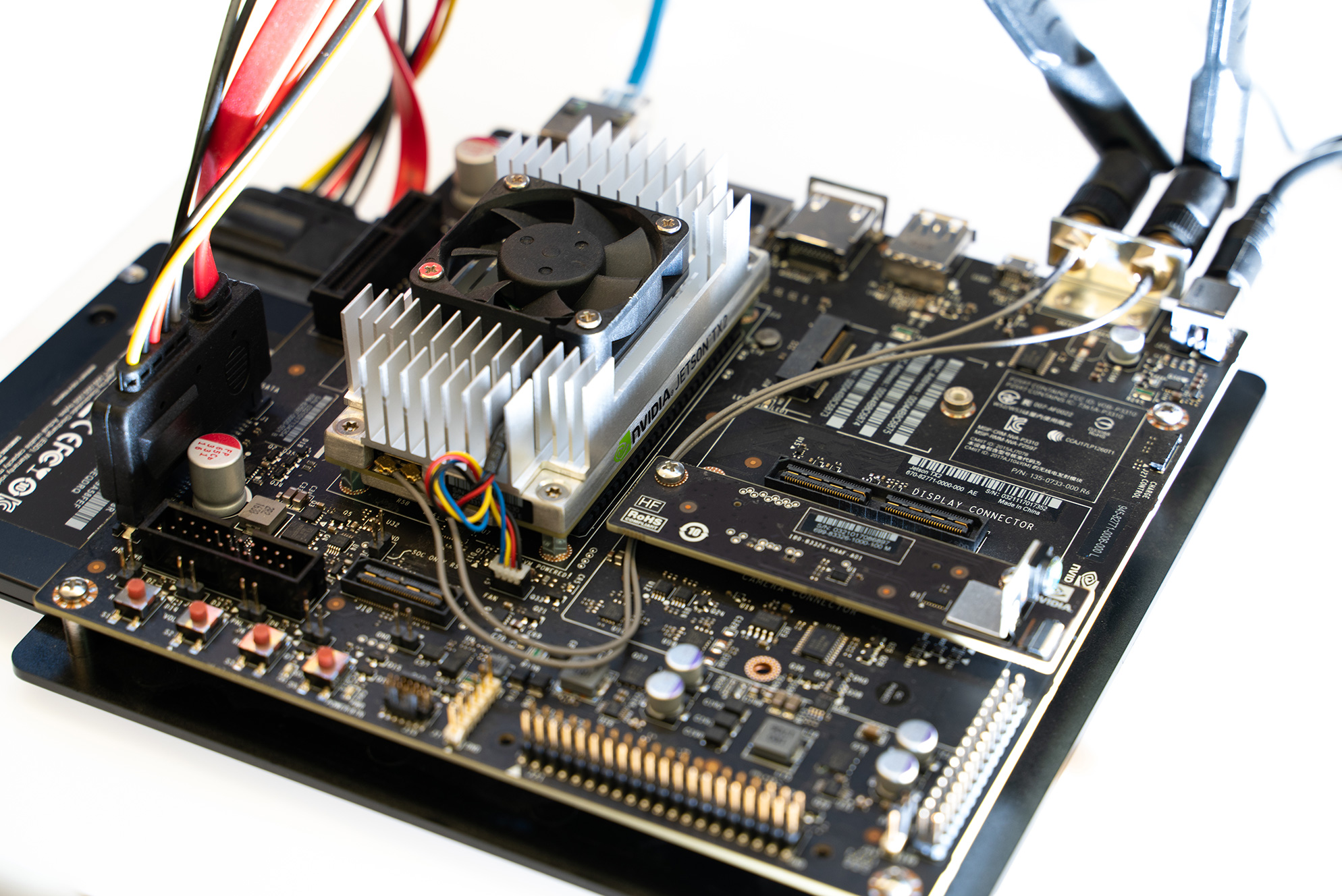}
    \caption{Jetson TX2 development board}
    \label{f:jetsontx2}
    \end{subfigure}
    \begin{subfigure}[!ht]{\linewidth}
    \centering
    \includegraphics[width=\textwidth]{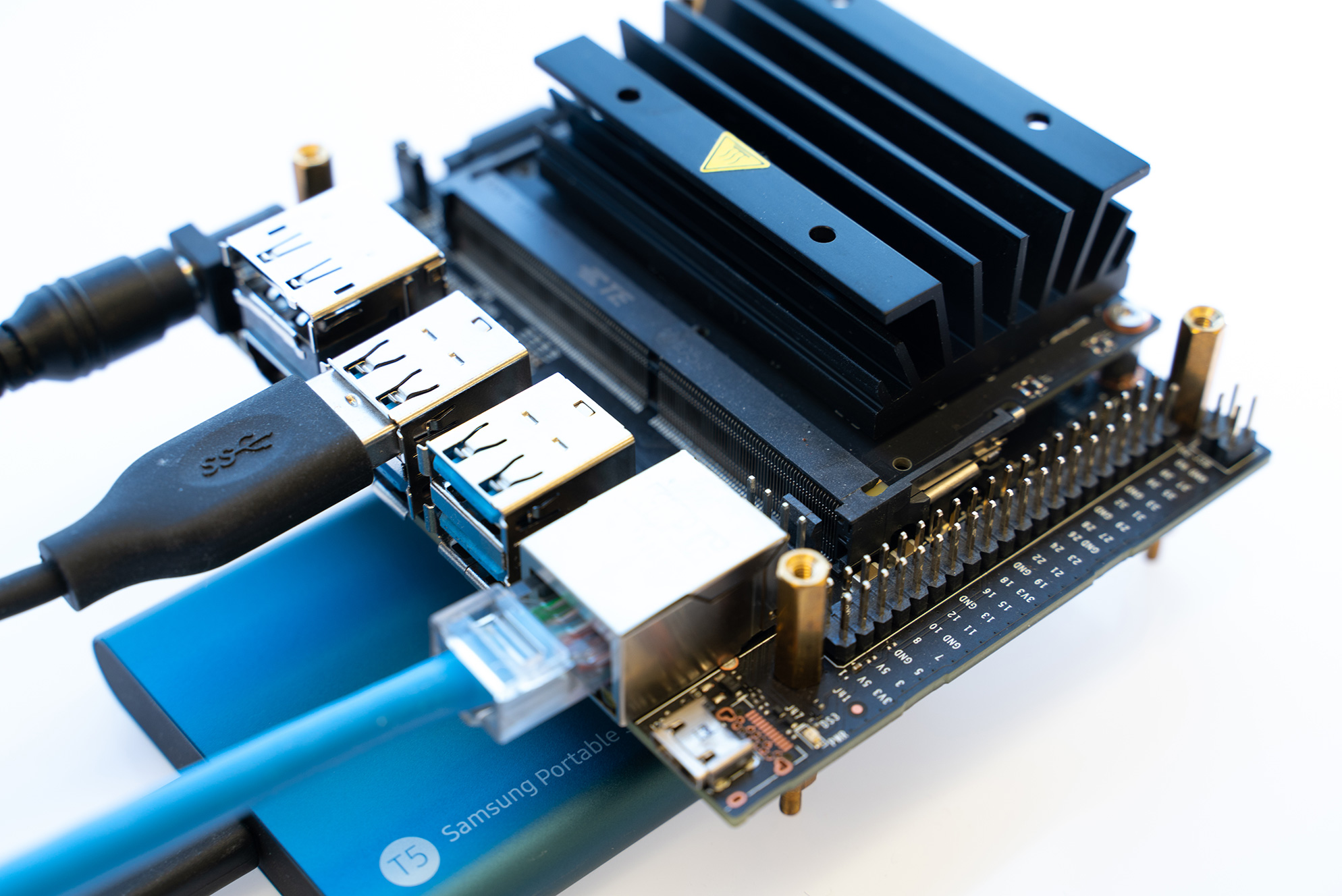}
    \caption{Jetson nano development board} \label{f:jetsonnano}
    \end{subfigure}    
    \caption{NVIDIA Jetson development boards. Photograph credits: Hsu-Kang Dow.}
    \label{f:jetsons}
\end{figure}

Jetson TX2 development board (Fig. \ref{f:jetsontx2}) is composed of a hexa-core ARM processor, 256 GPU cores, 8GB of memory (shared RAM for CPU and GPU) and 32 GB eMMC integrated storage. A Samsung 1TB SSD drive was connected to the  Jetson TX2 development board using the SATA interface. The system was running on Ubuntu 16.04.

Jetson Nano development (Fig. \ref{f:jetsonnano}) is composed of a quad-core ARM processor, 128 GPU cores and 4GB of memory (shared RAM for CPU and GPU). A Sandisk Extreme 64GB microSD (A2 rating) and a  Samsung 512 GB external SSD USB drive were used as the storage. This system was running on Ubuntu 18.04.

The evaluation was performed using the same methylation workflow used for the Rock64-cluster in section \ref{s:rock64-software-implementation}, with the use of \textit{f5c} CPU-GPU version instead of the CPU-only version being the only difference.

\subsection{Datasets}

A Nanopore MinION dataset of the T778 cancer cell-line of the human genome was used for the  evaluations  presented in section \ref{s:integration-res}. This dataset contained 771,325 reads with 11,393 and 194,983 as the average and maximum of read lengths. The total yield was 8.78 Gbases and total sizes of \textit{FAST5} and \textit{FASTQ} files were 845GB and 17GB respectively. The FAST5 files were of the single-FAST5 format. The dataset consisted of 198 batches of reads with each batch having 4000 reads.

\section{Results}\label{s:integration-res}

\subsection{Rock64-cluster}

For the aforementioned T778 MinION dataset, the complete methylation calling workflow consumed 5.88 hours on the Rock64-cluster. These 5.88 hours include the total processing time (processing steps in Fig. \ref{f:pipeline-rock64}) and all the overheads, i.e., overheads due to scheduling, file transfer to/from the NAS and tarball extraction. The tool used for methylation calling was \textit{f5c}. Note that, the time for the complete analysis on the Rock64-cluster (5.88 hours) is considerably lesser than the sequencing runtime on the MinION (typically 48 hours). 

During the analysis time (5.88 hours) mentioned above, five occasions of worker node freezes were recorded (worker node freezes explained in section \ref{s:f5p}). Four of the freezes resulted in watchdog time outs and eventually automatic restarts. However, the integrity of the analysis was not affected due to the failure handling mechanism of \textit{f5p} that reassigned the data batch once the worker node became alive (detailed in section \ref{s:f5p}). One freeze led to a totally dead device (not restarted by the watchdog), however, the analysis was continued with the remaining devices by \textit{f5p} as mentioned in section \ref{s:f5p}. Note that, the 5.88-hour analysis time includes the time lost due to these freezes and dead devices.

The summary of execution details discussed above for the T778 MinION dataset is listed in the first row of Table \ref{t:rock64-data-res}.  In Table \ref{t:rock64-data-res}, the first column describes the sample that is sequenced, the second column indicates the nanopore sequencer used (whether MinION, GrdION or PromethION), the third column lists the number of data batches in the dataset, the fourth column gives the typical sequencing runtime (48h for MinION/GridION and 64h for PromethION), the fifth column indicates the software used for methylation calling (\textit{f5c} or \textit{nanopolish}), the sixth column gives the total time for the execution of the methylation calling workflow on the Rock64-cluster, the seventh column gives the number of worker node freezes that were detected by the watchdog timer leading to automatic restarts and the eighth column gives the number of worker nodes retired due to complete freezes or consecutive failures. 

While the above T778 dataset was used for performing thorough evaluation and benchmarks, several other nanopore datasets were processed on the Rock64-cluster and their execution details are summarised in Table \ref{t:rock64-data-res} from the second row onwards. These details were collected while using the Rock64-cluster for in-house data processing of research data samples. Some datasets in Table \ref{t:rock64-data-res} have been processed using \textit{Nanopolish} instead of \textit{f5c} because those were processed before the development of \textit{f5c} as mentioned in section \ref{s:integration-exp}. The last two rows of Table \ref{t:rock64-data-res} are for the same dataset where one execution was using \textit{f5c} and the other using \textit{nanopolish}. Observe that \textit{f5c} performance is superior (45.08 hours for the complete workflow) compared to \textit{Nanopolish} (61.58 hours for the complete workflow), in spite of 3 worker nodes being retired in the \textit{f5c} execution compared to only 2 retired worker nodes in the \textit{Nanopolish} execution. Therefore, the processing time observed for the datasets processed using \textit{nanopolish} would be improved if executed using \textit{f5c}.

\begin{table}[!ht]
\caption{Execution results of several nanopore datasets of the human genome on the Rock64-cluster} \label{t:rock64-data-res}
\footnotesize
\begin{tabular}{|p{2.5cm}|p{1.8cm}|p{1.2cm}|p{1.7cm}|p{1.5cm}|p{1.5cm}|p{1cm}|p{1cm}|}
\hline
\textbf{Sample}                                          & \textbf{Sequencer}  & \textbf{Data batches} & \textbf{Sequencing run time (h)} & \textbf{\textit{f5c} or \textit{Nanopolish}} & \textbf{Processing time (h)} & \textbf{Node resets} & \textbf{Node retires} \\ \hline
T778 - lipsarcoma                     & MinION     & 198           & 48                          & \textit{f5c}            & 5.88                          & 4                                           & 1                         \\ \hline
MCF7 - breast cancer                   & GridION    & 727           & 48                          & \textit{Nanopolish}     & 18.08                         & 1                                           & 0                         \\ \hline
MCF7 - breast cancer                   & GridION    & 447           & 48                          & \textit{Nanopolish}     & 11.47                         & 1                                           & 1                         \\ \hline
NA12878                                         & PromethION & 2954          & 64                          & \textit{f5c}            & 50.27                         & 5                                           & 1                         \\ \hline
HCT116A - colon cancer             & PromethION & 1613          & 64                          & \textit{Nanopolish}     & 47.58                         & 5                                           & 1                         \\ \hline
Dk01 - prostate cancer                & PromethION & 2659          & 64                          & \textit{Nanopolish}     & 85.28                         & 5                                           & 2                         \\ \hline
LNCap - prostate cancer & PromethION & 2858          & 64                          & \textit{Nanopolish}     & 65.70                      & 1                                           & 2                         \\ \hline
PrEC - Prostate Epithelial                 & PromethION & 1256          & 64                          & \textit{Nanopolish}     & 45.52                      & 5                                           & 0                         \\ \hline
T778 - lipsarcoma                    & PromethION & 1210          & 64                          & \textit{Nanopolish}     & 35.63                      & 1                                           & 0                         \\ \hline
HBB20 PBMC                                      & PromethION & 2145          & 64                          & \textit{f5c}            & 45.08                      & 3                                           & 3                         \\ \hline
HBB20 PBMC                                      & PromethION & 2145          & 64                          & \textit{Nanopolish}     & 61.58                      & 4                                           & 2                         \\ \hline
\end{tabular}
\end{table}

\subsection{On NVIDIA Jetson SBCs}
\begin{figure}[!ht]
    \centering
    \includegraphics[width=\textwidth]{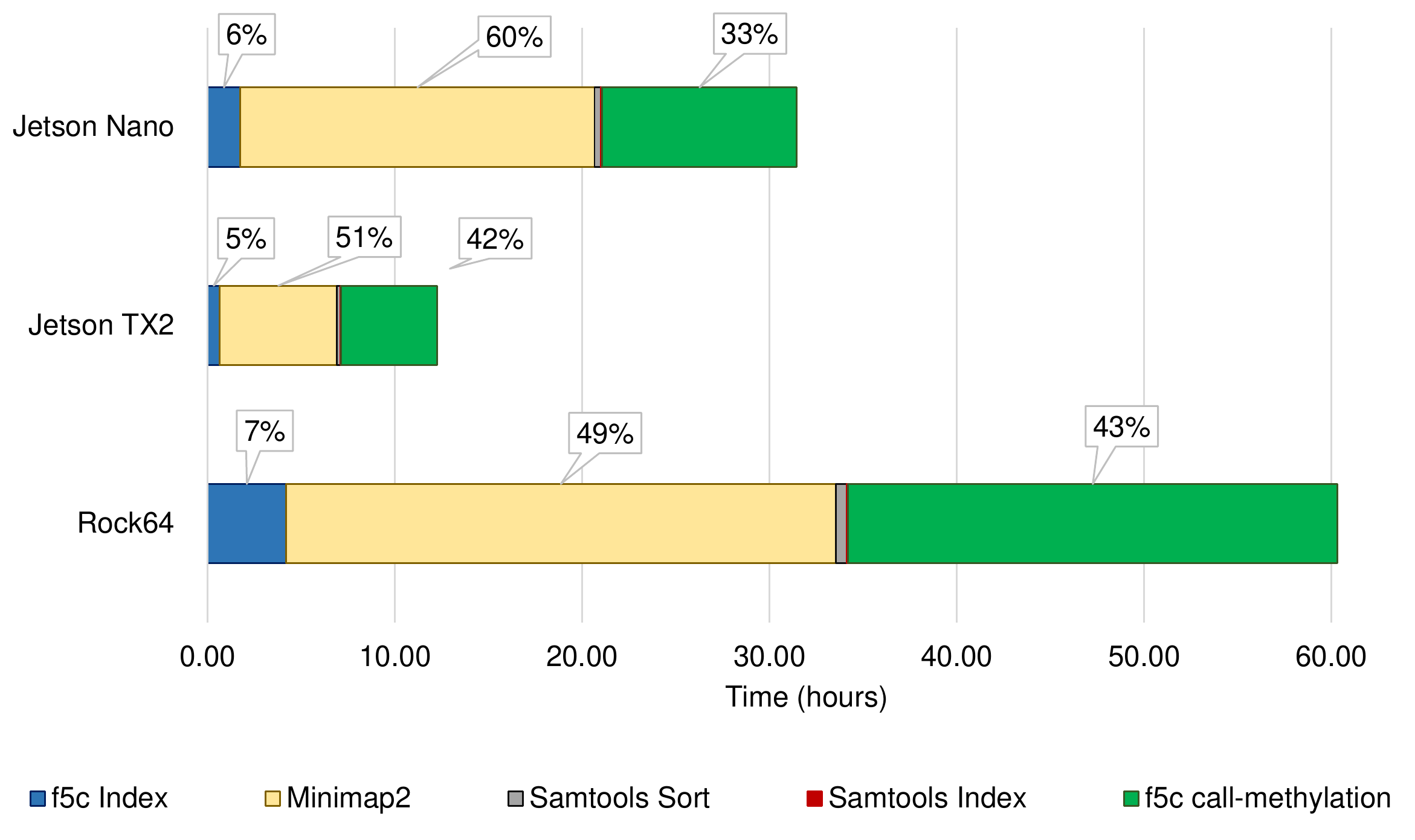}
    \caption{Comparison of Jetson TX2, Jetson Nano and Rock64 based on the single SBC execution times for the whole dataset}
    \label{f:perf-devices}
\end{figure}

The architecture proposed in section \ref{s:nanopore-cluster-arch} is generic, i.e., worker nodes are not limited to Rock64 SBCs.  Two alternative SBCs to Rock64 were benchmarked, namely, Jetson TX2 and Jetson Nano. Both Jetson SBCs are from NVIDIA and are equipped with GPUs making it possible to fully harness the GPU accelerated component of \textit{f5c}. The benchmarking was performed only on a single Jetson TX2 and a single Jetson Nano due to prohibitive costs. The performance of the methylation calling workflow on each of these SBcs is compared to the performance on a single Rock64 SBC in Fig. \ref{f:perf-devices}.

The x-axis of the horizontal bar chart in Fig. \ref{f:perf-devices} denotes the time in hours. The bars denote the sum of execution times for all the 198 batches of the T778 MinION dataset and different colours denote the breakdown of the execution time for each step in the workflow. Jetson TX2 was the fastest, consuming 12.27 hours, followed by Jetson Nano consuming 31.44 hours. Rock64 was the slowest consuming 60.31 hours. Thus, a single Jetson TX2 was around 5 times faster when compared to a single Rock64 SBC and a single Jetson Nano was around 2 times faster than a single Rock64 SBC.

On each SBC, the major portion of the time was contributed by the \textit{Minimap2} alignment step (49\%-60\%) followed by the methylation calling step (33\%-43\%). Methylation calling was performed using \textit{f5c}, the CPU-only version on Rock64 and the CPU-GPU version on the Jetson SBCs. \textit{f5c index} contributed less than 7\% of the total time and the times for \textit{Samtools sort} and \textit{Samtools index} were very small (around 1\%).

\begin{figure}
\centering
    \begin{subfigure}[!ht]{0.9\linewidth}
    \centering
    \includegraphics[width=\textwidth]{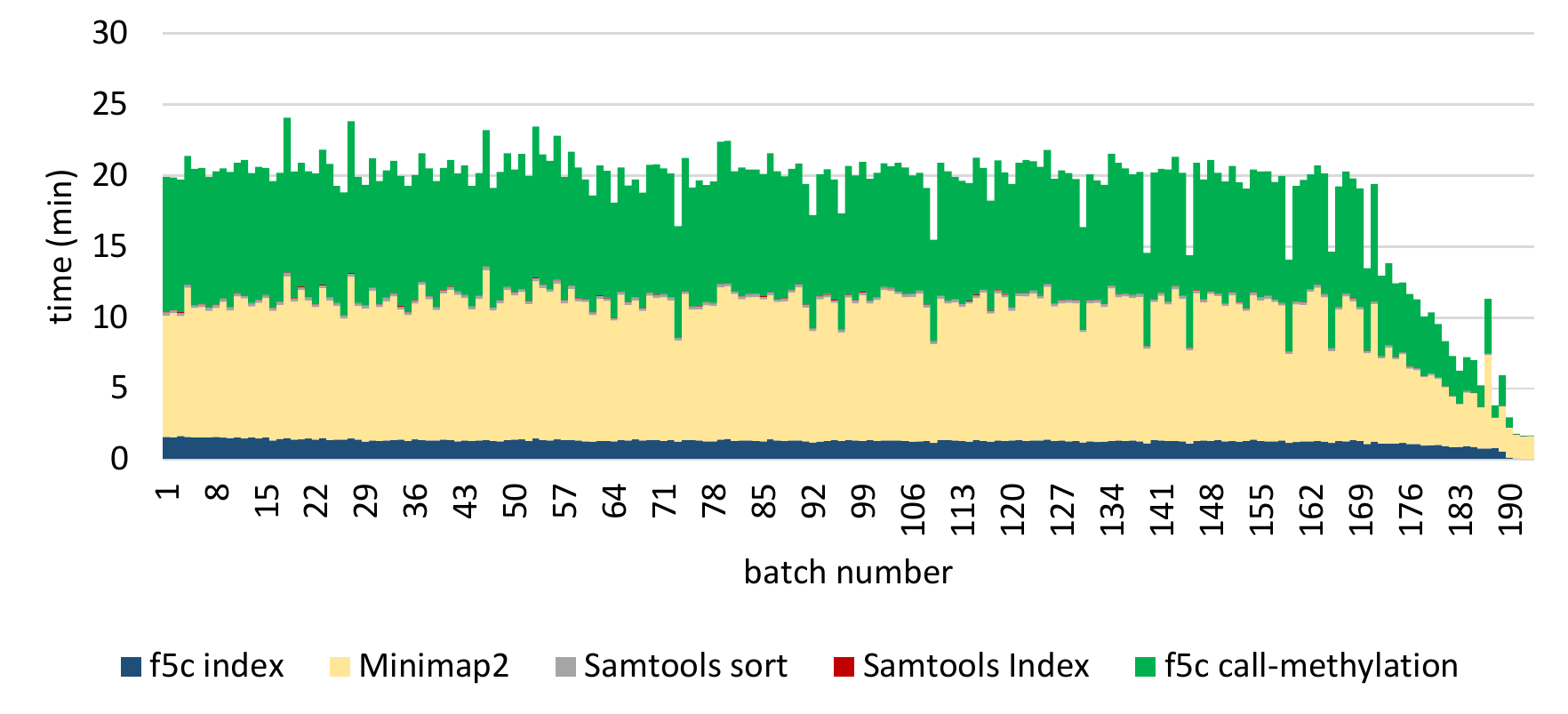}
    \caption{On a single Rock64 development board}
    \label{f:rock64-778-runtime}
    \end{subfigure}
    \begin{subfigure}[!ht]{0.9\linewidth}
    \centering
    \includegraphics[width=\textwidth]{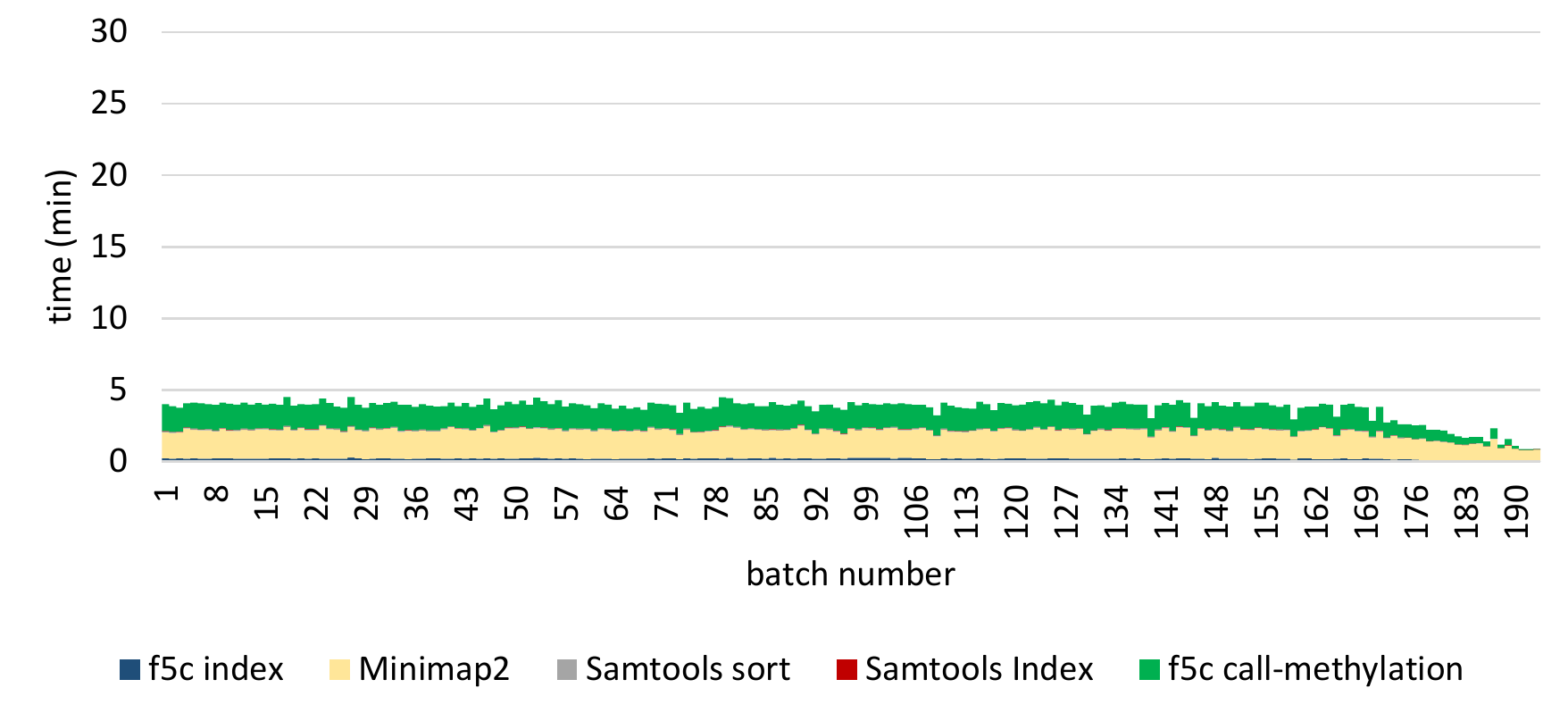}
    \caption{On a single Jetson TX2 development board} \label{f:tx2-778-runtime}
    \end{subfigure}    
    \begin{subfigure}[!ht]{0.9\linewidth}
    \centering
    \includegraphics[width=\textwidth]{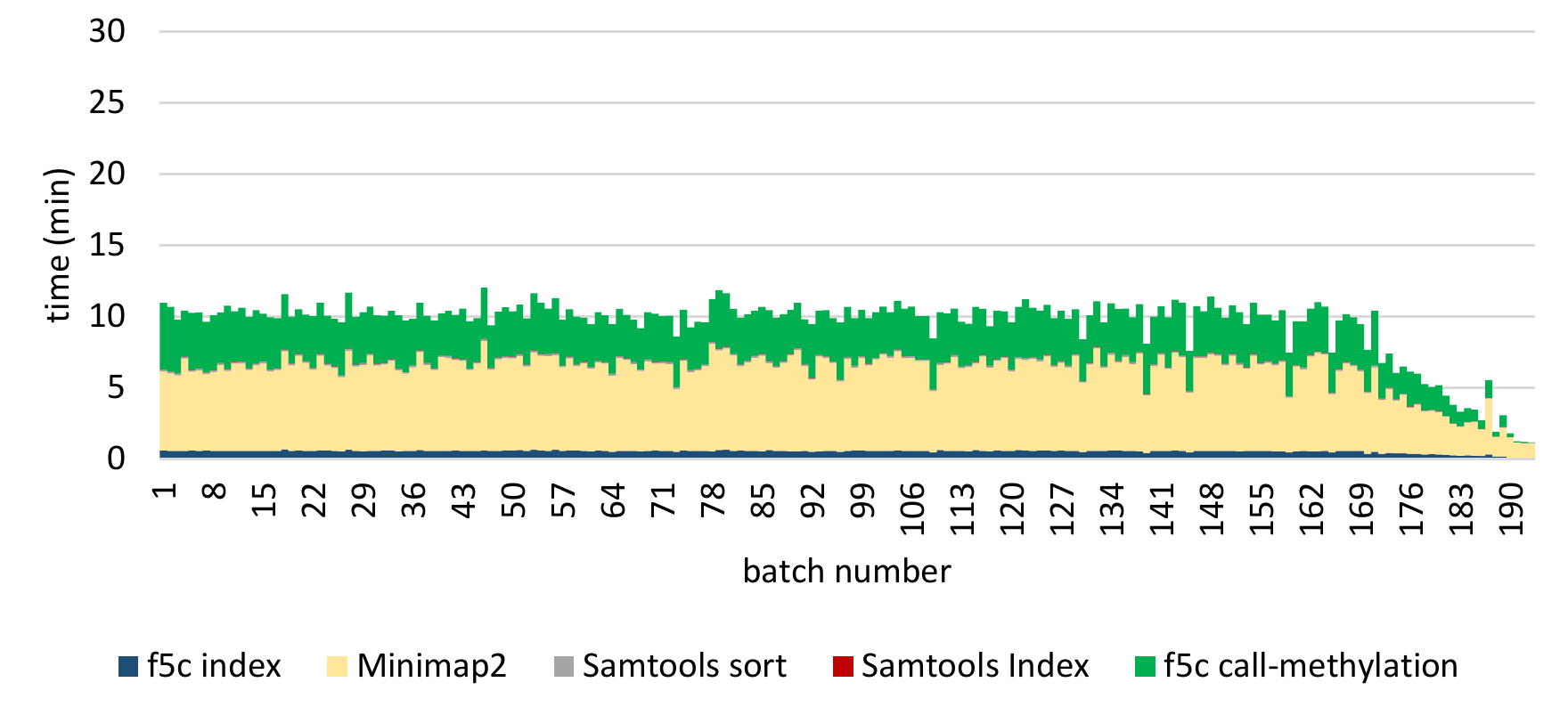}
    \caption{On a single Jetson nano development board} \label{f:nano-778-runtime}
    \end{subfigure}     
    \caption{Execution time on individual SBCs per each batch in the dataset}
    \label{f:778-runtimes}
\end{figure}

Fig. \ref{f:778-runtimes} shows the workflow execution time spent on each data batch of the 198 data batches of the T778 dataset. Fig. \ref{f:rock64-778-runtime} is for a single Rock64, Fig. \ref{f:tx2-778-runtime} is for a single Jetson TX2 and Fig. \ref{f:nano-778-runtime} is for a single Jetson Nano. The x-axis of each bar chart denotes the data batch number ranging from 1 to 198. The y-axis is the time in minutes where the bars represent the time spent on each data batch for the methylation calling workflow. Different colours in bars denote the breakdown of the time for different steps in the workflow. The average time for processing a data batch was 18.35 minutes for the Rock64 SBC, 3.72 minutes for the Jetson TX2 and 6.39 minutes for the Jetson Nano. Out of the 18.35 minutes for the Rock64, the major portion was consumed by \textit{Minimap2} alignment (8.95 minutes) followed by \textit{f5c} methylation calling (7.90 minutes). Similarly, 1.89 and 1.56 minutes for Jetson TX2 and 5.73 and 3.14 minutes for Jetson Nano were recorded for \textit{Minimap2} and \textit{f5c}, respectively.

\subsection{Real-time Processing Capability}

%device by device runtime compare
As mentioned in section \ref{s:integration-intro}, nanopore sequencers are capable of streaming the sequencing data and thus it is possible to process data on-the-fly. This section demonstrates the proof of concept of performing data analysis on-the-fly (real-time) using the architecture presented in section \ref{s:nanopore-cluster-arch}.

\begin{figure}[!ht]
    \centering
    \includegraphics[width=\textwidth]{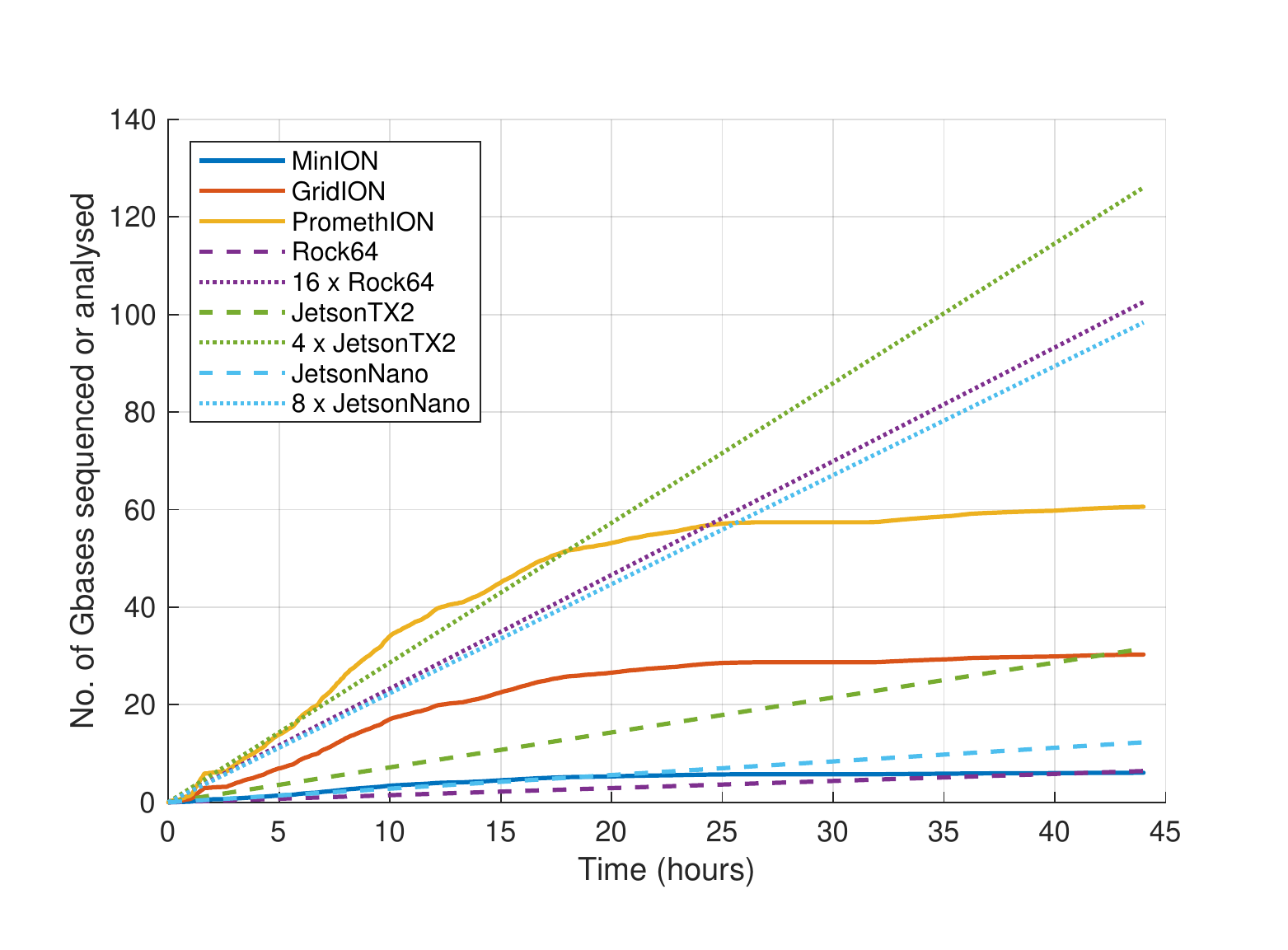}
    \caption{The comparison of the sequencing rate with the data analysis rate over the duration of the sequencing run}
    \label{f:yield-curve}
\end{figure}

How the sequencing rate varies over time is shown in Fig. \ref{f:yield-curve} for MinION (blue curve), GridION (orange curve) and PromethION (yellow curve). The x-axis denotes the time in hours and the y-axis denotes the cumulative number of bases sequenced (in Gbases) over time. Observe how the sequencing rate (gradient of the curve) is high at the beginning, when then slowly reduces and eventually becomes zero. 

Fig. \ref{f:yield-curve} also plots the cumulative number of bases possible to be processed using a single Rock-64 SBC (purple dashed line), a single Jetson TX2 (blue dashed line) and a single Jetson Nano (green dashed line). For these plots, the y-axis is now the number of gigabases analysed. The gradient of each plot is calculated by dividing the total workflow execution time in Fig. \ref{f:perf-devices} for the corresponding SBC by the total number of bases in the dataset. Observe that a single Rock64 device is barely adequate to keep up with the MinION curve. At first, the analysis lags when the sequencing rate is high, which then catches up when the sequencing rate drops. Observe that, a single Jetson Nano can easily keep up with a MinION and a Jetson TX2 is barely adequate to keep up with the GridION curve. 
Fig. \ref{f:yield-curve} also plots the cumulative number of bases possible to be processed using clusters made of each SBC., i.e., 16 Rock64 devices (purple dotted line), 4 Jetson TX2s (green dotted line) and 8 Jetson Nanos (blue dotted line). The gradients of these lines are equal to the product of the gradient for a single SBC and the number of devices. The number of SBCs in a cluster has been selected so that the cluster can more than adequately keep up with a PromethION flowcell. Such an extra margin between the analysis and the sequencing yield curves will smooth on-the-fly processing while allowing for disruptions such as device freezes. 

Note that the x-axis in Fig. \ref{f:yield-curve}  shows only the first 45 hours of the sequencing run as the curve is almost flat by this time, despite the sequence run of a MinION/GridION being 48 hours and PromethION being 64 hours. Also, the curves for the sequencers in Fig. \ref{f:yield-curve}  are based on typical average values. The exact curve can vary based on factors such as the quality of the sample and flow cell and also will change with technology improvements. Nevertheless, the presented proof of concept technique for estimating the number of SBCs for analysing data on-the-fly remains unaffected.

\subsection{Comparison with HPC}

\begin{figure}[!ht]
    \centering
    \includegraphics[width=\textwidth]{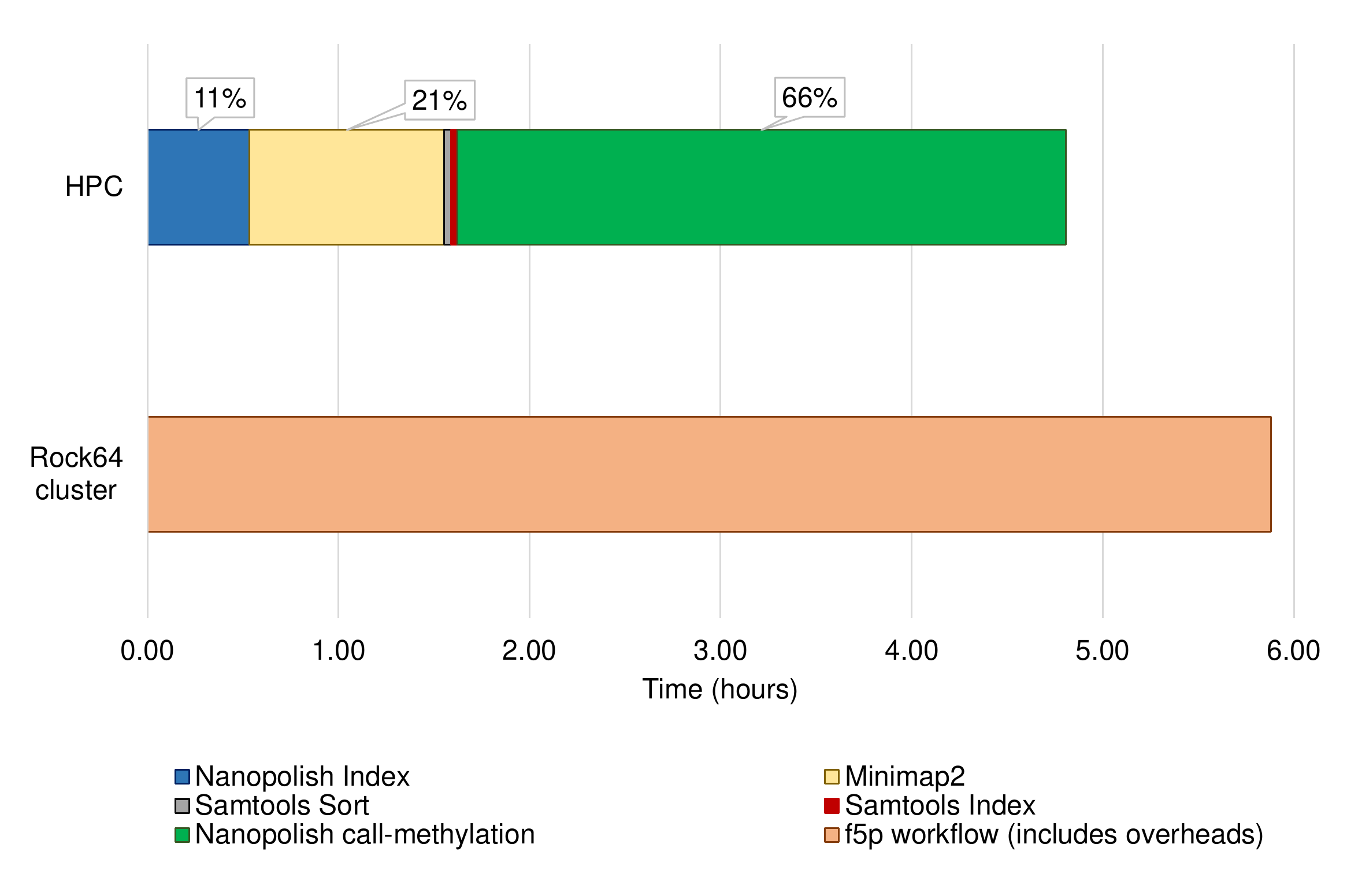}
    \caption{Comparison of proposed architecture on the Rock64-cluster with the original pipeline running on an HPC}
    \label{f:compare-with-hpc-rock64}
\end{figure}

The performance of the Rock64-cluster is compared to the performance of an HPC in Fig. \ref{f:compare-with-hpc-rock64}. The Rock64-cluster performed the methylation calling pipeline using \textit{f5p} as explained in section \ref{s:integration-exp}. The HPC was a server with 28 Xeon E5-2680 cores, 512GB RAM and 10 NVMe SSD drives in RAID configuration. The HPC executed the same methylation calling workflow as in Fig. \ref{f:pipeline-rock64} with original \textit{Minimap2} and original \textit{Nanopolish}. Observe that the time spent on the Rock64-cluster (5.88 hours) is comparable to the time consumed on HPC (4.81 hours). Importantly, the time for Rock64-cluster includes the overheads for copying to/from the NAS and extracting tarballs whereas the HPC processed the dataset that was already placed on the fast local SSD RAID drives.

Comparing the cost and size of the Rock64-cluster with that of the server (about 10 times approximately), it is surprising that the performance is similar. Further analysis of this surprising phenomenon is %in Appendix \ref{a:rock64-compare} and 
in chapter \ref{c:ioopti}.

\section{Discussion}

\subsection{Implementation for On-the-fly Processing}

The evaluation results presented on this chapter were based on datasets that were already residing on the NAS (datasets of previously sequenced samples), i.e., the whole dataset (all data batches) was available on the NAS when the workflow execution was started on the Rock64-cluster. While being adequate to demonstrate the proof of concept for on-the-fly (real-time) data processing, the future work could focus on implementing the scripts that automates the data transfer in real-time from the sequencer to the NAS. In fact, this implementation work is currently under progress as an undergraduate student project (\url{https://github.com/sashajenner/realf5p}) and is not claimed as a part of this thesis.

\subsection{Mobile Phone Cluster}

The proposed architecture in this chapter can also applicable to a cluster of mobile phones connected through Wi-Fi. The feasibility of performing the methylation calling workflow on an Android mobile phone was evaluated in an experimental environment (Fig. \ref{f:mobile-phone}) as described in Appendix \ref{a:onphone}. The development of a proper Android Application was undertaken by an undergraduate project group and is described in the pre-print at  \cite{samarakoon2020f5n}. Also, the Wi-Fi cluster implementation is under progress by the same group. The development of the Android application or the Wi-Fi cluster is not claimed under the thesis.

\begin{figure}[!ht]
    \centering
    \includegraphics[width=\textwidth]{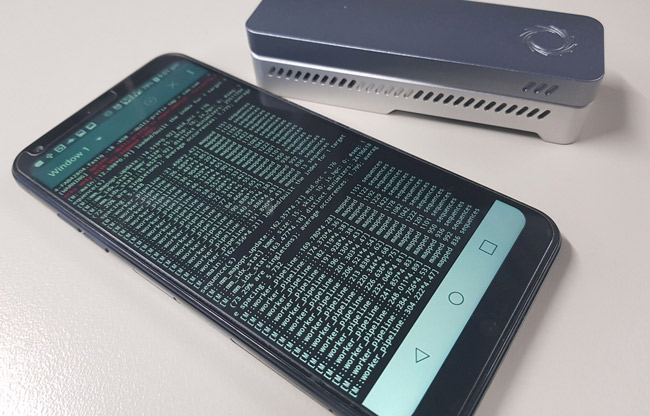}
    \caption{Methylation calling workflow on an Android mobile phone}
    \label{f:mobile-phone}
\end{figure}

\subsection{Potential Diagnostic Applications}

\begin{figure}[!ht]
    \centering
    \includegraphics[width=\textwidth]{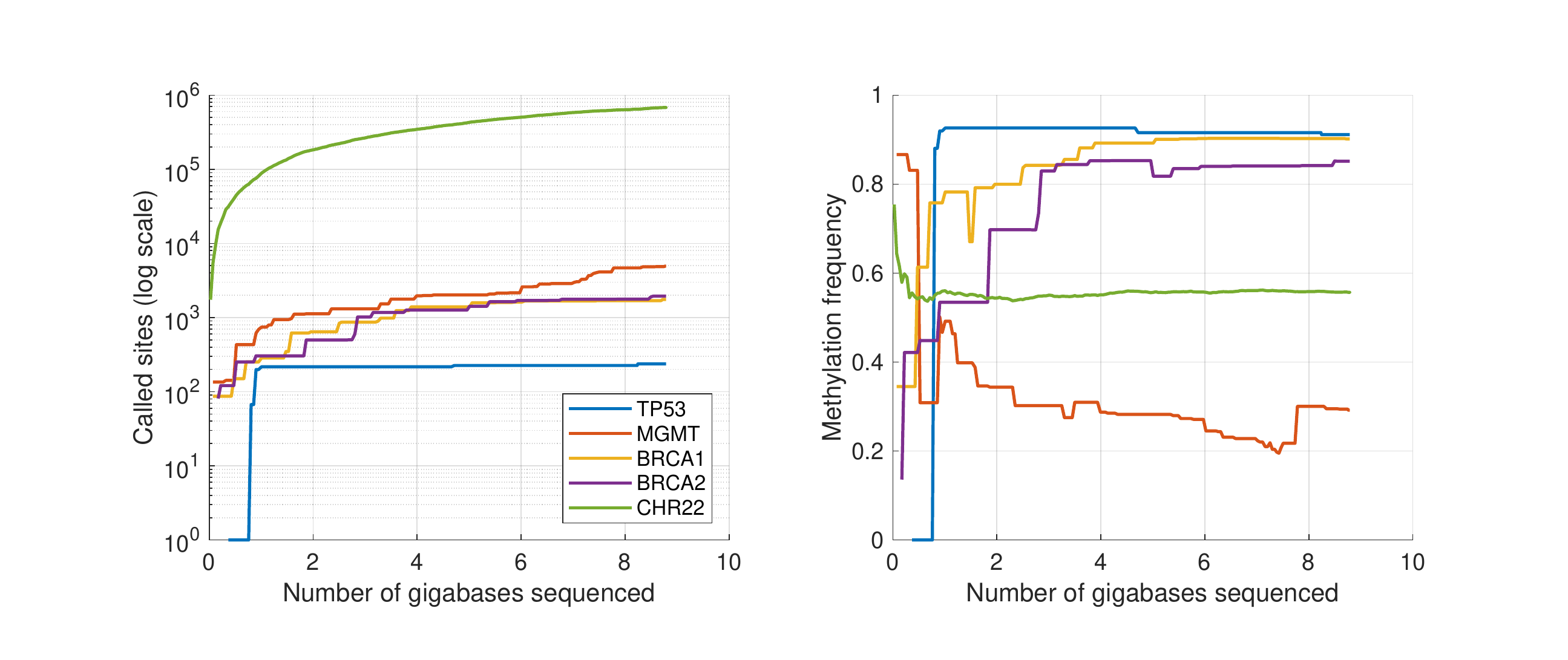}
    \caption[Potential applications of real-time methylation calling]{Potential applications of real-time methylation calling. Left graph - the variation of the number of called sites and the methylation frequency over the number of gigabases sequenced}
    \label{f:rock64-appli}
\end{figure}

The proposed architecture that realises portable real-time nanopore-based methylation detection systems has potential applications such as tissue classification,
diagnostic tests, environment, age, etc.  The basis for such an application is in Fig. \ref{f:rock64-appli} that shows how the methylation frequency changes with the number of bases sequenced for five different loci on the human genome (TP53, MGMT, BRCA1 and BRCA2 that are genes and  chromosome 22).  The number of bases sequenced (x-axis) is indicative of the time. As observed in the left plot in Fig. \ref{f:rock64-appli}, most of the CpG sites in the genomic locus are covered after around 2 gigabases of sequencing data (the gradient of the curves decreases). The right plot in Fig. \ref{f:rock64-appli}  shows that the methylation frequency across various loci stabilises after around 2 Gbases of sequencing data.

\section{Summary}

A system architecture was proposed for performing a popular DNA methylation detection workflow on a prototype embedded system. The workflow was realised on the proposed architecture by integrating the optimised software versions from previous chapters. The proposed architecture was evaluated using off-the-shelf single-board computers and was demonstrated that performing real-time analysis of nanopore sequencing is possible on an embedded system. Also, it was shown that the performance of the prototype embedded system is surprisingly similar to the performance on an HPC. The prototype system is fully functional and is integrated into the nanopore sequencing facility at the Garvan Institute of Medical research for performing methylation calling of the samples. The system architecture and the associated software for building a replica of the prototype are open-sourced at \url{https://github.com/hasindu2008/nanopore-cluster} and \url{https://github.com/hasindu2008/f5p}. 

%% file: 8-io-opti/main.tex
\chapter[I/O Optimisations]{Optimisation of Nanopore Sequence Analysis for Many-core CPUs}\label{c:ioopti}

\rule{\textwidth}{0.4pt} 
This chapter is prepared to be submitted as a publication in an ACM/IEEE journal/conference: \textbf{H. Gamaarachchi}, H. Saadat, S. Parameswaran, "Optimisation of Nanopore Sequence Analysis Software for Many-core CPUs", to be submitted [in progress], 2020.\\
\rule{\textwidth}{0.4pt}

Nanopore sequencing is a third generation (the latest) genome sequencing technology.  These modern advances in computational genomics are reshaping healthcare through life-saving applications in medicine and epidemiology, where quick turn-around time of results is critical. Nanopore sequence analysis software tools are inefficient in utilising the computing power offered by modern High Performance Computing systems equipped with many-core CPUs and RAID systems. In this chapter, we present a systematic experimental analysis to identify the potential bottlenecks, which reveals that the primary bottleneck is the thread-inefficient HDF5 library used to load nanopore data. We propose multiple optimisation strategies suitable for different practical scenarios to alleviate the bottleneck: 1) a new file format that offers up to  ${\sim}42\times$ I/O performance improvement; and, 2) a multi-process based solution for the scenario when using a new file format is not possible, that offers up to ${\sim}32\times$ I/O performance improvement.

We demonstrate the efficacy of our optimisations by integrating them to the popular \textit{Nanopolish} toolkit. Our experiments using a representative nanopore dataset demonstrate that the proposed optimisations enable improved  scaling of overall-performance with the number of threads (${\sim}6.5\times$ for 4 vs. 32 threads). Moreover, they also lead to overall-performance improvement (${\sim}2\times$ for 4 threads and ${\sim}6.5\times$ for 32 threads) and improved CPU utilisation (from 69\% to 99\% for 4 cores and from 22\% to 85\% for 32 threads)  for a given number of threads, when compared to the original \textit{Nanopolish}.

\section{Introduction}\label{s:io-intro}

% previously
%Nanopore sequencing has become mainstream with more and researchers increasingly adopting the technology. However, processing raw Nanopore signal data is extremely time consuming and is a well known fact amongst the bioinformatics community. It might be tempting for biologists to acquire more and more powerful resources, for instance, many CPU cores to process this data as fast as possible. Unfortunately, existing software tools are not necessarily capable of efficiently utilising a larger amount of cores. Hence, the overall execution time on such an expensive server might be the same as on a less expensive workstation. We present a motivational example based on a popular state-of-the-art Nanopore raw data analysis application called \textit{Nanopolish} \cite{simpson2017detecting} (Nanopolish Methylation Calling in particular).

Computational genomics has turned a new chapter in medical sciences and epidemiology \cite{scholz2016strain,riba2019big}. It enables promising applications such as accurate disease diagnosis, identifying genetic predisposition, and precision medicine \cite{computational-genomics}.
\emph{Genome sequencing} converts the genetic and biological information encoded in DNA molecules into computer readable data, which is typically hundreds or thousands of gigabytes in size. {\em Nanopore sequencing} is a leading third-generation (the latest) genome sequencing technology \cite{lu2016oxford}. Computational genomics software tools analyse the huge amount of data generated by genome sequencing to extract meaningful information for the above applications. 

Quick turn-around time of results in such applications is highly desirable. For instance, quick diagnostics can instantiate immediate treatments. Moreover, rapid results would enable faster tracking of disease spreading in epidemiological applications such as the ongoing Corona virus outbreak~\cite{de2020first}. However, to analyse the enormous amount of data with high speed, genomic computation software tools demand massive computing time. Thus, scientists typically use High Performance Computing (HPC) systems to run these software tools \cite{schmidt2017next}.

A modern HPC system offers significant computational power through many-core CPUs that are to be exploited through parallelism. The major advantage in such systems when compared to an ordinary personal computer is the availability of number of cores in the CPUs. Moreover, HPC systems have RAID storage composed of many disks for higher I/O throughput with the added benefit of reliability \cite{bland2013contemporary}.
%It might be tempting for biologists to acquire more and more powerful resources, for instance, many CPU cores to process this data as fast as possible. 
{\em Unfortunately, the existing software tools for nanopore sequencing are generally not capable of efficiently utilising the large number of cores available in many-core HPC systems, and thus fail to take maximal advantage of the available computing power.}
Consequently, the overall execution time of the applications on an expensive HPC system may not improve significantly when compared to its execution on a less expensive workstation or a personal computer (refer to chapter \ref{c:integration}). 
{\em In this chapter, we present software optimisations in nanopore software tools to enable them to take maximal advantage of the computing power offered by modern many-core HPC systems. }

To demonstrate the problem mentioned above, we present a motivational example using \textit{Nanopolish} \cite{simpson2017detecting}, which is a popular state-of-the-art nanopore raw data analysis toolkit \cite{nanopolish-popular}.

\begin{figure}
    % \centering
    % \includegraphics[width=\columnwidth]{io-opti/figs/iobottleneck.pdf}
    % \caption{Variation of execution time of original Nanopolish with the number of threads}
  \centering
    \begin{subfigure}[!ht]{0.49\linewidth}
      \centering
        \includegraphics[width=\textwidth]{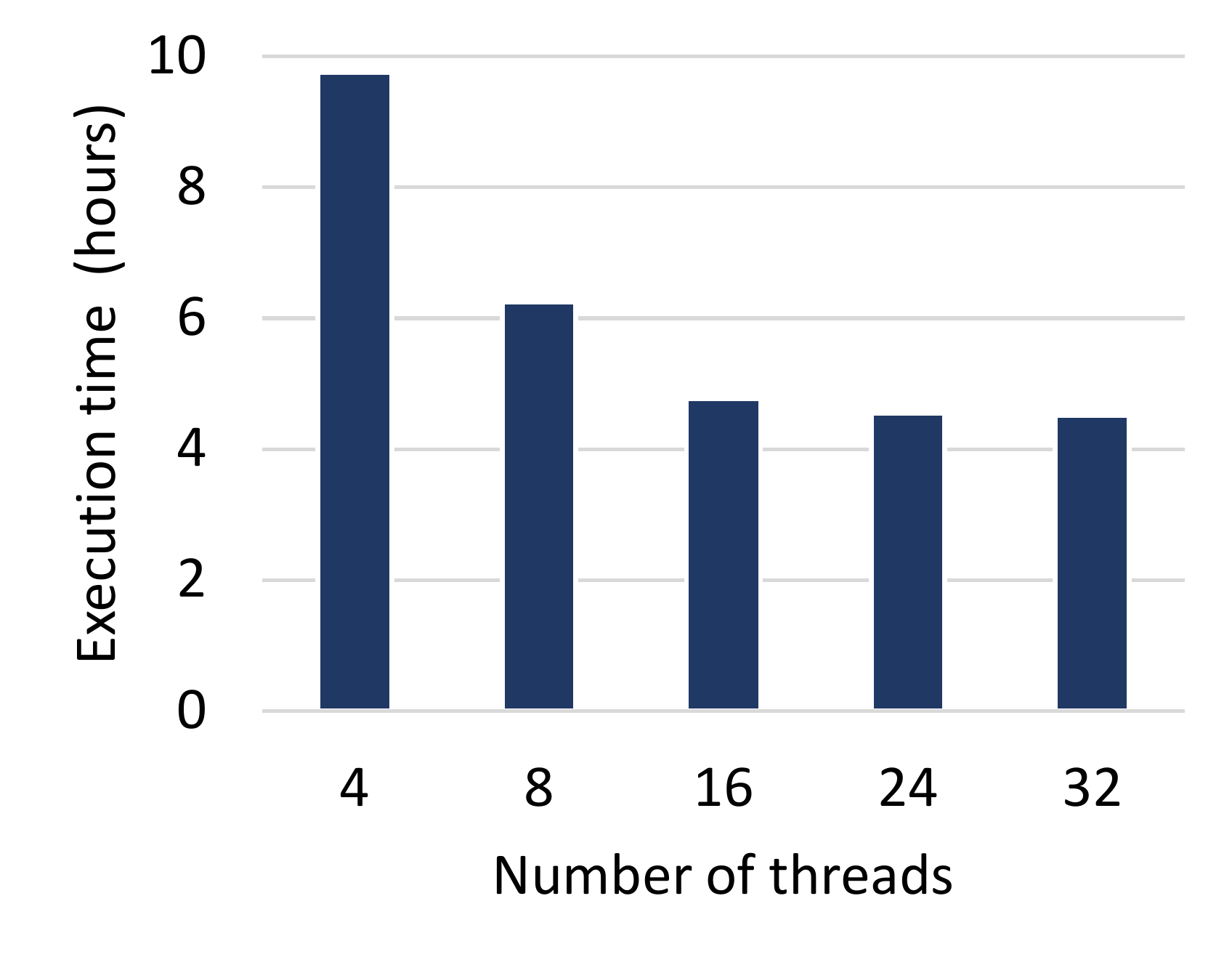}
        \caption{Execution time} 
        \label{f:nanopolish-orig-runtimea}
    \end{subfigure}
    \begin{subfigure}[!ht]{0.49\linewidth}
      \centering
        \includegraphics[width=\textwidth]{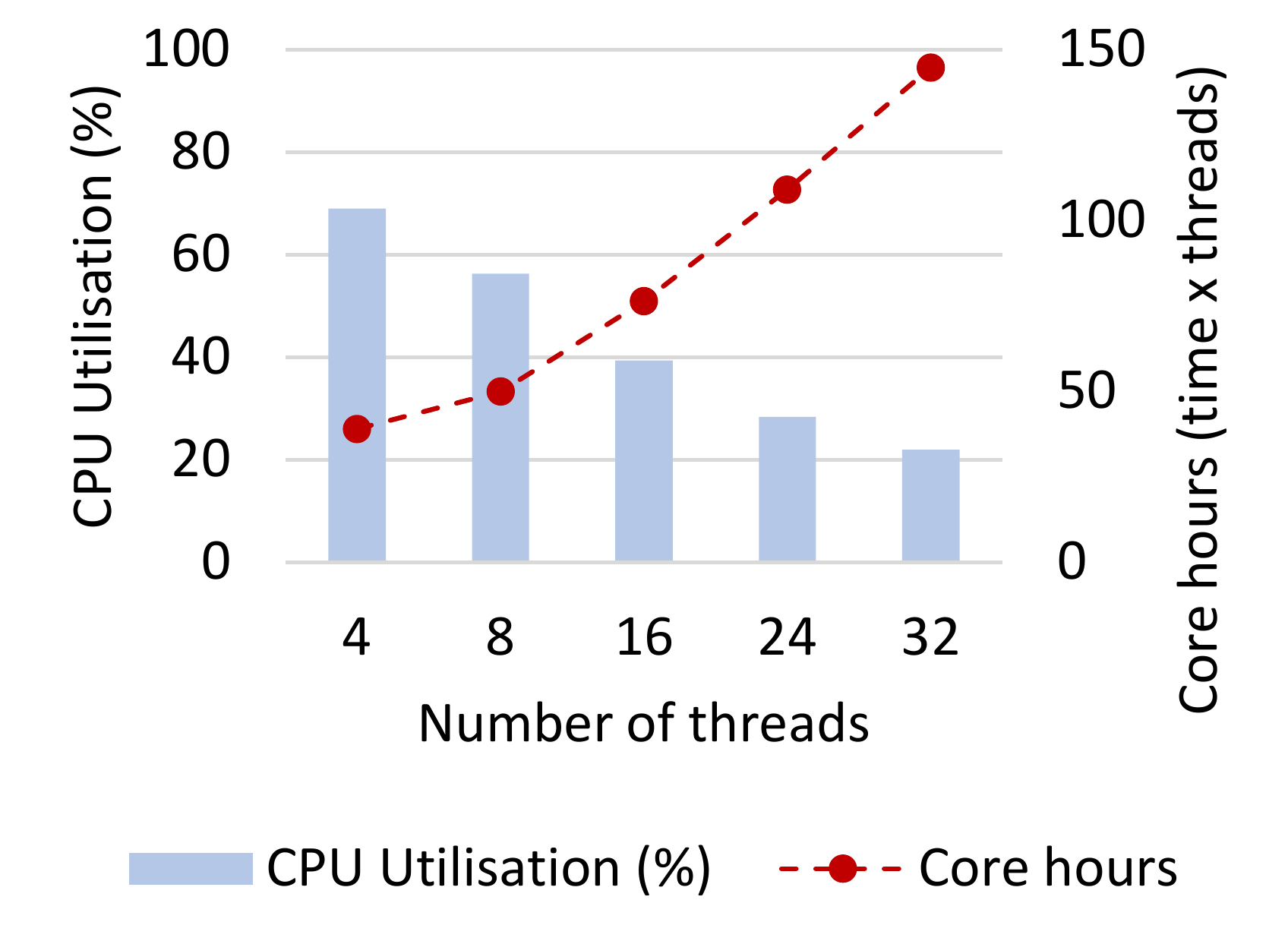}
        \caption{CPU utilisation \& core-hours} 
        \label{f:nanopolish-orig-runtimeb}
    \end{subfigure}
        \caption{Variation of (a) execution time, (b)  CPU utilisation and core-hours in original \textit{Nanopolish} with the number of data processing threads.} 
        \label{f:nanopolish-orig-runtime}    
\vspace{-3mm}
\end{figure}

\textbf{Motivational Example:\footnote{Refer to appendix \ref{a:ioopti} for another example on another dataset}}
We executed the \textit{call-methylation} tool in \textit{Nanopolish} toolkit on a representative dataset{\footnote{See the experimental setup under results for details of the dataset.}}. The experiment was performed on a high-end HPC system with 36 Intel Xeon cores{\footnote{See system S1 in Table \ref{t:systems-slow5} for the specification of the HPC system.}} using different number of threads. The graph in Fig. \ref{f:nanopolish-orig-runtime}a plots the execution time (y-axis) for \textit{Nanopolish}  against the number of threads (x-axis). We observe that when the tool is executed with four threads, the execution-time is nearly 10 hours. The execution-time does not improve significantly with increasing number of threads, and there is little improvement beyond 16 threads.
%LAPTOP%Moreover, the improvement when compared with a personal computer (PC) is less than 2X even with 32 threads.

To analyse further, Fig. \ref{f:nanopolish-orig-runtime}b plots the CPU utilisation{\footnote {CPU utilisation is calculated as in results.} (left y-axis) 
and the core-hours{\footnote{Core-hours is inspired by the common term \textit{man-hour}. It is equal to the product of the number of hours and the number of cores/threads \cite{core-hours}.}} (right y-axis)
for each case in the above experiment.
The CPU utilisation, for execution with four threads, is less than ideal (69\%). Moreover, as the number of threads increase, the CPU utilisation decreases significantly. Specifically, when executed with 32 threads, the CPU utilisation is as low as 22\%. We also observe that the core-hours (which should be constant with the number of threads in an ideal case) increase significantly, and hence depicting that employing greater number of threads is inefficient and not highly beneficial.}

%***************************************************************************
%***************************************************************************

Thus, procuring an HPC system with a higher number of CPU cores might not be beneficial for achieving quick turn-around time of results for nanopore software tools, and there is a need for software optimisations in nanopore software tools to exploit the available resources in HPC systems.  %In this chapter, we aim to address this challenge.
  {\em To this end, in this chapter, we first present a systematic experimental analysis to identify the potential bottlenecks that hinder the efficient utilisation of CPU resources in nanopore software tools. Then we present multiple optimisations--suitable for different practical scenarios--to overcome these bottlenecks and enable performance improvements.} 
Our experiments using the state-of-the-art \textit{Nanopolish} toolkit on HPC systems demonstrate that our proposed optimisations  enable improved CPU utilisation and hence improved performance scaling with the number of threads (Fig.~\ref{f:slow5-overall} and \ref{f:iop-overall}).
Moreover, they also enable improved performance for a given number of threads with respect to the original \textit{Nanopolish}. For example, for 32 threads, the CPU utilisation increases up to ${\sim}$85\% (which was 22\%), and a ${\sim}6.5\times$ speed up is achieved when compared to the original \textit{Nanopolish}. We believe that such improved performance will facilitate fast diagnostics and rapid epidemic response.

\smallskip
\textbf{Contributions:}
The key novel contributions of this chapter can be summarised as follows.
\begin{itemize} [leftmargin=*]
\item We, for the first time, present a  systematic analysis to identify the potential  bottlenecks in nanopore software tools. The analysis reveals that the primary bottleneck is caused by a limitation in an underlying library (HDF5) that serialises disk accesses from multiple threads
%, consequently  limiting the benefit of multiple disks \todo{(and essentially multiple cores)} in an high performance computer 
(Section~\ref{s:bottleneck}).
\item We propose an alternate file format (SLOW5) that alleviates the bottleneck by allowing random accesses from multiple parallel threads.  The proposed file format is designed by exploiting the domain knowledge of nanopore sequencing (Section~\ref{s:slow5}).
\item In some scenarios, it may not be practically possible to use a new file format. Therefore, we present a second solution based on multi-processes. This solution alleviates the bottleneck without requiring any modification to the existing file format (Section~\ref{s:multiproc}).
\item We demonstrate that the new multi-FAST5 file format--which is projected as a replacement of the existing FAST5 file format by the research community--also suffers from the same bottleneck, thus our proposed SLOW5 format is superior.
Moreover, our multi-process based solution is also effective in alleviating the bottleneck in multi-FAST5 (Section~\ref{s:comparisonnew}).
\end{itemize}

\textbf{chapter Organisation:}
Section~\ref{s:io-related} discusses the background and related work. Section~\ref{s:bottleneck} elaborates our analysis for identifying bottlenecks and its explanation. Our proposed optimisations and solutions are presented in Section~\ref{s:opti}. Section~\ref{s:res} presents our experimental setup and results. Finally, Section~\ref{s:io-discussion} is the discussion and the chapter is concluded in Section~\ref{s:io-conclusion}.

%==========================================================================================
\section{Background} \label{s:io-related}
%===========================================================================

\subsection{I/O}

Two types of I/O: synchronous I/O (blocking I/O) and asynchronous I/O (non-blocking I/O) are in the context of random accesses (opposed to sequential/streaming access) are discussed in subsections \ref{s:syncio} and \ref{a:syncio}, respectively.

\subsubsection{Synchronous I/O}\label{s:syncio}

Synchronous I/O is convenient to be programmed, and such programmed code are legible. Thus synchronous I/O is the most popular and predominantly used amongst typical programmers. Following is a simplified account of how random disk requests are served in a modern operating system. 

Consider a single-threaded program that requests I/O using standard \textit{read} or \textit{write} system calls (buffered read/write API calls such as \textit{fwrite}, \textit{fread}, \textit{fprintf}, \textit{getline}, etc are eventually mapped to these system calls). These system calls are synchronous calls which return when the requested data is read from the disk.  

\begin{figure}
    \centering
    \includegraphics[width=\textwidth]{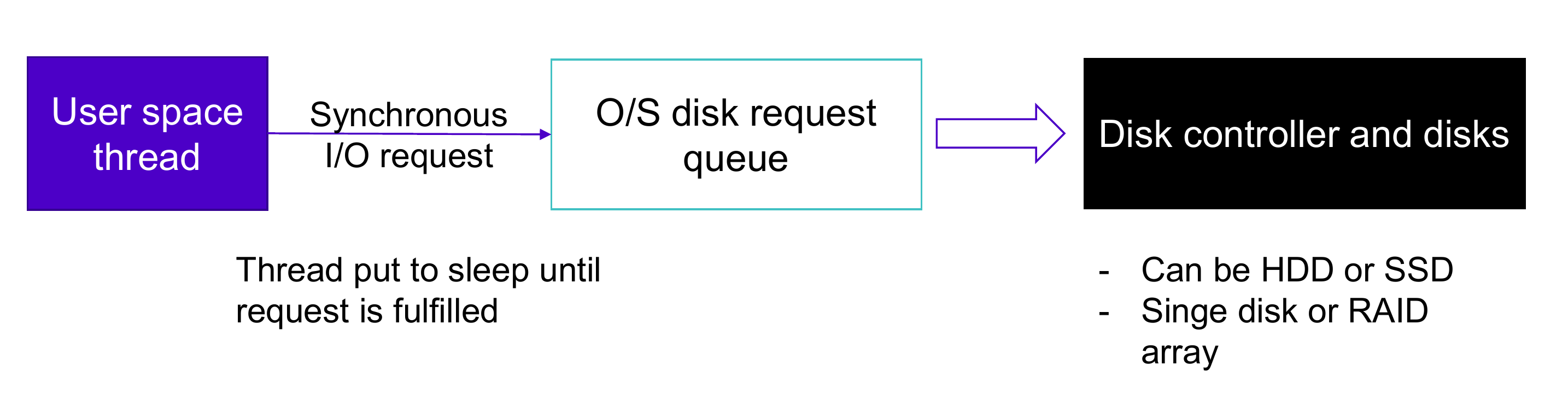}
    \caption{Elaboration of synchronous I/O}
    \label{f:sync_io}
\end{figure}

In Fig. \ref{f:sync_io}, the user-space thread is performing a synchronous I/O request. The operating system receives the system call and queues the disk request in its disk request queue. Momentarily, the user-space thread is put to sleep by the operating system, since a disk request is expected to take hundreds of thousands of CPU clock cycles. The operating system will schedule the disk request (assign to the disk controller) based on policies and priority levels imposed. The disk controller will perform the operation and the operating system will wake up the thread, once requested data reading is completed. If the disk system has a single disk, effectively one request can be served at a time\footnote{as the discussion is about random accesses, disk request merge operations are infrequent}. If the disk system has $K$ disks, up to $K$ requests may be served simultaneously, depending on the RAID level; i.e. $K$ simultaneous parallel reads are possible on a RAID 0 system with $K$ disks.

Consider a program with a single thread requesting  I/O as shown in Fig. \ref{f:sync_io}. Let $t$ be the average disk request service time (from the time of the system call to when the thread is woken up). Let a single thread be requesting $n$ synchronous disk reads sequentially. Despite the number of requests $n$, the total time spent on disk reading $T$ is : $ T = t \times n $. 

\begin{figure}
    \centering
    \includegraphics[width=\textwidth]{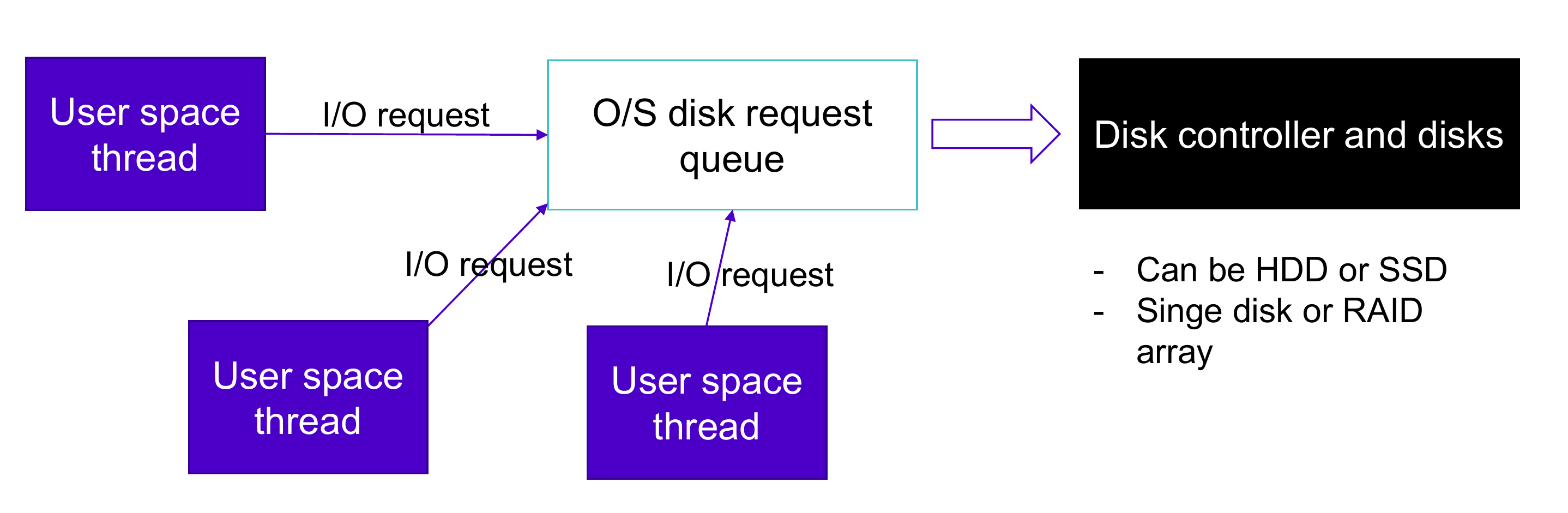}
    \caption{Elaboration of multi threaded synchronous I/O}
    \label{f:sync_io_mt}
\end{figure}

Now consider a program with multiple threads requesting I/O (I/O threads) as shown in Fig. \ref{f:sync_io_mt}. One thread put to sleep due to an I/O request, does not limit other threads from requesting I/O . Thus, if we launch $K$ I/O threads and if the disk controller can serve $K$ requests in parallel, the total time for disk reading is $T'$ : $ T' = t \times \frac{n}{K}$

The scenario in Fig. \ref{f:sync_io_mt} is achieved by programs where threads are having an independent code path - where each processing thread independently performs disk accesses on demand. However, in programs that perform data processing batch by batch, where one single thread reads a batch of data from the disk and assigns to multiple processing threads to be processed in parallel, it is the scenario in Fig. \ref{f:sync_io}.

%Thus, to exploit multiple disks in parallel, asynchronous I/O which we discuss below is needed (note that we are focusing on random accesses, not streaming disk accesses). 

\subsubsection{Asynchronous I/O}\label{a:syncio}

Asynchronous I/O is pertinent to highly responsive applications like web servers and database servers. In asynchronous I/O, the system call that requests I/O will return immediately. The user-space thread won't be put to sleep and this can continue to submit another I/O request or execute some other task while the disk request is being served. 

\begin{figure}
    \centering
    \includegraphics[width=\textwidth]{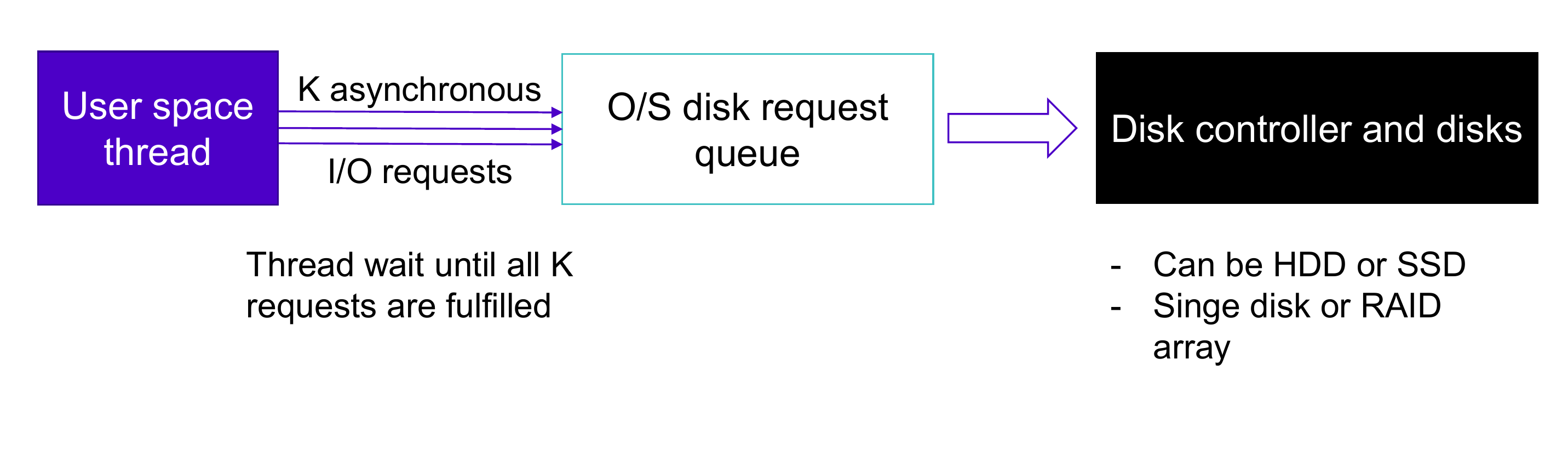}
    \caption{Elaboration of asynchronous I/O}
    \label{f:async_io}
\end{figure}

Consider the asynchronous I/O example in  Fig. \ref{f:async_io} where a single thread submits multiple I/O requests to the operating system simultaneously. In Fig. \ref{f:async_io}, the single user space thread submits K I/O requests in parallel. Then the thread can either poll or wait for a notification from the operating system for I/O request completion. Assume we have $n$ total disk requests to be performed. If the disk system can perform $K$ accesses in  parallel and if the time for a single disk accesses  is $t$ ($K$ parallel accesses take $t$ as well), the total time $ T'= t \times \frac{n}{K}$. Note that the time is the same as for Fig. \ref{f:sync_io_mt}. 

This type of asynchronous I/O is suitable when a program performs reading data and processing batch by batch where one thread performs I/O and then assigns multiple threads or to an accelerator card (eg: GPU to be processed). This is in contrast to independently processing threads we discussed under synchronous I/O above, as threads need to converge in this case.

Asynchronous I/O can be performed by: (1) native synchronous I/O system calls in the operating system or (2)  a library that emulates asynchronous I/O through a thread pool that use synchronous I/O system calls in the operating system. 

From the two methods above, (1) allows `real' asynchronous I/O, but only if supported by the operating system. Early Linux kernels (before version 2.5) did not have native asynchronous I/O systems calls, however, they are available in modern Linux kernels starting from version 2.5 onwards \cite{fixing-async}. Despite that, asynchronous I/O implementation in the Linux kernel has been a controversial topic amongst Linux developers \cite{add-support}, is complicated \cite{fixing-async}, have various drawbacks \cite{epoll-alternative,fixing-async} and does not support certain file systems such as NFS \cite{nonblocking-async-io}. GNU C Library (Glibc) does not provide wrapper functions for asynchronous I/O system calls \cite{iosetup}. Instead the programmers must use low-level system calls which are not easy and non-portable (Architecture specific). Third party libraries such as \textit{libaio} \cite{libaio} which uses Linux native asynchronous I/O system calls have attempted to provide an abstract interface. 

An example of the method (2) above is the current Portable Operating System Interface (POSIX) compatible asynchronous I/O (AIO) library provided by Glibc. POSIX AIO implementation in GlibC is provided in the user-space and uses multiple threads \cite{aio-glibc}. The developer of POSIX AIO have admitted that their approach is expensive and  have scalability issues which are expected to be fixed in the future through a state-machine-based implementation of asynchronous I/O \cite{aio-glibc}. Further, POSIX AIO is not implemented in all systems (eg: Windows subsystem for Linux) 

While the POSIX AIO is suitable for typical I/O loads, the programmers can also spawn multiple I/O threads per batch and assign the disk accesses amongst them. This is suitable if the batch size is big and the thread spawning time is small compared to the I/O time of the batch.

\subsection{Nanopore raw data analysis}

\textbf{Genomics:} 
DNA is a molecule composed of a long strand of millions of units called \textit{nucleotide bases} (or simply called \textit{bases}). Genome sequencers read a DNA strand in relatively smaller fragments (around 10,000 bases long in nanopore) and converts them into digitised data, termed as \textit{reads}\cite{lu2016oxford} in the domain of bioinformatics. In this chapter we refer to them explicitly as \textit{genomic reads} to avoid confusion with disk reads. 

\textbf{Nanopore Sequence Analysis:} Nanopore sequencing is a leading third-generation (the latest) sequencing technology~\cite{lu2016oxford}.
Oxford Nanopore Technologies (ONT) is the company that produces nanopore sequencers. A nanopore sequencer is composed of an array of pico-ampere range current sensors that measure the ionic current disruptions when DNA fragments pass through  nanometer scale protein pores \cite{lu2016oxford,jain2016oxford}. The raw sensor output for a \textit{genomic read} is a time series current signal and is referred to as \textit{raw signal} or \textit{raw data}. ONT stores the raw signal and other metadata (e.g., sampling frequency) in a file format called FAST5~\cite{fast5}. FAST5 is essentially a Hierarchical Data Formats 5 (HDF5)~\cite{hdf5} file with a specific schema defined by ONT.  The only existing library for accessing HDF5 format is the official library developed and maintained by the non-profit organisation HDF Group \cite{hdfsucks1,hdfsucks2}.

\textbf{\textit{Nanopolish}:}
Raw data analysis toolkits analyse the sequencer outputs using complex algorithms and extract meaningful information.
\textit{Nanopolish} is currently a popular state-of-the-art nanopore raw data analysis toolkit. 
\textit{Nanopolish} is used in a number of genomic workflows such as methylation detection~\cite{simpson2017detecting}, variant detection~\cite{leija2019evaluation}, draft genome polishing~\cite{loman2015complete,jain2018nanopore} and real-time molecular epidemiology for the ongoing Corona Virus outbreak~\cite{de2020first}. \textit{Nanopolish} is written in C/C++ and supports multi-threaded execution through OpenMP. It is an open-source toolkit with a large and complex codebase~\cite{simpson2017detecting,nanopolishcode}.

\textbf{Previous Work on Optimising \textit{Nanopolish}:}  
Nanopore sequence analysis is a relatively new field that only emerged in the last decade. Thus, optimisation efforts to improve performance of  nanopore software tools are rare. 
In \textit{Nanopolish}, calculation of log likelihood ratio is a predominantly used CPU intensive computation kernel \cite{simpson2017detecting}. To reduce CPU time for log likelihood computation, \textit{Nanopolish} authors have already employed a fast table-driven log-sum implementation established elsewhere in \cite{eddy2011accelerated}. 
%Also, a compute-intensive core component in \textit{Nanopolish} has been accelerated using GPUs \cite{gamaarachchi2019gpu}. 
However, none of the existing works have focused on  improving the overall performance of \textit{Nanopolish} on HPC systems with many-cores and Redundant Array of Independent Disks (RAID). Our proposed optimisations are orthogonal to the methods discussed above.

%Other genomics file format such as SAM/BAM \cite{li2009sequence}, CRAM \cite{fritz2011efficient} etc that exploits locality. Random accesses to genomic coordinates in a genome using faidx \cite{}. Block based compressed for gzip \cite{}. If there are any HDF5 related optimisations can be out here.

\textbf{Previous Work on Optimising Sequence Analysis:}
Several optimisation efforts exist for the second generation sequencing software (also known as \textit{Next Generation Sequencing}) ~\cite{al2016workflow,kawalia2015leveraging,kathiresan2015optimization,kathiresan2017accelerating,gamaarachchi2018cache,fritz2011efficient}. However, software used for nanopore sequencing (third-generation) is distinct from the first and second generations \cite{amarasinghe2020opportunities}.
Nanopore technology involves processing raw signal data, which is not the case for first and second generation. 

In this chapter, we for this first time, identify the major causes behind the inefficient resource-utilisation by nanopore software tools and present multiple optimisations to alleviate  those issues.

%===========================================================================
\section{Identification and Explanation of bottleneck} \label{s:bottleneck}
%===========================================================================

The motivational example in Section \ref{s:io-intro} revealed that \textit{Nanopolish} is unable to efficiently utilise multiple cores in the system. There can be two reasons for an application to be unable to utilise parallel resources. These are: 1) data processing bottleneck; and/or 2) I/O bottleneck. In this section, we identify and explain that the primary reason of the under-utilisation is I/O bottleneck.

\subsection{Identification of the Bottleneck}
We employed performance monitoring and profiling tools in the motivational example setup, to hypothesise the causes of inefficient resource-utilisation and performance.

{\em Hypothesis-1: The performance of the software tool is bounded by file I/O.}
 We observed through \textit{htop} utility in Linux that majority of \textit{Nanopolish} threads are in the `D' state. The `D' state is defined as the `state of the process for disk sleep (uninterruptible)'~\cite{htop}.  This leads to our first hypothesis that the software tool is bounded by file I/O. In fact, \textit{Nanopolish} incurs a large number of random disk accesses when reading millions of FAST5 files (based on HDF5) in a nanopore dataset\footnote{A nanopore dataset of a single genome sample contains millions of \textit{genomic reads} (fragments of DNA), and each \textit{genomic read} is stored in a separate FAST5 file. Thus, accessing millions of such \textit{genomic reads} incurs millions of random disk accesses (opposed to sequential/streaming access)}.

 {\em Hypothesis-2: The file I/O bottleneck is caused by the HDF5 library and not by the limitation of physical disks.}
We observed the disk usage statistics through the {\em iostat} utility to find that disk system is not fully utilised (i.e., the observed number of disk reads per second was around 100, while the particular disk system could handle more than 1000 IOPS). This implies that I/O bottleneck is not due to the limitation of physical disks to serve data fast enough to saturate the processor. To investigate further, we profiled \textit{Nanopolish} with \textit{Intel Vtune} under \textit{concurrency profiling}. It reveals that the majority of the `wait time' is due to a conditional variable (synchronisation primitive) in the underlying library called HDF5 library (Hierarchical Data Format 5--used to access raw nanopore data stored in FAST5 file). A closer look into the HDF5 library revealed that the thread-safe version of the HDF5 library serialises the calls for disk read requests~\cite{hdf-thread-efficiency}. Thus, we hypothesise that the reduced CPU utilisation is caused by the disk requests being serialised by the HDF5 library, consequently causing the bottleneck and limiting the utility of a multi-disk RAID system.

\begin{figure}[!t]
  \centering
\includegraphics[width=\linewidth]{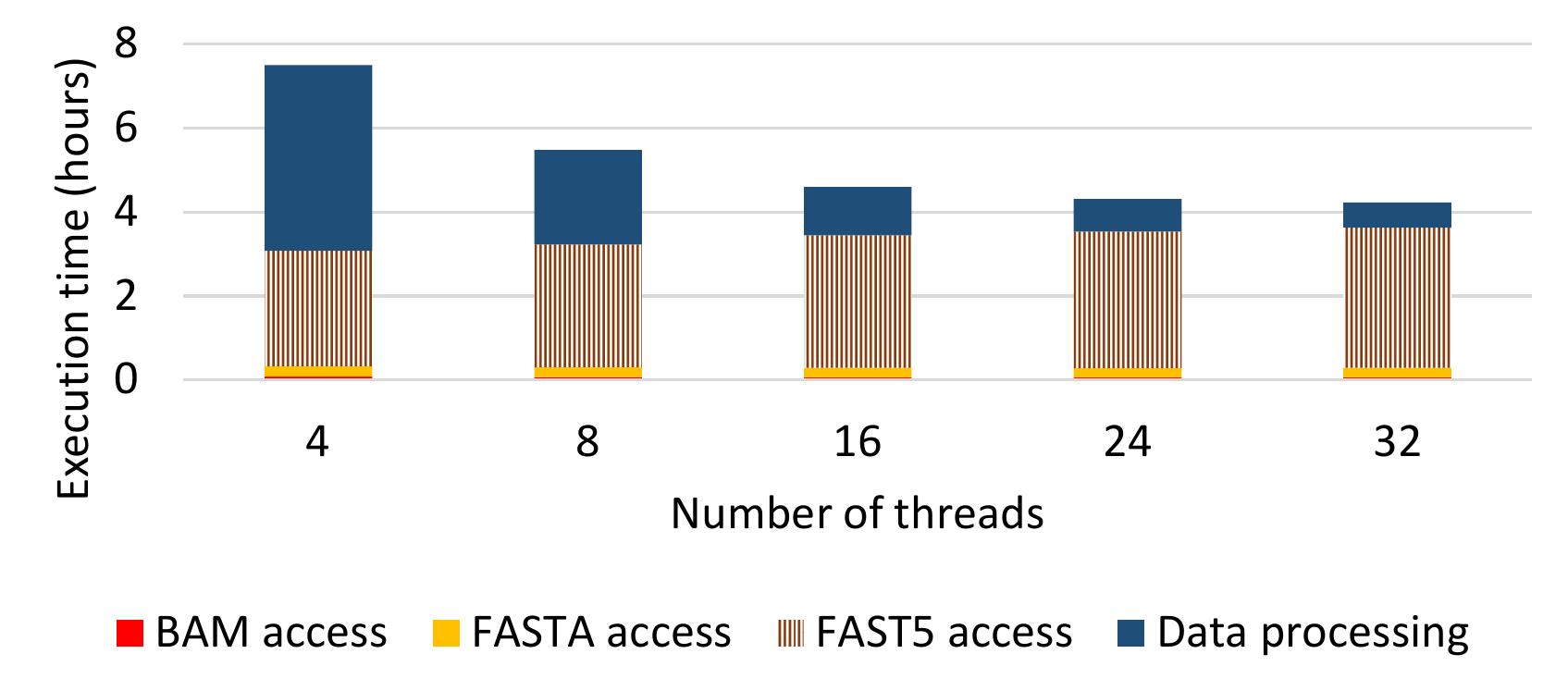}

    \caption{Decomposition of time for individual components in restructured \textit{Nanopolish}.} 
    \label{f:hdf5_thread_inefficiency}
\vspace{-3mm}    
\end{figure}
%*****************************************
%-------------------------------------

\subsection{Verification of the Identified Bottleneck}
%-------------------------------------

To verify the above identified cause of the bottleneck, we performed a deeper analysis.
For this, we first restructured \textit{Nanopolish} such that wall-clock time spent on I/O operations and data processing can be separately measured to determine the time spent on individual components in the program.

We run the restructured \textit{Nanopolish} with various number of threads (for FAST5 access and data processing). 
The results are presented in Fig. \ref{f:hdf5_thread_inefficiency}. The x-axis in the figure represents the number of threads used and the y-axis represents the total execution time. Different colours in the bars (see legend) denotes the decomposition of the total execution time into different components\footnote{FASTA access refers to random access to reference genome (stored in FASTA file format) performed using \textit{faidx} component in \textit{htslib} library. BAM access refers to sequential access performed through \textit{htslib} library to the genomic alignment records (stored in BAM file format)}.

We observe from Fig. \ref{f:hdf5_thread_inefficiency} that: 
1) the contribution by the BAM access and FASTA access to the overall execution time is negligible;  
2) a major portion of the time is consumed by FAST5 access (patterned brown); 
3) time consumption of FAST5 access (patterned brown) does not improve with increasing number of threads; and,
4) data processing time improves with increasing number of threads (solid blue).
This confirms that the bottleneck is caused by file I/O and not because of any data processing bottlenecks.

\subsection {Detailed Explanation of the Identified Bottleneck}

In this subsection, we explain the  major limitation in HDF5 library  that prevents efficient parallel accesses, consequently causing the  bottleneck.

\textbf{HDF5 Library and its Limitations in Thread Efficiency:} 
HDF5 library uses synchronous I/O calls and even the latest HDF5 implementation (HDF5-1.10) does not support asynchronous I/O{\footnote{In synchronous I/O calls, the OS, upon receiving the call puts the user-space thread to sleep and the thread can no longer submit I/O requests until the disk reading is completed and woken by the OS. Conversely, asynchronous I/O system calls return immediately without the thread being put to sleep and the thread can continue to submit another asynchronous request.}}. This, by itself, is not an issue as multiple synchronous I/O operations can be performed in parallel using multiple I/O threads to exploit the high throughput of RAID systems in HPC systems. 
Fig. \ref{f:sync_io_mt} demonstrates how multiple I/O threads can be used to perform parallel disk accesses using synchronous I/O. Suppose the disk system has $K$ disks, up to $K$ requests may be served simultaneously depending on the RAID level; i.e., $K$ simultaneous parallel reads are possible on a RAID 0 system with $K$ disks. 
Let $t$ be the average disk request service time (from the time of the system call to when the thread is woken up). 
For a program that launches $K$ I/O threads and if the disk controller can serve $K$ requests in parallel, the total time for $n$ disk reads is $T' = t \times \frac{n}{K}$.
%*************************************************

However, the HDF group (that maintains the HDF5 library) mentions that the thread-safe version of the HDF5 library is not thread efficient and that it effectively serialises the calls for disk read requests~\cite{hdf-thread-efficiency}. The global lock in the thread safe version of the HDF5 library creates limitations.
Following is an extract from the FAQ section of the HDF web site~\cite{hdf-thread-efficiency}.
``Users are often surprised to learn that (1) concurrent access to different datasets in a single HDF5 file and (2) concurrent access to different HDF5 files both require a thread-safe version of the HDF5 library. Although each thread in these examples is accessing different data, the HDF5 library modifies global data structures that are independent of a particular HDF5 dataset or HDF5 file. HDF5 relies on a semaphore around the library API calls in the thread-safe version of the library to protect the data structure from corruption by simultaneous manipulation from different threads. Examples of HDF5 library global data structures that must be protected are the freespace manager and open file lists.''
 
\begin{figure}
    \centering
    \includegraphics[width=\columnwidth]{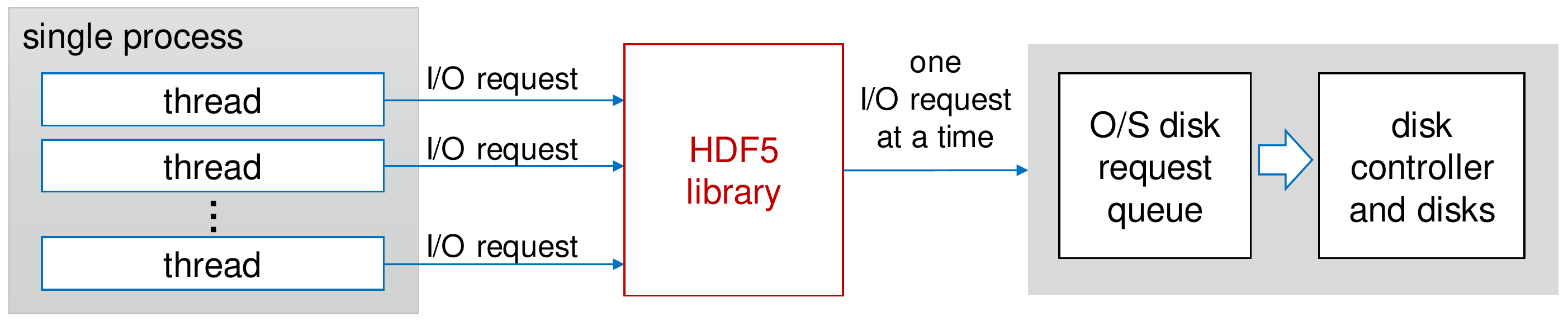}
    \vspace{-3mm}  
    \caption{Elaboration of the limitation in HDF library.}
    \label{f:hdf_hell}
    %\vspace{-3mm}  
\end{figure}

 Thus, in spite of having multiple I/O threads, I/O requests for HDF5 files have to go through the HDF5 library. Fig. \ref{f:hdf_hell} demonstrates this, where  $K$ I/O threads are requesting I/O from the HDF5 library in parallel. However, the lock inside the HDF5 serialises the parallel requests, effectively issuing only one request at a time to the operating system disk request queue. The operating system will put the thread to sleep and this is equivalent to a single I/O thread. Thus, the total time spent on disk accesses $T$ will be $T = t \times n$, and essentially, the high throughput capability of multiple disks in a RAID configuration is under-utilised.

\section{Proposed Optimisations}\label{s:opti}

In this section, we present two types of solutions to overcome the bottleneck in nanopore software tools. The first approach is to use an alternate file format (Section \ref{s:slow5}). 
However, current nanopore software tools have been developed on top of the FAST5 (HDF5) format because of its adoption by Oxford Nanopore technologies as the file format for storing the raw signal. Thus, using a new file format may not always be practical. For such scenarios, we present a second solution that uses multi-processes instead of multi-threads for I/O operations (Section \ref{s:multiproc}). This second solution does not require any changes to FAST5 format or the HDF5 library. Furthermore, we also present a few more optimisations to \textit{Nanopolish} that enable further speed-up (Section \ref{s:restructuring}).

%===================================================================
\subsection{Alternate File Format (SLOW5)}\label{s:slow5}

We propose a new file format called SLOW5\footnote{{The name SLOW5 is ironical to FAST5}} for storing nanopore raw signal data as an alternate to FAST5. We considered the domain knowledge from nanopore sequence analysis and the characteristics of disk accesses to design the new file format.

\textbf{SLOW5 File Format:} 
We design our proposed SLOW5 file format by extending the simple and well-known tab-separated values (TSV) format, using inspiration from the gold standard genomic file formats such as SAM \cite{li2009sequence} and VCF \cite{danecek2011variant}. An example of the proposed file format is shown in Table \ref{t:slow5-format}.
The structure of the file is explained below.

The first set of lines of the SLOW5 file comprises the file header. Each header line starts with the character $\#$. 
Generic metadata such as the file version and global metadata of the sequenced genome sample are also stored in the header. The global metadata  is common to all the \textit{genomic reads} and contains information such as the sequencing flow-cell identifier, and sequencing run identifier, etc.
The last line in the header gives the column names of the upcoming data, which are tab-separated. Note that, not all metadata and data fields are depicted in Table \ref{t:slow5-format} for the sake of brevity.

The header is followed by data where each line (row) represents a single \textit{genomic read}. In other words, for {\em N} \textit{genomic reads}, there are {\em N} data lines in the file. 
The \textit{genomic read} information fields (e.g., read-identifier,  number of signal samples, and the raw signal)  are tab separated and are in the same order as defined in the last line of the header. The \textit{raw\_signal} column contains the current signal values separated by commas. Note that all data corresponding to a single \textit{genomic read} are placed contiguously in the same row, thus facilitating locality in disk accesses.

\textbf{Working Explanation:} Random accesses to the \textit{genomic read} records in a SLOW5 file are facilitated by an index called the SLOW5 index. The SLOW5 index is another tab-separated file as shown in Table \ref{t:slow5-index}. Each line corresponds (except the first header line) to a \textit{genomic read}. The first column is the read-identifier of the \textit{genomic read}, the second column is the file offset to the corresponding SLOW5 record, and the third column is the size of the corresponding SLOW5 record in bytes (including the new line character). For performing random disk accesses to SLOW5, the SLOW5 index is first loaded to a hash table in RAM where the read-identifier serves as the hash table key and the rest of the data is used as hash table values. For a given read-identifier, the file offset and the record length is obtained from this hash table and the program can move the file pointer to the offset (i.e. using \textit{lseek}) and load the record to the memory. {Multiple random accesses to SLOW5 can be performed in parallel either through synchronous I/O calls with multiple threads, or through asynchronous I/O if supported by the operating system. 
Note that the raw signal data is read-only during nanopore sequence analysis.
Thus, SLOW5 is inherently thread-safe without any need of global locks.}

\subsection{Multi-process based Solution \label{s:multiproc}}

The original \textit{Nanopolish} runs a single process with multiple threads.
We propose a multi-process based solution for scenarios where the existing FAST5 cannot be replaced. 
Multiple threads in a single process share same address space and thus the lock in HDF5 library affects multiple threads.
Multiple threads are typically used to run sub-tasks in parallel while conveniently sharing data amongst the threads. In contrast, multiple processes have their own independent address spaces and are typically used to run isolated tasks in parallel.
We exploit the presence of independent address spaces in multiple processes to circumvent the lock in HDF5 library.
%*************************************************
\begin{figure}[!t]
    \centering
    \includegraphics[width=\linewidth]{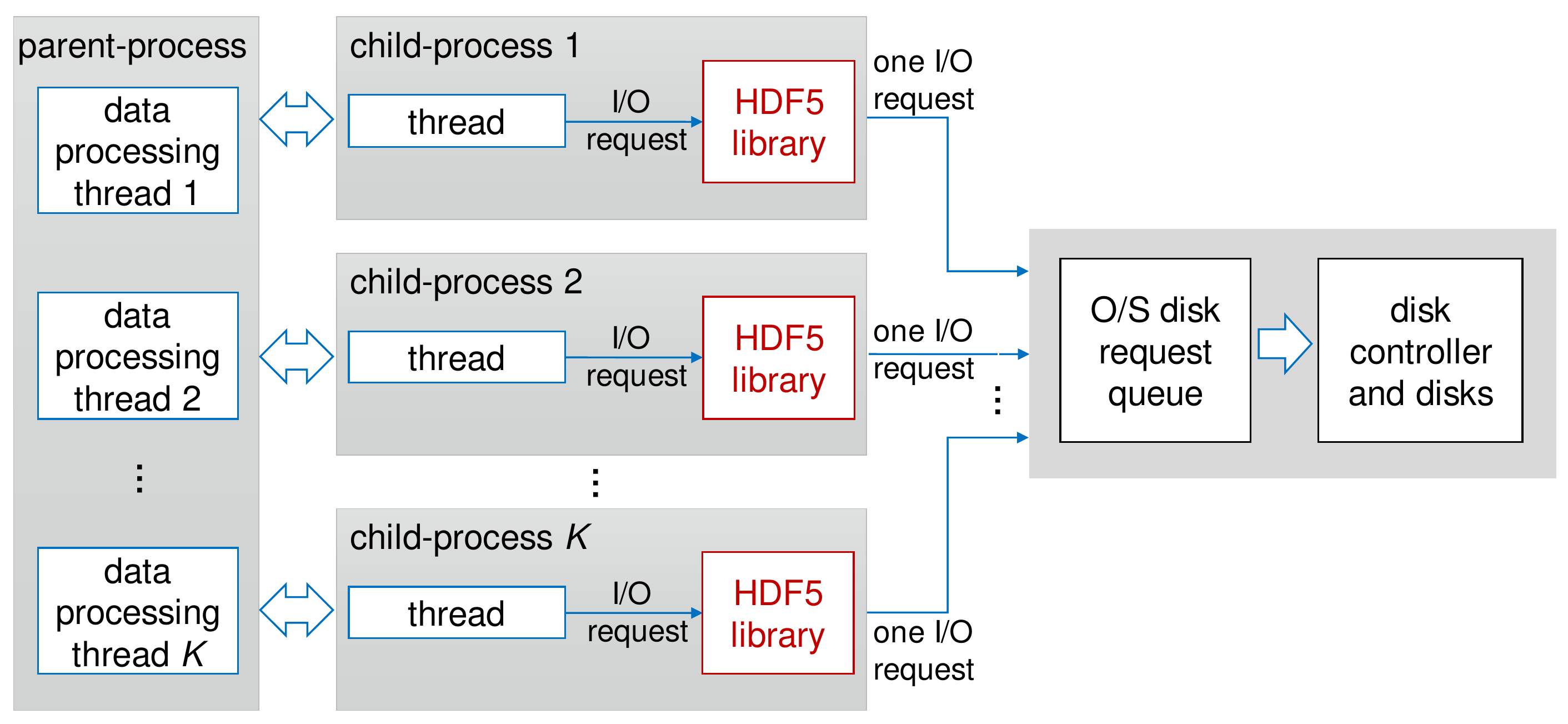}
    \caption{The proposed multi-process based solution.}
    \label{f:hdf5_multiproc}
    \vspace{-4mm}  
\end{figure}
%**************************************************

\textbf{Overview:} 
Our proposed multi-process based solution is elaborated in~ Fig.~\ref{f:hdf5_multiproc}.
We use multi-threads in the single parent-process for data processing and multiple child-processes for~I/O.
The parent-process performs data processing using multiple threads in parallel.
Each {child-}process has its own instance of the HDF5 library, as a consequence of independent address spaces. Moreover, each {child-}process has only a single thread that requests I/O. 
Thus, a single instance of the HDF5 library gets only one request at a time. 
In effect, there are multiple instances of the HDF5 library that can submit multiple I/O requests in parallel to the operating system (as opposed to the situation in Fig. \ref{f:hdf_hell}), thus benefiting from the high throughput offered by the RAID configuration. Formally, if there are $K$ processes and if the disk controller can serve $K$ requests in parallel, the total time spent on I/O operations will be $T' = t \times \frac{n}{K}$ (similar to the case in Fig. \ref{f:sync_io_mt}). 

\textbf{Details:} The proposed multi-process based solution can be adopted for Nanopore data processing using a pool of processes that performs FAST5 I/O. Multiple processes are spawned at the beginning of the program using the \textit{fork} system call. These forked child-processes form a pool of processes that exist until the lifetime of the parent-process, solely performing I/O of FAST5 files. The data processing can be performed by multiple threads spawned by the parent-process as usual. 
The parent-process when it requires to load signal data of $N$ reads (FAST5 accesses), first splits the list of reads to $K$ parts where $K$ is the number of child-processes. Then, each part is assigned to each child-process, which performs the assigned FAST5 accesses. When data is loaded, the child-processes send  data to the parent-process. 
The communication (data transfer) between the parent-process and child-processes can be implemented relatively easily using unnamed pipes out of the available  Inter Process Communication (IPC) techniques (still not easy as threads that share the same memory space). 

\textit{Note-1:} A fork-join model for multi-processes (as could be done for the multi-threading) is unsuitable to be used instead of 
the process pool model presented above. Firstly, creating a process can be very expensive and could easily become the biggest bottleneck than the file reading itself. Secondly, forking in the middle of a program could double the memory usage and is usually problematic.

\textit{Note-2: }
We propose the use of multi-threads in the single parent-process for data processing and multiple child-processes for I/O. The possibility of using separate processes for both data processing and I/O is discussed in Section~\ref{s:othersolutions} with its caveats.

\begin{figure}[!t]
    \centering
    \includegraphics[width=\columnwidth]{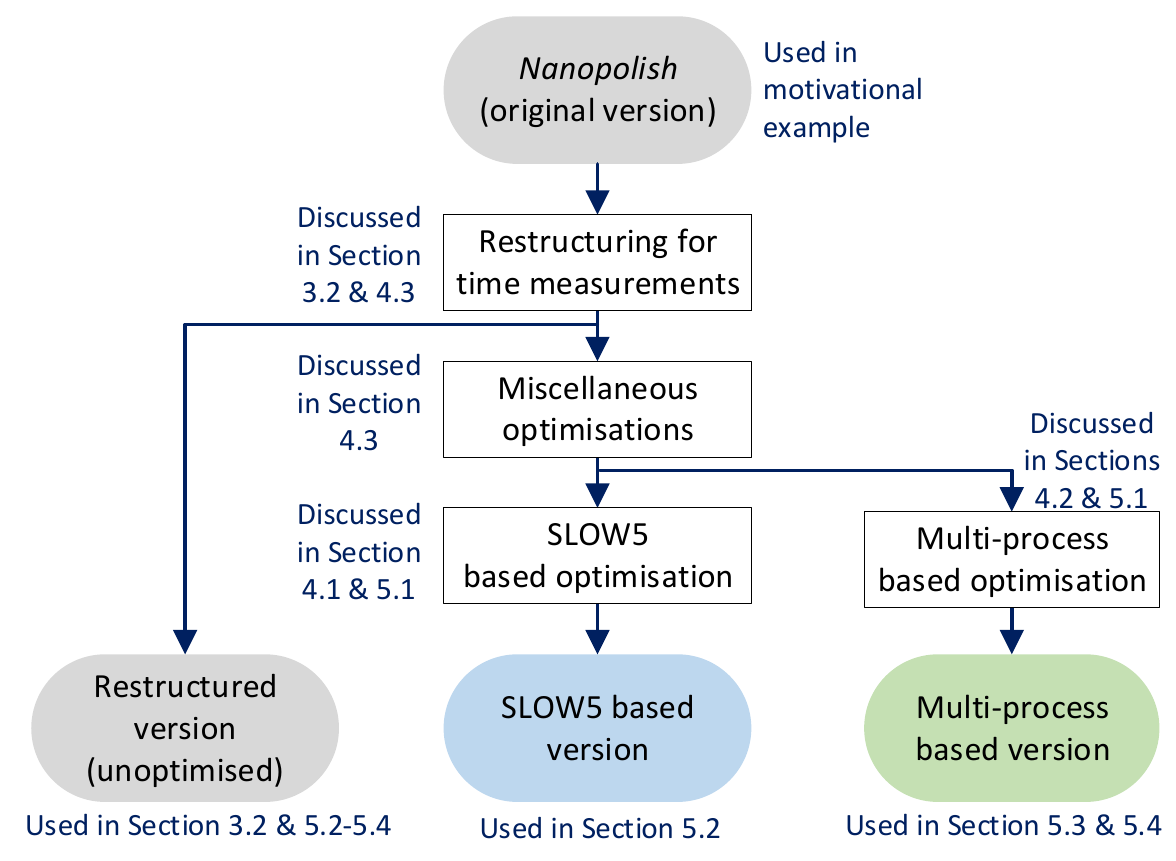}
    \caption{\small Flow diagram depicting modifications to \textit{Nanopolish}}
    \label{f:impletree}
    \vspace{-3mm}  
\end{figure}

\subsection{Restructuring \& Miscellaneous Optimisations} \label{s:restructuring}

In addition to the above I/O related optimisations, we also performed restructuring and a few other software optimisations with respect to multi-threading and memory. 
Our restructuring allows us to measure the execution time separately for I/O (including the execution time breakdown for different file formats) and data processing, without significant effect on performance, whereas, the software optimisation improve the processing time.

The original \textit{Nanopolish} implementation uses openMP for multi-threading. We restructured \textit{Nanopolish} to perform multi-threading using a lightweight fork-join model with work-stealing implemented using POSIX Threads (\textit{pthreads}).
Moreover, the restructured \textit{Nanopolish} performs I/O operations and data processing batch by batch (batch of \textit{genomic reads}), i.e., a batch of \textit{genomic reads} are loaded from the disk and the batch is then processed, subsequently, results of the batch are written to disk. 
I/O operations are interleaved with data processing, i.e. when the first batch is being processed, the second batch will be loaded from the disk. 

The restructured \textit{Nanopolish} was further optimised with strategies such as: reducing the number of memory allocations (\textit{malloc}) for dynamic 2D arrays by allocating a 1D array, an appropriate batch size that fits the available RAM, and a better load-balancing between multi-threads, etc.  
While space limits our ability to explain each of these optimisations, the details can be found in the open sourced code of this research project. For the sake of clarity, the overview of our restructuring and optimisations to original \textit{Nanopolish} and various resulting versions with their usage, are shown in~Fig.~\ref{f:impletree}.

\begin{table*}[!b]
\centering
\caption{Example of SLOW5 file format}\label{t:slow5-format}
\footnotesize
\begin{tabular}{|llllllc|}

\hline

\multicolumn{7}{c}{\textbf{SLOW5 file format}} \\
\cline{1-7}
\multicolumn{7}{|l|}{\#fileformat: slow5v1.0}                                                                                                             	\\
\multicolumn{7}{|l|}{\#exp\_start\_time:   2020-01-01T00:00:00Z}                                                                                  			\\
\multicolumn{7}{|l|}{\#run\_id: 855cdb4b26948}                                                                                                              \\
\multicolumn{7}{|l|}{\#flow\_cell\_id:   FAH00000}                                                                                                			\\
\#read\_id & n\_samples & digitisation & offset     & range      & sampling\_rate & \multicolumn{1}{l|}{raw\_signal}                             			\\
read-0     & 123456         & 8192         & 6          & 1467.6     & 100000           & \multicolumn{1}{l|}{498,492,501,508,503,505,509,…} 				\\
{read-1} & {2000} & {8192}   & {5} & {1467.6} & {4000}     & {400,401,500,403,407,478,510,…}                                                  				\\
\textbf{.} & \textbf{.} & \textbf{.}   & \textbf{.} & \textbf{.} & \textbf{.}     & \textbf{.}                                                  			\\
\textbf{.} & \textbf{.} & \textbf{.}   & \textbf{.} & \textbf{.} & \textbf{.}     & \textbf{.}                                                  			\\
\textbf{.} & \textbf{.} & \textbf{.}   & \textbf{.} & \textbf{.} & \textbf{.}     & \textbf{.}                                                  			\\
read-\textit{N}     & 10000        & 8192         & 3          & 1467.6     & 4000           & \multicolumn{1}{l|}{559,545,560,551,550,565,701,…}         	\\               
\cline{1-7} 

\end{tabular}
\end{table*}

\begin{table*}[!b]
\centering
\caption{Example of SLOW5 index}\label{t:slow5-index}
\begin{tabular}{|lll|}
\hline
\multicolumn{3}{c}{\textbf{SLOW5 index}}\\
\cline{1-3}
\#read\_id & file\_offset & rec\_length 				\\
read-0     & 67           & 500000                		\\
read-1  & 500001   & 1000000        					\\
\textbf{.} & \textbf{.}   & \textbf{.}         		\\
\textbf{.} & \textbf{.}   & \textbf{.}         		\\
\textbf{.} & \textbf{.}   & \textbf{.}         		\\
read-\textit{N}     &  364459005610  &    1580072      \\
\cline{1-3}

\end{tabular}
\end{table*}

%====================================================================================
\section{Experiment and Results} \label{s:res}
%====================================================================================
\subsection{Experimental Setup}
%----------------------------------------

\textbf{Implementation of the Alternate File Format:} 
We implemented a C program to convert FAST5 (HDF5) files into our SLOW5 format. The program also constructs a SLOW5 index as per the description in section \ref{s:slow5}. The restructured and optimised \textit{Nanopolish} (discussed in section \ref{s:restructuring}) was modified to support reading from SLOW5 format (Fig. \ref{f:impletree}). At the beginning of the program, the SLOW5 index is loaded onto a hash table that resides in RAM. For each \textit{genomic read} in a batch, the start position of the corresponding SLOW5 record (file offset) and size of the record (in bytes) is obtained from the index. Then, that information for all the \textit{genomic reads} in the batch are submitted as I/O requests. 
The \textit{POSIX AIO} library in \textit{glibc} is used for performing asynchronous I/O.

\textbf{Implementation of the Multi-process Pool:} 
The restructured and optimised \textit{Nanopolish} was modified such that FAST5 files are loaded using a multi-process pool as per the description in Section \ref{s:multiproc} (Fig. \ref{f:impletree}).  
At the beginning of the program, $K$ child-processes are spawned using \textit{fork} system call. Then, during the execution of the program, the parent-process divides the batch of \textit{genomic reads} into $K$ parts and assigns each part to a child-process. Child-processes performs FAST5 file reading (through HDF5 library) in parallel. After completion of reading by the child-processes, the data is collected by the parent-process. 
The inter-process communication is implemented using \textit{unnamed pipes} in Linux.

\textbf{Datasets and Computer Systems:} 

A representative nanopore dataset of the human genome was used for the evaluation and the details are in Table \ref{t:datasets-slow5}. This dataset is a complete nanopore MinION dataset of the T778 cancer cell-line \cite{stratford2012characterization,garsed2014architecture}. 
The computer systems used for the experiments and their specification are given in Table~\ref{t:systems-slow5}. Unless otherwise stated, the experiments in the chapter have been performed on system S1. System S2 was used for limited number of experiments due to the limited availability.  For the experiments associated with Network File System (NFS), the NFS storage on system S3 was mounted on system S1. For NFS, default parameters for the NFS server and client in Linux were used. Note that the operating system disk cache on S3 was also cleared before any NFS experiment.

%*****************************************
\begin{table}[!t]
\caption{Dataset} \label{t:datasets-slow5}
%\vspace{-3mm}  
\centering
\begin{tabular}{ccccccc}
\hline
\textbf{ID} &
\textbf{Sample} & \textbf{\begin{tabular}[c]{@{}c@{}}No. of \\ Gbases\end{tabular}} & \textbf{\begin{tabular}[c]{@{}c@{}}No. of \\ reads\end{tabular}} & \textbf{\begin{tabular}[c]{@{}c@{}}Average \\ read length\end{tabular}} & \textbf{\begin{tabular}[c]{@{}c@{}}Max \\ read length\end{tabular}} & \textbf{\begin{tabular}[c]{@{}c@{}}FAST5 \\ file size\end{tabular}} \\
\hline
\hline
D1 &T778            & 8.787                                                             & 771 325                                                          & 11 393                                                                  & 194 983                                                             & 845GB \\
\hline
\end{tabular}
\end{table}

\textbf{Measurements and Calculations:}  
The measurement and calculations for our results are performed as follows.

\textit{1) The Overall execution time} (wall-clock time) and \textit{the CPU time} (user mode + kernel mode) of the program (all version shown in Fig.~\ref{f:impletree}) were measured by running the program through \textit{GNU time} utility in Linux. 

\textit{2) The CPU utilisation percentage} is computed as in equation \ref{e:cpuusagecompute}. Note that this CPU utilisation percentage is a normalised value based on the number of data processing threads that which the program was executed with. 

\begin{equation}\label{e:cpuusagecompute}
    CPU\:utilisation = \frac{CPU\:time}{execution\:time \times number\:of\:threads} \times 100\%
\end{equation}
% \usepackage{booktabs}
\begin{table}[!t]
\caption{Computer systems} \label{t:systems-slow5}
%\vspace{-3mm}  
\footnotesize
\centering
\begin{tabular}{@{}lccc@{}}
\hline
\textbf{ System ID}          & S1                              & S2                                & S3 \\ 
\hline
\hline
\textbf{ Description}        & HPC with HDD RAID               & HPC with SSD RAID                 & NFS server \\
\textbf{ CPU}                & 2 $\times$ Intel Xeon Gold 6154 & 2 $\times$ Intel Xeon Gold 6148   & 4 $\times$ Intel Xeon X7560 \\
\textbf{ CPU cores}          & 36                              & 40                                & 32 \\
\textbf{ RAM}                & 384 GB                          & 768 GB                            & 256 GB \\
\textbf{ Disk System}        & 12$\times$10TB HDD drives       & 6$\times$4TB NVMe drives          & 10$\times$3TB HDD drives \\
\textbf{ File System}        & ext4                            & ext4                              & ext4 \\
\textbf{ RAID config.}       & RAID6                           & RAID0                             & RAID5 \\
\textbf{ OS}                 & Ubuntu 18.04.3 LTS              & CentOS 7.6.1810                   & Ubuntu 14.04.6 LTS \\ \hline
\end{tabular}
\end{table}
%*****************************************

\textit{3) Execution time for individual components (I/O operations and data processing)} in the restructured and/or optimised  \textit{Nanopolish} (three versions at the bottom of Fig.~\ref{f:impletree}) was measured by inserting \textit{gettimeofday} function calls into appropriate locations in the software source code.
To prevent the operating system disk cache affecting the accuracy of I/O results, we cleared the disk cache (\textit{pagecache}, \textit{dentries} and \textit{inodes}) each time before a program execution. Despite the effect of the hardware disk controller cache (${\sim}$8GB) being negligible due to the large dataset size (${\sim}$850GB), we still executed a mock program run prior to each experiment. Note that the operating system disk cache on S3 was also cleared before any NFS experiment.

\textit{4) Core-hours} is calculated as the product of the number of processing threads employed and the number of hours (wall-clock time) spent on the job. This metric is inspired by the metric man-hours used in labour industry and is used in Cloud Computing domain to calculate the data processing cost \cite{core-hours}. In an ideally parallel program, this metric remains constant with the number of  cores/threads.

%

%-----------------------------------------
\subsection{Results: Alternate File Format (SLOW5)}\label{s:resslow5}
%-----------------------------------------

\begin{figure}[!t]
  \centering
    \begin{subfigure}[!ht]{0.49\linewidth}
      \centering
        \includegraphics[width=\textwidth]{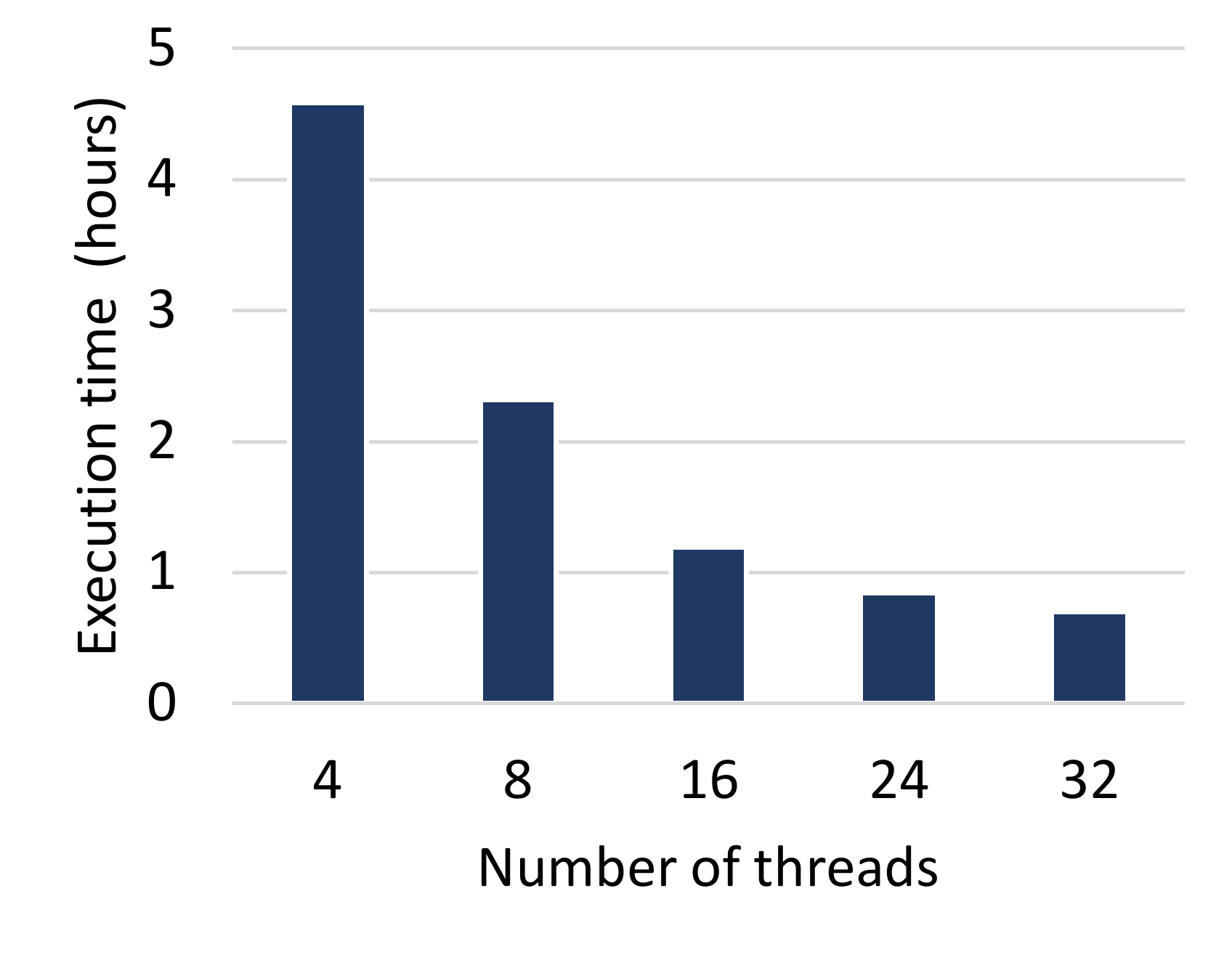}
        \caption{Execution time} 
        \label{f:slow5-overalla}
    \end{subfigure}
    \begin{subfigure}[!ht]{0.49\linewidth}
      \centering
        \includegraphics[width=\textwidth]{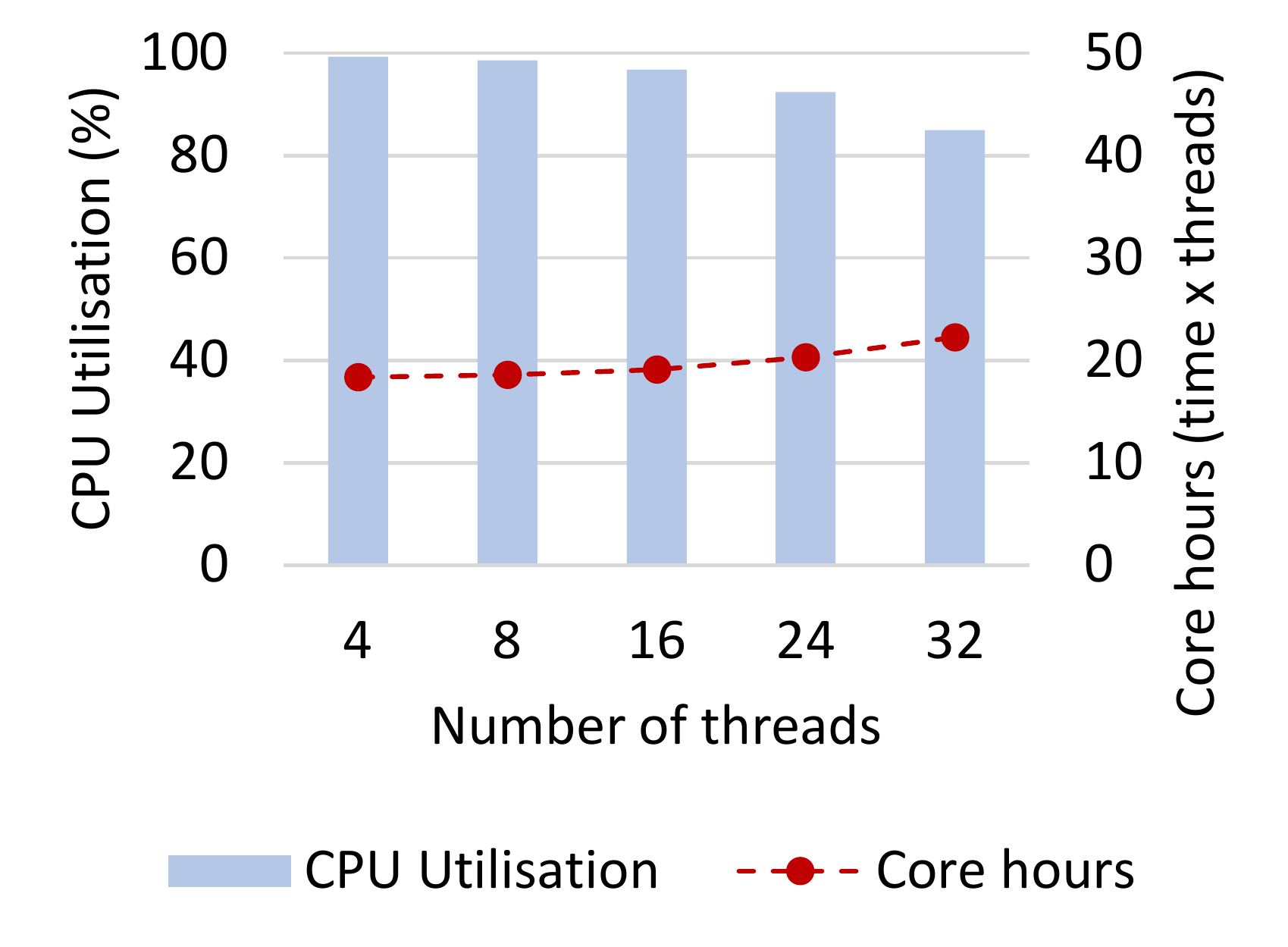}
        \caption{CPU utilisation} 
        \label{f:slow5-overallb}
    \end{subfigure}  
  
    \caption{Overall execution time and CPU utilisation when SLOW5 format is used} 
     \label{f:slow5-overall}
     \vspace{-3mm}  

\end{figure}

\textbf{Overall Execution Time and CPU Utilisation:}
The overall execution time when our proposed SLOW5 file format is used in the restructured and optimised \textit{Nanopolish} is shown in Fig. \ref{f:slow5-overall}a, while the CPU utilisation and the core-hours are depicted in Fig.~\ref{f:slow5-overall}b. The x-axis represents the number of data processing threads which the program was executed with. The number of I/O threads for \textit{glibc} POSIX AIO was also set to the same number of threads as the number of data processing threads.

\textit{Observation-1: Performance has improved w.r.t  original \textit{Nanopolish} for a given number of threads.}
To observe this, we compare Fig. \ref{f:nanopolish-orig-runtime}a with Fig. \ref{f:slow5-overall}a.
At 4 threads, execution time improved by ${\sim}$2$\times$ compared to original \textit{Nanopolish}. At 8, 16, and 24 threads speedups of ${\sim}$2.5$\times$, ${\sim}$4$\times$, ${\sim}$5.5$\times$ can be observed, respectively. At 32 threads, ${\sim}6.5\times$ speedup is observed. In other words, speedup of our optimised version over  original \textit{Nanopolish} increases with the number of threads.

\textit{Observation-2: CPU Utilisation has improved  with the number of threads when compared with  original \textit{Nanopolish}.} 
Comparing Fig.~\ref{f:nanopolish-orig-runtime}b with Fig. \ref{f:slow5-overall}b reveals that
CPU utilisation at 4 threads improved to 99\% which was 69\% for original \textit{Nanopolish}.  The CPU utilisation increases to 99\% from 56\%, 97\% from 39\%, and 92\% from 28\%, at 8, 16 and 24 threads, respectively w.r.t  original \textit{Nanopolish}.
 At 32 threads, an improvement of CPU utilisation to 85\% was observed which was as low as 22\% for  original \textit{Nanopolish}.

\textit{Observation-3: Performance scaling with number of threads is improved.} 
This is evident by the core-hours plot in Fig. \ref{f:slow5-overall}b, whose values are much smaller and almost constant when compared with its counter-part in
Fig.~\ref{f:nanopolish-orig-runtime}b.
For the original \textit{Nanopolish}, the execution time with 32 threads improved only by ${\sim}$2.1$\times$ compared to running with 4 threads (from 9.7h to ${\sim}$4.5h). Our optimised version improved by ${\sim}$6.5$\times$ at 32  threads compared to 4 threads (${\sim}$4.5h to ${\sim}$0.7h).

%----------------------------------

\begin{figure}[!b]
  \centering

\begin{subfigure}[!ht]{0.49\linewidth}
  \centering
    \includegraphics[width=\textwidth]{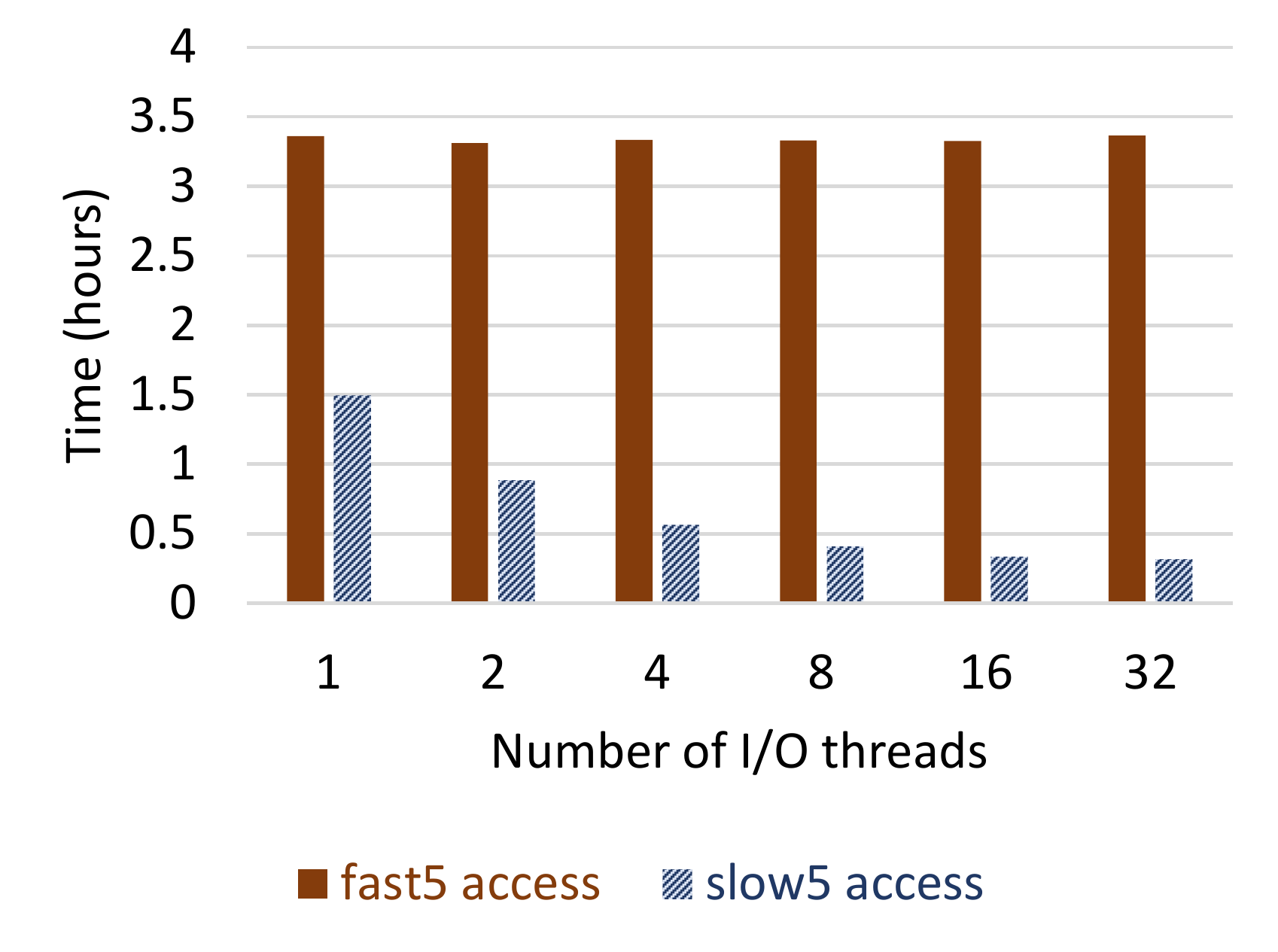}
    \caption{On system S1: HDD RAID} 
    \label{f:slow5a}
\end{subfigure}
\begin{subfigure}[!ht]{0.49\linewidth}
  \centering
    \includegraphics[width=\textwidth]{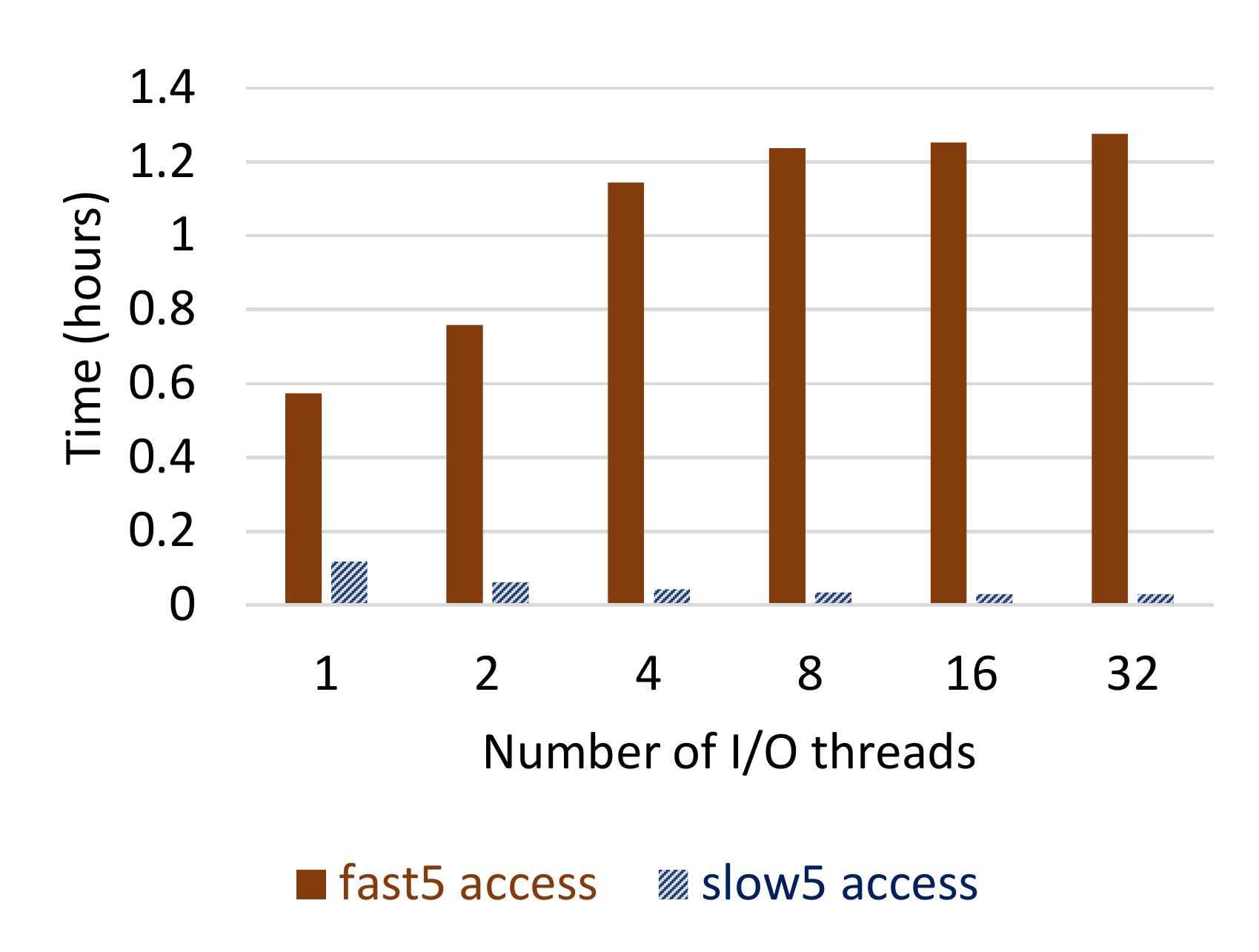}
    \caption{On system S2: SSD RAID} 
    \label{f:slow5b}
\end{subfigure}
    \caption{Comparison of FAST5 vs SLOW5 access} 
    \label{f:slow5}
\end{figure}

\textbf{I/O Time Consumption:}

It was discussed previously that the identified bottleneck is due to I/O. Therefore, to get more insight into the effectiveness of our proposed solution, we plot and compare the time spent in I/O operations. Specifically, the time spent for reading nanopore raw signal data on system S1 when using SLOW5 format is compared to when using FAST5 format in Fig. \ref{f:slow5a}. We make the following observations from the figure: 
1) there is no improvement in FAST5 access time (brown bars) despite increasing the number of threads used (due to the lock in HDF5 library); 2) in contrast, there is a significant improvement for the proposed SLOW5 access time (blue bars) with the increased number of threads; and,
3) even at a single thread, the proposed SLOW5 is ${\sim}2\times$ times faster than FAST5, and at 32 threads the improvement of SLOW5 compared to FAST5 is ${\sim}10\times$. 
The speed-up in I/O time for the single thread is contributed by the exploitation of locality (discussed in Section~\ref{s:slow5}) and our lightweight SLOW5 access implementation.

%-------------------------------------

\begin{figure}[!t]
  \centering
    \begin{subfigure}[!ht]{0.49\linewidth}
      \centering
        \includegraphics[width=\textwidth]{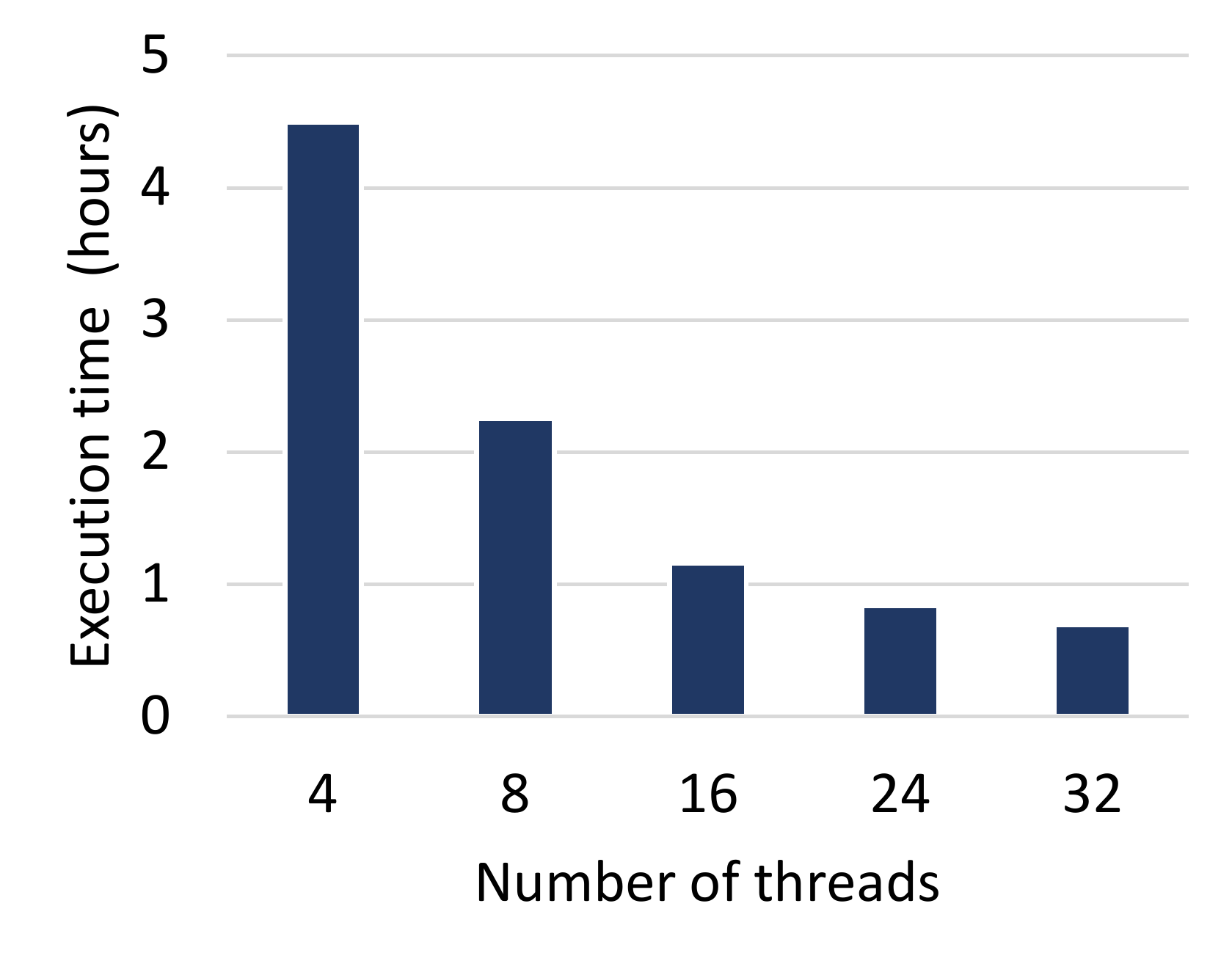}
        \caption{Execution time} 
        \label{f:iop-overalla}
    \end{subfigure}
    \begin{subfigure}[!ht]{0.49\linewidth}
      \centering
        \includegraphics[width=\textwidth]{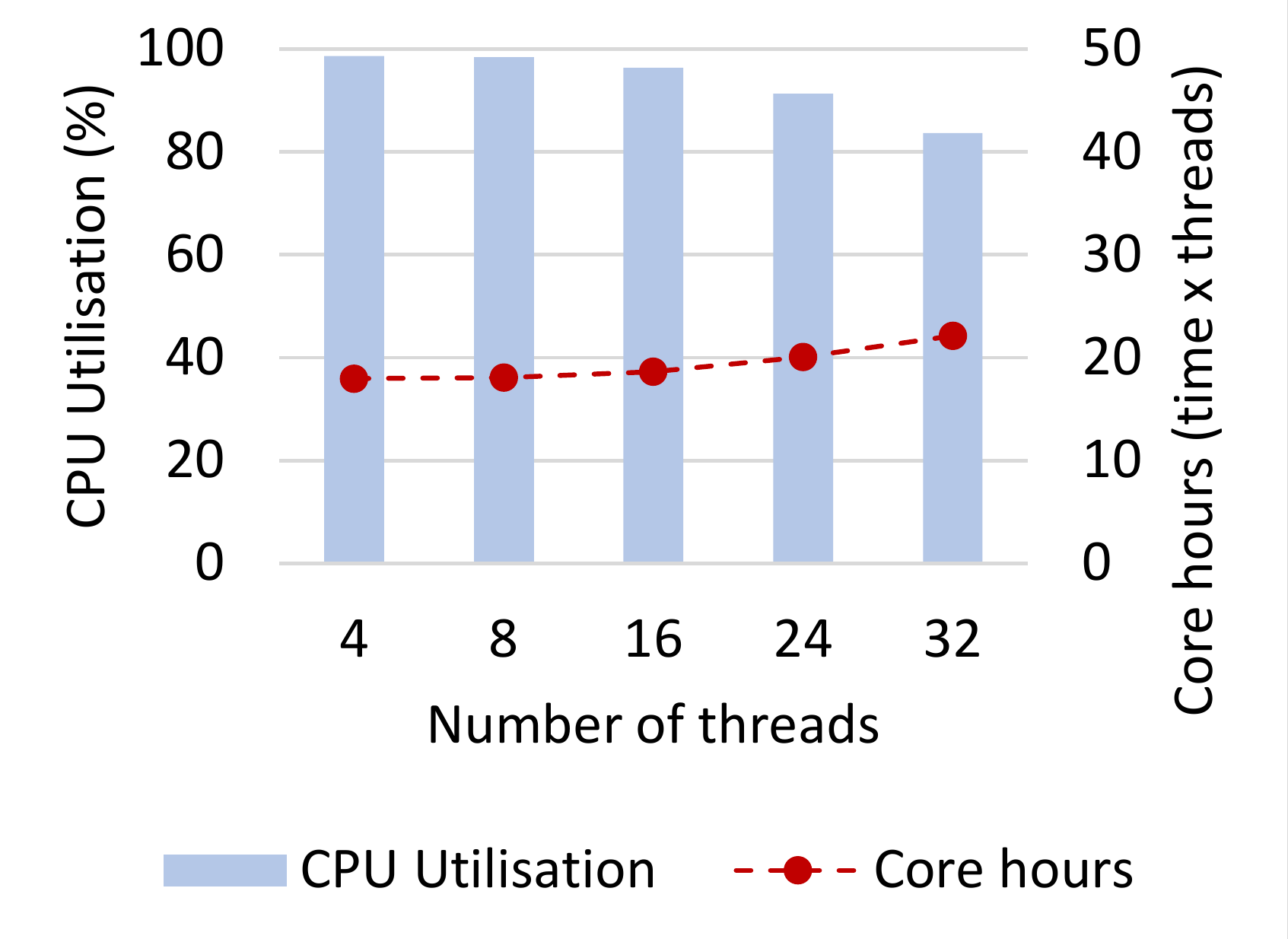}
        \caption{CPU utilisation} 
        \label{f:iop-overallb}
    \end{subfigure}    
  
    \caption{Overall results for multi-process pool} 
    \label{f:iop-overall}
\end{figure}

\begin{figure}[!b]
  \centering
\begin{subfigure}[!ht]{0.49\linewidth}
  \centering
    \includegraphics[width=\textwidth]{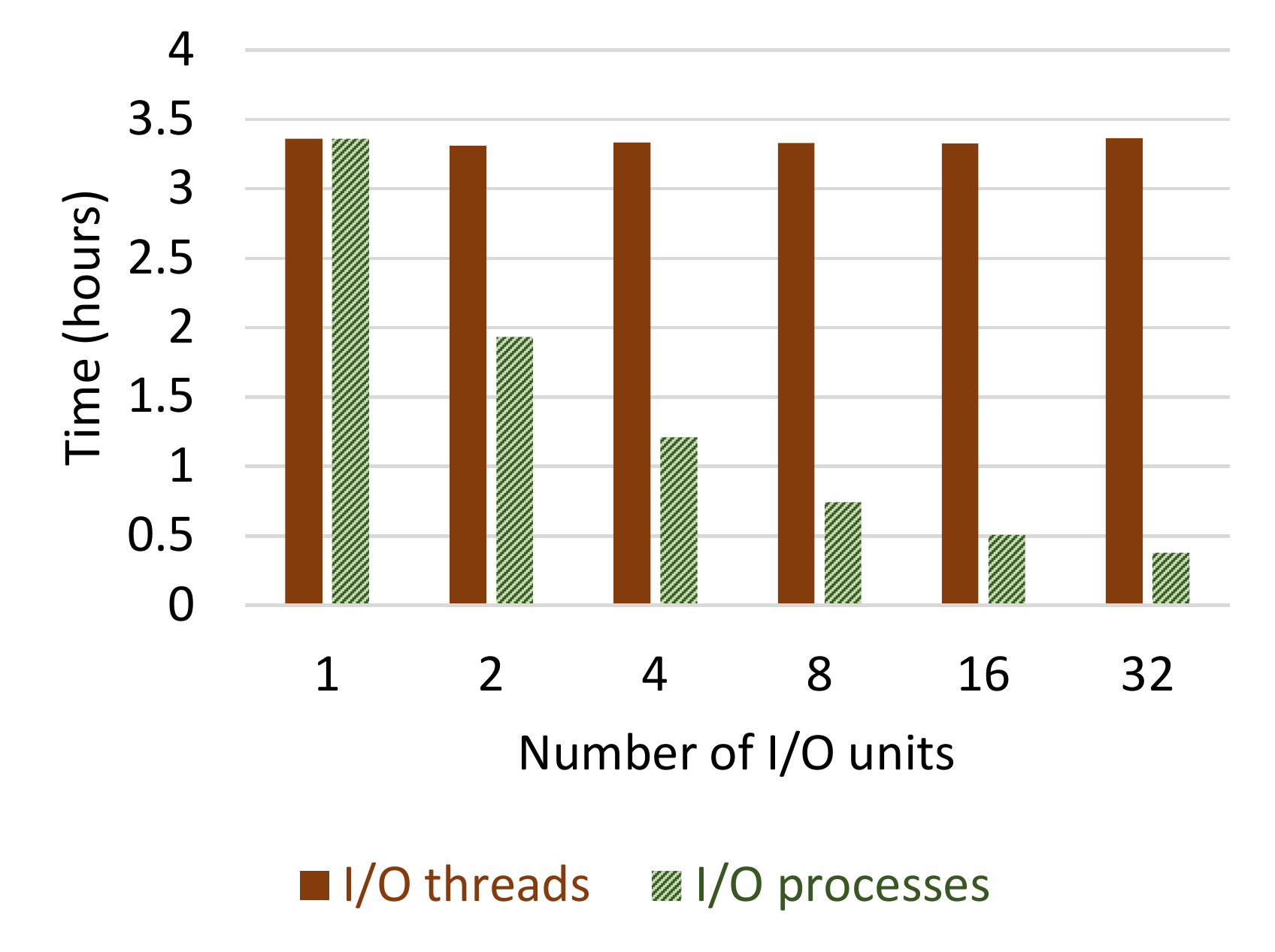}
    \caption{On system S1: HDD RAID} 
    \label{f:fast5-multiproca}
\end{subfigure}
\begin{subfigure}[!ht]{0.49\linewidth}
  \centering
    \includegraphics[width=\textwidth]{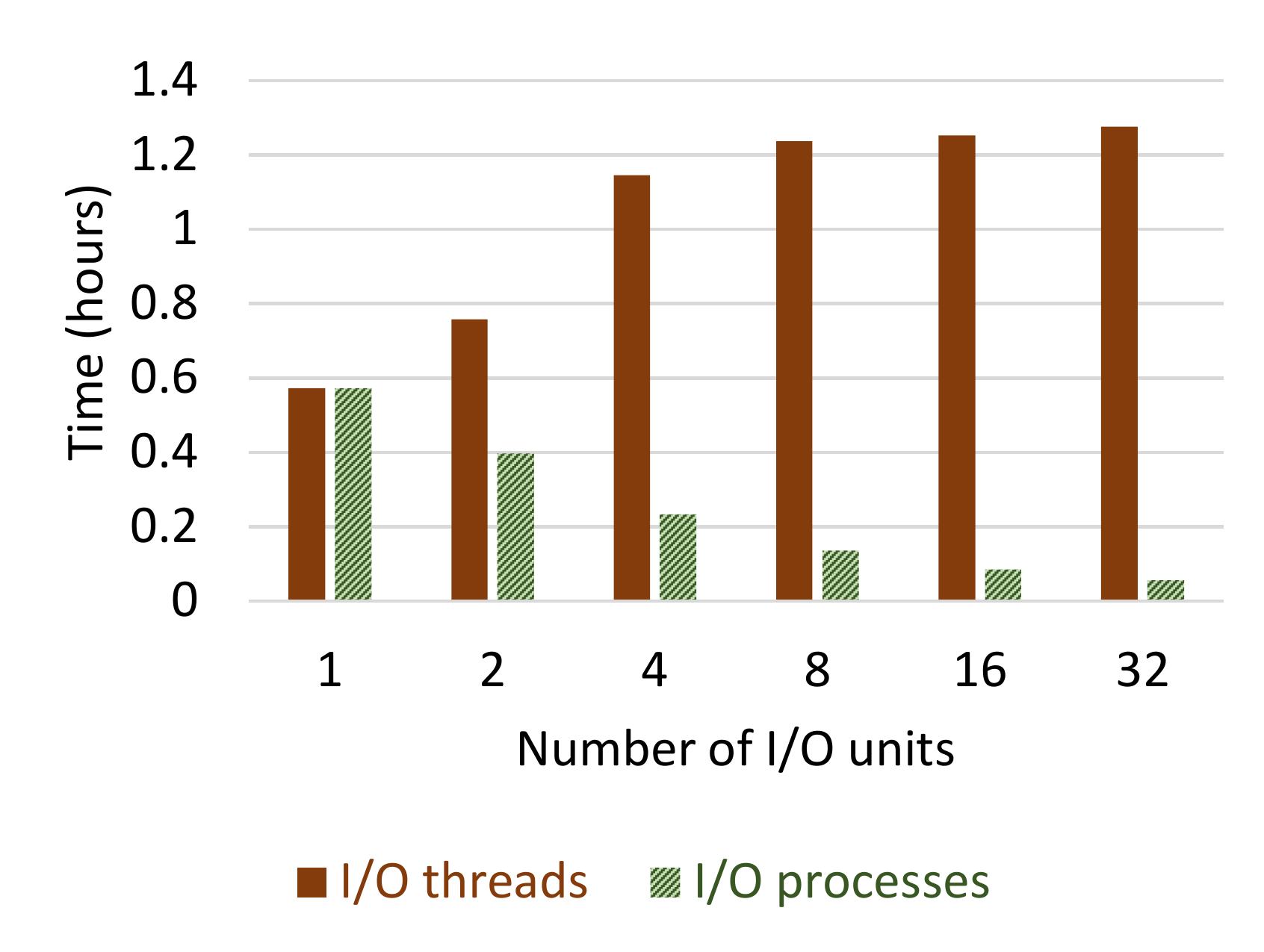}
    \caption{On system S2: SSD RAID} 
    \label{f:fast5-multiprocb}
\end{subfigure}
    \caption{FAST5 file access using multiple I/O threads vs I/O processes} 
    \label{f:fast5-multiproc}
\end{figure}

Above experiments on system S1 demonstrated that our proposed solution effectively improves performance of \textit{Nanopolish}. The S1 system consists of HDD RAID. Now, we demonstrate that our solution is also effective on SSD RAID using experiments on system S2. As discussed above, the I/O decomposition results are more insightful, therefore we present the I/O decomposition results on S2 system (SSD RAID based) for the sake of brevity of the manuscript.  Fig. \ref{f:slow5b} shows the comparison of FAST5 access time to SLOW5 access time, where similar observations can be made. In fact, FAST5 access time (brown bars) got worse with the number of threads, whereas SLOW5 access time (blue bars) improved with the number of threads. At 32 threads SLOW5 was ${\sim}42\times$ faster than FAST5 on SSD RAID. Thus, our proposed solution is effective for the HDD based RAIDs as well as the SSD based RAIDs.

\emph{Note:} 
The file sizes of the new SLOW5 format are comparable to the existing FAST5 format. Specifically, the dataset which was 845 GB in FAST5 format (Table \ref{t:datasets-slow5}), reduced to 340 GB when converted to SLOW5. The reduced size when converted to SLOW5 is due to storing global metadata in the header in SLOW5, instead of redundantly storing those for each read. SLOW5 index is quite small (47 MB) compared to gigabytes of RAM available on an HPC.
%\todo{[Will talking about the other dataset make things too confusing? For the large D2 dataset we ran with all cores available on server S1 and the overall execution time improved to 22.04 hours for restructured Nanopolish with SLOW5 which was 69.29 hours previously for FAST (mentioned in section \ref{s:deepnaly}). The FAST5 access time which was 62.72h earlier (mentioned in section \ref{s:deepnaly}) improved to \todo{3X} when our SLOW5 format was used.]}

%\todo{[Talk about IOPS and disk utilisation. Is it too much technical stuff?? See how the MB/s and IOPS change over time for disks]} 
%\todo{[Talk about lseek operations?]}

%\todo{[Talk about file sizes? :  for D1 fast5s were XX GB and in SLOW5 (uncompressed) was XX GB.  For D2 XX YY. Fast5 HDF5 are already compressed. The reduction in D1 is due to the removal of redundant data.]}

%-------------------------------------
\subsection{Results: Multi-process Pool}

{\textbf{Overall Execution Time and CPU Utilisation:}
Overall execution time for the restructured and optimised \textit{Nanopolish} when a multi-process pool is used for FAST5 access is shown in Fig. \ref{f:iop-overalla}, whereas the CPU utilisation and the core-hours are depicted in Fig.~\ref{f:iop-overallb}.
The x-axis of the figure corresponds to the number of data processing threads which is also equal to the  number of I/O processes. 
The results in Fig.~\ref{f:iop-overall} are similar to that of the SLOW5 solution discussed in previous subsection. 
The key observations in Fig. \ref{f:iop-overall} compared to original \textit{Nanopolish} are also similar to the first solution. These are:
1) improved performance w.r.t. the original \textit{Nanopolish} for a given number of threads;
2) improved CPU Utilisation; and, 3) better performance scaling with increasing number of threads, as depicted by the near-flat core-hour plot.

\textbf{I/O Time Consumption:}
Similar to the previous section, we evaluate the time spent in I/O operations. We compare the results for the multi-threaded and the multi-process based versions. The plots are presented in Fig.~\ref{f:fast5-multiproc}, with the x-axis denoting the number of processes/threads used. 
On HDD RAID (Fig. \ref{f:fast5-multiproca}), the FAST5 access time does not improve with increased I/O threads (brown bars), while it 
significantly improves with increased I/O processes (green bars). At 32 threads/processes the improvement was ${\sim}9\times$. 
On SSD RAID (Fig. \ref{f:fast5-multiprocb}), the FAST5 access time gets worse with increased I/O threads. In contrast, it significantly improves with increased I/O processes. Using 32 I/O processes is ${\sim}23\times$ faster than using 32 I/O threads on SSD RAID.

In summary, using processes instead of threads for I/O operations alleviates the I/O bottleneck, while using multiple-threads for data processing in a single parent-process avoids introduction of any additional significant bottlenecks, as depicted by the above results.

\subsection{Comparison of SLOW5 to FAST5 with Multi-process Pool}

Comparing the time for SLOW5 access in Fig. \ref{f:slow5} with I/O process based pool for FAST5 (Fig. \ref{f:fast5-multiproc}) shows that SLOW5 outperforms FAST5 even when multiple I/O processes are used especially at lower number of threads/processes. 

%======================================================
\subsection{Comparison with Multi-FAST5 Format}
\label{s:comparisonnew}
%======================================================
\begin{figure}[!t]
  \centering
\begin{subfigure}[!ht]{0.49\linewidth}
  \centering
    \includegraphics[width=\textwidth]{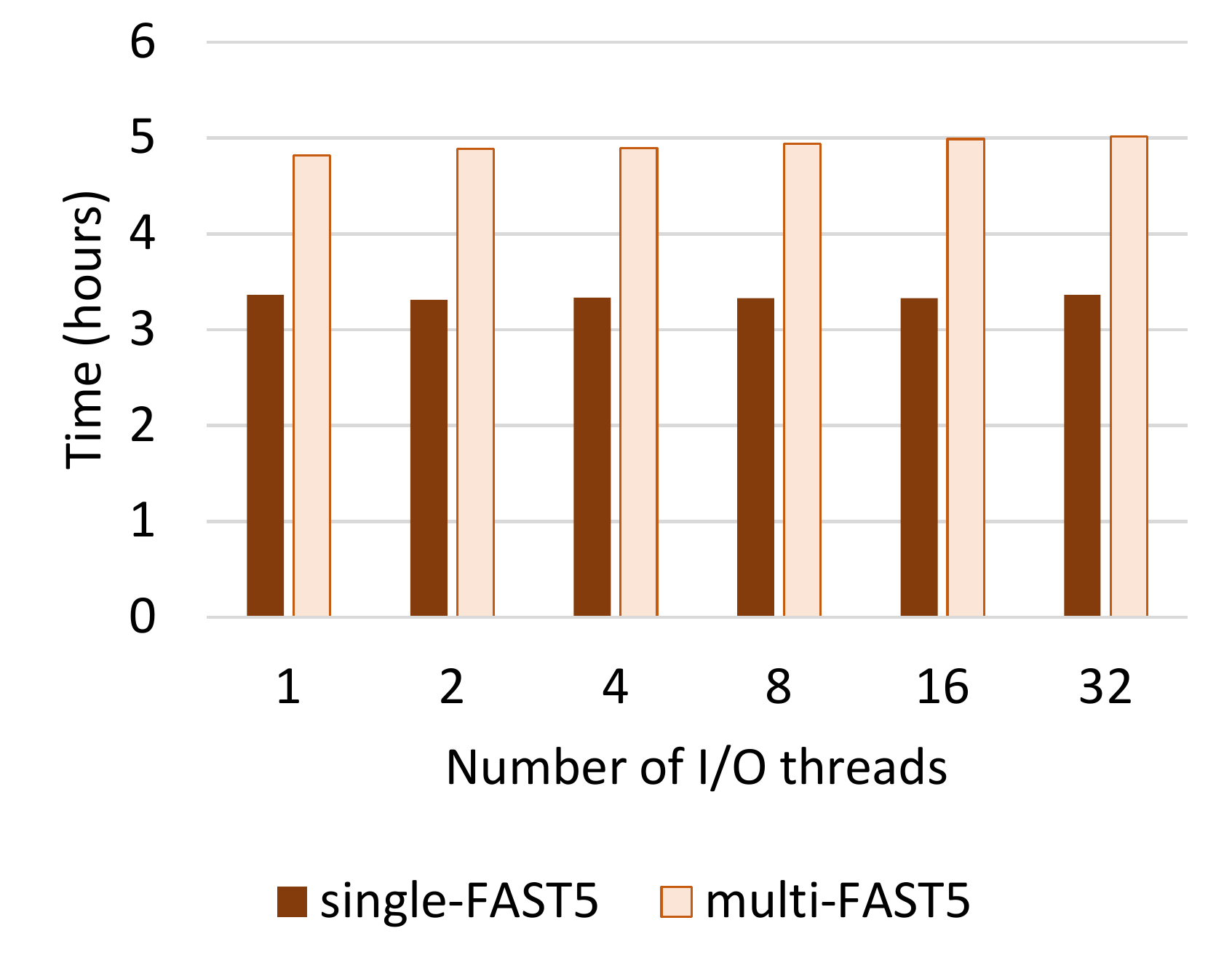}
    \caption{On system S1: HDD RAID} 
    \label{f:multifast5-threadsa}
\end{subfigure}
\begin{subfigure}[!ht]{0.49\linewidth}
  \centering
    \includegraphics[width=\textwidth]{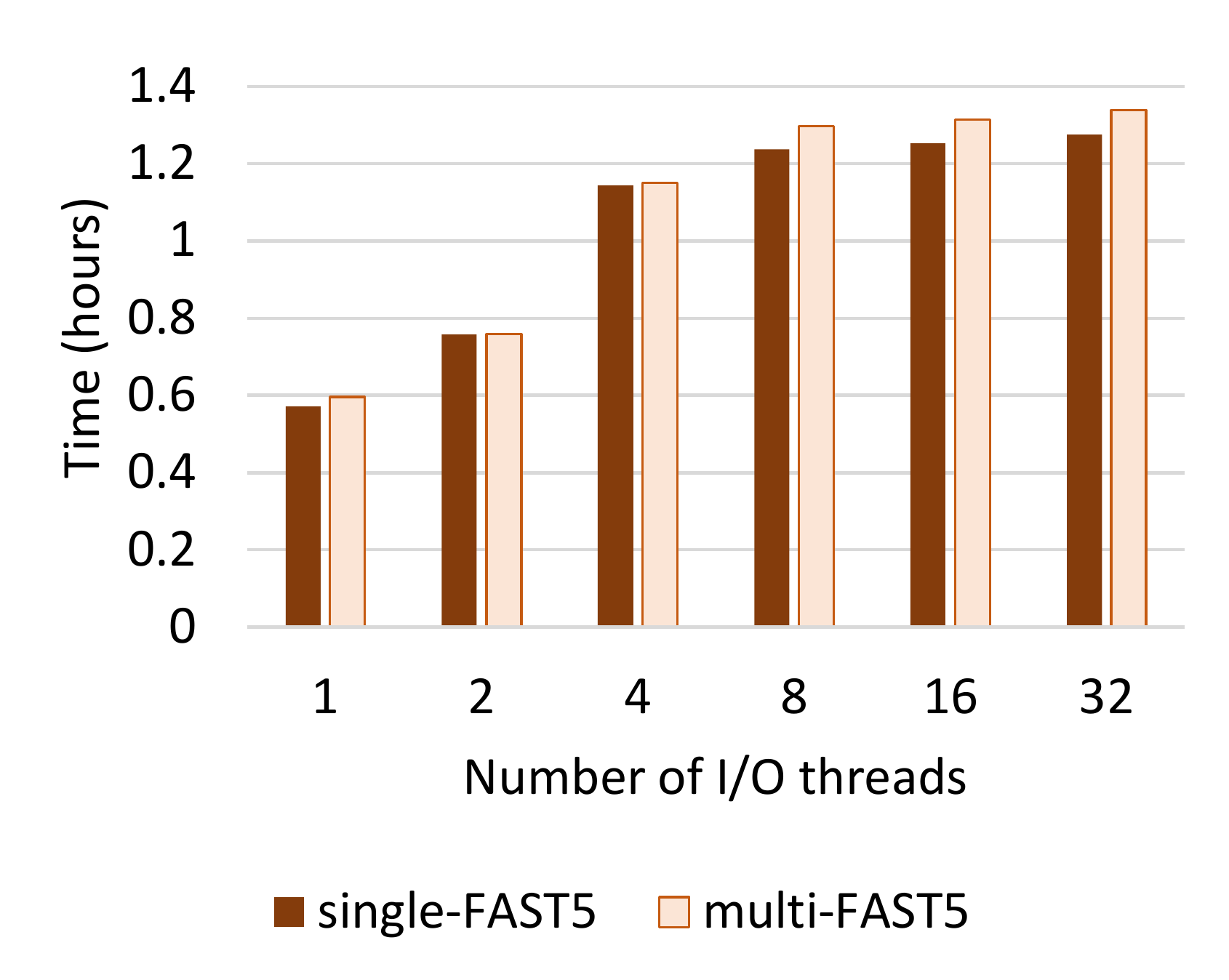}
    \caption{On system S2: SSD RAID} 
    \label{f:multifast5-threadsb}
\end{subfigure}
    \caption{Single-FAST5 vs Multi-FAST5 using I/O threads} 
    \label{f:multifast5-threads}
\end{figure}
ONT is recently working on a new file format: known as \textit{multi-FAST5}. It is projected to replace the existing FAST5 format in near future. 
The raw signals from multiple \textit{genomic reads} (by default 4000 \textit{genomic reads}) are packed into a FAST5 file and such files are termed as \textit{multi-FAST5}. 
Multi-FAST5 reduces the gigantic amount of small single-FAST5 files generated from a sequencing run, easing the file management (eg: copying/moving files, listing files). Multi-FAST5 files are also HDF5 files where the schema is an extended version for that of single-FAST5. 
%One would argue that the latest multi-FAST5 format from ONT would perform better than the single-FAST5 files (previous experiments used single-FAST5). Therefore, 
Next, we demonstrate that multi-FAST5 suffers from a similar bottleneck, and thus our proposed SLOW5 is superior to the new multi-FAST5 format. Moreover, our multi-process based solution is also applicable and effective for multi-FAST5 format.

%\textbf{ The new multi-FAST5 also suffers from a similar bottleneck.}
\textbf{Performance Bottleneck in Multi-FAST5:}
First, we compare the file access time in multi-FAST5 to single-FAST5 in Fig. \ref{f:multifast5-threads}.
Unfortunately, the access time does not improve by the use of multiple I/O threads on HDD RAID, similar to single-FAST5 (Fig. \ref{f:multifast5-threadsa}). In fact, multi-FAST5 performance is actually worse than that of single-FAST5. 
On SSD RAID (Fig. \ref{f:multifast5-threadsb}), the performance of multi-FAST5 and single-FAST5 are almost the same and gets gradually worse with the number of threads.

%\textbf{The proposed multi-process based solution improves the performance for Multi-FAST5 format.}
\textbf{Proposed Multi-process Solution on Multi-FAST5:}
Now we demonstrate that our multi-process based solution is also applicable and effectively improves the performance for the new  multi-FAST5 format. The access time for Multi-FAST5 and single-FAST5 with our multi-process solution for different number of threads are in Fig. \ref{f:multifast5-iop}.
 With our solution, the trend of multi-FAST5 access time is similar to that of single-FAST5 files (on both HDD RAID and SSD RAID), that is, it gets significantly better with the number of I/O processes used. Note that when the time with single-FAST5 is compared, multi-FAST5 takes more  time than single FAST5, visibly in the HDD RAID.

\begin{figure}[!t]
  \centering
\begin{subfigure}[!ht]{0.49\linewidth}
  \centering
    \includegraphics[width=\textwidth]{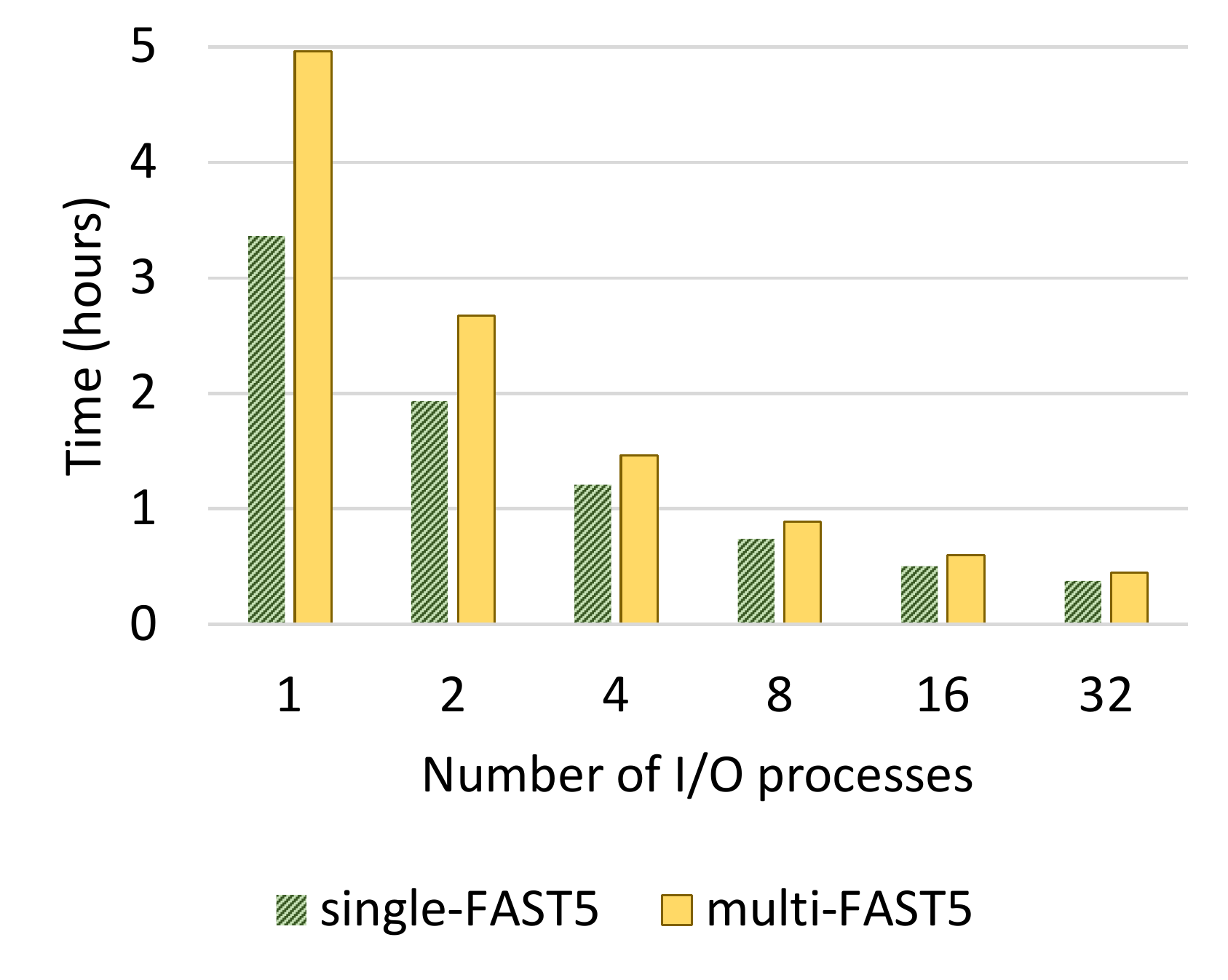}
    \caption{On system S1: HDD RAID} 
    \label{f:multifast5-iopa}
\end{subfigure}
\begin{subfigure}[!ht]{0.49\linewidth}
  \centering
    \includegraphics[width=\textwidth]{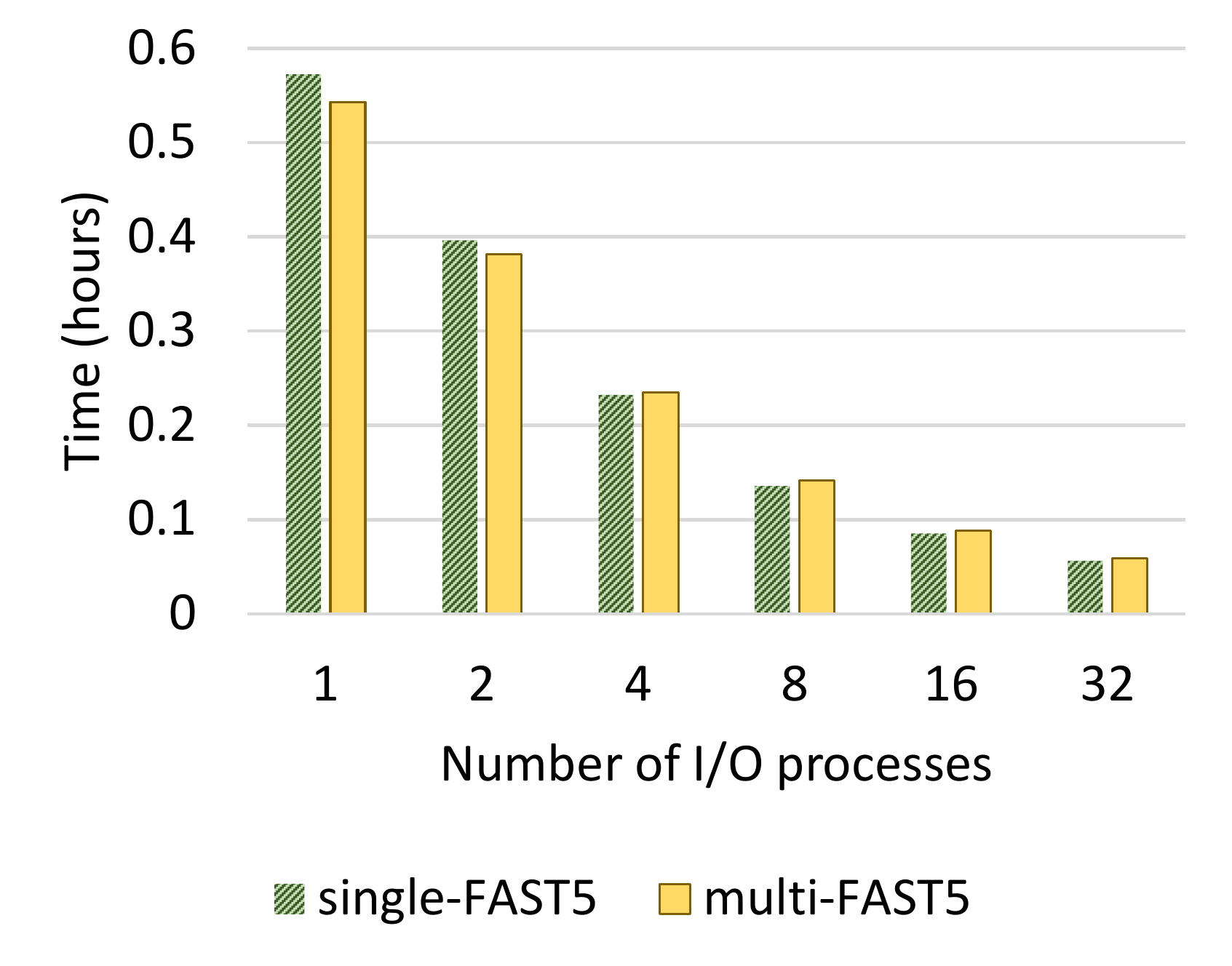}
    \caption{On system S2: SSD RAID} 
    \label{f:multifast5-iopb}
\end{subfigure}
    \caption{Single-FAST5 vs Multi-FAST5 using I/O processes} 
    \label{f:multifast5-iop}

\end{figure}

}

% %-------------------------------------
\subsection{On NFS}

\begin{figure}
\centering
    \begin{subfigure}[!ht]{0.49\linewidth}
    \centering
    \includegraphics[width=\textwidth]{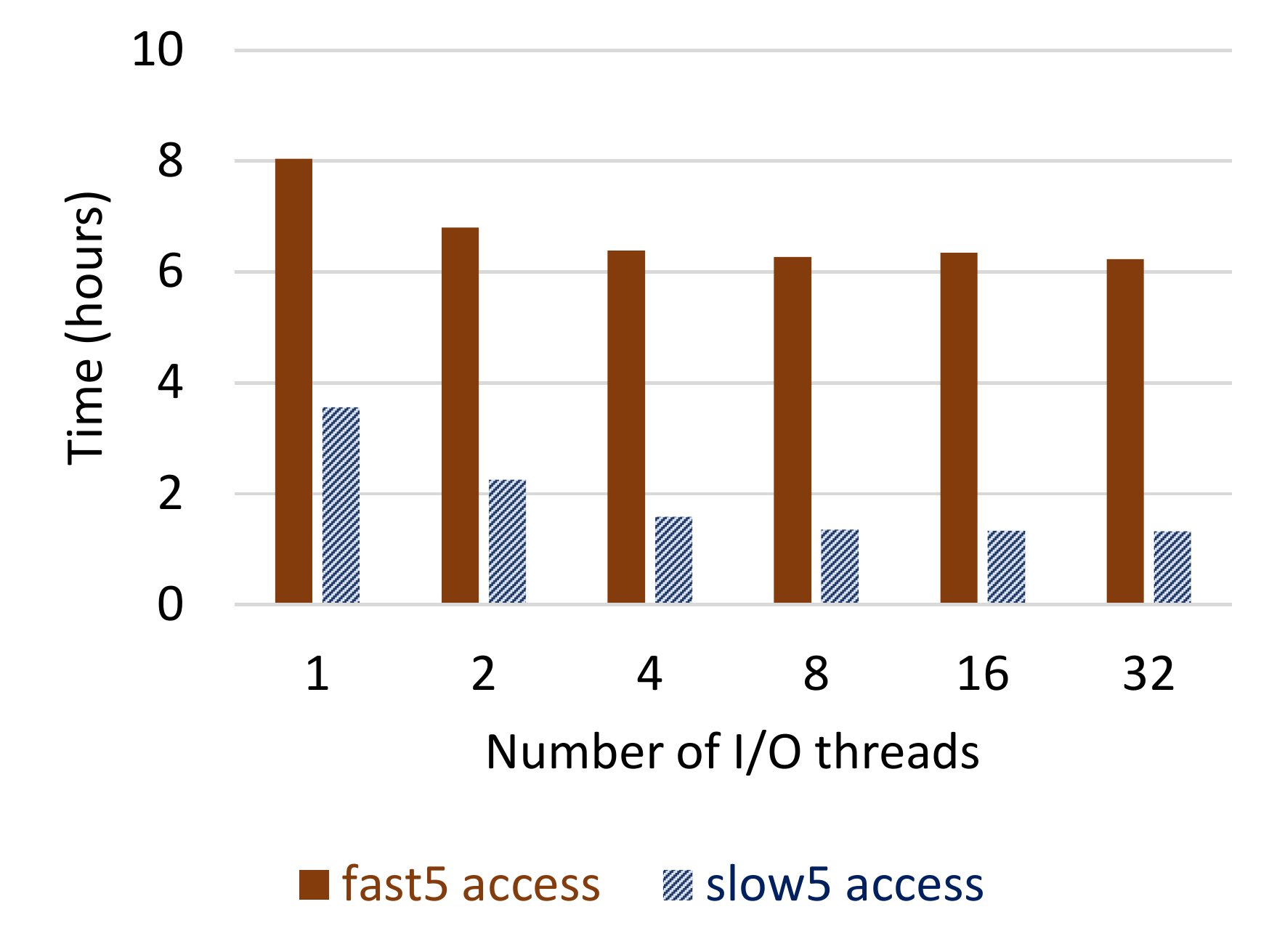}
    \caption{SLOW5 performance on NFS}
    \label{f:nfs-slow5}
    \end{subfigure}
    \begin{subfigure}[!ht]{0.49\linewidth}
    \centering
    \includegraphics[width=\textwidth]{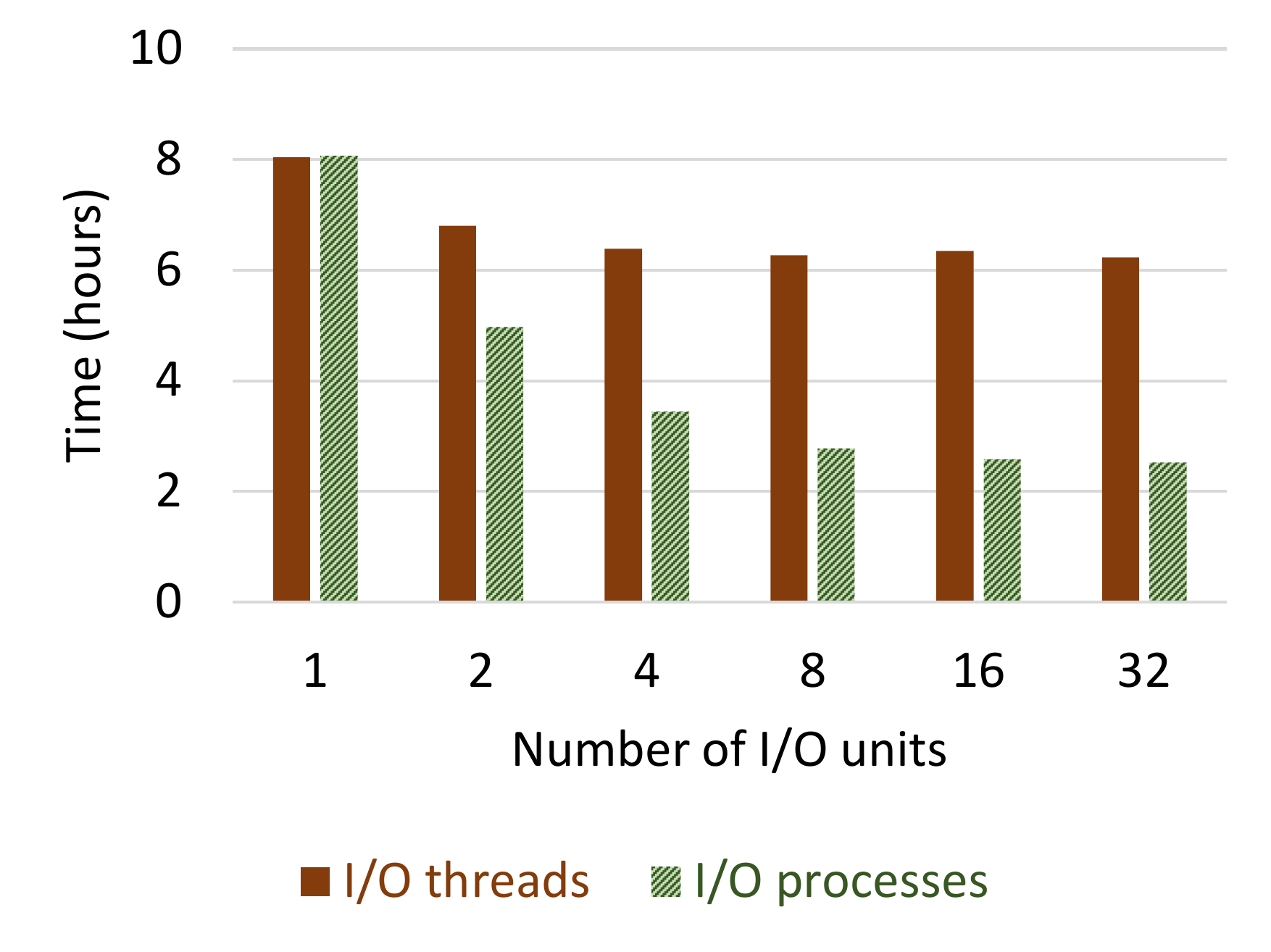}
    \caption{FAST5 performance on NFS} \label{f:nfs-fast5}
    \end{subfigure}    
    \caption{Performance on NFS}
    \label{f:nfs}
\end{figure}

Our proposed optimisations has the potential to benefit direct execution of nanopore data analysis tools on data residing on a network attached storage. HPC cluster environments predominantly use such network attached storage in addition to local RAID systems. We demonstrate the performance of our proposed methods on NFS in Fig. \ref{f:nfs}. 

Fig. \ref{f:nfs-slow5} compares our SLOW5 format with FAST5 on NFS over multiple I/O threads. Use of multiple I/O threads for accessing FAST5 files on NFS (brown bars), slightly improves the performance up to around 4 threads (unlike previously on local RAID), which then saturates. SLOW5 access (blue bars) is much faster than FAST5.  SLOW5 access time improves up to around 8 threads which then saturates.

Fig. \ref{f:nfs-fast5} compares our proposed process pool based method to using multiple I/O threads.  Use of multiple I/O processes (green bars) considerably improves the FAST5 access performance up to around 8 processes, which then slowly saturates, a similar trend to that with SLOW5. Comparing SLOW5 (blue bars in Fig. \ref{f:nfs-slow5}) to FAST5 access using multiple I/O processes (Fig. \ref{f:nfs-fast5}) shows that SLOW5 performance is superior.

%  %----------------------------------------------------------------------

Refer to appendix \ref{a:ioopti} for supplementary results and analyses.

%===============================================================
\section{Discussion}\label{s:io-discussion}
%===============================================================
%\HasComment{I restructured this section. I didnt modify any text in 6.1 and 6.3, but revised 6.2 more thoroughly with proper story}

\subsection{Other Possible Solutions}
\label{s:othersolutions}
In this chapter, we presented two solutions to overcome the I/O bottleneck caused by the FAST5 file format and demonstrated their efficacy using experiments. Additionally, there are few other possible solutions to the problem, as discussed below.

\textbf{Fixing HDF5 Library:}
As discussed above, using a new file format may not always be practical and we presented a multi-process based solution in such a scenario.
Another candidate solution is to fix (optimise) the HDF5 library to be thread efficient. However, HDF5 library is a complicated library with a large code base of $>$300,000 lines of C code and such a fix has to be potentially done by the HDF Group \cite{hdfsucks1,hdfsucks2}. The HDF5 Group mentions that the future plan to implement efficient multi-threaded access is currently hindered by inadequate resources \cite{hdf-thread-efficiency}. Therefore, such a fix is unlikely to happen in the near future. Moreover, there is no other alternate library to read HDF5 files\cite{hdfsucks1,hdfsucks2}.

\textbf{Naive Approaches of Multi-processing:}
Instead of using a process pool solely for FAST5 I/O and multi-threads for parallel data processing (as proposed in Section~\ref{s:multiproc}), programmers may use multi-processes for both the I/O operations and parallel data processing. This would be easier than implementing a pool of processes, however, this is only suitable for trivially parallel cases. If the application needs to share data among multiple processing units, processes are unsuitable due to the complexity that arise when performing inter-process communication.

Alternatively, the programmer may let users manually split data and launch multiple processes. Unfortunately, this method exerts additional burden on the  user, i.e., custom scripts must be written for data splitting, launching data processing and concatenating the result. Moreover, this is only suitable for trivially parallel applications where data can be easily split. Also, an expensive HPC system with dozens of cores is superfluous as the user could use a cluster of low cost networked computers (as shown in chapter \ref{c:integration}).

In summary, using processes instead of threads potentially solves the I/O bottleneck as we demonstrated in results. However, it is important to note that \textit{processes} in an operating system are meant for isolation whereas \textit{threads} are for sharing data. Inter-process communication requires system calls, while inter-thread communication involves sharing the same memory space. Further, processes are expensive to be spawned and are not lightweight (unlike threads). Thus, using processes as a replacement to threads makes the code relatively complicated. Therefore, we suggest that using the SLOW5 format is a superior solution than the multi-processes based solution. 

\subsection{Future Directions}

{As shown in section \ref{s:resslow5}, SLOW5 file size is smaller than FAST5 due to the efficient storage of metadata. SLOW5 file size can potentially be further reduced by using a binary encoding instead of ASCII and/or by applying block compression techniques such as BGZF that still allows random access \cite{li2011tabix}. Having both ASCII and binary formats is useful, where the former is human readable and the latter is space efficient. In fact, gold standard file formats in genomics such as SAM and VCF that are in ASCII have their binary counterparts BAM and BCF.}

After applying our proposed optimisations proposed in this chapter, the next bottleneck  in \textit{Nanopolish} could be the FASTA access (random access to reference genome) which is performed using the \textit{faidx} component in \textit{htslib} library. This \textit{faidx} is not currently thread-safe and thus only single threaded access is possible. However, FASTA is a simple ASCII based format and thus extending \textit{faidx} for thread efficiency is feasible as future work.

\subsection{Potential Impact on other Toolkits and Domains}

It is likely that the identified limitation in HDF5 libary is a primary bottleneck in several other nanopore software toolkits, which also use the HDF5 library such as \textit{Tombo}~\cite{stoiber2016novo}, \textit{NanoMod}~\cite{liu2019nanomod} and \textit{SquiggleKit}~\cite{ferguson2019squigglekit}.
Thus, our proposed optimisations are potentially useful in such toolkits.
Our work may also guide nanopore software developers to avoid the identified bottleneck in future. Furthermore, HDF5 is also used in other engineering domains such as physics, astronomy, weather forecasting \cite{folk2011overview}. Therefore, we believe that our work will inspire optimisations in those domains.

%=================================================================
\section{Summary}
\label{s:io-conclusion}
%=================================================================

In this chapter, we demonstrated with an example that nanopore software fail to take maximal advantage of the computing power offered by many-core processors in HPC systems, despite multi-threaded implementation. To address this problem, we presented a systematic experimental analysis to identify potential performance bottlenecks in nanopore software tools for running on many-core CPUs. We identified that the bottleneck is caused by inefficient file I/O associated with the HDF5 library used for loading nanopore raw data. The inefficiency in file I/O in HDF5 is due to a global lock which limits multiple threads requesting file accesses in parallel. Then, we proposed multiple optimisations to alleviate the bottleneck. We proposed a new file format that facilitates efficient file access using multiple threads. For the scenarios where the original format must be used, we presented a multi-process based solution. Thus, our proposed optimisations can be used as an alternative, or alongside the existing file-format.  Our experiments demonstrated that our optimisations not only enable improved performance for a given number of threads (${\sim}2\times$ for 4 threads and ${\sim}6.5\times$ for 32 threads), but also enable improved CPU utilisation (from 69\% to 99\% for 4 cores and from 22\% to 85\% for 32 threads) when compared to original \textit{Nanopolish}. Consequently, improved performance scaling with the number of threads was also achieved (${\sim}6.5\times$ for 4 vs. 32 threads).

%% file: 9-conclusion/main.tex
\chapter{Conclusion and Future Directions}\label{c:conclusion}

DNA sequencing is a revolutionary technology that is reshaping the field of medicine and healthcare. In addition, DNA sequencing has important applications in other fields such as epidemiology and forensics. Over the last two decades, the size of DNA sequencers has shrunk from the size of a fridge to that of a mobile phone and sequencing cost per genome has remarkably reduced by more than 1000 times. These remarkable improvements are expected to continue further. Unfortunately, hundreds to thousands of gigabytes of data output from today's ultra-portable sequencers is analysed on non-portable high-performance computers or cloud computers, which was the case even a couple of decades ago.

This thesis moved the DNA sequence analysis from high-performance computers to portable computing devices, a timely need for enhancing the use of ultra-portable sequencers in point-of-care or in-the-field. The objective was achieved using computer architecture-aware optimisation of complex DNA analysis workflows. Such optimisations enabled efficient mapping of the software to exploit complex features of modern computer hardware. Domain knowledge of both computer architecture and DNA sequence analysis was simultaneously used to achieve the twin goals of achieving efficient compute resource utilisation with no impact on accuracy. Therefore, this thesis is an attempt to bridge the two domains, DNA sequencing and computer architecture.

In this thesis, gold-standard DNA sequence analysis software tools were systematically examined for bottlenecks and architecture-aware optimisations were performed at I/O level, processor level, RAM level, cache level and at the register level. The optimised software tools were used to perform complete end-to-end analysis workflows on prototype embedded systems composed of single-board computers. The performance and accuracy were evaluated using real and representative datasets. The resultant embedded systems were fully functional with performance comparable to an unoptimised workflow on a high-performance computer. The constructed prototype embedded systems are currently being used for in-house data analysis at Garvan Institute of Medical Research, Sydney. Such low cost, energy-efficient, sufficiently fast and portable embedded system enables complete DNA analysis in point-of-care or in-the-field.

In addition to prototype embedded systems composed of single board computers, this thesis has also made it possible to run DNA analysis workflows on commodity portable computing devices such as laptops, tablets and mobile phones. The optimisations proposed in this thesis also benefit running DNA analysis workflows on high-performance computers through a magnitude of times faster performance. Optimised versions of software produced under the thesis are released as open-source software. The prototype embedded systems constructed under this thesis are fully functional that they are currently being used for in-house nanopore sequence data processing at Garvan Institute of Medical Research, Sydney. The open-source software produced under the thesis is being used by several research centres globally and users have surprised by the significant speedup achieved compared to existing software. The conclusion from each chapter of this thesis is given below.

A popular variant calling software for second-generation sequence data called \textit{Platypus} was optimised for efficient usage of the memory hierarchy. Systematically examining the steps in variant calling revealed that 60\% of the total variant calling time is consumed by \textit{de Bruijn} graph construction during the local re-assembly step. After carefully inspecting the data access patterns, optimisations were proposed to improve the locality of memory accesses both at cache level and register level.  The existing algorithm was modified to integrate the proposed optimisations, which in turn improved the efficient usage of faster cache memories and registers.  The results showed that these changes improve the performance of \textit{de Bruijn} graph construction by a factor of around two when implemented on a general-purpose processor. The modified algorithm opens the door to a much higher acceleration of local re-assembly on GPU, FPGA and ASIP. The implementation of the algorithm which is integrated into the \textit{Platypus} Variant Caller is publicly available at \url{https://github.com/hasindu2008/platycflr}.

The gold standard software for aligning long reads generated from third-generation high-throughput sequencers called \textit{Minimap2} was optimised for removed memory capacity. \textit{Minimap2} relies on a large hash table data structure (constructed out of the reference genome) stored in RAM for the alignment process. Large reference genomes such as the human genome require 11GB for the hash table alone. Mere parameter optimisation in \textit{Minimap2} cannot substantially reduce memory usage without considerably sacrificing alignment quality.
Memory capacity optimisations were proposed to substantially reduce memory usage. Memory capacity optimisations included partitioning an alignment index, saving the internal state, and merging the output \textit{a posteriori}. This strategy reduced the memory requirements for aligning long reads to the human reference genome from 11GB to less than 2GB, with minimal impact on accuracy. This work made it possible to perform read alignment to large reference genomes using computers with limited volatile memory.  The optimised version of Minimap2 is available as open-source at \url{https://github.com/hasindu2008/minimap2-arm} and is also integrated into the original \textit{Minimap2} software.

A popular signal analysis toolkit for analysing nanopore raw signal data called \textit{Nanopolish} was optimised for CPU-GPU heterogeneous systems.  Examining the methylation calling tool in the \textit{Nanopolish} toolkit revealed that around 70\% of the runtime is consumed by an algorithm called Adaptive Banded Event Alignment (ABEA).
Despite this algorithm being not embarrassingly parallel, an approach was proposed that made this algorithm efficiently execute on GPUs. The high variability of the read lengths was one of the main challenges, which was remedied through a number of memory optimisations and a heterogeneous processing strategy that uses both CPU and GPU. Proposed optimisations yielded around 3-5$\times$ performance improvement on a CPU-GPU system when compared to a CPU. CPU-GPU optimised ABEA was integrated back into a completely re-engineered version of the \textit{Nanopolish} methylation calling tool and this resultant new software was named \textit{f5c}.  It was demonstrated that \textit{f5c} is adequately capable of processing data from a portable nanopore sequencer in real-time using an embedded SoC equipped with an ARM processor (with six cores) and NVIDIA GPU (256 cores). \textit{f5c} not only benefits embedded SoC but also a wide range of systems equipped with GPUs from laptops to servers. \textit{f5c} was not only around 9$\times$ faster on an HPC but also reduced the peak RAM by around 6$\times$ times. The source code of \textit{f5c} is made available at \url{https://github.com/hasindu2008/f5c}.

A system architecture was proposed for performing a popular DNA methylation detection workflow on a prototype embedded system. The workflow was realised on the proposed architecture by integrating the optimised software versions from previous chapters. The proposed architecture was evaluated using off-the-shelf single-board computers and was demonstrated that performing real-time analysis of nanopore sequencing is possible on an embedded system. It was further demonstrated that the performance of the prototype embedded system is surprisingly similar to the performance on an HPC. The system architecture and the associated software for building a replica of the prototype are released and the open-source code is available at \url{https://github.com/hasindu2008/nanopore-cluster} and \url{https://github.com/hasindu2008/f5p}.

The cause behind the unexpected slow performance on an HPC was identified to be the Nanopore software failing to take maximal advantage of the computing power offered by many-core processors in HPC systems, despite its multi-threaded implementation. A systematic experimental analysis was conducted to identify potential performance bottlenecks in nanopore software tools for running on many-core CPUs. This analysis revealed that the bottleneck is caused by inefficient file I/O associated with the HDF5 library used for loading nanopore raw data. The inefficiency in file I/O in HDF5 was identified to be due to a global lock which limits multiple threads requesting file accesses in parallel. Multiple optimisations were proposed to alleviate the bottleneck: a new file format that facilitates efficient file access using multiple threads; and, a multi-process-based solution for the scenarios where the original format must be used. Thus, the proposed optimisations can be used as an alternative, or alongside the existing file-format. The experiments demonstrated that the optimisations not only enable improved performance for a given number of threads (${\sim}2\times$ for 4 threads and ${\sim}6.5\times$ for 32 threads) but also enable improved CPU utilisation (from 69\% to 99\% for 4 cores and from 22\% to 85\% for 32 threads) when compared to the original \textit{Nanopolish}. Consequently, improved performance scaling with the number of threads was also achieved (${\sim}6.5\times$ for 4 vs. 32 threads).

Conclusively, the architecture-aware optimisations presented in this are significant contributions that result in an ultra-portable DNA analysis system which additionally benefit the performance of DNA analysis workflow on an HPC.

\section{Future Directions}

In the upcoming decades, DNA sequencers will further miniaturise and the sequencing cost will be increasingly affordable. Consequently, DNA tests have the potential to be routine and decentralised as are today's blood tests. The realisation of this prospect requires sequence analysis devices also to be further miniaturised. This thesis has put the foundation by demonstrating functional complex DNA sequence analysis workflows on prototypical embedded systems constructed out of over-the-shelf embedded computing device. The goal was archived through architecture-aware optimisation of the analysis software, and with the future goal of building domain-specific architectures for DNA sequence analysis in mind.

The proposed architecture in this thesis was evaluated by a prototype constructed out of multiple single-board computers interconnected using Ethernet. The prototype is bulky, mainly due to many cables. However, designing a custom carrier board that can accommodate multiple off-the-shelf system-on-modules with integrated Ethernet and power delivery will produce a system that is many times smaller than the prototype. Besides, the proposed system can be miniaturised into a single chip by designing a multiprocessor system on a chip (MPSoC) composed of application-specific instruction-set processors (ASIP). Such an MPSoC will be a magnitude of times smaller, with superior performance and lower energy when compared to the current prototype. Such an MPSoC integrated into an ultra-portable sequencer will enable complete DNA analysis on the palmtop.

Orthogonal to the above directions, the currently developed embedded system can be extended to an end-to-end application of direct biological significance such as a diagnostic test. However, such a direction would require strong collaboration with biologists and clinicians. Furthermore, there are other branches in genomics that can be explored, for instance, Ribonucleic acid (RNA) workflows, meta-genomic analyses and de-novo assembly. As exemplified in this thesis for two DNA analysis workflows (genetic variant detection using second-generation sequencing and epigenetic modification detection using third-generation sequencing), the other workflows also will have significant room for improvement through architecture-aware optimisation alone. Increased collaboration between researchers from the two domains---computer architecture and DNA sequencing---will be favourable to efficiently reduce the gap between DNA sequencing analysis.

%% file: 9.5-appendices/1-minimap.tex
\chapter[Appendix: Featherweight Long Read Alignment]{Supplementary Materials - Featherweight Long Read Alignment using Partitioned Reference Indexes}\label{a:minimap-supps}

\rule{\textwidth}{0.4pt} 
This appendix is published as supplementary material of \cite{minimap2arm} in Nature Scientific Reports under Creative Commons CC BY license.\\
\rule{\textwidth}{0.4pt} 

\section[Supplementary Note 1]{Supplementary Note 1 - Detailed Methodology of the merging}\label{s:minimap2-supp1}

This supplementary note elaborates the merging method in detail together with some implementation details.

\subsection{Serialising (dumping) of the internal state}
\vspace{3mm}

For each part of the partitioned index, a separate intermediate file (which we refer to as a dump) is created in the binary format [refer line 36-44 in \url{https://github.com/hasindu2008/minimap2-arm/blob/v0.1-alpha/merge.c}].
After a read is aligned to the partition of the index currently in memory, all the intermediate states for its alignments are dumped into this binary file [line 501-506 in \url{https://github.com/hasindu2008/minimap2-arm/blob/v0.1-alpha/map.c}]. Binary format was preferred as it reduces the file size compared to ASCII. When the last read is mapped to the current partition of the index in memory, the dump will contain the intermediate state of the mappings for all the reads, in the same order as the reads in the input read set.  If the partitioned index had n partitions, at the end of the n\textsuperscript{th} partitions we will have n such dumps.

The dumped internal state includes; fifteen 32-bit unsigned integers (such as the reference ID, chaining scores, query and reference start and end), two 32-bit signed integers and one floating point value. All these information are inside a single structure in Minimap2 (called \textit{mm\_reg1\_t} in \textit{minimap.h}) which made the dumping convenient. The size required for a single alignment is around 80 bytes. 

If the user has requested Minimap2 to generate the base-level alignment, then the internal state for base-level alignment are also dumped. Base-level alignment information include; six 32-bit integers (such as the base level alignment score, number of CIGAR operations and a variable size flexible integer array for storing CIGAR operations. These information are stored inside another structure in Minimap2 (called \textit{ mm\_extra\_t}), which is only allocated if the base level alignment has been requested. The memory address to this structure is stored as a pointer in the previously mentioned \textit{mm\_reg1\_t} structure. When dumping, we flatten the information linearly (eliminate memory pointers) to the file.

In addition to the above, a quantity called \textit{replen} (sum of lengths of regions in the read that are covered by highly repetitive k-mers) is dumped. This is a per read quantity. We save the \textit{replen} to the same dump file that we discussed above, just after the information for each mapping.
For each read there will be a \textit{replen} for each part of the index, that is saved in the dump for that particular part of the partitioned index [line 495 of \url{https://github.com/hasindu2008/minimap2-arm/blob/v0.1-alpha/map.c}].
\vspace{3mm}

\subsection{Merging operation} 
 \vspace{3mm}
When alignment of all reads to all parts of the index completes, the merging operation is invoked [\textit{merge} function in \url{https://github.com/hasindu2008/minimap2-arm/blob/v0.1-alpha/merge.c}]. We simultaneous open the read file and the dump files for all parts of the partitioned index.
Reads are sequentially loaded while loading all the internal states for the alignments of that read. This includes the internal state for all its alignments (includes the base-level information if it had been requested) as well as the \textit{replen} from each dump file. The flattened data in the files are restored to their original structures when loading to the memory.

If no base-level alignments had been requested, the alignments are sorted based on the chaining score in descending order [function \textit{mm\_hit\_sort\_by\_score} in \url{https://github.com/hasindu2008/minimap2-arm/blob/v0.1-alpha/merge.c}]. If base-level alignment had been requested, they are sorted based on the base-level DP alignment score. 
Categorisation of primary and secondary chains is performed on the sorted alignments according to the same method done on Minimap2 (using \textit{mm\_set\_parent} function). 
This fixes the issue with the primary vs secondary flag. Then the alignment entries are filtered based on the user requested number of secondary alignments and the priority ratio (using \textit{mm\_select\_sub} function). This eliminates the issue of outputting secondary alignments for each part of the index that makes the output size huge. 
If the output has been requested in form of a SAM file, the best primary alignment is set to the primary flag while all other primary alignments are set to supplementary (using \textit{mm\_set\_sam\_pri} function). 

The mapping quality (MAPQ) estimation depends on the length of the read covered by repeat regions in the genome. To compute a perfect value for this quantity, the whole index needs to be in the memory which is the case for a single reference index. 
However, we estimate this quantity by taking the maximum out of the \textit{replen} values that were dumped for the particular read. The Spearman correlation of this estimated value to the perfect \textit{replen} was 0.9961. 
As the mapping quality is anyway an estimation, computing the mapping quality based on the estimated \textit{replen} does not affect the final results significantly.

\vspace{3mm}
\subsection{Emulated single reference index}
\vspace{3mm}
For memory efficiency, Minimap2 stores meta-data of reference sequences (such as the sequence name and sequence length) only in the reference index (refer to \textit{mm\_idx\_t} struct in \textit{minimap.h}). The order in which the sequences reside in the \textit{struct} array forms a unique numeric identifier for each reference sequence.

In the internal state for mappings only this numeric identifier is stored. The meta-data for the reference sequence are resolved using these numeric identifiers, only during the output printing. However, during merging we do not have the reference indexes in memory and the numeric identifiers cannot be resolved. Hence, we construct an emulated single reference index. For this, we save the meta-data of the reference sequences when each part of the partitioned index is loaded [line 47-54 in \url{https://github.com/hasindu2008/minimap2-arm/blob/v0.1-alpha/merge.c}]. These meta-data go to the beginning of the dump file for the particular part of the index. At the beginning of the merging, the meta-data is 
loaded back to form an emulated single reference index [line 164-173 in \url{https://github.com/hasindu2008/minimap2-arm/blob/v0.1-alpha/merge.c}]. However, the numeric identifiers in
the internal states from the dump files are incorrect (as numeric the identifier is an independent incrementing index for each part of the index). These are corrected to be compatible with the numeric identifiers in the emulated single reference index by adding the correct offset [line 254 in \url{https://github.com/hasindu2008/minimap2-arm/blob/v0.1-alpha/merge.c}].

%This is fixed during the merge stage by adding an offset. 
%However in a partitioned index, by default the ID numbers for chromosomes  are based on only what it in that corresponding part. 
 %- Dumping the sequence IDs, sequence names and the lengths of the sequences in the index mi. 
As a side effect of this emulated single reference index, a correct SAM header can be output even in the partitioned mode.
Further, the merging process which merges the mappings for a read at a time, outputs the mappings for a particular read ID adjacently. Hence, no additional sorting is required for any downstream analysis tools that require so.\\\\ 

\section[Chromosome Balancing]{Supplementary Note 2 - Detailed Methodology of the chromosome balancing}\label{s:minimap2-supp2}

\subsection{Memory efficiency for references with unbalanced lengths}
\vspace{3mm}

The existing partitioned index construction method in Minimap2, does not balance the size of index partitions when the reference genome has sequences (chromosomes) with highly varying lengths. This existing index construction method puts the reference sequences to the index in the order they exist in the reference genome. When constructing a partitioned index, it moves to the next part of of the index only when the user specified number of bases per index (by default 4 Gbases) is exceeded. When building a partitioned index for overlap finding, the parts would be approximately equal in size as the length of the longest read would be a few mega bases. However, in case of a reference genomes like the human genome where the chromosomes are of highly variable lengths, the size of the parts are unbalanced. The largest part of the index determines the peak memory. Hence, an unbalance will hinder the maximum efficiency for systems with limited memory.
For instance, consider a hypothetical genome (total length 700M) 
with following chromosomes and lengths in the order chr1 (300M), chr2 (320M), chr3 (60M), chr4 (20M).
Providing a value of 350M as the number of bases in a partition (with the intention of splitting into 2 parts), will create an unbalanced index as follows.

\begin{itemize}
\item part1 : chr1, chr2 : total length - 620M 
\item part2 : chr3, chr4 : total length - 80M 
\end{itemize}

We follow a simple partitioning approach to balance this out. Instead of the number of bases per partition, the number of partitions is taken as a user input.
The reference sequences are first sorted in descending order based on the sequence length (length without the ambiguous N bases). The sum of bases in each partition is initialised to 0. The, the sorted list in traversed  in order while assigning the current sequence into the partition with the minimum sum of bases. The sum of bases in that partition is updated accordingly. Using this strategy, we get a distribution as follows.

\begin{itemize}
\item part1 : 300M, 60M : total length - 360M
\item part2 : 320M, 20M : total length - 340M
\end{itemize}
\vspace{2cm}

\section{Supplementary Note 3 - Instructions to run the tools} \label{s:minimap2-supp3}

\subsection{Example}
\vspace{5mm}

\begin{enumerate}

\item
  Download and compile minimap2 that supports partitioned indexes and
  merging
  
\begin{lstlisting}[language=bash]
wget https://github.com/hasindu2008/minimap2-arm/archive/v0.1.tar.gz
tar xvf v0.1.tar.gz && cd minimap2-arm-0.1 && make
\end{lstlisting}

\item
  Download the human reference genome and create a partitioned index with 4 partitions

\begin{lstlisting}[language=bash]
wget -O hg38noAlt.fa.gz http://bit.ly/hg38noAlt && gunzip hg38noAlt.fa.gz
./misc/idxtools/divide_and_index.sh hg38noAlt.fa 4 hg38noAlt.idx ./minimap2 map-ont

\end{lstlisting}

Note : \url{http://bit.ly/hg38noAlt} redirects to 
\\
\url{ftp://ftp.ncbi.nlm.nih.gov/genomes/all/GCA/000/001/405/GCA_000001405.15_GRCh38/seqs_for_alignment_pipelines.ucsc_ids/GCA_000001405.15_GRCh38_no_alt_analysis_set.fna.gz} 

\item
  Download a Nanopore NA12878 dataset and run Minimap2 with merging

\begin{lstlisting}[language=bash]
wget -O na12878.fq.gz http://bit.ly/NA12878
./minimap2 -a -x map-ont hg38noAlt.idx na12878.fq.gz --multi-prefix tmp > out.sam
\end{lstlisting}

Note : \url{http://bit.ly/NA12878} redirects to 
\\
\url{http://s3.amazonaws.com/nanopore-human-wgs/rel3-nanopore-wgs-84868110-FAF01132.fastq.gz}
\end{enumerate}

Notes :

\begin{itemize}
\item
  To perform mapping without base-level alignment use:

\begin{lstlisting}[language=bash]
./minimap2 -x map-ont hg38noAlt.idx na12878.fq.gz --multi-prefix tmp > out.paf
\end{lstlisting}
\end{itemize}

\begin{itemize}
\item
  From Minimap2 version 2.12-r827
  [https://github.com/lh3/minimap2/blob/master/NEWS.md\#release-212-r827-6-august-2018]
  onwards, the merging functionality has been integrated into the main
  repository. This version additionally supports paired-end short reads
  and the merging operation is multi-threaded. Use
  \texttt{-\/-split-prefix} option instead of \texttt{-\/-multi-prefix}.
\end{itemize}

\vspace{5 mm}

\subsection{Index construction with chromosome size
balancing}

\vspace{3 mm}

\texttt{divide\_and\_index.sh} is the wrapper script for balanced index construction. It takes the reference genome and outputs a partitioned index optimised for reduced peak memory. Its usage is as follows:

\begin{lstlisting}[language=bash]
usage : ./divide_and_index.sh <reference.fa> <num_parts> <out.idx> <minimap2_exe> 
<minimap2_profile>

reference.fa - path to the fasta file containing the reference genome
num_parts - number of partitions in the index
out.idx - path to the file to which the index should be dumped
minimap2_exe - path to the minimap2 executable
minimap2_profile - minimap2 pre-set for indexing (map-pb or map-ont)

Example : ./divide_and_index.sh hg19.fa 4 hg19.idx minimap2 map-ont
\end{lstlisting}

Functionality of \texttt{divide\_and\_index.sh} is as follows. 
\begin{enumerate}
\item
Compiling \texttt{divide.c} using \texttt{gcc} to produce
\texttt{divide}. 
\item Calling the compiled binary \texttt{divide} to split
the reference genome into partitions such that the total length of
chromosomes in each partition are approximately equal. 
\item Calling the
minimap2 binary separately on each reference partition to produce a
separate index file for each partition. 
\item Combining all the index files
to produce a single partitioned index file.
\end{enumerate}
\vspace{5 mm}
\subsection{Running Minimap2 on a partitioned index with merging}

\vspace{3 mm}
To run minimap2 on an index created using the above method :

\begin{lstlisting}[language=bash]
minimap2 -x <profile> <partioned_index.idx> <reads.fastq> --multi-prefix <tmp-prefix>
\end{lstlisting}

\texttt{-\/-multi-prefix} which takes a prefix for temporary files,
enables the merging of the outputs generated through iterative mapping
to index partitions.

%% file: 9.5-appendices/2-f5c.tex
\chapter{Appendix: \textit{f5c} Documentation}\label{a:f5c-documentation}

\rule{\textwidth}{0.4pt} 
This appendix is based on the \textit{f5c} documentation available at \url{https://hasindu2008.github.io/f5c} associated with the GitHub repository at \url{https://github.com/hasindu2008/f5c}.\\
\rule{\textwidth}{0.4pt}

\section{Readme}

\textit{f5c} is an optimised re-implementation of the \emph{call-methylation} and
\emph{eventalign} modules in
\href{https://github.com/jts/nanopolish}{Nanopolish}. Given a set of
basecalled Nanopore reads and the raw signals, \emph{f5c
call-methylation} detects the methylated cytosine and \emph{f5c
eventalign} aligns raw nanopore DNA signals (events) to the base-called
read. \emph{f5c} can optionally utilise NVIDIA graphics cards for
acceleration.

First, the reads have to be indexed using \texttt{f5c index}. Then, invoke
\texttt{f5c call-methylation} to detect methylated cytosine bases.
Finally, you may use \texttt{f5c meth-freq} to obtain methylation
frequencies. Alternatively, invoke \texttt{f5c eventalign} to perform
event alignment. The results are almost the same as from \textit{nanopolish}
except for a few differences due to floating point approximations.

\emph{Full Documentation} :
\href{https://hasindu2008.github.io/f5c/docs/overview}{\url{https://hasindu2008.github.io/f5c/docs/overview}}

\emph{Pre-print} :
\href{https://www.biorxiv.org/content/10.1101/756122v1}{\url{https://doi.org/10.1101/756122}}

%\href{https://travis-ci.org/hasindu2008/f5c}{\includegraphics{https://travis-ci.org/hasindu2008/f5c.svg?branch=master}}

\subsection{Quick start}\label{quick-start}

If you are a Linux user and want to quickly try out, download the
compiled binaries from the
\href{https://github.com/hasindu2008/f5c/releases}{latest release}. For
example:

\begin{lstlisting}[language=bash]
VERSION=v0.4
wget "https://github.com/hasindu2008/f5c/releases/download/$VERSION/f5c-$VERSION-binaries.tar.gz" && tar xvf f5c-$VERSION-binaries.tar.gz && cd f5c-$VERSION/
./f5c_x86_64_linux        # CPU version
./f5c_x86_64_linux_cuda   # cuda supported version
\end{lstlisting}

Binaries should work on most Linux distributions and the only dependency
is \texttt{zlib} which is available by default on most distros.

\subsection{Building}\label{building}

Users are recommended to build from the
\href{https://github.com/hasindu2008/f5c/releases}{latest release} tar
ball. You need a compiler that supports C++11. Quick example for Ubuntu
:

\begin{lstlisting}[language=bash]
sudo apt-get install libhdf5-dev zlib1g-dev   #install HDF5 and zlib development libraries
VERSION=v0.4
wget "https://github.com/hasindu2008/f5c/releases/download/$VERSION/f5c-$VERSION-release.tar.gz" && tar xvf f5c-$VERSION-release.tar.gz && cd f5c-$VERSION/
scripts/install-hts.sh  # download and compile the htslib
./configure
make                    # make cuda=1 to enable CUDA support
\end{lstlisting}

The commands to install hdf5 (and zlib) \textbf{development libraries}
on some popular distributions :

\begin{lstlisting}[language=bash]
On Debian/Ubuntu : sudo apt-get install libhdf5-dev zlib1g-dev
On Fedora/CentOS : sudo dnf/yum install hdf5-devel zlib-devel
On Arch Linux: sudo pacman -S hdf5
On OS X : brew install hdf5
\end{lstlisting}

If you skip \texttt{scripts/install-hts.sh} and \texttt{./configure}
hdf5 will be compiled locally. It is a good option if you cannot install
hdf5 library system wide. However, building hdf5 takes ages.

Building from the Github repository additionally requires
\texttt{autoreconf} which can be installed on Ubuntu using
\texttt{sudo apt-get install autoconf automake}.

Other building options are detailed in section \ref{building-f5c}.Instructions
to build a docker image is detailed section \ref{title-docker-image}.

\subsubsection{NVIDIA CUDA support}\label{nvidia-cuda-support}

To build for the GPU, you need to have the CUDA toolkit installed. Make sure
nvcc (NVIDIA C Compiler) is in your PATH.

The building instructions are the same as above, except that you should
call \textit{make} as :

\begin{lstlisting}[language=bash]
make cuda=1
\end{lstlisting}

Optionally you can provide the CUDA architecture as :

\begin{lstlisting}[language=bash]
make cuda=1 CUDA_ARCH=-arch=sm_xy
\end{lstlisting}

If your CUDA library is not in the default location
/usr/local/cuda/lib64, point to the correct location as:

\begin{lstlisting}[language=bash]
make cuda=1 CUDA_LIB=/path/to/cuda/library/
\end{lstlisting}

Refer to section \ref{title-cuda-troubleshooting} for troubleshooting CUDA related problems.

\subsection{Usage}\label{usage}

\begin{lstlisting}[language=bash]
f5c index -d [fast5_folder] [read.fastq|fasta]
f5c call-methylation -b [reads.sorted.bam] -g [ref.fa] -r [reads.fastq|fasta] > [meth.tsv]
f5c meth-freq -i [meth.tsv] > [freq.tsv]
f5c eventalign -b [reads.sorted.bam] -g [ref.fa] -r [reads.fastq|fasta] > [events.tsv]
\end{lstlisting}

Refer to section \ref{title-commands-and-options} for all the commands and options.

\subsubsection{Example}\label{example}

Follow the same steps as in
\href{https://nanopolish.readthedocs.io/en/latest/quickstart_call_methylation.html}{Nanopolish
tutorial} while replacing \texttt{nanopolish} with \texttt{f5c}. If you
only want to perform a quick test of f5c :

\begin{lstlisting}[language=bash]
#download and extract the dataset including sorted alignments
wget -O f5c_na12878_test.tgz "https://f5c.page.link/f5c_na12878_test"
tar xf f5c_na12878_test.tgz

#index, call methylation and get methylation frequencies
f5c index -d chr22_meth_example/fast5_files chr22_meth_example/reads.fastq
f5c call-methylation -b chr22_meth_example/reads.sorted.bam -g chr22_meth_example/humangenome.fa -r chr22_meth_example/reads.fastq > chr22_meth_example/result.tsv
f5c meth-freq -i chr22_meth_example/result.tsv > chr22_meth_example/freq.tsv
#event alignment
f5c eventalign -b chr22_meth_example/reads.sorted.bam -g chr22_meth_example/humangenome.fa -r chr22_meth_example/reads.fastq > chr22_meth_example/events.tsv
\end{lstlisting}

\subsection{Acknowledgement}\label{acknowledgement}

This repository reuses code and methods from
\href{https://github.com/jts/nanopolish}{Nanopolish}. The event detection code is from Oxford Nanopore's
\href{https://github.com/nanoporetech/scrappie}{Scrappie
basecaller}. Some code snippets have been taken from
\href{https://github.com/lh3/minimap2}{Minimap2} and
\href{http://samtools.sourceforge.net/}{Samtools}.

\section{Building f5c}\label{building-f5c}

Note : Building from the Github repository requires \texttt{autoreconf}
which can be installed on Ubuntu using
\texttt{sudo apt-get install autoconf automake}.

Clone the git repository.

\begin{lstlisting}[language=bash]
git clone https://github.com/hasindu2008/f5c && cd f5c
\end{lstlisting}

Alternatively, download the
\href{https://github.com/hasindu2008/f5c/releases}{latest release}
tarball and extract. eg :

\begin{lstlisting}[language=bash]
VERSION=v0.4
wget "https://github.com/hasindu2008/f5c/releases/download/$VERSION/f5c-$VERSION-release.tar.gz" && tar xvf f5c-$VERSION-release.tar.gz && cd f5c-$VERSION/
\end{lstlisting}

While we have tried hard to avoid the dependency hell, three
dependencies (zlib, HDF5 and HTS) could not be avoided.

Currently 3 building methods are supported.

\begin{enumerate}
\itemsep1pt\parskip0pt\parsep0pt
\item
  Locally compiled HTS library and system wide HDF5 library
  (recommended).
\item
  Locally compiled HTS and HDF5 libraries (HDF5 local compilation -
  takes a bit of time).
\item
  System wide HTS and HDF5 libraries (not recommended as HTS versions
  can be old).
\end{enumerate}

\subsection{Method 1 (recommended)}\label{method-1-recommended}

Dependencies : Install the HDF5 (and zlib development libraries).

\begin{lstlisting}[language=bash]
On Debian/Ubuntu : sudo apt-get install libhdf5-dev zlib1g-dev
On Fedora/CentOS : sudo dnf/yum install hdf5-devel zlib-devel
On Arch Linux: sudo pacman -S hdf5
On OS X : brew install hdf5
\end{lstlisting}

Now build f5c.

\begin{lstlisting}[language=bash]
autoreconf              # skip if compiling a release, only required when building from github
scripts/install-hts.sh  # download and compiles htslib in the current folder
./configure
make                    # or make cuda=1 if compiling for CUDA
\end{lstlisting}

\subsection{Method 2 (time consuming)}\label{method-2-time-consuming}

Dependencies : Install the zlib development libraries.

\begin{lstlisting}[language=bash]
On Debian/Ubuntu : sudo apt-get install zlib1g-dev
On Fedora/CentOS : sudo dnf/yum install zlib-devel
\end{lstlisting}

Now build f5c.

\begin{lstlisting}[language=bash]
autoreconf                      # skip if compiling a release, only required when building from github
scripts/install-hts.sh          # download and compiles htslib in the current folder
scripts/install-hdf5.sh         # download and compiles HDF5 in the current folder
./configure --enable-localhdf5
make                            # or make cuda=1 if compiling for CUDA
\end{lstlisting}

\subsection{Method 3 (not recommended)}\label{method-3-not-recommended}

Dependencies : Install HDF5 and hts.

\begin{lstlisting}[language=bash]
On Debian/Ubuntu : sudo apt-get install libhdf5-dev zlib1g-dev libhts1
\end{lstlisting}

Now build f5c.

\begin{lstlisting}[language=bash]
autoreconf                      # skip if compiling a release, only required when building from github
./configure --enable-systemhts
make                            # or make cuda=1 if compiling for CUDA
\end{lstlisting}

\subsection{Docker Image}\label{title-docker-image}

To build a docker image:

\begin{lstlisting}[language=bash]
git clone https://github.com/hasindu2008/f5c && cd f5c
docker build .
\end{lstlisting}

Note down the image uuid and run f5c as:

\begin{lstlisting}[language=bash]
docker run -v /path/to/local/data/data/:/data/ -it :image_id  ./f5c call-methylation -r /data/reads.fa -b /data/alignments.sorted.bam -g /data/ref.fa
\end{lstlisting}

\subsection{CUDA
Troubleshooting}\label{title-cuda-troubleshooting}

\subsection{Compiling Issues}\label{compiling-issues}

\subsubsection{\texttt{make: nvcc: Command not found} error when I compile
with
\texttt{make cuda=1}}\label{make-nvcc-command-not-found-error-when-i-compile-with-make-cuda1}

Make sure that the NVIDIA CUDA toolkit is installed. See instruction at
the \href{https://docs.nvidia.com/cuda/}{official installation
guide}. If you still get this error after the toolkit installation,
then \texttt{nvcc} is probably not in your PATH. In that case, either
add the nvcc location to your PATH or manually specify the nvcc location
through a Makefile variable.

Example:

If you installed the CUDA toolkit through \texttt{apt} in Ubuntu,

\begin{lstlisting}[language=bash]
make cuda=1 NVCC=/usr/local/cuda/bin/nvcc
\end{lstlisting}

If you did the
\href{https://docs.nvidia.com/cuda/cuda-installation-guide-linux/index.html\#runfile}{Runfile
installation} on Ubuntu,

\begin{lstlisting}[language=bash]
make cuda=1 NVCC=/usr/local/cuda-<toolkit-version>/bin/nvcc
\end{lstlisting}

Note that the location of \texttt{nvcc} might be different depending on
your distribution and the installation method.

\subsubsection{Cannot find \texttt{-lcudart\_static}
error.}\label{cannot-find--lcudartux5fstatic-error.}

The default CUDA library path in the Makefile is set to be
\texttt{/usr/local/cuda/lib64}.

While this is the default path for an Ubuntu 64-bit system with the CUDA
toolkit installed using the package manager \texttt{apt}, it might be
different on your system. You can manually specify the path to the cuda
library when compiling.

Example:

If you did the
\href{https://docs.nvidia.com/cuda/cuda-installation-guide-linux/index.html\#runfile}{Runfile
installation} on Ubuntu,

\begin{lstlisting}[language=bash]
make cuda=1 CUDA_LIB=/usr/local/cuda-<toolkit-version>/lib64/
\end{lstlisting}

If you are using Ubuntu 32-bit,

\begin{lstlisting}[language=bash]
make cuda=1 CUDA_LIB=/usr/local/cuda/lib/
\end{lstlisting}

Note that the location of the CUDA library path might be different
depending on your distribution and the installation method.

\subsubsection{memcpy was not declared in this scope
error}\label{memcpy-was-not-declared-in-this-scope-error}

If you get an error like this:

\begin{lstlisting}[language=bash]
$ make cuda=1
nvcc -x cu -g -O2 -std=c++11 -lineinfo -Xcompiler -Wall -I./htslib -DHAVE_CUDA=1 -rdc=true -c src/f5c.cu -o build/f5c_cuda.o
/usr/include/string.h: In function 'void* __mempcpy_inline(void*, const void*, size_t)':
/usr/include/string.h:652:42: error: 'memcpy' was not declared in this scope
return (char *) memcpy (__dest, __src, __n) + __n;
^
Makefile:76: recipe for target 'build/f5c_cuda.o' failed
make: *** [build/f5c_cuda.o] Error 1
\end{lstlisting}

Compile with \texttt{-D\_FORCE\_INLINES} appended to
\texttt{CUDA\_CFLAGS} when calling \texttt{make}:

\begin{lstlisting}[language=bash]
CUDA_CFLAGS+="-D_FORCE_INLINES" make cuda=1
\end{lstlisting}

The issue is reported in
\href{https://github.com/hasindu2008/f5c/issues/36}{\#36}. And this fix
is suggested in
\href{https://github.com/opencv/opencv/issues/6500}{opencv/opencv\#6500}.

\subsection{Runtime Errors}\label{runtime-errors}

\subsubsection{CUDA driver version is insufficient for CUDA runtime
version}\label{cuda-driver-version-is-insufficient-for-cuda-runtime-version}

Check the following in order:

\begin{enumerate}
\item
  \textbf{Do you have an NVIDIA GPU / is your NVIDIA GPU recognised by the
  system?}\\On most distributions you can use the following command to
  verify:

\begin{lstlisting}[language=bash]
lspci | grep -i "vga\|3d\|display"
\end{lstlisting}

  It should list the NVIDIA GPU:

\begin{lstlisting}[language=bash]
01:00.0 3D controller: NVIDIA Corporation GP107M [GeForce GTX 1050 Ti Mobile] (rev a1)
\end{lstlisting}
\item
  \textbf{Have you installed the NVIDIA driver (not the open source nouveau
  driver)?}\\ On most distributions you can check your graphics card driver
  using

\begin{lstlisting}[language=bash]
lspci -nnk | grep -iA2 "vga\|3d\|display"
\end{lstlisting}

  If the kernel driver output contains \texttt{nvidia}, then you are
  using the correct driver.

\begin{lstlisting}[language=bash]
01:00.0 3D controller [0302]: NVIDIA Corporation GP107M [GeForce GTX 1050 Ti Mobile] [10de:1c8c] (rev a1)
       Kernel driver in use: nvidia
       Kernel modules: nvidiafb, nouveau, nvidia_396, nvidia_396_drm
\end{lstlisting}
\item
  \textbf{If you are using a Tegra GPU (e.g.~Jetson TX2), does the current user
  belong to the ``video'' user group?}\\Check the current group names
  with \texttt{group {[}user{]}}.\\
\item
  \textbf{Is the CUDA driver version too old for the toolkit that is used to
  compile with?}\\ 
  See the \href{https://docs.nvidia.com/deploy/cuda-compatibility/index.html\#binary-compatibility}{cuda binary compatibility guide}.\\The
  release CUDA binary that we provide is compiled using CUDA toolkit
  6.5. The CUDA runtime library is statically linked and therefore
  release CUDA binaries work on driver version \textgreater{}=
  340.21.\\Use \texttt{nvidia-smi} to check your driver version.

\begin{lstlisting}[language=bash]
$ nvidia-smi
+-----------------------------------------------------------------------------+
| NVIDIA-SMI 396.44                 Driver Version: 396.44                    |
\end{lstlisting}

  If you compiled the binary yourself, see the
  \href{https://docs.nvidia.com/deploy/cuda-compatibility/index.html\#binary-compatibility}{cuda binary compatibility guide}
  to check if your toolkit version and driver version match.
\end{enumerate}

\subsection{Commands and
options}\label{title-commands-and-options}

\subsection{Available f5c tools}\label{available-f5c-tools}

\begin{lstlisting}[language=bash]
Usage: f5c <command> [options]

command:
         index               Build an index mapping from basecalled reads to the signals measured by the sequencer (same as nanopolish index)
         call-methylation    Classify nucleotides as methylated or not (optimised nanopolish call-methylation)
         meth-freq           Calculate methylation frequency at genomic CpG sites (optimised nanopolish calculate_methylation_frequency.py)
         eventalign          Align nanopore events to reference k-mers (optimised nanopolish eventalign)
\end{lstlisting}

\subsubsection{Indexing}\label{indexing}

\begin{lstlisting}[language=bash]
Usage: f5c index [OPTIONS] -d nanopore_raw_file_directory reads.fastq
Build an index mapping from basecalled reads to the signals measured by the sequencer
f5c index is equivalent to nanopolish index by Jared Simpson

  -h, --help                           display this help and exit
  -v, --verbose                        display verbose output
  -d, --directory                      path to the directory containing the raw ONT signal files. This option can be given multiple times.
  -s, --sequencing-summary             the sequencing summary file from albacore, providing this option will make indexing much faster
  -f, --summary-fofn                   file containing the paths to the sequencing summary files (one per line)
\end{lstlisting}

\subsubsection{Calling methylation}\label{calling-methylation}

\begin{lstlisting}[language=bash]
Usage: f5c call-methylation [OPTIONS] -r reads.fa -b alignments.bam -g genome.fa
   -r FILE                    fastq/fasta read file
   -b FILE                    sorted bam file
   -g FILE                    reference genome
   -w STR[chr:start-end]      limit processing to genomic region STR
   -t INT                     number of threads [8]
   -K INT                     batch size (max number of reads loaded at once) [512]
   -B FLOAT[K/M/G]            max number of bases loaded at once [2.0M]
   -h                         help
   -o FILE                    output to file [stdout]
   --iop INT                  number of I/O processes to read fast5 files [1]
   --min-mapq INT             minimum mapping quality [30]
   --secondary=yes|no         consider secondary mappings or not [no]
   --verbose INT              verbosity level [0]
   --version                  print version
   --disable-cuda=yes|no      disable running on CUDA [no]
   --cuda-dev-id INT          CUDA device ID to run kernels on [0]
   --cuda-max-lf FLOAT        reads with length <= cuda-max-lf*avg_readlen on GPU, rest on CPU [3.0]
   --cuda-avg-epk FLOAT       average number of events per kmer - for allocating GPU arrays [2.0]
   --cuda-max-epk FLOAT       reads with events per kmer <= cuda_max_epk on GPU, rest on CPU [5.0]
   -x STRING                  profile to be used for optimal CUDA parameter selection. user-specified parameters will override profile values
advanced options:
   --kmer-model FILE          custom k-mer model file
   --skip-unreadable=yes|no   skip any unreadable fast5 or terminate program [yes]
   --print-events=yes|no      prints the event table
   --print-banded-aln=yes|no  prints the event alignment
   --print-scaling=yes|no     prints the estimated scalings
   --print-raw=yes|no         prints the raw signal
   --debug-break [INT]        break after processing the specified batch
   --profile-cpu=yes|no       process section by section (used for profiling on CPU)
   --skip-ultra FILE          skip ultra long reads and write those entries to the bam file provided as the argument
   --ultra-thresh [INT]       threshold to skip ultra long reads [100000]
   --write-dump=yes|no        write the fast5 dump to a file or not
   --read-dump=yes|no         read from a fast5 dump file or not
   --meth-out-version [INT]   methylation tsv output version (set 2 to print the strand column) [1]
   --cuda-mem-frac FLOAT      Fraction of free GPU memory to allocate [0.9 (0.7 for tegra)]
\end{lstlisting}

\subsubsection{Calculate methylation
frequency}\label{calculate-methylation-frequency}

\begin{lstlisting}[language=bash]
Usage: meth-freq [options...]

  -c [float]        Call threshold. Default is 2.5.
  -i [file]         Input file. Read from stdin if not specified.
  -o [file]         Output file. Write to stdout if not specified.
  -s                Split groups
\end{lstlisting}

\subsubsection{Aligning events}
\begin{lstlisting}[language=bash]
Usage: f5c eventalign [OPTIONS] -r reads.fa -b alignments.bam -g genome.fa
   -r FILE                    fastq/fasta read file
   -b FILE                    sorted bam file
   -g FILE                    reference genome
   -w STR[chr:start-end]      limit processing to genomic region STR
   -t INT                     number of threads [8]
   -K INT                     batch size (max number of reads loaded at once) [512]
   -B FLOAT[K/M/G]            max number of bases loaded at once [2.0M]
   -h                         help
   -o FILE                    output to file [stdout]
   --iop INT                  number of I/O processes to read fast5 files [1]
   --min-mapq INT             minimum mapping quality [30]
   --secondary=yes|no         consider secondary mappings or not [no]
   --verbose INT              verbosity level [0]
   --version                  print version
   --disable-cuda=yes|no      disable running on CUDA [no]
   --cuda-dev-id INT          CUDA device ID to run kernels on [0]
   --cuda-max-lf FLOAT        reads with length <= cuda-max-lf*avg_readlen on GPU, rest on CPU [3.0]
   --cuda-avg-epk FLOAT       average number of events per kmer - for allocating GPU arrays [2.0]
   --cuda-max-epk FLOAT       reads with events per kmer <= cuda_max_epk on GPU, rest on CPU [5.0]
   -x STRING                  profile to be used for optimal CUDA parameter selection. user-specified parameters will override profile values
advanced options:
   --kmer-model FILE          custom k-mer model file
   --skip-unreadable=yes|no   skip any unreadable fast5 or terminate program [yes]
   --print-events=yes|no      prints the event table
   --print-banded-aln=yes|no  prints the event alignment
   --print-scaling=yes|no     prints the estimated scalings
   --print-raw=yes|no         prints the raw signal
   --debug-break [INT]        break after processing the specified batch
   --profile-cpu=yes|no       process section by section (used for profiling on CPU)
   --skip-ultra FILE          skip ultra long reads and write those entries to the bam file provided as the argument
   --ultra-thresh [INT]       threshold to skip ultra long reads [100000]
   --write-dump=yes|no        write the fast5 dump to a file or not
   --read-dump=yes|no         read from a fast5 dump file or not
   --summary FILE             summarise the alignment of each read/strand in FILE
   --sam                      write output in SAM format
   --print-read-names         print read names instead of indexes
   --scale-events             scale events to the model, rather than vice-versa
   --samples                  write the raw samples for the event to the tsv output
   --cuda-mem-frac FLOAT      Fraction of free GPU memory to allocate [0.9 (0.7 for tegra)]
\end{lstlisting}

%% file: 9.5-appendices/3-f5c-misc.tex
\chapter[Appendix: \textit{f5c}]{Supplementary Materials - \textit{f5c}}\label{a:f5c-supps}

\section{Why Nanopolish had to be re-engineered?}

There are three reasons why \textit{Nanopolish} had to be completely re-engineered into \textit{f5c} for a successful GPU implementation.

\begin{itemize}
    \item  \textit{Nanopolish} performs on-demand loading of signal data from file (a CPU thread assigned to the particular read invokes a file access just prior to signal alignment). However, transferring read by read to the GPU will incur a massive penalty and thus a batch of reads have to be transferred at once. Thus, we had to re-write the Nanopolish processing framework in such a way that loading and processing of a batch are performed batch wise.  In \textit{f5c}, we read a batch of data to the RAM and then bulk transfer to GPU memory, a batch of \textit{n} reads at a time. 

    \item Nanopolish thread model un-suitable for GPU acceleration---a thread is dynamically assigned to a read using openMP, thus each read has its own code path. However, offloading a batch of reads to the GPU for signal alignment requires code paths of all the reads in the batch to have converged before the GPU kernel is invoked. In addition, accurately measuring time, benchmarking and profiling of individual algorithmic components is hindered by such divergent code paths. \textit{pthread} based approach that interleaves input reading, processing and output.

    \item Nanopolish is not optimised for efficient resource utilisation (eg: marginal performance improvement beyond 16 threads on servers and heavy-weight for embedded systems due to spurious \textit{malloc} calls). A comparison of such a version with the GPU would result in an apparent high speedup, which is unfair.
    
\end{itemize}

\section{Additional advantages of \textit{f5c} over \textit{Nanopolish}}

In addition to the GPU acceleration of ABEA, \textit{f5c} has many additional advantages over original \textit{Nanopolish}.

\begin{itemize}
\item I/O and processing are interleaved in \textit{f5c}: the I/O latency is considerably minimised.
\item Our CPU version alone is around 1.5X-2X faster than the \textit{Nanopolish} call methylation implementation and is very lightweight - suitable for embedded systems due to the careful use of  data structures and algorithms.
\item \textit{f5c} is capable of detecting load balance problems between CPU and GPU, and report user with suggestion for appropriate parameters.
\item  \textit{f5c} works with package manager's system wide installations of HDF5 (no need of thread-safe build of HDF5), hence no need locally compile HDF5.
\item Dependency hell has been minimised for both CPU and GPU versions. Compatible with
g++ 4.8 or higher, and CUDA toolkit 6.5 or higher.  
\item \textit{f5c} has suggestive error message for troubleshooting, especially the issues with respect to GPU. 
\item \textit{Pthread} based thread framework written in C that interleaves I/O with processing is very lightweight and can be a starting point for future Nanopore tools.
\item \textit{f5c} allows benchmarking section by section to identify the bottlenecks in performance.
\item \textit{f5c} framework is suitable for the acceleration of core kernels through other methods such as FPGA.
\end{itemize}

\chapter[Appendix: Portable Binaries]{Generating portable binaries for ONT tools}\label{a:potablebin}

\rule{\textwidth}{0.4pt} 
This appendix is based on a blog article published at \url{https://hasindu2008.github.io/portable-binary}.\\
\rule{\textwidth}{0.4pt}

Compiling software can sometimes be a nightmare due to numerous
dependencies. This is specifically the case for bioinformatics tools
that utilise signal level data from Oxford Nanopore (ONT) sequencers.
According to my experience, the major cause behind compilation troubles
in ONT tools is the Hierarchical Data Format 5 (HDF5)
library\footnote{Currently, the raw signal data from the ONT sequencers are stored in HDF5 file format. Possibly due to the complexity of HD5, there are no alternate library  implementations than the official library from the HDF Group. Compiling the HDF5 library takes time. Luckily, package managers' versions of HDF5 library exists, but there seem to be some inconsistencies across various distributions. Thus, a software developed on one Linux distribution will rarely compile without any trouble on a different system. For example, the header file \textit{<hdf5.h>} resides directly \textit{include} directory on certain systems, while in some other system it can be \textit{<hdf5/hdf5.h>} or \textit{<hdf5/serial/hdf5.h>}.}. While a system admin may enjoy tedious compilations,
it is not the case for users of bioinformatics tools. What if the tool
developers release pre-compiled binaries? Some would object this as it
is not a perfect solution. Nevertheless, I believe that it is far better
than releasing an unusable tool due to users giving up at the
compilation stage. Further, pre-compiled binaries are less bulky
compared to docker images. Generating a ``portable binary'' that runs on
numerous Linux distributions/version is tricky, but possible with some
additional work from the developer's side.

Explained below is a recipe (or probably some key points) to generate
``portable binaries'' for ONT tools. In summary, this strategy uses a
combination of static linking and dynamic linking to generate a
``portable binary''. Dynamically linking all libraries means that the
user would have to install the exact version of the library as the
developer. On the other end, statically linking everything is also not
ideal\footnote{Static linking is not ideal due to libraries such as GNU libc being non-portable (see \url{http://stevehanov.ca/blog/?id=97}. But, many reasons why statically linked binaries are good for bioinformatics tools are detailed in \url{http://lh3.github.io/2014/07/12/about-static-linking}.
}. Thus, a hybrid static and dynamic linking strategy is
the way to go. However, this recipe is limited to C/C++ based tools. In
addition, portability here means that a binary compiled for a particular
operating system would work on other distributions (and versions) of the
same operating system. For example, the binary for Linux on x86\_64
architecture would run despite the distribution (whether Ubuntu, Debian,
Fedora or Red hat) or the version. Portability here does NOT mean that
the Linux version will run on Windows or that x86\_64 version will run
on ARM.

\section{Key points}\label{key-points}

The key points for a successful portable binary are:

\begin{enumerate}
\item
  Avoid unnecessary dependencies as much as possible.
\item
  Identify libraries which causes problems when dynamically linked. Such
  libraries are good candidates for linking statically. Examples are:

    \begin{itemize}
    \item
      libraries which do not honour backward compatibility
    \item
      non-mature libraries which the API frequently change
    \item
      the name/location of the shared object (.so) file installed by the
      package manager is different on different distributions.
    \end{itemize}

\item
  Identify libraries which are better to be left dynamically linked. The
  best example is glibc which is not recommended to be statically
  linked. Luckily glibc sufficiently maintains backward
  compatibility and can be left dynamically linked.
\item
  Generate the binaries on a machine (a virtual machine is sufficient)
  with an old Linux distribution (eg: Ubuntu 14 or even better if Ubuntu
  12) installed with older libraries. For instance, glibc which we
  decided to be left dynamically linked is NOT
  forward-compatible\footnote{binaries compiled for an older glibc version will run on a system with a newer glibc (glibc is backward compatible). However, binaries compiled for newer glibc versions will not always work with an older glibc (glibc is not forward compatible).
}.
\item
  Try to avoid package manager's version for libraries when statically
  linking. Instead, compile those libraries yourself with minimal
  features that you require for your tool. For instance, statically
  linking HDF5 package manager's version, also require linking
  additional libraries such as \emph{libsz} (a lossless compression
  library for scientific data) and \emph{libaec} (Adaptive Entropy
  Coding library). Those can be avoided if we compile HDF5 ourselves
  without those features (if your tool does not require those additional
  features).
\end{enumerate}

\section{A case study with \textit{f5c}}\label{a-case-study-with-f5c}

Now let's go through the above points with reference to
\href{https://github.com/hasindu2008/f5c/{]}}{f5c}, a tool which we are
currently developing that utilises ONT raw data.

\begin{enumerate}
\item
  We tried our best to avoid dependencies. However, three external
  dependencies HDF5, HTSlib (high-throughput sequencing), zlib (data
  compression library) and obviously standard libraries (such as glibc,
  pthreads) could not be avoided. We generate the binaries for f5c on
  Ubuntu 14.
\item
  HDF5 and HTSlib are statically linked. The location of the .so file of
  HDF5 is not consistent across distributions and even different
  versions in the same distribution. HTSlib that comes with the package
  manager is an older version and f5c required a newer version to
  support long reads. Thus, we statically link HDF5 and HTSlib. For the
  CUDA supported version of f5c, we statically link the CUDA runtime
  library as well, which is explained later.
\item
  zlib and other standard libraries are dynamically linked. Executing
  the command \emph{ldd} on a release binary of f5c (``portable
  binary'') gives the list of dynamically linked libraries shown below.
  Note that HD5F and HTSlib were statically linked and thus not seen in
  the \emph{ldd} output.
   \begin{lstlisting}[language=bash]
   $ldd ./f5c
   linux-vdso.so.1 =>  (0x00007fffc91fb000)
   libdl.so.2 => /lib/x86_64-linux-gnu/libdl.so.2 (0x00007f61550d0000)
   libpthread.so.0 => /lib/x86_64-linux-gnu/libpthread.so.0 (0x00007f6154eb0000)
   libz.so.1 => /lib/x86_64-linux-gnu/libz.so.1 (0x00007f6154c90000)
   libstdc++.so.6 => /usr/lib/x86_64-linux-gnu/libstdc++.so.6 (0x00007f61548f0000)
   libm.so.6 => /lib/x86_64-linux-gnu/libm.so.6 (0x00007f61545e0000)
   libgcc_s.so.1 => /lib/x86_64-linux-gnu/libgcc_s.so.1 (0x00007f61543c0000)
   libc.so.6 => /lib/x86_64-linux-gnu/libc.so.6 (0x00007f6153fe0000)
   /lib64/ld-linux-x86-64.so.2 (0x00007f6155400000)
   \end{lstlisting}
\item
  We compile on a virtual machine with Ubuntu 14.
\item
  To highlight why compiling external libraries ourselves with minimal
  features, let us see the additional output of \emph{ldd} when f5c was
  dynamically linked with the package managers' HDF5 (see below).
  Observe that now in addition to the actual HDF5 library
  (\emph{libhdf5\_serial.so}) we have got two additional dependencies
  (\emph{libsz.so} and \emph{libaec.so}). Thus, if one is to statically
  link this package manager's HDF5 version, then \emph{libsz} and
  \emph{libaec} also would have to be statically linked. Compiling HDF5
  ourselves let us drop these features which we do not want.

   \begin{lstlisting}[language=bash]
   $ldd ./f5c
   ...
   libhdf5_serial.so.10 => /usr/lib/x86_64-linux-gnu/libhdf5_serial.so.10 (0x00007f1b21f30000)
   libsz.so.2 => /usr/lib/x86_64-linux-gnu/libsz.so.2 (0x00007f1b20c30000)
   libaec.so.0 => /usr/lib/x86_64-linux-gnu/libaec.so.0 (0x00007f1b20800000)
   ...
   \end{lstlisting}
\end{enumerate}
 
\subsection{Note on CUDA libraries}\label{note-on-cuda-libraries}

CUDA runtime is not both forward and backward compatible and requires
the exact version to be installed. Hence dynamically linked CUDA runtime
is of no much use. Luckily CUDA runtime library has been designed to
support static linking. In fact, NVIDIA recommends statically compiling
the CUDA runtime library (refer the
\href{https://docs.nvidia.com/cuda/cuda-c-best-practices-guide/index.html}{CUDA
best practices guide}) and the default behaviour of the CUDA C compiler
(nvcc) 5.5 or is to statically link the CUDA runtime. However, CUDA
runtimes are coupled with CUDA driver versions. NVIDIA states that CUDA
Driver API is backward compatible but not forward compatible (see
\href{https://docs.nvidia.com/cuda/cuda-c-best-practices-guide/index.html\#cuda-compatibility-and-upgrades}{here})
and thus CUDA Runtime compiled against a particular Driver will work on
later driver releases, but may not work on earlier driver versions. As a
result, generating the binary should better be done with an old CUDA
toolkit version. Otherwise, the users will have to install the latest
drivers to run this binary. For f5c we installed the CUDA 6.5 toolkit
version on the Ubuntu 14 virtual machine to generate CUDA binaries.

\subsection{Example commands}\label{example-commands}

Assume we have HDF5 and HTSlib locally compiled and the static libraries
(libhdf5.a and libhts.a) are located in ./build/lib/. These libraries
are statically linked as :

\begin{lstlisting}[language=bash]
<gcc/g++> [options] <object1.o> <object2.o> <...> build/lib/libhdf5.a -ldl build/lib/libhts.a  -lpthread -lz  -o binary
\end{lstlisting}

To statically link the CUDA runtime when using gcc or g++:

\begin{lstlisting}[language=bash]
<gcc/g++> [options]   <object1.o> <object2.o> <...> build/lib/libhdf5.a build/lib/libhts.a  -L/usr/local/cuda/lib64 -lcudart_static -lpthread -lz  -lrt -ldl -o binary
\end{lstlisting}

Alternatively if CUDA toolkit 5.5 higher NVIDIA C compiler \emph{nvcc}
links the CUDA runtime statically by default:

\begin{lstlisting}[language=bash]
nvcc [options] <object1.o> <object2.o> <...> build/lib/libhdf5.a build/lib/libhts.a  -lpthread -lz  -lrt -ldl -o binary
\end{lstlisting}

After generating the binary issue the \emph{ldd} command to verify if
the intended ones are statically linked. The output of \emph{ldd} lists
the dynamically linked libraries and the statically linked libraries
should NOT appear in this output.

\begin{lstlisting}[language=bash]
ldd ./binary
\end{lstlisting}

% \begin{center}\rule{3in}{0.4pt}\end{center}

% Credits to \href{https://github.com/danielltb}{@danielltb} for sharing
% valuable knowledge about static linking and compatibility.

% \begin{center}\rule{3in}{0.4pt}\end{center}

%% file: 9.5-appendices/4-nanocluster-f5p.tex
\chapter{Appendix: Rock64-cluster and \textit{f5p}}\label{a:nanocluster}

\rule{\textwidth}{0.4pt} 
This appendix is based on the documentation associated with the GitHub repositories at \url{https://github.com/hasindu2008/nanopore-cluster} and \\ \url{https://github.com/hasindu2008/f5p}.\\
\rule{\textwidth}{0.4pt} 

\section{Rock64-cluster}\label{nanopore-cluster}

\subsection{Required Hardware}\label{required-hardware}

A cluster of computers connected to each other using Gigabit Ethernet. We built our cluster using 16 Rock64 single board computers and the list of items we used are:

\begin{itemize}
\item
  Rock64 single board computers (4GB RAM)
\item
  Rock64 heat sinks
\item
  64 GB eMMC modules
\item
  USB to type H barrel 5V DC power cables
\item
  Orico DUB-8P-BK USB charging stations (as power supplies for Rock64 devices)
\item
  Copper Cylinders for Raspberry Pi
\item
  HPE OfficeConnect 1950 24G 2SFP+ 2XGT switch
\item
  Ethernet cables
\item
  USB adaptor for eMMC modules (to flash eMMC)
\end{itemize}

\subsection{Connecting nodes
together}\label{connecting-nodes-together}

\begin{itemize}
\item
  Build the cluster.
\item
  Connect the nuts and bolts (using copper cylinders).
\item
  Flash Linux distributions onto eMMCs (we flashed Ubuntu).
\item
  Plug eMMCs and heat sinks to the Rock64s.
\item
  Connect Rock64s onto the switch and power supplies.
\item
  Configure the switch.
\item
  Assign IP addresses to the Rock64 devices.
\item
  Provide Internet to the Rock64 devices.
\end{itemize}

\subsection{Setting up the \emph{head
node}}\label{setting-up-the-head-node}

The node which will be used to control, connect and assign work to the
other \emph{worker nodes} is referred to as the \emph{head node}. This
\emph{head node} can be a Rock64 itself or any other computer. We used
an old PC as the \emph{head node}. On the head node you may want to do
the following.

\begin{itemize}
\item
  Install and configure
  \href{https://docs.ansible.com/ansible/latest/index.html}{\emph{ansible}}.
  \emph{Ansible} will be used to launch commands on all \emph{worker
  nodes}, centrally from the \emph{head node}.
\item
  Install \emph{ansible}. On Ubuntu:
\begin{lstlisting}[language=bash]
    sudo apt-add-repository ppa:ansible/ansible
    sudo apt update
    sudo apt install ansible
\end{lstlisting}
\item
  Configure \emph{ansible}. You need to edit your
  \texttt{/etc/ansible/ansible.cfg} and\\ \texttt{/etc/ansible/hosts}. Our
  sample config files are at \url{scripts/sample_config/ansible}.
\item
  Create an SSH key on the head node. You can use the command
  \texttt{ssh-keygen}. This key is needed for password-less access to
  the \emph{worker nodes}.
\item
  Mount the network attached storage. You can add an entry to the
  \texttt{/etc/fstab} for persist across reboots.
\item
  Optionally, you can install \emph{ganglia} to monitor various metrics
  of the nodes. In Ubuntu you may use:
\begin{lstlisting}[language=bash]
  sudo apt-get install ganglia-monitor rrdtool gmetad ganglia-webfrontend
  sudo cp /etc/ganglia-webfrontend/apache.conf /etc/apache2/sites-enabled/ganglia.conf
 \end{lstlisting}
\end{itemize}

You have to edit the configuration files
\texttt{/etc/ganglia/gmetad.conf} and\\ /etc/\texttt{ganglia/gmond.conf}.
Our sample configuration files are at \url{scripts/sample_config/ganglia}. In
summary, commands are:

You may refer to the tutorial \href{https://hostpresto.com/community/tutorials/how-to-install-and-configure-ganglia-monitor-on-ubuntu-16-04/}{[here]} on installing and configuring ganglia.

\begin{itemize}
\item
  Optionally, you can configure \emph{rsyslog} and
  \href{https://loganalyzer.adiscon.com/}{LogAnalyzer} to centrally view
  the logs through a web browser.
\item
  Add path of scripts/system to PATH.
\end{itemize}

\subsection{Compiling software and preparing the folder
structure}\label{compiling-software-and-preparing-the-folder-structure}

On one of your nodes (rock64 devices):

\begin{itemize}
\item
  Compile the software. We compiled minimap2, nanopolish, samtools and
  f5c for ARM architecture.
\item
  Create folder named \texttt{nanopore} under \texttt{/}.
\item
  Put compiled binaries to a folder named \texttt{/nanopore/bin}.
\item
  Put the reference genome and a minimap2 index under
  \texttt{/nanopore/reference}.
\item
  Create a folder named \texttt{/nanopore/scratch} for later use.
\end{itemize}

The directory structure should look like bellow :
\begin{lstlisting}[language=bash]
  nanopore
  |__ bin
  |   |__ f5c
  |   |__ minimap2-arm
  |   |__ nanopolish
  |   |__ samtools
  |__ reference
  |   |__ hg38noAlt.fa
  |   |__ hg38noAlt.fa.fai
  |   |__ hg38noAlt.idx
  |__ scratch
\end{lstlisting}

\subsection{Setting up \emph{woker
nodes}}\label{setting-up-woker-nodes}

On the \emph{worker node}:

\begin{itemize}
\item
  Change device name.
\item
  Change the time zone (and configure ntp).
\item
  Perform apt update and package installation eg: nfs-common ganglia-monitor.
\item
  Mount the network attached storage.
\item
  Create a swap space.
\item
  Copy the binaries and the folder structure we constructed before.
\end{itemize}

A shell script that perform the above is available at \url{scripts/new_workernode_setup/run_on_workernode.sh}.

On the \emph{head node}:
\begin{itemize}
\item
  Copy ssh-key to the \emph{worker node}.
\item
  Copy \emph{ganglia} configuration files to the \emph{worker node}.
\item
  Copy \emph{rsyslog} configuration files to the \emph{worker node}.
\end{itemize}

A shell script that perform the above is available at \url{scripts/new_workernode_setup/run_on_headnode.sh}.

\section{\textit{f5p}}\label{a:f5p}

\textit{f5p} is a lightweight job scheduler and daemon for nanopore data processing on a nanopore mini-cluster.

\subsection{Pre-requisites}\label{pre-requisites}

\begin{itemize}
\item
  A compute-cluster composed of devices running Linux connected to each
  other preferably using Ethernet.
\item
  One of the devices will act as the \emph{head node} to issue commands
  to other \emph{worker nodes}.
\item
  A shared network mounted storage for storing data.
\item
  SSH key based access from \emph{head node} to \emph{worker nodes}.
\item
  Optionally you may configure
  \href{https://docs.ansible.com/ansible/latest/index.html}{ansible} to
  automate configuration tasks.
\end{itemize}

\subsection{Getting started}\label{getting-started}

\subsection{Building and initial
configuration}\label{building-and-initial-configuration}

\begin{enumerate}
  
\item
  First build the scheduling daemon (\emph{f5pd}) and client
  (\emph{f5pl}).

\begin{lstlisting}[language=bash]
make
\end{lstlisting}

\item
  Scheduling client (\emph{f5pl}) is destined for the \emph{head node}.
  Copy the scheduling daemon (\emph{f5pd}) to all \emph{worker nodes}.
  If you have configured ansible, you adapt the following command.

\begin{lstlisting}[language=bash]
ansible all -m copy -a "src=./f5pd dest=/nanopore/bin/f5pd mode=0755"
\end{lstlisting}

\item
  Run the scheduling daemon (\emph{f5pd}) on all \emph{worker nodes}.
  You may want to add (\emph{f5pd}) as a
  \emph{\href{http://manpages.ubuntu.com/manpages/cosmic/man5/systemd.service.5.html}{systemd
  service}} that runs on the start-up. See
  \href{https://github.com/hasindu2008/f5_pipeline/blob/master/scripts/f5pd.service}{scripts/f5pd.service}
  for an example \emph{systemd configuration} and
  \href{https://github.com/hasindu2008/f5_pipeline/blob/master/scripts/install_f5pd_service.sh}{scripts/install\_f5pd\_service.sh}
  for an example script.
\item
  On the \emph{head node} create a file containing the list of IP
  addresses of the \emph{worker nodes}, one IP address per line. An
  example is in
  \href{https://github.com/hasindu2008/f5_pipeline/blob/master/data/ip_list.cfg}{data/ip\_list.cfg}.
\item
  Optionally, you may install a web server on the \emph{head node} and
  host the scripts under
  \href{https://github.com/hasindu2008/f5_pipeline/tree/master/scripts/front}{scripts/front}
  to view the log on a web-browser. You will need to edit the paths in
  these scripts to point to the log location. Note that these scripts
  are not probably safe to be hosted on a public server.
\end{enumerate}

\subsection{Running for a dataset}\label{running-for-a-dataset}

\begin{enumerate}
  
\item
  Modify the shell script
  \href{https://github.com/hasindu2008/f5_pipeline/blob/master/scripts/fast5_pipeline.sh}{scripts/fast5\_pipeline.sh}
  for your use-case. This script is to be called on \emph{worker nodes}
  by (\emph{f5pd}), each time a data unit is assigned. The example
  script:
    \begin{itemize}
    \item
      takes a location of a tar file on the network mount (which contains a
      batch of \emph{fast5} files) as the argument;
    \item
      deduce the location of \emph{fastq} file on the network mount
      associated to the tar file;
    \item
      copy the tar file and \emph{fastq} file to the local storage;
    \item
      runs a methylation-calling pipeline that uses the tools
      \emph{minimap2}, \emph{samtools} and \emph{nanopolish}; and,
    \item
      copy the results back to the network mount.
    \end{itemize}
Note that this scripts should exit with a non zero status if any thing
went wrong. After modifying the script, copy it to the \emph{worker
nodes} to the location\\ \texttt{/nanopore/bin/fast5\_pipeline.sh}

\item
  On the \emph{head node} create a file containing the list of tar files
  (each tar file contains a fast5 batch), one tar file per line. An
  example is in
  \href{https://github.com/hasindu2008/f5_pipeline/blob/master/data/file_list.cfg}{data/file\_list.cfg}.
\item
  Launch the \emph{f5pl} with the IP list and the tar file list you
  previously created as the arguments.
\end{enumerate}
\begin{lstlisting}[language=bash]
./f5pl data/ip_list.cfg data/file_list.cfg
\end{lstlisting}
You may adapt the script
\href{https://github.com/hasindu2008/f5_pipeline/blob/master/scripts/run.sh}{scripts/run.sh}
which performs a run discussed above.

%% file: 9.5-appendices/5-phone.tex
\chapter[Appendix: Bioinformatics on Mobile Phone]{Getting Command line Bioinformatics Tools Working on Android}\label{a:onphone}
\rule{\textwidth}{0.4pt} 
This appendix is based on the blog articles published at \url{https://hasindu2008.github.io/linux-tools-on-phone} and \url{https://hasindu2008.github.io/linux-tools-on-phone2}.\\
\rule{\textwidth}{0.4pt}

This is a very hacky method and is solely for testing out. In summary, we generate a completely statically linked binary on an ARM based single board computer running Linux. The method is only for tools written in C/C++. I will show steps for four examples, namely minimap2, samtools, f5c and nanopolish.

Then statically linked binaries which we generated can be downloaded from \url{http://bit.ly/2INNeRv}.
The sample data\footnote{Contains chromosome 22, a small set of \href{https://github.com/nanopore-wgs-consortium/NA12878}{NA12878 Nanopore reads} and some E.coli Nanopore reads from the \href{https://nanopolish.readthedocs.io/en/latest/quickstart_consensus.html}{Nanopolish tutorial}.} for the following examples can be downloaded from \url{http://bit.ly/2XOK1Yg}

\section{Requirements}

\begin{itemize}
\item A mobile phone running Android. Does not require rooting\footnote{At the time of writing Android (tested on Android 7 and 8) seem to allow executing binaries from `/data/local/tmp' through the Android Debug Bridge (ADB). As long as this is not blocked in the future versions, the method should work.}. My phone used for testing was a cheap \href{https://www.gsmarena.com/lg_q6-8756.php}{LG Q6} phone running Android 7.
\item An ARM based single board computer (will call it SBC here onwards) running Linux. We used an \href{https://wiki.odroid.com/odroid-xu4/odroid-xu4}{Odroid XU4} running Ubuntu 16.04.4 LTS.
\item A USB cable to connect your phone. Optionally a host computer (laptop or a desktop) to connect the phone. Even the SBC can be used as the host.
\end{itemize}

You might wonder if the mobile phone and the SBC) should have the same ARM architecture (i.e. ARMv7 or ARMv8). Not necessarily. The LG Q6 mobile phone had an ARMv8 (Octa-core 1.4 GHz Cortex-A53) processor architecture while the Odroid XU4 had ARMv7 (Cortex-A15 2Ghz and Cortex-A7 Octa core). However, the mobile phone despite its ARMv8 64-bit processor, was still running a 32-bit version of the OS, thus running `cat /proc/cpu' on the phone through Android Debug Bridge (ADB) output the following (a similar outcome to that on a latest Raspberry Pi with ARMv8 processor running the 32-bit Raspbian).

\begin{lstlisting}[language=bash]
mh:/data/local/tmp $ cat /proc/cpuinfo
processor       : 0
model name      : ARMv7 Processor rev 4 (v7l)
BogoMIPS        : 38.40
Features        : half thumb fastmult vfp edsp neon vfpv3 tls vfpv4 idiva idivt vfpd32 lpae evtstrm aes pmull sha1 sha2 crc32
CPU implementer : 0x41
CPU architecture: 7
CPU variant     : 0x0
CPU part        : 0xd03
CPU revision    : 4
\end{lstlisting}

In case your mobile is running a 64-bit OS, you might need an SBC running 64-bit as well.

\section{Steps}

\begin{enumerate}
\item Setup the Android Debug Bridge (ADB)

	You have to setup your host computer to be able to connect to your phone through ADB. There are a number of tutorials for this on the Internet which you can follow. For example see \url{https://devsjournal.com/download-minimal-adb-fastboot-tool.html}. Note that this step might slightly vary for different Android phones. This is the summary of what we did:

	\begin{itemize}
	\item Installed the minimal version of ADB. The ADB command line tool comes with the Android SDK, but we preferred the minimal version of ADB as it is light weight. For Windows, you can download minimal ADB from \url{https://forum.xda-developers.com/showthread.php?t=2317790}. For Linux, you may use the package manager (eg : `sudo apt-get install android-tools-adb android-tools-fastboot').

	\item Installed the USB drivers for the phone. We used the OEM version (through the manufacturer website given at \url{https://developer.android.com/studio/run/oem-usb#Drivers}. Even the \href{https://adb.clockworkmod.com}{Universal ADB driver} should work for most phones.

	\item Enabled developer options on Android and then allowed USB debugging.

	\item  Connected the phone through USB to the computer. Opened a command line on the computer and issued `adb devices' command. If everything is successful, the phone connected to the computer should be listed.

	\begin{lstlisting}[language=bash]
	C:\Program Files (x86)\Minimal ADB and Fastboot>adb devices
	List of devices attached
	LGM70059258dab  device
	\end{lstlisting}

	If your phone is not listed (usually it happens to me most of the time due to incompatible driver or ADB versions etc) you will have to do a bit of playing around with some patience.
	\end{itemize}

\item Download the source code of the tool onto the SBC and compile with `-static' option to generate a statically linked binary. See examples in the next section.

\item Copy the static binary to the location `/data/local/tmp' of the mobile phone using the `adb push` command. This location `/data/local/tmp' allows us setting executable permissions and running a binary through ADB. This location works up till Android 8.1.0 version. Hopefully will not be restricted in the future versions.

	\begin{lstlisting}[language=bash]
	C:\Program Files (x86)\Minimal ADB and Fastboot>adb push "/path/to/binary" /data/local/tmp/
	\end{lstlisting}

\item Launch an `adb shell' (will give us a shell on the phone) and set executable permission to the binary we just copied. Then you can execute the binary on the phone.

	\begin{lstlisting}[language=bash]
	C:\Program Files (x86)\Minimal ADB and Fastboot>adb shell
	mh:/ $ cd /data/local/tmp
	mh:/data/local/tmp $ chmod +x binaryname
	mh:/data/local/tmp $ ./binaryname
	\end{lstlisting}

\end{enumerate}

\section{Examples}
\subsection{minimap2}

\begin{figure}[!t]
  \centering
\begin{subfigure}[!ht]{0.49\linewidth}
  \centering
    \includegraphics[width=\textwidth]{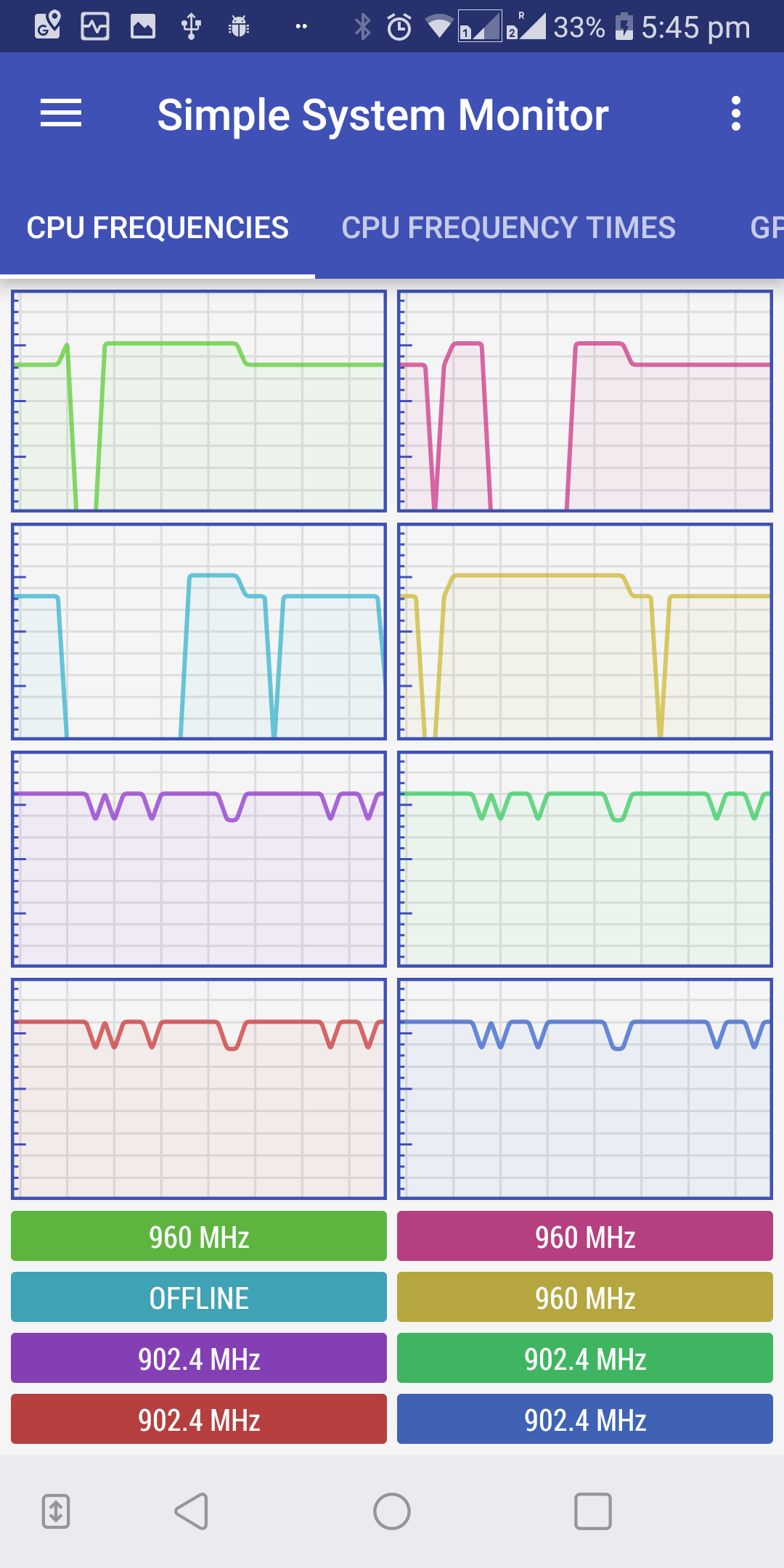}
    \caption{CPU Usage} 
\end{subfigure}
\begin{subfigure}[!ht]{0.49\linewidth}
  \centering
    \includegraphics[width=\textwidth]{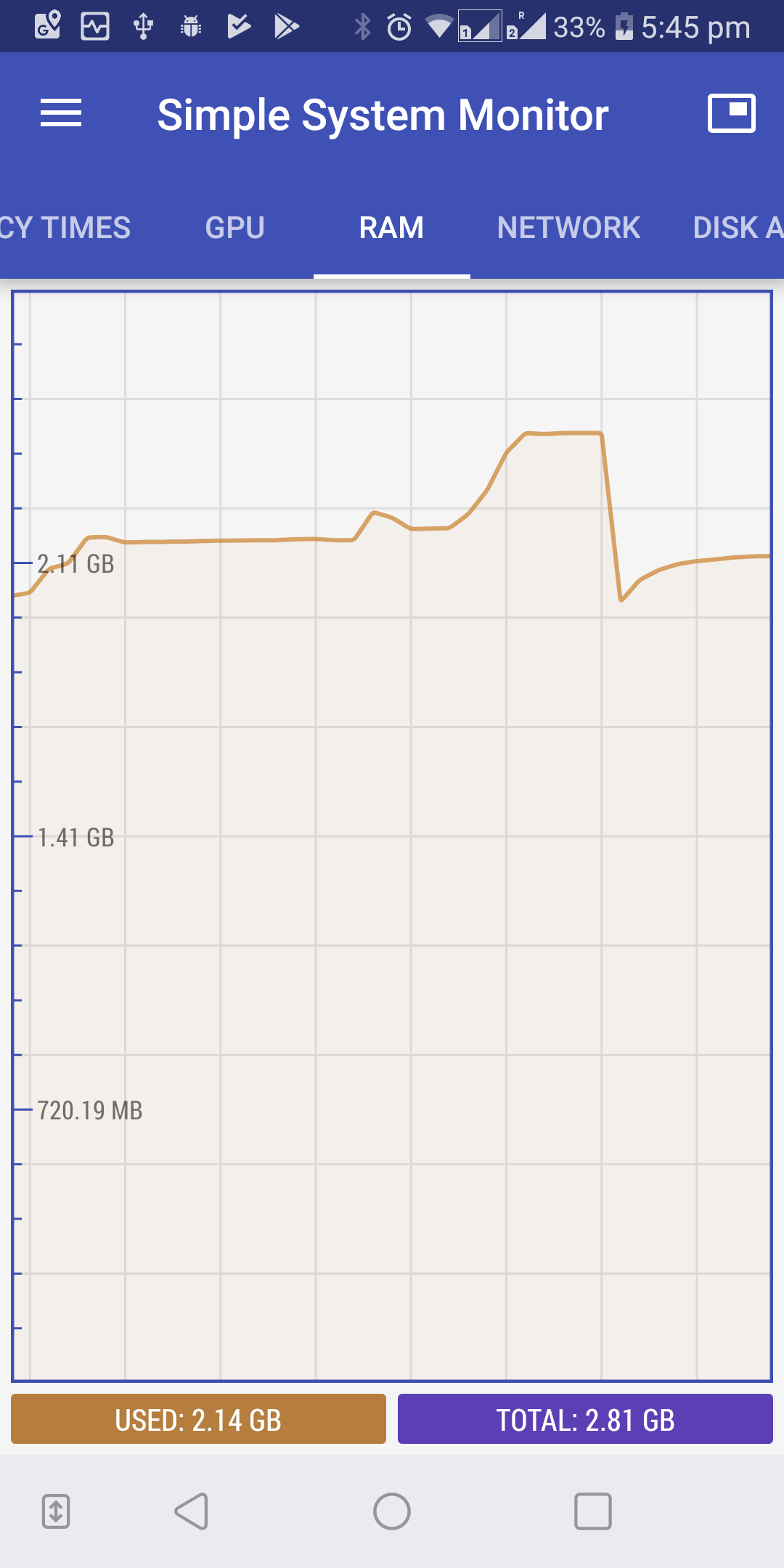}
    \caption{RAM usage} 
\end{subfigure}
    \caption{CPU and RAM usage} 
    \label{f:cpu-ram-usage-on-phone}
\end{figure}

\begin{enumerate}
\item First, download the minimap2 source code on to the SBC. This example uses my fork of minimap2 which was patched to support ARM. You may also use version 2.7 or higher from the original minimap2 repository at \url{https://github.com/lh3/minimap2} which supports ARM.
	\begin{lstlisting}[language=bash]
	wget -O minimap2-arm.tar.gz "https://github.com/hasindu2008/minimap2-arm/archive/v0.1.tar.gz" && tar xvf minimap2-arm.tar.gz && cd minimap2-arm-0.1/
	\end{lstlisting}

\item Open the \textit{Makefile} (located inside the extracted source code directory) using a text editor and get rid of \textit{getopt.o} by changing line 35 and 36 in \textit{Makefile} from:

	\begin{lstlisting}[language=bash]
	make
	minimap2:main.o getopt.o libminimap2.a
			$(CC) $(CFLAGS) main.o getopt.o -o $@ -L. -lminimap2 $(LIBS)
	\end{lstlisting}
	to
	\begin{lstlisting}[language=bash]
	make
	minimap2:main.o libminimap2.a
			$(CC) $(CFLAGS) main.o -o $@ -L. -lminimap2 $(LIBS)
	\end{lstlisting}

	This is to prevent the potential compilation error in the next step (i.e. multiple definition of `getopt' due to that in \textit{getopt.c} in current folder and the one in \textit{libc}). Note that in latest minimap2 versions, \textit{getopt.c} has been changed to \textit{ketopt.c} and this step is not required.

\item Compile with the `-static' option by passing the `CC' variable in Make as `gcc -static'. You will need to have the dependency \textit{zlib development files} installed (package manager can be used, eg: `sudo apt-get install zlib1g-dev').

	\begin{lstlisting}[language=bash]
	$ make arm_neon=1 CC="gcc -static"
	\end{lstlisting}

\item Make sure that the generated binary is statically linked.

	\begin{lstlisting}[language=bash]
	$ ldd ./minimap2
			not a dynamic executable
	\end{lstlisting}

\item Copy this binary to your mobile phone through ADB. We first copied the binary from the SBC to the laptop and then issued:

	\begin{lstlisting}[language=bash]
	C:\Program Files (x86)\Minimal ADB and Fastboot>adb push "C:\Users\hasindu\Desktop\minimap2" /data/local/tmp/
	C:\Users\hasindu\Desktop\minimap2: 1 file pushed. 14.0 MB/s (1470676 bytes in 0.100s)
	\end{lstlisting}

\item Provide executable permissions and launch minimap2 without arguments on your phone to see the usage message.

	\begin{lstlisting}[language=bash]
	C:\Program Files (x86)\Minimal ADB and Fastboot>adb shell
	mh:/ $ cd /data/local/tmp
	mh:/data/local/tmp $ ls -l minimap2
	-rw-rw-rw- 1 shell shell 1470676 2019-06-15 17:31 minimap2
	mh:/data/local/tmp $ chmod +x minimap2
	mh:/data/local/tmp $ ./minimap2
	Usage: minimap2 [options] <target.fa>|<target.idx> [query.fa] [...]
	....
	\end{lstlisting}

\item Copy a reference genome and set of reads on to the storage of your mobile phone. We copied chr22  and a set of nanopore reads file onto \textit{/sdcard/genome/} on my phone (chr22.fa and 740475-67.fastq in our test dataset (available at \url{http://bit.ly/2XOK1Yg}). You can use `adb push' or the Windows Explorer based phone browser.

\item Now align some reads to the reference. We ran with 4 threads instead of 8 threads as the phone otherwise got laggy. The `-K5M' option to limit the batch size to cap the peak memory (my phone had only 3GB of RAM). Note that chr22 reference is small and fits adequately to 2GB RAM. If you want to align to a full human genome on a limited memory system see chapter \ref{c:minimap} and appendix \ref{a:minimap-supps}.

	\begin{lstlisting}[language=bash]
	127|mh:/data/local/tmp $ ./minimap2 -x map-ont -a /sdcard/genome/chr22.fa /sdcard/genome/740475-67.fastq -t4 -K5M > /sdcard/genome/reads.sam
	[M::mm_idx_gen::8.923*0.99] collected minimizers
	[M::mm_idx_gen::10.035*1.27] sorted minimizers
	[M::main::10.035*1.27] loaded/built the index for 1 target sequence(s)
	[M::mm_mapopt_update::10.394*1.26] mid_occ = 136
	[M::mm_idx_stat] kmer size: 15; skip: 10; is_hpc: 0; #seq: 1
	[M::mm_idx_stat::10.617*1.26] distinct minimizers: 4817802 (89.47% are singletons); average occurrences: 1.368; average spacing: 7.784
	[M::worker_pipeline::18.912*2.37] mapped 493 sequences
	[M::worker_pipeline::52.854*2.01] mapped 413 sequences
	[M::worker_pipeline::64.470*2.33] mapped 443 sequences
	[M::worker_pipeline::109.708*2.07] mapped 457 sequences
	[M::worker_pipeline::151.487*2.32] mapped 454 sequences
	[M::worker_pipeline::162.448*2.42] mapped 317 sequences
	[M::worker_pipeline::174.190*2.52] mapped 410 sequences
	[M::worker_pipeline::183.692*2.59] mapped 496 sequences
	[M::worker_pipeline::190.814*2.63] mapped 301 sequences
	[M::main] Version: 2.11-r797
	[M::main] CMD: ./minimap2 -x map-ont -a -t4 -K5M /sdcard/genome/chr22.fa /sdcard/genome/740475-67.fastq
	[M::main] Real time: 190.969 sec; CPU: 501.840 sec
	\end{lstlisting}

\item Optionally, observe the CPU and RAM usage by installing a system monitor application on your phone (Fig. \ref{f:cpu-ram-usage-on-phone}). We used \href{https://play.google.com/store/apps/details?id=com.dp.sysmonitor.app&hl=en_AU}{simple system monitor}.

\item When everything is done, check the output file.

	\begin{lstlisting}[language=bash]
	mh:/data/local/tmp $ ls -l /sdcard/genome/740475-67.fastq
	-rw-rw---- 1 root sdcard_rw 85784776 2018-06-29 19:39 /sdcard/genome/740475-67.fastq
	\end{lstlisting}

\end{enumerate}

\subsection{Samtools}

\begin{enumerate}
\item Download samtools source code.
	\begin{lstlisting}[language=bash]
	wget -O samtools.tar.gz "https://github.com/samtools/samtools/releases/download/1.9/samtools-1.9.tar.bz2" && tar -xvf samtools.tar.gz && cd samtools-1.9/
	\end{lstlisting}

\item Compile with `-static'. You need to have dependencies installed or else disable unwanted components through flags to ./configure. See official Samtools installation documentation  at \url{http://www.htslib.org/download}.

	\begin{lstlisting}[language=bash]
	./configure CC="gcc -static" --without-curses
	make
	\end{lstlisting}

\item Verify if statically linked.

	\begin{lstlisting}[language=bash]
	$ ldd ./samtools
			not a dynamic executable
	\end{lstlisting}

\item Copy the binary to your phone.

	\begin{lstlisting}[language=bash]
	C:\Program Files (x86)\Minimal ADB and Fastboot>adb push "C:\Users\hasindu\Desktop\samtools" /data/local/tmp/
	C:\Users\hasindu\Desktop\samtools: 1 file pushed. 9.3 MB/s (4859024 bytes in 0.496s)
	\end{lstlisting}

\item Set executable permissions and run. Output from minimap2 above (reads.sam) is sorted and then indexed in the example below.

	\begin{lstlisting}[language=bash]
	C:\Program Files (x86)\Minimal ADB and Fastboot>adb shell
	mh:/ $ cd  /data/local/tmp/
	mh:/data/local/tmp $ chmod +x samtools
	mh:/data/local/tmp $ ./samtools sort /sdcard/genome/reads.sam  > /sdcard/genome/reads.bam
	mh:/data/local/tmp $ ./samtools index /sdcard/genome/reads.bam
	\end{lstlisting}
\end{enumerate}

\subsection{F5C}

\begin{enumerate}

\item Download the source code and compile statically as follows. Library compilation will take time, bare with patience.

	\begin{lstlisting}[language=bash]
	wget -O f5c.tar.gz https://github.com/hasindu2008/f5c/releases/download/v0.1-beta/f5c-v0.1-beta-release.tar.gz && tar xvf f5c.tar.gz && cd f5c-v0.1-beta/
	scripts/install-hts.sh          # download and compiles htslib in the current folder
	scripts/install-hdf5.sh         # download and compiles HDF5 in the current folder
	./configure --enable-localhdf5
	make  CXX="g++ -static"            
	\end{lstlisting}

\item Copy the binary to the phone as in previous examples. Also, copy a set of Nanopore data including fast5 files (ecoli\_2kb\_region in our test dataset available at \url{http://bit.ly/2XOK1Yg}). Then index and perform  methylation calling using f5c as below.

	\begin{lstlisting}[language=bash]
	#indexing
	1|mh:/data/local/tmp $ ./f5c index -d /sdcard/genome/ecoli_2kb_region/fast5_files/ /sdcard/genome/ecoli_2kb_region/reads.fasta
	[readdb] indexing /sdcard/genome/ecoli_2kb_region/fast5_files/
	[readdb] num reads: 112, num reads with path to fast5: 112

	#f5c for mthylation calling
	1|mh:/data/local/tmp $./f5c call-methylation -r /sdcard/genome/ecoli_2kb_region/reads.fasta -g /sdcard/genome/ecoli_2kb_region/draft.fa -b /sdcard/genome/ecoli_2kb_region/reads.bam   >  /sdcard/genome/ecoli_2kb_region/ref.tsv

	[meth_main::1.595*0.98] 125 Entries (0.7M bases) loaded
	[pthread_processor::11.151*6.09] 125 Entries (0.7M bases) processed

	[meth_main] total entries: 125, qc fail: 0, could not calibrate: 0, no alignment: 0, bad fast5: 0
	[meth_main] total bases: 0.7 Mbases
	[meth_main] Data loading time: 1.419 sec
	[meth_main]     - bam load time: 0.021 sec
	[meth_main]     - fasta load time: 0.353 sec
	[meth_main]     - fast5 load time: 1.041 sec
	[meth_main]         - fast5 open time: 0.195 sec
	[meth_main]         - fast5 read time: 0.818 sec
	[meth_main] Data processing time: 9.555 sec

	[main] CMD: ./f5c call-methylation -r /sdcard/genome/ecoli_2kb_region/reads.fasta -g /sdcard/genome/ecoli_2kb_region/draft.fa -b /sdcard/genome/ecoli_2kb_region/reads.bam
	[main] Real time: 11.417 sec; CPU time: 68.170 sec; Peak RAM: 0.143 GB
	\end{lstlisting}
\end{enumerate}

\subsection{Nanopolish}

\begin{enumerate}

\item Download the source code and compile statically as follows. Library compilation will take time, bare with patience. This example uses my fork of nanopolish patched for ARM support. You may also use v0.11.0 or higher from the original nanopolish repository at \url{https://github.com/jts/nanopolish} that supports ARM.

	\begin{lstlisting}[language=bash]
	git clone --recursive  https://github.com/hasindu2008/nanopolish-arm && cd nanopolish-arm
	git checkout v0.1
	make -j8          			#let HDF5 and htslib compile
	make clean							#clean only the nanopolish related object files (leaving compiled HDF5 and htslib instact)
	make  CC="gcc -static" CXX="g++ -static"
	\end{lstlisting}

\item Copy the binary to the phone as in previous examples. The launch nanopolish.

	\begin{lstlisting}[language=bash]

	#indexing
	1|mh:/data/local/tmp $ ./nanopolish index -d /sdcard/genome/ecoli_2kb_region/fast5_files/ /sdcard/genome/ecoli_2kb_region/reads.fasta

	#variant calling
	1|mh:/data/local/tmp $ ./nanopolish variants -r /sdcard/genome/ecoli_2kb_region/reads.fasta -b /sdcard/genome/ecoli_2kb_region/reads.bam -g /sdcard/genome/ecoli_2kb_region/draft.fa -t4  -w "tig00000001:200000-202000" -p1 > /sdcard/genome/ecoli_2kb_region/variants.vcf
	[post-run summary] total reads: 101, unparseable: 0, qc fail: 0, could not calibrate: 0, no alignment: 0, bad fast5: 0

	#methylation calling
	1|mh:/data/local/tmp $ ./nanopolish call-methylation -r /sdcard/genome/ecoli_2kb_region/reads.fasta -g /sdcard/genome/ecoli_2kb_region/draft.fa -b /sdcard/genome/ecoli_2kb_region/reads.bam   >  /sdcard/genome/ecoli_2kb_region/ref.tsv
	[post-run summary] total reads: 143, unparseable: 0, qc fail: 0, could not calibrate: 0, no alignment: 0, bad fast5: 0

	\end{lstlisting}
\end{enumerate}

\section{Running Directly on Phone}

The section above shows how Linux command line bioinformatics tools (such as minimap2) can be run on an Android mobile phone through Android Debug Bridge. That method required us to issue commands to the phone from the host PC via USB. This section shows how we can make it a bit fancier, by issuing commands directly from the mobile phone. In summary, we will install a virtual terminal app to the phone and issue commands from there.

This post assumes that the binaries have been already copied to `/data/local/tmp' on your mobile phone by following the steps in the previous section.

\subsection{On Android 7.0 or before}

\begin{figure}[!t]
  \centering
\begin{subfigure}[!ht]{0.8\linewidth}
  \centering
    \includegraphics[width=\textwidth]{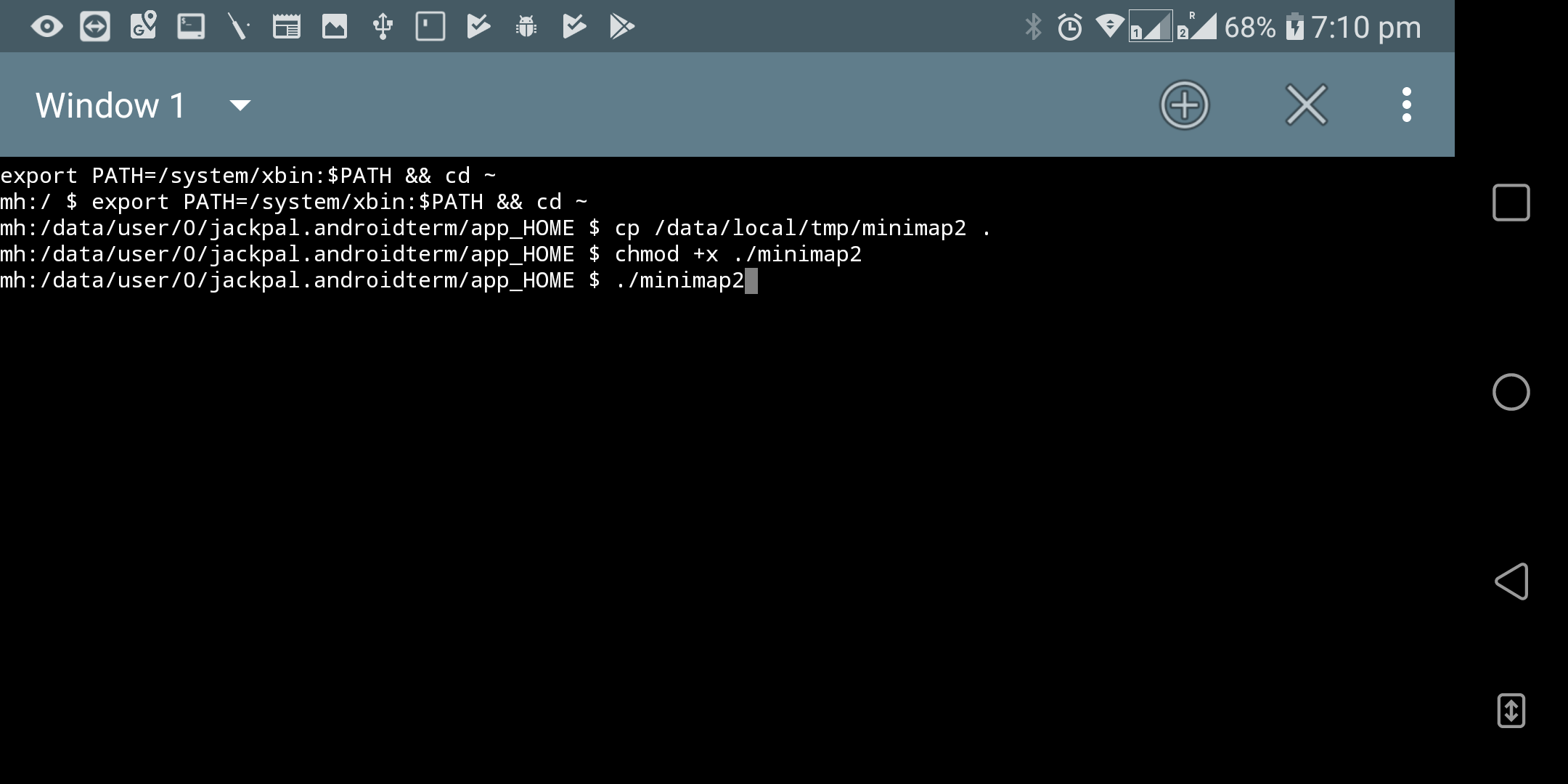}
    \caption{Entering Commands} 
\end{subfigure}

\begin{subfigure}[!ht]{0.4\linewidth}
  \centering
    \includegraphics[width=\textwidth]{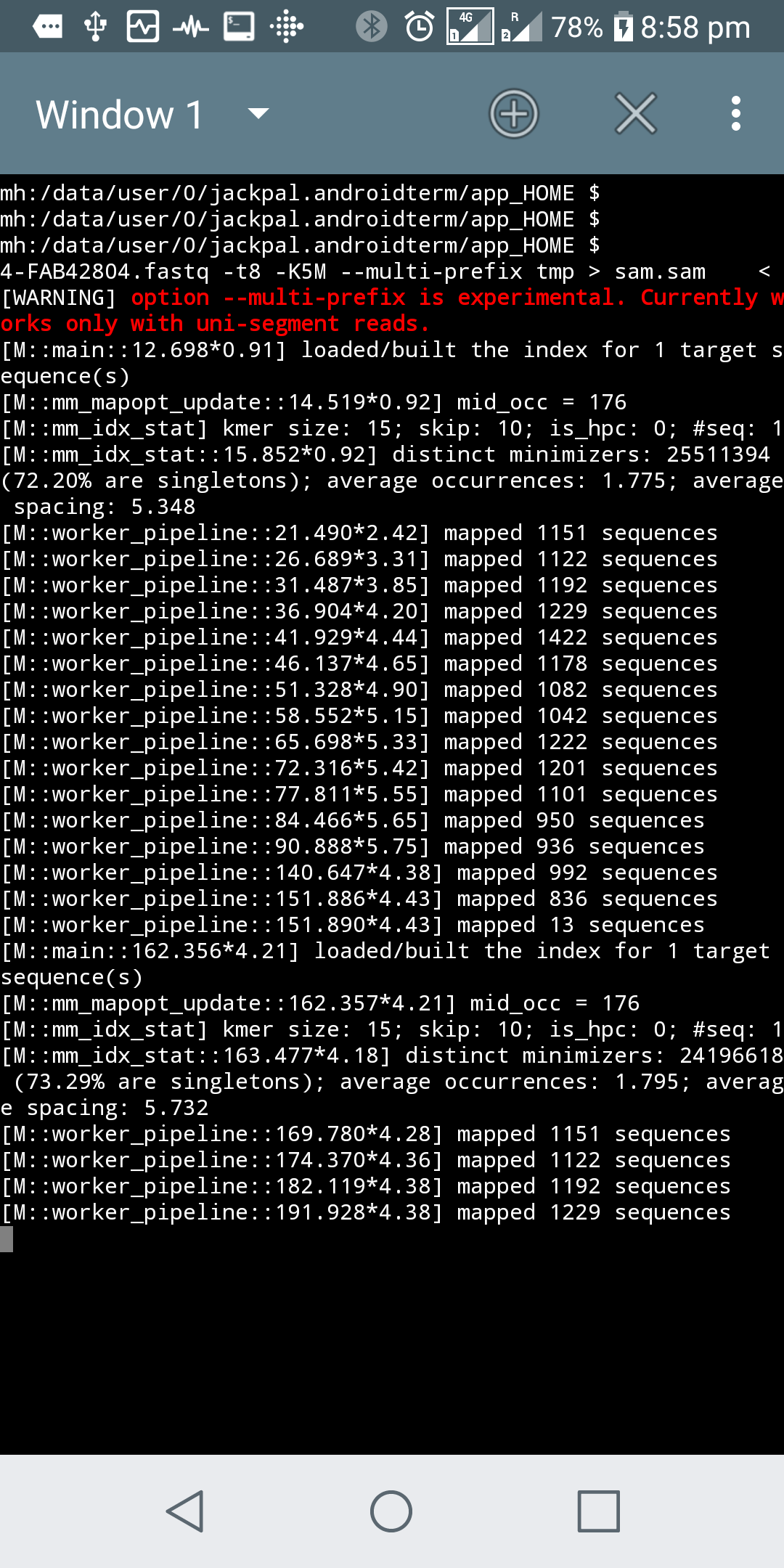}
    \caption{Minimap2 execution} 
\end{subfigure}
\begin{subfigure}[!ht]{0.4\linewidth}
  \centering
    \includegraphics[width=\textwidth]{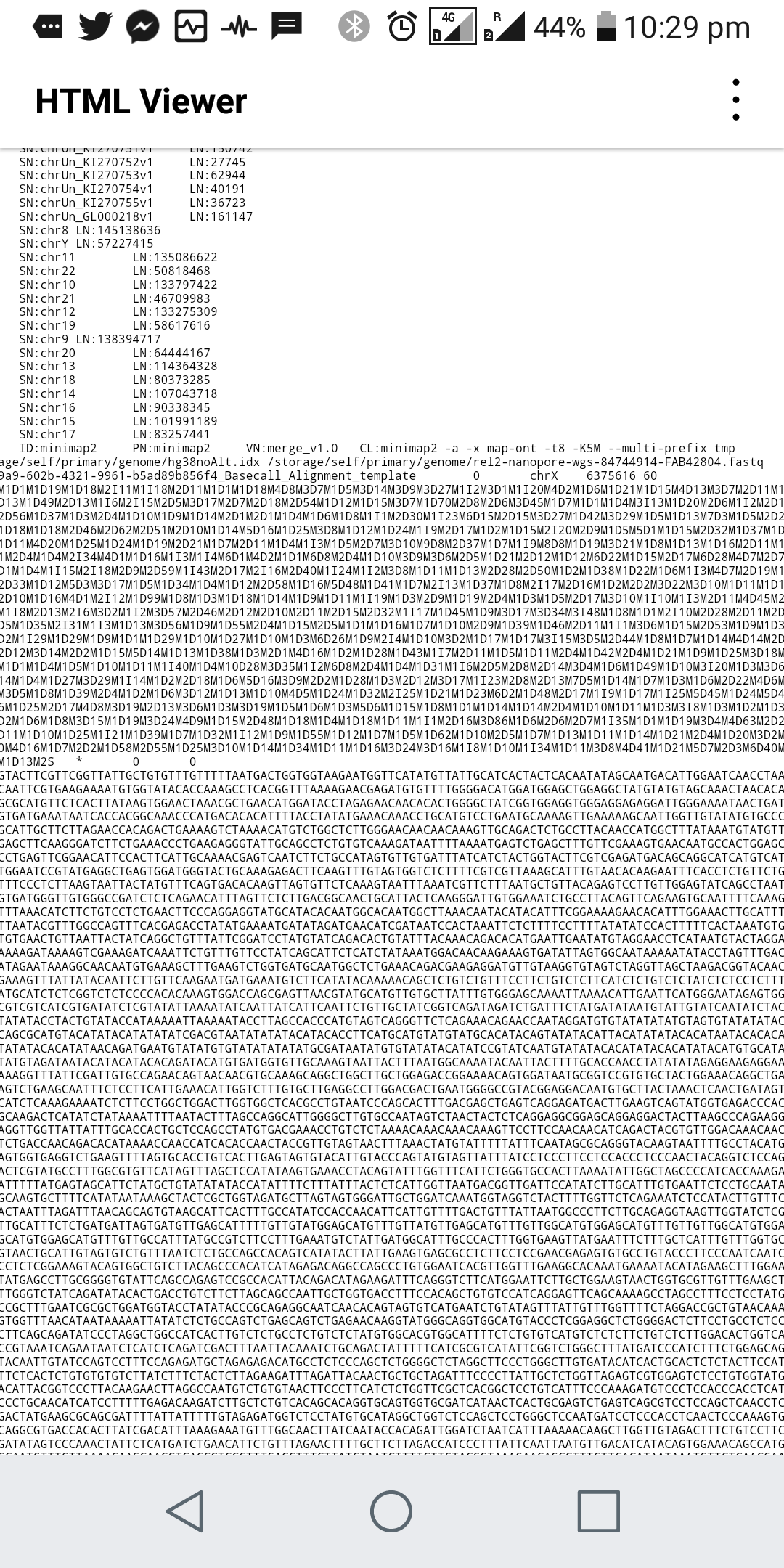}
    \caption{SAM output} 
\end{subfigure}
    \caption{Executing \textit{Minimap2} using terminal emulator} 
    \label{f:minimap2e-on-phone}
\end{figure}

\begin{itemize}
\item Install a terminal emulator on your Android phone, for instance, \href{https://play.google.com/store/apps/details?id=jackpal.androidterm&hl=en_AU}{Terminal Emulator for Android}.

\item Launch the terminal emulator app.

\item On the terminal emulator append `/system/xbin' to `PATH' (the location of tools such as `cp' - might vary on your phone). Then change the current directory to the home, copy the binary, give executable permission and then launch the tool. An example for minimap2 is below (and Fig. \ref{f:minimap2e-on-phone}).

	\begin{lstlisting}[language=bash]
	export PATH=/system/xbin:$PATH && cd ~
	cp /data/local/tmp/minimap2 .
	chmod +x minimap2
	./minimap2
	\end{lstlisting}

\end{itemize}

\subsection{On Android 8.x}

The Above method, unfortunately, will not work on the latest Android 8. You may get a ``Bad System Call'' error when you attempt to run a binary using the terminal emulator. This is due to the \href{https://android-developers.googleblog.com/2017/07/seccomp-filter-in-android-o.html}{seccom filter} introduced in Android 8.0. If you have a rooted phone surely you can get over this by running as sudo. But luckily, still there is a way for non rooted phones - use an app that emulates the ADB client, for instance, \href{https://play.google.com/store/apps/details?id=com.cgutman.androidremotedebugger&hl=en}{Android Remote Debugger}. Limitation of this method is you need a host PC (with ADB configured) to initially launch ADB server on the phone.

\begin{figure}[!t]
  \centering
\begin{subfigure}[!ht]{0.49\linewidth}
  \centering
    \includegraphics[width=\textwidth]{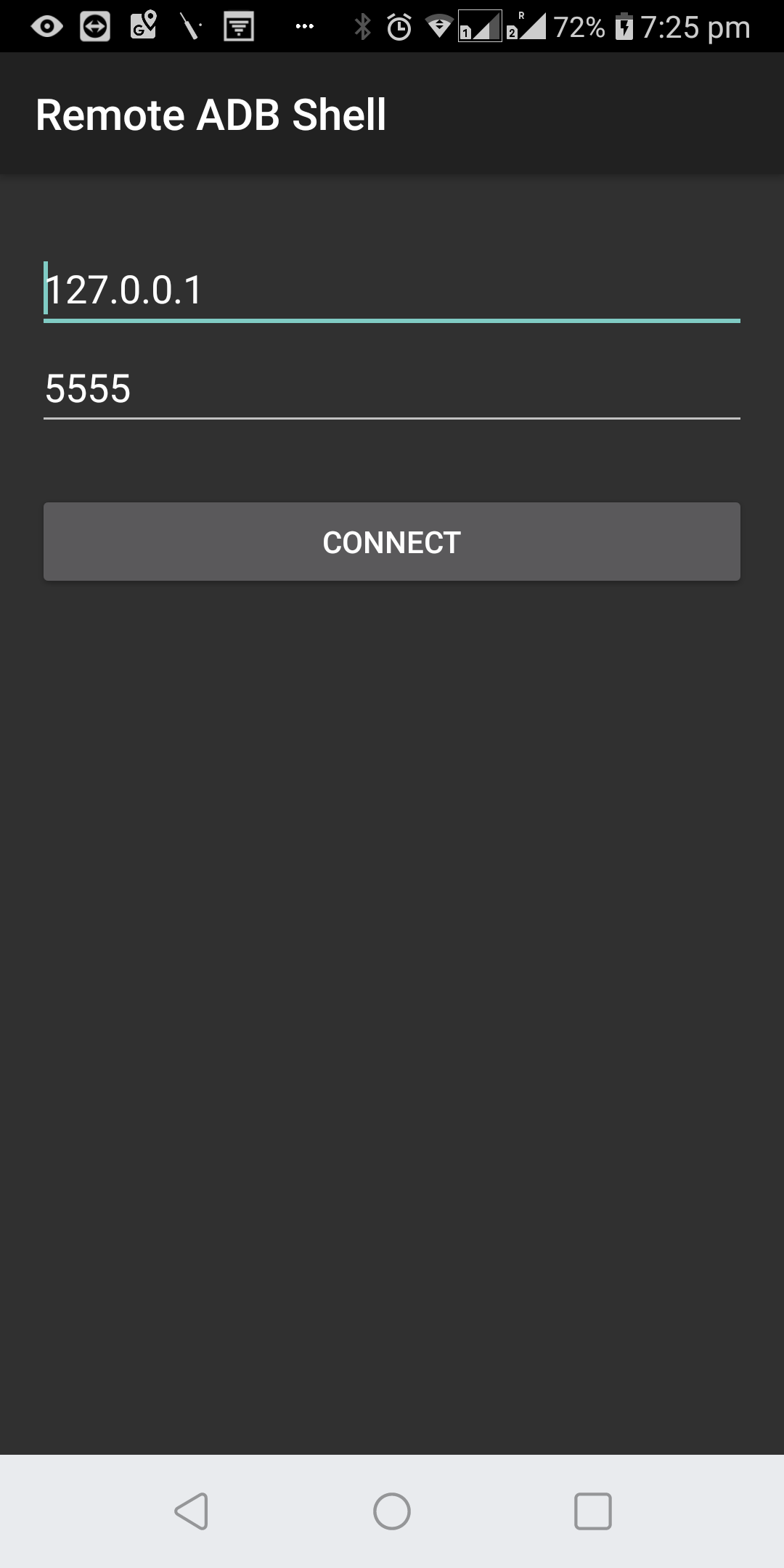}
\end{subfigure}
\begin{subfigure}[!ht]{0.49\linewidth}
  \centering
    \includegraphics[width=\textwidth]{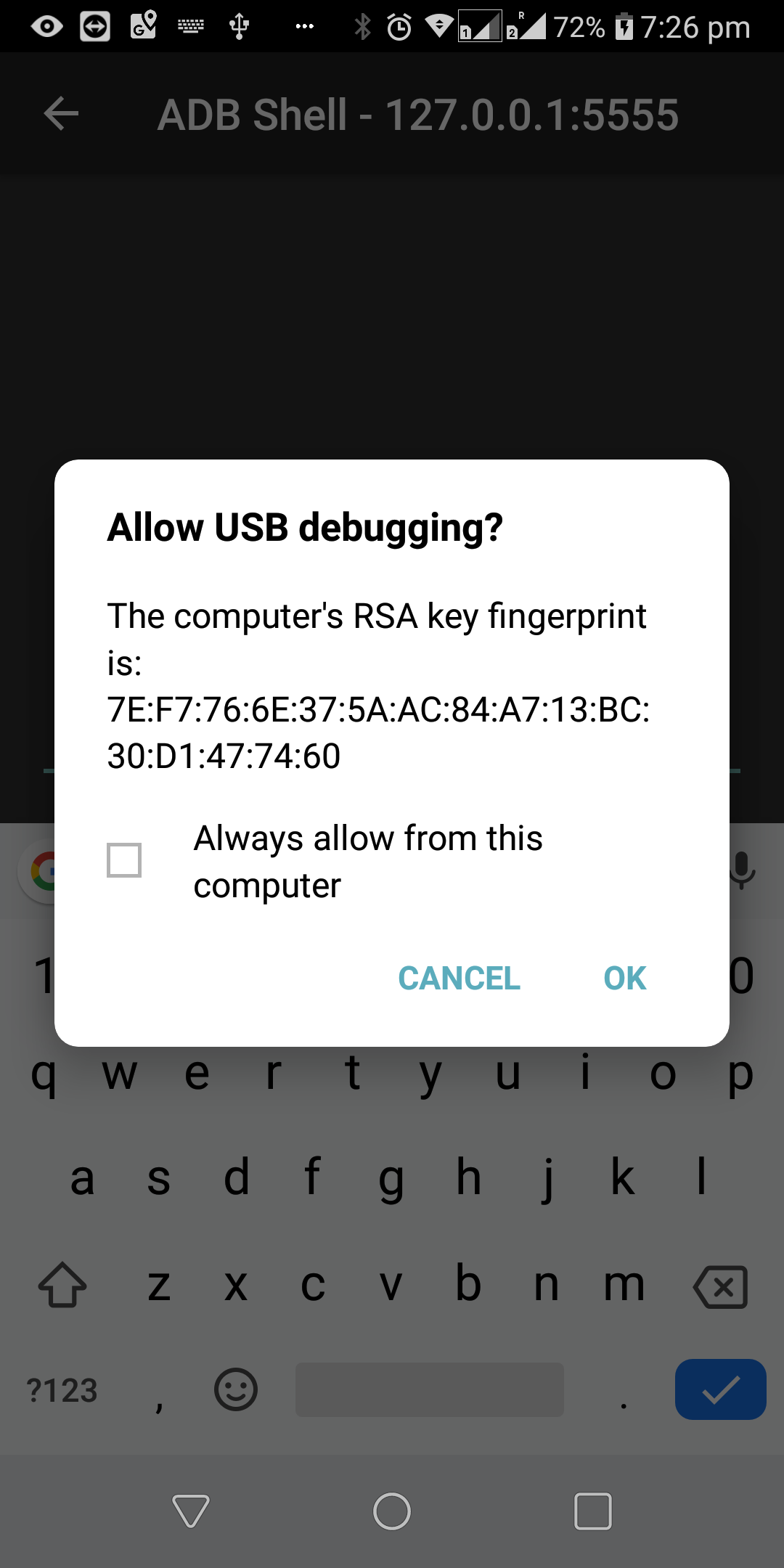}
\end{subfigure}
    \caption{Remote ADB} 
    \label{f:remote-adb-on-phone}
\end{figure}

\begin{figure}[!t]
  \centering
    \includegraphics[width=\textwidth]{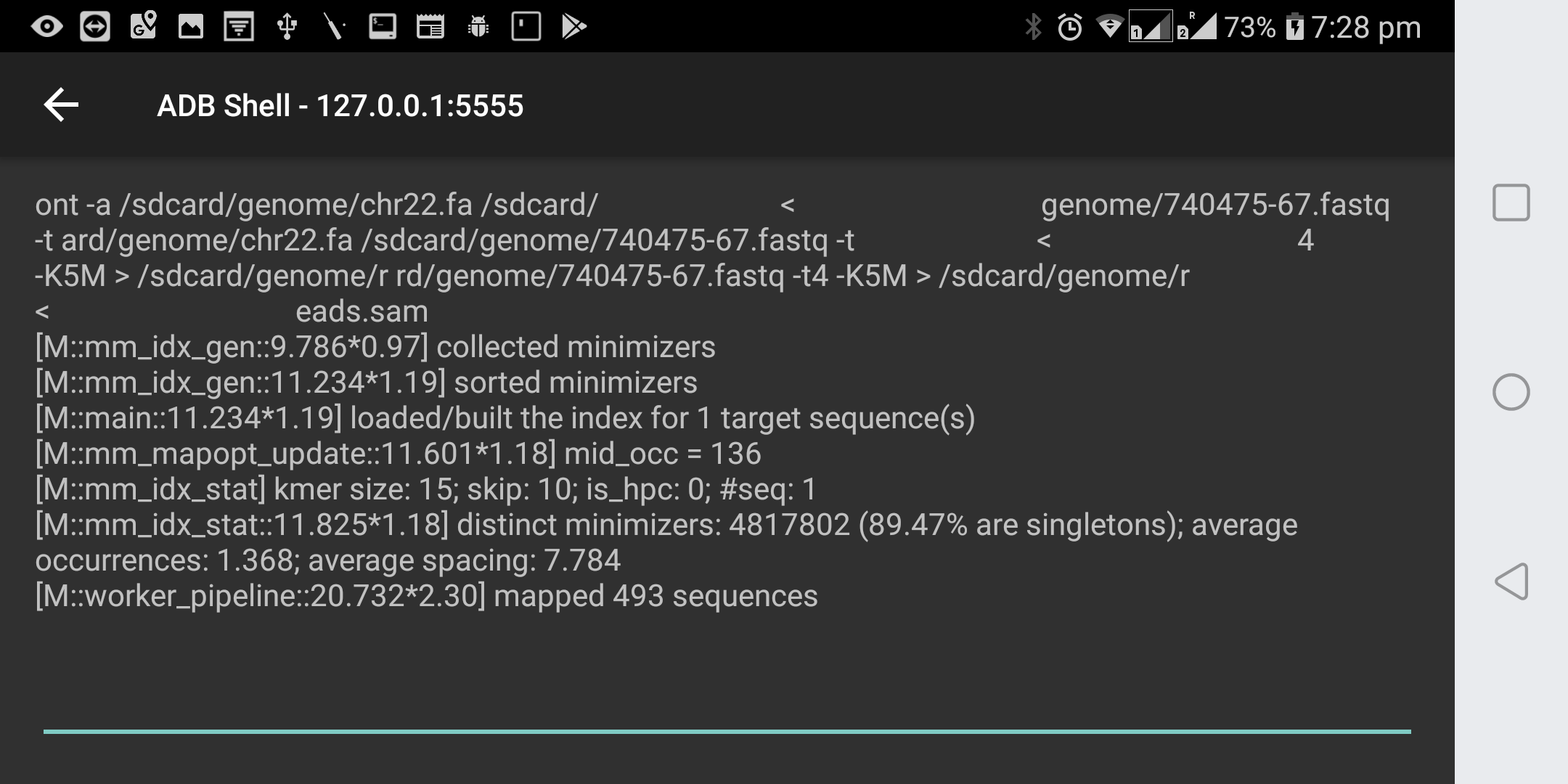}
    \caption{Execution using remote ADB} 
    \label{f:remote-adb2-on-phone}
\end{figure}

\begin{enumerate}
\item Install \href{https://play.google.com/store/apps/details?id=com.cgutman.androidremotedebugger&hl=en}{Android Remote Debugger} on your phone

\item Connect the phone through USB to the host computer (need to have ADB configured as we did in the previous section and on a command prompt issue the following.

	\begin{lstlisting}[language=bash]
	C:\Program Files (x86)\Minimal ADB and Fastboot>adb tcpip 5555
	restarting in TCP mode port: 5555
	\end{lstlisting}

	You can disconnect from the computer after launching the server as above. However, you will need to perform this step every time you reboot your phone.

\item Launch the Android Remote Debugger app and connect to the localhost (127.0.0.1) on port 5555 (Fig. \ref{f:remote-adb-on-phone}).

\item Now change directory to `/data/local/tmp' and execute the binary (Fig. \ref{f:remote-adb2-on-phone}).

\end{enumerate}

\subsection{Is there a proper way?}

All the methods above are hacky and suitable only in a development environment. While I have not myself investigated proper ways, here are some thoughts.

\begin{itemize}
\item Compile the binaries and link against \textit{bionic}, the standard C library for Android (opposed to static linking). We have to use a cross compiler for this, i.e. gcc-arm-linux-androideabi. however the dependencies (such as \textit{zlib}) have to be compiled ourselves using the cross compiler (cannot use the versions from \textit{apt}). However, additional requirements such as mandated position independent executables and \href{https://android.googlesource.com/platform/bionic/+/master/android-changes-for-ndk-developers.md}{restrictions on text relocations} will further complicate the compilation. After getting it compiled, you would make an Android application that acts as a wrapper that calls the compiled binaries, for instance, what is suggested at \url{https://stackoverflow.com/questions/5583487/hosting-an-executable-within-android-application}.

\item The most proper way (but a lot of work for sure) would be to use the \href{https://developer.android.com/ndk/guides}{Android NDK} to compile the C codes into native libraries (might require a restructuring of the source code) which then can be called through an Android app through JNI.
\end{itemize}

%% file: 9.5-appendices/5.5-misc-open-source.tex
\chapter{Appendix: Open-source Contributions}\label{a:opensource}

\subsection{User comments for \textit{f5c}}

\textit{``I have just had the first sample finish after placing them on faster storage.  I have to say the speed has left me speechless and in shock.  I was expecting an improvement in speed, but this is something much more than an "improvement". With iops at 16 and the drives on faster disks it took just 13 hours for a 40x human sample.  That is really impressive!''} --- a \textit{f5c} user

\begin{figure}
    \centering
    \includegraphics[width=\textwidth]{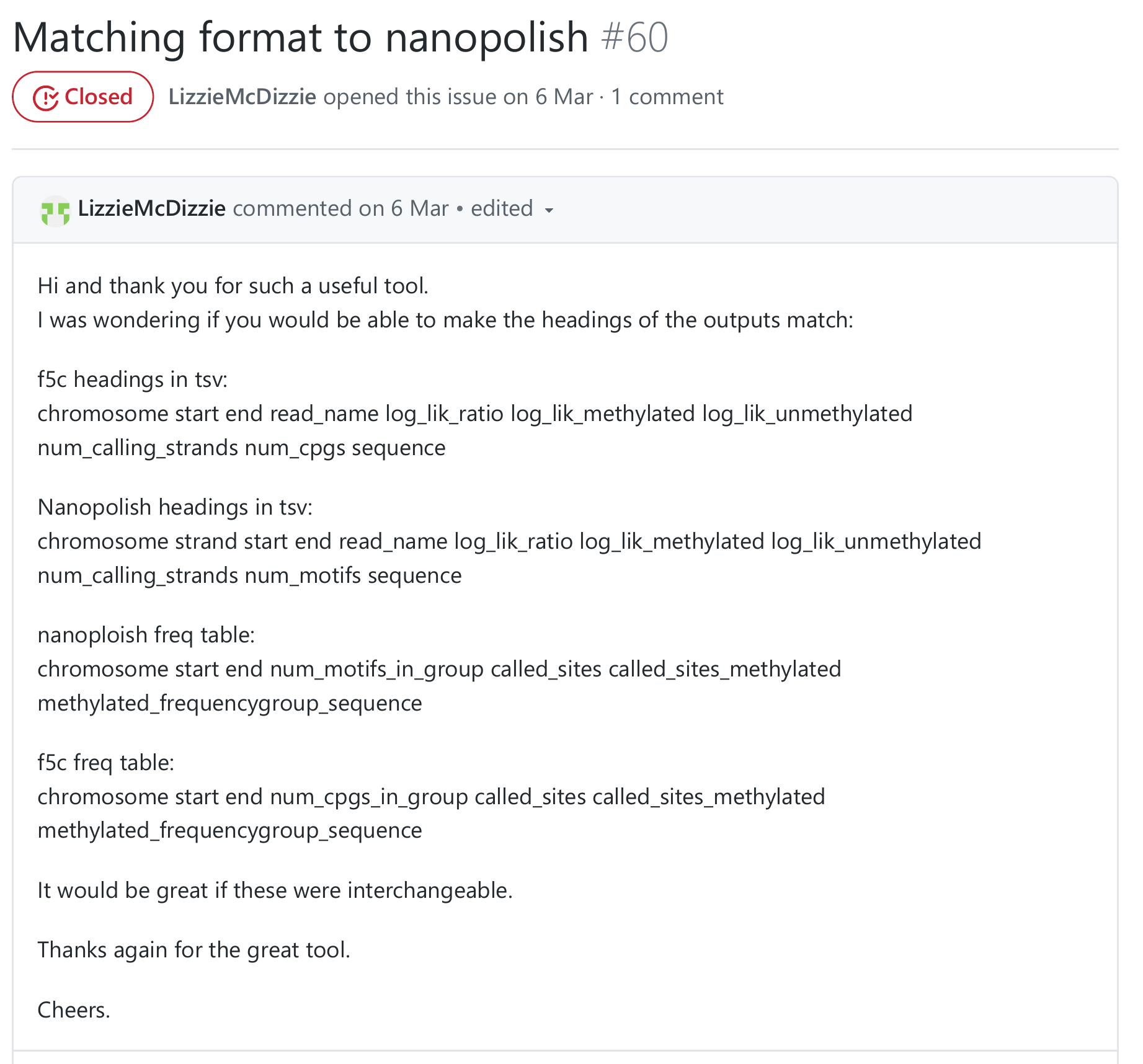}
    %\caption{Caption}
    %\label{fig:my_label}
\end{figure}

\clearpage

\subsection{Contributions to \textit{Minimap2}}

\begin{figure}[H]
    \centering
    \includegraphics[width=\textwidth]{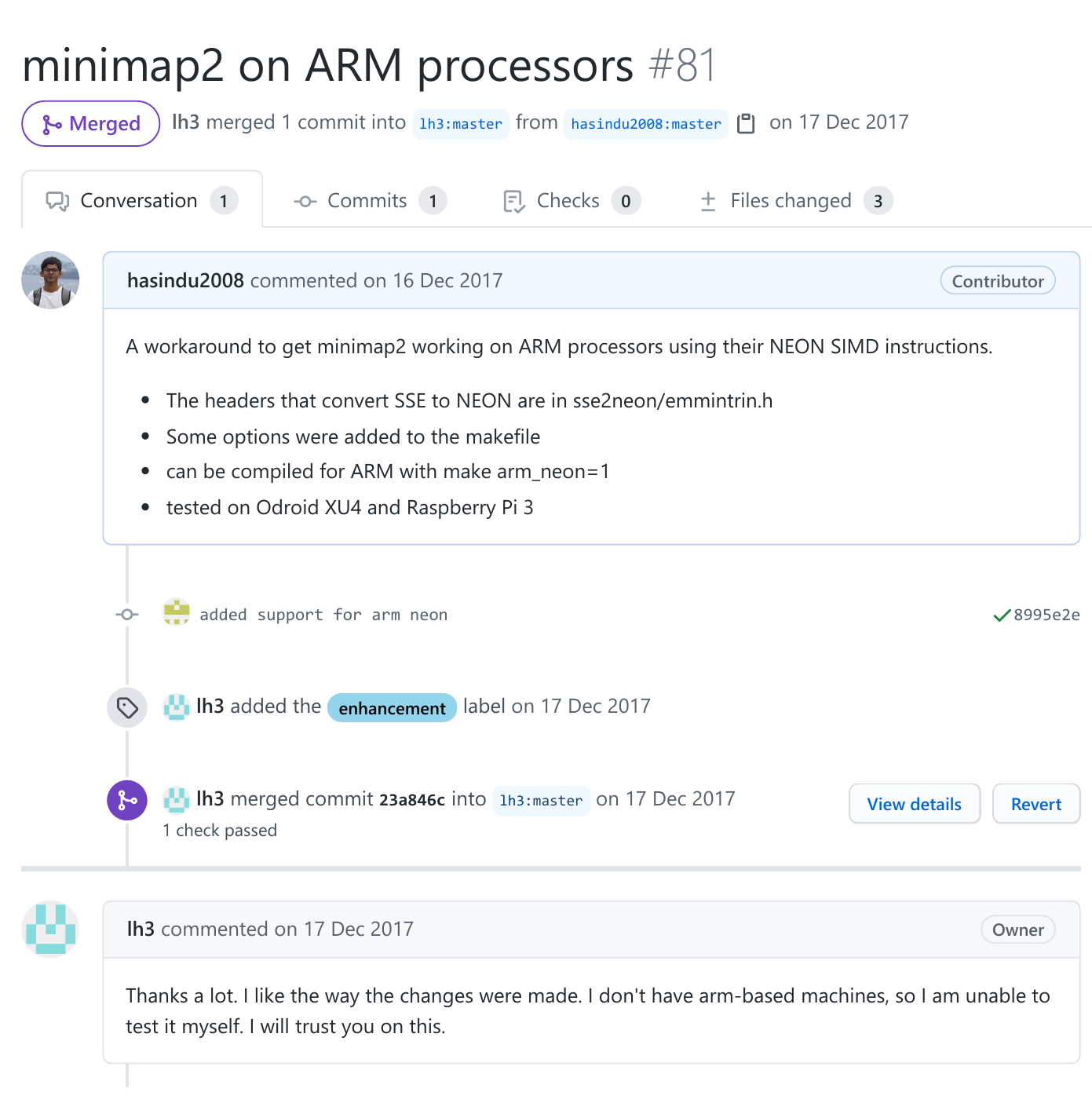}
    %\caption{Caption}
    %\label{fig:my_label}
\end{figure}

\begin{figure}
    \centering
    \includegraphics[width=\textwidth]{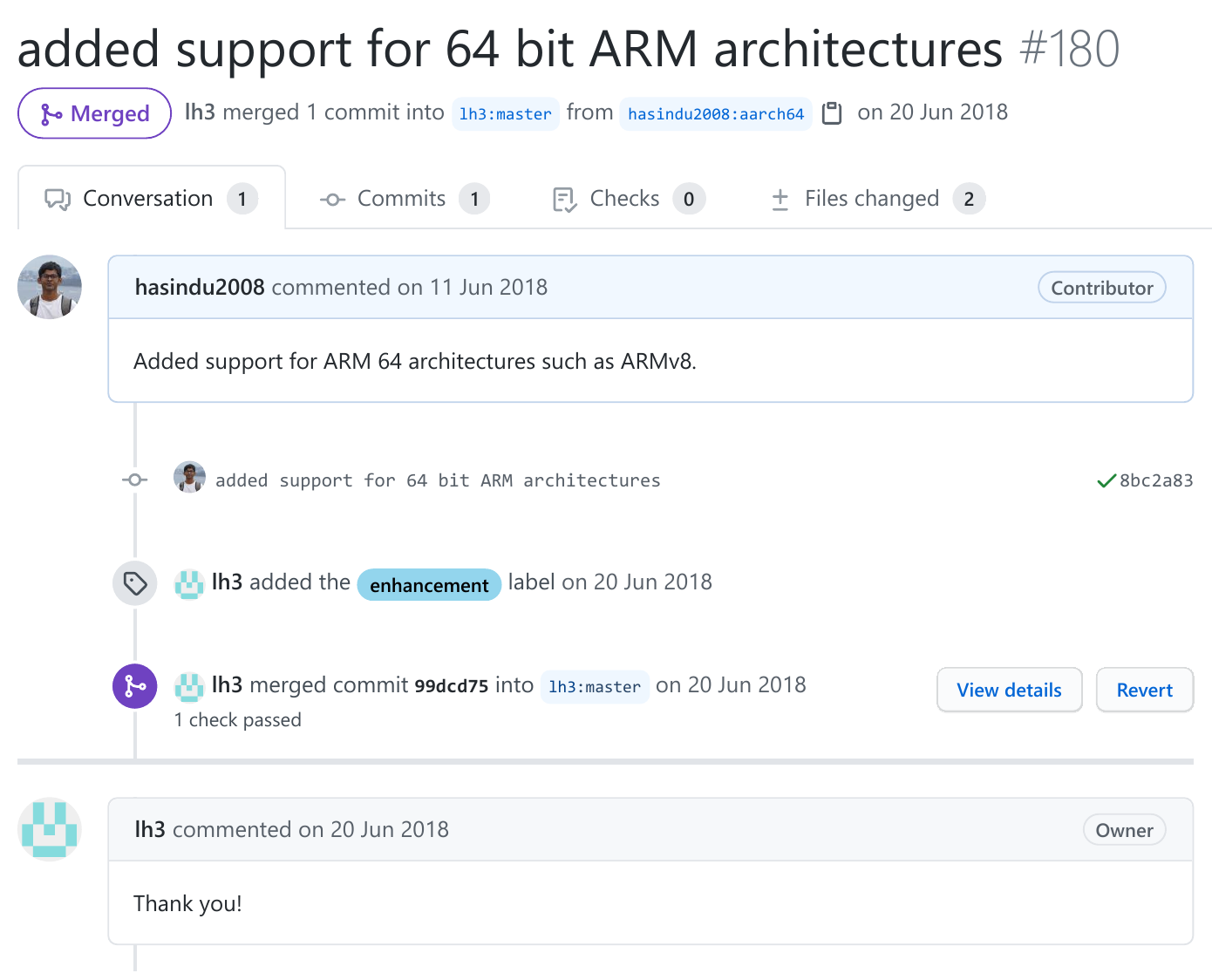}
    %\caption{Caption}
    %\label{fig:my_label}
\end{figure}

\begin{figure}
    \centering
    \includegraphics[width=\textwidth]{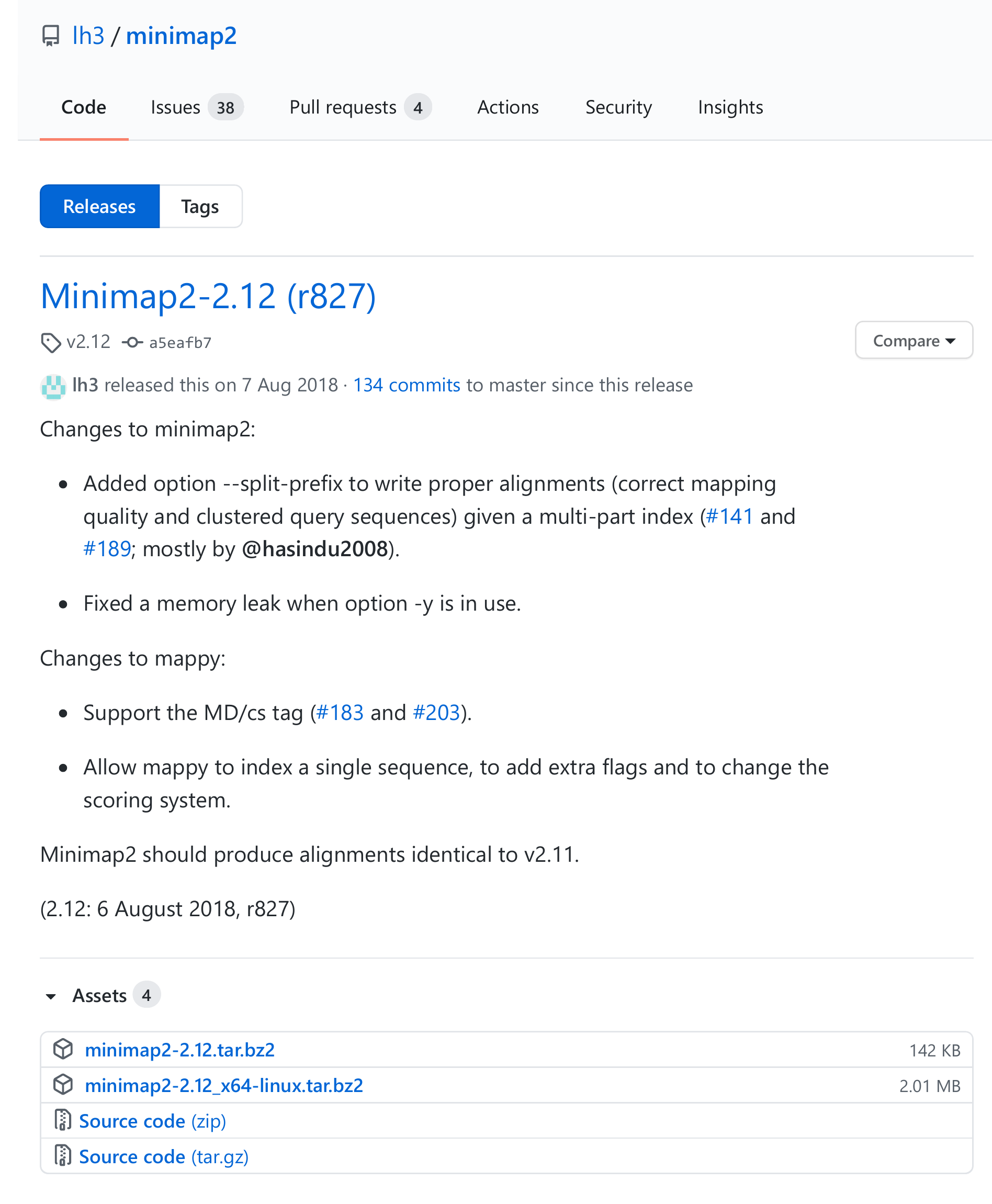}
    %\caption{Caption}
    %\label{fig:my_label}
\end{figure}

\clearpage

\subsection{Contributions to \textit{Nanopolish}}

\begin{figure}[H]
    \centering
    \includegraphics[width=\textwidth]{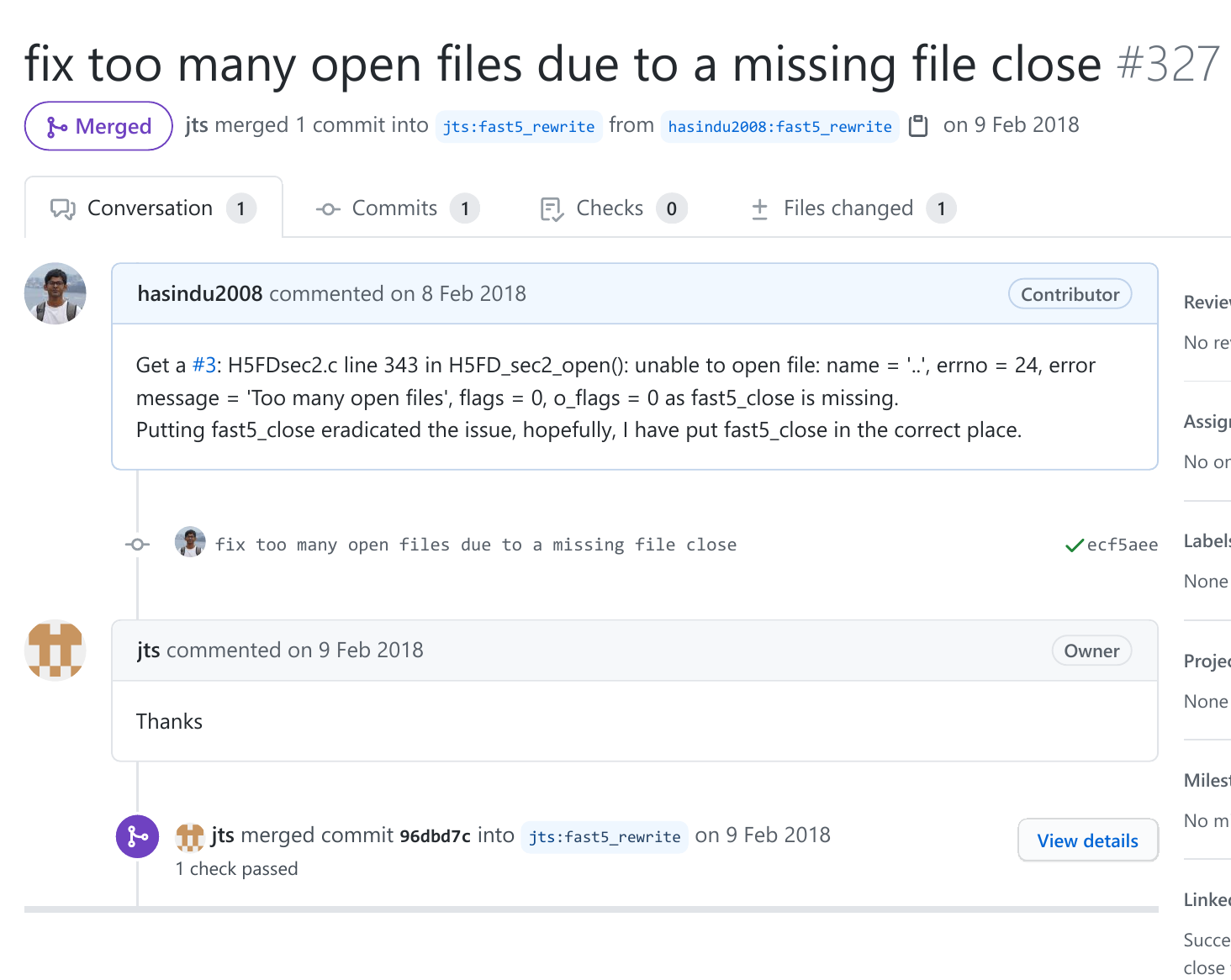}
    %\caption{Caption}
    %\label{fig:my_label}
\end{figure}

\begin{figure}
    \centering
    \includegraphics[width=0.8\textwidth]{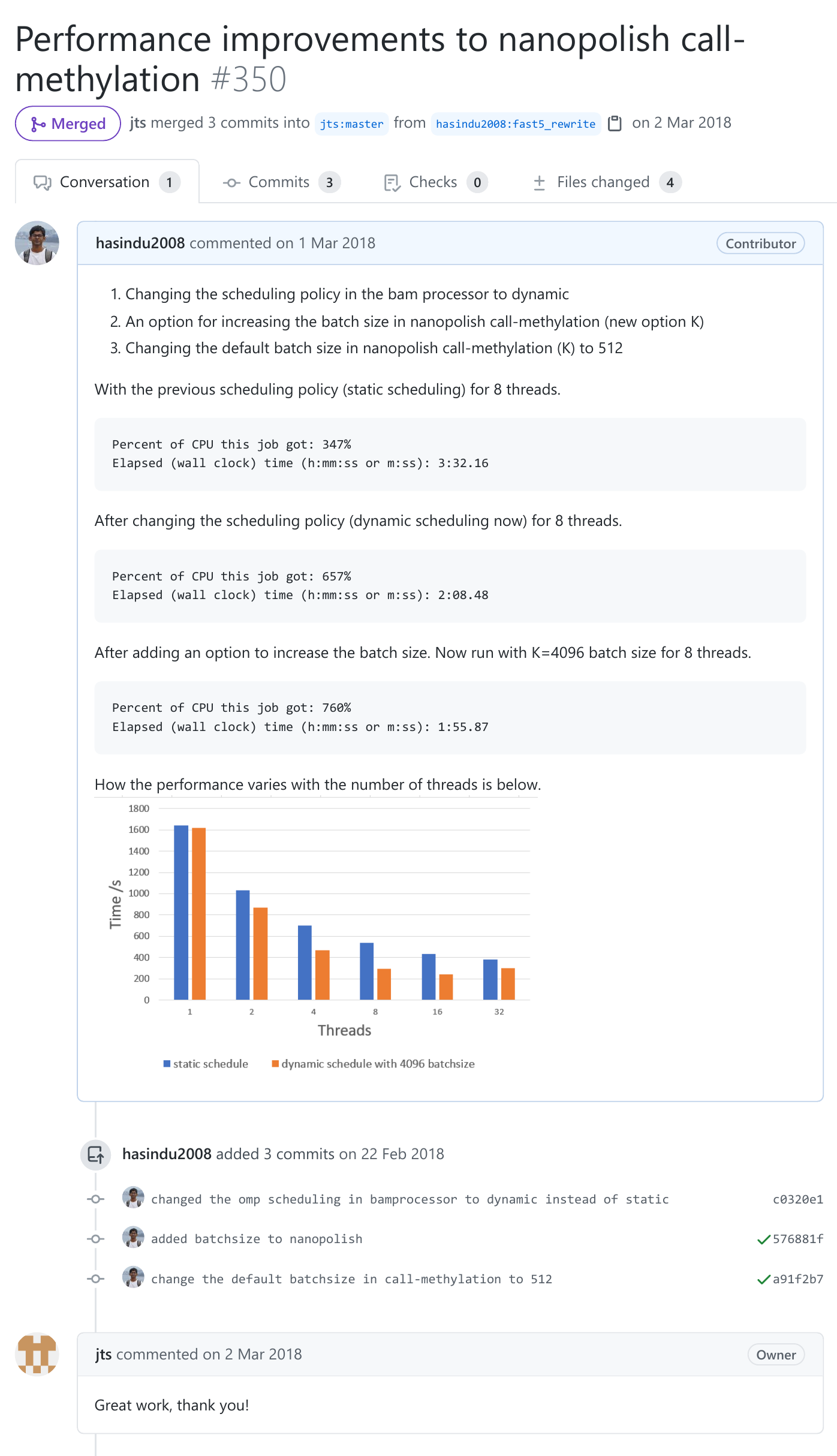}
    %\caption{Caption}
    %\label{fig:my_label}
\end{figure}

\begin{figure}
    \centering
    \includegraphics[width=\textwidth]{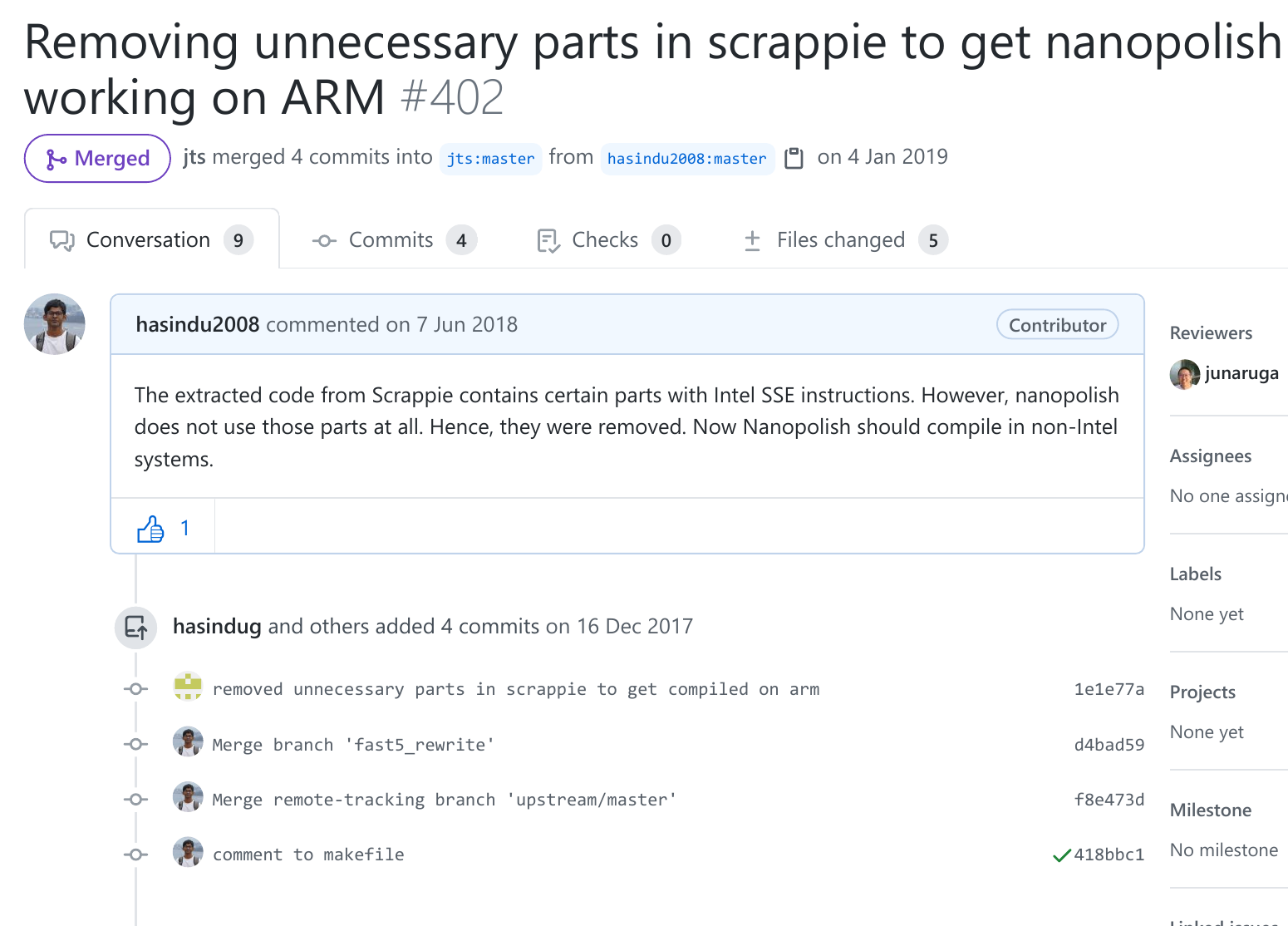}
    %\caption{Caption}
    %\label{fig:my_label}
\end{figure}

\begin{figure}
    \centering
    \includegraphics[width=0.8\textwidth]{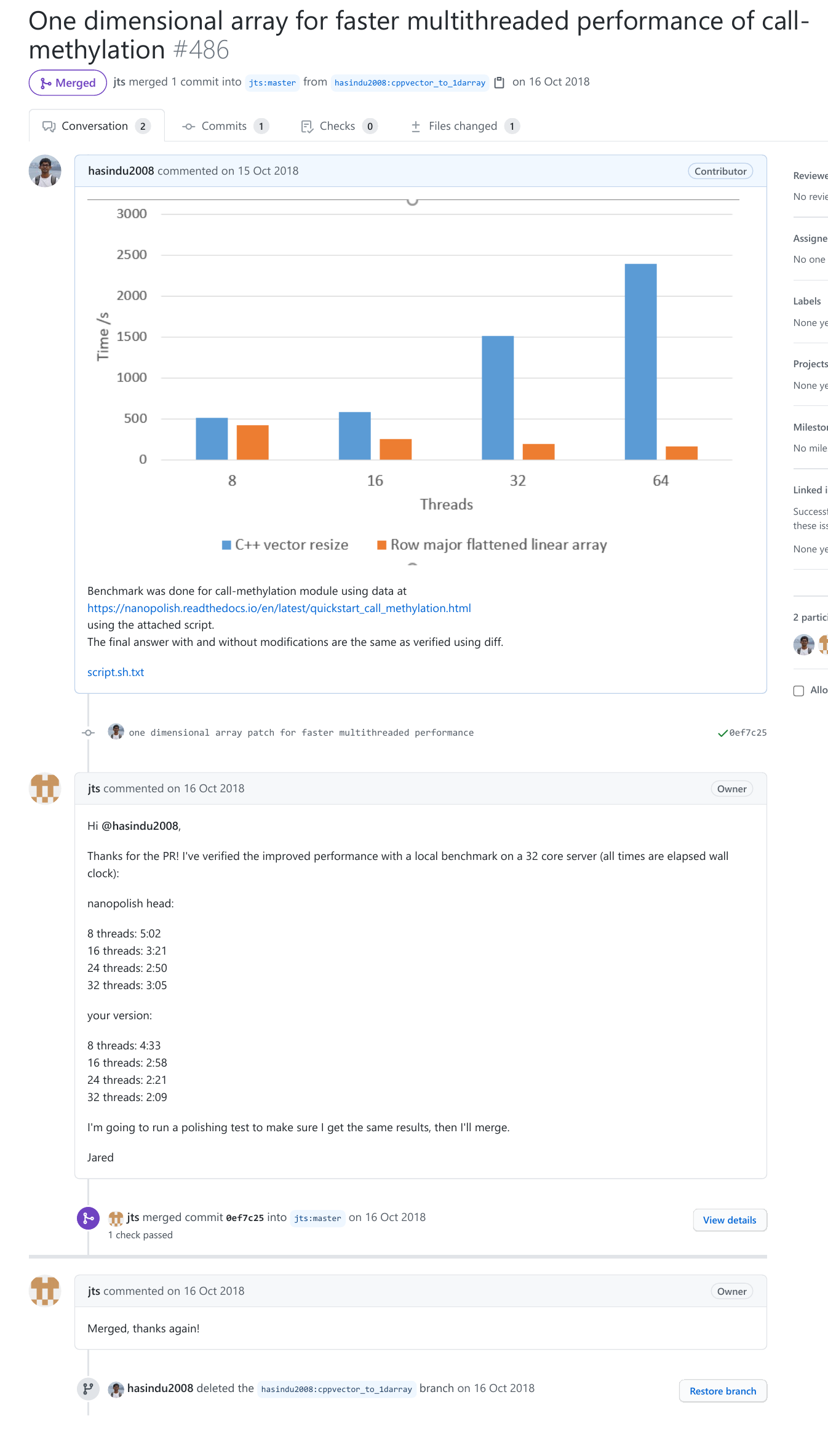}
    %\caption{Caption}
    %\label{fig:my_label}
\end{figure}

\clearpage

%% file: 9.5-appendices/6-io.tex
\chapter[Appendix: I/O Optimisations]{Supplementary Materials - Optimisation of Nanopore Sequence Analysis for Many-core CPUs}\label{a:ioopti}

\section{Extended Motivational Example on another System} \label{s:extended-motivational}

\begin{figure}
    \centering
    \includegraphics[width=\textwidth]{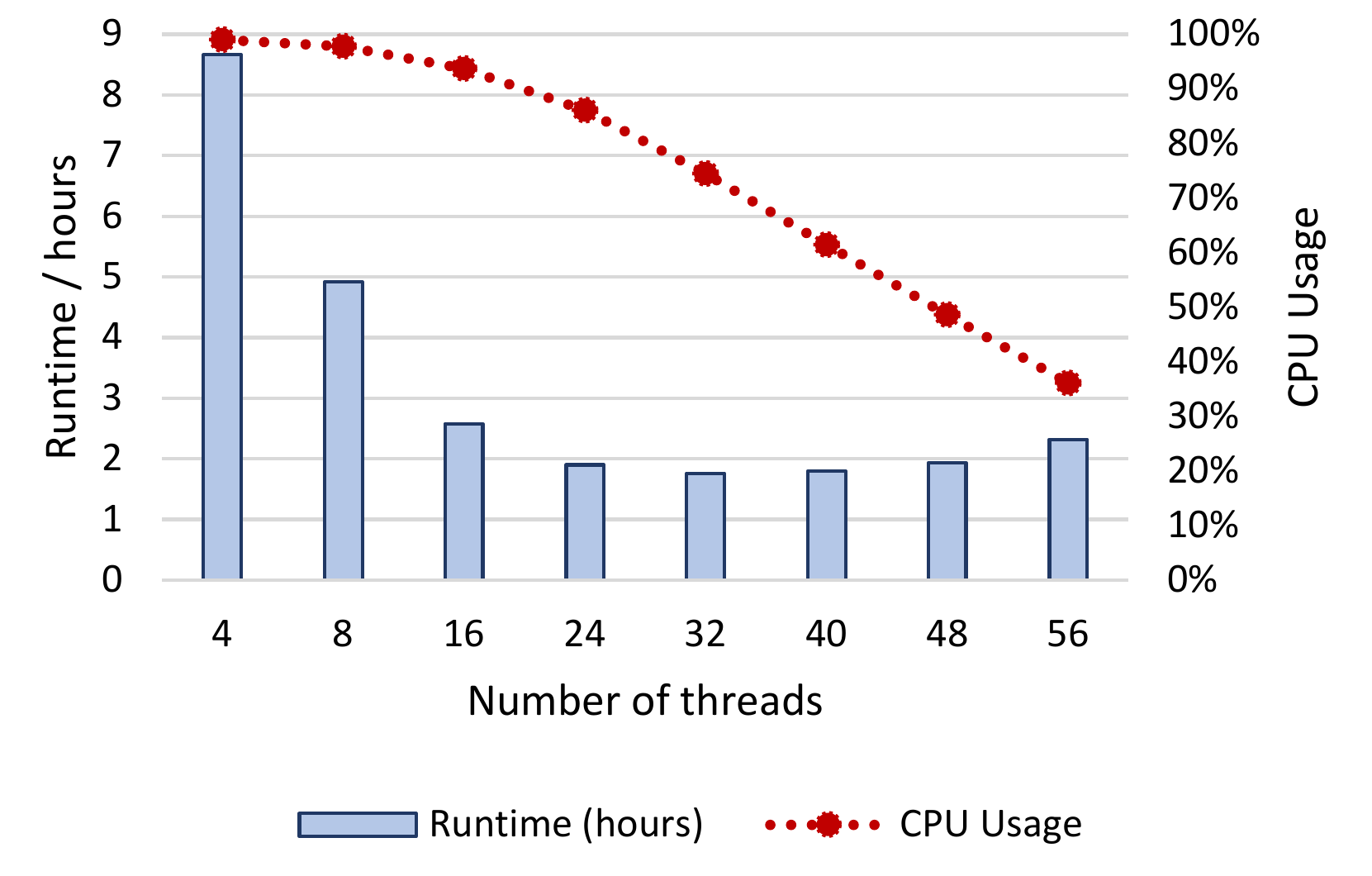}
    \caption{Variation of runtime of original Nanopolish with the number of threads}
    \label{f:a-nanopolish-orig-runtime}
\end{figure}

To further demonstrate that the resource usage inefficiency, we executed the methylation calling tool Nanopolish \cite{simpson2017detecting} on another high-end server with 28 Intel Xeon cores (56 logical cores or hyperthreads) and a Redundant Array of Independent Disks (RAID) storage composed of multiple Non-Volatile Memory Express (NVMe)  Solid-State Drives (SSD). See system S4 in Table \ref{t:a-systems-slow5} for full information of the server used. The graph in Fig. \ref{f:a-nanopolish-orig-runtime} plots the runtime for Nanopolish (left y-axis) when run with a different number of threads (x-axis). The graph in Fig. \ref{f:a-nanopolish-orig-runtime} also plots the CPU usage for each case under the right y-axis, where CPU utilisation is calculated as explained in experimental setup:

At four threads, the CPU utilisation was closer to 100\%, meaning that 4 CPU cores \footnote{More accurately four virtual (logical) cores as Intel Processors employ hyper threading.} were fully used. However, when called with 56 threads, the CPU utilisation was 30\% meaning that out of the 56 CPU cores, only around 20 CPU cores were really used. This observation confirms that procuring a server with a higher number of CPU cores would not be beneficial for similar cases.

\begin{figure}
    \centering
    \includegraphics[width=\textwidth]{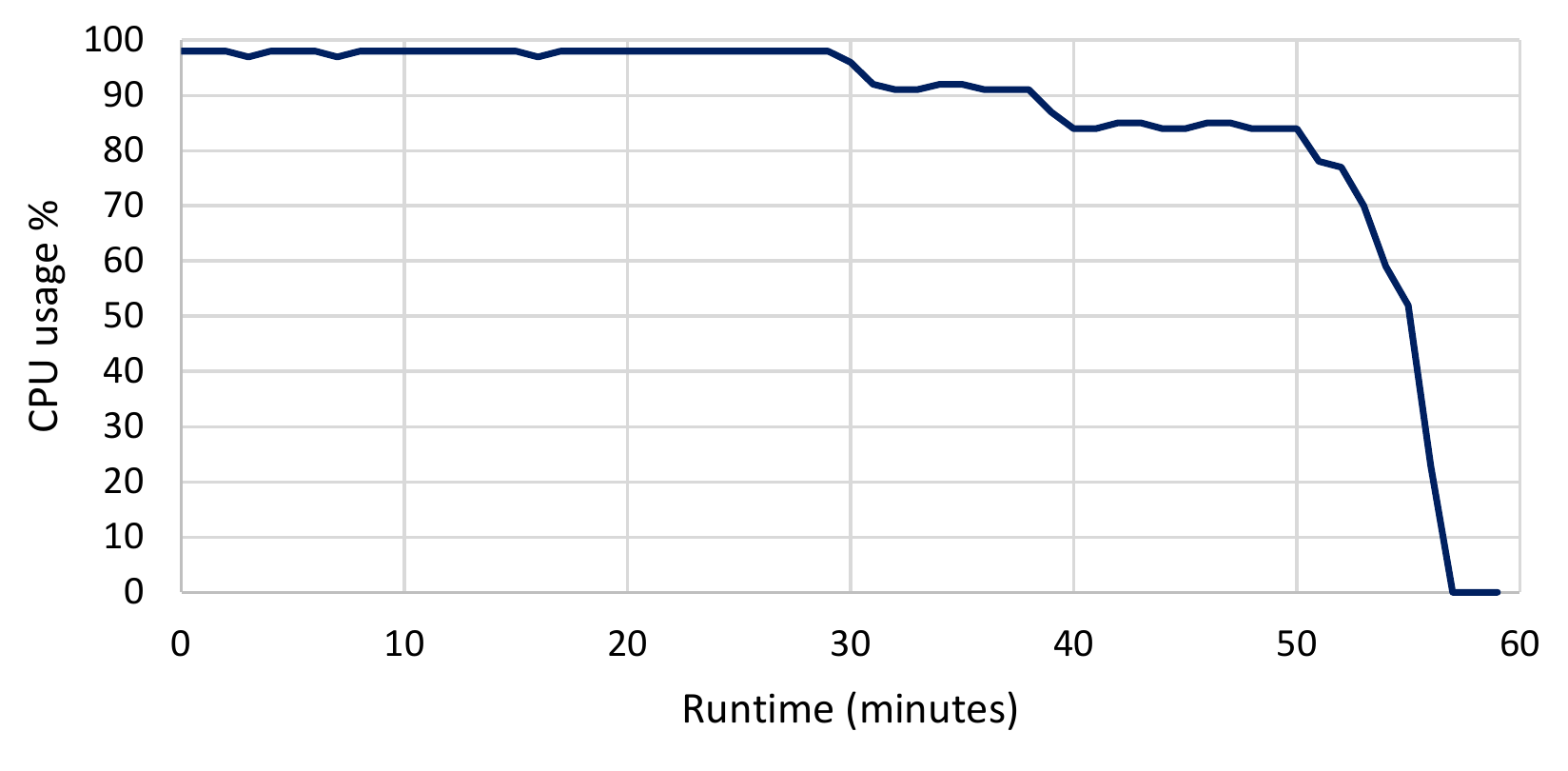}
    \caption{CPU utilisation with the runtime}
    \label{f:a-nanopolish-multi-proc}
\end{figure}

To verify the above mentioned limitation of the HDF5 library, we manually split the dataset into 14 roughly equal parts and then separately (but in parallel) launched 14 \textit{Nanopolish} processes. Each process was launched with 4 threads, thus the 56 threads are expected to be active to be able to fully utilise all 56 cores. This 4 thread and 14 process configuration was selected since at 4 threads the CPU utilisation was around $\sim$100\% (Fig. \ref{f:a-nanopolish-orig-runtime}). Since each process has its own address space and a synchronisation primitives in one process does not affect the other processes. How the CPU utilisation (calculated as in equation \ref{e:cpuusagecompute} out of all 56 CPU threads) varied with the time is in Fig. \ref{f:a-nanopolish-multi-proc}.

Observe that the CPU utilisation is closer to 100\% up to around 30 minutes. After 30 minutes the drop in CPU utilisation is due to certain processors finishing earlier than the others. Nevertheless, this plot shows that the underlying disk system could service faster than the CPU could process and confirms that the low CPU utilisation when running one Nanopolish process with 56 threads was due to the limitation of underlying software (HDF5 library in this case).

\section[Extended Bottleneck using Another Dataset]{Extended Deeper analysis of the Bottleneck using Another Dataset}\label{s:deepnaly}

We restructured Nanopolish such that time spent on I/O and processing can be separately measured. It is such that a batch of reads (several thousand) are read from the disk using a single thread (to mimic the behaviours of the lock in HDF5) and the batch is assigned to multiple threads equal to the maximum number of cores to be processed. This repeats until all reads are processed. On system S1 (Table \ref{t:systems-slow5}) for the dataset D2 (Table \ref{t:a-datasets-slow5}), the execution time was 72.19 hours (3 days!). Data loading ridiculously contributed took 95.96\% of the total time (69.27) and only 4.04\% for processing (2.92 h). Out of the time for data loading, reading of \textit{FAST5 files} (HDF5) ridiculously contributed to 90.54\% (62.72h). Random access to FASTQ/FASTA files performed using \textit{faidx} in \textit{htslib} took 8.58\% (5.95) and sequential access to the BAM file performed through \textit{htslib} took only 0.88\% (0.58h).

\begin{figure}[!ht]
  \centering
\begin{subfigure}[!ht]{0.49\textwidth}
  \centering
    \includegraphics[width=\textwidth]{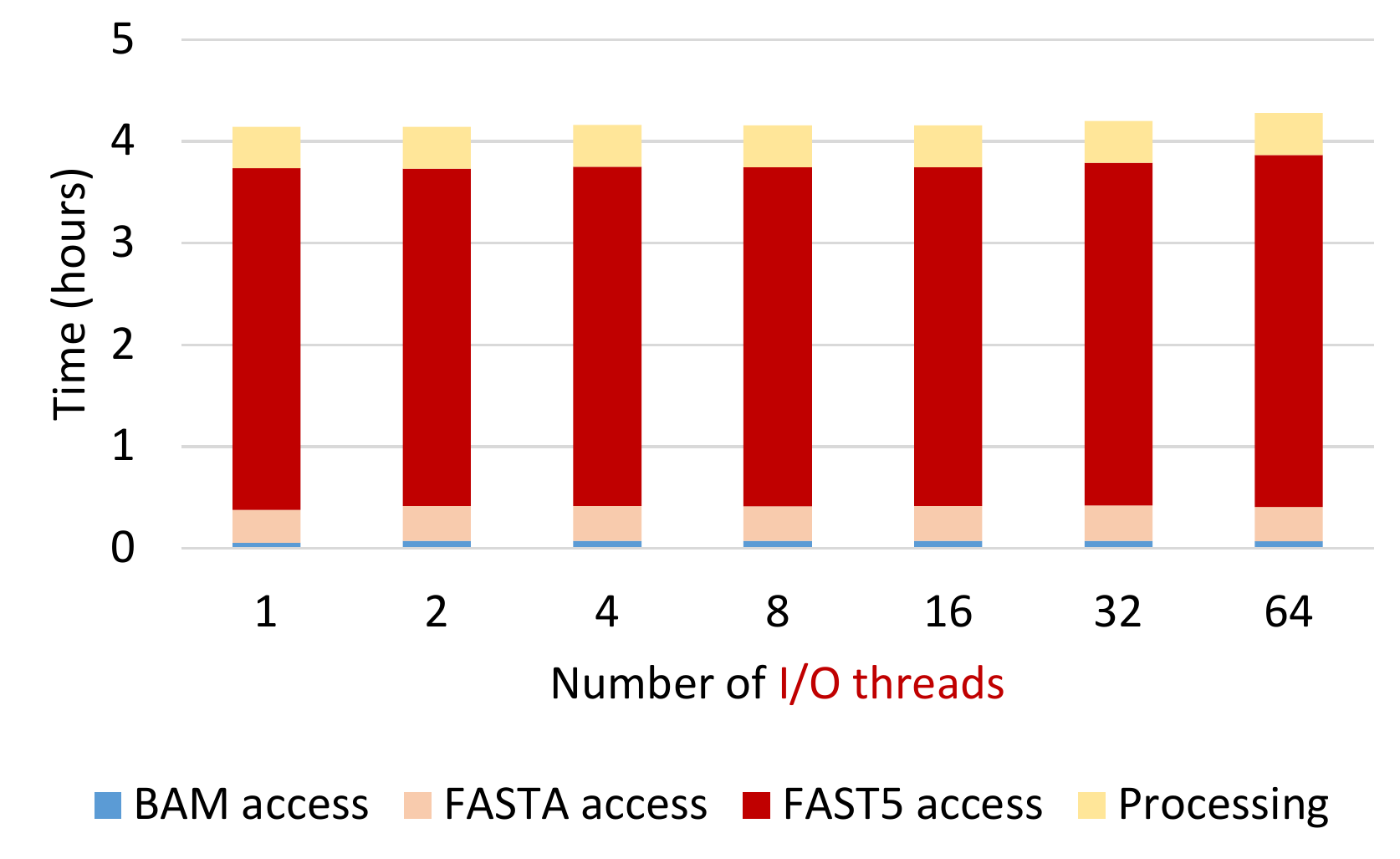}
    \caption{on system S1 comprising HDD RAID} 
    %\label{f:mem_balance}
\end{subfigure}
\begin{subfigure}[!ht]{0.49\textwidth}
  \centering
    \includegraphics[width=\textwidth]{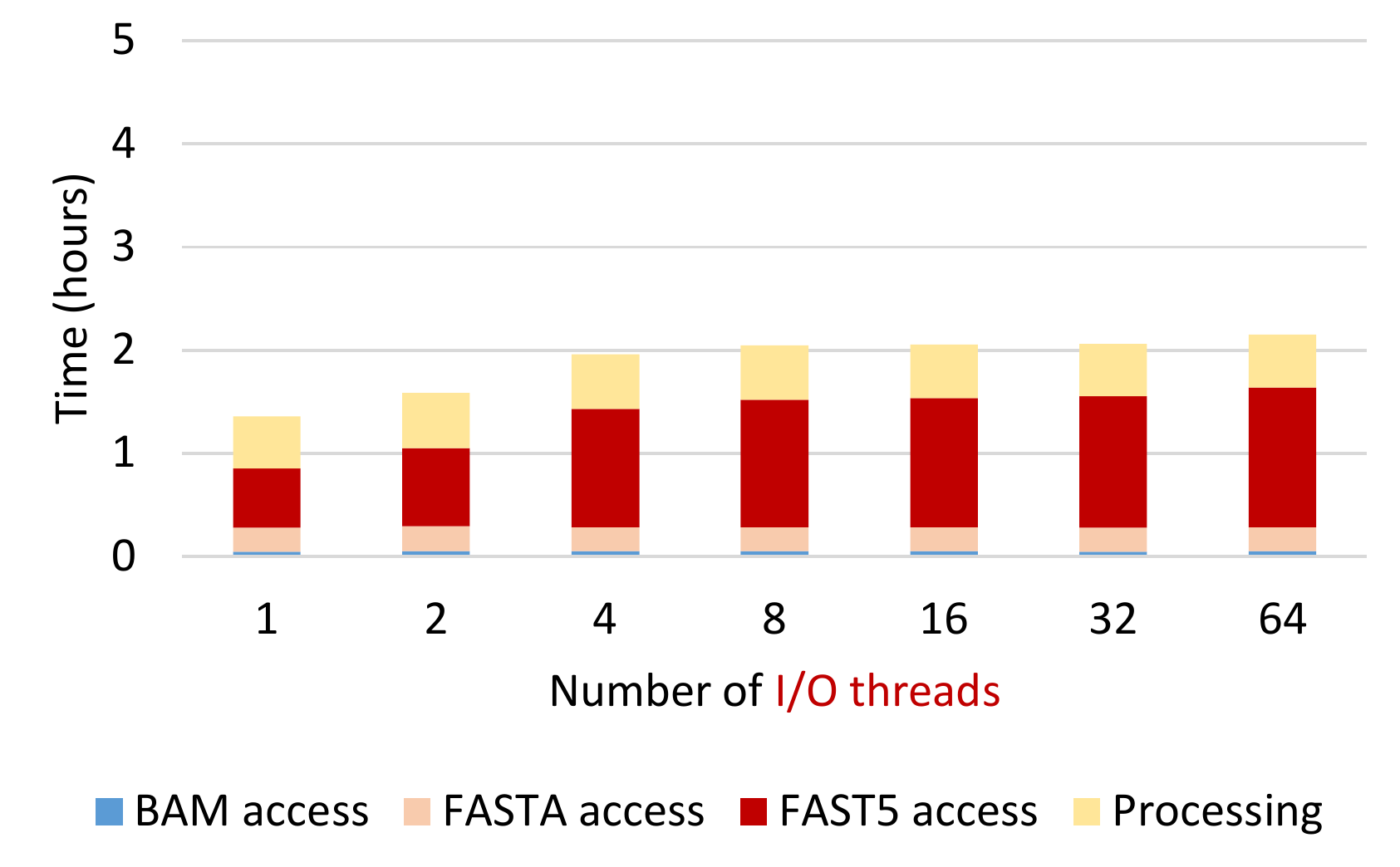}
    \caption{on system S2 comprising SSD RAID} 
    %\label{f:load_balance}
\end{subfigure}
    \caption{Inefficiency of multi-threaded I/O to the HDF5 library} 
    \label{f:a-hdf5_thread_inefficiency}
\end{figure}

Instead of a single thread performing FAST5 reading, now we ran the experiment with multiple threads to further consolidate our explanation of the HDF5 bottleneck. Dataset D1 (manageable runtime than D2 for extensive testing) was used and the experiment was conducted on two systems, S1 that contains an HDD RAID and S2 with an SSD RAID (refer to Tables \ref{t:systems-slow5}, \ref{t:datasets-slow5}  and \ref{t:a-datasets-slow5} for detailed information of the systems and the datasets). As expected, the time for reading FAST5 did not improve with the number of I/O threads as shown in the Fig. \ref{f:a-hdf5_thread_inefficiency}. On the system with SSD RAID FAST5 access time even got worse with multiple threads. Similar to the dataset D2 above, processing time (performed using all available cores), FASTA access and BAM access (performed using 1 thread) took significantly lower time compared to FAST5 access for D1 as exemplified in Fig. \ref{f:a-hdf5_thread_inefficiency}.

%====================================================================================
\section{Extended Results}
%====================================================================================
%\subsection{Additional Datasets and Systems}
%----------------------------------------

%D1 was used  for extensive testing, as the execution time was manageable ($<$12 hours).  Throughout the paper, unless otherwise stated dataset D1 is involved.

%*****************************************
\begin{table*}[!ht]
\caption[An Oxford Nanopore PromethION dataset]{\textbf{Dataset D2}. D2 is an Oxford Nanopore PromethION dataset. D2 was only used for a limited number of experiments due to the massive execution time (up to 72 hours).} \label{t:a-datasets-slow5}
\begin{tabular}{|l|p{1.6cm}|p{1.5cm}|p{1.9cm}|p{1.5cm}|p{1.5cm}|p{1.3cm}|p{1.3cm}|}
\hline
\textbf{ID} & \textbf{Sample}      & \textbf{No. of Gbases} & \textbf{No. of reads} & \textbf{average read length} & \textbf{max read length} & \textbf{FASTQ file size} & \textbf{FAST5 file size}\\ \hline
%D1 & T778     & 8.787         & 771 325      & 11 393              & 194 983    & 17GB &   845GB  \\ \hline
D2 & NA12878 & 66.093        & 23 654 340   &     2794.13     &  1 034 523                & 127GB & 1.6TB \\ \hline
\end{tabular}
\end{table*}
%778-5000ng and NA12878 + cancerquin (LXBAB132606)
%*****************************************

%*****************************************
\begin{table*}[!ht]
\caption{System used for experiment in section \ref{s:extended-motivational}} \label{t:a-systems-slow5}
\begin{tabular}{|l|p{2cm}|p{2cm}|p{1cm}|l|p{1.5cm}|p{1.5cm}|p{2cm}|}
\hline
\textbf{ID} & \textbf{Description}                       & \textbf{CPU}                      & \textbf{CPU cores} & \textbf{RAM}        & \textbf{Disk System} & \textbf{RAID configuration}                          & \textbf{OS}                 \\ \hline
%S1 & server with HDD RAID & 2 $\times$ Intel Xeon Gold 6154 & 36             & 384 GB     & 12$\times$10TB HDD drives & RAID6 & Ubuntu 18.04.3 LTS \\ \hline
%S2 & server with SSD RAID & 2 $\times$ Intel Xeon Gold 6148 & 40             & 768 GB     & 6$\times$4TB NVMe drives & RAID0    & CentOS 7.6.1810    \\ \hline
%S3 & NFS server           & 4 $\times$ Intel Xeon X7560     & 32             & 256 GB     & 10$\times$3TB HDD drives &    RAID5    & Ubuntu 14.04.6 LTS  \\ \hline
S4 & server with SSD RAID array & Intel Xeon CPU E5-2680   & 28             & 512 GB     & 10 NVMe drives     & RAID10  & Centos 7            \\ \hline
\end{tabular}
\end{table*}
%*****************************************

%-----------------------------------------
\subsection{Results from Alternate File Format (SLOW5) for Another Dataset}
%-----------------------------------------

For the large D2 dataset we ran with all cores available on server S1 and the overall execution time improved to 22.04 hours for restructured Nanopolish with SLOW5 which was 69.29 hours previously for FAST (mentioned in section \ref{s:deepnaly}). The FAST5 access time which was around 62 hours earlier (mentioned in section \ref{s:deepnaly}) improved to  around 6 hours when our SLOW5 format was used.

%\todo{[Talk about file sizes? :  for D1 fast5s were 845 GB (GZ compression filter for raw signal, but large due to some intermediate data during basecalling eventtable) and in SLOW5 (uncompressed) was 340 GB. SLOW5 index was 47MB. For D2 1.6 TB for fast5 and 4.2TB for SLOW5 and 1.4 GB for SLOW5 index. Fast5 HDF5 are already compressed. The reduction in D1 is due to the removal of redundant data.]}

%-------------------------------------
%\subsection{Extended Results from Multi-process Pool}
%-------------------------------------

% \subsection{Impact of Proposed Solutions on the System Calls}

% The number of system calls were recorded by executing the applications through the \textit{strace} utility in Linux.

% \todo{[Talk about lseek operations?]}

\subsection{Impact of Proposed Solutions on Disk IOPS}

\begin{figure}
    \centering
    \includegraphics[width=\textwidth]{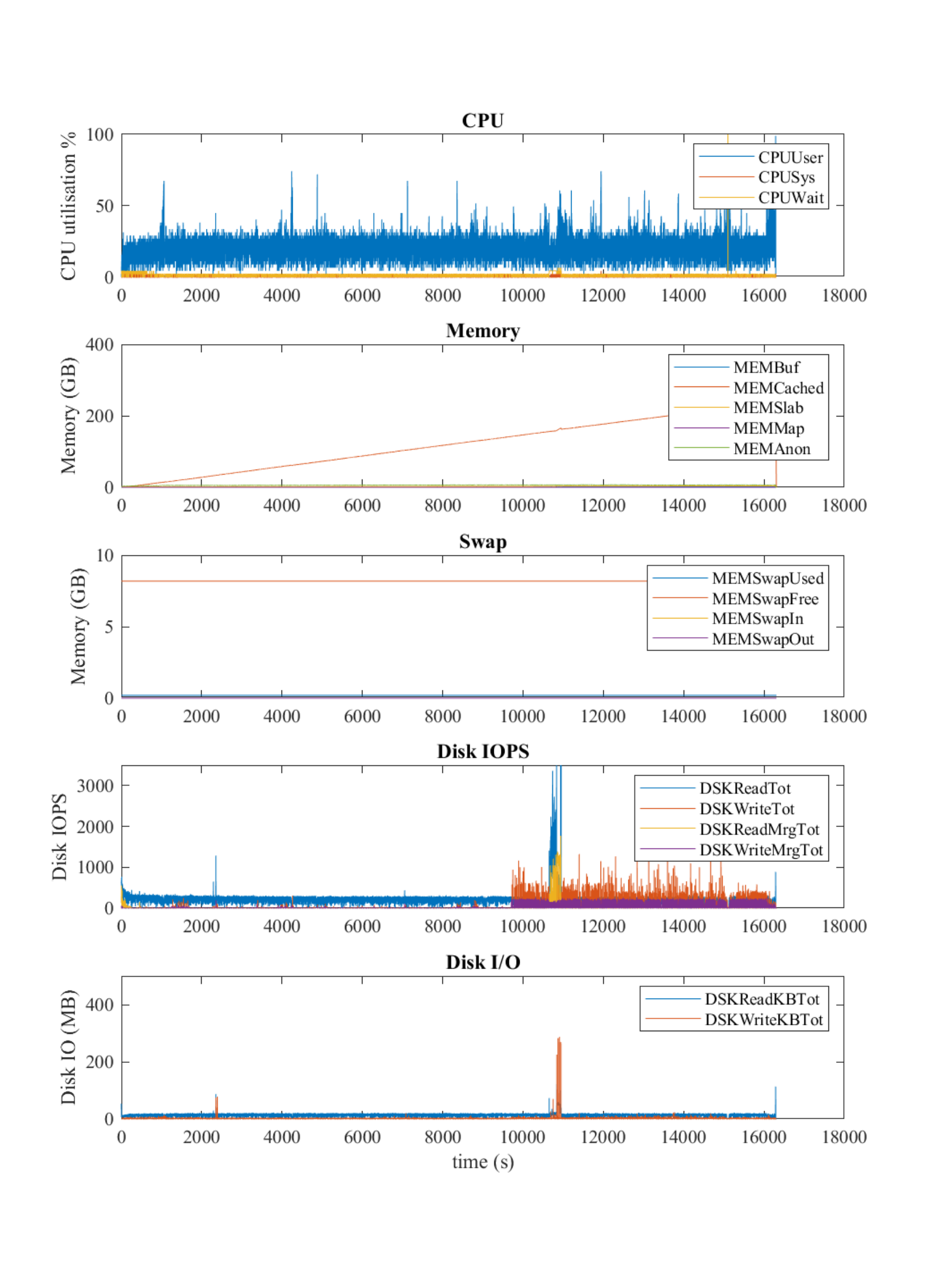}
    \caption{Statistics collected using \textit{collectl} for \textit{Nanopolish}}
    \label{f:iopsnano}
\end{figure}

\begin{figure}
    \centering
    \includegraphics[width=\textwidth]{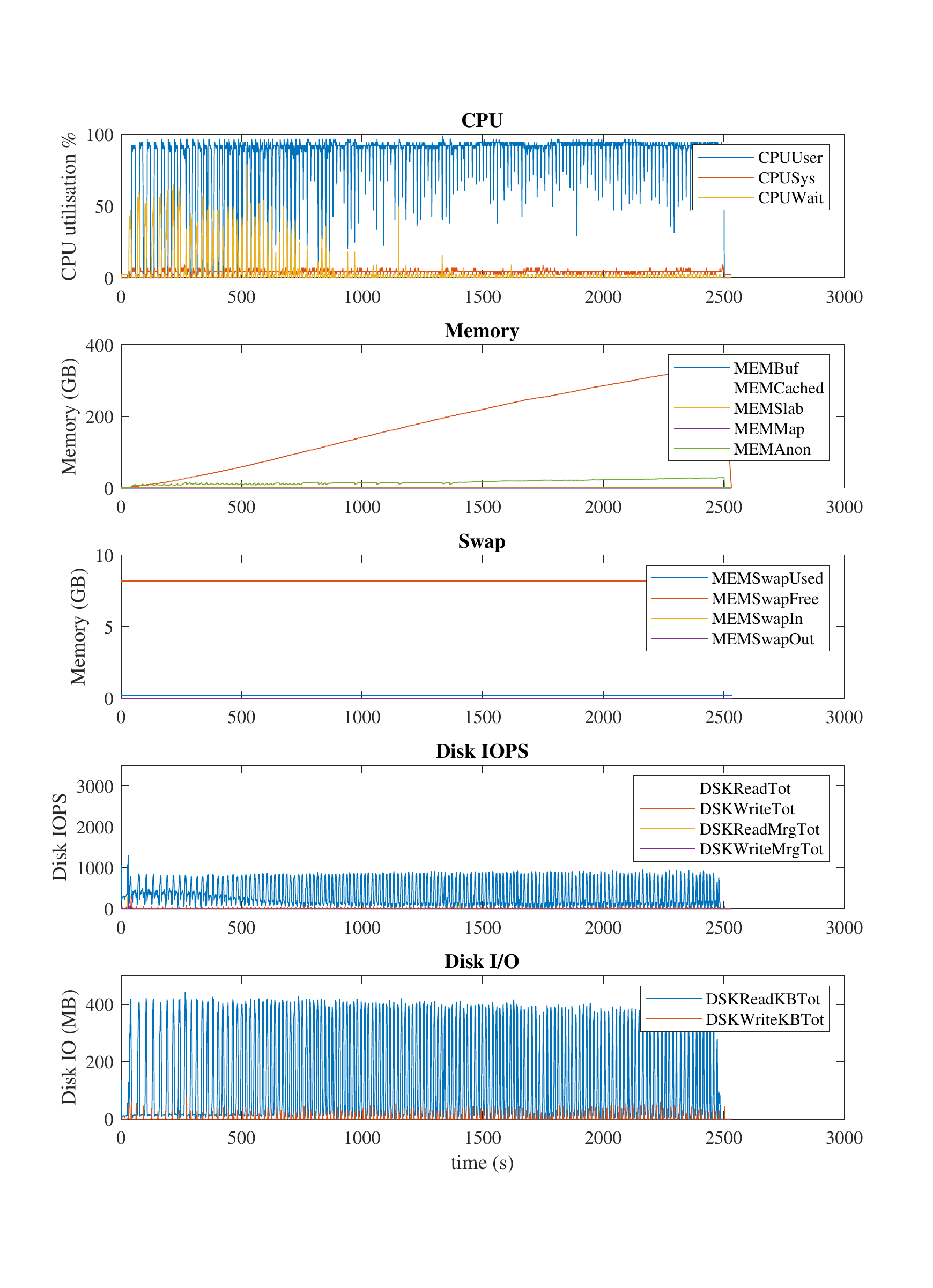}
    \caption{Statistics collected using \textit{collectl} for \textit{SLOW5}}
    \label{f:iopsslow5}
\end{figure}

\begin{figure}
    \centering
    \includegraphics[width=\textwidth]{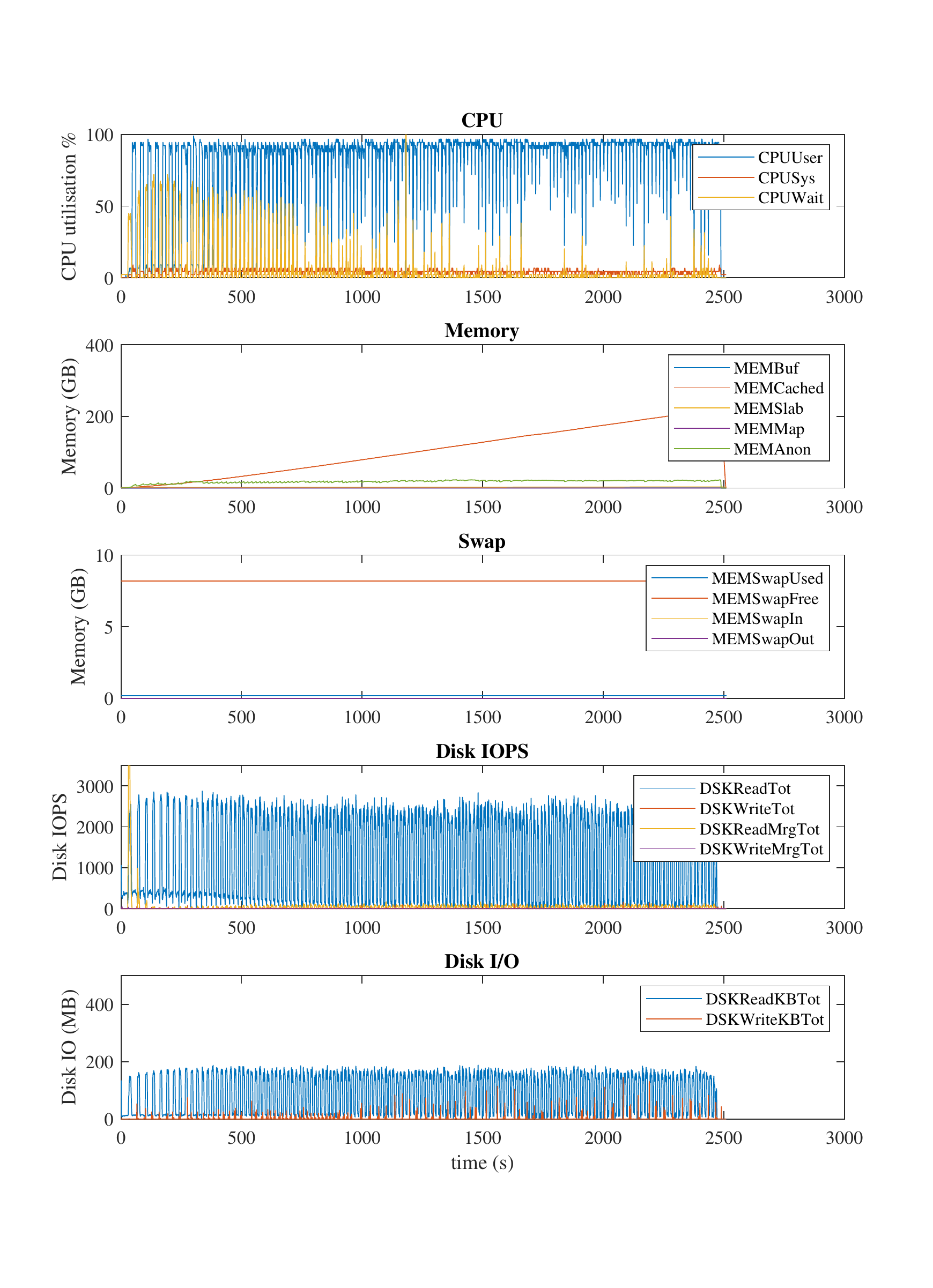}
    \caption{Statistics collected using \textit{collectl} for \textit{FAST5} multi-process pool}
    \label{f:iopsiops}
\end{figure}

System resource utilisation statistics for original \textit{Nanopolish}, optimised \textit{Nanopolish} with \textit{SLOW5} format and optimised \textit{Nanopolish} with FAST5 multi-process pool are in Fig. \ref{f:iopsnano}, Fig. \ref{f:iopsslow5} and Fig. \ref{f:iopsiops}, respectively. The statistics were collected using the \textit{collectl} utility in Linux while each application was executing with 32 threads on System S1.

Observe that disk Input/output operations per second (IOPS) for \textit{SLOW5} format is lesser than that for the \textit{FAST5} multi-process pool while the disk I/O (amount of data read per second) for \textit{SLOW5} format is larger than for  the \textit{FAST5} multi-process pool. This observation demonstrates the efficacy of \textit{SLOW5} format that exploits the locality in data for efficient disk access. Storing all the data and metadata associated with a \textit{genomic-read} in a single contiguous record reduces the random disk access operations (IOPS) while increasing the amount of data read per second.

%\todo{[For large datset? Compare IOS with SLOW5]}

\chapter[Appendix: Poster Presentations]{Poster Presentations}\label{a:posters}

\begin{figure}
    \centering
    \includegraphics[width=\textwidth]{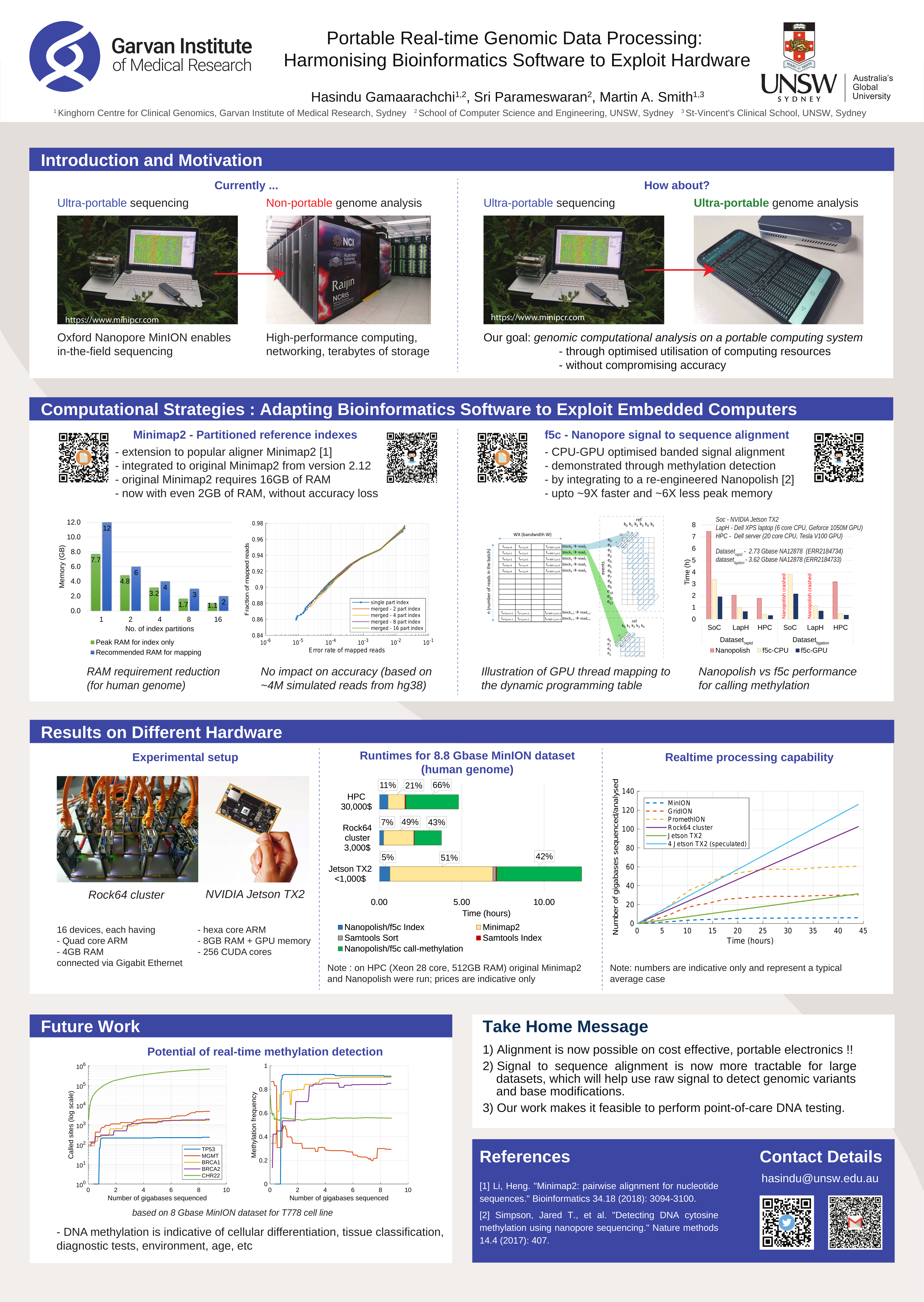}
    \caption{Poster presented at AGTA 2019}
    \label{f:poster-agta}
\end{figure}

The poster presented at Australasian Genomic Technologies Association (AGTA) Conference 2019 that attracted the best student poster award is in Fig. \ref{f:poster-agta}

\begin{figure}
    \centering
    \includegraphics[width=\textwidth]{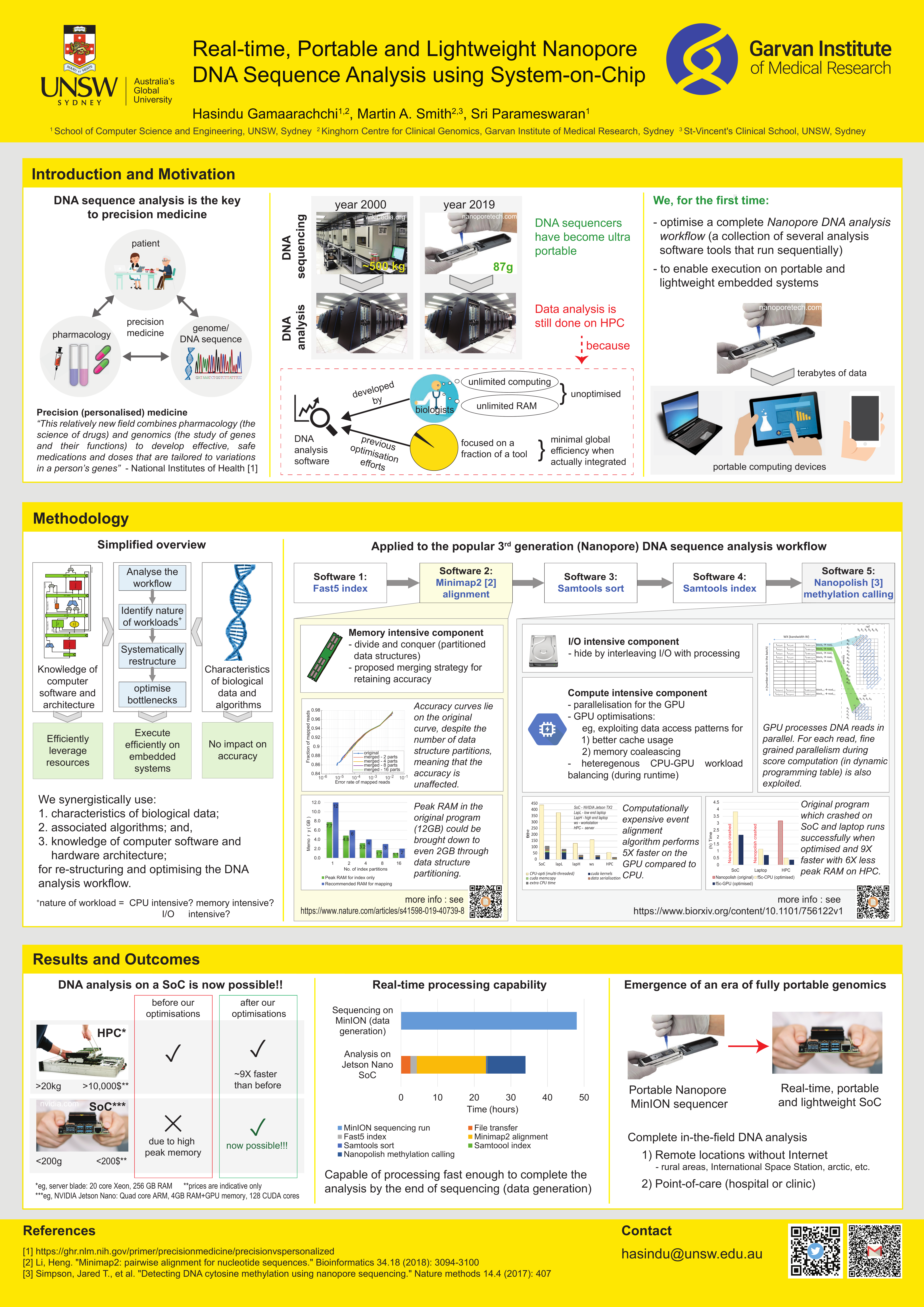}
    \caption{Poster presented at ACM SRC at ESWEEK 2019}
    \label{f:poster-src}
\end{figure}

The poster presented at ACM SRC at ESWEEK 2019 that was shortlisted to the next level in the competition is in Fig. \ref{f:poster-src}

